\documentclass[11pt,oneside]{book} 

\usepackage{
	amsmath, 
	amsthm, 
	amssymb,
	fancyhdr, 
	setspace, 
}
\usepackage[
	paper=letterpaper,
	left=1in,
	top=1in,
	bottom=1in,
    right=1in,
	includehead,
	includefoot
]
{geometry}

\usepackage{titlesec}
\titleformat{\chapter}[display]
  {\normalfont\Large\bfseries\centering} 
  {\chaptertitlename\ \thechapter}       
  {15pt}                                 
  {\Large}                               

\usepackage{xcolor}
\usepackage[linktocpage=true]{hyperref}
\usepackage[
    backend=bibtex, 
    natbib=true,
    style=phys,
    biblabel=brackets,
    giveninits=true,
    abbreviate=false,
    doi=false, url=false, isbn=false, 
    block=space,
    maxnames=3,          
    minnames=1,          
    uniquelist=false,    
    backref=false,       
    backrefstyle=none    
]{biblatex}

\renewbibmacro*{title}{%
  \ifboolexpr{
    test {\ifentrytype{article}}
    or
    test {\ifentrytype{inbook}}
    or
    test {\ifentrytype{incollection}}
    or
    test {\ifentrytype{inproceedings}}
    or
    test {\ifentrytype{thesis}}
    or
    test {\ifentrytype{unpublished}}
  }
  {\printtext[title]{\thefield{title}}}
  {\printtext[title]{\thefield{title}}}}
\renewbibmacro*{booktitle}{%
  \printtext[booktitle]{\thefield{booktitle}}}

\usepackage{physics}
\usepackage{slashed}

\usepackage{enumitem}

\newcommand{\subheading}[1]{%
    \vspace{0.5\baselineskip}
    \noindent\textbf{#1}
    \vspace{0.5\baselineskip}
    \newline
}

\usepackage{subcaption}

\usepackage[version=4,arrows=pgf-filled,
textfontname=sffamily,mathfontname=mathsf]{mhchem}

\usepackage{xspace}
\newcommand{\enscat}{$e+N\rightarrow e+N$\xspace}
\newcommand{\ld}{\textit{l}\ce{D2}\xspace}
\newcommand{\lh}{\textit{l}\ce{H2}\xspace}
\newcommand{\rqe}{$R^{QE}$\xspace}
\newcommand{\rsf}{$R_{n/p}^{sf}$\xspace}
\newcommand{\q}{$Q^2$\xspace}

\newcommand{\w}{$W^2$\xspace}

\newcommand{\ep}{$\epsilon$\xspace}
\newcommand{\gmn}{E12-09-019\xspace}
\newcommand{\ntpe}{E12-20-010\xspace}
\newcommand{\ebbcal}{E^{\text{BBCAL}}_{e'}\xspace}
\newcommand{\qeq}[1]{$Q^2 = #1$ \SI{}{(GeV/c)^2}\xspace}
\newcommand{\deep}{$D(e,e'p)$\xspace}
\newcommand{\deen}{$D(e,e'n)$\xspace}
\newcommand{\deeN}{$D(e,e'N)$\xspace}
\newcommand{\heep}{$H(e,e'p)$\xspace}

\newcommand{\sect}{Section\xspace}
\newcommand{\eqn}{Equation\xspace}

\newcommand{\tab}{Table\xspace}
\newcommand{\fig}{Figure\xspace}

\newcommand{\sfig}{0.7\columnwidth}
\newcommand{\thpq}{$\theta_{pq}$\xspace}
\newcommand{\dx}{$\Delta{x}$\xspace}
\newcommand{\dy}{$\Delta{y}$\xspace}
\newcommand{\dxb}{\bm{$\Delta{x}$}\xspace}
\newcommand{\dyb}{\bm{$\Delta{y}$}\xspace}
\newcommand{\xhob}{$x_{HCAL}^{obs}$\xspace}
\newcommand{\yhob}{$y_{HCAL}^{obs}$\xspace}
\newcommand{\xhex}{$x_{HCAL}^{exp}$\xspace}
\newcommand{\xhexp}{${\left[x_{HCAL}^{exp}\right]}^p$\xspace}
\newcommand{\yhex}{$y_{HCAL}^{exp}$\xspace}
\newcommand{\qvect}{$\vb{q}$\xspace}
\newcommand{\eovp}{$E_{BBCAL}/p$\xspace}
\newcommand{\eovpb}{\bm{$E_{BBCAL}/p$}\xspace}
\newcommand{\eove}{$E_{BBCAL}/E'_{e}$\xspace}
\newcommand{\gfsbs}{g4sbs\xspace}
\newcommand{\geantf}{\textit{Geant4}\xspace}
\newcommand{\simc}{SIMC\xspace}
\newcommand{\hsptab}{\hspace{0pt}}
\newcommand{\hsptabn}{\hspace{-2.5pt}}
\newcommand{\he}{$^{\ce{3}}$\ce{He}\xspace}

\usepackage{todonotes}

\usepackage[]{caption} 
\usepackage{amsmath}
\usepackage{amsthm}
\theoremstyle{plain}

\theoremstyle{definition}

\makeatletter
\@addtoreset{equation}{chapter}
\@addtoreset{figure}{chapter}
\@addtoreset{table}{chapter}
\makeatother
\renewcommand*{\thefootnote}{\arabic{footnote}}
\makeatletter
\@addtoreset{footnote}{chapter}  
\makeatother

\usepackage{multirow}

\usepackage{array}

\usepackage{siunitx} 

\usepackage{bm} 

\usepackage{placeins}  

\usepackage{tikz} 

\usepackage{tablefootnote}

\newcommand{\mythesistitle}{Precision Measurements of the Neutron Magnetic Form Factor to High Momentum Transfer using Durand's Method}
\newcommand{\myname}{Provakar Datta}
\newcommand{\myyear}{2024}
\newcommand{\degreethree}{MS, University of Connecticut, 2020}
\newcommand{\degreetwo}{MSc, Indian Institute of Technology Madras, 2017}
\newcommand{\degreeone}{BSc, St. Xavier's College Kolkata, 2015}
\newcommand{\majoradvisor}{Andrew J. R. Puckett}
\newcommand{\associateone}{Thomas C. Blum}
\newcommand{\associatetwo}{Richard T. Jones}
\newcommand{\associatethree}{Mark K. Jones}

\begin{document}

\frontmatter

\pagestyle{fancy}      
\fancyhead[C]{Provakar Datta -- University of Connecticut, 2024}       
\fancyfoot{}

\thispagestyle{empty}
\begin{center}
{\Large
{\bf\mythesistitle}
}
\vspace{.6in}

\myname, PhD\\
University of Connecticut, \myyear

\vspace{.6in}

{\large\bf ABSTRACT}
\end{center}

\addcontentsline{toc}{chapter}{Abstract}

Protons and neutrons, collectively known as nucleons, along with electrons, constitute the fundamental building blocks of the visible universe. Understanding their internal structure is crucial for addressing key scientific questions about our origin and existence. Elastic electron-nucleon scattering provides insights into the spatial distributions of charge and current within nucleons through their electromagnetic form factors. Accurate knowledge of these form factors over a broad range of \q, the squared four-momentum transfer in the scattering process, reveals details about the nucleon's internal structure. However, high-\q data of the nucleon electromagnetic form factor is scarce due to the challenges associated with such measurements.

This thesis reports preliminary results from high-precision measurements of the neutron magnetic form factor ($G_M^n$) to unprecedented \q using Durand's method, also known as the ``ratio" method. Systematic errors are greatly reduced by extracting $G_M^n$ from the ratio of neutron-coincident (\deen) to proton-coincident (\deep) quasi-elastic electron scattering from deuteron. The scattered electrons were detected in the BigBite spectrometer, which features multiple Gas Electron Multiplier (GEM) layers with large active area for high-precision tracking at very high rates. Simultaneous nucleon detection was performed by the Super BigBite spectrometer, which utilizes a dipole magnet with large solid angle acceptance at forward angles and a novel hadron calorimeter with very high and comparable detection efficiencies for both protons and neutrons. This setup could handle very high luminosity, making high-\q measurements feasible. Data were collected at five \q points: $3$, $4.5$, $7.4$, $9.9$, and $13.6$ (GeV/c)$^2$. Preliminary results are reported for all, with the lowest two \q points in good agreement with existing world data, while the higher points significantly extend the \q range in which $G_M^n$ is known accurately. The precision of the highest \q point is expected to remain unmatched for years to come.


\newpage
\pagestyle{plain}

\setcounter{page}{1}
\begin{center}
{\Large
{\bf\mythesistitle}
}
\vspace{.6in}

\myname

\vspace{.6in}

\degreeone \\
\degreetwo \\
\degreethree

\vfill
A Dissertation \\
Submitted in Partial Fulfillment of the \\
Requirements for the Degree of \\
Doctor of Philosophy \\
at the \\
University of Connecticut

\vspace{.25in}

\myyear

\end{center}
\newpage

\phantom{skip}
\vspace{.5in}
\begin{center}
Copyright by

\vspace{.2in} 
\myname

\vspace{6in} 
\myyear
\end{center}
\newpage


\begin{center}
{\large\bf APPROVAL PAGE}

\vspace{.5in}
Doctor of Philosophy Dissertation

\vspace{.5in}
{\Large\bf\mythesistitle}

\vspace{.5in}
Presented by\\
\myname, BSc, MSc, MS 
\vspace{.75in}
\end{center}

\begin{center}
\begin{minipage}{4.5in}
Major Advisor \hfill$\underset{\hbox{\majoradvisor}}{\rule{3in}{1pt}}$\\[15pt]

Associate Advisor \hfill$\underset{\hbox{\associateone}}{\rule{3in}{1pt}}$\\[15pt]

Associate Advisor \hfill$\underset{\hbox{\associatetwo}}{\rule{3in}{1pt}}$\\[15pt]

Associate Advisor \hfill$\underset{\hbox{\associatethree}}{\rule{3in}{1pt}}$
\end{minipage}

\vspace{1in}
University of Connecticut\\
\myyear
\end{center}

\newpage

\thispagestyle{plain}

\begin{center}
To my parents\\
Barun C. Datta and Anjona Datta \\
\vspace{1em}
and my sister \\
Sonia Datta Ghosh
\end{center}

\newpage
\thispagestyle{plain}
\begin{center}
{\large\bf ACKNOWLEDGMENTS}
\end{center}
\doublespacing
\phantomsection
\addcontentsline{toc}{chapter}{Acknowledgments}
This dissertation marks the achievement of my childhood dream of becoming a scientist. It is very close to my heart, as are the people whose love, good wishes, guidance, support, and encouragement have made this a reality. As a child, I thought becoming a scientist only meant pursuing research to become a world expert in that subject. However, what I didn't appreciate then is that it is a journey that pushes one out of their comfort zone, where they meet brilliant people, share both hardships and joys, and ultimately become a better person. Although it is impossible to acknowledge everyone I have met along the way, I would like to take this opportunity to reflect on this incredible journey and give credit to as many individuals as possible.

I grew up in a small town named Chakdah in West Bengal, a state in eastern India, with my parents, Barun Datta and Anjona Datta, and my wonderful sister, Sonia Datta Ghosh. While my father stayed far from home to earn a living, my mother took care of my sister and me. The sacrifices they have made over the years, which are impossible to convey in words, are why I am here today. My grandmother Anjali Bose, uncle Ashim Bose, brother-in-law Subhajit Ghosh, cousins Pijush Sen, Anik Sarkar, Shubhankar Bose, Debashish Dutta, family friends Madhumita Patra, Kajal Sanyal—who mean more than family—and childhood friend Kaustav Sanyal have showered me with unwavering support and blessings during both easy and challenging times.

Growing up, my knack for mathematics and science was shaped and nurtured by my teachers: Surajit Mondal, Bibhas Das, Rohit Pal, and Anish Ghosh. They encouraged me to pursue a career in core science rather than follow the ongoing trend of pursuing a degree in engineering or medicine. I vividly remember the day I told my father that I wanted to pursue core science. In response, he said, ``Do what you want. I have full faith in you, and I know that you will succeed in whatever you choose to do." That simple sentence, filled with trust and respect, bestowed upon me a great sense of responsibility and helped me stay on track all these years. In 2012, I began a physics major at St. Xavier's College Kolkata (SXCCAL), marking the beginning of a new era in my life.

The knowledgeable and caring professors at SXCCAL and my passionate cohorts contributed tremendously to my growth as a physicist. During this time, I made some of my lifelong friends: Sudarshana Laha, Ishani Ganguly, Satabdwa Majumdar, Vinayak Lahiri, Dipan Majumdar, Debmallya Chanda, and Madumita Sarkar, who have supported me all along. After three years of my bachelor’s degree, I moved far from home to Chennai to pursue my master’s at the Indian Institute of Technology (IIT) Madras. This experience opened up a new world for me and gave me my first taste of research. There, I had the privilege of learning from some of the finest physicists in India: V. Balakrishnan, Arul Lakshminarayan, MV Satyanarayana, and Dawood Kothawala. Their profound knowledge, passion, and ability to explain complex ideas simply not only enriched my understanding of physics but also fueled my desire to pursue it further. During this time, I encountered ``Introduction to Elementary Particles" by Griffiths, fell in love with the subject immediately, and decided to pursue a PhD in this related field.

During the final semester of my master’s program, my father passed away. It was devastating, and I was ready to give up, but I couldn't, thanks to my mother, my sister, and the family and friends mentioned above. With their endless support and encouragement, I mustered the courage to fly thousands of miles away to pursue my dream. In 2018, I joined UConn as a graduate student and met my colleagues: Sebastian Seeds, Zachery Harris, Jonathan Feliz, Kaitlin Lyszak, Bren Backhaus, and Gabe Kovacs. We endured classes and prelims together and created all sorts of fun, forming bonds for a lifetime. UConn gave me the opportunity to learn from professors like Gerald Dunne, who is undoubtedly the best physics teacher I have ever had. During this time, I met Ayush and Sayani Sengupta, and Tapan and Subhra Chaudhuri, who became my family away from home.

In 2019, I joined Andrew Puckett's group and began my dissertation research. Andrew accepted me despite my lack of prior research experience in this field. He has been an excellent mentor, guiding me throughout with his profound knowledge and unwavering passion for this field. He deserves credit for most of my knowledge about software and data analysis. Most importantly, Andrew has always been present. Whenever I got stuck with anything, he was there to help, sometimes even on weekends. I cannot express how grateful I am for that. It suffices to say that without his continuous support, this work would not have been possible.

In 2021, I moved to JLab to actively participate in the installation and operation of my thesis experiment and became involved with the BigBite Calorimeter (BBCAL) group. The group was led by Mark Jones, the current Hall A and C leader, who has the reputation of knowing everything. Having heard so much about him, I was a bit intimidated to meet him at first. However, after meeting him, I realized that he is a wealth of knowledge wrapped in humility. I haven't stopped asking him questions since then! I regard Mark as my onsite advisor and give him credit for most of the hardware and operational skills I have gained over the years.

Through the BBCAL group, I also met Arun Tadepalli, who has been guiding me like a big brother since then. We spent days and nights in the experimental hall laying hundreds of cables, installing connectors, and testing modules for BBCAL installation during COVID. It was this unspeakable effort that resulted in a lifelong bond. Sebastian, with whom I joined UConn and Andrew's group, has been a partner in this effort as well. Together, we have worked countless hours, spent numerous sleepless nights in the counting house, hiked many miles, discussed philosophy for hours, and watched multiple classics over the years. I cherish each and every one of these moments. This journey would not have been the same without him. Sebastian is one of the kindest people I know, and his companionship has made me a better person, too.

The most time-consuming part of my thesis work was data analysis. It was arduous, involved, and filled with roadblocks. However, I had the privilege of working with a brilliant group of people whose hard work, insights, and guidance helped me surmount each and every obstacle. This group consisted of my fellow thesis students: Sebastian Seeds, Anuruddha Rathnayake, Maria Satnik, John Boyd, and Zeke Wertz; postdoc Eric Fuchey; and professors and scientists: Andrew Puckett, Bogdan Wojtsekhowski, David Armstrong, Nilanga Liyanage, and Arun Tadepalli. I am grateful to each and every one of them.

Beyond my immediate cohort, JLab has brought me into a wonderful community of fellow students—Casey Morean, Sean Jeffas, Kate Evans, Vimukthi Gamage, Bhasitha Dharmasena, Cameron Cotton, and Faraz Chahili; and staff scientists—Steve Wood, Brad Sawatzky, Bill Henry, Chandan Ghosh, and Scott Barcus—whose support made this long, challenging journey bearable. We traveled to conferences together, went on hikes, and played board games, among other activities. Brad and Steve have been invaluable sources of knowledge from whom I have learned a tremendous amount. Sean introduced me to tennis, which quickly became a beloved hobby. Bill and I would play whenever we could—morning or evening, winter or summer—helping me refine my skills while also providing an important outlet for stress relief. Together, we all created countless memories that I will cherish for the rest of my life.

None of this would have been possible without the success of the experiment. It took the dedicated efforts of countless individuals to make that happen. I would like to thank the US Department of Energy Office of Science for funding this work, the accelerator division for ensuring the delivery of the beam during the experiment, the entire Hall A collaboration, and anyone else who contributed to the success of \gmn. Last but not least, I would like to extend my gratitude to the SBS collaboration for allowing me to be a part of this incredible effort.

\vspace{4em}
\hspace{-1em}Provakar Datta \\
Newport News, VA \\
October 2024

\newpage  
\onehalfspacing
{\hypersetup{hidelinks,linktoc=all} 
\tableofcontents
}
\clearpage  
\onehalfspacing
\clearpage
\phantomsection
\addcontentsline{toc}{chapter}{List of Figures}
\listoffigures
\clearpage
\phantomsection
\addcontentsline{toc}{chapter}{List of Tables}
\listoftables
\mainmatter
\doublespacing 


\chapter{Introduction}
The quest to understand the nature of matter is ancient yet incomplete. Over the past two centuries, sophisticated scientific methods have been developed to tackle this fundamental question, culminating in the establishment of the Standard Model of elementary particles in the 1970s. Nucleons, the building blocks of atomic nuclei and the only stable baryons, have played a central role in this progress. Once thought to be point-like particles, advancements in experimental nuclear physics in the mid-20th century uncovered their complex internal structure. Explaining this structure continues to be a key challenge for both experimental and theoretical physicists.

One of the most powerful tools for probing the internal structure of nucleons is scattering them with point-like leptonic probes, such as electrons. This technique, pioneered by Robert Hofstadter in the 1950s, remains as relevant today as it was 70 years ago. This thesis focuses on a new measurement that uses this approach to enhance our understanding of the neutron's internal structure.

In this chapter, we briefly review the discovery of nucleons and their internal structure, which led to the development of Quantum Chromodynamics (QCD), a cornerstone of the Standard Model. We will then introduce the quantum mechanical formulation of elastic electron-nucleon ($eN$) scattering, leading to the nucleon electromagnetic form factors, which connect the $eN$ scattering cross-section to the internal structure of nucleons.
\section{Nucleons to QCD: A Historical Perspective}
\label{sec:ch1:nucleontoQCD}
The first scientifically sound approach to explaining the nature of matter emerged in the early 19th century with the concept of atoms as indivisible units. However, in 1897, J. J. Thomson's discovery of the electron revealed that atoms were not indivisible but had internal structures. In 1909, the gold foil experiment conducted by Hans Geiger and Ernest Marsden under the supervision of Ernest Rutherford led to the discovery, in 1911, of a small, dense, positively charged nucleus at the center of the atom \cite{GegierMarsden1909,Rutherford1911,Geiger1913LXITL}. Rutherford then identified the proton as a fundamental component of this nucleus in 1917 \cite{doi:10.1080/14786440608635919}. Finally, in 1932, James Chadwick completed the picture with his discovery of the neutron \cite{Chadwick:1932ma}.

The nearly identical masses of the proton and neutron led to the concept of isospin \cite{Heisenberg1932}, a quantum number arising from SU(2) symmetry that treats them as two states of a single particle, the "nucleon", similar to how an electron's spin up and spin down states are viewed as one. In 1935, Hideki Yukawa proposed the concept of the strong nuclear force between nucleons, mediated by a particle he called the meson, which was later identified as the pion, to explain the stability of the nucleus. This strong force, recognized as one of the fundamental forces of nature, later became the foundation of Quantum Chromodynamics (QCD), a key component of the Standard Model of elementary particles and interactions.

In 1928, shortly before the discovery of the neutron, Paul Dirac proposed his revolutionary relativistic wave equation to describe all massive spin-1/2 point-like elementary particles, known as "Dirac particles" \cite{99d27fb6-4a63-3df7-9f70-2b4182168f59}. This equation predicted the magnetic moment of such a particle to be:
\begin{equation}
    \mu = g \left(\frac{e}{2m}\right) \frac{\hbar}{2}
\end{equation}
where $g=2$ is the Land\'e $g$-factor, $e$ and $m$ are the charge and mass of the particle, and $\hbar$ is Planck's constant divided by $2\pi$. Therefore, protons and neutrons, which were believed to be Dirac particles, were expected to have magnetic moments of $1\mu_N$ and $0\mu_N$, respectively\footnote{$\mu_N$ is the nuclear magneton, defined as $\mu_N=\frac{e\hbar}{2M_N}$ with $M_N$ being the nucleon mass.}. 

However, in 1933, Otto Stern's measurement of the proton magnetic moment ($\mu_p$) revealed a value of $2.79\mu_N$ \cite{Estermann1933}, a drastic deviation from the prediction. Later, in 1940, Alvarez and Bloch measured the free neutron's magnetic moment ($\mu_n$) to be $-1.91\mu_N$ \cite{PhysRev.57.111}, strongly suggesting the existence of internal structure within nucleons. A decade later, in the 1950s, the pioneering electron scattering experiments on atomic nuclei conducted by Robert Hofstadter and his team firmly established that the proton has an extended charge distribution and measured its size \cite{PhysRev.98.217,PhysRev.102.851,RevModPhys.28.214}. Stern, Alvarez, Bloch, and Hofstadter each received the Nobel Prize for their groundbreaking contributions to unraveling nuclear structure, laying the foundation for modern nuclear physics.

In 1947, the discovery of the pion from cosmic rays marked the beginning of a shift in particle physics, as the nucleon lost its unique role. Advancements in cosmic ray detection and particle accelerator technologies subsequently revealed numerous new subatomic particles with varying masses, charges, and lifetimes, leading to what became known as the ``particle zoo." Some of these particles, such as the kaon and lambda, exhibited unusually long lifetimes for strong interactions. To explain this, Murray Gell-Mann and Kazuhiko Nishijima independently introduced the concept of ``strangeness" in 1953, a new additive quantum number that must be conserved in strong and electromagnetic interactions \cite{PhysRev.92.833,10.1143/PTP.10.581}. This idea extended the SU(2) isospin group to an approximate SU(3) symmetry group, known as ``flavor SU(3)."

\begin{table}[h!]
    \caption[Properties of quarks]{Properties of quarks \cite{PDGQUARKMASS}.}
    \label{tab:ch1:quarkproperties}
    \centering
    \begin{tabular}{>{\hsptab}l<{\hsptab}>{\hsptab}c<{\hsptab}>{\hsptab}c<{\hsptab}>{\hsptab}c<{\hsptab}>{\hsptab}c<{\hsptab}>{\hsptab}c<{\hsptab}>{\hsptab}c<{\hsptab}}   
    \hline\hline
       \multirow{2}{*}{Properties}  &  \multicolumn{6}{c}{Quark Families} \\ \cline{2-7} \vspace{-1em} \\
                   & Up (u) & Down (d) & Charm (c) & Strange (s) & Top (t) & Bottom (b)\vspace{0.2em} \\ \hline \vspace{-1em} \\
       Charge      & $+\frac{2}{3}$e & $-\frac{1}{3}$e & $+\frac{2}{3}$e & $-\frac{1}{3}$e & $+\frac{2}{3}$e & $-\frac{1}{3}$e\\ \vspace{-1em} \\
       Spin        & $\frac{1}{2}$ & $\frac{1}{2}$ & $\frac{1}{2}$ & $\frac{1}{2}$ & $\frac{1}{2}$ & $\frac{1}{2}$\\ \vspace{-1em} \\
       Isospin (I$_3$)  & $\frac{1}{2}$ & $-\frac{1}{2}$ & $0$ & $0$ & $0$ & $0$\\
       Strangeness & $0$ & $0$ & $0$ & $-1$ & $0$ & $0$\\
       Mass (MeV)  & $2.16^{+0.49}_{-0.26}$ & $4.67^{+0.48}_{-0.17}$ & $1270\pm20$ & $93.4^{+8.6}_{-3.4}$ & $172690\pm300$ & $4180^{+30}_{-20}$\\ \vspace{-1em} \\
       \hline\hline
    \end{tabular}
\end{table}
Building on this, Murray Gell-Mann and George Zweig independently proposed the quark model in 1964, introducing quarks as fundamental constituents with three flavors: up (u), down (d), and strange (s) \cite{Gell-Mann:1964ewy,Zweig:1964jf}. Gell-Mann's ``Eightfold Way," based on this framework, elegantly organized the strongly interacting subatomic particles, collectively known as hadrons, and greatly simplified the particle zoo. In the Eightfold Way, hadrons are classified into two main categories based on their quark content:
\begin{itemize}
    \item \textit{Mesons}, obtained from the flavor SU(3) decomposition of a quark-antiquark pair:
    \begin{equation*}
        3 \otimes \Bar{3} = 8 \oplus 1,
    \end{equation*}
    are bosons with integer spin.
    \item \textit{Baryons}, obtained from the flavor SU(3) decomposition of three quarks:
    \begin{equation*}
        3 \otimes 3 \otimes 3 = 10 \oplus 8 \oplus 8 \oplus 1,
    \end{equation*}    
    are fermions with half-integer spin. The nucleons are included in this category.
\end{itemize}
Two u quarks each with charge +2/3e and isospin 1/2 and a d quark with charge -1/3e and isospin -1/2 were combined (uud) to create a proton with charge 1e and isospin 1/2. The neutron, represented by ddu, has a charge of 0 and isospin -1/2. Based on this framework, in the limit $m_u=m_d=m$, the calculation of the nucleon magnetic moment predicted the following:
\begin{equation*}
    \begin{aligned}
        \mu_n = -2\frac{e\hbar}{2m}, \,\, \mu_p = 3\frac{e\hbar}{2m}  \Rightarrow  \frac{\mu_n}{\mu_p} = -0.67
    \end{aligned}
\end{equation*}
which closely aligned with the experimental observation of $\frac{\mu_n}{\mu_p}=-0.68$. The quark model also predicted new particles which were later discovered. More experimental confirmations for the presence of quark mounted in the late 1960s and early 1970s. Notably, deep inelastic scattering (DIS) experiments at Stanford Linear Accelerator Center (SLAC), where electrons were scattered off protons at high energies, revealed a point-like substructure inside protons, consistent with quarks. This was reinforced by Bjorken scaling, observed in these experiments, where certain ratios of scattering cross-sections remained constant with increasing energy, suggesting the existence of smaller, indivisible components within hadrons. 

Despite its success, the quark model had several shortcomings. The contribution of up (u) and down (d) quarks, later known as the ``valence quarks," fell significantly short in explaining the nucleon mass and total spin. Additionally, the quark model could not account for the fact that quarks are never found in isolation and could not explain the existence of baryons such as the $\Delta^{++}$ particle, which consists of three identical quarks. This configuration, which results in a totally symmetric ground state, is forbidden by Fermi-Dirac statistics and Pauli exclusion principle.

These limitations of the quark model led to the development of Quantum Chromodynamics (QCD) in the early 1970s. QCD, the theory of the strong force, is grounded in gauge invariance under the SU(3) color group, an exact symmetry group known as ``color SU(3)." In QCD, quarks carry a type of charge called color charge, which comes in three varieties: red, blue, and green. The strong force is mediated by gluons, massless gauge bosons that, unlike photons in Quantum Electrodynamics (QED), also carry color charge and can self-interact. Furthermore, the strength of the strong coupling constant, $\alpha_s$, varies significantly with energy\textemdash this is known as the running of the coupling constant. These distinctive features of QCD lead to phenomena more complex than those encountered in QED:
\begin{itemize}
    \item \textbf{Color Confinement:} The strong coupling constant increases with distance, preventing quarks and gluons from existing in isolation. The energy required to separate quarks exceeds their mass, leading to the creation of additional quarks and colorless hadrons instead of isolated quarks. The large value of $\alpha_s$ at low energies also makes strong interactions impossible to calculate using standard perturbative methods, especially for lighter hadrons like nucleons.
    \item \textbf{Asymptotic Freedom:} At high energies, the strong coupling constant decreases at short distances, allowing quarks and gluons to become asymptotically free. This enables perturbative calculations of strong interactions at high energies. Perturbative QCD (pQCD) accurately predicted the scaling behavior of the proton structure function \(F_2\), observed in deep inelastic scattering experiments, confirming the existence of asymptotic freedom.
\end{itemize}

These properties of QCD reveal a more intricate internal structure of nucleons than previously suggested by the quark model. In addition to the ``valence quarks" that make up nucleons, QCD predicts the presence of ``sea quarks," arising from gluon-induced quantum fluctuations, where gluons momentarily split into quark-antiquark pairs. Sea quarks contribute significantly to the nucleon's properties, including its mass and spin, and their existence has been confirmed in deep inelastic scattering experiments. Alongside this development, three additional quark flavors\textemdash charm (c), bottom (b), and top (t)\textemdash were proposed and subsequently discovered, bringing the total number of quark flavors to six (see \tab \ref{tab:ch1:quarkproperties}).

Despite their complexity, nucleons remain the best testbed for studying the strong force due to their abundance. While a complete, first-principles description of nucleon structure from QCD is still lacking, effective field theories using mesons and baryons as degrees of freedom have proven highly successful. A central challenge in nuclear physics is bridging the gap between low-energy meson-baryon models and high-energy perturbative QCD (pQCD). Precise measurements of nucleon electromagnetic form factors via elastic electron-nucleon scattering to very high 4-momentum transfer will test theoretical predictions in this transition region, and such a measurement is the focus of this thesis.

\section{Elastic $eN$ Scattering and Nucleon Form Factors}
\label{sec:ch1:eNscattering}
Scattering a beam of charged particles off a target to probe its electromagnetic structure is a foundational technique in physics, underpinning all major discoveries related to hadronic structure. Electron scattering off nucleon targets, pioneered by Robert Hofstadter at SLAC in the 1950s, has become one of the most effective methods for probing the nucleon's electromagnetic structure. As leptons, electrons have precisely calculable electromagnetic interactions within Quantum Electrodynamics (QED), making the scattering cross-section entirely dependent on the hadronic vertex, which encodes the nucleon's internal structure. Furthermore, the stability of electrons and the relative ease of their production, acceleration, and steering facilitate the generation of high-energy, high-intensity beams, making electrons unparalleled for precision measurements of the nucleon's electromagnetic structure.

A variety of electron-nucleon ($eN$) scattering processes are studied to understand different aspects of the nucleon's internal structure. The simplest and most fundamental among these is elastic $eN$ scattering, where the 4-momentum of the $eN$ system is conserved, and the scattered nucleon remains intact in its ground state. In this section, we will derive an expression for the unpolarized elastic $eN$ scattering cross-section from first principles in terms of the nucleon electromagnetic form factors, which will aid in interpreting the measurements presented in this thesis. For the derivation, the mass of the electron $m_e$ will be neglected, and the natural units, i.e., $\hbar=c=1$, will be assumed unless otherwise specified.

\begin{figure}[h!]
    \centering
    \includegraphics[width=0.7\columnwidth]{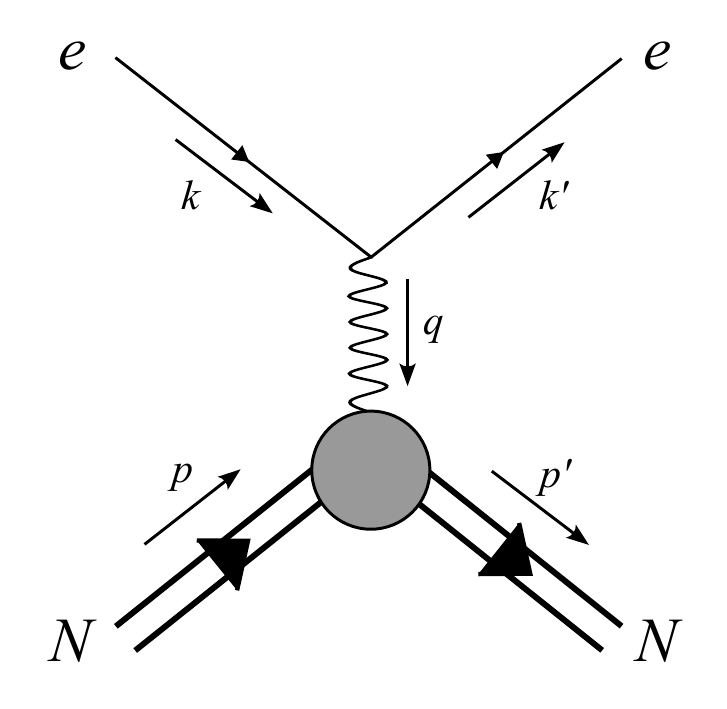} 
    \caption{\label{fig:ch1:feynmandiagram} Tree level Feynman diagram for elastic $eN$ scattering.}
\end{figure}
\subsection{QED Formalism}
\fig \ref{fig:ch1:feynmandiagram} depicts the Feynman diagram for elastic electron-nucleon ($eN$) scattering, 
\begin{equation*} 
    e(k) + N(p) \rightarrow e(k') + N(p'), 
\end{equation*} 
at leading order in the fine-structure constant $\alpha$. In this approximation, known as the one-photon exchange (OPE) approximation\footnote{Also referred to as the Born approximation, after Max Born who first proposed it.}, a single virtual photon is exchanged between the electron and the nucleon. Given the small value of $\alpha$ ($\approx \frac{1}{137}$), contributions from higher-order terms can be safely neglected in calculating the cross-section. 

According to Feynman rules for QED, the Lorentz-invariant scattering amplitude $\mathcal{M}$ for this process is given by:
\begin{equation}
    -i\mathcal{M} = \Bar{u}^{s'}(k')(ie\gamma^\mu)u^s(k) \left(-\frac{ig_{\mu\nu}}{q^2}\right) \Bar{u}^{r'}(p')(-ie\Gamma^\nu)u^r(p)
\end{equation}
where $\gamma^\mu$ are the Dirac matrices, $u^s$ is the free particle Dirac spinor with spin state $s$ and $\Bar{u}\equiv u^{\dagger}\gamma^0$ is its adjoint, $k$ and $k'$ ($p$ and $p'$) are the 4-momentum of the incoming and outgoing electrons (nucleons), respectively. The term $ie\gamma^\mu$ is the electron vertex factor. The photon propagator is given by $-\frac{ig_{\mu\nu}}{q^2}$, where $g_{\mu\nu}$ is the Minkowski metric tensor and $q=k-k'=p'-p$ is the 4-momentum transfer. The nucleon vertex factor, $-ie\Gamma^\nu$, describes the photon-nucleon coupling, where $\Gamma^\mu$ encapsulates the nucleon form factors.

\subheading{Expression for $\Gamma^\mu$}
The form of $\Gamma^\mu$ is not explicitly known (as indicated by the blob in the Feynman diagram); however, we can infer its most general expression based on the following considerations:
\begin{itemize}
    \item To the lowest order, $\Gamma^\mu$ must reduce to $\gamma^\mu$.
    \item $\Gamma^\mu$ can only depend on $\gamma^\mu$, $p$, $p'$, and $q$ (along with scalars and constants), as no other quantities contribute according to the Feynman rules. However, since $q=p-p'$, only three of these vectors are independent. For convenience, we can express them as $\gamma^\mu$, $(p-p')$, and $(p+p')$.
    \item The presence of terms involving $\gamma^5$ are forbidden as electromagnetic interaction preserves parity.
    \item $\Gamma^\mu$ must transform as a vector, i.e. in the same way as $\gamma^\mu$, to preserve Lorentz invariance. Therefore, it must be expressed as a linear combination of the independent vectors in the problem:
    \begin{equation}
    \label{eqn:ch1:gammageneral}
        \Gamma^\mu = \gamma^\mu A + (p'^{\mu} + p^\mu) B + (p'^{\mu} - p^\mu) C
    \end{equation}
\end{itemize}
The coefficients $A$, $B$, and $C$ depends solely on $q^2$, the only non-trivial scalar in the problem. Here, $q^2=-2p\cdot p' + 2M^2$ with $M$ being the mass of the nucleon.

Now, by means of the Ward identity, the current conservation at the nucleon vertex can be ensured as:
\begin{equation*}
\begin{aligned}    
    & q_{\mu} \Gamma^{\mu} = 0 \\
    \Rightarrow\,\, & (p'_{\mu} - p_\mu)\gamma^\mu A + (p'_{\mu} - p_\mu)(p'^{\mu} + p^\mu) B + (p'_{\mu} - p_\mu)(p'^{\mu} - p^\mu) C = 0
\end{aligned}    
\end{equation*}
The first term on the left-hand side vanishes when sandwiched between Dirac spinors due to the Dirac equation, which requires:
\begin{equation}
    \begin{aligned}
        \Bar{u}^{r'}(p')\slashed{p'} &= \Bar{u}^{r'}(p') M \\
        \slashed{p}u^{r}(p) &= M u^{r}(p)
    \end{aligned}
\end{equation}
where $\slashed{p}\equiv\gamma^\mu p_\mu$. The second term vanishes because, for elastic scattering, $p'^{\mu}p'_{\mu}=p^\mu p_\mu=M^2$. The third term does not vanish trivially, which necessitates that $C=0$. Further simplification of the expression can be achieved using the Gordon identity, defined as:
\begin{equation*}
\label{eqn:ch1:gordonidentity}
    \Bar{u}^{r'}(p')\gamma^\mu u^r(p) = \Bar{u}^{r'}(p') \left[ \frac{p'^{\mu} + p^\mu}{2M} + \frac{i\sigma^{\mu\nu}q_{\nu}}{2M} \right] u^r(p)
\end{equation*}
Thus, the resulting expression for $\Gamma^\mu$ is:
\begin{equation}
    \Gamma^\mu(p,p') = \gamma^\mu F_1(q^2) + \frac{i\sigma^{\mu\nu}q_{\nu}}{2M} F_2(q^2)
\end{equation}
Applying the Gordon identity in its alternate form: $\frac{i\sigma^{\mu\nu}q_{\nu}}{2M}=\gamma^\mu-\frac{p'^{\mu} + p^\mu}{2M}$, another commonly used expression for $\Gamma^\mu$ can be obtained:
\begin{equation}
\label{eqn:ch1:nucleonvertex}
    \Gamma^\mu(p,p') = \gamma^\mu (F_1(q^2)+F_2(q^2)) - \frac{p'^{\mu} + p^\mu}{2M} F_2(q^2)
\end{equation}
Here, $F_1$ and $F_2$ are the nucleon form factors, known as the Dirac and Pauli form factors, respectively. The exact functional forms of $F_1$ and $F_2$ are unknown, but they are real functions of $q^2$ and can be measured experimentally.

\subheading{Spin Averaged Squared Amplitude}
With the expression for $\Gamma^\mu$ now established, the next step is to compute the absolute square of the amplitude, $|\mathcal{M}|^2$, as the cross-section is proportional to it. Since we are dealing with unpolarized scattering, it is necessary to average over the initial spins and sum over the final spins. The spin-averaged squared amplitude is given by:
\begin{equation}
\label{eqn:ch1:msquared}
    \langle |\mathcal{M}|^2 \rangle = \frac{e^4}{q^4} L^{\mu\nu}_e W^N_{\mu\nu}
\end{equation}
where $L^{\mu\nu}_e$ and $W^N_{\mu\nu}$ represent the leptonic and hadronic tensors, respectively. We will compute each of these terms separately, for convenience. The leptonic tensor simplifies to:
\begin{align}
\label{eqn:ch1:leptonictensor}
    L^{\mu\nu}_e &= \frac{1}{2} \sum_{s',s} [\Bar{u}^{s'}(k')(ie\gamma^\mu)u^s(k)] [\Bar{u}^{s'}(k')(ie\gamma^\nu)u^s(k)]^* \notag \\
        &= \frac{1}{2}\,\text{Tr}[\gamma^\mu(\slashed{k} + m_e)\gamma^\nu(\slashed{k'} + m_e)] \notag \\
        &= 2\,[k^\mu k'^{\nu} + k'^{\mu} k^{\nu} - g^{\mu\nu}(k\cdot k' - m_e^2)] \notag\\
        &= 2\,(k^\mu k'^{\nu} + k'^{\mu} k^{\nu} - g^{\mu\nu}k\cdot k') \qquad[\text{imposing}\,\,m_e=0]
\end{align}
The factor of $\frac{1}{2}$ on the first line accounts for the averaging of the initial spins. We obtained the second line by applying Casimir's trick, and used several standard trace identities to derive the third line from the second (see \sect \ref{sec:appena:identities}). The term involving $m_e$ has been dropped in the third line since its contribution is negligible at the energy scales relevant to our analysis. The hadronic tensor can be simplified in a similar fashion as follows:
\begin{align}
\label{eqn:ch1:hadronictensor}
    W^N_{\mu\nu} &= \frac{1}{2} \sum_{r',r} [\Bar{u}^{r'}(p')(-ie\Gamma_\mu)u^r(p)] [\Bar{u}^{r'}(p')(-ie\Gamma_\nu)u^r(p)]^* \notag \\
        &= \frac{1}{2}\,\text{Tr}[\Gamma_\mu(\slashed{p} + M)\Gamma_\nu(\slashed{p'} + M)] \notag \\
        &= \frac{1}{2}\,\text{Tr} \left[ \left(\gamma_\mu (F_1+F_2) - \frac{p'_{\mu} + p_\mu}{2M} F_2\right) (\slashed{p} + M) \right. \notag\\
        &\qquad\qquad\times \left. \left(\gamma_\nu (F_1+F_2) - \frac{p'_{\nu} + p_\nu}{2M} F_2\right) (\slashed{p'} + M) \right] \notag\\
        &= 2\,\bigg\{ (F_1+F_2)^2 [p_\mu p'_{\nu} + p'_{\mu} p_{\nu} - g_{\mu\nu}(p\cdot p' - M^2)] \bigg. \notag\\
        &\qquad\qquad - (F_1+F_2)F_2 (p'_{\mu}+p_\mu)(p'_{\nu}+p_\nu) \notag\\
        &\qquad\qquad + \left.\frac{F_2^2}{4M^2} (p'_{\mu}+p_\mu)(p'_{\nu}+p_\nu) (p\cdot p' + M^2) \right\}
\end{align}
In the second line, we applied Casimir's trick, just as we did in the evaluation of the leptonic tensor. The third line introduces the explicit form of the hadronic tensor from \eqn \ref{eqn:ch1:nucleonvertex}. Several steps are involved in transitioning from line three to line four, which are detailed in \sect \ref{sec:appena:derivehadronictensor}. The hadronic tensor can be further simplified in the lab frame, which is defined below.

\subsection{Lab Frame Kinematics}
\label{ssec:ch1:labframekine}
The lab frame is defined with the electron beam along the positive Z-axis, the positive Y-axis pointing vertically upward toward the roof of the hall, and the positive X-axis directed beam-left when looking downstream, forming a right-handed coordinate system. In this frame, the target nucleon is at rest, and for simplicity, we assume the scattering occurs in the X-Z plane, setting the azimuthal scattering angle, $\phi_N$. Under these conditions, the 4-momenta of the particles involved in the scattering process are given by:
\begin{equation}
\begin{aligned}
    k^\mu &\equiv (E_e, 0, 0, E_e) \\
    k'^{\mu} &\equiv (E'_e, E'_e\sin\theta_e, 0, E'_e\cos\theta_e) \qquad[\text{imposing}\,\,m_e=0] \\
    p^\mu &\equiv (M, 0, 0, 0) \\
    p'^{\mu} &\equiv (E'_N, p'\sin(-\theta_N), 0, p'\cos(-\theta_N)) \\
             &= (M+\nu, -p'\sin\theta_N, 0, p'\cos\theta_N)
\end{aligned}    
\end{equation}
Here, $\theta_e$ is the polar scattering angle and $\nu$ is defined as:
\begin{equation}
    \nu = E_e - E'_e
\end{equation}
Defining some useful identities:
\begin{align}
    k\cdot k &= k'\cdot k' = 0 \\
    k\cdot k' &= E_eE'_e(1 - \cos\theta_e) = 2E_eE'_e\sin^2\frac{\theta_e}{2} \label{seqn:kkprime}\\
    p\cdot p' &= M(M+\nu) \label{seqn:ppprime} 
\end{align}
Useful identities involving the momentum transfer vector,
\begin{equation}
    q \equiv (\nu,\vb{q}) = k - k' = p' - p,
\end{equation}
are listed below: 
\vspace{-1.5em}
\begin{align}
    q^2 &= (k-k')^2 \notag\\
        &= -2k\cdot k' \label{seqn:kkprimeq2}\\
        &= -4E_eE'_e\sin^2\frac{\theta_e}{2}  \qquad [\text{using}\,\,\eqref{seqn:kkprime}]  \label{seqn:q2squared}\\
    q^2 &= (p-p')^2 \notag\\
        &= 2M^2 - 2p\cdot p' \label{seqn:q2m2pprime}\\
        &= -2M\nu \qquad [\text{using}\,\,\eqref{seqn:ppprime}] \\
    k\cdot q  &= -k'\cdot q = k\cdot(k-k') = - k\cdot k' = \frac{q^2}{2} \hfill \qquad [\text{using}\,\,\eqref{seqn:kkprimeq2}] \\
    k\cdot p  & = ME_e \label{seqn:kdotp} \\
    k\cdot p' &= k\cdot p + k\cdot q = ME_e + \frac{q^2}{2} \\
    k'\cdot p  & = ME'_e = ME_e + \frac{q^2}{2} \qquad [\text{using}\,\,\eqref{seqn:q2m2pprime}]\\
    k'\cdot p' &= k'\cdot p + k'\cdot q = ME_e \\
    p\cdot q  &=  p\cdot(p'-p) = M(M+\nu) - M^2 = 2M\nu \label{seqn:pdotq} 
\end{align}
Additionally,
\begin{align}
    Q^2 &= -q^2 = 4E_eE'_e\sin^2\frac{\theta_e}{2} \qquad [\text{using}\,\,\eqref{seqn:q2squared}] \\
    W^2 &= (p+q)^2 = M^2 + 2M\nu - Q^2 \qquad [\text{using}\,\,\eqref{seqn:pdotq}] \\
    E'_e &= \frac{E_e}{1+2\frac{E_e}{M}\sin^2\frac{\theta_e}{2}} = \frac{E_e}{1+\frac{E_e}{M}(1-\cos\theta_e)} \qquad [\text{using}\,\,\eqref{seqn:q2m2pprime},\,\eqref{seqn:q2squared},\,\&\,\eqref{seqn:ppprime}] \\
    E'_N &= E_e - E'_e + M = \frac{E_e^2}{M} \frac{1-\cos\theta_e}{1+\frac{E_e}{M}(1-\cos\theta_e)}
\end{align}

\subsection{Differential Cross-section}
The cross-section is given by Fermi's Golden rule:
\begin{equation}
\label{eqn:ch1:fermigoldenrule}
    d\sigma = \langle |\mathcal{M}|^2 \rangle \frac{d\Phi}{F}
\end{equation}
In the lab frame, for two-body scattering, the flux factor $F$ and the differential phase space $d\Phi$ are given by:
\begin{align}
    F &= 4\sqrt{(k\cdot p)^2 - m_e^2M^2} = 4E_eM \qquad [\text{using}\,\,\eqref{seqn:kdotp}] \label{seqn:flux}\\
    d\Phi &= (2\pi)^4 \delta^4 (k + p - k' - p') \frac{d^3\vb{k'}}{(2\pi)^32E'_e} \frac{d^3\vb{p'}}{(2\pi)^32E'_N} = \frac{1}{16\pi^2M} \frac{E'^{2}_e}{E_e} d\Omega \label{seqn:dphi}
\end{align}
The steps to simplify the expression for $d\Phi$ are detailed in \sect \ref{sec:appena:derivephasespace}. Substituting Equations \ref{seqn:flux} and \ref{seqn:dphi} into \ref{eqn:ch1:fermigoldenrule}, we obtain,
\begin{equation}
\label{eqn:ch1:differentialcrossection}
    \frac{d\sigma}{d\Omega} = \frac{\langle |\mathcal{M}|^2 \rangle}{64\pi^2M^2} \left(\frac{E'_e}{E_e}\right)^2
\end{equation}
This provides the differential scattering cross-section in the lab frame, expressed in terms of the invariant amplitude. 
\subsection{The Rosenbluth Formula and Sachs Form Factors}
\label{ssec:ch1:rosenbluthformula}
In the lab frame, the expression for the hadronic tensor $W^N_{\mu\nu}$ from \eqn \ref{eqn:ch1:hadronictensor} simplifies to (see \sect \ref{sec:appena:derivehadronictensor}):
\begin{equation}
\begin{aligned}
\label{eqn:ch1:hadronictensorsimplified}
    W_{\mu\nu}^N &= (p'_\mu + p_\mu)(p'_\nu + p_\nu)\left(F_1^2 - \frac{q^2}{4M^2}F_2^2\right) \\
    &\qquad - [(p'_\mu - p_\mu)(p'_\nu - p_\nu) + 2g_{\mu\nu} (p\cdot p' - M^2)] (F_1 + F_2)^2
\end{aligned}
\end{equation}
Substituting Equations \ref{eqn:ch1:leptonictensor} and \ref{eqn:ch1:hadronictensorsimplified} into \ref{eqn:ch1:msquared}, we obtain,
\begin{align}
\label{eqn:ch1:squaredamplitudeintermediate}
    \langle |\mathcal{M}|^2 \rangle &= \frac{e^4}{q^4} L^{\mu\nu}_e W^N_{\mu\nu} \notag\\
    &= \frac{2e^4}{q^4} (k^\mu k'^{\nu} + k'^{\mu} k^{\nu} - g^{\mu\nu}k\cdot k') \notag\\
    &\quad\qquad \times \bigg\{\bigg.(p'_\mu + p_\mu)(p'_\nu + p_\nu)\left(F_1^2 - \frac{q^2}{4M^2}F_2^2\right) \notag\\
    &\quad\qquad\qquad - [(p'_\mu - p_\mu)(p'_\nu - p_\nu) + 2g_{\mu\nu} (p\cdot p' - M^2)] (F_1 + F_2)^2 \bigg.\bigg\} \notag\\
    &= \frac{2e^4}{q^4} \left[ \mathcal{A} \left(F_1^2 - \frac{q^2}{4M^2}F_2^2\right) - \mathcal{B} (F_1 + F_2)^2 \right]
\end{align}
The coefficients $\mathcal{A}$ and $\mathcal{B}$ can be significantly simplified using lab frame kinematics, as detailed in \sect \ref{sec:appena:deriveamplitude}, leading to the final expression:
\begin{equation}
\label{eqn:ch1:squaredamplitudefinal}
    \langle |\mathcal{M}|^2 \rangle = \frac{e^4M^2 \cos^2\frac{\theta_e}{2}}{E_eE'_e\sin^4\frac{\theta_e}{2}} \left[ \left(F_1^2 - \frac{q^2}{4M^2}F_2^2\right) - \frac{q^2}{2M^2} (F_1 + F_2)^2 \tan^2\frac{\theta_e}{2}\right]
\end{equation}

Finally, the desired expression for the differential scattering cross-section in the One-Photon Exchange (OPE) approximation for unpolarized electron-nucleon scattering in the lab frame can be derived by substituting \eqn \ref{eqn:ch1:squaredamplitudefinal} into \ref{eqn:ch1:differentialcrossection}:
\begin{equation}
\label{eqn:ch1:rosenbluth}
    \frac{d\sigma}{d\Omega} =  \frac{\alpha^2 \cos^2\frac{\theta_e}{2}}{4E^2_e\sin^4\frac{\theta_e}{2}} \frac{E'_e}{E_e} \left[ \left(F_1^2 + \frac{Q^2}{4M^2}F_2^2\right) + \frac{Q^2}{2M^2} (F_1 + F_2)^2\tan^2\frac{\theta_e}{2} \right] ,
\end{equation}
where $\alpha=\frac{e^2}{4\pi}$ is the fine-structure constant. This is known as the \textit{Rosenbluth formula}, named after Marshall Rosenbluth, who derived it in 1950 \cite{PhysRev.79.615}. 

In the case of scattering from a structureless target with spin, such as a muon, where $\Gamma^\mu = \gamma^\mu$, implying $F_1 = 1$ and $F_2 = 0$, the Rosenbluth cross-section simplifies to:
\begin{equation}
\label{eqn:ch1:rosenbluthnostructure}
    \frac{d\sigma}{d\Omega} =  \frac{\alpha^2 \cos^2\frac{\theta_e}{2}}{4E^2_e\sin^4\frac{\theta_e}{2}} \frac{E'_e}{E_e} \left( 1 + \frac{Q^2}{2M^2}\tan^2\frac{\theta_e}{2} \right) 
\end{equation}
which, in the limit $Q^2 \rightarrow 0$, further reduces to:
\begin{equation}
\label{eqn:ch1:mott}
    \sigma_{\text{Mott}} = \left.\frac{d\sigma}{d\Omega}\right|_{Q^2 \rightarrow 0} =  \frac{\alpha^2 \cos^2\frac{\theta_e}{2}}{4E^2_e\sin^4\frac{\theta_e}{2}} \frac{E'_e}{E_e} 
\end{equation}
This is the well-known Mott cross-section, which describes elastic electron scattering from a point-like, spinless target.

\subheading{Introducing Sachs Form Factors}
The Sachs electric ($G_E$) and magnetic ($G_M$) form factors are expressed as linear combinations of the Dirac ($F_1$) and Pauli ($F_2$) form factors \cite{PhysRev.126.2256,RevModPhys.35.335}:
\begin{equation}
\label{eqn:ch1:sachsff}
    \begin{aligned}
        G_E(Q^2) &= F_1(Q^2) - \tau F_2(Q^2) \\
        G_M(Q^2) &= F_1(Q^2) + F_2(Q^2)        
    \end{aligned}
\end{equation}
where $\tau=\frac{Q^2}{4M^2}$. Substituting Equations \ref{eqn:ch1:sachsff} and \ref{eqn:ch1:mott} into \ref{eqn:ch1:rosenbluth}, yields the Rosenbluth formula in terms of the Sachs form factors:
\begin{equation}
\label{eqn:ch1:rosenbluthsachs}
    \frac{d\sigma}{d\Omega} =  \sigma_{\text{Mott}} \left[ \frac{G_E^2 + \tau G_M^2}{1+\tau} + 2\tau G_M^2 \tan^2\frac{\theta_e}{2} \right]     
\end{equation}
The term multiplied by $\sigma_{\text{Mott}}$ represents the deviation of the scattering cross-section due to the internal structure (including spin) of the nucleon. To isolate this part, a reduced cross-section $\sigma_{Red}$ can be defined as the ratio of the Rosenbluth cross-section to $\sigma_{\text{Mott}}$:
\begin{equation}
\label{eqn:ch1:sigmareduced}
    \sigma_{Red} = \epsilon(1+\tau)\frac{\frac{d\sigma}{d\Omega}}{\sigma_{\text{Mott}}} = \epsilon G_E^2 + \tau G_M^2
\end{equation}
where the longitudinal polarization of the virtual photon, $\epsilon$, is given by: 
\begin{equation}
    \epsilon = \frac{1}{1+ 2(1+\tau) \tan^2 \frac{\theta_e}{2}}
\end{equation}

Using the Sachs form factors instead of $F_1$ and $F_2$ offers several advantages. When expressed in terms of $G_E$ and $G_M$, the Rosenbluth cross-section avoids cross terms involving products of the form factors, simplifying the analysis. Moreover, the Sachs form factors offer a more intuitive physical interpretation. For instance, in the non-relativistic limit, $\tau \rightarrow 0$, $G_E$ and $G_M$ can be interpreted as the 3D Fourier transforms of the spatial distributions of charge and current within the nucleon, respectively.

\subheading{Limiting Behavior of the Form Factors}
\label{shead:limitingbehavior}
In the low $Q^2$ limit, $Q^2 \rightarrow 0$, the virtual photon lacks the resolution needed to probe inside the nucleons. As a result, the nucleons appear as spin-$\frac{1}{2}$ point particles with charge and magnetization equal to their respective electric charge and magnetic moment. In this limit, the Dirac ($F_1$) and Pauli ($F_2$) form factors reduce to: 
\begin{equation}
\begin{aligned}
    F^p_1(0) &= 1, \qquad F^p_2(0) = \kappa_p = \mu_p - 1 = 1.79\mu_N \\
    F^n_1(0) &= 0, \qquad F^n_2(0) = \kappa_n = \mu_n = -1.91\mu_N
\end{aligned}    
\end{equation}
where $\kappa_N$ is the anomalous magnetic moment of the nucleon. Similarly, the limiting behavior of the Sachs form factors can be derived using \eqn \ref{eqn:ch1:sachsff} as:
\begin{equation}
\begin{aligned}
    G_E^p(0) &= 1, \qquad G_M^p(0) = \mu_p = \kappa_p + 1 = 2.79\mu_N \\
    G_E^n(0) &= 0, \qquad G_M^n(0) = \mu_n = \kappa_n = -1.91\mu_N
\end{aligned}    
\end{equation}
These values provide the proper normalization for the respective form factors, ensuring consistency in the calculations. 
%

In the non-relativistic limit, $\tau \rightarrow 0$, the Rosenbluth cross-section simplifies to $\frac{d\sigma}{d\Omega} =  \sigma_{\text{Mott}} G_E^2$. In this limit, an approximate expression for the nucleon charge radius can be derived in terms of $G_E$ by expressing it as the Fourier transform of the spatial charge distribution $\rho(\vb{r})$ within the nucleon:
\begin{align}
    G_E &= \int_V \rho(\vb{r}) e^{i\vb{q}\vdot\vb{r}} d^3\vb{r} \notag\\
        &= 2\pi \int_0^\infty \rho(r)r^2dr \int_0^{\pi} \left( 1 + i|\vb{q}|r\cos\theta - \frac{1}{2}|\vb{q}|^2r^2\cos^2\theta + ... \right) \sin\theta d\theta \notag\\
        &= 1 - \frac{1}{6}|\vb{q}|^2\langle r^2\rangle + ...
\end{align}
On the second line, we have Taylor-expanded the exponential function and assumed a spherically symmetric charge distribution $\rho(r=|\vb{r}|)$. This clearly shows that, in the non-relativistic limit, the Sachs electric form factor $G_E$ is directly related to the mean squared (r.m.s.) charge radius of the nucleon, which can be expressed as:
\begin{equation}
\label{eqn:ch1:chargeradius}
    \langle r_C^2\rangle = -6 \left. \frac{dG_E(Q^2)}{dQ^2} \right|_{Q^2\rightarrow0}
\end{equation}
A similar expression can be derived for the mean squared magnetic radius, $\langle r^2_M \rangle$, of the nucleon in relation to the magnetic form factor $G_M$. However, its measurement is more challenging as the contribution of $G_M^2$ to $\sigma_{Red}$ is suppressed by a factor of $\tau$ (see \eqn \ref{eqn:ch1:sigmareduced}). Numerous experiments over several decades have measured the nucleon charge radius, yet the field remains active. Ongoing efforts are focused on resolving discrepancies between different experimental results and enhancing measurement precision (see \sect \ref{ssec:ch2:protonff}).

\chapter{Nucleon Form Factors}
In the previous chapter, we introduced the nucleon elastic electromagnetic form factors, which will henceforth be referred to as nucleon form factors for brevity, as fundamental quantities that encapsulate the internal structure of nucleons within the framework of elastic electron-nucleon ($eN$) scattering. Over the past seven decades, numerous experiments have measured these form factors using various techniques and experimental capabilities, resulting in a wealth of data. This extensive body of experimental evidence has contributed to the development and refinement of numerous nucleon models, thereby deepening our understanding of the nucleon's internal structure.

While a comprehensive review of all relevant experiments and theoretical models is beyond the scope of this work, this section provides a brief overview of the key experimental observables related to nucleon form factor measurements, the current status of experimental data, and their theoretical interpretations, with a particular focus on the high-$Q^2$ regime. We conclude with a brief overview of the ongoing Super BigBite Spectrometer (SBS) program, which aims to extend high-precision measurements of $G_M^n$, $G_E^n$, and $G_E^p$ to unprecedented $Q^2$ ranges. This discussion will serve as a segue into the introduction and detailed analysis of the \gmn experiment, which focuses on the high-$Q^2$ measurement of $G_M^n$, the primary subject of this thesis.

\section{Experimental Observables}
Nucleon form factors are probed through elastic electron-nucleon ($eN$) scattering, the formalism of which is discussed in detail in \sect \ref{sec:ch1:eNscattering}. However, the measurement techniques vary depending on whether the scattering is polarized or unpolarized. In this section, we will briefly review the corresponding experimental observables.
\subsection{Cross-section Measurements}
\label{ssec:ch2:rosenbluthseparation}
The Rosenbluth formula, as derived in \sect \ref{ssec:ch1:rosenbluthformula}, provides the unpolarized elastic $eN$ scattering cross-section in the laboratory frame within the one-photon exchange approximation, expressed in terms of the Sachs electromagnetic form factors as:
\begin{equation}
    \frac{d\sigma}{d\Omega} = \sigma_{\text{Mott}} \frac{\sigma_{Red}}{\epsilon(1+\tau)}, 
\end{equation}
where the reduced cross-section, $\sigma_{\text{Red}}$, is defined as:
\begin{equation}
    \begin{aligned}
        \sigma_{Red} = \tau G_M^2(Q^2) + \epsilon G_E^2(Q^2) = \sigma_T + \epsilon \sigma_L 
\end{aligned}
\end{equation}
with $\sigma_T$ and $\sigma_L$ representing the transverse and longitudinal components of the cross-section. The linear dependence of $\sigma_{\text{Red}}$ on $\epsilon$, the longitudinal polarization of the virtual photon, allows the separation of the Sachs electric ($G_E$) and magnetic ($G_M$) form factors by measuring the cross-section at the same $Q^2$ but with different values of $\epsilon$, corresponding to varying electron scattering angles and beam energy. This method is known as the \textit{Rosenbluth separation} or \textit{longitudinal/transverse (L/T) separation} technique. The reduced cross-section $\sigma_{Red}$ is commonly expressed in the following form:
\begin{equation}
    \sigma_{Red} = G_M^2(Q^2)\left( \tau + \frac{\text{RS}(Q^2)}{\mu_N^2}\,\epsilon \right)
\end{equation}
where $\text{RS}(Q^2) = \frac{\mu_N^2 G_E^2(Q^2)}{G_M^2(Q^2)}$ is the Rosenbluth slope.

The extraction of neutron form factors using the Rosenbluth separation technique is particularly challenging due to the absence of a stable free neutron target. Several measurement techniques have been adapted to overcome these challenges. Among them, \textit{Durand's method}, also known as the \textit{ratio method}, is considered the most reliable for measuring the neutron magnetic form factor $G_M^n$ and has been adapted for the measurements presented in this thesis. For a detailed description of this method, refer to Sections \ref{sssec:ch2:durandsmethod} and \ref{sec:measurementtechnique}.
%
\subsection{Double-polarization Measurements}
In polarized elastic $eN$ scattering, the interaction between the magnetic moments of the electron and the nucleon introduces spin-dependent effects in the scattering amplitude. These effects give rise to polarization observables that are sensitive to the nucleon form factor ratio, $G_E/G_M$, providing an alternative or complementary method to cross-section measurements for determining these quantities. This approach, motivated by the reduced sensitivity of the cross-section to ${G_E^p}^2$ at high $Q^2$ \cite{Akhiezer:1968ek,Akhiezer:1973xbf}, is significantly less prone to systematic uncertainties than the Rosenbluth separation technique and is considered superior for high $Q^2$ measurements of the nucleon electric form factors.

\subheading{Recoil Polarization/Polarization Transfer}
In the one-photon exchange  (OPE) approximation, the transferred polarization components in elastic scattering of a longitudinally polarized electron beam off an unpolarized target ($\overrightarrow{e}N\rightarrow e\overrightarrow{N}$) can be expressed in terms of the Sachs form factors as follows:
\begin{equation}
    \begin{aligned}
        I_0P_n &= 0 \\
        I_0P_l &= hP_e\frac{E_e + E'_e}{M}\sqrt{\tau(1+\tau)} \tan^2\frac{\theta_e}{2} G_M^2 \\
        I_0P_t &= -h P_e 2\sqrt{\tau(1+\tau)} \tan\frac{\theta_e}{2} G_E G_M     
    \end{aligned}
\end{equation}
Here, $I_0 = G_E^2 + \frac{\tau}{\epsilon}G_M^2$ is a normalization factor, $P_n$, $P_l$, and $P_t$ represent the normal, longitudinal, and transverse polarization transfer components, respectively, $h = \pm1$ indicates the beam helicity states, and $P_e$ is the magnitude of the electron beam polarization. The remaining variables retain their usual definitions as described in \sect \ref{ssec:ch1:labframekine}. A simple rearrangement yields the form factor ratio in terms of $P_t$ and $P_l$:
\begin{equation}
\label{eqn:ch2:recoilpolarization}
    \frac{G_E}{G_M} = - \frac{P_t}{P_l} \frac{E_e + E'_e}{2M} \tan\frac{\theta_e}{2}
\end{equation}
This method, known as \textit{recoil polarization} or the \textit{polarization transfer} technique, is primarily used for high-$Q^2$ measurements of the proton electric form factor, $G_E^p$.

\begin{figure}[h!]
    \centering
    \includegraphics[width=0.5\columnwidth]{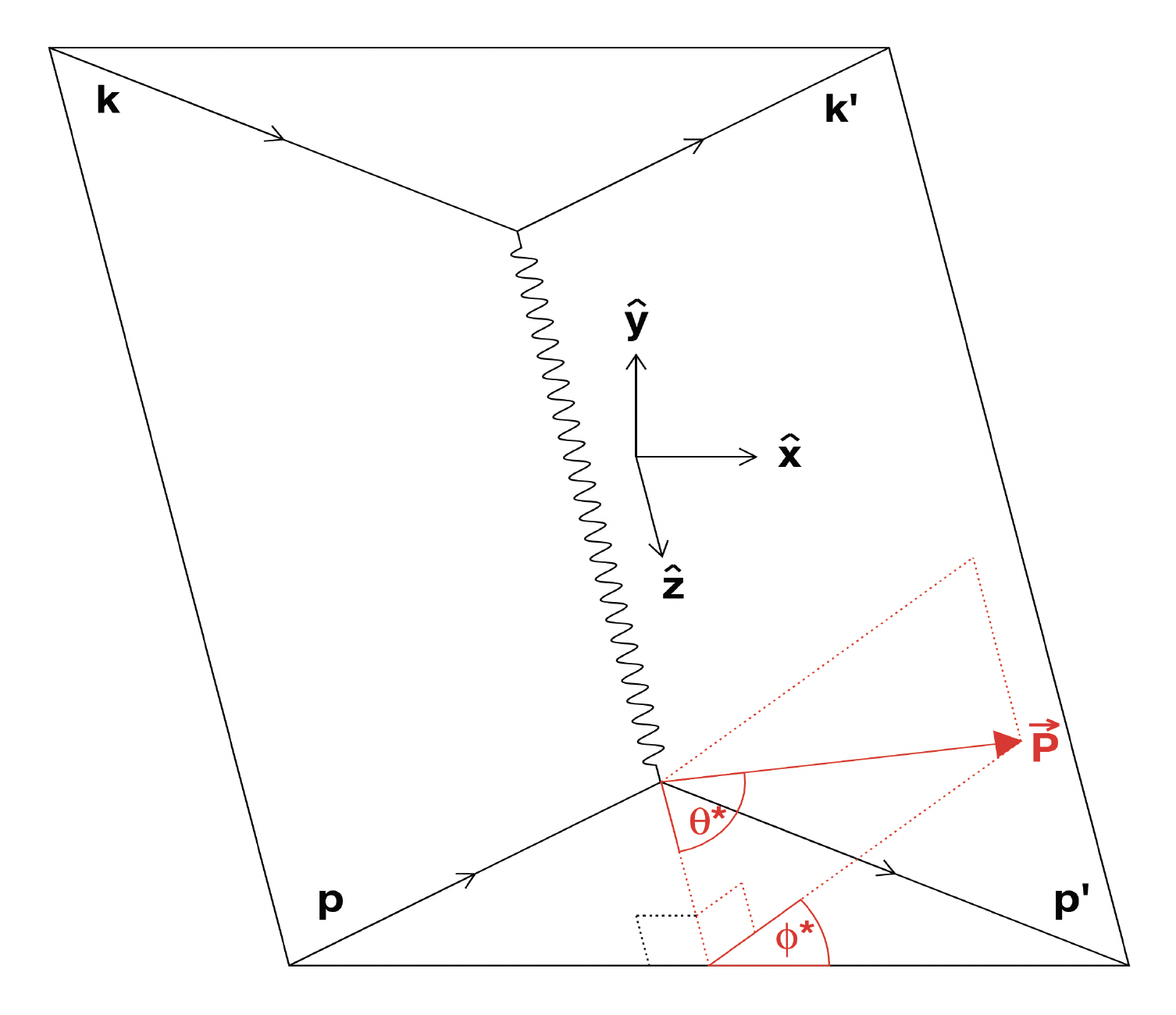}
    \caption{\label{fig:ch2:beamassymetry} Illustration of the polarized elastic $eN$ scattering process for beam-target double-spin asymmetry measurement ($\overrightarrow{e}\overrightarrow{N} \rightarrow eN$), showing the definitions of the angles $\theta^*$ and $\phi^*$ in terms of the target polarization vector $\va{P}$ and the momentum transfer vector $\vb{q} = \vb{k} - \vb{k'}$. In this setup, the $\vu{z}$ axis is aligned with $\vu{q}$, the $\vu{x}$ axis lies in the reaction plane, and the $\vu{y} = \vu{q} \times \vu{k}$ axis is perpendicular to the reaction plane, forming a right-handed coordinate system. This figure is adapted from \cite{Gross2023}.}
\end{figure}

\subheading{Beam-Target Double-Spin Asymmetry}
For neutron form factor measurements, the recoil polarization technique is significantly more challenging due to the poorly known neutron polarimetry. However, the relative ease of controlling and maintaining the polarization of $^{\ce{3}}$\ce{He} for extended periods allows the use of the \textit{beam-target double-spin asymmetry} technique at higher $Q^2$. In this method, a longitudinally polarized electron beam is scattered off a polarized target ($\overrightarrow{e}\overrightarrow{N}\rightarrow eN$), and the asymmetry in the cross-section between positive ($\sigma_+$) and negative ($\sigma_-$) beam helicity states ($h$) is measured in OPE approximation as:
\begin{equation}
\begin{aligned}
    A &= hP_e P_N\, \frac{\sigma_+ - \sigma_-}{\sigma_+ + \sigma_-} \\
    &= -hP_e P_N\,\frac{2\sqrt{\tau(1+\tau)}\tan\frac{\theta_e}{2}}{\left(\frac{G_E}{G_M}\right)^2 + \frac{\tau}{\epsilon}} \left( \frac{G_E}{G_M} \sin\theta^* \cos\phi^* + \sqrt{\tau\left(1 + (1+\tau)\tan^2\frac{\theta_e}{2}\right)}\cos\theta^* \right)
    %
\end{aligned}
\end{equation}
where $P_e$ and $P_N$ represent the electron beam and target polarization, respectively, and $\theta^*$ and $\phi^*$ are angles defined in \fig \ref{fig:ch2:beamassymetry}. From this equation, it is evident that the maximum value of the double-spin asymmetry $A$ in relation to the form factor ratio occurs when $\theta^* = \pi/2$ and $\phi^* = 0$ or $\pi$, meaning the target polarization is perpendicular to the momentum transfer vector $\vb{q}$ and parallel to the scattering plane. Under these conditions, the asymmetry simplifies to:
\begin{equation}
\label{eqn:ch2:beamtargetasy}
    A_{\perp} = -hP_e P_N\,\frac{2\sqrt{\tau(1+\tau)}\tan\frac{\theta_e}{2}}{\left(\frac{G_E}{G_M}\right)^2 + \frac{\tau}{\epsilon}} \frac{G_E}{G_M}
\end{equation}
It is important to note that this discussion applies strictly to elastic electron scattering from free nucleons. Appropriate nuclear corrections are necessary for quasi-elastic scattering from polarized bound nuclei, such as deuteron ($D$) or \he.

\section{Existing Experimental Data}
\label{sec:ch2:formfactordata}
Early cross-section measurements of the Sachs form factors at low-\q found that:
\begin{equation}
    G_E^p \approx \frac{G_M^p}{\mu_p} \approx \frac{G_M^n}{\mu_n} \approx G_D,
\end{equation}
where the dipole form factor $G_D$ is defined as:
\begin{equation}
    G_D = \left( 1 + \frac{Q^2}{0.71} \right)^{-2}.
\end{equation}
This behavior is consistent with the expected limiting behavior, as discussed in \sect \ref{shead:limitingbehavior}. However, higher \q measurements have revealed significant deviations from these trends, along with several unexpected findings that have deepened our understanding of the nucleon's internal structure. This section provides a brief summary of the empirical data available at the time of writing for proton and neutron form factor measurements, using both cross-section and double-polarization techniques.

\subsection{Proton Form Factor}
\label{ssec:ch2:protonff}
\begin{figure}[h!]
    \centering
    \includegraphics[width=1\columnwidth]{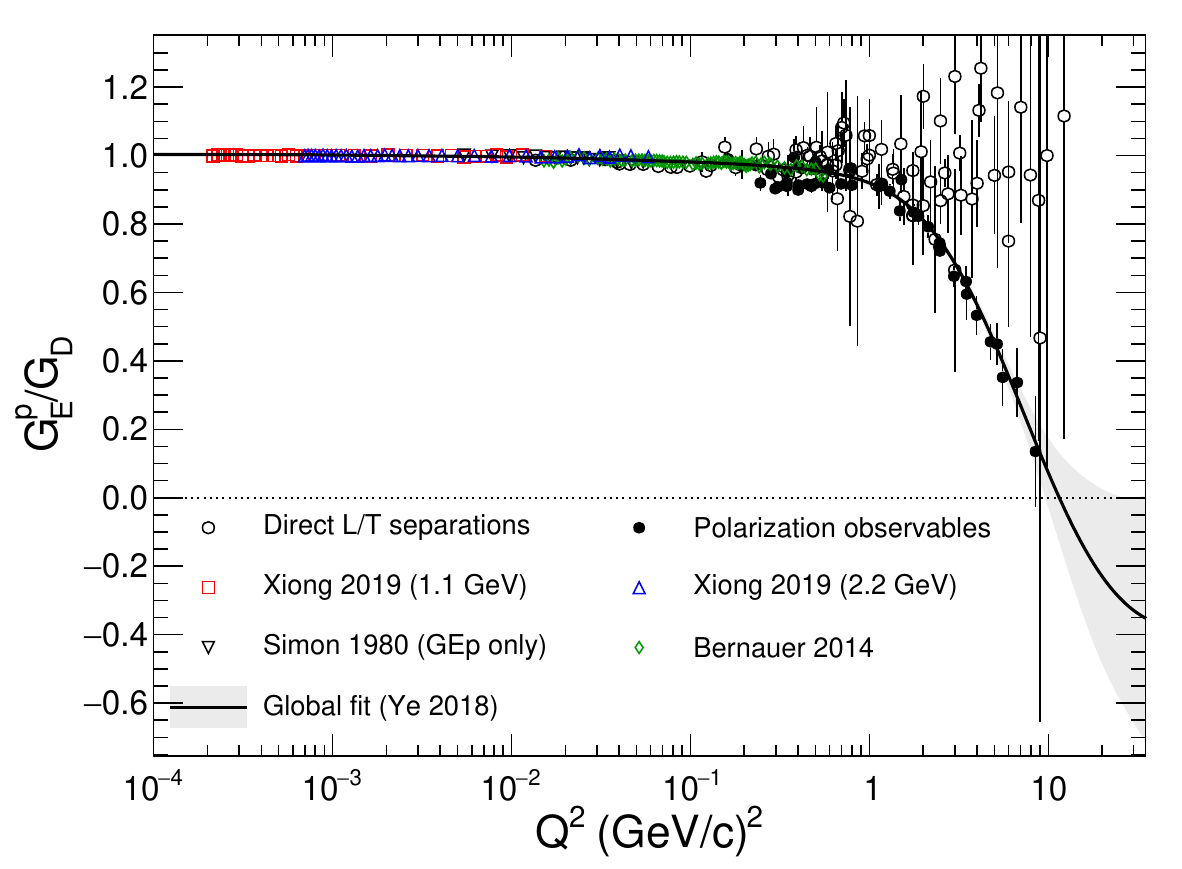}
    \caption{\label{fig:ch2:gepdata} World data for the proton electric form factor $G_E^p/G_D$. Open shapes represent cross-section measurements \cite{PhysRevLett.128.102002,PhysRev.142.922,Bartel:1973rf,BERGER197187,PhysRevD.4.45,BORKOWSKI1974269,Walker:1993vj,PhysRevD.50.5491,PhysRevLett.94.142301,PhysRevC.70.015206}, while filled circles denote double-polarization measurements, including recoil polarization \cite{PhysRevLett.80.452,PhysRevLett.84.1398,PhysRevLett.88.092301,PhysRevC.71.055202,PhysRevC.64.038202,Pospischil2001,MACLACHLAN2006261,PhysRevLett.104.242301,PhysRevC.85.045203,PhysRevLett.106.132501,PhysRevC.96.055203,PhysRevC.84.055204,ZHAN201159,PhysRevLett.105.072001} and beam-target double-spin asymmetry \cite{PhysRevC.74.035201,PhysRevLett.98.052301,SANE:2018cub}. Among the cross-section data, green diamonds represent the Mainz A1 dataset \cite{PhysRevC.70.015206}, while red squares and blue triangles indicate datasets from PRad experiment \cite{Xiong2019}. The global fit is sourced from \cite{YE20188}. See the text for more details. This figure is adapted from \cite{Gross2023}.}
\end{figure}

\begin{figure}[h!]
    \centering
    \includegraphics[width=1\columnwidth]{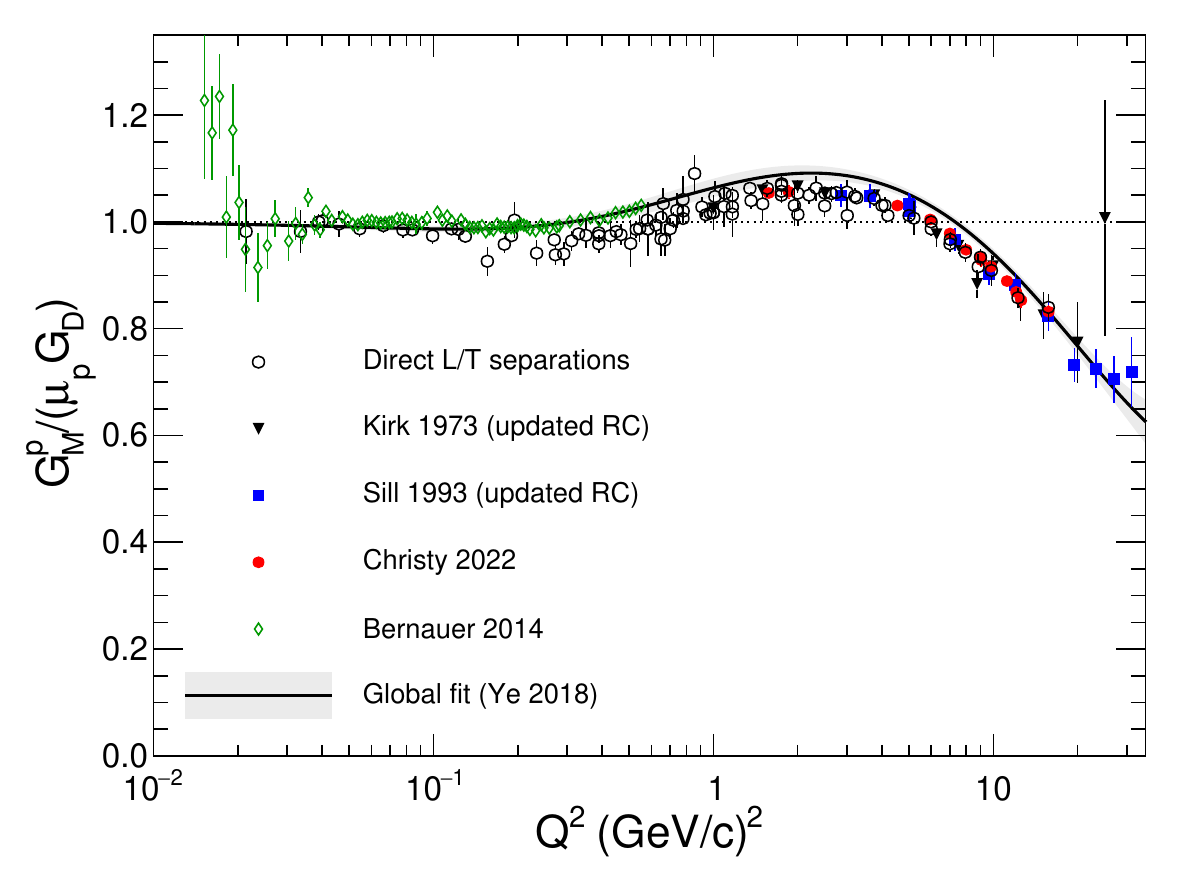}
    \caption{\label{fig:ch2:gmpdata} World data for the proton magnetic form factor $G_M^p/(\mu_pG_D)$. Red filled circles represent the most recent cross-section data from Jefferson Lab \cite{PhysRevLett.128.102002}. Black filled triangles and blue filled squares denote older SLAC data \cite{PhysRevD.48.29,PhysRevD.8.63} reanalyzed with state-of-the-art radiative corrections by the authors of \cite{PhysRevLett.128.102002}. Green open circles represent the Mainz A1 dataset \cite{PhysRevC.70.015206}. The global fit is sourced from \cite{YE20188}. See the text for more details. This figure is adapted from \cite{Gross2023}.}
\end{figure}

A nearly complete collection of existing proton electric ($G_E^p$) and magnetic ($G_M^p/\mu_p$) form factor data, normalized to $G_D$, is shown in Figures \ref{fig:ch2:gepdata} and \ref{fig:ch2:gmpdata}, respectively, covering both cross-section \cite{PhysRevLett.128.102002,PhysRev.142.922,Bartel:1973rf,BERGER197187,PhysRevD.4.45,BORKOWSKI1974269,Walker:1993vj,PhysRevD.50.5491,PhysRevLett.94.142301,PhysRevC.70.015206} and double-polarization \cite{PhysRevLett.80.452,PhysRevLett.84.1398,PhysRevLett.88.092301,PhysRevC.71.055202,PhysRevC.64.038202,Pospischil2001,MACLACHLAN2006261,PhysRevLett.104.242301,PhysRevC.85.045203,PhysRevLett.106.132501,PhysRevC.96.055203,PhysRevC.84.055204,ZHAN201159,PhysRevLett.105.072001,PhysRevC.74.035201,PhysRevLett.98.052301,SANE:2018cub} measurements. In these plots, the open circles represent published proton form factor data extracted using the Rosenbluth separation technique. The filled circles in \fig \ref{fig:ch2:gepdata} represent proton form factors extracted from polarization observables, which directly measure the ratio $G_E^p/G_M^p$. The ratio has been converted to $G_E^p$ using the value of $G_M^p$ obtained from the global fit provided in \cite{YE20188} (also shown in the same figure). A brief historical overview of the collection of these data, along with key achievements and surprises, is discussed below.

Proton form factor measurements from cross-section data have a long history, dating back to the 1950s with the pioneering work of Robert Hofstadter \cite{PhysRev.98.217,PhysRev.102.851,RevModPhys.28.214}. Early experiments struggled with precision, but significant improvements were made throughout the 1960s. In the 1970s, several precise data sets covering a broad range of $Q^2$ became available, marking a significant advancement. However, as higher-$Q^2$ data became available, inconsistencies began to emerge. For example, data from a SLAC experiment \cite{PhysRevD.4.45} suggested that $G_E^p/G_D$ exceeded unity for $Q^2$ between $1$ and $3.8$ (GeV/c)$^2$, contradicting the downward trend observed in other experiments \cite{BERGER197187,Bartel:1973rf}. At the same time, experiments at Mainz \cite{BORKOWSKI1974269} provided precise measurements of $G_E^p$ and $G_M^p$ at very low $Q^2$ ($0.014-0.12$ (GeV/c)$^2$), achieving a significant milestone in the high-precision measurement of $G_E^p$ and the determination of the proton charge radius. Around the same time, SLAC experiments \cite{PhysRevD.8.63} pushed the limits of $ep$ cross-section measurements to $Q^2 = 25$ (GeV/c)$^2$, revealing a drop-off in $G_M^p/\mu_pG_D$ beyond $Q^2 = 7$ (GeV/c)$^2$. This finding was confirmed in the 1990s by another SLAC experiment \cite{PhysRevD.48.29}, which measured the $ep$ cross-section up to $Q^2 = 30$ (GeV/c)$^2$ with improved statistical precision\footnote{Reanalysis of \cite{PhysRevD.8.63} $\&$ \cite{PhysRevD.48.29} datasets with the ``state-of-the-art" radiative correction (RC) is presented in \fig \ref{fig:ch2:gmpdata}.}. In the early 2000s, several Jefferson Lab (JLab) experiments contributed significantly to this effort by measuring proton form factors via Rosenbluth separation. Notably, some JLab experiments \cite{PhysRevLett.94.142301} detected the elastically scattered proton (e,p), instead of the more commonly used electron (e,e'), and found excellent agreement with the existing (e,e') data. Since the systematics for proton detection differ from those for electron detection, such agreement reinforced confidence that the experimental systematics are well understood. The most recent high-precision cross-section data for $G_M^p$ up to \qeq{15.75} also come from JLab  \cite{PhysRevLett.128.102002}. The data is shown as red-filled circles in \fig \ref{fig:ch2:gmpdata}.

The measurement of proton form factors using polarization observables began with early experiments in the 1970s at SLAC \cite{PhysRevLett.37.1258}, using the beam-target double-spin asymmetry technique (see \eqn \ref{eqn:ch2:beamtargetasy}). The recoil polarization (see \eqn \ref{eqn:ch2:recoilpolarization}) technique, first applied at MIT-Bates \cite{PhysRevLett.80.452}, successfully measured the $G_E^p/G_M^p$ ratio, particularly at low $Q^2$, demonstrating its potential for future studies. Experiments at facilities like MAMI \cite{Pospischil2001} and JLab \cite{PhysRevC.64.038202,MACLACHLAN2006261,PhysRevLett.105.072001} refined this method, providing more precise measurements of $G_E^p/G_M^p$, especially at low to moderate $Q^2$ values. These early results were consistent with Rosenbluth separation measurements but offered significantly improved precision.

\subheading{Form Factor Ratio Puzzle}
A breakthrough occurred in the late 1990s and early 2000s when JLab experiments \cite{PhysRevLett.84.1398,PhysRevC.71.055202,PhysRevLett.88.092301,PhysRevC.85.045203,PhysRevLett.104.242301} extended $G_E^p/G_M^p$ measurements using polarization transfer method to higher $Q^2$ values (up to $8.5$ (GeV/c)$^2$). These results revealed a sharp decline in the $\mu_p G_E^p/G_M^p$ ratio, contradicting previous expectations from cross-section measurements, which suggested that the ratio would remain near unity. The data showed that $G_E^p$ decreases faster than $G_M^p$ as $Q^2$ increases, indicating differences in the spatial distributions of charge and magnetization at short distances. This finding, confirmed by multiple JLab experiments, significantly reshaped our understanding of the nucleon’s internal structure.

The observed discrepancy between cross-section and polarization observables at high-\q is thought to be \cite{PhysRevLett.91.142303,PhysRevLett.91.142304} due to ``hard" two-photon exchange (TPE), a next-to-leading order process in QED radiative corrections to elastic $eN$ scattering. In this process, both exchanged photons carry ``large" momentum, making it impossible to calculate model-independently and thus neglected in standard radiative correction prescriptions. Cross-section measurements generally require large radiative corrections ($10-30\%$ at modest-to-large \q \cite{Gross2023}), whereas double-polarization observables do not, due to the nearly identical radiative effects in the polarization observables, which tend to cancel out in the ratio. Several experiments have studied the $\epsilon$ dependence of elastic $ep$ cross-sections, confirming the significance of the hard TPE effect.

When the discrepancy was first observed, theorists such as Maximon and Tjon \cite{PhysRevC.62.054320} refined the conventional radiative correction (RC) prescription based on the work of Tsai \cite{PhysRev.122.1898} and Mo and Tsai \cite{RevModPhys.41.205}. Reanalysis of older SLAC data \cite{PhysRevLett.128.102002,PhysRevC.93.055201} using updated RC prescription \cite{PhysRevC.62.054320} has shown that the magnitude and significance of the discrepancy between cross-section and polarization observables are reduced but not eliminated. The most recent results \cite{PhysRevLett.128.102002} of these reanalyses are shown in \fig \ref{fig:ch2:gmpdata}. Notably, the global fit \cite{YE20188} shown in the same figure (and also in \fig \ref{fig:ch2:gepdata}) includes phenomenological hard TPE corrections, which are not applied to the displayed data but tend to increase $G_M^p$ by roughly $2-3\%$ \cite{Gross2023} in the region where the discrepancy between cross-section and polarization observables is largest.

\subheading{Proton Radius Puzzle}
\label{subh:ch2:protonradiuspuzzle}
A major surprise in low-$Q^2$ elastic $ep$ cross-section measurements arose in the context of determining the mean squared charge radius of the proton $r_C^p$ (see \eqn \ref{eqn:ch1:chargeradius}). In 2010, the CREMA collaboration published an extremely precise extraction of the proton charge radius using Lamb shift measurements in muonic hydrogen \cite{Pohl2010}, yielding a value of $0.84$ fm—approximately seven standard deviations smaller than the previously accepted proton radius of $0.88$ fm derived from $ep$ scattering measurements and ordinary hydrogen spectroscopy. 

Reevaluation of existing $ep$ scattering data highlighted issues with the extraction of $r_C^p$, as well as ``overly optimistic" treatment of systematic uncertainties in some experiments \cite{Punjabi2015}. To address these concerns, the Mainz A1 collaboration performed $1422$ $ep$ cross-section measurements \cite{PhysRevC.70.015206} in a very low-$Q^2$ region (0.0038 to 0.98 (GeV/c)$^2$) over a wide range of beam energies and scattering angles, achieving unprecedented accuracy. The obtained cross-sections were fitted using advanced techniques, resulting in a proton charge radius of $r_C^p \approx 0.879 \pm 0.005_{\text{stat}} \pm 0.004_{\text{syst}}$, consistent with existing $ep$ scattering data.\footnote{While the MAMI extraction of $G_E^p$ was in agreement with global data, their $G_M^p$ extraction showed a slower-than-expected falloff with $Q^2$, suggesting a smaller proton magnetic radius. This trend is clearly visible in Figure \ref{fig:ch2:gmpdata}} 

More recently, in 2019, the PRad collaboration conducted a new $ep$ cross-section measurement \cite{Xiong2019}, reaching a $Q^2$ as low as $Q^2 = 0.0002$ (GeV/c)$^2$. Using a novel method that involved precision calorimetry, a windowless gas target, and simultaneous Møller scattering measurements to constrain the absolute cross-section normalization, they obtained a proton radius of $r_C^p \approx 0.83 \pm 0.01$ fm. This marked the first agreement between $ep$ scattering and muonic hydrogen measurements of the proton charge radius.

\subsection{Neutron Form Factor}
\label{ssec:ch2:neutronff}
\begin{figure}[h!]
    \centering
    \includegraphics[width=1\columnwidth]{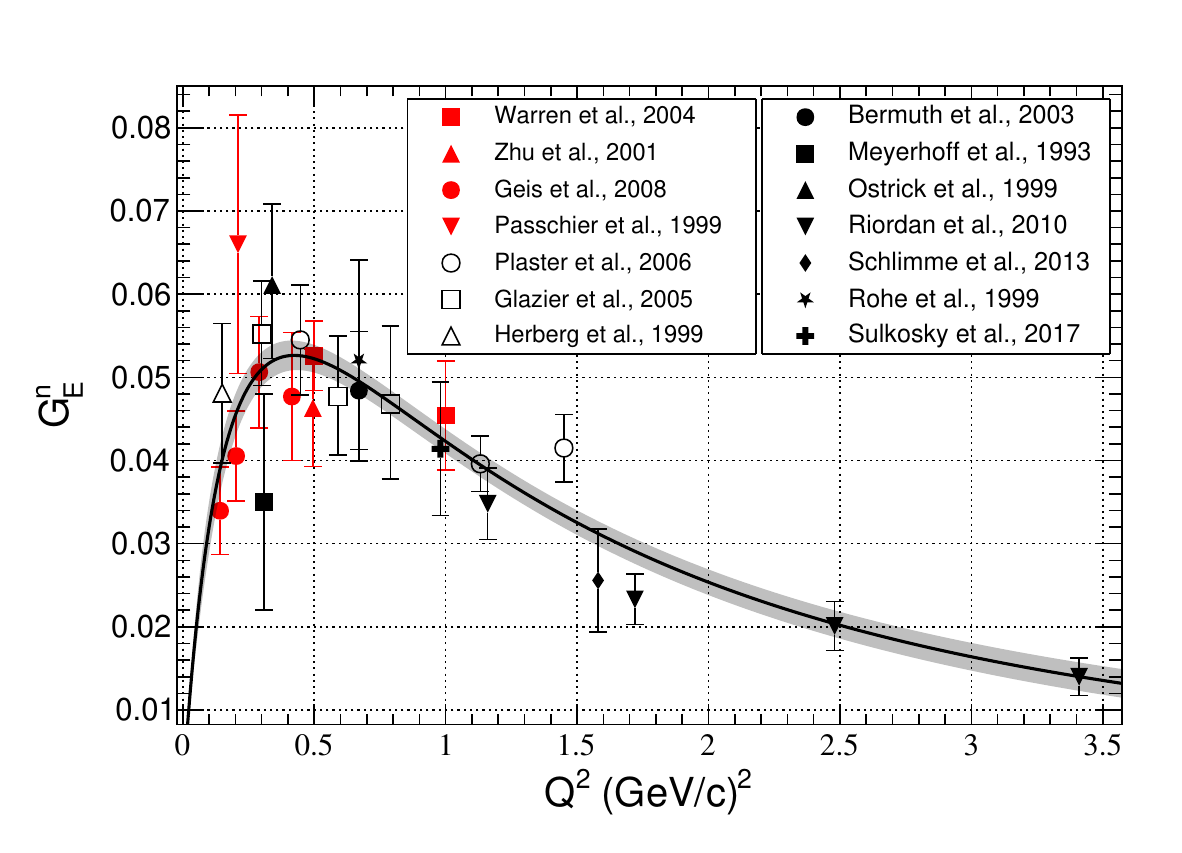}
    \caption{\label{fig:ch2:gendata} World data for the neutron electric form factor $G_E^n$. Black filled shapes represent beam-target double-spin asymmetry measurement data using a polarized $^{\ce{3}}$\ce{He} target \cite{PhysRevLett.105.262302,PhysRevC.96.065206,PhysRevLett.111.132504,BERMUTH2003199,Becker1999}. Red filled shapes denote beam-target double-spin asymmetry data with a polarized deuteron target \cite{PhysRevLett.101.042501,PhysRevLett.92.042301,PhysRevLett.87.081801,PhysRevLett.82.4988}, while open shapes represent measurements obtained via recoil neutron polarization \cite{Glazier:2004ny,Herberg1999,PhysRevC.73.025205}. The global fit is sourced from \cite{YE20188}. See the text for more details.}
\end{figure}

\begin{figure}[h!]
    \centering
    \includegraphics[width=1\columnwidth]{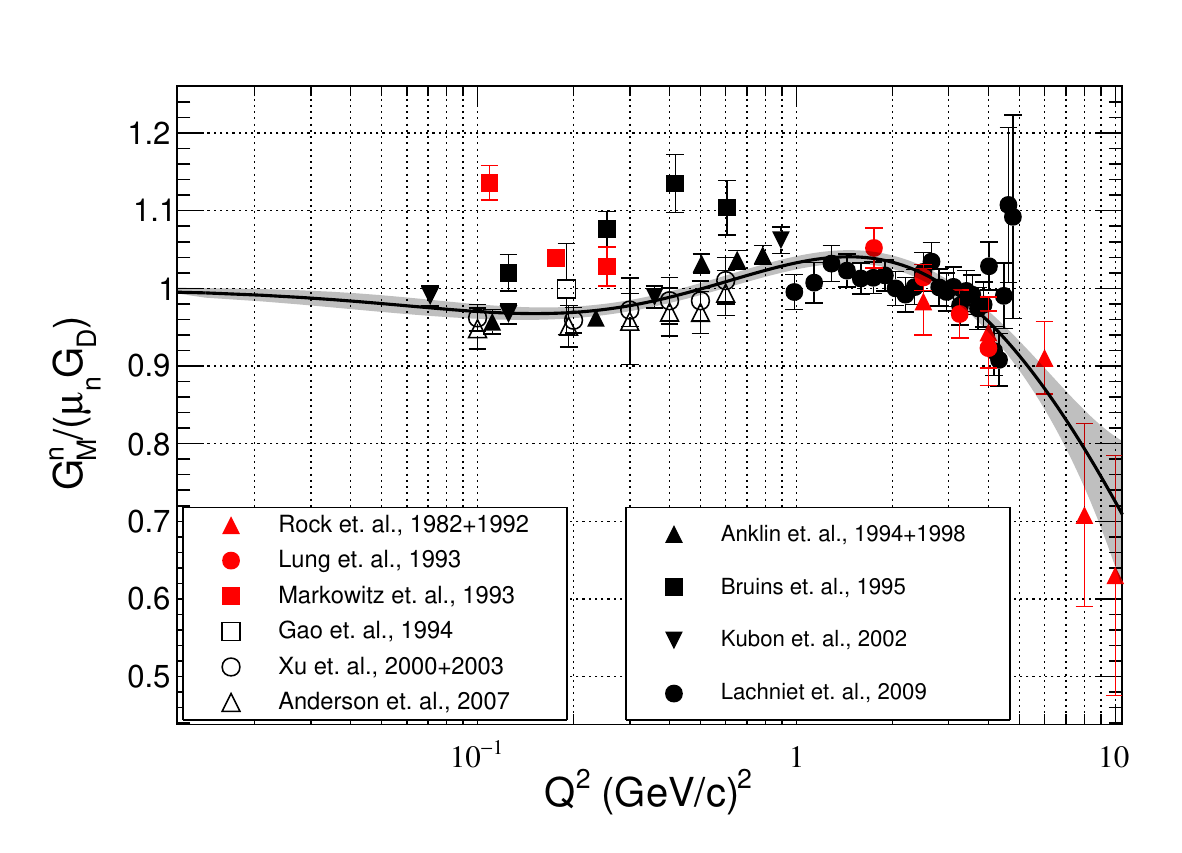}
    \caption{\label{fig:ch2:gmndata} World data for the neutron magnetic form factor $G_M^n/(\mu_nG_D)$. Black filled shapes represent measurements obtained using the ratio (or Durand's) method \cite{PhysRevLett.102.192001,ANKLIN1994313,ANKLIN1998248,PhysRevLett.75.21,KUBON200226}, while red filled shapes denote absolute cross-section measurements \cite{PhysRevLett.70.718,PhysRevD.46.24,PhysRevC.48.R5}. Black open shapes indicate beam-target double-spin asymmetry measurements \cite{PhysRevLett.85.2900,PhysRevC.67.012201,PhysRevC.75.034003,PhysRevC.50.R546}. The global fit is sourced from \cite{YE20188}. See the text for further details.} 
\end{figure}

A nearly complete collection of neutron electric ($G_E^n$) and magnetic ($G_M^n/\mu_n$) form factor data is shown in Figures \ref{fig:ch2:gendata} and \ref{fig:ch2:gmndata}, respectively. The relatively sparse data presented in these plots, compared to the proton form factor data shown in Figures \ref{fig:ch2:gepdata} and \ref{fig:ch2:gmpdata}, highlights the fact that measuring neutron form factors is significantly more challenging than measuring their proton counterparts. The primary difficulty stems from the lack of a stable free neutron target, necessitating electron scattering measurements from bound nuclei, such as deuteron ($D$) or $^{\ce{3}}$\ce{He}. Extracting cross-sections from these nuclei requires model-dependent nuclear corrections. Additionally, the charge neutrality of neutrons makes them difficult to detect with high efficiency, complicating coincidence measurements.

Traditionally, cross-section measurements to extract neutron form factors have been performed on a deuteron ($D$) target. Absolute cross-sections have been measured either in quasi-elastic single-arm $D(e,e')$ or coincidence $D(e,e'p)n$ reactions, or in a few cases, in elastic $eD$ reactions. Since the neutron is electrically neutral, $G_E^n$ vanishes at $Q^2=0$ and remains near zero—much smaller than the proton contribution to the quasi-elastic $eD$ cross-section—over almost the entire measured \q range. As a result, the magnetic term dominates, making these measurements primarily sensitive to $G_M^n$. For these reasons, all reliable $G_E^n$ data with reasonable precision come from polarization observables. Both recoil polarization (see \eqn \ref{eqn:ch2:recoilpolarization}) and beam-target double-spin asymmetry (see \eqn \ref{eqn:ch2:beamtargetasy}) techniques have been used to measure $G_E^n$, as discussed below.

The measurement of $G_E^n$ using the recoil polarization technique began in the late 1980s at MIT-Bates \cite{PhysRevC.50.R1749}, using the $D(e,e'pn)$ reaction at a \qeq{0.255}. They measured the transverse polarization component with a known beam polarization. This initial experiment demonstrated the feasibility of the technique, which was further refined at MAMI in the 1990s. At MAMI, both the transverse and longitudinal polarization components were measured \cite{Herberg1999}, extending $G_E^n$ measurements to a \q of up to $0.8$ (GeV/c)$^2$. Later, measurements at JLab \cite{PhysRevC.73.025205} further advanced the technique, achieving precise $G_E^n$ data at higher \q values up to $1.45$ (GeV/c)$^2$ with reduced uncertainties. These advancements provided critical tests for theoretical models, deepening our understanding of neutron and proton form factors.

Despite the success of the recoil polarization technique for $G_E^n$ measurements, the challenges of neutron polarimetry persisted, motivating the use of the beam-target double-spin asymmetry technique. The first such measurement was conducted at NIKHEF in 1999, at \qeq{0.21}, using a polarized electron beam and a vector-polarized deuterium gas target. This experiment \cite{PhysRevLett.82.4988} provided important constraints on $G_E^n$ up to \qeq{0.7} when combined with earlier data from \cite{PhysRevLett.74.2427} and \cite{PLATCHKOV1990740}. Later, JLab experiments in Hall C \cite{PhysRevLett.87.081801,PhysRevLett.92.042301} measured $G_E^n$ at higher \q values ($0.5$ and $1.0$ (GeV/c)$^2$) using a solid polarized deuterated ammonia ($ND_3$) target, marking a significant step forward in polarized target use at larger \q. In the mid-2000s, an experiment at MIT-Bates Lab \cite{PhysRevLett.101.042501} measured $G_E^n/G_M^n$ using a longitudinally polarized beam and a vector-polarized $D$ target, achieving excellent agreement with Vector Meson Dominance (VMD) based models \cite{PhysRevC.64.035204} and meson-cloud calculations \cite{PhysRevC.66.032201} (see \sect \ref{sec:ch2:theory}). A significant development was the use of polarized $^{3}$He, which effectively acts as a polarized neutron target. Early experiments at MIT-Bates \cite{PhysRevC.44.R571,PhysRevLett.68.2901}, done at lower \q (up to $0.2$ (GeV/c)$^2$), were unable to produce useful results due to significant spin-dependent effects. However, it was noted \cite{PhysRevLett.68.2901} that at higher \q, such effects would diminish significantly. In the 1990s, several experiments at MAMI \cite{BERMUTH2003199,Becker1999} successfully extracted $G_E^n$ at higher \q, ranging from $0.31$ to $0.67$ (GeV/c)$^2$. In 2006, the GEN(1) experiment at JLab \cite{Riordan:2010id} marked a significant leap by measuring the neutron form factor ratio $G_E^n/G_M^n$ up to \qeq{3.41}.

The beam-target double-spin asymmetry technique using a polarized $^{3}$He target was also employed for measuring $G_M^n$ in the low-\q regime ($0-0.6$ (GeV/c)$^2$) by several experiments \cite{PhysRevLett.85.2900,PhysRevC.67.012201,PhysRevC.75.034003,PhysRevC.50.R546}. A precise extraction of $G_M^n$ from this technique requires fully relativistic three-body calculations, which are quite challenging. Due to the superior sensitivity to $G_M^n$ of quasi-elastic $eD$ cross-sections, as discussed above, its early measurements, dating back to 1965, at low \q were performed using quasi-elastic single-arm $D(e,e')$ reactions \cite{PhysRev.139.B458,PhysRevD.8.753}. However, these measurements suffered from sizable theoretical uncertainties, primarily due to large final state interactions (FSI). To reduce theoretical uncertainties, several experiments have utilized coincidence cross-section measurements in the $D(e,e'pn)$ reaction for $G_M^n$ extraction \cite{PhysRev.173.1357,PhysRev.141.1286}. The primary challenge they faced was the accurate determination of neutron detection efficiency. A few experiments, including a recent JLab experiment \cite{PhysRevLett.132.162501}, took advantage of the similarity of the ground-state wave functions of the mirror nuclei $^{3}$H and $^{3}$He to extract $G_M^n$ from the ratio of quasi-elastic ${}^3H(e,e')$ and ${}^3He(e,e')$ cross-sections with minimal nuclear corrections. However, accurately modeling the inelastic background remained challenging.

\subsubsection*{Durand's Method/Ratio Method}
\label{sssec:ch2:durandsmethod}
The method least sensitive to uncertainties for $G_M^n$ extraction is the one proposed by Durand \cite{PhysRev.115.1020}, which involves simultaneous measurements of neutron-tagged (\deen) and proton-tagged (\deep) quasi-elastic scattering cross-sections from $eD$ scattering. This approach, commonly known as the ratio method, extracts $G_M^n$ from the ratio of these two cross-sections by using the well-known proton cross-section and subtracting a small term involving $G_E^n$. Refer to \sect \ref{sec:measurementtechnique} for a detailed description of the measurement technique. Detecting nucleons in coincidence provides a powerful means to suppress inelastic background, even at large \q. Systematic uncertainties related to the electron beam, target, electron detection and reconstruction, trigger, and data acquisition cancel out in the ratio. Additionally, uncertainties in the deuteron wave function, final state interactions, meson exchange contributions, and radiative effects largely cancel out since they are nearly identical between neutron-tagged and proton-tagged events. This leaves the relative detection efficiency of protons and neutrons as the primary source of uncertainty in this method. The near-cancellation of the aforementioned uncertainties, which plague other measurement techniques, makes this the most suitable method for $G_M^n$ extraction, allowing for measurements at very high \q.

The earliest experiments using the ratio method were conducted at DESY in the 1970s, reaching \q values up to $1.5$ (GeV/c)$^2$ \cite{PhysRevLett.16.592}. In 1995, measurements at Bonn \cite{PhysRevLett.75.21} extended this technique to lower \q values, in the range $0.13$ to $0.61$ (GeV/c)$^2$, with careful calibration of neutron detection efficiency using the $H(\gamma,\pi^-)n$ reaction. Further refinements came from a series of experiments conducted at NIKHEF \cite{ANKLIN1998248} and MAMI \cite{ANKLIN1994313}, spanning the \q range of $0.61$ to $0.78$ (GeV/c)$^2$. A subsequent MAMI experiment extended the \q range further, from $0.71$ to $0.89$ (GeV/c)$^2$, achieving a statistical precision of $1.5\%$. Throughout this period, neutron detection efficiency remained a critical focus. Neutron detectors were taken to the Paul Scherrer Institute (PSI) for efficiency measurements. The NIKHEF and MAMI experiments showed excellent agreement for matching \q values; however, their obtained $G_M^n$ values were smaller than those obtained by Bonn and MIT-Bates \cite{PhysRevC.48.R5}. It was suggested \cite{PhysRevLett.79.5186} that the Bonn group had overestimated their neutron efficiency by neglecting pion electroproduction contributions, although the Bonn team defended their approach \cite{PhysRevLett.79.5187}, arguing that the contribution was negligible for their kinematic conditions.

The most recent high-\q measurement of $G_M^n$ \cite{PhysRevLett.102.192001} using the ratio method came from the CLAS collaboration at JLab Hall B. $G_M^n$ was measured in fine \q bins across the range of $1$ to $4.8$ (GeV/c)$^2$. They employed an innovative dual-cell design, with liquid hydrogen and deuterium cells separated by $4.7$ cm, for the simultaneous in-situ calibration of neutron detection efficiency using the $H(e,e'\pi^+)n$ reaction. These results aligned well with previous SLAC data \cite{PhysRevD.46.24} extracted from quasi-free $ed$ cross-sections.

The work presented in this thesis extends the high-precision measurement of $G_M^n$ to an unprecedented \q value of 13.6 (GeV/c)$^2$ using the ratio method, a detailed description of which is provided in the following chapters.

%

%

\section{Theoretical Interpretation}
\label{sec:ch2:theory}
This section provides a brief overview of the theoretical interpretation of nucleon form factor data, with a focus on high-\q regime. Topics such as lattice QCD calculations, which are currently more precise at low-\q, and low momentum transfer chiral perturbation theory are omitted, as they are primarily relevant to low-\q measurements. For a more comprehensive review, readers are encouraged to consult \cite{Punjabi2015}, \cite{Gross2023}, and the references therein.

\subsection{Perturbative QCD (pQCD)}
In the perturbative QCD (pQCD) framework, the nucleon is treated as a system of three quarks moving in parallel. During elastic scattering, the momentum of the virtual photon is absorbed by one of these quarks and is subsequently redistributed to the remaining two quarks via two hard gluon exchanges. This approach, which is valid at high-\q due to asymptotic freedom, successfully predicts the dominance of the helicity-conserving Dirac form factor $F_1$ over the helicity non-conserving Pauli form factor $F_2$, where $F_1$ scales as $Q^{-4}$ and $F_2$ scales as $Q^{-6}$.

Belitsky, Ji, and Yuan \cite{PhysRevLett.91.092003,} refined this picture by including non-zero quark orbital angular momentum in the nucleon’s light-cone wave function and derived the high-\q behavior of the ratio $F_2/F_1$  as:
\begin{equation*}
    \frac{F_2}{F_1} \propto \frac{\ln^2{\frac{Q^2}{\Lambda^2}}}{Q^2},
\end{equation*}
where $\Lambda$ is a non-perturbative mass scale. This ``double-logarithmic" behavior, with $\Lambda\approx0.2-0.3$ GeV, was later confirmed in available proton data for $Q^2>3$ (GeV/c)$^2$, offering further support for the validity of pQCD in describing nucleon form factors at high momentum transfers. However, it is worth noting that similar agreement was not observed for neutron $F^n_2/F^n_1$ data, obtained using $G_E^n/G_M^n$ data up to \qeq{3.4} and $G_M^n$ data up to \qeq{4.8}, at least for the corresponding values of $\Lambda$ \cite{PhysRevLett.106.252003}.

\subsection{Vector Meson Dominance}
\begin{figure}[h!]
    \centering
    \includegraphics[width=0.4\columnwidth]{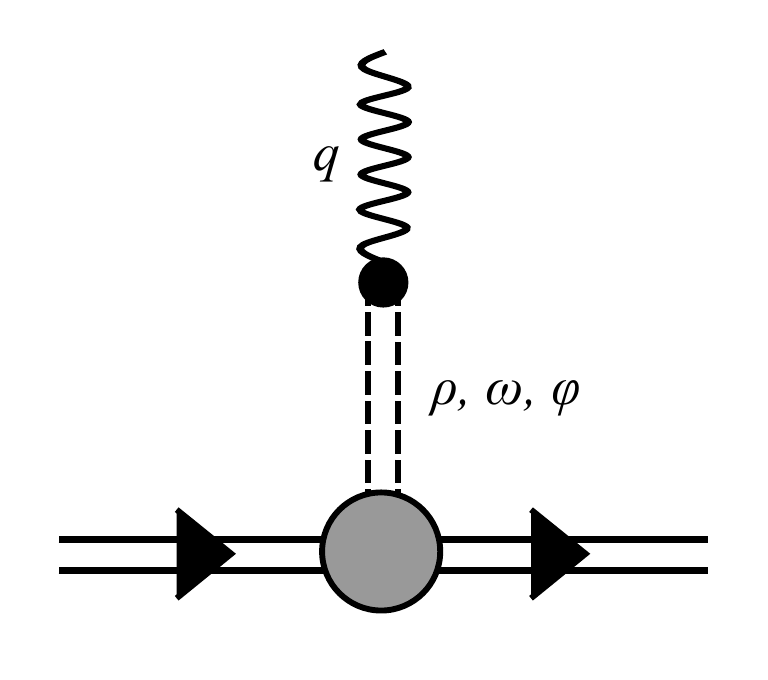}
    \caption{\label{fig:ch2:vmdpicture} Vector meson dominance (VMD) picture of virtual photon coupling to nucleon through intermediate vector meson.}
\end{figure}

The Vector Meson Dominance (VMD) model is one of the earliest to successfully describe the global features of nucleon form factors and predict the falloff of the ratio $G_E^p/G_M^p$. In VMD, a type of dispersion analysis, the form factors are treated as complex functions of $q^2$, which can be analytically continued into the timelike region ($q^2>0$). In this framework, the virtual photon couples to a low-lying vector meson—such as the $\rho(770)$, $\omega(782)$, or $\phi(1020)$—which shares the same spin and parity as the photon, and the vector meson subsequently interacts with the nucleon (see \fig \ref{fig:ch2:vmdpicture}). Notably, the existence of the $\rho$, $\omega$, and $\phi$ mesons was predicted by this framework \cite{PhysRev.106.1366,PhysRevLett.2.365} and later confirmed through experiments in reactions like $\pi N\rightarrow \pi\pi N$ or $e^+ e^- \rightarrow\,\text{pions}$.

The propagator for single vector meson exchange with simple couplings is expressed as:
\begin{equation*}
    \frac{m_{\text{V}}^2}{m_{\text{V}}^2 - q^2},
\end{equation*}
where $m_{\text{V}}$ is the mass of the vector meson. By assigning the vector meson a form factor, one can obtain the $Q^{-4}$ falloff of form factors at large $Q^2$, in line with data and predictions from pQCD. One early example of such a fit was provided by Iachello, Jackson, and Lande \cite{IACHELLO1973191} in 1973, which predicted the falloff in the $G_E^p/G_M^p$ ratio decades before the polarization transfer experiments. Gari and Krümpelmann \cite{GARI1992159} later refined this approach, achieving $Q^{-4}$ and $Q^{-6}$ falloff for $F_1$ and $F_2$, respectively, providing a better match to experimental data and pQCD predictions at large $q^2$. Subsequent work by Lomon \cite{PhysRevC.64.035204,Lomon2006}, incorporating a second $\rho$ meson (the $\rho'(1450)$) and a second $\omega$ meson (the $\omega'(1419)$), resulted in a robust parametrization that successfully described all nucleon form factors.

\subsection{Constituent Quark Model}
Constituent quark models (CQMs) were among the first models to successfully explain the static properties of hadrons, such as magnetic moments, as discussed in \sect \ref{sec:ch1:nucleontoQCD}. These models played a critical role in the development of QCD, particularly by highlighting the need for the color quantum number to preserve the Pauli exclusion principle in baryonic states like the $\Delta^{++}$, whose spin-flavor-orbital wave function is totally symmetric under the exchange of quarks. CQMs represent a class of nucleon models that describe nucleons as the quantum mechanical ground state of three constituent quarks confined within a potential.

Early CQMs, such as those developed by De R\'ujula, Georgi, and Glashow \cite{PhysRevD.12.147}, and by Isgur and Karl \cite{PhysRevD.18.4187}, treated baryonic wave functions as non-relativistic. In the latter model, the confining potential was assumed to be a long-range harmonic oscillator potential, supplemented by a one-gluon exchange potential to account for hyperfine splitting. This additional gluon exchange led to some $D$-state admixture in the baryon ground state, implying a slightly non-spherical charge distribution. This was confirmed by observing non-zero electric quadrupole ($E2$) and Coulomb quadrupole ($C2$) amplitudes in the $N\rightarrow \Delta(1232)$ transition.

However, to accurately calculate nucleon form factors at both low and high-\q, CQMs require a relativistic treatment. The main challenge lies in transforming the wave function from the nucleon’s rest frame to a moving frame. Paul Dirac proposed three approaches to handle this problem: instant, point, and light-front forms of dynamics \cite{RevModPhys.21.392}. These representations differ based on which Poincar\'e group generators are kinematical (interaction-independent) or dynamical (interaction-dependent).

The light-front form is particularly useful for form factor calculations due to the relative ease of transforming states between frames. Calculations start with a wave function based on CQMs, which is then transformed into the light-front form by applying a Melosh (or Wigner) rotation to the Dirac spinors of each quark. Initial work by Chung and Coester \cite{PhysRevD.44.229} using Gaussian wave functions predicted a falling ratio of $G_E^p/G_M^p$ at large \q, but underestimated the observed rate. A few years later, Schlumpf \cite{PhysRevD.47.4114} resolved this discrepancy by using a wave function with a power-law falloff, significantly improving the results. Extending this, French, Jennings, and Miller \cite{PhysRevC.54.920,PhysRevC.65.065205} predicted a zero-crossing of the $G_E^p/G_M^p$ ratio at \qeq{5\,\,\text{and}\,\,6}.

CQMs can be further improved by incorporating pionic degrees of freedom, as pions, being light, dominate the long-distance behavior of nucleon wave functions. In chiral quark models, nucleon form factors are calculated by treating pion effects perturbatively. One such model is the Light-Front Cloudy Bag Model proposed by Miller \cite{PhysRevC.66.032201}, where pion cloud effects were calculated through one-loop diagrams using the Schlumpf wave function, achieving good agreement with form factor data across both low and high-\q regimes. Further improvements were made by introducing constituent quark form factors and dynamically dressing them with mesons. The Clo\"et-Miller model \cite{PhysRevC.86.015208}, which combined quark and pion cloud effects, successfully explained the quark spin fraction of the proton’s spin and predicted a zero-crossing of $G_E^p$ at \qeq{12.3}.
\subsection{GPDs and the Transverse Densities}
Generalized Parton Distributions (GPDs), introduced independently by Ji \cite{PhysRevLett.78.610} and Radyushkin \cite{RADYUSHKIN1996417}, represent the probability amplitude for removing a quark with a certain momentum fraction and reinstating it with a different momentum fraction, while also encoding the spatial separation between the initial and final states. GPDs bridge the gap between parton distribution functions (PDFs) and form factors, allowing for a 3D imaging of the nucleon's internal structure. GPDs can be extracted from measurements of deeply virtual Compton scattering (DVCS) and other hard exclusive processes. In such extractions, precision nucleon form factor data play a crucial role by helping to determine the Bethe-Heitler contribution. Additionally, form factor data impose powerful constraints on the GPD moments, which enter the Ji sum rule \cite{PhysRevLett.78.610} for nucleon spin decomposition.

Alternatively, nucleon form factors can be derived from GPDs. Although experimental GPD data are limited due to the challenges associated with such measurements, good GPD models, such as the one proposed by Guidal \textit{et al} \cite{PhysRevD.72.054013}, have been successful in explaining the high-\q behavior of form factor data. When the longitudinal momentum transfer $\xi =0$, the integration of the GPD moments, $H$ and $E$, over the longitudinal momentum fraction $x$ of the nucleon carried by the quark, yields the nucleon form factors:
\begin{equation}
    \begin{aligned}
        F_1(Q^2) &=  \sum_q e^q \int_{-1}^1 dx H^q(x,0,Q^2), \\
        F_2(Q^2) &=  \sum_q e^q \int_{-1}^1 dx E^q(x,0,Q^2), 
    \end{aligned}
\end{equation}
where $e^q$ is the quark charge.

As discussed in \sect \ref{ssec:ch1:rosenbluthformula}, in the non-relativistic limit, the nucleon electromagnetic form factor represents the 3D Fourier transform of the spatial distributions of charge and magnetization within the nucleon. However, at higher \q, the presence of different wave functions for the initial and final nucleons invalidates a straightforward probability or density interpretation. Nevertheless, in the infinite momentum frame (IMF), where the observer approaches the nucleon at the speed of light or vice versa, one can exploit the relationship between GPD moments and the impact parameter ($b$) parton distribution functions (PDFs). This allows for a model-independent determination of the parton charge density of nucleons in transverse space, expressed in terms of the nucleon electromagnetic form factor.

If the Z-axis, $\vu{z}$, is chosen to lie along $(p + p')/2$, where $p$ and $p'$ represent the incoming and outgoing nucleon 4-momenta, respectively, and the frame is further arranged such that the virtual photon 4-momentum satisfies $q^+ = 0$ and has a transverse component, $\vb{q}_\perp$, in the X-Y plane such that $q^2 = -\vb{q}_\perp^2 = -Q^2$, then the transverse (X-Y plane) charge density of an unpolarized nucleon can be derived according to Miller's work \cite{PhysRevLett.99.112001} as:
\begin{equation}
    \rho^N_0(b) = \int_0^\infty \frac{dQ}{2\pi^2} QJ_0(bQ)F_1^N (Q^2) 
\end{equation}
where $J_n$ is the cylindrical Bessel function of order $n$. Building on this work, Carlson and Vanderhaeghen \cite{PhysRevLett.100.032004} derived the transverse charge density for a nucleon polarized along the $\vu{x}$ direction in the following form:
\begin{equation}
    \rho^N_T(\vb{b}) = \rho^N_0(b) - \sin\phi_b \int_0^\infty \frac{dQ}{2\pi}\frac{Q^2}{2M_N} J_1(bQ) F_2^N(Q^2),
\end{equation}
where $\vb{b} = b(\cos\phi_b\vu{x} + \sin\phi_b\vu{y})$. The unpolarized transverse charge density of the neutron reveals a negative core, in contrast to earlier beliefs that the neutron has a positive core surrounded by a negative pion cloud.

\section{The SBS Program}
\begin{figure}[h!]
    \centering
    \includegraphics[width=1\columnwidth]{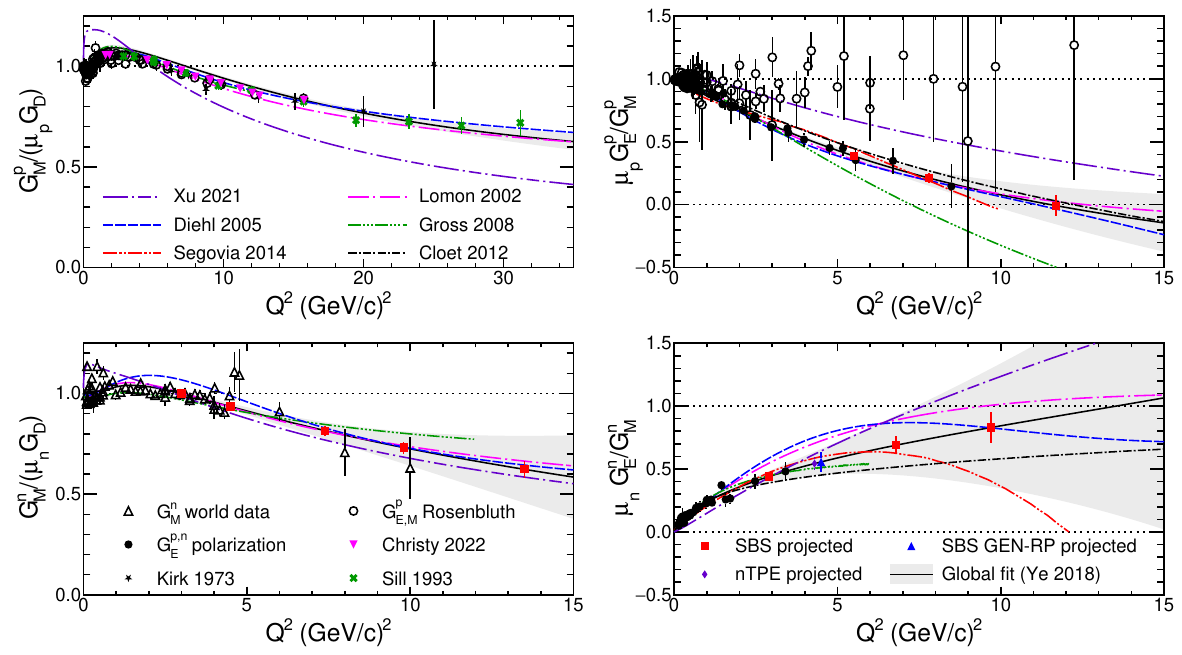}
    \caption{\label{fig:ch2:sbsprojection} The projection of SBS high-\q form factor data. This figure is adapted from \cite{Gross2023}.}
\end{figure}

In the previous section, we highlighted the far-reaching significance of high-precision nucleon form factor measurements across a wide range of \q. However, our review of the existing nucleon form factor data (see \sect \ref{sec:ch2:formfactordata}) clearly shows a lack of high-precision measurements at large \q, except for $G_M^p$. This issue is particularly pronounced for the neutron form factors. To address this gap, the Super BigBite Spectrometer (SBS) collaboration at Jefferson Lab (JLab) proposed a series of experiments aimed at high-precision measurements of $G_M^n$, $G_E^n$, and $G_E^p$, extending to unprecedented \q ranges. As illustrated in \fig \ref{fig:ch2:sbsprojection}, these measurements are expected to provide high-precision data for all nucleon form factors up to or beyond \qeq{10}, representing a significant advancement in our understanding of the nucleon's internal structure.

%
As of this writing, eight SBS experiments have been fully approved, with five having successfully completed data collection since the spectrometers were commissioned in the fall of 2021 at JLab's experimental Hall A. A brief description of the SBS high-\q form factor experiments is provided below:
\begin{enumerate}
    \item \textbf{\gmn}, or SBS-GMn, was the first SBS experiment, running from October 2021 to February 2022. It extended high-precision measurements of the neutron magnetic form factor $G_M^n$ over the range of \qeq{4\,\,\text{to}\,\,13.6}. The simultaneous measurement of quasi-elastic neutron-tagged (\deen) and proton-tagged (\deep) scattering cross-sections allowed for the use of Durand's (or ratio) method of measurement. The \gmn experiment is the subject of this thesis and will be discussed in detail in the following chapters.
    %
    \item \textbf{\ntpe}, or SBS-nTPE, was a short but high-impact experiment that ran parasitically with \gmn. It performed the first Rosenbluth separation of the neutron form factors at high \q ($4.5$ (GeV/c)$^2$) by taking measurements at two \ep points ($0.51$ and $0.80$). This experiment also used Durand's method to extract the elastic neutron-to-proton cross-section ratio at each \ep point.
    %
    \item \textbf{E12-09-016}, or GEnII, collected data in two run groups between October 2022 and October 2023. This experiment extended high-precision measurements of the neutron electric form factor $G_E^n$ over the range of \qeq{3.4\,\,\text{to}\,\,10} via a beam-target double-spin asymmetry measurement on a polarized \he target. Notably, the novel design of this target achieved a record-breaking figure of merit (FOM), enabling the experiment to reach such high \q.
    %
    \item \textbf{E12-17-004}, or GEnRP, measured the neutron electric form factor $G_E^n$ via recoil neutron polarization ($D(\overrightarrow{e},e'\overrightarrow{n})$) at \qeq{4.5}, which is approximately $3$ (GeV/c)$^2$ higher than the previous such measurement. This experiment, completed most recently in April-May 2024, will significantly enhance our understanding of neutron polarimetry. Comparing its results with those from \ntpe will provide valuable insights into the validity of the one-photon exchange (OPE) approximation in elastic $en$ scattering.
    %
    \item \textbf{E12-07-109}, or GEpV, is scheduled to collect data in the spring of 2025. This experiment will extend high-precision measurements of the proton electric form factor $G_E^p$ in the range of \qeq{8.5\,\,\text{to}\,\,12} using recoil proton polarization ($H(\overrightarrow{e},e'\overrightarrow{p})$).
\end{enumerate}

Several challenges are associated with high-\q form factor measurements, like those pursued by the SBS collaboration. For instance, the elastic $eN$ scattering cross-section decreases as $Q^{-12}$, necessitating very high luminosity. Additionally, as \q increases, the virtual photon angle decreases, causing the particles of interest to be detected at forward angles with high momentum. To address these challenges, spectrometers with large solid angle acceptance, capable of handling high luminosity at forward angles, are essential. Neutron form factor measurements, in particular, require the detection of high-energy neutrons with high efficiency, which presents an additional challenge.

To achieve the required high luminosity, the SBS collaboration takes advantage of the upgraded Continuous Electron Beam Accelerator Facility (CEBAF) at JLab, which can deliver up to 12 GeV electron beams with unparalleled intensity and precision (see \sect \ref{sec:ch3:cebafelectronbeam}). In combination with high-density, cryogenically cooled liquid deuterium (\ld) and hydrogen (\lh) targets available at JLab (see \sect \ref{ssec:ch3:cryotarget}), the luminosity required for SBS experiments ($\approx10^{38}$ cm$^{-2}$s$^{-1}$) is achieved\footnote{For the E12-09-016 experiment, gaseous \he targets with state-of-the-art FOM were provided by the University of Virginia (UVA).}.

To address other challenges, three primary detector subsystems were built. These include a large solid-angle acceptance dipole magnet with sufficient field strength to separate high-energy nucleons by charge, coupled with the capability to reach forward angles, achieved by cutting the magnet yoke to accommodate the beamline (see \sect \ref{ssec:ch3:sbsmagnet}). A hadron calorimeter with a large active area was developed to detect high-energy nucleons with high efficiency, alongside excellent spatial and temporal resolution (see \sect \ref{ssec:ch3:hcal}). Additionally, multiple gas electron multiplier (GEM) detectors with very large active areas, excellent position resolution, and the ability to withstand high rates were employed (see \sect \ref{ssec:ch3:bbgem}).
\chapter{The {\gmn} Experiment: Design and Configuration}


The \gmn experiment is the inaugural experiment of the Super BigBite Spectrometer (SBS) collaboration's high-\q nucleon electromagnetic form factor (EMFF) measurement program. It aims to extend the high-precision measurement of the neutron magnetic form factor, $G_M^n$, from $Q^2=4$ to \SI{13.6}{(GeV/c)^2}, leading to a significant enhancement of our understanding of the neutron's internal structure.
\begin{figure}[h!]
	\centering
	\fboxsep=0.75mm
    \fboxrule=1pt
	\fcolorbox{gray}{lightgray}{\includegraphics[width=0.9\columnwidth]{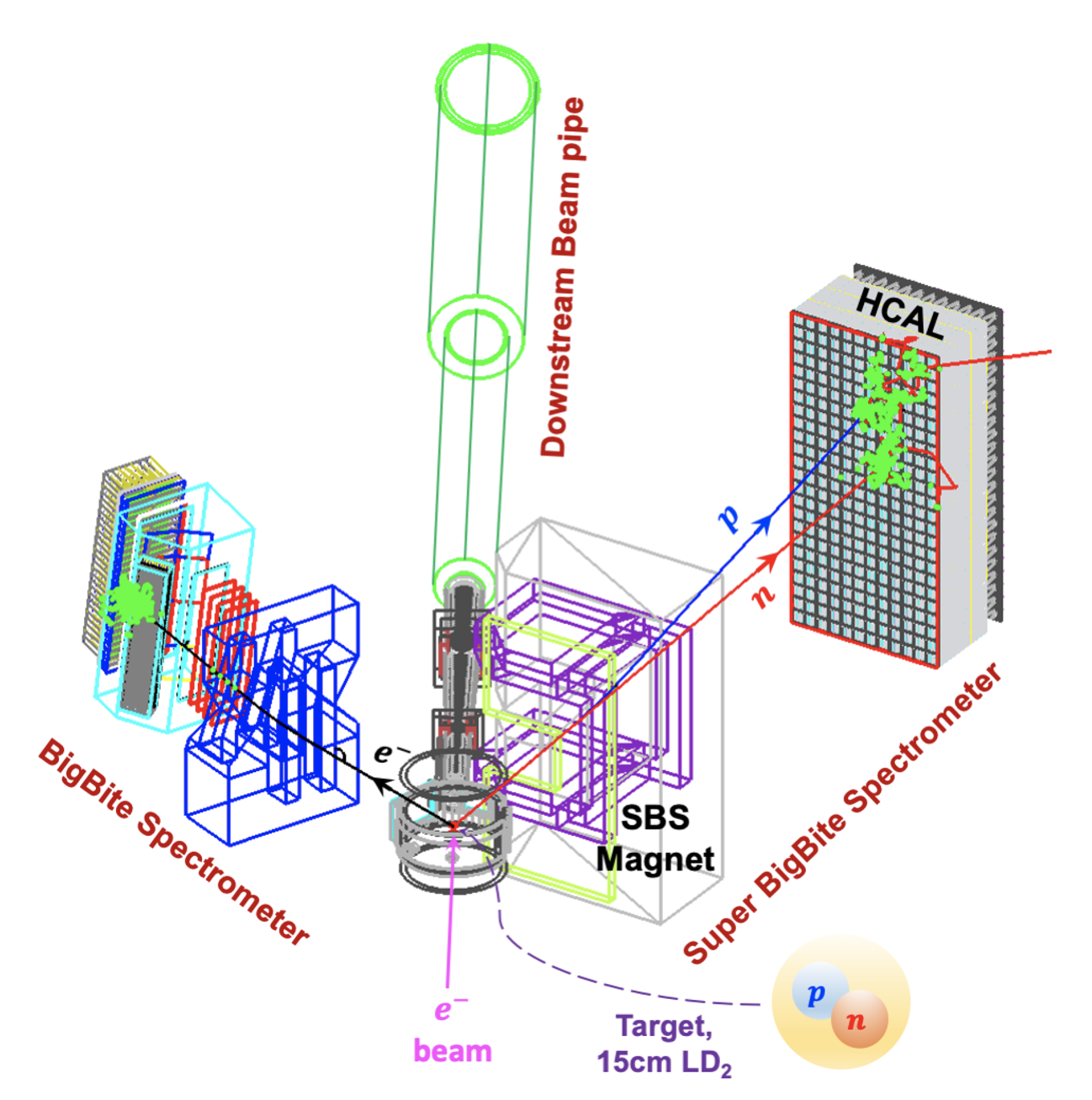}}
	\caption{\label{fig:gmnsetup} Experimental setup of \gmn in Geant4.}
\end{figure}

The installation of the spectrometers for the SBS program at Jefferson Lab's Hall A began in the summer of 2021. Data collection for \gmn commenced immediately after completion and continued for nearly five months, from October 14, 2021, to February 8, 2022. Figure \ref{fig:gmnsetup} illustrates a schematic of the \gmn experimental setup. A high-energy electron beam produced by the accelerator collided with the deuterium target in the scattering chamber. The scattered electrons and nucleons were then simultaneously detected by the upgraded BigBite Spectrometer (BBS) and the new Super BigBite Spectrometer (SBS), respectively. This simultaneous detection allowed the use of ``Durand's Method" of measurement, which is insensitive to many systematic uncertainties that affect other measurement techniques.

Within the BBS, electrons were bent upwards by the BigBite dipole magnet based on their momentum, tracked with high precision by five Gas Electron Multiplier (GEM) detectors, and finally stopped by the BigBite calorimeter (BBCAL) for high-resolution energy reconstruction. Additionally, a Cherenkov detector and a timing hodoscope within the BBS facilitated particle identification and precise time measurements. In the SBS, the scattered nucleons were first separated by charge by the SBS dipole magnet and then detected by the hadron calorimeter (HCAL) with very high and comparable efficiencies for proton and neutron.

In this chapter, we will discuss in detail the measurement technique, kinematics, and various parts of the experimental setup of \gmn.

\section{Measurement Technique}
\label{sec:measurementtechnique}
The ``Durand method", also known as the ``ratio method" \cite{PhysRev.115.1020}, was used by the \gmn experiment to determine $G_{M}^n$ from electron-deuteron scattering measurements. This method assumes that the electron-deuteron scattering cross-section arises solely from scattering by the struck nucleon. This assumption holds if the proton-neutron interference term in the underlying cross-section is suppressed, achievable by restricting the scattering region to the forward cone around the momentum transfer vector (\qvect) direction (defined in \sect \ref{ssec:ch4:enkinecorr}). The resulting events are the neutron-tagged (\deen) and proton-tagged (\deep) quasi-elastic (QE) scattering events from deuteron—a loosely bound system of a proton and neutron in perpetual ``Fermi motion". With simple nuclear corrections, QE scattering from deuteron at high \q can be viewed as a sum of scattering from free neutron and proton targets.

The ``ratio method" requires the simultaneous detection of the scattered electron and nucleons \cite{ANKLIN1994313,PhysRevLett.70.718,PhysRevLett.102.192001} to form the quasi-elastic scattering cross-section ratio \rqe, defined as:
\begin{equation}
    R^{QE} = \frac{\dv{\sigma}{\Omega}|_{D(e,e'n)}}{\dv{\sigma}{\Omega}|_{D(e,e'p)}}
\end{equation}

The ratio \rqe is insensitive to many systematic effects, such as scattered electron detection and reconstruction efficiencies, DAQ deadtime, and electron trigger efficiency, which often complicate other measurement techniques. Moreover, as a coincidence measurement, a simple cut requiring electron-nucleon coincidence significantly reduces inelastic background, enabling measurements at high \q. 

The elastic neutron-tagged to proton-tagged cross-section ratio, $R$, can be extracted from \rqe with corrections for final-state interactions and other nuclear effects using the following equation:
\begin{equation}
\label{eqn:rprime}
    R = \frac{\dv{\sigma}{\Omega}|_{n(e,e')}}{\dv{\sigma}{\Omega}|_{p(e,e')}} = \frac{R^{QE}}{1+f_{corr}} 
\end{equation}
The advantage of the ``ratio method" lies in the near cancellation of these corrections in the ratio since they are almost identical for neutrons and protons. Calculations show that the residual correction on the ratio, $f_{corr}$, will be under $0.1\%$ in the \gmn region of interest.

$R$, representing the $en$ to $ep$ Born cross-section ratio, is the most fundamental model-independent observable of the \gmn experiment. It can be expressed using the Rosenbluth formula as follows:
\begin{equation}
\label{eqn:ch3:rrosenbluth}
    R = \frac{\frac{\sigma_{\text{Mott}}}{\epsilon_n(1+\tau_n)} \left( \epsilon_n{G^n_{E}}^2 + \tau_n {G^n_{M}}^2 \right)}{\frac{\,\sigma_{\text{Mott}}}{\epsilon_p(1+\tau_p)} \left( \epsilon_p{G^p_{E}}^2 + \tau_p {G^p_M}^2 \right)}
\end{equation}
This expression can be greatly simplified as the factors $\sigma_{\text{Mott}}$, $\epsilon$, and $\tau$ are approximately the same for $en$ and $ep$ scattering.

Finally, by inverting \eqn \ref{eqn:ch3:rrosenbluth}, $G_M^n$ can be expressed in terms of $R$ as:
\begin{equation}
\label{eqn:ch3:gmnfromr}
    G_M^n = -\left[ \frac{1}{\tau_n}\frac{\epsilon_{n}(1+\tau_{n})}{\epsilon_{p}(1+\tau_{p})}\sigma^p_{Red}R - \frac{\epsilon_n}{\tau_n}{G_E^n}^2 \right]^{\frac{1}{2}},
\end{equation}
where the minus sign is assumed due to the neutron's negative magnetic dipole moment. The proton reduced cross-section, $\sigma^p_{Red}$, and $G_E^n$ are derived from the parametrization of nucleon electromagnetic form factors based on existing global data. As improved models become available, more accurate $G_M^n$ values can be extracted from the measured $R$ values.

\section{Kinematics}
\label{sec::kine}
Table \ref{tab:sbsconfig} outlines the kinematic points of the \gmn experiment. Data were collected at six unique combinations of \q and \ep values. The lowest two {\q} points overlap with the existing CLAS e5 measurements of $G_M^n$ \cite{PhysRevLett.102.192001} while the remaining higher {\q} points greatly extend the region in which $G_M^n$ is known with high precision. Data recorded at two \ep points at \qeq{4.5} will be used by the \ntpe experiment to do high precision Rosenbluth separation (see \sect \ref{ssec:ch2:rosenbluthseparation}) of the neutron form factors to shed some light on the two-photon exchange contribution in elastic $en$ scattering. 
\begin{table}[h!]
\caption{\label{tab:sbsconfig}Kinematics of \gmn. SBS conf. is a dummy index assigned to each unique experimental configuration, \q is the central {\q}, \ep is the longitudinal polarization of the virtual photon, $E_{beam}$ is the beam energy, $\theta_{BB}(d_{BB})$ is the BigBite central angle (target-magnet distance), $\theta_{SBS}(d_{SBS})$ is the Super BigBite central angle (target-magnet distance), $\theta_{HCAL}(d_{HCAL})$ is the HCAL central angle (target-HCAL distance), $\Bar{E_{e}}$ is the average scattered electron energy,  $\Bar{p}_{N}$ is the average scattered nucleon momentum. Data taken at two \ep points for $Q^2=4.5$ \SI{}{(GeV/c)^2} will be used by {\ntpe} experiment to do Rosenbluth separation to shed light on the two-photon exchange contribution in elastic $en$ scattering. Additional important kinematic specific parameters can be found in \tab \ref{tab:sbsconfig2}.}
\centering
\begin{tabular}{>{\hsptab}c<{\hsptab}>{\hsptab}c<{\hsptab}>{\hsptab}c<{\hsptab}>{\hsptab}c<{\hsptab}>{\hsptab}c<{\hsptab}>{\hsptab}c<{\hsptab}>{\hsptab}c<{\hsptab}>{\hsptab}c<{\hsptab}>{\hsptab}c<{\hsptab}}
\hline\hline\vspace{-1.1em} \\ 
SBS   & $Q^{2}$         &\multirow{2}{*}{\ep}& $E_{beam}$& $\theta_{BB}$/$d_{BB}$& $\theta_{SBS}$/$d_{SBS}$& $\theta_{HCAL}$/$d_{HCAL}$& $\Bar{E_{e}}$    & $\Bar{p}_{N}$ \\ 
conf. &\SI{}{(GeV/c)^2} &   & (GeV)     & (\SI{}{deg/m})        & (\SI{}{deg/m})          & (\SI{}{deg/m})            & \SI{}{(GeV)}& (\SI{}{GeV/c}) \vspace{0.2em} \\ \hline \vspace{-1.1em} \\ 
4  & 3.0  & 0.72 & 3.73 & 36.0/1.79 & 31.9/2.25 & 31.9/11.0 & 2.12 & 2.4 \\ 
9  & 4.5  & 0.51 & 4.03 & 49.0/1.55 & 22.5/2.25 & 22.0/11.0 & 1.63 & 3.2 \\ 
8  & 4.5  & 0.80 & 5.98 & 26.5/1.97 & 29.9/2.25 & 29.4/11.0 & 3.58 & 3.2 \\ 
14 & 7.4  & 0.46 & 5.97 & 46.5/1.85 & 17.3/2.25 & 17.3/14.0 & 2.00 & 4.8 \\ 
7  & 9.9  & 0.50 & 7.91 & 40.0/1.85 & 16.1/2.25 & 16.1/14.0 & 2.66 & 6.1 \\ 
11 & 13.6 & 0.41 & 9.86 & 42.0/1.55 & 13.3/2.25 & 13.3/14.5 & 2.67 & 8.1 \\ 
\hline\hline
\end{tabular}
\end{table}
%
%
\begin{table}
    \caption{\label{tab:sbsconfig2}Important experimental parameters for \gmn kinematics (sorted chronologically). This table is an extension of \tab \ref{tab:sbsconfig}. ``SBS conf." is a dummy index assigned to each unique experimental configuration, \q is the central \q value, \ep is the longitudinal polarization of the virtual photon, ``Run range" is the range of good data runs taken during the configuration, including all targets, $I_{SBS}$ is the SBS magnet current used during data collection for the corresponding target, expressed as a percentage of \SI{2100}{A} (* denotes the SBS current used for production), $B_{SBS}^{max}$ is the maximum SBS dipole field strength, estimated from observed proton deflection (see \sect \ref{sssec:ch4:dxdycorrcut}), $\Sigma$ is the total charge collected on \ld target at the production SBS field settings, ``GEM conf." is the GEM detector configuration (see \tab \ref{tab:gemconfig}). The \ce{Al} shield column notes the presence or absence of the \ce{Al} shield in front of the scattering chamber window for electrons (see \sect \ref{sec:ch3:exptarget}).}
    \centering
    \begin{tabular}{>{\hsptabn}c<{\hsptabn}>{\hsptabn}c<{\hsptabn}>{\hsptabn}c<{\hsptabn}>{\hsptabn}c<{\hsptabn}>{\hsptabn}c<{\hsptabn}>{\hsptabn}c<{\hsptabn}>{\hsptabn}c<{\hsptabn}>{\hsptabn}c<{\hsptabn}>{\hsptabn}c<{\hsptabn}>{\hsptabn}c<{\hsptabn}} 
    \hline\hline\vspace{-1.1em} \\ 
        SBS & \q & \multirow{2}{*}{\ep} & \multirow{2}{*}{Run range} & \multicolumn{2}{c}{$I_{SBS}$ $(\%)$} & $B_{SBS}^{max}$ & $\Sigma$ & GEM &  \ce{Al} \\ \vspace{-1.1em} \\ \cline{5-6} \vspace{-1.1em} \\
        conf. & (GeV/c)$^2$ &  &  & \ld & \lh & (T) & (C) & conf. & shield \vspace{0.2em} \\ \hline \vspace{-1.1em} \\
        4  & 3.0  & 0.72 & 11436-11616 & 0,30*,50     & 0,30,50     & 1.780 & 0.05 & 0 & No  \\
        7  & 9.9  & 0.50 & 11965-12073 & 85*          & 85          & 1.275 & 0.59 & 0 & No  \\
        11 & 13.6 & 0.41 & 12278-13064 & 0,100*       & 0,100       & 1.275 & 10.5 & 1 & Both \\
        14 & 7.4  & 0.46 & 13193-13404 & 70*          & 0,70        & 1.275 & 1.69 & 2 & Yes \\
        8  & 4.5  & 0.80 & 13435-13620 & 0,50,70*,100 & 0,50,70,100 & 1.230 & 0.84 & 2 & Yes \\
        9  & 4.5  & 0.51 & 13656-13799 & 70*          & 70          & 1.275 & 2.87 & 2 & Yes \\
    \hline\hline
    \end{tabular}
\end{table}

The BB and Super BB spectrometers, each weighing on the order of several tens of tons, required manual movement during configuration changes, followed by a geodetic survey to align the detector subsystems and determine their precise locations. This task was further complicated by the unavailability of the Hall A overhead crane and additional complexities from the ongoing COVID-19 pandemic. As a result, setting up for data collection at a specific kinematic point typically required an entire day shift.

Addressing unexpected issues during the \gmn experiment led to significant hardware changes between kinematics, such as the removal of some SBS magnet coils to reduce stray fields, replacing faulty GEM layers, and installing an \ce{Al} shield in front of the scattering chamber window to reduce background rates in the front GEM layers. Keeping track of these changes, summarized in \tab \ref{tab:sbsconfig2}, is crucial for accurate event reconstruction and realistic simulation of the experimental setup.

The \gmn production data was collected on the \ld target at specific SBS field settings designed to maximize neutron-proton separation while keeping upward-deflected protons within the HCAL acceptance. The production SBS field settings and the collected total charge for each \gmn kinematic are listed in \tab \ref{tab:sbsconfig2}. The unexpectedly low statistics at \qeq{9.9}, compared to the other high-\q datasets, resulted from unforeseen experimental downtime. However, the data collected was sufficient for the desired statistical precision. In addition to production data, ample \lh data was recorded at various field settings for calibration, as listed in the same table. Dedicated data on optics and dummy targets were also taken at each kinematic.

\section{The Electron Beam}
\label{sec:ch3:cebafelectronbeam}
The Continuous Electron Beam Accelerator Facility (CEBAF) at Jefferson Lab is a racetrack-shaped electron beam accelerator located in a tunnel 25 feet below ground. It is capable of delivering a continuous wave (CW) electron beam of up to 12 GeV energy with unparalleled intensity and precision. CEBAF features four experimental halls—A, B, C, and D—where fixed-target electron scattering experiments are conducted to probe the electromagnetic structure of hadrons. This section provides a brief overview of the process by which a high-energy electron beam is produced at CEBAF, including its generation, acceleration, and delivery to the experimental halls, followed by a brief description of the Hall A beamline.

%
%
\begin{figure}[ht!]
    \centering
    \includegraphics[width=0.9\columnwidth]{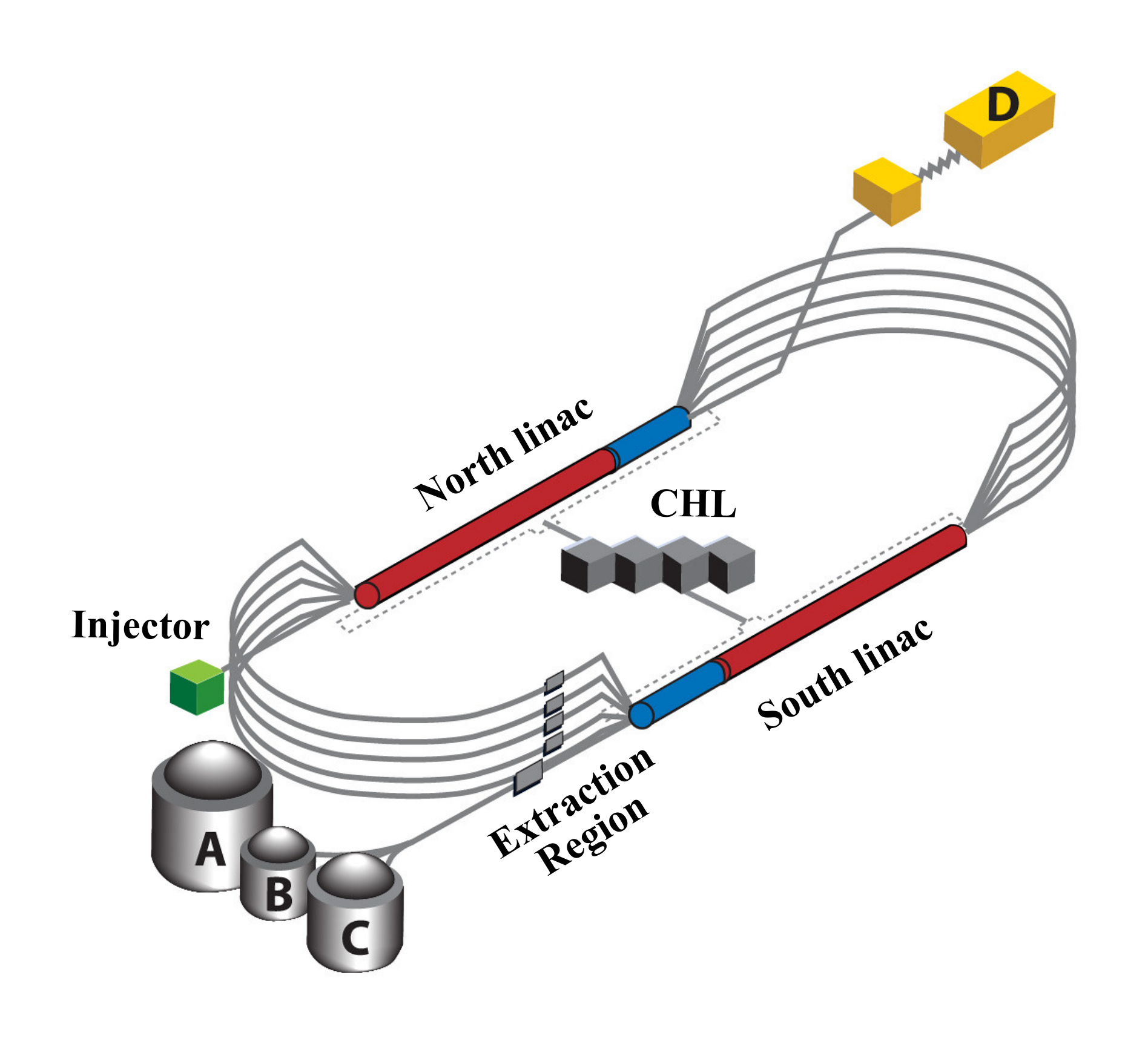}
    \caption{\label{fig:ch3:cebafschematic} Schematic of the \SI{12}{GeV} CEBAF.}
\end{figure}
\subsection{Generation} 
Electrons are produced via laser-induced photoemission from a superlattice gallium arsenide (\ce{GaAs}) photocathode \footnote{\ce{GaAs}/\ce{GaAsP} strained-layer superlattice photocathode, consisting of 14 periodic layers of \ce{GaAs} and \ce{GaAsP}, capable of producing $85\%$ polarization at 780 nm. QE: $1\%$, 6 mA/W}. The photocathode is housed in a 100 keV electron gun operated under ultra-high vacuum ($10^{-11}-$\SI{E-12}{Torr}). Four independent RF-gain-switched lasers, each capable of producing circularly polarized 780 nm light at a pulse rate of 499/249.5 MHz, are used to generate unique electron beams for each experimental hall. The frequency and phase of each laser beam are ingeniously selected to produce a 1497 MHz train of electrons, enabling simultaneous operation of all four experimental halls.

For example, when only Halls A, B, and C are operational, their respective laser sources are pulsed at 499 MHz, with a $120^{\circ}$ RF-phase shift introduced between them. These three independent beam bunches are then interlaced to create a single 1497 MHz beam bunch train, which conserves the duty factor. This frequency matches the fundamental RF frequency of the linacs. The properties of each bunch, particularly the charge, are preserved in the formation of the bunch train, allowing for precise control over the simultaneous operation of the halls. The beam bunch train is then injected into the north linac for acceleration \footnote{When Hall D is operational, the laser frequencies are adjusted to accommodate a fourth beam in the bunch train. One possible configuration is to send a 249.5 MHz beam to Halls A and D, and a 499 MHz beam to Halls B and C, as was the case during the \gmn experiment.}.
\begin{figure}[ht!]
    \centering
    \includegraphics[width=1\columnwidth]{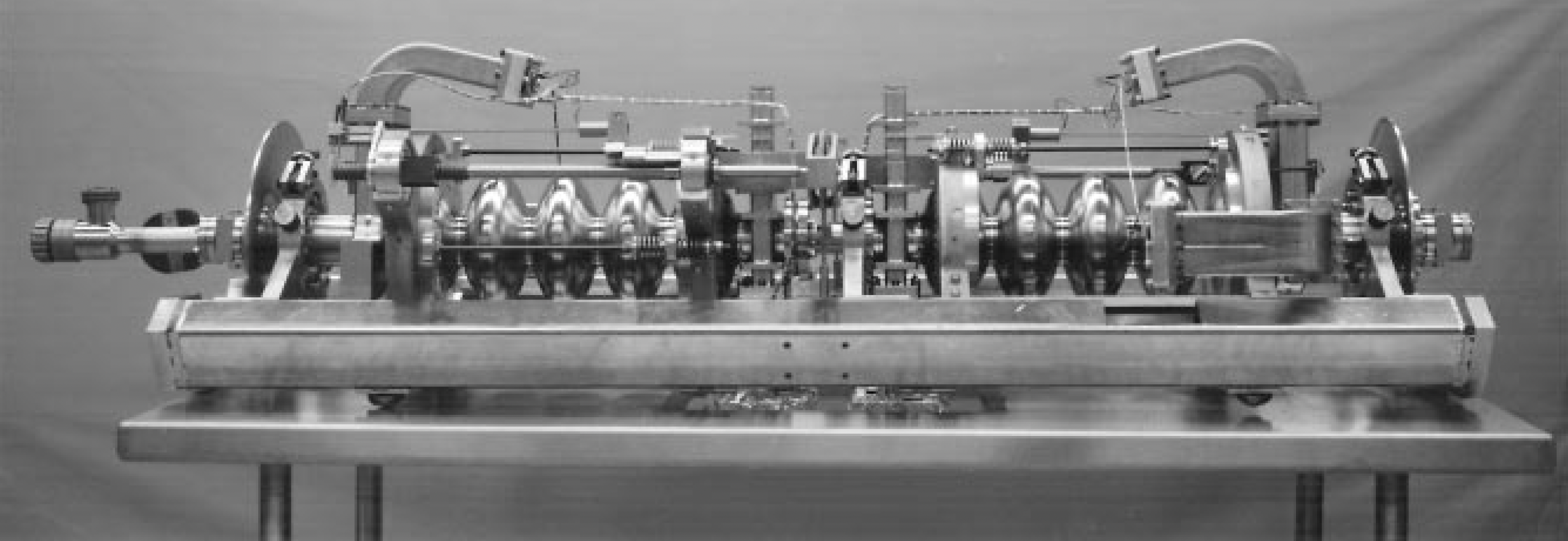}
    \caption[A CEBAF cryounit]{\label{fig:ch3:cebafcavity} A CEBAF cryounit, consisting of two hermetically paired five-cell SRF cavities \cite{Leemann2001}.}
\end{figure}
\subsection{Acceleration} The CEBAF consists of two antiparallel linear accelerators (linacs), each capable of producing \SI{1.1}{GeV} energy gradient. One of the ingenious features of CEBAF's design is the use of ten, five at each end, magnetic arcs to link the linacs enabling multipass beam recirculation. A total of five and a half passes are needed to reach the highest energy \footnote{Hall D exclusively receives the highest energy beam due to its physical location, while the other halls receive a maximum beam energy of \SI{11}{GeV}. During \gmn, operational constraints limited Hall A to a maximum beam energy of \SI{9.9}{GeV}.}. This ``race-track" design dramatically reduced the cost of building CEBAF by significantly reducing the number of Superconducting Radio-Frequency (SRF) cavities - another very important innovation in CEBAF design.  Twice the accelerating gradients and negligible losses due to resistive heating make SRFs immensely cost-effective compared to room-temperature RFs. 

SRF cavities are the building blocks of CEBAF's linacs. There are two types of SRF cavities: one composed of five elliptically shaped niobium cells and the other of seven elliptically shaped niobium cells\footnote{The development of seven-cell cryomodules was one of the key innovations that enabled CEBAF's energy upgrade from $6$ to $12$ GeV in 2017.}. The five-cell cavity is powered by a 5 kW klystron and can produce an energy gain of 3.75 MeV per pass, while the seven-cell cavity is powered by a 13 kW klystron and can produce an energy gain of 13.5 MeV per pass. Each cavity operates at a frequency of 1497 MHz, with optimal performance achieved by maintaining the cavity walls at 2 K and sustaining an ultra-high vacuum ($\approx$ \SI{E-12}{Torr}) inside the cavity.

\renewcommand{\thefootnote}{\fnsymbol{footnote}}
\begin{table}[h!]
    \centering
    \caption{Specifications of the C20 and C100 cryomodules installed in the 12 GeV CEBAF \cite{Leemann2001,PhysRevAccelBeams.27.084802}.}
    \label{tab:ch3:cryomodule}    
    \begin{tabular}{lp{3em}l}
    \hline\hline
       \multirow{2}{*}{Quantity} & \multicolumn{2}{c}{Cryomodule Type} \\ \cline{2-3} \vspace{-1em} \\
                                           & C20             &  C100 \\ \hline \vspace{-1em} \\
       Total number in the injector        & 1+$\frac{1}{4}$ & 1    \\ \vspace{-1em} \\
       Total number/linac                  & 20\tablefootnote{\label{myfootnote}According to \cite{PhysRevAccelBeams.27.084802}, between 2006 and 2019, some C20 cryomodules were replaced with refurbished C50 modules, although the exact number of these modules in each linac is not explicitly mentioned. C50 modules are five-cell modules with an energy gain of 6.25 MeV per pass, approximately $67\%$ higher than that of the C20 modules.}              & 5 \\
       SRF cavity/cryomodule               & 8               & 8 \\
       Energy gain/SRF cavity/pass (MeV)   & 3.75            & 13.5 \\
       Niobium cells/SRF cavity            & 5               & 7 \\
       SRF cavity length (m)               & 0.5             & 0.7 \\ \vspace{-1em} \\
       \hline\hline
    \end{tabular}
\end{table}
Two hermetically paired SRF cavities of the same type constitute one ``cryounit" (see \fig \ref{fig:ch3:cebafcavity}), and four cryounits form a ``cryomodule." Cryomodules containing five-cell and seven-cell SRF cavities are referred to as ``C20" and ``C100" cryomodules, respectively \cite{PhysRevAccelBeams.27.084802}. Twenty C20\hyperref[myfootnote]{\footnotemark[\value{footnote}]} cryomodules, along with five C100 cryomodules, make up one CEBAF linac, capable of producing an energy gain of 1.1 GeV per pass. The cryomodules are immersed in a liquid helium bath circulated by the Central Helium Liquefier (CHL) plant via two parallel cooling loops to maintain the 2 K operational temperature. \tab \ref{tab:ch3:cryomodule} provides a summary of the specifications for both types of cryomodules.
\renewcommand{\thefootnote}{\arabic{footnote}}

\subsection{Delivery to Halls} The CEBAF infrastructure for beam transportation consists of six main subsystems: the injector, linacs, spreaders, recombiners, recirculation arcs, and extraction regions. The purpose and function of the injector and linacs have already been discussed. In this section, the remaining subsystems will be the focus.

For the racetrack design to work efficiently, the transport of the electron beam from one linac to the next through the recirculation arcs, while preserving its properties, is crucial. After exiting a linac, the beam is spread by a spreader via differential vertical bending according to its energy and directed to the appropriate energy-tuned recirculation arc. At the other end of the arc, a recombiner, which is optically similar and essentially a mirror image of the spreader, phase matches the individual beams before sending them into the next linac.

The ten recirculation arc beamlines—five at each end—are designed to avoid any phase space dilution caused by error sensitivity, synchrotron radiation excitation, or optical aberrations. The final step is to extract the correct beam bunch, with the desired energy, from the beam bunch train and direct it to the corresponding experimental hall. For Halls A, B, and C, the extraction regions are designed to achieve this by performing a ``two-beam" split using subharmonic RF separators. These separators, located at the end of the south linac in each recirculation arc, are phased to provide maximum deflection to the desired beam bunch, which is then extracted into the transport lines of the corresponding hall. The remaining out-of-phase beam bunches are deflected in the opposite direction with the correct amplitude and angle, ensuring their continued forward propagation into the recirculation arc.

\subsection{Hall A Beamline}
\begin{figure}[ht!]
    \centering
    \includegraphics[width=1\columnwidth]{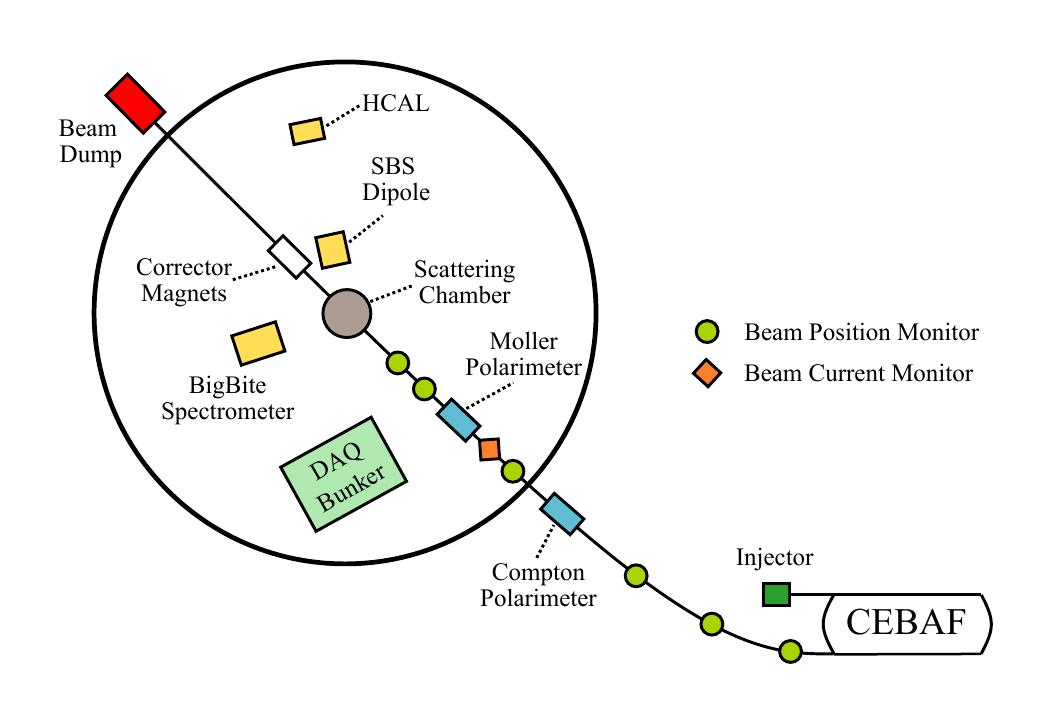}
    \caption[Schematic of the Hall A beamline]{\label{fig:ch3:hallabeamline} Schematic of the Hall A beamline (not to scale). Only key components are shown.}
\end{figure}

Hall A is a cylindrical concrete structure with a radius of approximately 27 meters and a domed top with a maximum height of 24 meters, located underground. The Hall A beamline runs through the center, mounted 3.05 meters above the floor \cite{osti_1844333}. It carries the extracted Hall A beam bunch from downstream of the extraction region into the hall, where it terminates at the beam dump located at the end of Hall A. The key components of the Hall A beamline relevant to the \gmn experiment include beam current monitors, beam position monitors, beam raster magnets, superharps, ion chambers, and downstream corrector magnets. \fig \ref{fig:ch3:hallabeamline} presents a schematic of the Hall A beamline, illustrating the approximate locations of some of these components.

\subsubsection{Beam Current Monitors}
The Hall A beamline is designed to handle beam intensities ranging from 1 to 120 $\mu$A. However, due to constraints such as the target's melting point, data acquisition (DAQ) livetime, trip limits of the gas electron multiplier (GEM) detectors, and the accelerator's maximum power limit, the maximum current during the \gmn experiment was restricted to a significantly lower value of 15 $\mu$A. The beam current was continuously measured and monitored throughout the experiment using beam current monitors (BCMs), an essential part of the beam diagnostic system. The setup includes an electromagnetically shielded and thermally insulated parametric current transformer (PCT) known as the ``Unser," along with two stainless-steel resonant radio-frequency (RF) cavities, one located upstream (US) and the other downstream (DS) of the Unser monitor.

The Unser monitor produces a signal with amplitude proportional to the beam intensity as the electron beam passes through the transformer coils inside it. The proportionality constant, determined via calibration with a beam of known current, allows the calculation of the actual beam current from the Unser signal amplitude. Although the Unser is highly linear, its precision degrades over time due to signal drift. Therefore, it is used primarily for periodic calibration of the US and DS BCMs but not for continuous monitoring throughout the experiment.

The US and DS RF cavities are tuned to the fundamental CEBAF linac frequency of 1497 MHz. The signal from each cavity is proportional to the beam intensity and is stable over long durations. These signals are downconverted from 1497 MHz to 1 MHz for easier processing and then fed into RMS-to-DC converters, which provide a DC signal proportional to the root-mean-square (RMS) beam current. The DC output from these converters is split into multiple channels (two for the US BCM and four for the DS BCM), each with a different amplification level\footnote{Different amplifications allow accurate measurements across a wide range of beam currents. For instance, at low beam currents, higher amplification provides more reliable readings, while at high beam currents, lower amplification is preferable.}. These amplified signals are then read by voltmeters to provide a direct measurement of the beam current.


\subsubsection{Beam Position Monitors}
The Hall A beam position monitoring (BPM) system is a crucial part of the beam diagnostic infrastructure, responsible for continuously tracking the position and trajectory of the electron beam as it moves through the beamline. BPMs are strategically placed along the beamline, particularly after beam-steering magnets and other focusing elements, to provide precise, real-time feedback on the beam’s position in both the horizontal (X) and vertical (Y) directions. This ensures that the beam remains precisely aligned with the target, helping to avoid various failure modes associated with beam excursion.

\begin{figure}[ht!]
    \centering
    \includegraphics[width=0.6\columnwidth]{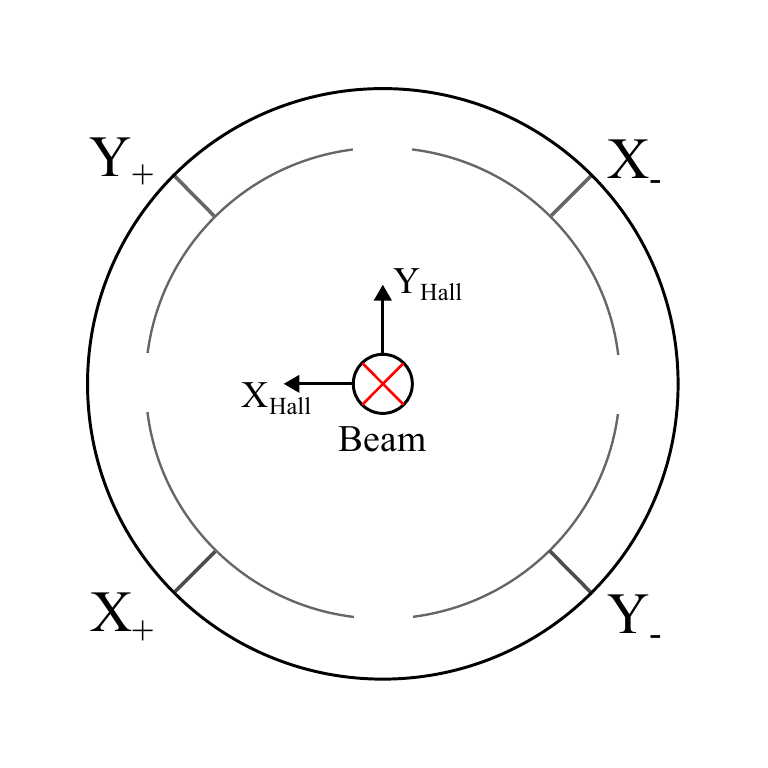}
    \caption{\label{fig:ch3:bpm} Transverse cross-section of a beam position monitor (BPM), reproduced from \cite{osti_1821604}.}
\end{figure}
%
Each BPM consists of four wire antennas arranged symmetrically at 45$^\circ$ angles relative to the hall coordinate system, as shown in \fig \ref{fig:ch3:bpm}. As the beam passes through the BPM, it generates signal in the antennas, the strength of which is proportional to the distance between the beam and the antenna. The beam position coordinates $(x_{rot},y_{rot})$ in the ``rotated" coordinate system are given by
\begin{equation}
    x_{rot} = C\,\frac{x_+ - x_-}{x_+ + x_-},\qquad y_{rot} = C\,\frac{y_+ - y_-}{y_+ + y_-},
\end{equation}
where $C=18.76$ mm is a calibration constant \cite{Silwal2012} and $x_+$, $x_-$, $y_+$, and $y_-$ are integrated ADC signals measured by the corresponding antennas. The beam position coordinates $(x,y)$ in the hall coordinate system are then calculated using a straightforward coordinate transformation of the following form:
\begin{equation}
    \begin{pmatrix}
        x\\
        y
    \end{pmatrix} 
    = 
    \begin{pmatrix}
        \cos45^\circ & -\sin45^\circ\\
        \sin45^\circ & \cos45^\circ
    \end{pmatrix}
    \begin{pmatrix}
        x_{rot}\\
        y_{rot}
    \end{pmatrix}    
\end{equation}
The Hall A BPMs operate in auto-gain mode, where the system adjusts the gain of each antenna based on beam intensity to maintain a constant signal integral. 
\subsubsection{Beam Energy Measurements}
Precise knowledge of the beam energy is essential for reconstructing the 4-momentum of the incident electron beam, making it a critical component of physics analysis. CEBAF is designed to provide a beam with highly stable energy, achieving an energy spread on the order of $10^{-4}$ GeV or less across all five energy passes. Real-time monitoring of the beam energy is conducted by multiple BPMs positioned at the beginning, middle, and end of the Hall A arc, where the extracted CEBAF beam is bent into Hall A by eight dipole magnets. These BPMs operate in fixed-gain mode, unlike the standard Hall A BPMs, allowing their measured signals to be used in determining the deviation of the beam energy from the nominal value. This measurement, however, is relative and requires multiplication by a calibration constant—determined through absolute energy measurement—to obtain the ``true" beam energy. Following the CEBAF 12 GeV upgrade, this calibration constant has been determined to be 1.003 \cite{Santiesteban:2021lot}.

Throughout the \gmn experiment, real-time beam energy readings were recorded in the data stream via the Experimental Physics and Industrial Control System (EPICS), with a frequency of one entry every 5–8 seconds. During analysis, the beam energy per run was determined by calculating the arithmetic mean of all entries recorded for that run and then multiplying by the calibration constant. The average beam energy for each kinematic setting, as listed in \tab \ref{tab:sbsconfig}, was computed by taking the mean of the beam energies across all runs for that kinematic setting.

\subsubsection{Beam Raster}
Upon entering the hall, the electron beam is rastered to distribute the heat load on the target and ensure uniform irradiation. This process is especially important when using thin or cryogenic targets, which are vulnerable to damage from localized heating caused by the intense beam. Rastering is achieved using a set of dipole magnets that deflect the beam at a rate of 25 kHz, following a pattern chosen by the user. During the \gmn experiment, a square raster with dimensions of 2 mm × 2 mm was used.

\begin{figure}[ht!]
    \centering
    \includegraphics[width=0.7\columnwidth]{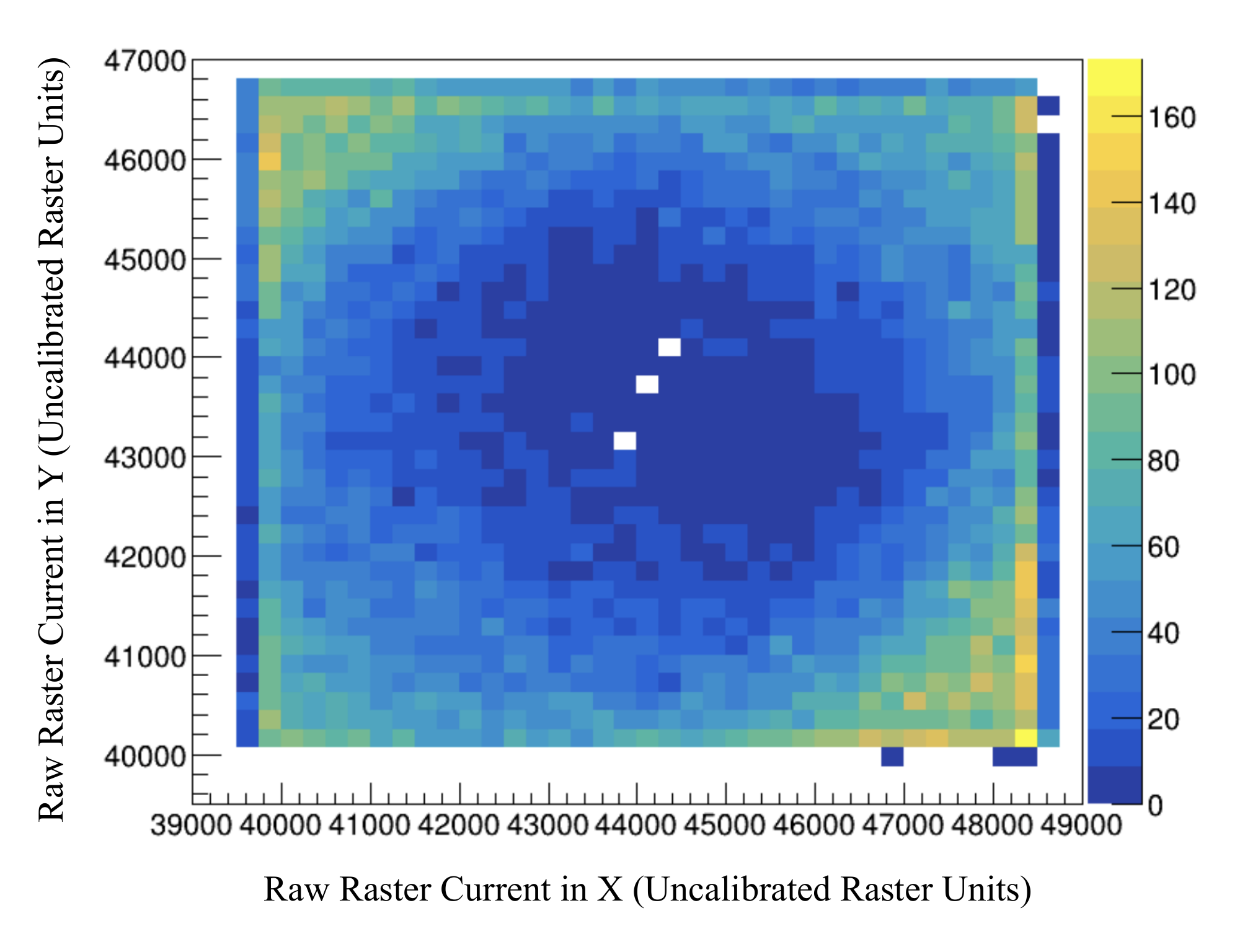}
    \caption{\label{fig:ch3:raster} Raster map on a carbon foil target with a 2 mm hole at its center, using data from run 13181 taken during \qeq{7.4} kinematics.}
\end{figure}
When establishing the beam in the hall—either at the beginning of each kinematic setting or after a long downtime—the raster size was verified by directing the rastered beam onto a thin carbon foil target with a 2 mm hole in the center. A well-centered, visible hole in the resulting raster map, as shown in \fig \ref{fig:ch3:raster}, confirmed proper alignment. Additionally, the intrinsic beam spot size before rastering was optimized by performing a ``harp scan," which involved passing an array of conducting wires through the low-intensity ($<5\mu$A) and low-duty-cycle unrastered beam.


%

%
 
\section{Experimental Targets}
\label{sec:ch3:exptarget}
\begin{figure}[h!]
     \centering
     \begin{subfigure}[b]{0.51\textwidth}
         \centering
         \includegraphics[width=\textwidth]{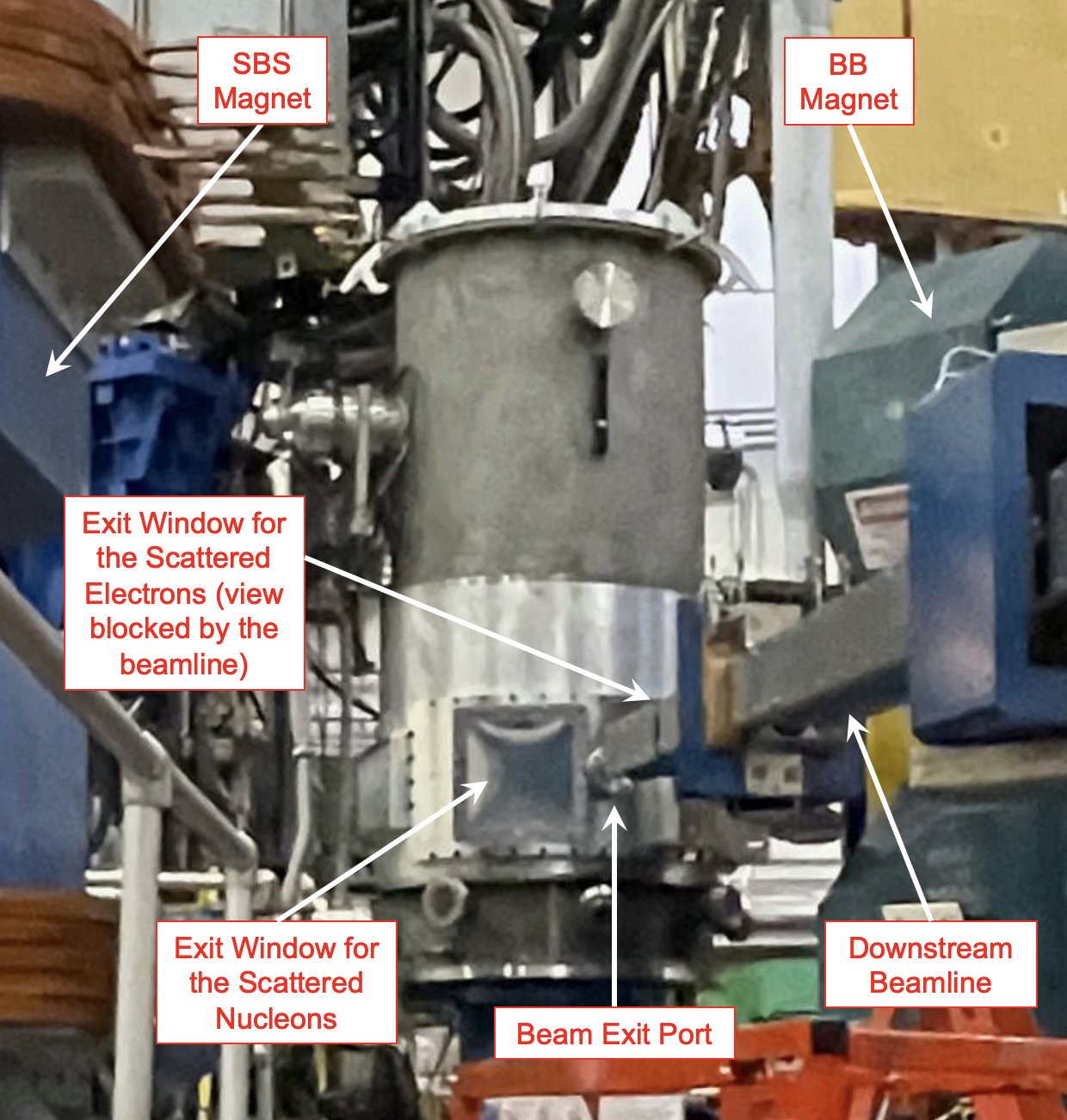}
         \caption{}
         \label{sfig:tgtscatcham}
     \end{subfigure}
     \hfill
     \begin{subfigure}[b]{0.4\textwidth}
         \centering
         \includegraphics[width=\textwidth]{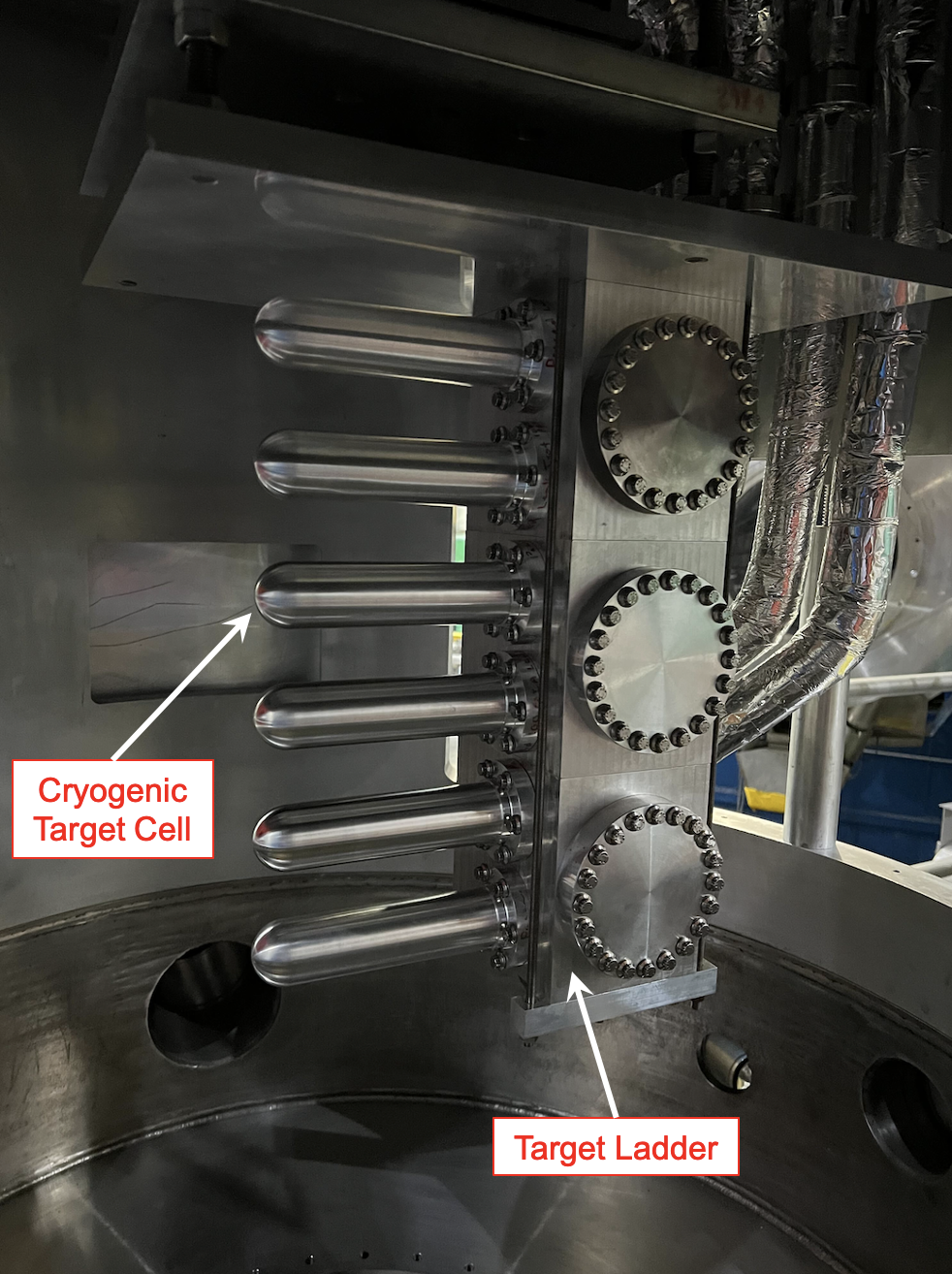}
         \caption{}
         \label{sfig:tgtcell}
     \end{subfigure}
     \caption{Components of the experimental target system used in the \gmn experiment. (a) The outside view of the scattering chamber when looking upstream. (b) The inside view of the scattering chamber showing the target ladder.}
     \label{fig:tgt}
\end{figure}

The \gmn experimental targets used for production and calibration were housed in the scattering chamber, as shown in \fig \ref{fig:tgt}. These targets were primarily categorized as cryogenic or solid. Positioned on a ladder for precise vertical motion, they intercepted the electron beam path. Upon interaction, some electrons scattered off the target while others continued to the beam dump. 

Scattered particles exited through thin windows on either side of the chamber and were eventually detected by the BigBite and Super BigBite spectrometers. The scattering chamber, an aluminum cylinder with \SI{2}{inch} thick walls and a \SI{41}{inch} internal diameter, included exit windows, beam entrance and exit ports, electrical feedthroughs, vacuum pump ports, and cryogenic target plumbing. Maintained in a high vacuum ($<$ \SI{E-6}{Torr}), it was connected directly to the evacuated accelerator beamline to minimize secondary interactions before and immediately after scattering. During data collection at \qeq{13.6}, a \SI{0.125}{in} thick \ce{Al} shield\footnote{The first data run with the \ce{Al} shield in place was run $12675$. The \ce{Al} plate replaced a \SI{10}{mm} thick polyethylene bar, which was briefly installed for runs $12556$ to $12674$. A small but significant portion of data collected at the beginning of this \q had no extra shielding.} was installed in front of the exit window for scattered electrons to reduce background in the front Gas Electron Multiplier (GEM) detectors, which remained in place for the remainder of the experiment. 

\subsection{Cryogenic Targets}
\label{ssec:ch3:cryotarget}
The cryogenic target, or cryotarget, system used in the \gmn experiment comprised liquid deuterium (\ld) and liquid hydrogen (\lh) loops. One of the major challenges associated with the neutron form factor measurement is the absence of a stable free neutron target. However, the deuteron, being a loosely bound system of a proton and a neutron with a binding energy of approximately \SI{2.2}{MeV}, serves as an excellent substitute for a free neutron target. Consequently, \ld was used as the production target for the experiment, while data taken on \lh were utilized for detector calibrations.  
\begin{table}[h!]
\caption{\label{tab:tgtcelldim} Density and cell dimensions of the cryotargets used in \gmn \cite{Meekins2022}. Note: Target densities have been computed using the National Institute of Standards and Technology (NIST) Chemistry WebBook, based on their temperature and pressure during the experiment (see Figure \ref{fig:tgtgui}). The cell diameters have been obtained from Dave Meekins through private communication.}
\centering
\begin{tabular}{cccccc}
\hline\hline
Target & Density    & Diameter & Entrance window & Exit window    & Wall  \\ 
loop & (\SI{}{g/ml})& (mm)     & thickness (mm)  & thickness (mm) & thickness (mm) \\ \hline 
\SI{15}{cm} \ld & $0.16694$ & $40.64$ & $0.119\pm0.003$ & $0.155\pm0.008$ & $0.137\pm0.015$\\ 
\SI{15}{cm} \lh & $0.07248$ & $40.64$ & $0.132\pm0.004$ & $0.152\pm0.009$ & $0.136\pm0.009$\\ 
\hline\hline
\end{tabular}
\end{table}

The cryotarget cells are aluminum (Al 7075) cans with a shape of a ``cigar-tube", as depicted in Figure \ref{sfig:tgtcell}. Their dimensions are summarized in Table \ref{tab:tgtcelldim}. These cells are connected to their corresponding recirculation loops for the continuous and stable flow of the target material. The \ld flowing through the target cell is maintained at \SI{22}{K} and \SI{24}{psiA}, while the \lh is maintained at an operational temperature and pressure of \SI{19}{K} and \SI{25}{psiA}, respectively. Each recirculation loop comprises a heat exchanger for cooling the target material to its operational temperature and a high-speed pumping fan, operating at \SI{60}{Hz}, to propel the liquid flow through the loop. The cooling at the heat exchanger is facilitated by a continuous flow of helium coolant supplied by an End Station Refrigerator (ESR) at approximately \SI{14}{K} and \SI{12}{atm}, returning at around \SI{20}{K} and \SI{3}{atm}.

\begin{figure}[h!]
    \centering
    \includegraphics[width=0.85\columnwidth]{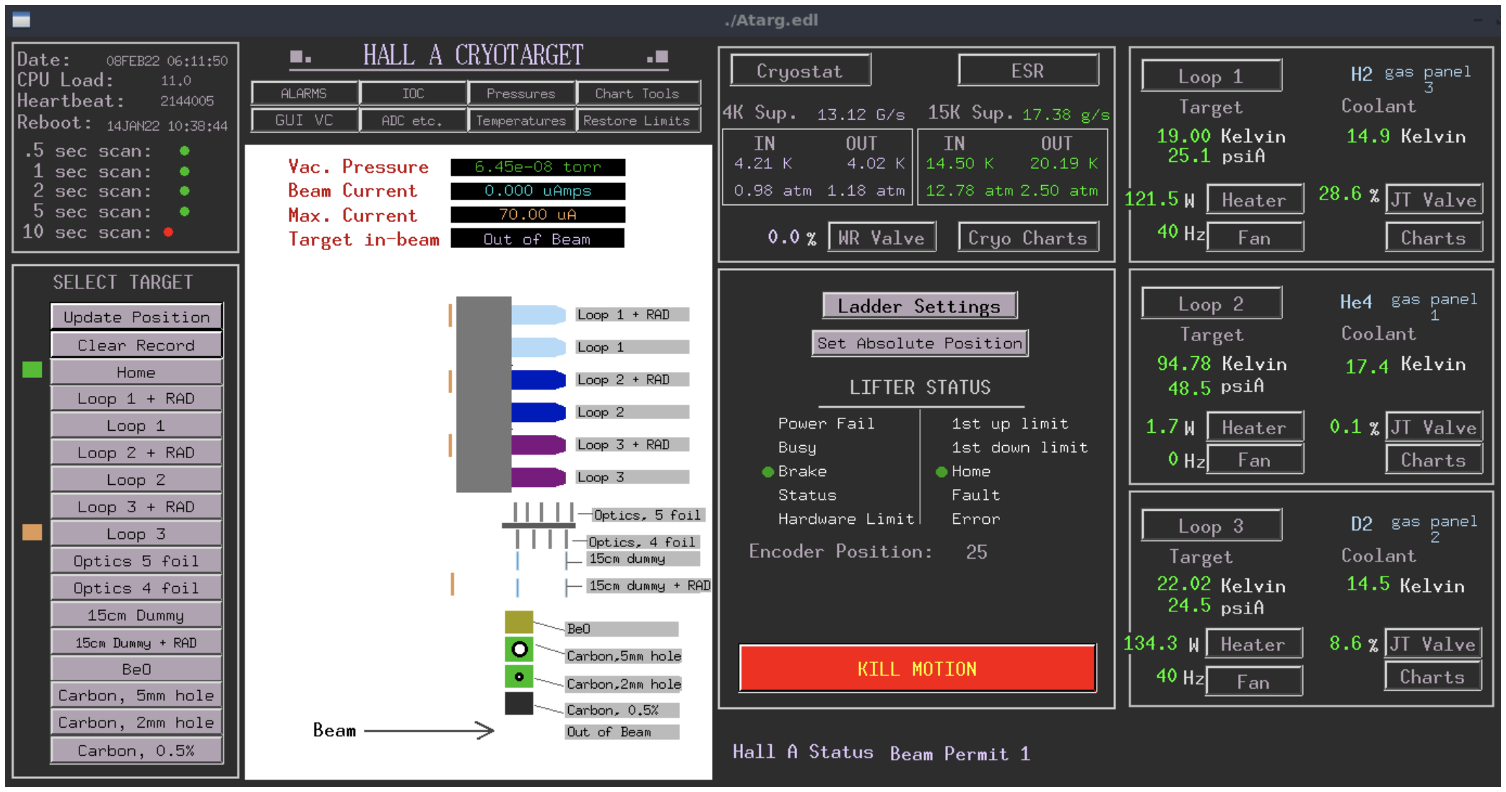}
    \caption{\label{fig:tgtgui} A screenshot of the target operator GUI captured during \gmn on February 8, 2022, displaying a detailed view of the target ladder along with the temperature and pressure readings of the cryotargets.}
\end{figure}
The incident electron beam deposits a significant amount of heat in the target, ranging from $5-80$ \SI{}{J/s} during the \gmn experiment, proportional to the beam current. The flow of helium coolant to the heat exchanger can be regulated by a Joule-Thompson (JT) valve, ensuring that the heat generated by the beam is compensated for, thereby preventing an increase in the target temperature within the cell, a phenomenon known as target boiling. For safety, it is customary to adjust the JT valve such that there is a buffer of at least 75 J/s beyond the anticipated beam heating. The heat needed to compensate for this extra cooling is provided by a variable high-power heater ($P_{max}=$ \SI{1}{kW}) and a variable low-power heater installed at the entrance and exit of the heat exchanger, respectively. These heaters operate on PID feedback circuits to maintain the operational temperature of the target within $\pm$ \SI{0.01}{K} by continuously receiving inputs from three strategically placed thermometers: at the entrance and exit of the target cell, as well as at the entrance of the heat exchanger.

\subsection{Solid Targets}
\label{ssec:ch3:solidtarget}
In addition to the cryotargets, several solid targets were available during the experiment, as depicted in Figure \ref{fig:tgtgui}. These encompassed the dummy target employed for background estimation resulting from scattering from the cryotarget shell, single and multi-foil optics targets utilized in BB optics program, and carbon hole targets utilized for aligning the beam spot at the target. A summary detailing the thicknesses and positions of these solid targets can be found in Table \ref{tab:tgtsolid}.
\begin{table}[h!]
\caption{\label{tab:tgtsolid} Thickness and position of the solid targets used in \gmn \cite{Meekins2022}. Note: Both dummy foils are at the same (vertical) position on the target ladder.}
\centering
\begin{tabular}{cccc}
\hline\hline
Solid target & Material    & Thickness (\SI{}{g/cm^2}) & Position in $z$ (\SI{}{cm}) \\ \hline 
Carbon, 0.5\% (1 foil) & $99.95\%$ Carbon & $0.044\pm0.001$ & 0\\ 
Optics 4 foil & Carbon           & $0.044\pm0.001$ & $\pm2.5$, $\pm7.5$ \\
Optics 5 foil & Carbon           & $0.044\pm0.001$ & 0, $\pm5$, $\pm10$ \\
\SI{15}{cm} Dummy (Upstream) & Al 7075 & $0.350\pm0.0003$ & $-7.5$ \\
\SI{15}{cm} Dummy (Downstream) & Al 7075 & $0.349\pm0.0003$ & $+7.5$ \\
\hline\hline
\end{tabular}
\end{table}
\section{The BigBite Spectrometer}
\begin{figure}[ht!]
    \centering
    \includegraphics[width=0.9\columnwidth]{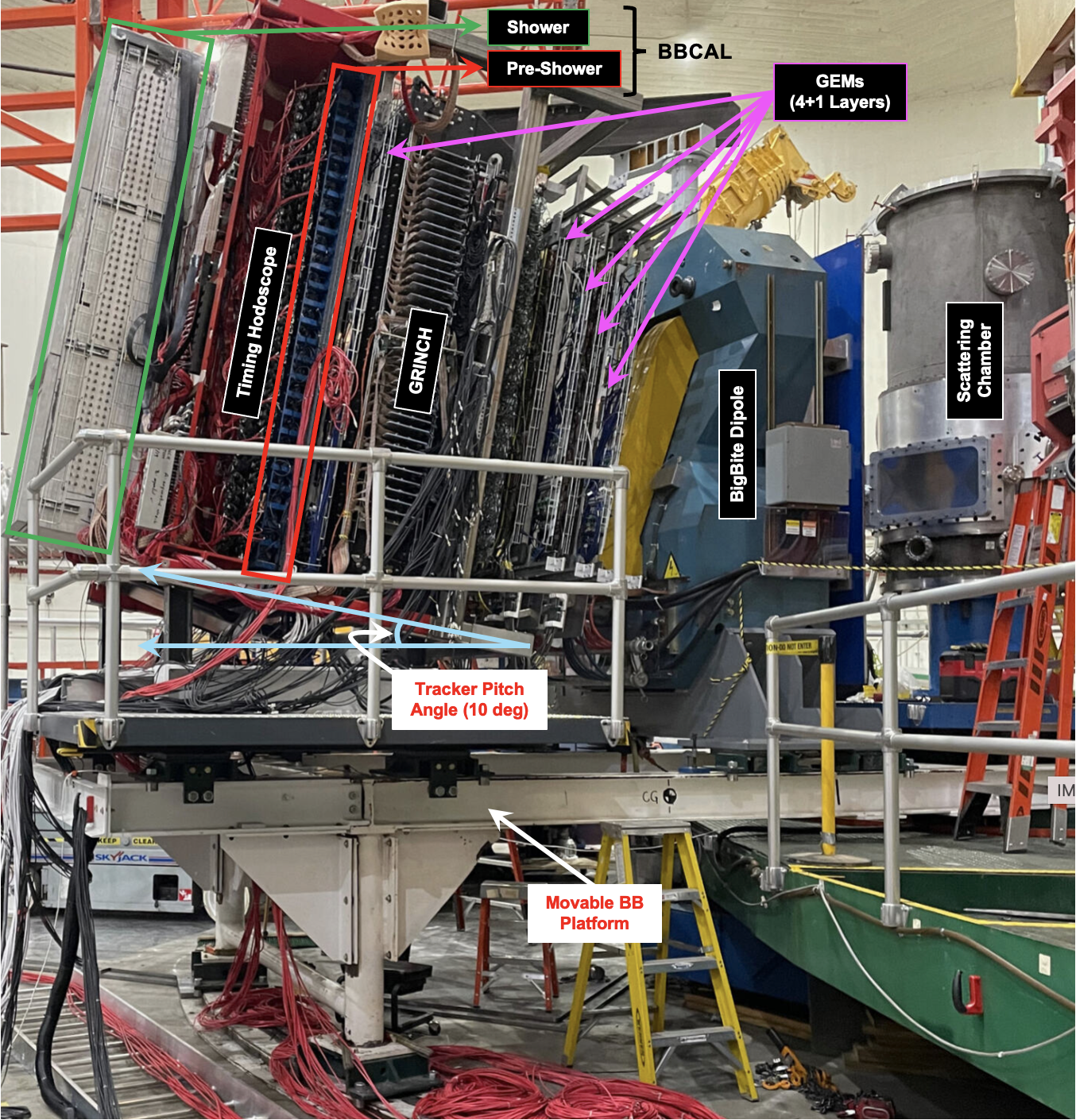}
    \caption{\label{fig:bbinhall} The BigBite Spectrometer in Hall A.}
\end{figure}

The BigBite Spectrometer, or BB, is a non-focusing spectrometer located on the left side of the beamline when looking downstream from the scattering chamber. It serves as the electron arm of the {\gmn} experiment, capable of detecting high-energy ($1-4$ \SI{}{GeV}) scattered electrons with an angular resolution of $1-2$ \SI{}{mrad} and a relative momentum resolution of $\approx 1-1.5\%$. The BB spectrometer features a large-aperture dipole magnet followed by the BigBite detector package, which comprises many sub-detectors as depicted in Figure \ref{fig:bbinhall}. This section will discuss the design and operation of the various sub-systems of the BB spectrometer.
\subsection{BB Dipole Magnet}
The BB dipole magnet, also known as the BB magnet, is a large-aperture, non-focusing, water-cooled, H-shaped dipole magnet crucial for reconstructing the momentum of high-energy ($1-4$ \SI{}{GeV}) scattered electrons entering the BB spectrometer. Positioned at the entrance of the BB spectrometer, it is followed by the BB detector package, situated on a platform inclined upward by $10^{\circ}$ relative to the ground to maximize acceptance, as depicted in Figure \ref{fig:bbinhall}. 

\begin{figure}[ht!]
     \centering
     \begin{subfigure}[b]{0.45\textwidth}
         \centering
         \includegraphics[width=\textwidth]{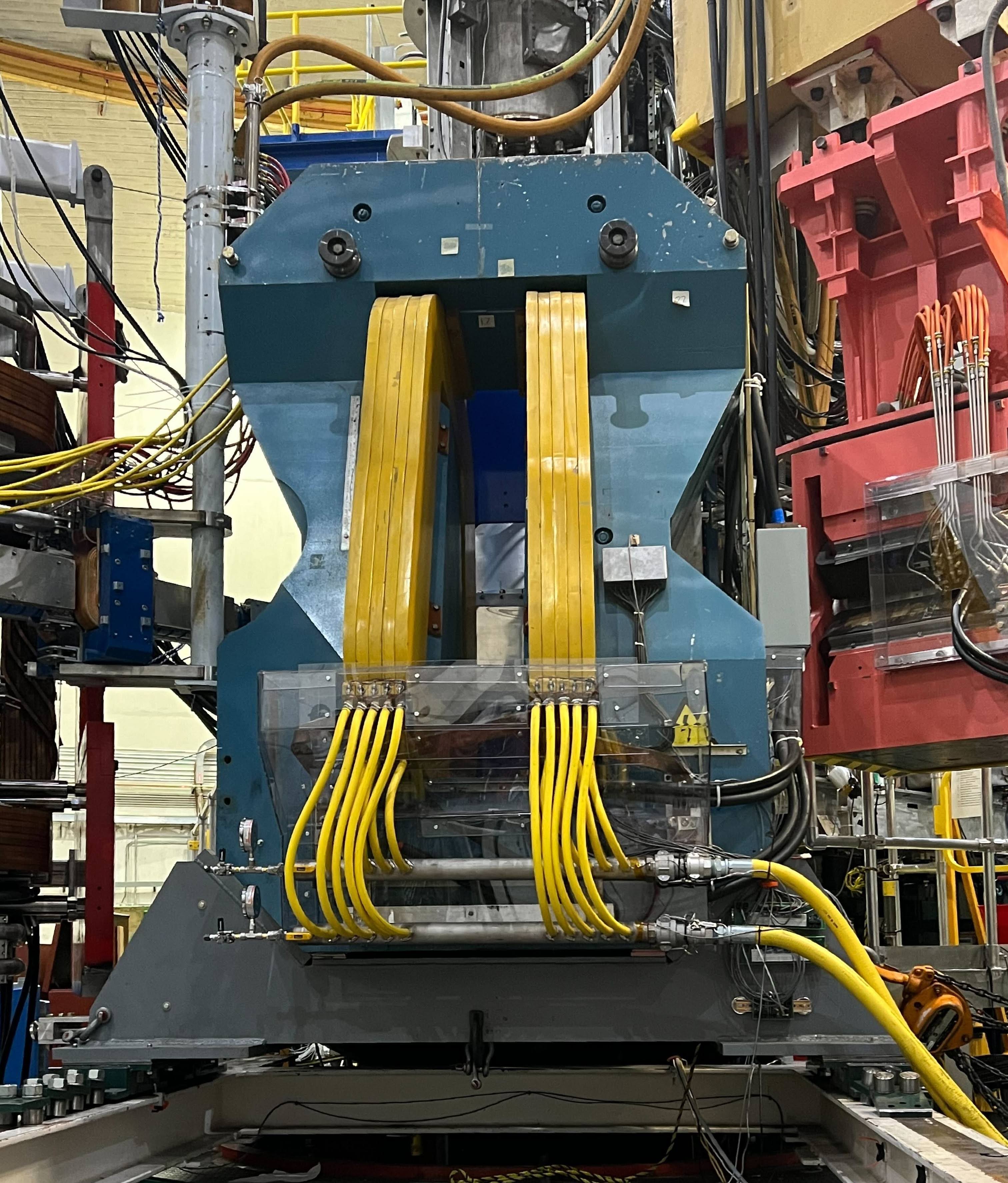}
         \caption{}
     \end{subfigure}
     \hfill
     \begin{subfigure}[b]{0.4\textwidth}
         \centering
         \includegraphics[width=\textwidth]{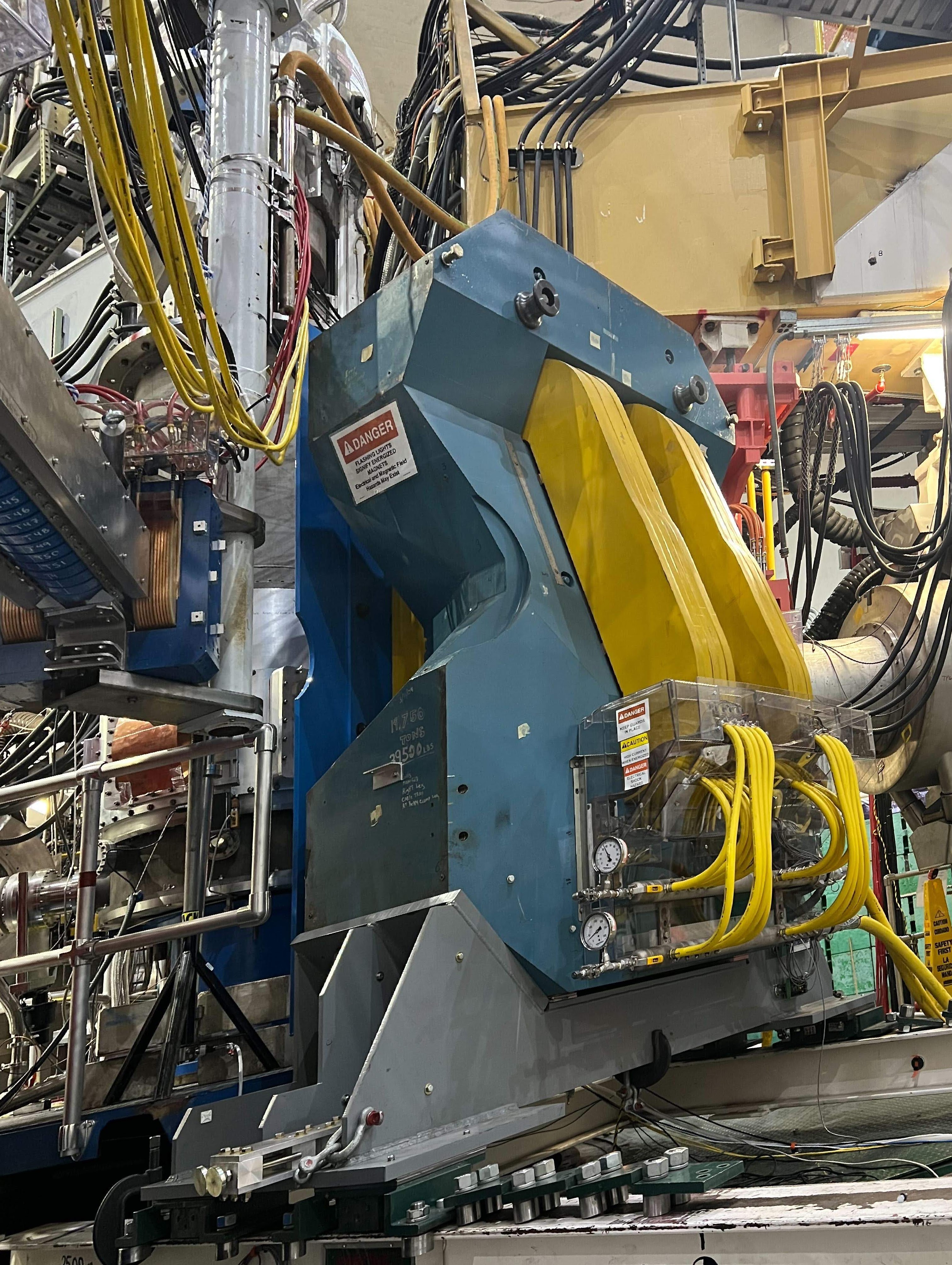}
         \caption{}
     \end{subfigure}
     \caption{The BigBite dipole magnet in Hall A. (a) Back view. (b) Side view.}
     \label{fig:ch3:bbmag}
\end{figure}
The BB magnet was designed and built at the Internal Target Facility of the AmPS ring at NIKHEF in collaboration with the Budker Institute for Nuclear Physics in Novosibirsk \cite{DELANGE1998182}. Subsequently, it was acquired by Jefferson Lab and first used in the E01-015 experiment in Hall A \cite{HALLASTNDRDEQUIP}. Its entrance face is perpendicular to the central trajectory, while the exit face has a pole face rotation of $20^{\circ}$ (see \fig \ref{fig:ch3:bbmag}), adjusting the field integral for particles entering the upper or lower region \cite{MIHOVILOVIC201220}. This ensures a uniform dispersion across the spectrometer's acceptance. Weighing approximately \SI{20}{tons}, the BB magnet is positioned on a movable platform in the Hall to facilitate adjustments of spectrometer angles and the target-to-magnet distance to desired values, as listed in Table \ref{tab:sbsconfig}.

The BB magnet has an opening of $95\times25$ \SI{}{cm^2} with a central yoke length of \SI{71}{cm}. The dipole field in the gap is in the horizontal direction perpendicular to the motion of the scattered particle. Throughout the experiment, the magnet was operated at its maximum operational current of \SI{750}{A} with negative polarity, causing electron tracks to bend upwards. The resulting field integral of approximately \SI{1}{Tm} was sufficient to provide an average deflection of $4.6^{\circ}$ to \SI{3.6}{GeV} electron tracks ensuring high resolution momentum reconstruction across all \gmn kinematics.
%
\subsection{Gas Electron Multiplier}
\label{ssec:ch3:bbgem}
The addition of Gas Electron Multiplier (GEM) based trackers represents the most significant upgrade to the BigBite Spectrometer compared to its older counterpart. The following features of the GEMs make them indispensable for the SBS high-{\q} EMFF measurement program including the {\gmn} experiment:
\begin{itemize}
    \item Capability to handle very high background rates of the order of hundreds of \SI{}{kHz/cm^2}
    \item Excellent spatial resolution of approximately $70$ $\mu$m for a single hit
    \item Large acceptance, covering up to $60\times200$ \SI{}{cm^2} active area
\end{itemize}
There are $5$ GEM layers located in the BigBite spectrometer. The first four layers, also known as the front trackers, are stacked one after another in parallel downstream of the BigBite magnet. The fifth layer, or the back tracker, sits further downstream sandwiched between the Pre-Shower calorimeter and the GRINCH. Combining precise measurements of particle hit coordinates from each GEM layer with the BigBite optics model allows for the reconstruction of scattered electron tracks. The process of BB track reconstruction is involved and will be discussed in \sect \ref{ssec:trackreconst}. In this section we will delve into the design and operation of the SBS GEMs.   
\subsubsection{Design and Operation}
\label{sssec:gemdesop}
\begin{figure}[ht!]
    \centering
    \includegraphics[width=0.55\columnwidth]{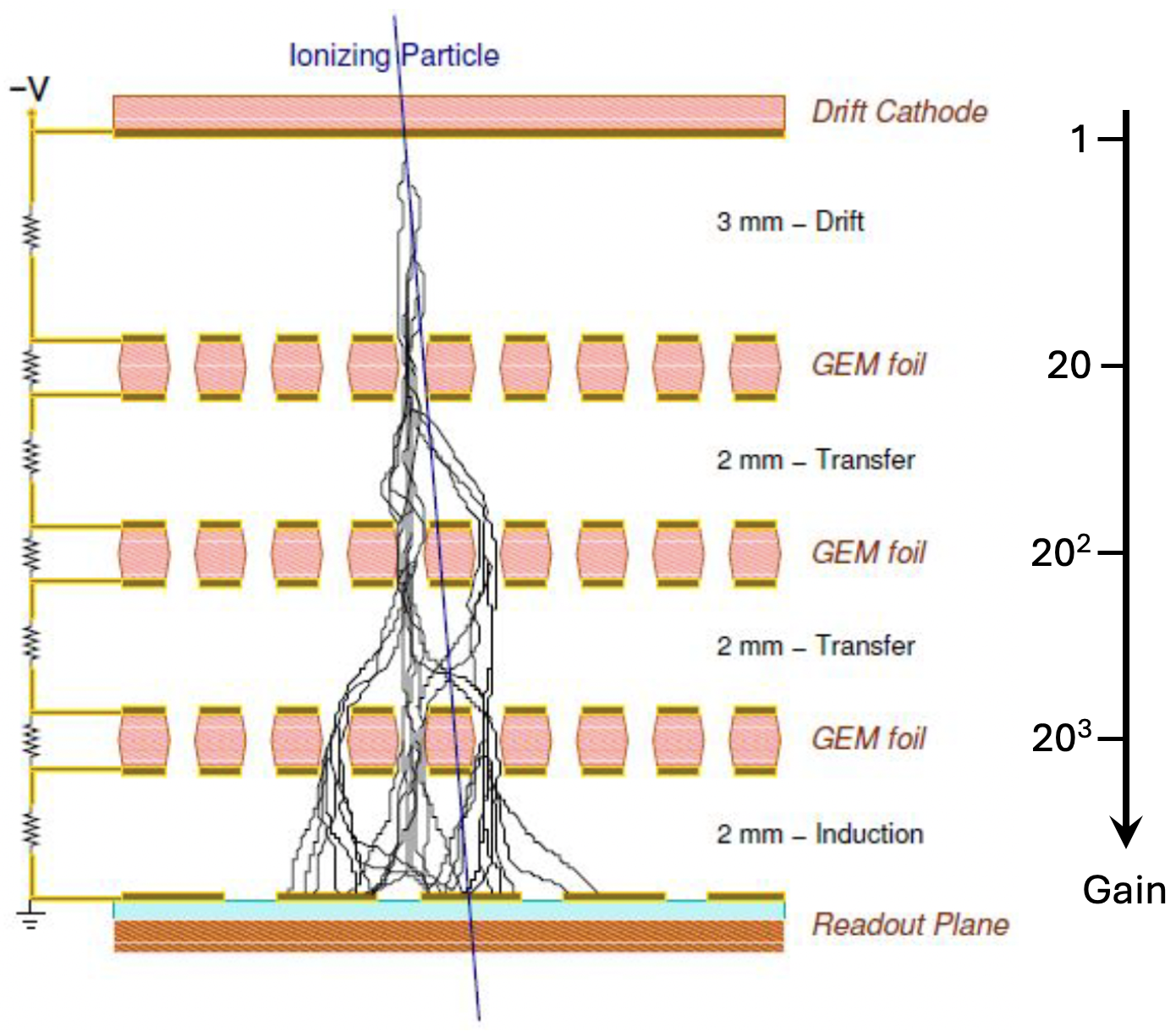}
    \caption{\label{fig:gem3foil} Schematic of a ``triple-foil" GEM detector.}
\end{figure}

The design of the SBS GEMs is derived from those used in the COMPASS experiment, which were the first to be employed under high-rate conditions. A GEM detector typically consists of a drift cathode foil, one or more GEM foils, and an electronics readout board (ROB). The effective gain of a GEM detector can be optimized for a given high voltage (HV) by adjusting the number of GEM foils. The SBS GEM detectors employ a ``triple-foil" design, also known as the ``triple-GEM" design, incorporating three GEM foils between the drift cathode and the ROB. Figure \ref{fig:gem3foil} illustrates the working principle of a ``triple-foil" GEM detector. 

\subheading{Gas Mixture}
The GEM detectors are filled with gas that ionizes upon interaction with high-energy electrons, creating an avalanche and generating recognizable signals recorded by the ROB. Choosing the gas for the GEM detector is crucial, considering factors such as cost-effectiveness, chemical inertness, and low ionization energy. While noble gas like \ce{Ar} fits these criteria well, using pure \ce{Ar} has drawbacks, including energy loss via photon emission instead of secondary electron production upon excitation. To suppress such spurious photon-induced effects, \ce{Ar} is mixed with \ce{CO2}, a polyatomic gas, which absorbs photon energy through rotational and vibrational modes. Following gain and detection efficiency optimization, a gas mixture of $75\%$ \ce{Ar} and $25\%$ \ce{CO2} was chosen for the SBS GEMs, supplied to all five GEM detectors through a gas distribution system.

\subheading{Drift Cathode and GEM Foil}
The drift cathode, a thin copper foil, is maintained at the highest potential relative to the ROB, which is grounded as the anode. The \SI{3}{mm} thick space between the drift cathode and the first GEM foil constitutes the initial ionization region, known as the ``drift region". A GEM foil consists of a $50$-micron-thick layer of polyimide with a $5$-micron-thick copper coating on both sides. Polyimide, a dielectric material with outstanding thermal stability, mechanical resilience, and chemical resistance, insulates the two metal sides while serving simultaneously as an anode on one side and cathode on the other when potential is applied across them. 

\begin{figure}[ht!]
     \centering
     \begin{subfigure}[b]{0.5\textwidth}
         \centering
         \includegraphics[width=\textwidth]{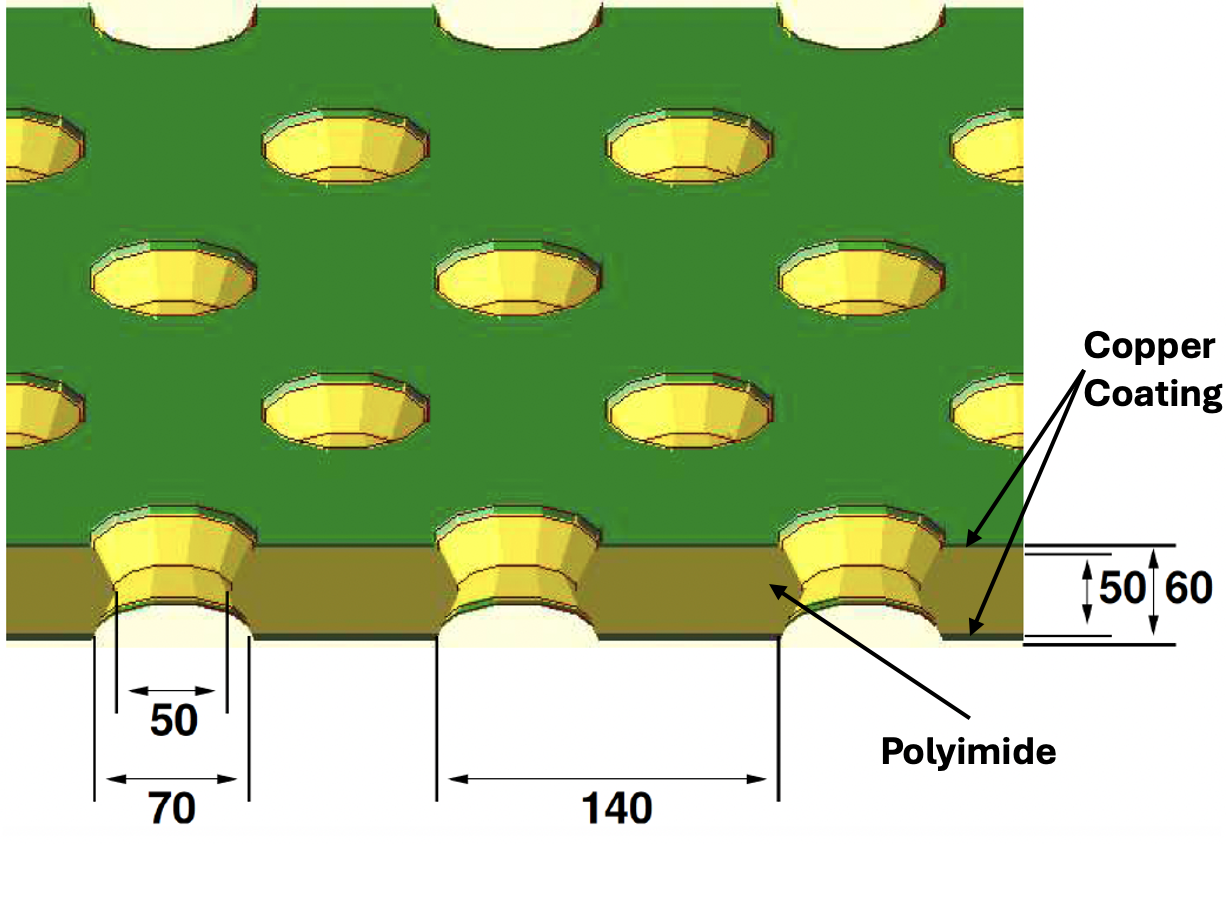}
         \caption{}
         \label{sfig:gemfoildim}
     \end{subfigure}
     \hfill
     \begin{subfigure}[b]{0.4\textwidth}
         \centering
         \includegraphics[width=\textwidth]{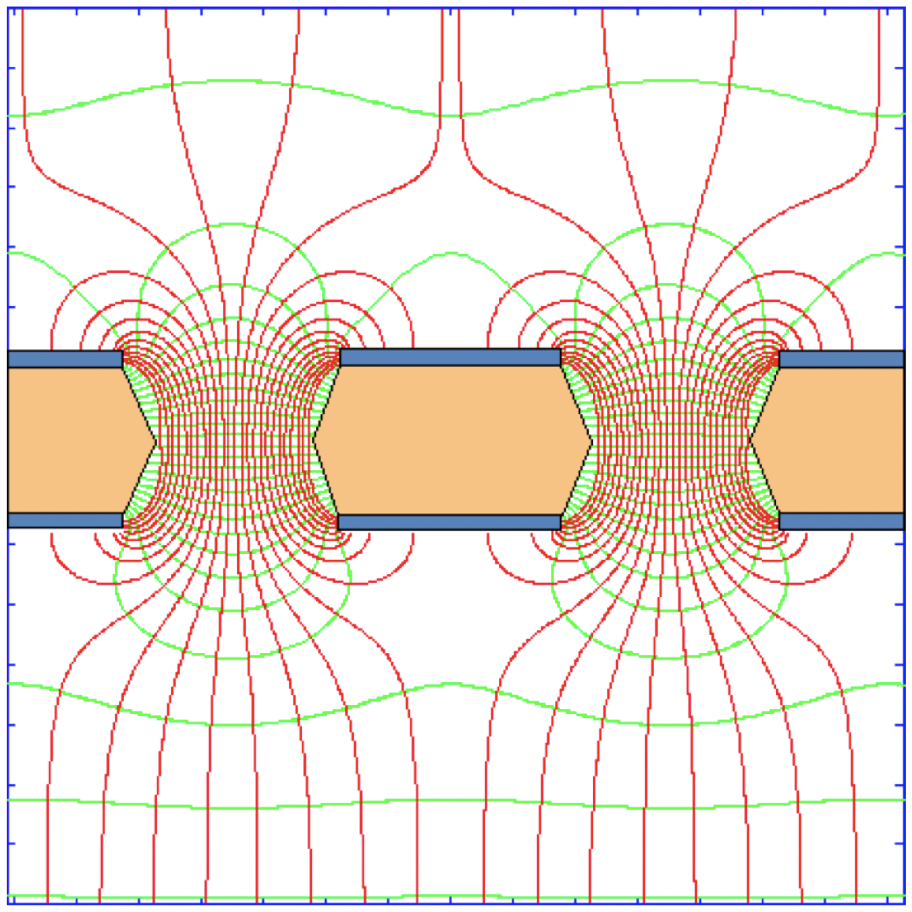}
         \caption{}
         \label{sfig:gemfoilfield}
     \end{subfigure}
     \caption{Design of a GEM foil. (a) Various design parameters of a GEM foil \cite{thesisLH}. (b) Electric field lines within GEM holes.}
\end{figure}

Charged particles traverse the GEM foil through hourglass-shaped pores of very high density, featuring an inner diameter of $50$ microns and an outer diameter of $70$ microns with a pitch of \SI{140}{microns} as depicted in Figure \ref{sfig:gemfoildim}. The application of a very high voltage (approximately \SI{400}{V}) across the GEM foil generates a substantial electric field within these small hole regions, accelerating electrons and ions and inducing ionization collisions with the gas atoms in the detector volume. This process leads to an electron avalanche. Each avalanche produced by a single GEM foil results in a $20$x gain, making the effective gain of a ``triple-foil" GEM detector $20^3$x. Figure \ref{sfig:gemfoilfield} illustrates a simulation of the electric field lines within a GEM hole. As mentioned above, a ``triple-foil" GEM detectors such as SBS GEMs have three GEM foils and the \SI{2}{mm} gap between two consecutive GEM foils is known as the "transfer region".

\begin{figure}[ht!]
     \centering
     \begin{subfigure}[b]{0.4\textwidth}
         \centering
         \includegraphics[width=\textwidth]{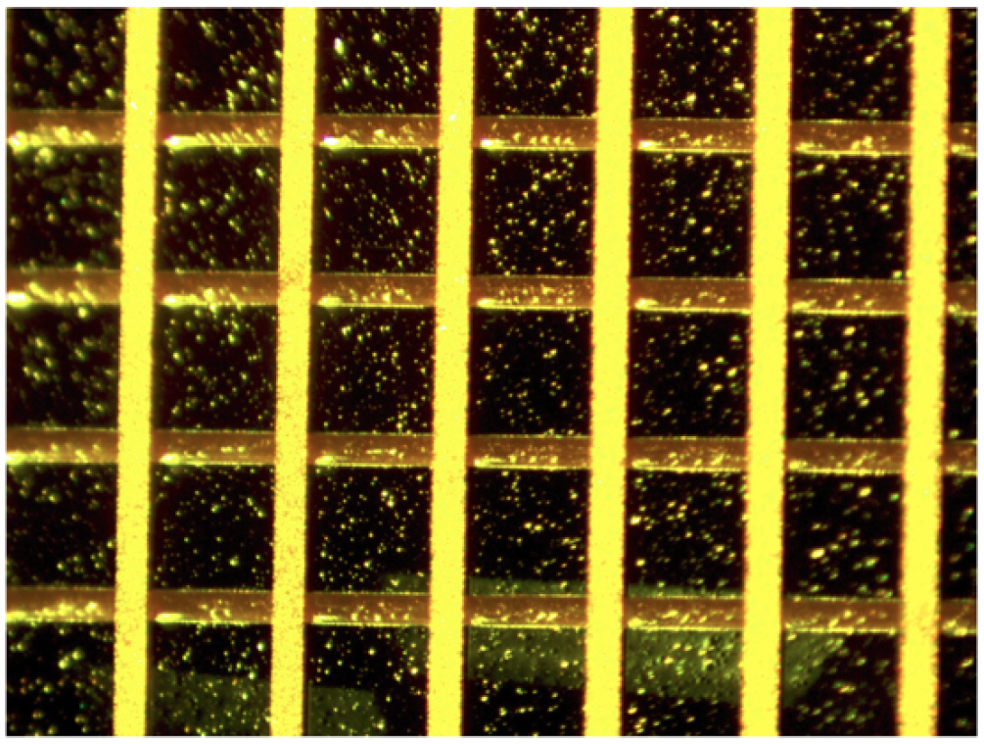}
         \caption{}
         \label{sfig:gemstripXY}
     \end{subfigure}
     \hfill
     \begin{subfigure}[b]{0.4\textwidth}
         \centering
         \includegraphics[width=\textwidth]{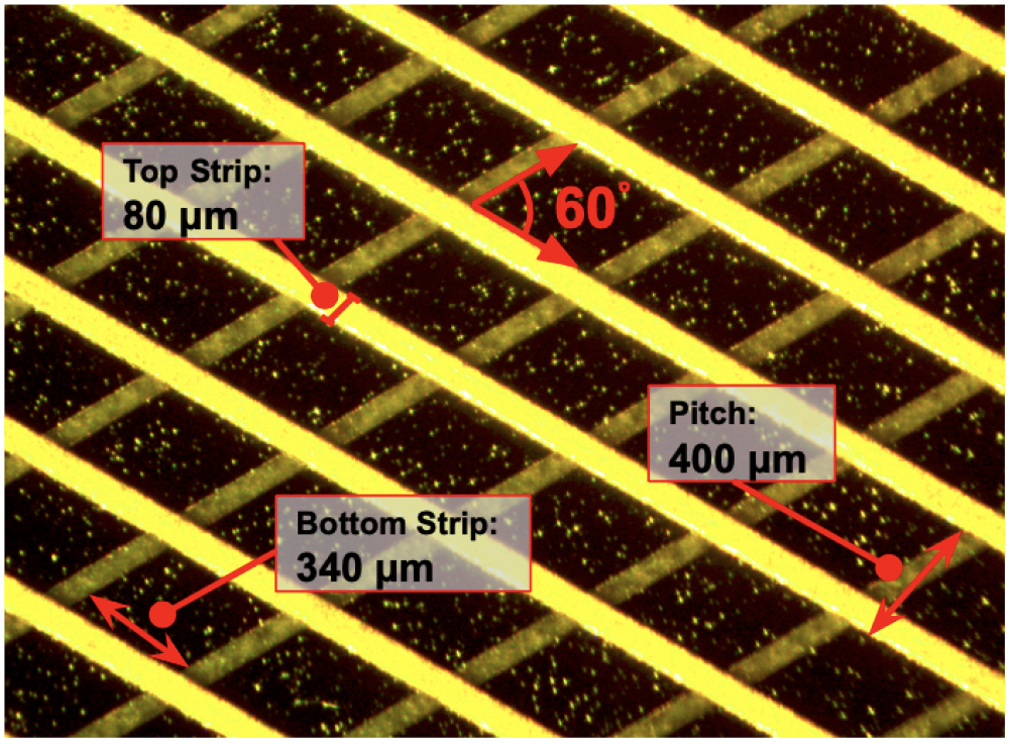}
         \caption{}
         \label{sfig:gemstripUV}
     \end{subfigure}
     \caption{Two types of GEM readout strip orientation used in SBS GEMs. (a) XY Cartesian: relative orientation of $90^{\circ}$. (b) Stereo angle UV: relative orientation of $60^{\circ}$.}
     \label{fig:gemstriporien}
\end{figure}

\subheading{Readout Board}
The space between the third GEM foil and the readout board (ROB) is \SI{3}{mm} thick and known as the ``induction region". The ROB comprises two sets of thin parallel copper strips separated by a thin layer of polyimide. The widths of the strips in the upper ($80$ microns) and lower ($340$ microns) layers are unequal to account for the discrepancy in their charge sharing induced by the gap between them. Combining the charge collected by the strips of one of these two layers provides one coordinate of the position of the track. Two such coordinates measured by the two layers associated with a single track give the precise location of the track in the GEM detector. The relative orientation of the strips in the upper and lower levels defines the readout coordinate system. Two different types of readout coordinate systems have been used in the SBS GEMs: XY Cartesian and stereo angle UV. While the strips in the two layers are orthogonal in the XY Cartesian design, their relative orientation is $60^{\circ}$ in the stereo angle UV design, as depicted in Figure \ref{fig:gemstriporien}.

\subheading{Readout Electronics}
The readout electronics of a GEM detector comprise Analog Pipeline Voltage 25 (APV25) chips and their associated backplanes, VME-based Multi-Purpose Digitizers (MPDs), and VXS Trigger Processors (VTPs). The APV cards are directly mounted onto the ROB and connected to the readout strips. Signals collected by the APV cards are transmitted to MPDs for high-speed digitization via HDMI cables. Digitized signals from the MPDs are then conveyed to VTPs through optical fibers. The VTPs are tasked with data reduction via common mode subtraction and zero suppression, enabling online analysis. While the APV cards and their backplanes are radiation-hardened, the MPDs and VTPs are not. Therefore, the MPDs are housed in a shielded area adjacent to the detector stack, while the VTPs are located in the DAQ bunker. 
\begin{figure}[ht!]
     \centering
     \begin{subfigure}[b]{0.3\textwidth}
         \centering
         \includegraphics[width=\textwidth]{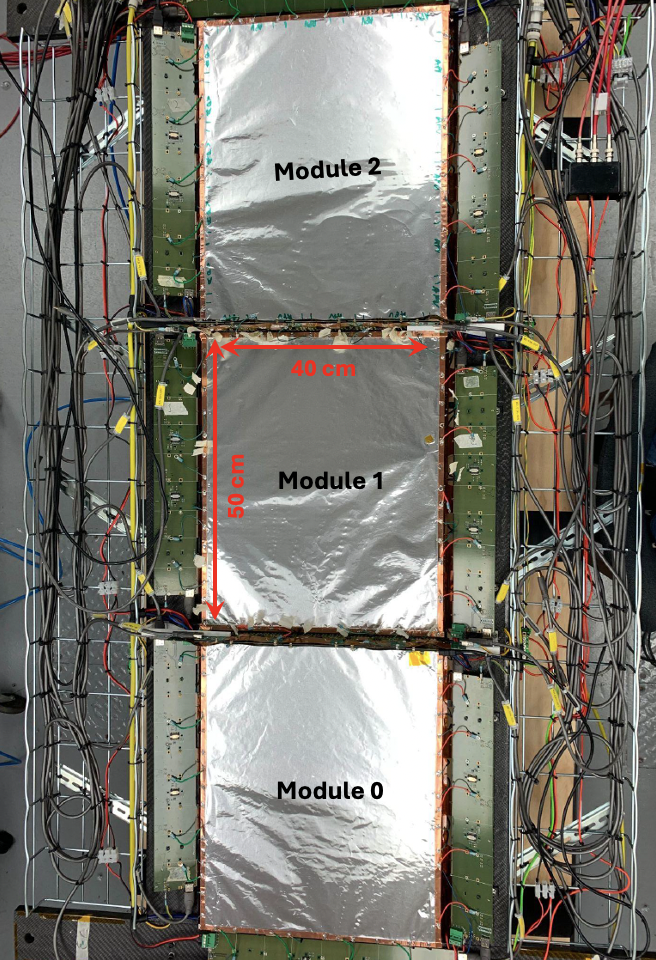}
         \caption{}
         \label{sfig:gemINFN}
     \end{subfigure}
     \hfill
     \begin{subfigure}[b]{0.255\textwidth}
         \centering
         \includegraphics[width=\textwidth]{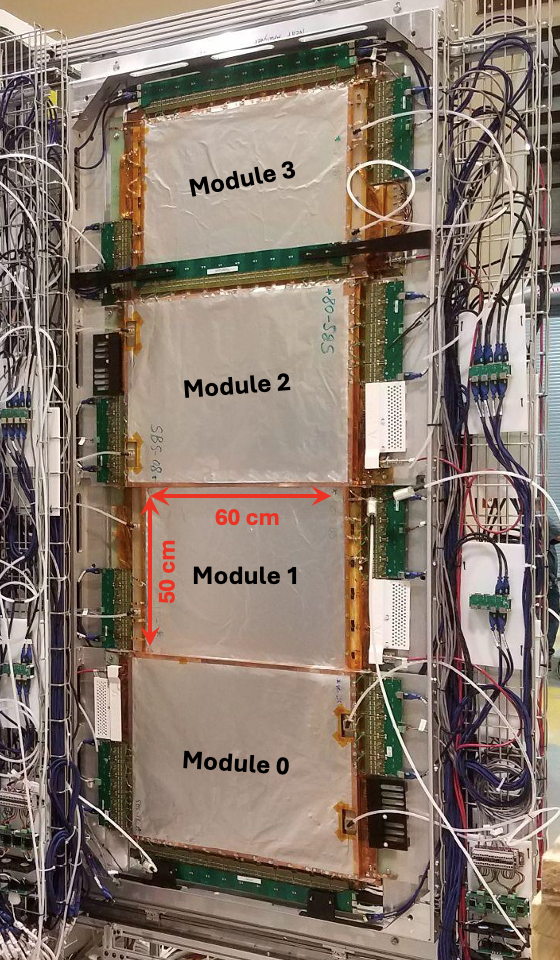}
         \caption{}
         \label{sfig:gemUVAXY}
     \end{subfigure}
     \hfill
     \begin{subfigure}[b]{0.3\textwidth}
         \centering
         \includegraphics[width=\textwidth]{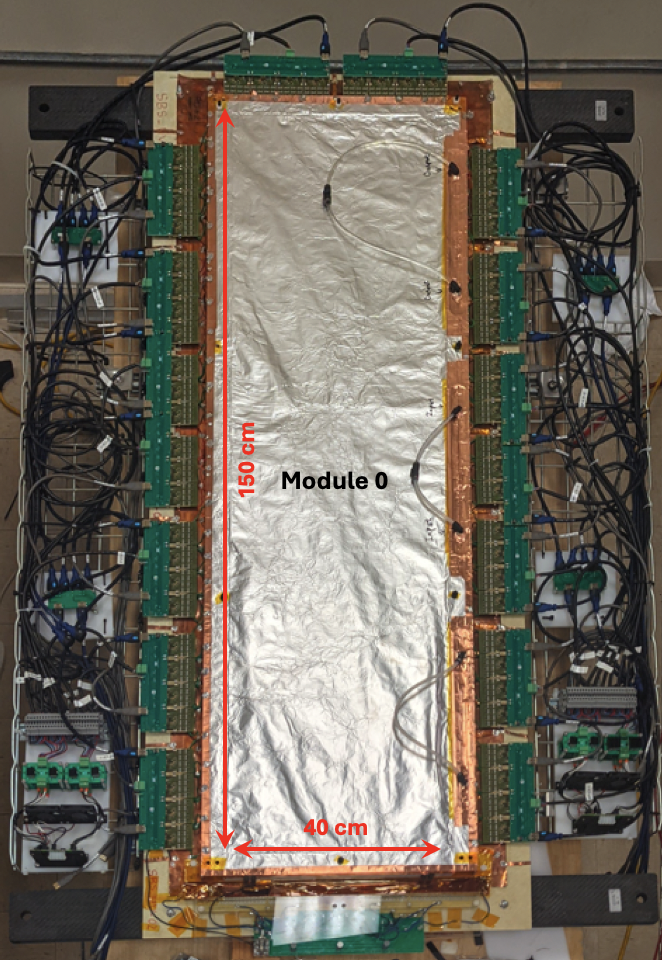}
         \caption{}
         \label{sfig:gemUVAUV}
     \end{subfigure}
     \caption{Types of SBS GEMs. (a) INFN XY GEM: $3$ modules each with a dimension of $40\times50$ \SI{}{cm^2}. (b) UVA XY GEM: $4$ modules each with a dimension of $60\times50$ \SI{}{cm^2}. (c) UVA UV GEM: $1$ module with a dimension of $40\times150$ \SI{}{cm^2} \cite{thesisJB}.}
     \label{fig:gemconfigs}
\end{figure}

\subheading{Voltage Supply}
Both high voltage (HV) and low voltage (LV) supplies are necessary for the operation of a GEM detector. As evident from the above discussion, the operation of a GEM detector is enabled by applying and maintaining a stable high voltage across the detector (\SI{3.7}{kV} at $745$ $\mu$A) and the individual GEM foils ($\approx$ \SI{400}{V}). HV supplied by the W-IE-NE-R MPOD EHS 8060n power supply is distributed to each GEM foil by a PCB fitted to the GEM detector. Additionally, the APV25 cards are powered by a \SI{5}{V} DC voltage supplied by a \SI{120}{V} power supply through a low voltage regulator board. 

\subsubsection{SBS GEMs: INFN and UVA}
\label{sssec:sbsgemtypes}
The SBS GEMs are manufactured by the University of Virginia (UVA) and the Istituto Nazionale di Fisica Nucleare (INFN). While the basic design remains consistent, there are differences between the GEMs produced by these two institutions. INFN produces only one type of SBS GEM, known as the INFN XY GEM. These GEMs feature an XY Cartesian readout coordinate system and are divided into three GEM modules, each with dimensions of $40\times50$ \SI{}{cm^2}. However, UVA produces two types of SBS GEMs - the UVA XY GEM and the UVA UV GEM. Similar to the INFN XY GEM, the UVA XY GEMs feature an XY Cartesian readout coordinate system but are divided into four GEM modules, each with dimensions of $60\times50$ \SI{}{cm^2}. As the name suggests, the UVA UV GEMs feature a stereo angle UV readout coordinate system and consist of a single GEM module with dimensions of $40\times150$ \SI{}{cm^2}. Figure \ref{fig:gemconfigs} illustrates images of all three types of GEMs mentioned above. 
\begin{table}[ht!]
\caption{\label{tab:gemconfig} BigBite (BB) GEM configuration during {\gmn}. Layer 0 is the first BB GEM layer located downstream of the BB magnet and layer 4 is the back tracker located between the Pre-Shower and the GRINCH.}
\centering
\begin{tabular}{ccccccc}\hline\hline
   GEM & \multicolumn{5}{c}{BigBite GEM Type in Layer} & Run \\ \cline{2-6}
   Configuration & 0 & 1 & 2 & 3 & 4 & Range \\ \hline
   0 & UVA UV & INFN XY & UVA UV & INFN XY & UVA XY & 11180-12073 \\
   1 & UVA UV & INFN XY & UVA UV & UVA UV & UVA XY & 12078-13086 \\
   2 & UVA UV & UVA UV & UVA UV & UVA UV & UVA XY & 13095-13799 \\ \hline\hline
\end{tabular}
\end{table}

During the {\gmn} experiment, all three types of GEMs were utilized in the BigBite (BB) Spectrometer. Each of the five GEM detectors in the BB spectrometer constitutes a GEM layer, with layer numbering starting from $0$. For instance, the first GEM layer downstream of the BB magnet is layer $0$, and the last GEM layer, or the back tracker situated between the Pre-Shower and the GRINCH, is layer $4$. While the back tracker, a UVA XY GEM, remained constant throughout the experiment, several front trackers had to be replaced. The {\gmn} experiment commenced with two UVA UV GEMs in layers $0$ and $2$, two INFN XY GEMs in layers $1$ and $3$, and a UVA XY GEM in layer $4$, labeled as GEM configuration $0$. However, within two months of operation, both INFN XY GEMs were substituted with UVA UV GEMs due to various issues, such as dead modules and instability. Detailed descriptions of the BB GEM configurations during the \gmn experiment can be found in Table \ref{tab:gemconfig}.

\subsection{Gas Ring Imaging Cherenkov} 
\label{ssec:ch3:grinch}
The Gas Ring Imaging Cherenkov detector (GRINCH), positioned between the fourth and fifth GEM layers, identifies electron tracks among background particles like charged pions. When electrons enter the gaseous medium inside GRINCH, they emit Cherenkov radiation in a cone shape. This radiation is reflected by a mirror placed along the particle's trajectory and directed toward a PMT array for detection. The resulting detection cluster, a cross-section of the cone perpendicular to the trajectory, forms a ring, giving the detector its name. The angle of Cherenkov radiation, $\theta_{\text{c}}$, emitted by a charged particle traveling with velocity $v$ in a medium with refractive index $n$, is given by the equation: 
\begin{equation} 
    \cos\theta_{\text{c}} = \frac{c}{nv}, 
\end{equation} 
where $c$ is the speed of light in vacuum. Since pions are heavier than electrons, they travel at lower velocities for the same momentum, leading to a larger value of $\cos\theta_{\text{c}}$ and, therefore, a smaller Cherenkov angle $\theta_{\text{c}}$. As a result, in detectors like GRINCH, only electrons produce visible Cherenkov light clusters, enabling effective pion rejection.

\begin{figure}[h!]
    \centering
    \includegraphics[width=1\columnwidth]{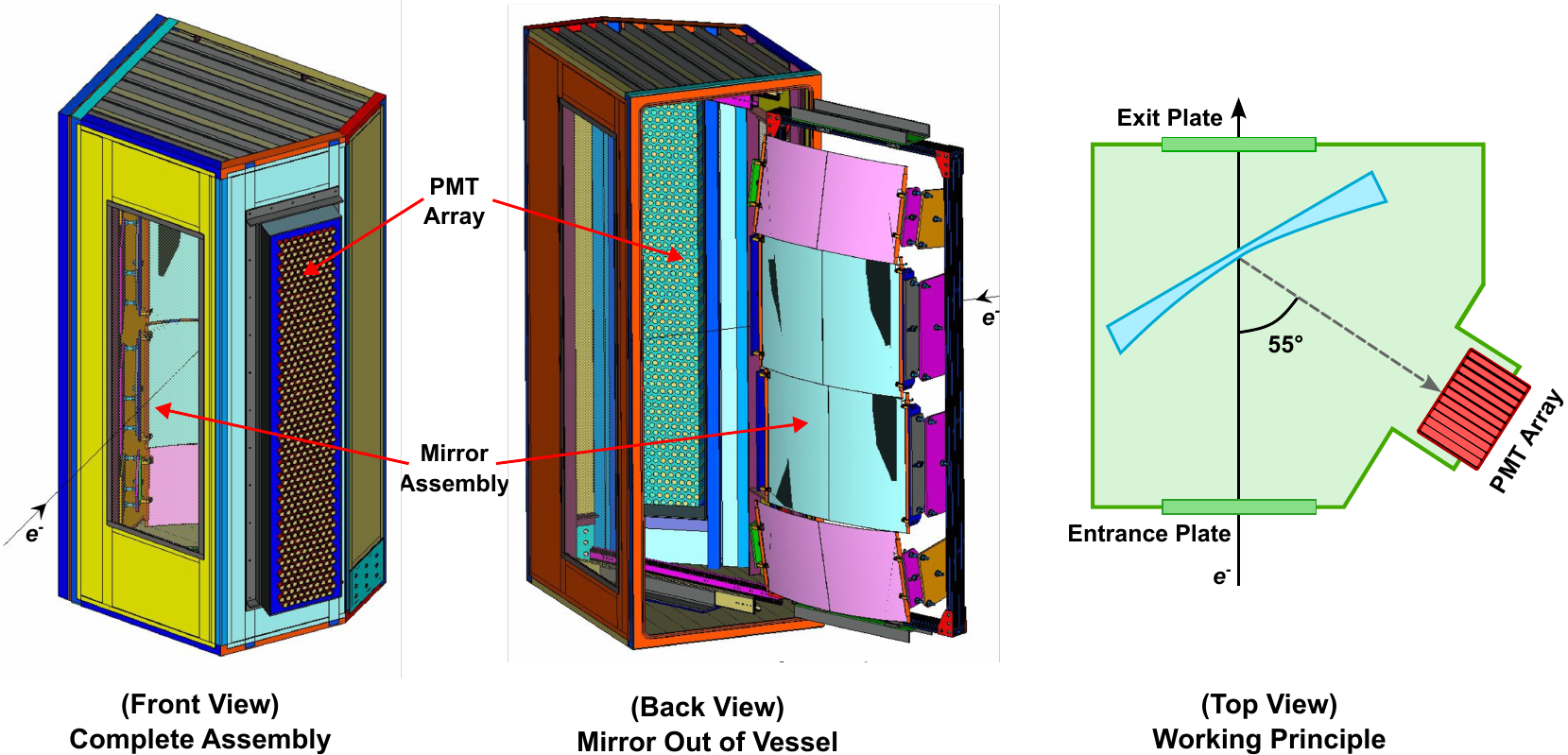}
    \caption{\label{fig:ch3:grinch} Design and working principle of the Gas Ring Imaging Cherenkov (GRINCH) detector.}
\end{figure}
\fig \ref{fig:ch3:grinch} shows CAD drawings of the GRINCH design. GRINCH has three main components:
\begin{enumerate}
    \item \textbf{Vessel:} An air-sealed container, \SI{88.9}{cm} deep, houses the detector assembly and heavy gas medium needed for Cherenkov radiation. For the GMn experiment, the vessel was filled with \ce{C4F8} heavy gas, which has a refractive index of $1.00132$ at \SI{405}{nm} and a pion threshold of \SI{2.7}{GeV}. Pions exceeding this energy threshold will form clusters in the GRINCH, reducing its pion rejection capabilities since the PMT array lacks the resolution to distinguish between clusters formed by pions and electrons.
    \item \textbf{Mirror Assembly:} Four highly reflective mirrors, each with an incident angle of $55^{\circ}$, are arranged in a staggered configuration along a vertical plane that intersects the spectrometer axis at $117.5^{\circ}$. The middle mirrors ($75.02\times62.54$ cm$^2$) are slightly larger than the edge mirrors ($75.02\times42.54$ cm$^2$), which are inclined ($\approx 10^{\circ}$) about the backplane toward the spectrometer axis to maximize acceptance.
    \item \textbf{PMT Array:} The Cherenkov light reflected by the mirrors focuses onto the PMT array, located in front of the mirror assembly at a $55^{\circ}$ angle relative to the spectrometer axis. The array contains $510$ 9125B 29 mm PMTs arranged in a honeycomb pattern. Signals from the PMTs are transmitted to NINO ASIC-based amplifier/discriminator cards attached to the detector frame. These processed signals are then sent to the DAQ bunker via \SI{30}{m} long 34-pin ribbon cables to be digitized and recorded by VETROC modules, which capture leading and trailing edge times along with time-over-threshold (TOT) information per PMT signal.
\end{enumerate}

Due to several issues, useful GRINCH data is not available for all GMn kinematics. GRINCH was only filled with \ce{C4F8} gas midway through the experiment; before that, it contained \ce{CO2}, which has a lower-than-optimal refractive index. This affected the two highest-\q kinematics, where additional pion rejection was most needed, and the lowest-\q kinematics. While data taken at \qeq{7.4} included the heavy gas, it was compromised by malfunctioning VETROC modules. However, the GRINCH data quality for the remaining \qeq{4.5} datasets is good. Still, the energy of many generated charged pions at high \ep kinematics exceeds the pion threshold of \ce{C4F8}, reducing the effectiveness of pion rejection. As a result, the most effective performance of GRINCH was observed during \qeq{4.5} low-\ep kinematics.


\subsection{BigBite Calorimeter} 
\label{ssec:ch3:bbcal}
The homogeneous electromagnetic calorimeter in the BigBite spectrometer, known as the BigBite Calorimeter or BBCAL, consists of two segments: the Pre-Shower (PS) and Shower (SH) calorimeters. The PS is positioned downstream of the back-tracker, followed by a timing hodoscope (TH), while the SH calorimeter is located downstream of the TH, thus completing the BigBite spectrometer setup. 

The PS detector consists of 52 lead-glass (LG) modules arranged in 26 rows and 2 columns, while the SH detector includes 189 LG blocks arranged in 27 rows and 7 columns. Primary electrons lose energy in the LG modules primarily through bremsstrahlung and ionization, which trigger a cascade of electron-positron pair production and further bremsstrahlung processes, collectively known as an electromagnetic shower. The Cherenkov radiation, emitted by both the primary electron and the secondary charged particles in the visible to UV spectrum, is then detected by BBCAL PMTs. On average, approximately $25-30\%$ of the scattered electron's energy is absorbed by the PS, with the remainder captured by the SH. Together, the PS and SH reconstruct nearly $100\%$ of the incident electron's energy, providing excellent energy resolution ($5.4\%$ at \SI{3.6}{GeV}). 

The Minimum Ionizing Particles (MIPs) like pions deposit only a small fraction of their energy in the PS, allowing for easy identification and filtering from analysis, which effectively reduces inelastic background. The high energy resolution of BBCAL enables the formation of a stable and efficient electron-based trigger, which served as the main trigger for the \gmn experiment, as detailed in \sect \ref{ssec:bbtrig}. Additionally, the BBCAL cluster energy and position provides the initial position and slope\footnote{Determining the slope requires additional input from the BigBite optics model and assumes that the electron track originates from the target, as detailed in \sect \ref{ssec:trackreconst}.} of the scattered electron track, enabling track reconstruction under high-rate conditions, such as those encountered in \gmn.

%
\subsubsection{Detector Assembly}
\label{ssec:bbcaldetassembly}
Although the basic components are more or less the same, there are significant differences in the design of the SH and PS calorimeters. This section will detail the important aspects of their design, pointing out the key differences. 

\begin{figure}[h!]
	   \centering
	   \fboxsep=0.75mm
       \fboxrule=1pt
       \includegraphics[width=1\columnwidth]{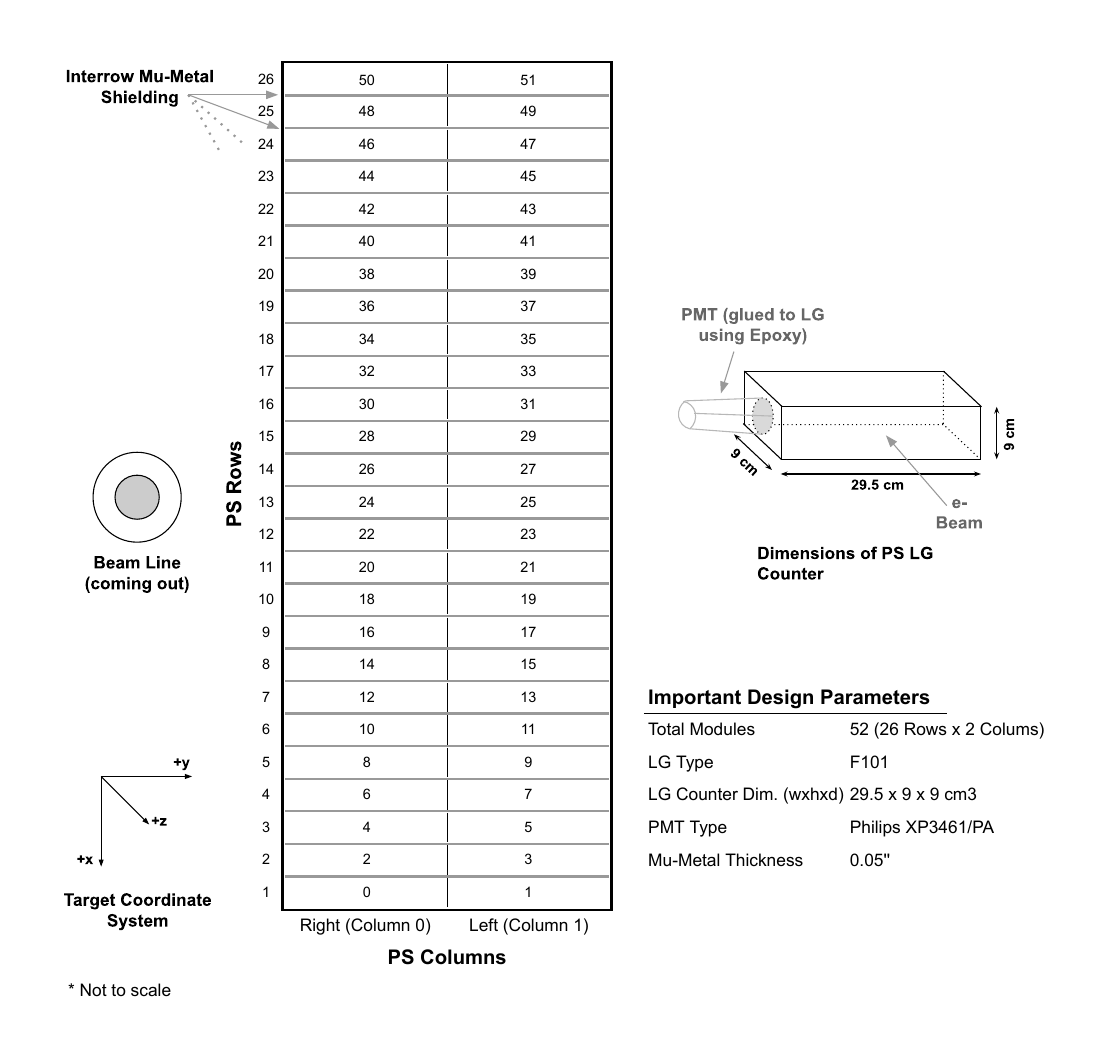}
       \caption{\label{fig:pssketch} Pre-Shower (PS) detector map (back view) with a summary of the important design parameters.}
    \end{figure}

    \subheading{Pre-Shower Detector} The Pre-Shower (PS) detector installed in the upgraded BigBite Spectrometer, often designated as the ``New" PS detector, was constructed in the Fall of 2020 at the Jefferson Lab. This construction involved the replacement of counters and PMTs from the ``Old" PS detector \footnote{As a part of the old BigBite spectrometer, The ``Old" PS detector has been used in multiple Jefferson Lab experiments, including E-02-013, E-06-010, etc., during the \SI{6}{GeV} era.} with refurbished counterparts sourced from the electromagnetic calorimeter employed in the HERMES experiment \cite{AVAKIAN199869}. The primary improvements of the ``New" PS detector over the ``Old" one are as follows:
    \begin{enumerate}
        \item More radiation-hardened: Achieved by replacing TF1 LG counters with F101 LG counters. 
        \item Better shielded against stray magnetic field: Achieved by installing mu-metal plates in between two consecutive rows. 
    \end{enumerate}
    The remaining part of this section will be dedicated to describing the design of the ``New" PS detector. For brevity, the term ``New" will be dropped. Figure \ref{fig:pssketch} has a map of the PS detector along with a summary of the important design parameters.
    %
    %

    The PS detector consists of $52$ F101 LG counters, each with dimensions of $29.5\times9\times9$ \SI{}{cm^3}. Table \ref{table:f101prop} summarizes the chemical composition and important properties of F101 LG. The counters are arranged in $26$ rows of $2$ columns, positioned facing each other and perpendicular to the spectrometer axis. Signals generated in each counter are read out by a Philips XP3461/PA PMT. A PS module, consisting of an LG counter and a PMT, serves as the basic unit of the PS calorimeter.
    \begin{table}[h!]
    \caption{\label{table:f101prop}Chemical composition and important properties of F101 LG \cite{AVAKIAN1996155}.}
    \centering
    \begin{tabular}{c c} 
        \hline\hline
        Chemical Composition & weight (\%) \\
            \ce{Pb3O4} & 51.23 \\ 
            \ce{SiO2}  & 41.53 \\
            \ce{K2O}   & 7.0 \\
            \ce{Ce}    & 0.2 \\
        Density           & \SI{3.86}{gcm^{-3}} \\ 
        Refractive Index  & \SI{1.65}{}  \\
        Radiation Length  & \SI{2.78}{cm} \\
        Moli\`ere Radius  & \SI{3.28}{cm} \\ 
        Critical Energy   & \SI{17.97}{MeV} \\ [0.5ex]
        \hline\hline
    \end{tabular}
    \end{table}
    
    The PS detector frame is not light-tight. It is completely covered from the front and back but remains open on both sides to allow passage for the attachment of signal and HV cables to the PMTs. Such a design made it necessary to wrap each PS counter in the following two layers:
    \begin{enumerate}
        \item Inner layer: Constructed from aluminized mylar foil, with the mylar side facing inward, thus making contact with the polished surfaces of the LG blocks. Such a layer improves light collection. 
        \item Outer layer: Constructed from black Tedlar film to provide optical isolation from outside light.
    \end{enumerate}

    The scattered particles must traverse at least \SI{9}{cm} of F101 LG material while passing through the PS, which corresponds to approximately $3$ radiation lengths. This thickness is insufficient to stop high-energy electron tracks, which are fully absorbed by the SH calorimeter further downstream.  
    
        
    %
    %
    \begin{figure}
	   \centering
	   \fboxsep=0.75mm
       \fboxrule=1pt
       \includegraphics[width=1\columnwidth]{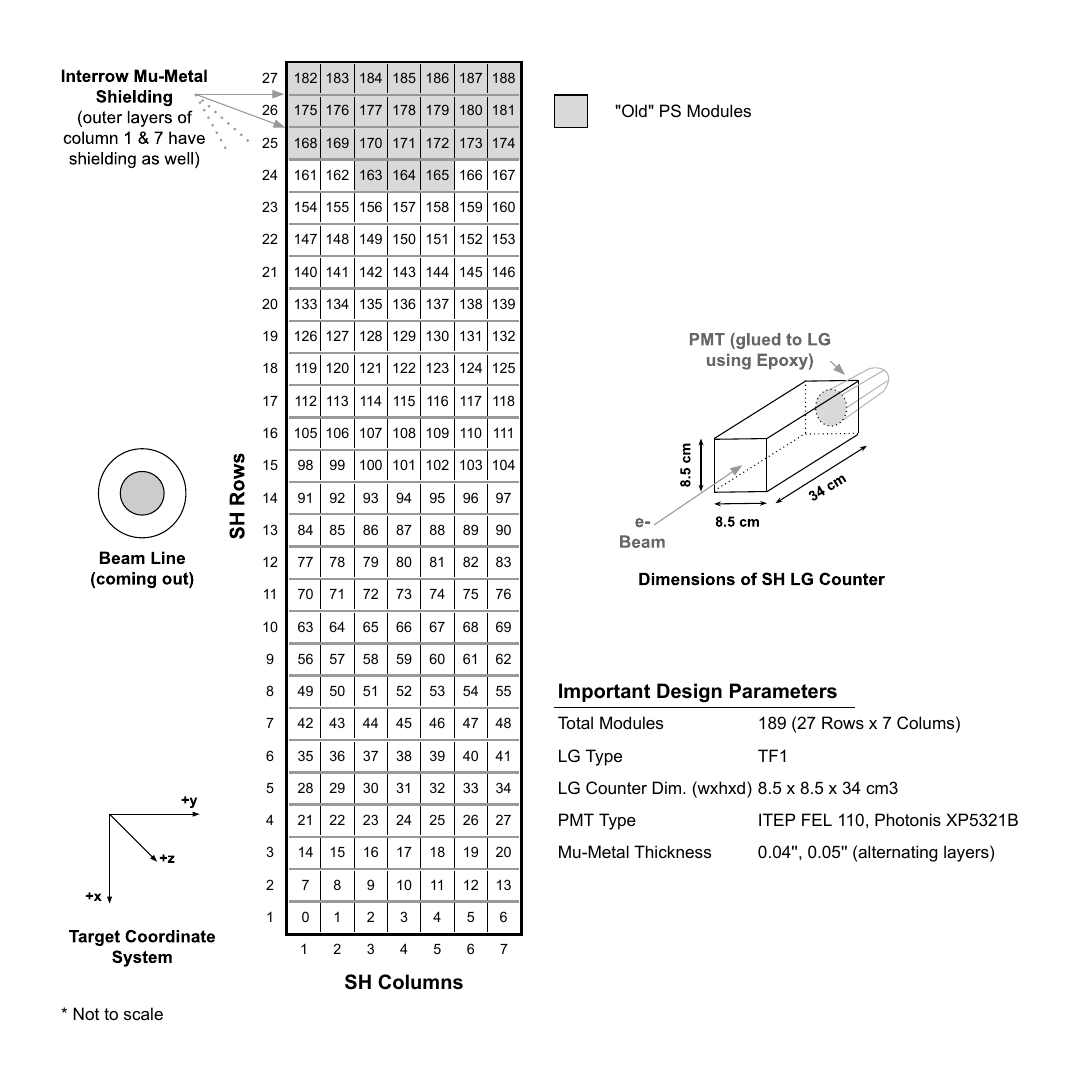}
	   \caption{\label{fig:shsketch} Shower (SH) detector map (back view) with a summary of the important design parameters.}
    \end{figure}
    %
    \subheading{Shower Detector} Similar to the Pre-Shower (PS) detector, the Shower (SH) detector used in the upgraded BigBite Spectrometer is a refurbished version of its predecessor. Each SH module, constituting the building block of the SH calorimeter, consists of a lead-glass (LG) counter and a PMT. As part of the refurbishment process, inefficient modules from different sections of the SH calorimeter were replaced with modules from its top four rows. The missing modules in the top four rows were then filled with the ``Old" PS modules. Figure \ref{fig:shsketch} has a map of the SH detector along with a summary of the important design parameters.

    There are $189$ TF1 LG counters in the SH calorimeter, each with dimensions of $8.5\times8.5\times34$ \SI{}{cm^3}. Table \ref{table:tf1prop} summarizes the chemical composition and important properties of TF1 LG. The counters are stacked in $27$ rows of $7$ columns facing the spectrometer axis. Signals generated in each counter are read out by a PMT attached to it.
    \begin{table}[h!]
    \caption{\label{table:tf1prop}Chemical composition and important properties of TF1 LG \cite{BALATZ2005114,ASTVATSATUROV1973105}.}
    \centering
    \begin{tabular}{c c} 
        \hline\hline
        Chemical Composition & weight (\%) \\
            \ce{PbO}   & 51.2 \\ 
            \ce{SiO2}  & 41.3 \\
            \ce{K2O}   & 7.0 \\
            \ce{As2O3} & 0.5 \\
        Density           & \SI{3.86}{gcm^{-3}} \\ 
        Refractive Index  & \SI{1.65}{}  \\
        Radiation Length  & \SI{2.5}{cm} \\
        Moli\`ere Radius  & \SI{3.5}{cm} \\ 
        Critical Energy   & \SI{15}{MeV} \\ [0.5ex]
        \hline\hline
    \end{tabular}
    \end{table}

    Since some of the SH modules were replaced with the ``Old" PS modules as part of the refurbishment, there are currently two types of PMTs in the SH calorimeter: ITEP FEL 110 and Photonis XP5321B. To enhance light collection by each PMT and screen the counters from each other, each LG counter was wrapped in aluminized mylar foil before installation. All the SH modules are kept in a light-tight box for optical isolation from external light sources. Additionally, mu-metal sheets have been installed in between two SH rows and at the outer layers of the first and last SH columns to provide shielding against stray magnetic fields.

    The SH calorimeter adds \SI{34}{cm} TF1 LG material to the path of the scattered particles, which is approximately $14$ radiation lengths. Together the SH and PS calorimeters constitute a depth of approximately $17$ radiation lengths, providing sufficient material to fully contain high-energy scattered electrons enabling high-resolution energy reconstruction.  
    

    %
%
    
\subsubsection{Signal Circuit}
\label{sssec:bbcalcircuit}
\begin{figure}[h!]
    \centering
    \includegraphics[width=0.9\columnwidth]{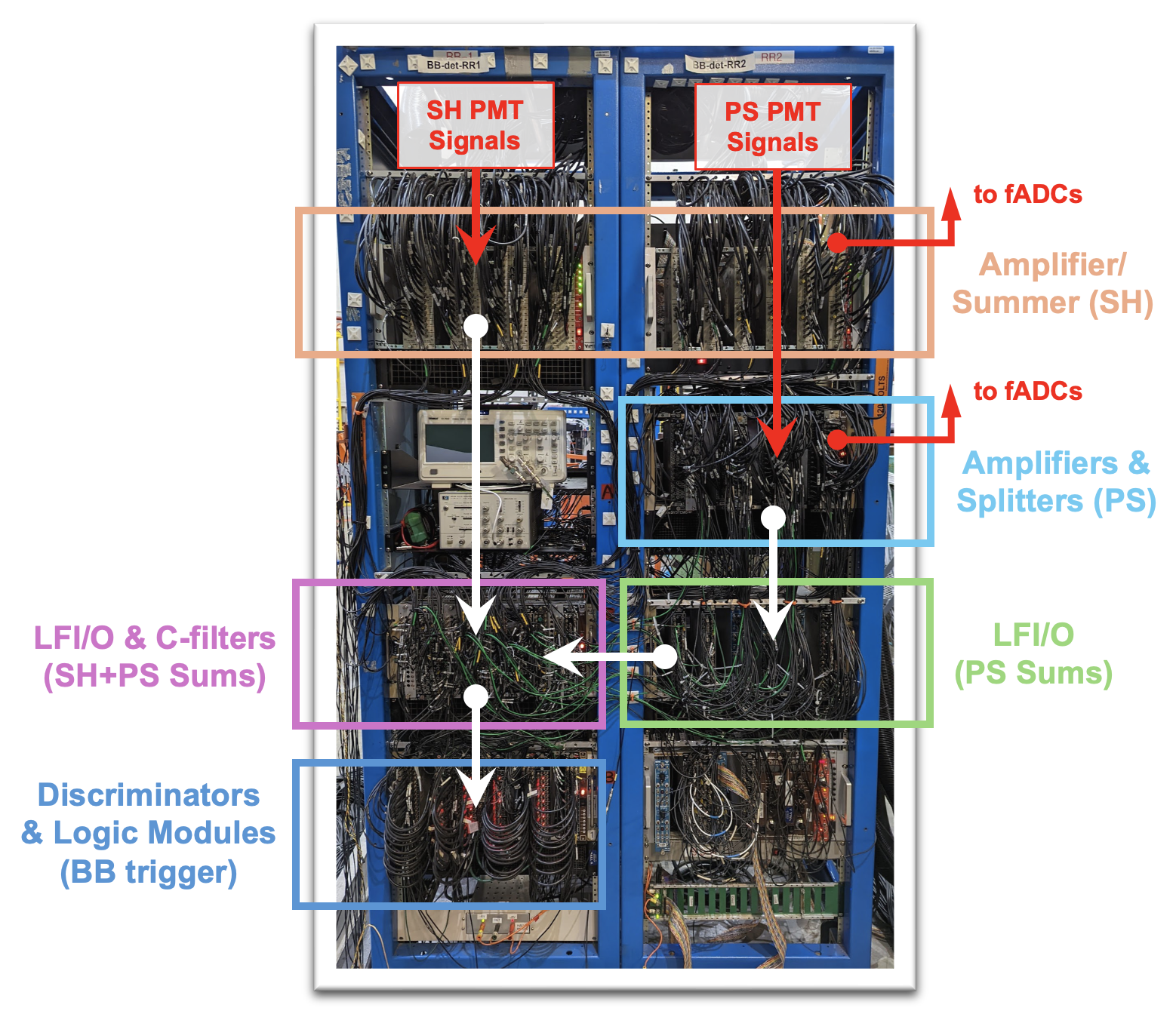}
    \caption{\label{fig:bbcalfend} The BigBite calorimeter front-end electronics and the implementation of BB electron trigger. See text for more details.}
\end{figure}

The raw PMT signals from the SH and PS detectors were processed at the BBCAL front-end located near the detector in the experimental hall. The front-end consisted of various NIM modules assembled in a weldment, as illustrated in \fig \ref{fig:bbcalfend}. The following outlines the path traversed by the PMT signals through these modules:

SH PMT signals, after generation, were sent directly into a custom-made Summer/Amplifier (S/A) module via \SI{12.5}{m} RG174 coaxial cables, where they were split into two copies:
\begin{enumerate}
    \item One copy, exiting the back of the module, was amplified by approximately 5x and then transmitted to the DAQ bunker via a \SI{50}{m} long signal cable for digitization and recording by the fADC 250 module.
    \item \label{itm:SHfour} The other copy, amplified by approximately 3.5x, was summed with signals from six other SH modules in the same row. The summed output was then routed into a quad of the Phillips Scientific (PhSc) 740 linear fan-in/fan-out (LFI/O) module, where it was combined with signals from overlapping SH and PS detectors to form the trigger logic, as discussed in \sect \ref{ssec:bbtrig}.
\end{enumerate}

The signals from the PS PMTs followed a slightly different path. After generation, they were routed into a 2-output 10x PhSc amplifier via \SI{12.5}{m} long signal cables. One copy of the amplified signal was sent to the DAQ bunker via a \SI{50}{m} long signal cable for digitization and recording by the fADC 250 module. The other copy was sent into a custom-made 2-output splitter module. Each output of the splitter module, with an effective gain of 5x the original PS signal, was routed into a quad of an LFI/O module to sum signals from overlapping PS rows. The outputs of these quads were then routed to another set of LFI/Os, where the final summing of signals from overlapping SH and PS rows occurred to form the BBCAL trigger logic (see Step \ref{itm:SHfour} above).

\begin{table}[h]
\centering
\caption{Saturation level of electronic modules in the BBCAL signal circuit.}
\begin{tabular}{>{\hsptab}l<{\hsptab}>{\hsptab}c<{\hsptab}>{\hsptab}c<{\hsptab}>{\hsptab}c<{\hsptab}>{\hsptab}c<{\hsptab} }
    \hline\hline
      & Total & Input Chan & Output Chan & Max Allowed Input  \\
      & Modules      & per Module     & per Module 
     & Signal Amp (mV) \\
    \hline
    Summer/Amplifier (S/A) & $14$ & $14$ & $14+6$ & $200$ \\
    PhSc 776 Amplifier   & $4$ & $16$ & $16$ & $300$ \\
    PhSc 740 LFI/O       & $10$& $16$ & $16$ & $1200$ \\
    LeCroy 428F LFI/O  & $3$ & $16$ & $16$ & $700$ \\
    fADC 250           & $16$& $16$ & N/A  & $2000$ \\
    \hline\hline
\end{tabular}
\label{tab:satlev}
\end{table}
The BBCAL front-end was initially assembled in 2006 for the E-02-013 experiment. Similar to the SH and PS detectors, the BBCAL front-end used in the {\gmn} experiment is a refurbished version of the original. Any faulty modules were either repaired or replaced, and a comprehensive characterization of all electronic modules was conducted. Table \ref{tab:satlev} lists the various types of modules used in the BBCAL signal circuit, along with their respective saturation levels. The determination of the saturation level of each type of module was crucial to decide at what amplitude every PS and SH PMT signal should be aligned at the trigger level via PMT gain-matching. A detailed description of the BBCAL trigger logic and its implementation can be found in Section \ref{ssec:bbtrig}.
\subsubsection{HV System}
\label{ssec:bbcalhv}
The High Voltage (HV) distribution system for the BBCAL PMTs comprises of two LeCroy 1458 HV crates with built-in Raspberry Pi (rpi)-based software control, twenty-one LeCroy 1461N HV cards, and $482$ SHV cables. The HV crates are installed in a rack, one on top of the other, in the DAQ bunker to avoid radiation damage. 

Each LeCroy 1458 HV crate can hold sixteen type 1461N HV cards, each equipped with twelve HV supply channels. Such a design necessitates a total of twenty-one HV cards to be installed in two HV crates to accommodate all $241$ BBCAL PMTs. Among the twenty-one cards, eleven of them are installed in the upper crate to supply HV to all the PMTs of the right column of PS and rows 1 through $12$ of SH. The remaining five HV cards in the upper crate are used to supply HV to GRINCH PMTs, and the six empty slots in the bottom crate are kept as spares. HV from each 1461N HV card channel to the corresponding BBCAL PMT is carried by a \SI{50}{m} long SHV cable, assembled by joining one \SI{40}{m} long and another \SI{10}{m} long SHV cable.

The rpi installed in each LeCroy 1458 HV crate enables software control by running the HV server. The upper and lower crates run the ``rpi17" and ``rpi18" servers, respectively. Once the servers are online, a JAVA-based Graphical User Interface (GUI) is used to set the HV value for individual channels and monitor the corresponding voltage and current read-backs. It also offers convenience features such as saving and loading an HV settings file containing the HV values for all the BBCAL PMTs. One can also set the voltage and current trip limits for individual HV channels to trigger an alarm in case of any failure. Similar to other sub-systems, three EPICS Process Variables (PVs) are assigned to each BBCAL HV channel to keep track of the set voltage (V0Set), voltage read-back (VMon), and current read-back (IMon). These variables are automatically logged in the Hall A electronic logbook (HALOG) at the start of every production run.

\subsection{Timing Hodoscope}
The Timing Hodoscope, or TH, in the BB spectrometer is a highly segmented scintillation detector positioned between the Shower and Pre-Shower calorimeters, covering their entire active area. Its main purpose is to provide a high-precision ($\sigma_{t} \approx$ \SI{200}{ps}) timing reference for the scattered electron tracks across the full range of electron momentum detected by BB. It is designed to operate under high-rate conditions to avoid imposing significant limitations on the experiment's desired luminosity. This section will be dedicated to discussing the design and operation of the TH. 
%

\subsubsection{Design}
\begin{figure}[ht!]
     \centering
     \begin{subfigure}[b]{0.47\textwidth}
         \centering
         \includegraphics[width=\textwidth]{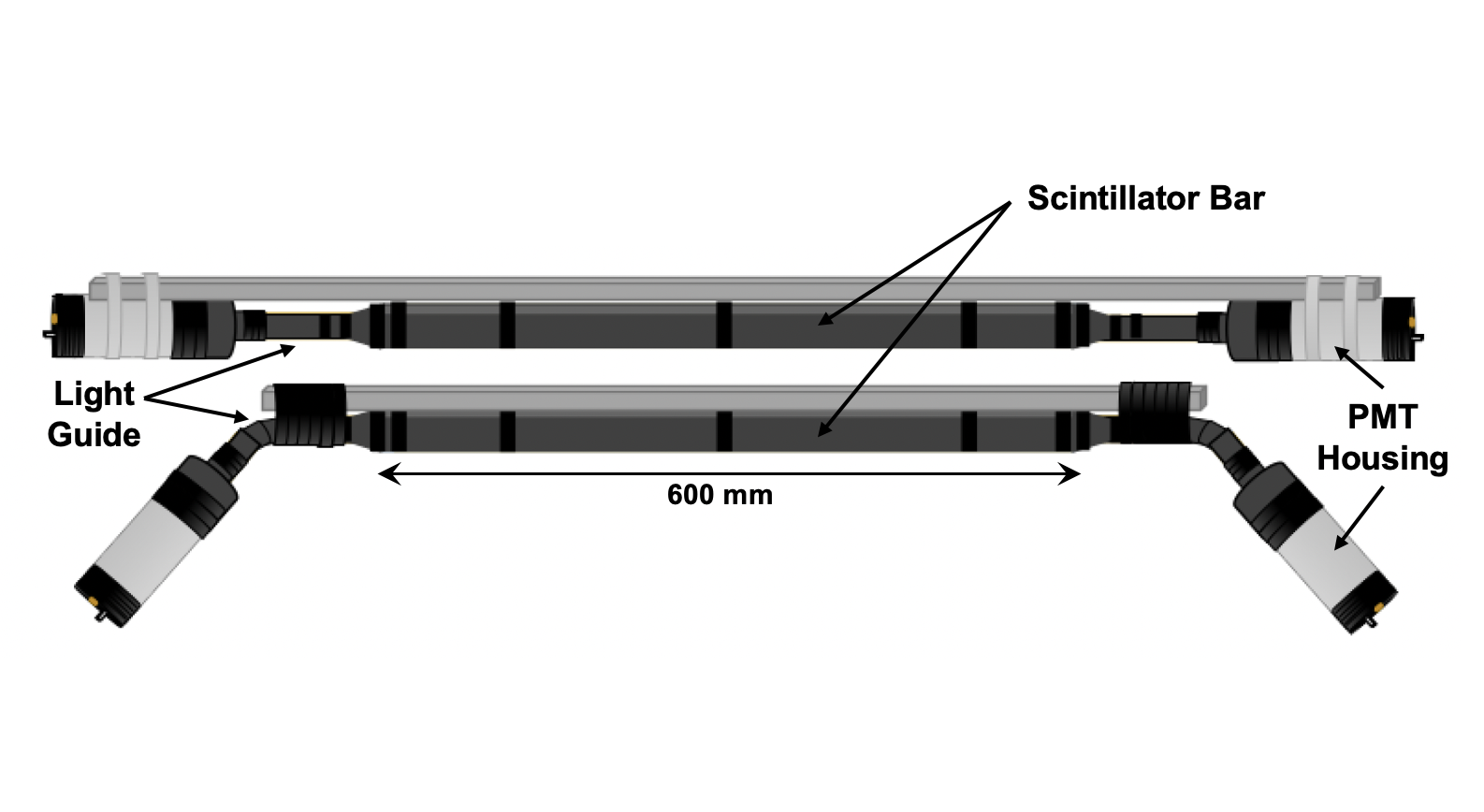}
         \caption{}
     \end{subfigure}
     \hfill
     \begin{subfigure}[b]{0.47\textwidth}
         \centering
         \includegraphics[width=\textwidth]{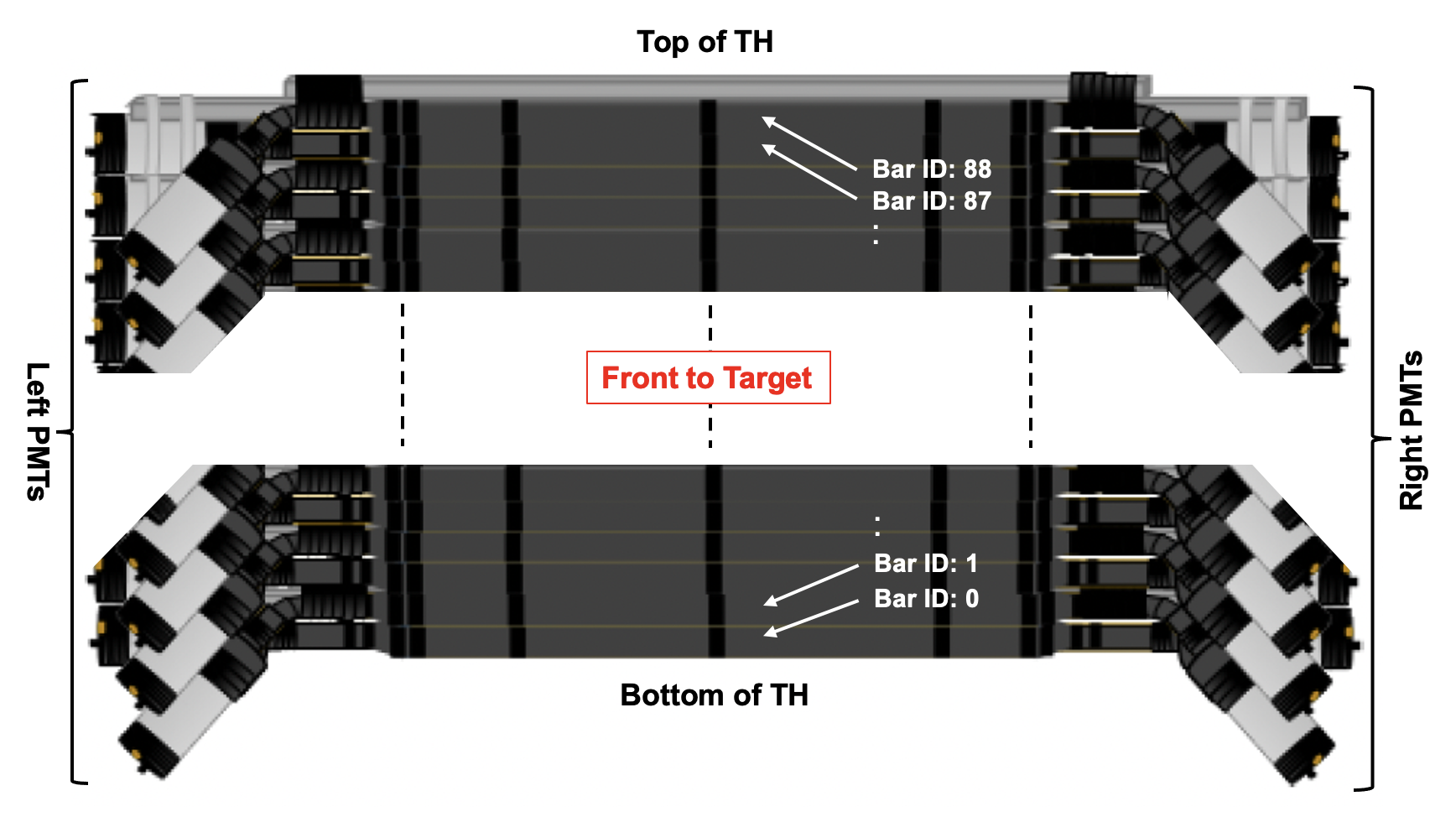}
         \caption{}
     \end{subfigure}
     \caption{Timing Hodoscope (TH) detector map and geometry \cite{thesisRalphM}\protect\footnotemark. (a) TH bars with straight (top) and curved (bottom) light-guides. (b) Stacking of the TH bars.}
     \label{fig:thbar}
\end{figure}

The TH comprises $89$ vertically stacked Eljen Technologies EJ200 plastic scintillator bars, positioned perpendicular to the spectrometer axis. Each bar, measuring $600\times25\times25$ \SI{}{mm^3}, is wrapped in black industrial paper for light-tightness. Electron Tubes ET 9142 single-channel PMTs are mounted at both ends of each bar, totaling $178$ readout channels. The PMTs are housed in light-tight enclosures and connected to the bars via Eljen Technologies UVT acrylic rod light guides, with a diameter of \SI{24}{mm}. Exposed ends of the light guides are inserted into the PMT housing, sealed with foam material and black electrical tape. Due to spatial constraints, the light guides alternate between straight and curved geometries as they move vertically through the stack of scintillators, as shown in Figure \ref{fig:thbar}. High-energy scattered electrons deposit energy in the EJ200 plastic scintillators, which is then re-emitted as light with a peak wavelength of \SI{425}{nm} and collected by the associated PMTs. 
\footnotetext{The original design of the TH included $90$ scintillator bars, as mentioned in the citation. However, due to space constraints, the $90^{\text{th}}$ bar was never installed.}

\subsubsection{Readout Electronics}
Signals from the PMTs are transmitted to $12$ custom-made $16$-channel amplifier / discriminator cards, each based on the NINO Application-Specific Integrated Circuit (ASIC). These cards are attached to the detector, with six on either side, and connected to the PMTs via \SI{1.5}{m} long type RG 174 coaxial signal cables. The NINO cards are carefully tuned to achieve fast and low-noise processing of the PMT signals, which is essential for high-precision time measurements. Each card can output two sets of processed signals through a pair of $34$-pin ribbon cable connectors:
\begin{enumerate}
    \item One set of output signals from all $12$ NINO cards is transmitted to two CAEN V1190A multi-hit TDCs located in the DAQ bunker via twelve \SI{30}{m} long $34$-pin ribbon cables, to record signal time and NINO time-over-threshold (TOT) information.
    \item Another set of NINO card outputs, from only $4$ NINO cards (two on either side), is directed to four $16$-channel fADC modules located in the same crate as the CAEN V1190A TDCs. These modules record the amplitude and integral of the associated PMT signals for calibration purposes.
\end{enumerate}

\subsubsection{Voltage Supply}
All the PMTs and NINO cards in the TH are powered by high-voltage (HV) and low-voltage (LV) distribution systems, respectively. For the HV system, four 48-channel A1932A HV distributors, housed in a CAEN SY1527LC mainframe, supply HVs to all $178$ TH PMTs. These HVs are transmitted from the DAQ bunker to four HV distribution boxes attached to the detector frame via four \SI{60}{m} long braided $48$-channel multi-way cables. Subsequently, HV is delivered to individual PMTs through $178$ \SI{4}{m} long HV cables. Throughout the {\gmn} experiment, the operational HV values for TH PMTs ranged between \SI{-700}{V} to \SI{-1100}{V} \cite{thesisRalphM}.

Similarly, the LV distribution system for powering the NINO cards follows a comparable structure. LV from a KEYSIGHT N5744A DC module is distributed from the DAQ bunker to the LV distribution box attached to the detector frame via two \SI{60}{m} long 8-AWG power cables. From there, twelve \SI{5}{m} long 20-AWG power cables connect the NINO cards to the LV distribution box, ensuring a stable \SI{5}{V} DC voltage supply to individual NINO cards, each with a current draw of \SI{1.3}{A}.
\section{The Super BigBite Spectrometer}
\begin{figure}[ht!]
    \centering
    \includegraphics[width=0.9\columnwidth]{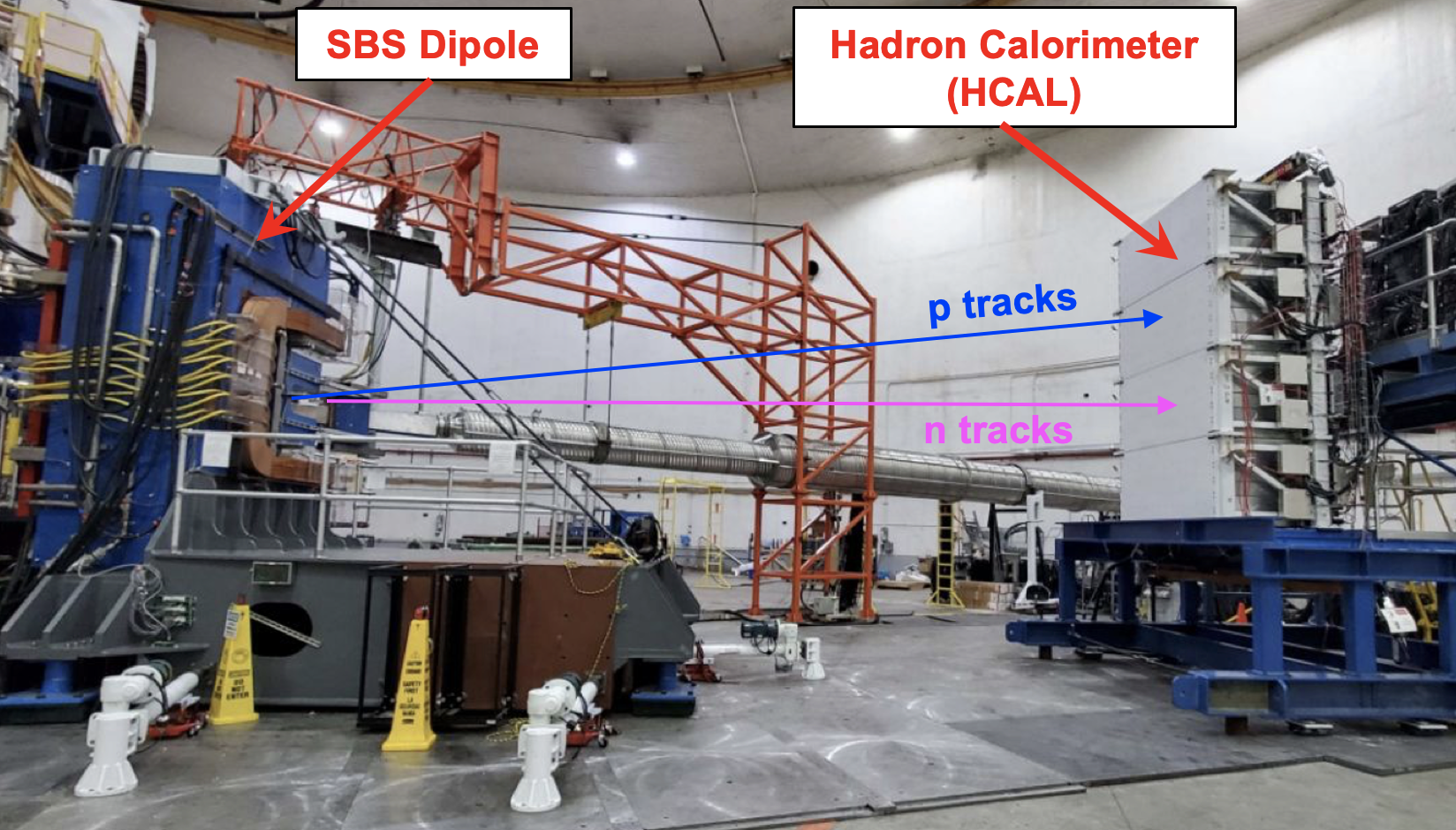}
    \caption{\label{fig:sbsinhall} The Super BigBite Spectrometer in Hall A.}
\end{figure}

The Super BigBite Spectrometer, or SBS, is located on the right side of the beamline when looking downstream from the scattering chamber and constitutes the hadron arm of the {\gmn} experiment. It consists of a large-aperture ($48\times122$ \SI{}{cm^2}) dipole magnet capable of separating high-energy ($1-9$ \SI{}{GeV}) scattered nucleons by charge, and a large-acceptance ($1.8\times3.6$ \SI{}{m^2}) hadron calorimeter capable of detecting nucleons with very high ($> 90\%$) and comparable efficiencies. Figure \ref{fig:sbsinhall} depicts the SBS in Hall A. This section will be dedicated to discussing the design and operation of the SBS dipole magnet and the hadron calorimeter.
\subsection{SBS Dipole Magnet}
\label{ssec:ch3:sbsmagnet}
The SBS dipole magnet, also known as the SBS magnet or 48D48, is a large-aperture, non-focusing, water-cooled dipole magnet capable of separating high energy ($1-9$ \SI{}{GeV}) nucleons by charge. It marks the entrance to the Super BigBite Spectrometer, with the hadron calorimeter directly following. Its approximately \SI{50}{msr} solid angle acceptance, up to \SI{1.8}{T} dipole field strength, and the capability to reach a forward angle of $10^{\circ}$ make it indispensable for the SBS high-{\q} EMFF measurement program.  

\begin{figure}[ht!]
     \centering
     \begin{subfigure}[b]{0.476\textwidth}
         \centering
         \includegraphics[width=\textwidth]{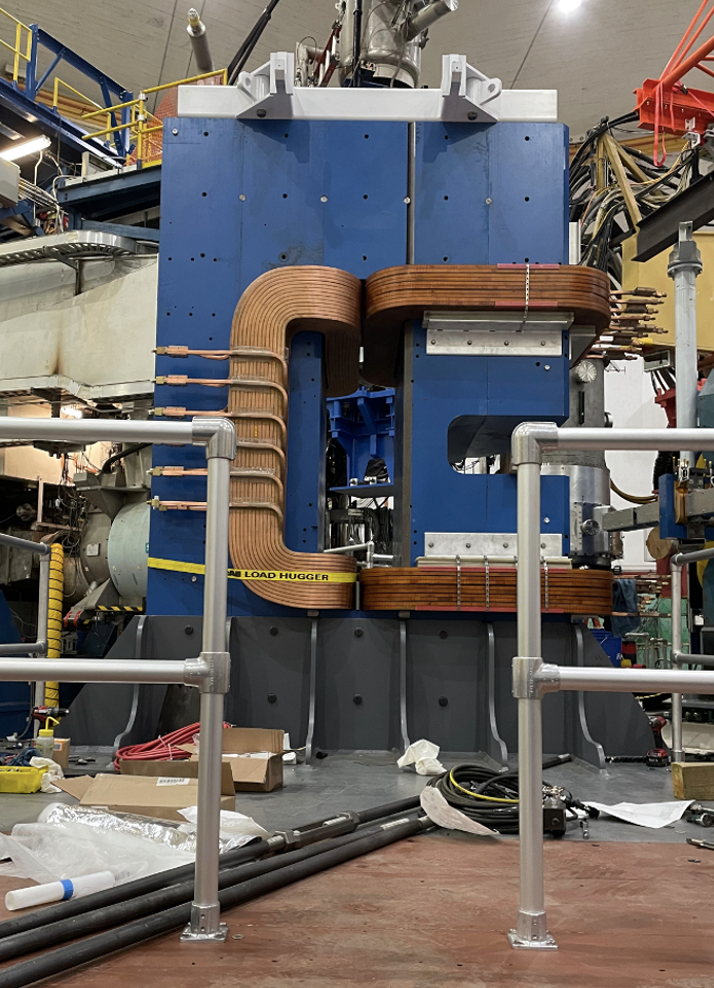}
         \caption{}
     \end{subfigure}
     \hfill
     \begin{subfigure}[b]{0.45\textwidth}
         \centering
         \includegraphics[width=\textwidth]{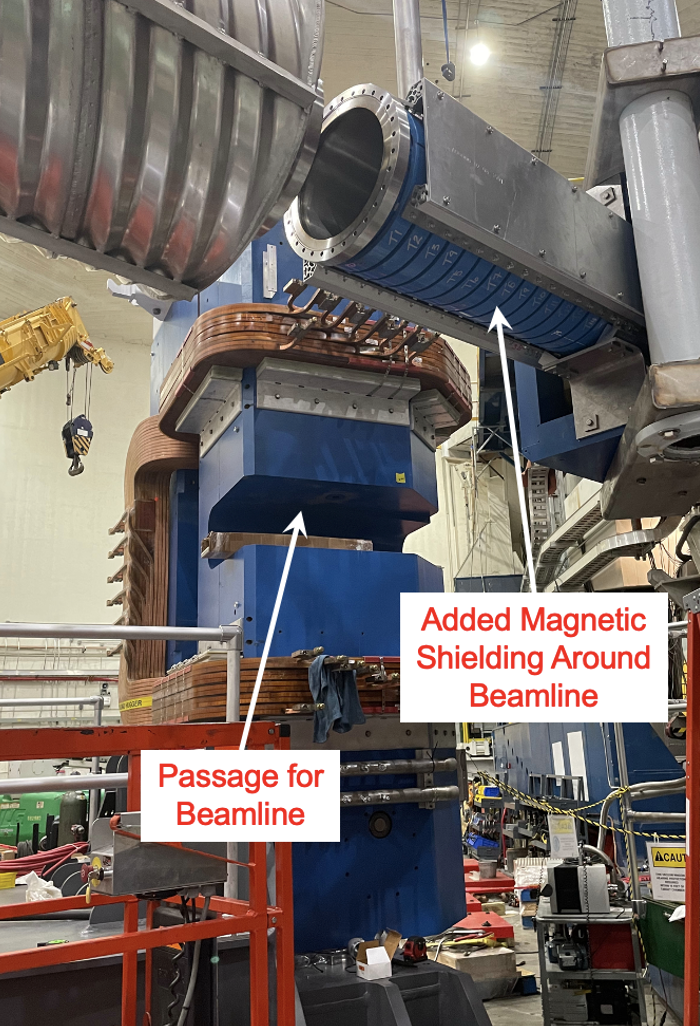}
         \caption{}
     \end{subfigure}
     \caption{The SBS magnet in Hall A. (a) Back view. (b) Side view.}
     \label{fig:sbsmaginhall}
\end{figure}

The Brookhaven National Laboratory (BNL) had four 48D48 magnets in storage, from which two were acquired for the SBS program \cite{SBSMAGROBIN}. One of them was transformed into the SBS magnet with several necessary upgrades, while the iron from the other was repurposed as a counterweight to enhance the stability of the SBS magnet \cite{PCBOGDAN}. The most significant upgrade to the BNL 48D48 magnet was the addition of a slot in its yoke (see Figure \ref{fig:sbsmaginhall}) to allow passage for the downstream beamline, enabling access to very forward angles ($\ge 10^{\circ}$), essential to reach high-\q. Weighing more than \SI{100}{tons}, the SBS magnet is positioned on a movable platform in the Hall for ease of adjusting spectrometer angles to desired values, as listed in Table \ref{tab:sbsconfig2}. However, the distance from the target to the magnet remained constant at \SI{2.25}{m} throughout the experiment.
%
%
%

The SBS magnet has an aperture of $48\times122$ \SI{}{cm^2} and a yoke length of \SI{1.22}{m}. It is energized by two coils, each with $120$ turns, generating a horizontal dipole field perpendicular to the motion of the scattered particles. During the experiment, the magnet was operated with positive polarity, causing proton tracks to bend upwards. The maximum operating current was \SI{2100}{A}, corresponding to a maximum field strength, $B_{SBS}^{max}$, which decreased significantly from \SI{1.78}{T} to \SI{1.28}{T} (see \tab \ref{tab:sbsconfig2}) due to modifications made between the \qeq{3\,\,\&\,\,9.9} kinematics to reduce stray fields. Despite this reduction, the field strength still provided an average deflection of $3.3^{\circ}$ to \SI{8.1}{GeV/c} proton tracks, ensuring adequate separation between \deen and \deep events at the HCAL, crucial for extracting \rqe.

The SBS field settings for production data at each kinematic, listed in \tab \ref{tab:sbsconfig2}, were chosen to maximize the separation of neutron and proton tracks while maintaining the upward-bent proton tracks within the HCAL acceptance. Greater separation minimizes systematic uncertainty due to nucleon misidentification from final-state interactions and/or the deuteron wave-function's long tail. Additionally, sufficient \ld and \lh data were collected at various SBS field strengths to evaluate the uniformity of the HCAL proton detection efficiency across the detector's active area, which is vital for determining the systematic uncertainty related to nucleon detection efficiency.

\subsection{Hadron Calorimeter} 
\label{ssec:ch3:hcal}
The sampling hadron calorimeter in the Super BigBite spectrometer, also known as the HCAL, is designed for the detection of high energy nucleons ($1-$\SI{10}{GeV}) with very high and comparable efficiencies. It is capable of detecting both neutrons and protons with excellent position and time resolutions. The energy resolution, however, is degraded significantly due to the sampling of the deposited energy.

The HCAL is located behind the SBS magnet, covering a large active area of $2 \times 3.8$ \SI{}{m^2}. Each HCAL module consists of $80$ interleaved layers of iron and scintillator plates, creating an effective depth of \SI{1}{m}. This design provides sufficient interaction length for protons and neutrons across the momentum range of interest for the SBS program, resulting in a very high detection efficiency (greater than $90\%$) for both particles. When nucleons interact with the iron plates, they produce hadronic showers of secondary particles, some of which generate signals in the associated PMTs by interacting with the scintillator plates. Approximately $8\%$ of the incident particle's kinetic energy is sampled in this process, defining HCAL's sampling fraction. 

\begin{figure}[ht!]
     \centering
     \begin{subfigure}[b]{0.478\textwidth}
         \centering
         \includegraphics[width=\textwidth]{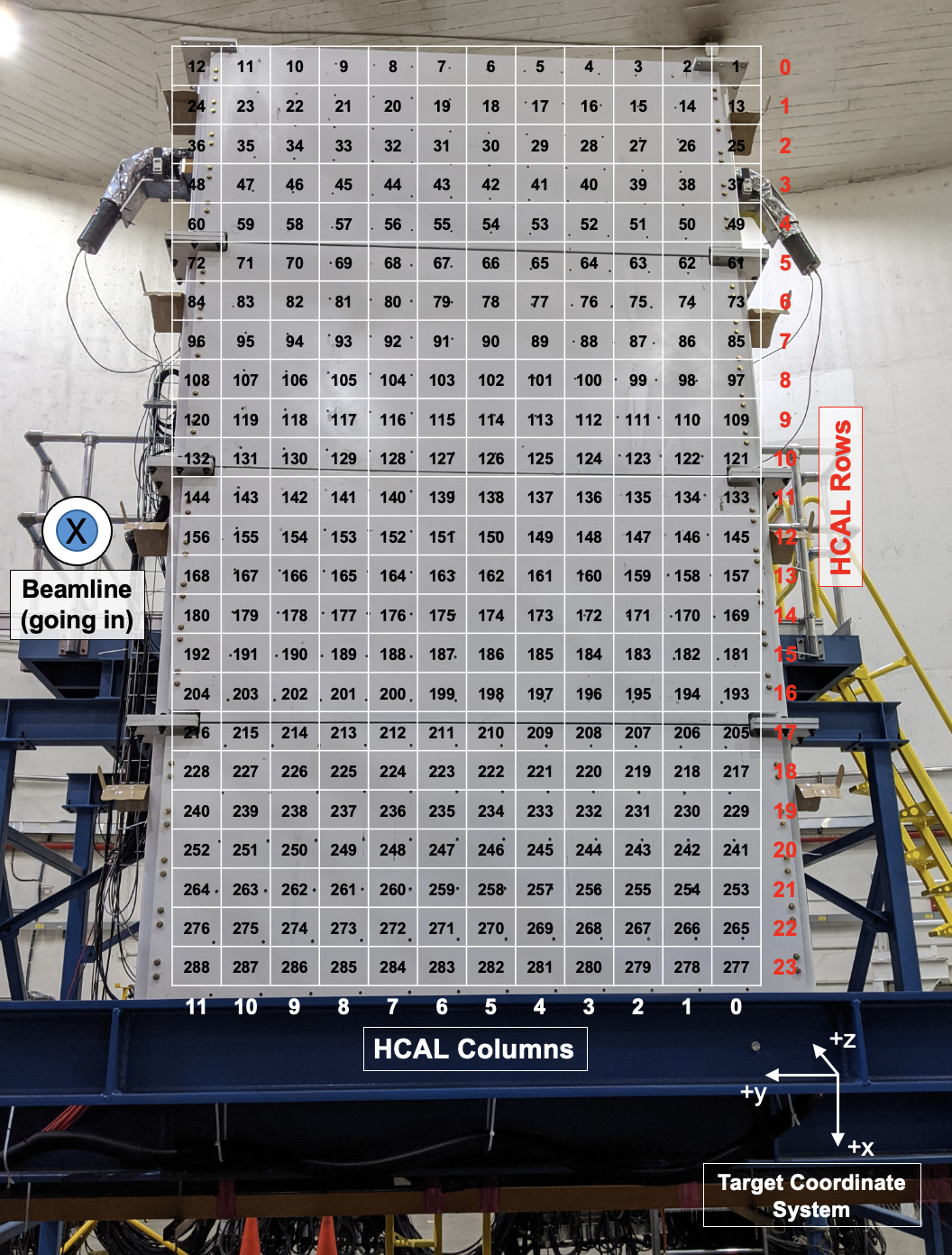}
         \caption{}
         \label{sfig:hcalfview}
     \end{subfigure}
     \hfill
     \begin{subfigure}[b]{0.47\textwidth}
         \centering
         \includegraphics[width=\textwidth]{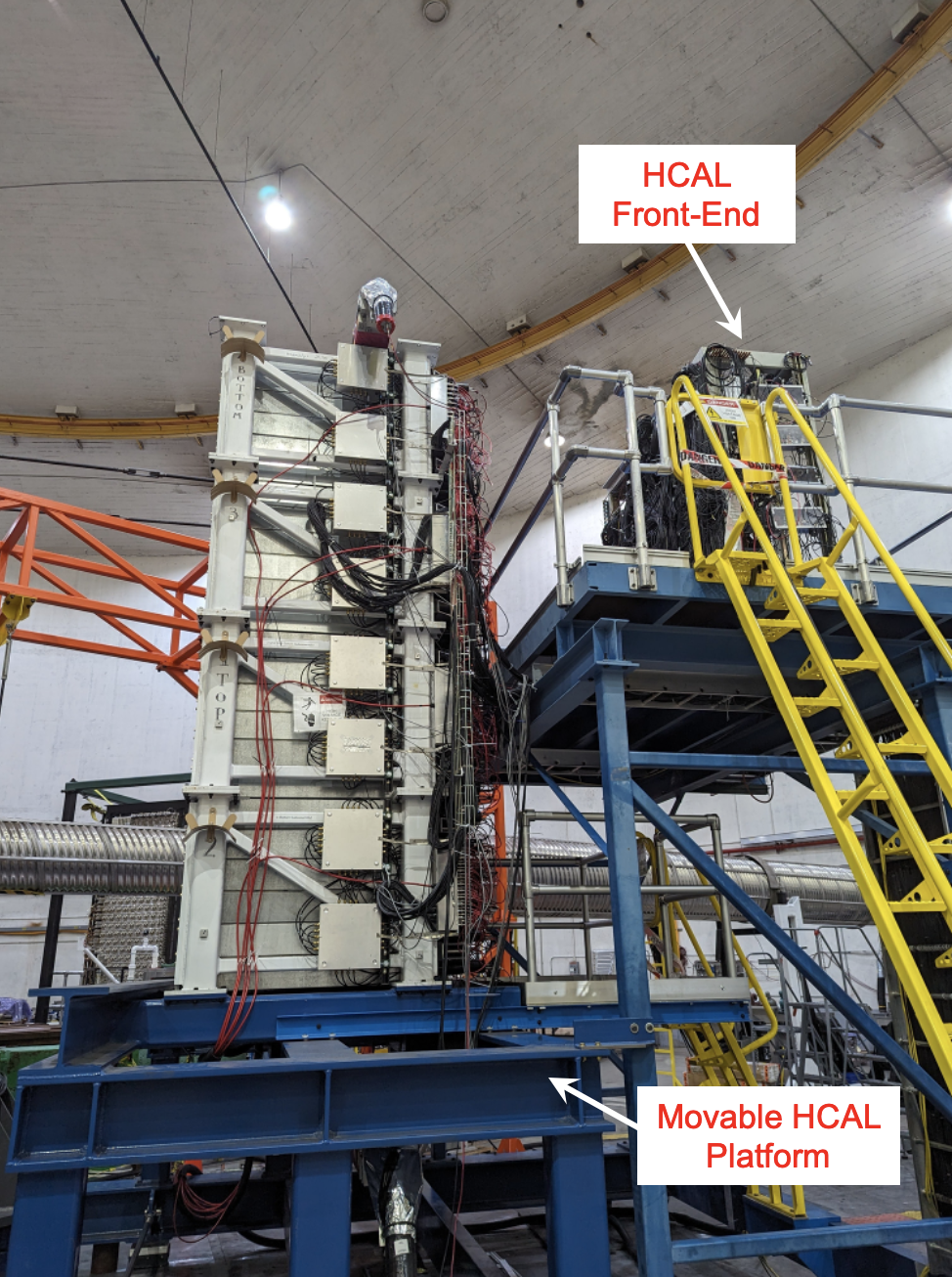}
         \caption{}
         \label{sfig:hcalsview}
     \end{subfigure}
     \caption{The hadron calorimeter (HCAL) in Hall A. (a) Front view of HCAL when looking downstream from the target. A detector map has been overlaid on the image, indicating the position of the detector modules with their corresponding module number (black digits), row number (red digits), and column number (white digits). NOTE: The association may not be exact in some areas due to imperfect alignment. (b) Side view of HCAL along with its front-end electronics.}
     \label{fig:hcaldetmap}
\end{figure}
While this low sampling fraction leads to poor energy resolution, the high segmentation of HCAL's active area and the broad spread of hadronic showers from high-energy nucleons provide excellent position resolution (5-6 cm), enabling precise separation of \deen and \deep events essential for extracting \rqe. Additionally, the excellent time resolution allows for strict electron-nucleon coincidence time cuts, effectively reducing accidental background. Although a nucleon-based trigger was created by logically summing HCAL signals, as discussed in Section \ref{ssec:sbstrig}, to be used in coincidence with the electron trigger, it wasn't employed during \gmn due to poor efficiency.

\subsubsection{Detector Assembly}
The design of the HCAL is based on the COMPASS HCAL1 calorimeter\cite{Vlasov2006} at CERN. It comprises 288 detector modules, each measuring $15\times15\times100$ \SI{}{cm^3}, arranged in $24$ rows of $12$ columns, forming the primary structure. These modules are distributed among four craneable sub-assemblies, each containing $6$ rows and $12$ columns, allowing for relocation of this \SI{40}{ton} detector.

\begin{figure}[ht!]
     \centering
     \begin{subfigure}[b]{0.478\textwidth}
         \centering
         \includegraphics[width=\textwidth]{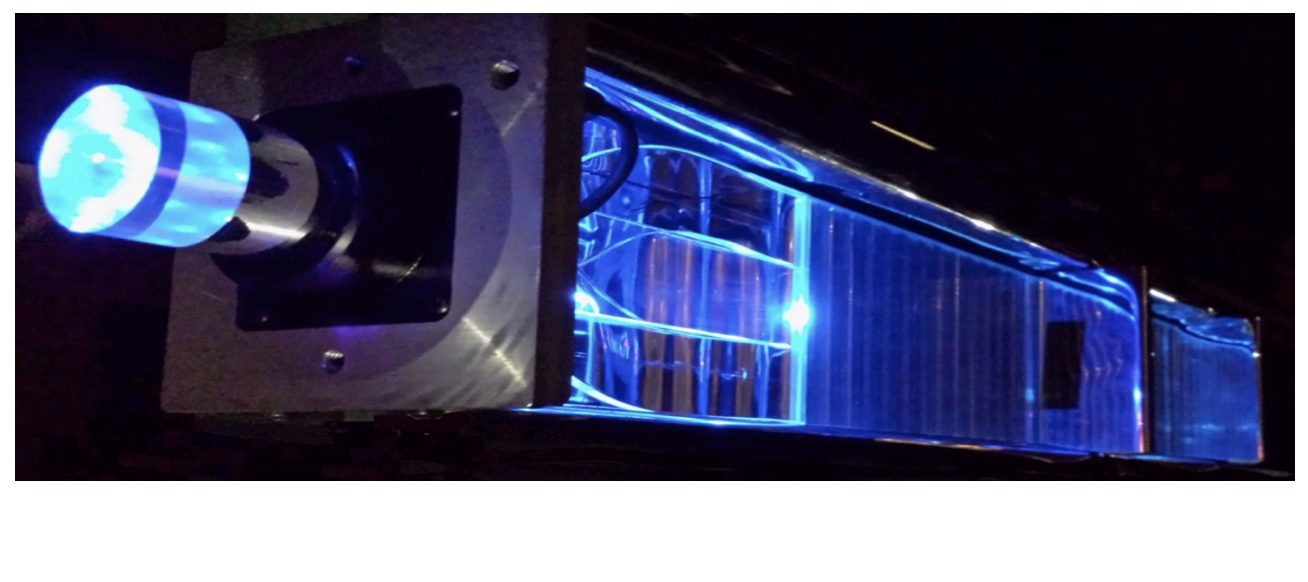}
         \caption{}
     \end{subfigure}
     \hfill
     \begin{subfigure}[b]{0.47\textwidth}
         \centering
         \includegraphics[width=\textwidth]{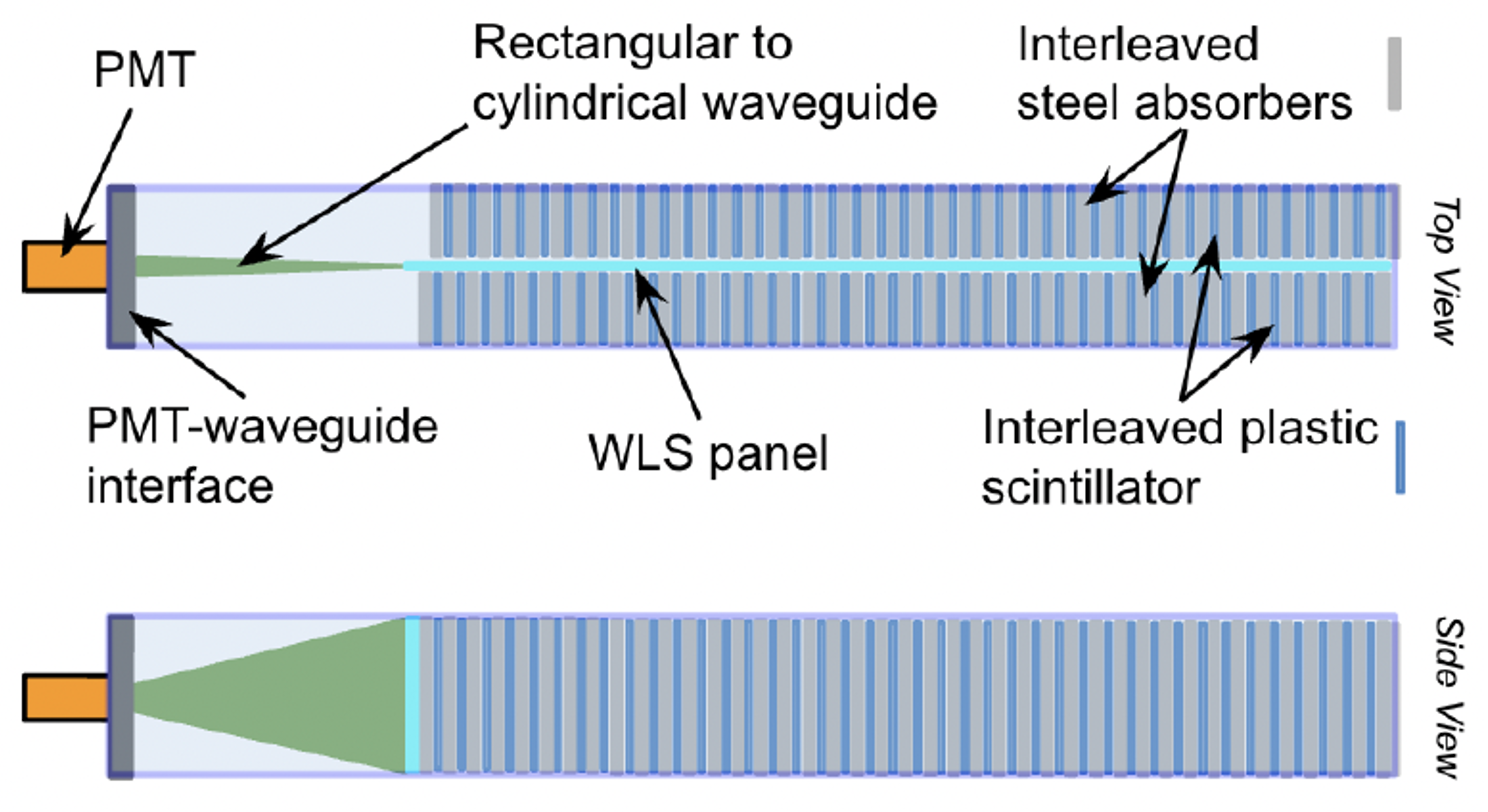}
         \caption{}
     \end{subfigure}
     \caption{Anatomy of a Hadron Calorimeter (HCAL) Module. (a) Photograph of an actual HCAL module, (b) Schematic representation of its cross-section \cite{thesisSSeeds}.}
     \label{fig:hcalmodule}
\end{figure}
In the center of each HCAL module lies a St. Gobain BC-484 wavelength shifter (WLS) bar, linked to a PMT through a rectangular-to-cylindrical light guide, as illustrated in \fig \ref{fig:hcalmodule}. Flanking the WLS are $40$ layers of \SI{1.5}{cm} thick iron absorbers, alternating with $40$ layers of \SI{1}{cm} thick 2,5-Diphenyloxazole (PPO) plastic scintillators. These layers are slightly offset from one another to enhance light output. The output spectrum of the PPO peaks at \SI{385}{nm}, aligning closely with the peak absorption wavelength of the St. Gobain BC-484 WLS, thus optimizing light collection. The generated light is collected by the two types of PMTs installed in HCAL - $192$ Photonis XP2262 2" 12-stage PMTs and $96$ Photonis XP2282 2" 8-stage PMTs. 
\begin{table}
    \caption{Important design parameters of the Hadron Calorimeter (HCAL). The angles of HCAL and its distance from the target used during different \gmn kinematics can be found at \tab \ref{tab:sbsconfig}. Additionally, HCAL detector map can be found in \fig \ref{fig:hcaldetmap}.}
    \label{tab:ch3:hcaldesignparam}
    \centering
    \begin{tabular}{lc}
    \hline\hline \\ \vspace{-2.2em} \\
       Module dimensions (wxh)  & $15.24\times15.24$ \SI{}{cm^2} \\
       Vertical gap between modules     & \SI{0.635}{cm} \\
       Horizontal gap between modules   & \SI{0.254}{cm} \\
       Segmentation (rowsxcolumns)      & $24\times12$ \\
       Vertical offset of HCAL above beamline  & \SI{75}{cm} \\
    \hline\hline
    \end{tabular}
\end{table}

A summary of the key design parameters of HCAL is provided in \tab \ref{tab:ch3:hcaldesignparam}. These parameters are crucial for the event reconstruction and simulation purposes.

\subsubsection{Signal Circuit}
\label{ssec:hcalsigcircuit}
The front-end electronics of HCAL are positioned on a mezzanine directly behind the detector, ensuring consistent access to modules across the top and bottom rows. Raw PMT signals are then transmitted from the detector to the front-end and subsequently to the DAQ bunker via \SI{100}{m} long RG-58 A/U coaxial cables. 
Here are the sequential steps individual HCAL PMT signals undergo at the front-end:
\begin{enumerate}
    \item After reaching the front-end, the signal is directed to a $16$-channel 10x Phillips Scientific 776 (PhSc 776) Amplifier. Each input channel of the PhSc 776 amplifier produces two identical copies of the amplified PMT signal. A total of $18$ such modules are installed to process signals from $288$ HCAL PMTs.
    \begin{enumerate}
        \item One copy of the amplified signal goes into the fADC module located at the DAQ bunker to get digitized and then recorded. 
        \item Another copy goes into custom-made $16$-channel splitter module.
    \end{enumerate}
    \item As the splitter splits the output of 10x amplifier into two copies, the effective gain of each copy becomes 5x. 
    \begin{enumerate}
        \item One of these copies undergoes dual discrimination using two sets of LeCroy 2313 Discriminators. Eighteen such discriminators are positioned at the front-end, while another set is situated at the DAQ bunker. Employing dual discrimination aims to mitigate any distortion of logic pulses over extended cable lengths. Subsequently, the discriminated signals are fed into five JLab-manufactured f1TDC modules, also housed in the DAQ bunker, for digitization and recording. 
        \item The remaining copy goes into the the $32$-channel UVA-120 summing modules for the formation of the nucleon trigger logic as discussed in Section \ref{ssec:sbstrig}. 
    \end{enumerate}
\end{enumerate}

\subsubsection{HV System}
The High Voltage (HV) distribution system for the HCAL PMTs follows a structure similar to the one used for BBCAL PMTs, as detailed in Section \ref{ssec:bbcalhv}. Twenty-four $12$-channel 1461N HV cards are evenly distributed between two LeCroy 1458 HV crates, accommodating HV channels for all $288$ HCAL PMTs. SHV cables exiting the LeCroy HV crates connect to twelve 24-channel HV boxes, where they are bundled into larger HV cables, each \SI{75}{m} long. These cables transport the bundled HV signals from the DAQ bunker to twelve similar HV boxes attached to the HCAL. The HV boxes attached to the HCAL then split back the bundled HV signals and transport them to individual HCAL PMTs.

The HCAL HV crates are situated in a rack within the DAQ bunker, stacked one on top of the other. The upper and lower crates run ``rpi20" and ``rpi21" HV servers, respectively, enabling software controls identical to those available for BBCAL. This includes automatic logging of the EPICS PVs associated to HCAL HV parameters in the HALOG at the start of every data run. Throughout the \gmn experiment, the operational HV values for HCAL PMTs ranged between \SI{-1200}{V} to \SI{-2300}{V}.

%
\section{Trigger and Data Acquisition}
\label{sec:trigndaq}
The \gmn experiment used a trigger-based data acquisition system. The electron trigger for the BigBite (BB) spectrometer was formed by BBCAL signals, while the nucleon trigger for the Super BigBite Spectrometer (SBS) was formed by HCAL signals. Both of these triggers were included in the Trigger Supervisor (TS). However, the BB single-arm trigger was primarily used for production. The high efficiency and stability of the BB electron trigger allowed for the application of a higher threshold, keeping the data acquisition (DAQ) rate feasible. During the commissioning phase, subsystem-specific cosmic counter-based triggers were used. Additionally, the GRINCH and HCAL had in-built LED-based trigger systems for gain-matching and the continuous monitoring of the gain stability of the PMTs. Table \ref{tab:gmntriggers} lists the triggers available during \gmn.
\begin{table}[h!]
\caption{\label{tab:gmntriggers} List of available triggers during \gmn. ``TS input" refers to the input channel number at Level 1 of the trigger supervisor (see Section \ref{ssec:daq}). ``Trigger bits" are the corresponding decimal bits. The ``prescale" values used during production reduced the trigger rate by a factor of $2^{\text{Prescale}-1}$ (a prescale of -1 disables the trigger). The ``rate" column shows the range of prescaled trigger rates observed during production. For more details on the BB electron trigger and the SBS nucleon trigger, refer to the text.}
\centering
\begin{tabular}{c c c c c c}
\hline\hline
TS    & Trigger & Prescale & Rate  & Description   & Purpose\\ 
input & bits    &          & (kHz) & of the trigger&        \\ \hline
1 & 1  & 0  & 1-6  & BB electron (BBHI) & Physics \\ 
2 & 2  & -1 & 0    & SBS nucleon        & Calibration \\ 
3 & 4  & -1 & 0    & BB-SBS coincidence & Unused \\ 
4 & 8  & -1 & 0    & Left HRS           & Unused \\ 
5 & 16 & 5  & 0.01 & GRINCH LED         & Calibration \\ 
6 & 32 & 0  & 0.01 & HCAL LED           & Calibration \\ 
7 & 64 & -1 & 0    & BB electron (BBLO) & Unused \\ 
\hline\hline
\end{tabular}
\end{table}

\subsection{BB Electron Trigger}
\label{ssec:bbtrig}
The BB electron trigger is implemented at the BBCAL front-end using various NIM modules, as shown in Figure \ref{fig:bbcalfend}. Signals summed over individual SH and PS rows are combined logically to form the final trigger and the underlying trigger logic is as follows.
\begin{enumerate}
    \item Both PS and SH rows are divided into $25$ groups each. The groups are numbered from $1$ to $25$ for both SH and PS, running from the bottom to the top of the detectors.  
    \item All the PS groups consist of two consecutive rows. However, only the first and last $7$ SH groups consist of two rows, while the remaining groups contain three. For instance, PS group $1$ (or PG-1) is the sum of the signals from PS rows $1$ and $2$, and SH group $1$ (or SG-1) is the sum of the signals from SH rows $1$ and $2$, and so on. Similarly, PG-8 is the sum of PS rows $8$ and $9$, but SG-8 is the sum of SH rows $8$, $9$, and $10$ (see \tab \ref{tab:bbcaltrigsum}).
    \item The corresponding SH and PS groups are then summed together to form $25$ trigger sums. For instance, sum of SG-1 and PG-1 forms SC 1-2, the first trigger sum, and SG-25 and PG-25 are combined to form SC 25-26, the last trigger sum.
    \item Finally, a global OR of the $25$ trigger sums form the BB electron trigger.
\end{enumerate}

\begin{table}[h!]
    \caption{\label{tab:bbcaltrigsum}List of BigBite electron trigger sums formed by the Shower (SH) and Pre-Shower (PS) rows. See text for details on implementation.}
    \centering
    \begin{tabular}{lll}
    \hline\hline\vspace{-1.1em} \\ 
        Trigger Sums & SH $\&$ PS Groups & Associated SH $\&$ PS Rows  \vspace{0.2em} \\ \hline \vspace{-1.1em} \\
        SC 1-2 & SG-1 + PG-1 & SH-1 + SH-2 + PS-1 + PS-2\\
        SC 2-3 & SG-2 + PG-2 & SH-2 + SH-3 + PS-2 + PS-3\\
        SC 3-4 & SG-3 + PG-3 & SH-3 + SH-4 + PS-3 + PS-4\\
        SC 4-5 & SG-4 + PG-4 & SH-4 + SH-5 + PS-4 + PS-5\\
        SC 5-6 & SG-5 + PG-5 & SH-5 + SH-6 + PS-5 + PS-6\\
        SC 6-7 & SG-6 + PG-6 & SH-6 + SH-7 + PS-6 + PS-7\\
        SC 7-8 & SG-7 + PG-7 & SH-7 + SH-8 + PS-7 + PS-8\\
        SC 8-9 & SG-8 + PG-8 & SH-8 + SH-9 + SH-10 + PS-8 + PS-9\\
        SC 9-10 & SG-9 + PG-9 & SH-9 + SH-10 + SH-11 + PS-9 + PS-10\\
        SC 10-11 & SG-10 + PG-10 & SH-10 + SH-11 + SH-12 + PS-10 + PS-11\\
        SC 11-12 & SG-11 + PG-11 & SH-11 + SH-12 + SH-13 + PS-11 + PS-12\\
        SC 12-13 & SG-12 + PG-12 & SH-12 + SH-13 + SH-14 + PS-12 + PS-13\\
        SC 13-14 & SG-13 + PG-13 & SH-13 + SH-14 + SH-15 + PS-13 + PS-14\\
        SC 14-15 & SG-14 + PG-14 & SH-14 + SH-15 + SH-16 + PS-14 + PS-15\\
        SC 15-16 & SG-15 + PG-15 & SH-15 + SH-16 + SH-17 + PS-15 + PS-16\\
        SC 16-17 & SG-16 + PG-16 & SH-16 + SH-17 + SH-18 + PS-16 + PS-17\\
        SC 17-18 & SG-17 + PG-17 & SH-17 + SH-18 + SH-19 + PS-17 + PS-18\\
        SC 18-19 & SG-18 + PG-18 & SH-18 + SH-19 + SH-20 + PS-18 + PS-19\\
        SC 19-20 & SG-19 + PG-19 & SH-20 + SH-21 + PS-19 + PS-20\\
        SC 20-21 & SG-20 + PG-20 & SH-21 + SH-22 + PS-20 + PS-21\\
        SC 21-22 & SG-21 + PG-21 & SH-22 + SH-23 + PS-21 + PS-22\\
        SC 22-23 & SG-22 + PG-22 & SH-23 + SH-24 + PS-22 + PS-23\\
        SC 23-24 & SG-23 + PG-23 & SH-24 + SH-25 + PS-23 + PS-24\\
        SC 24-25 & SG-24 + PG-24 & SH-25 + SH-26 + PS-24 + PS-25\\
        SC 25-26 & SG-25 + PG-25 & SH-26 + SH-27 + PS-25 + PS-26\\ \vspace{-1.2em} \\
    \hline\hline
    \end{tabular}
\end{table}
The inclusion of three SH rows instead of two in the $11$ trigger sums formed by the middle rows of the SH and PS detectors gives more weight to the events generated within the BB acceptance. Additionally, such a design accounts for the slight mismatch in alignment between SH rows and their corresponding PS ones.
\subsubsection{Implementation}
\label{ssec:bbtrigimplement}
As outlined in Section \ref{sssec:bbcalcircuit}, signals from the SH and PS PMTs are amplified and split into two copies at the BBCAL front-end. First copies of the signals are recorded via fADCs located at the DAQ bunker, while the second copies are logically summed together to form the SH and PS groups, as discussed above. Following steps are then taken to sum the SH and PS groups to create the final electron trigger (see \fig \ref{fig:ch3:bbtrigschematic}).
\begin{enumerate}
    \item Phillips Scientific model 740 (PhSc 740) Quad Linear Fan In/Fan Out (LFI/O) modules are used to sum associated SH and PS groups. Summed signals from all the SH rows corresponding to an SH group and the signal from the associated PS group are given as inputs to one quad of an LFI/O module. The output of this quad is nothing but the corresponding trigger sum. For instance, an LFI/O quad outputs trigger sum SC 1-2 when the corresponding inputs are the summed signals from SH rows $1$ and $2$, i.e., SG-1, and the signal from PG-1. A total of $7$ PhSc 740 LFI/O modules are used to from all $25$ trigger sums. Each quad has four identical output channels out of which three are used:
    \begin{enumerate}
        \item One copy of the output gets carried to the DAQ bunker via a \SI{50}{m} long signal cable to be recorded through a fADC 250 channel. A total of two fADC modules, which are located in the BB timing hodoscope VXS crate, are used to record signals from all $25$ trigger sums. This is a part of the trigger performance monitoring system.
        \item Remaining two copies of the LFI/O outputs are used to form two identical sets of BB electron triggers - BBCAL High (BBHI) and BBCAL Low (BBLO) \footnote{Two identical sets of triggers were implemented to allow setting up a veto between them while keeping one of the triggers at a higher threshold (BBCAL High) than the other (BBCAL Low). This would help achieve a cleaner selection of pion events for E12-21-005 \& E12-20-008 experiments.}. Among these two, only BBHI was used during {\gmn} experiment. Hence, going forward, we will focus exclusively on the implementation of the BBHI trigger set.
    \end{enumerate}
\begin{figure}[ht!]
    \centering
    \includegraphics[width=0.8\columnwidth]{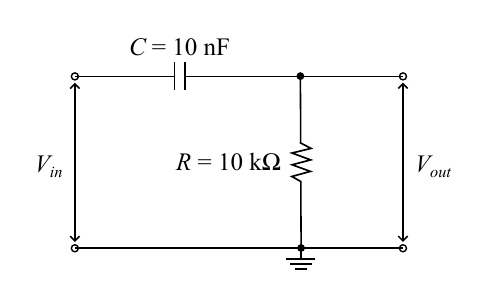}
    \caption{\label{fig:ch3:cfilter} Circuit diagram of the high-pass filter installed in the BB trigger circuit.}
\end{figure}
    \item Significant DC offsets and baseline fluctuations were noticed in some of the LFI/O outputs during the commissioning phase, affecting the stability of the trigger. Summer/Amplifier (S/A) modules sitting upstream of the signal circuit and the overheating of the NIM crates were identified as the main sources of these issues. High-pass filters were fabricated for each LFI/O output and installed in LEMO connector-based 16-channel NIM patch panels to filter out any DC offset and related fluctuations. \fig \ref{fig:ch3:cfilter} shows a circuit diagram of the filter that is used. 
    \item The filtered trigger sum signals are processed by two $16$-channel PhSc 706 Discriminators, which have been modified in-house to allow for remote threshold adjustments. Users can adjust the discriminator thresholds within a range of \SI{-10}{mV} to \SI{-1}{V} using a Python-based GUI. This GUI interacts with a DAC module located in one of the HCAL VXS crates at the DAQ bunker, connected to the discriminators via \SI{50}{m} long signal cables. Additionally, an rpi-based read-back system is implemented to continuously monitor the threshold values through the same GUI. The pulse widths of both discriminators, however, were kept constant at \SI{40}{ns}, throughout the experiment.
\begin{figure}
    \centering
    \includegraphics[width=1\columnwidth]{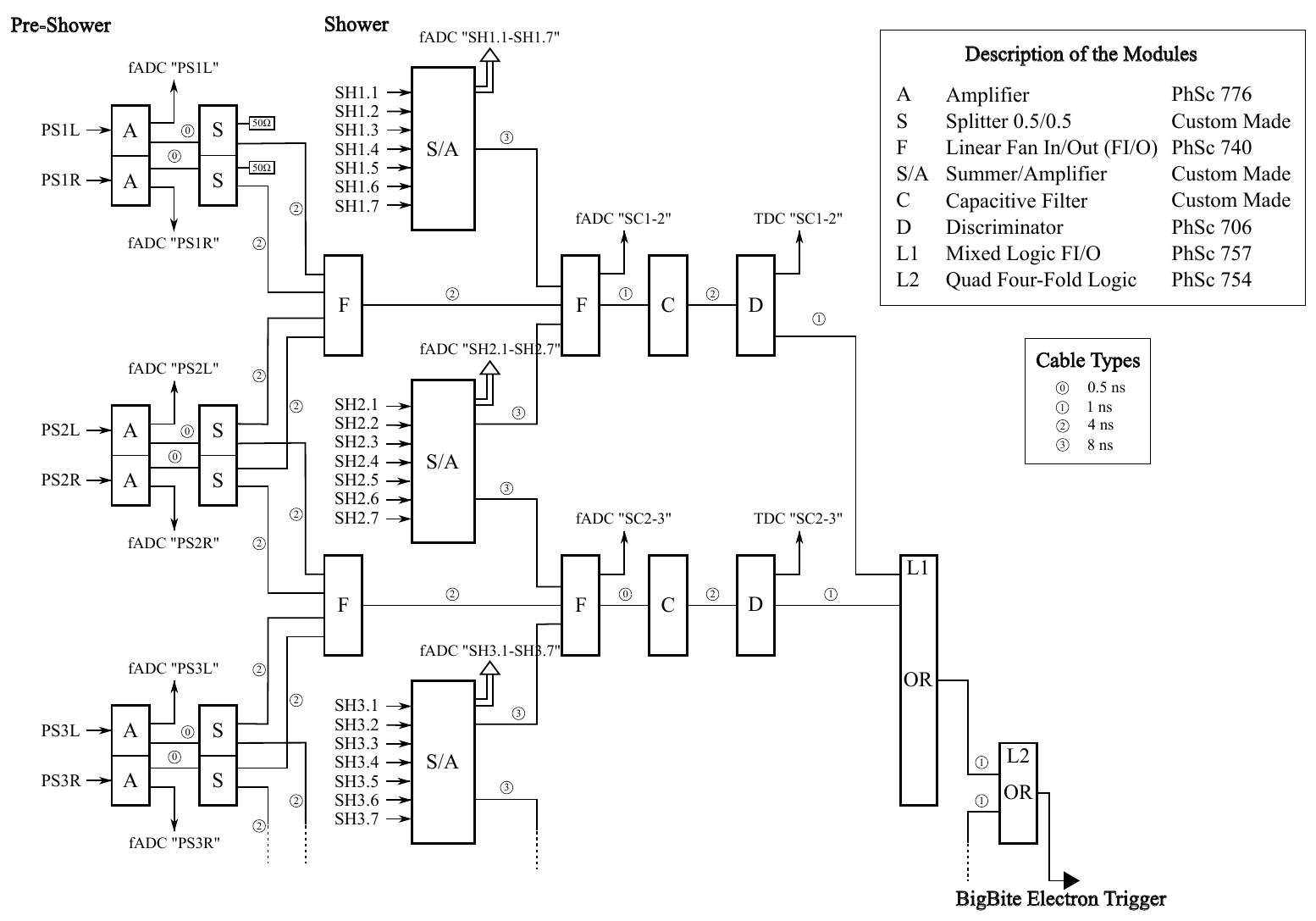}
    \caption{\label{fig:ch3:bbtrigschematic} Schematic of BigBite electron trigger logic used for \gmn. See text for details. Note: ``PhSc" is used as an abbreviation for Phillips Scientific instead of the standard ``PS" to avoid confusion with Pre-Shower.}
\end{figure}
    \item Each channel of a PhSc 706 discriminator has two identical outputs:
    \begin{enumerate}
        \item One copy of the output is carried to the DAQ bunker to record the trigger counts and time via a series of electronics including two $16$-channel PhSc 726 Level Translators, one $32$-channel SIS3820 Multi Purpose Scaler, and one $128$-channel CAEN V1190A Multihit TDC. 
        \item The second copy of the discriminated trigger sum signals go into two PhSc 757 Mixed Logic FI/O modules, which are operated in octal mode to compute logical OR of all the input signals. Outputs of these two modules are then ORed by feeding them into a PhSc 754 Quad Four-Fold Logic Unit to form the BB electron trigger.
    \end{enumerate}
    \item A copy of the final trigger is made available at the DAQ bunker via \SI{50}{m} long signal cable for the following purposes:
    \begin{enumerate}
        \item Inclusion in the Trigger Supervisor (TS) to enable triggering data acquisition by all the detector sub-systems.
        \item Recording trigger counts and time via various VME modules.
    \end{enumerate}
\end{enumerate}
%
\subsubsection{Calibration}
\label{ssec:trigcalib}
A PMT can operate within a range of gains by adjusting the HV supplied to it, and those installed in BBCAL are no exception. Therefore, if not properly gain-matched, the amplitude (and integral) of signals coming from different BBCAL PMTs can vary for the same amount of energy deposition in the corresponding lead glass (LG) counters. Summing such signals via LFI/O modules will result in a highly biased and inefficient trigger. Avoiding this situation by ensuring proper gain-matching of all BBCAL PMTs is the primary goal of the BB electron trigger calibration process. 

\subheading{BBCAL PMT Gain-Matching}
High-energy cosmic-ray muons lose energy as Minimum Ionizing Particles (MIPs) in LG material. The total energy deposition in a single SH (or PS) LG counter due to a nearly vertical cosmic event is approximately \SI{72}{MeV}\footnote{This is a rough estimate, solely based on experimental observations.}, irrespective of its position in the detector. Such uniform energy deposition across the detector during a cosmic run allows us to calculate an optimized set of HV values to match the PMT gains by aligning the corresponding signal amplitudes to a target value. 

First of all, good cosmic events are selected for analysis by requiring verticality cuts. A cosmic event in a given SH counter is considered good if:
\begin{enumerate}
    \item the top two and the bottom two neighbors of that module have good hits, while
    \item its immediate horizontal neighbors do not have good hits.
\end{enumerate}
For the PS modules, however, no cuts on the horizontal neighbors are applied, as the PS detector has only two columns. As is clear from the verticality cut definition, the modules at the edges of the SH and PS detectors are slightly less constrained than those inside. 

\begin{figure}[ht!]
     \centering
     \begin{subfigure}[b]{0.496\columnwidth}
         \centering
         \includegraphics[width=\textwidth]{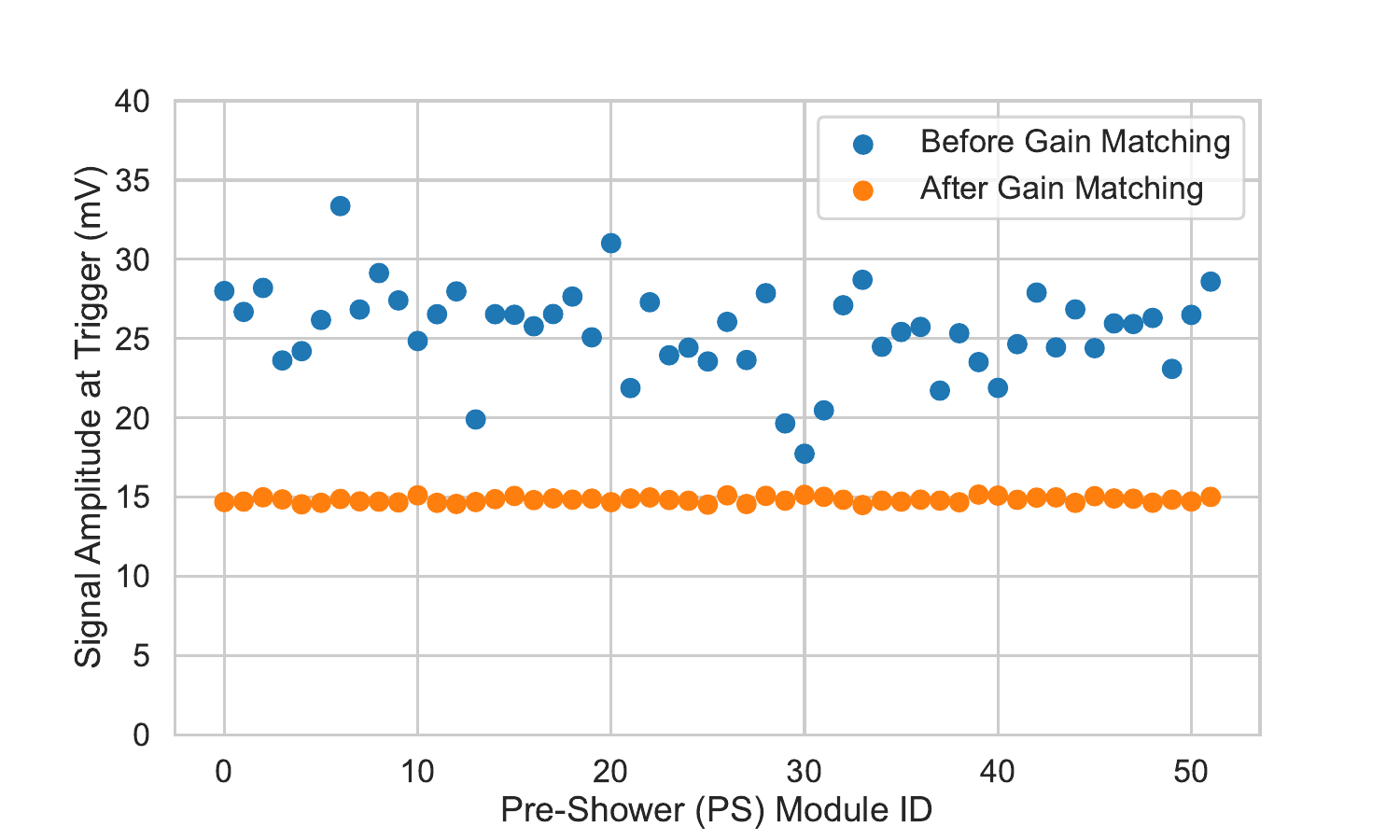}
     \end{subfigure}
     \hfill
     \begin{subfigure}[b]{0.496\columnwidth}
         \centering
         \includegraphics[width=\textwidth]{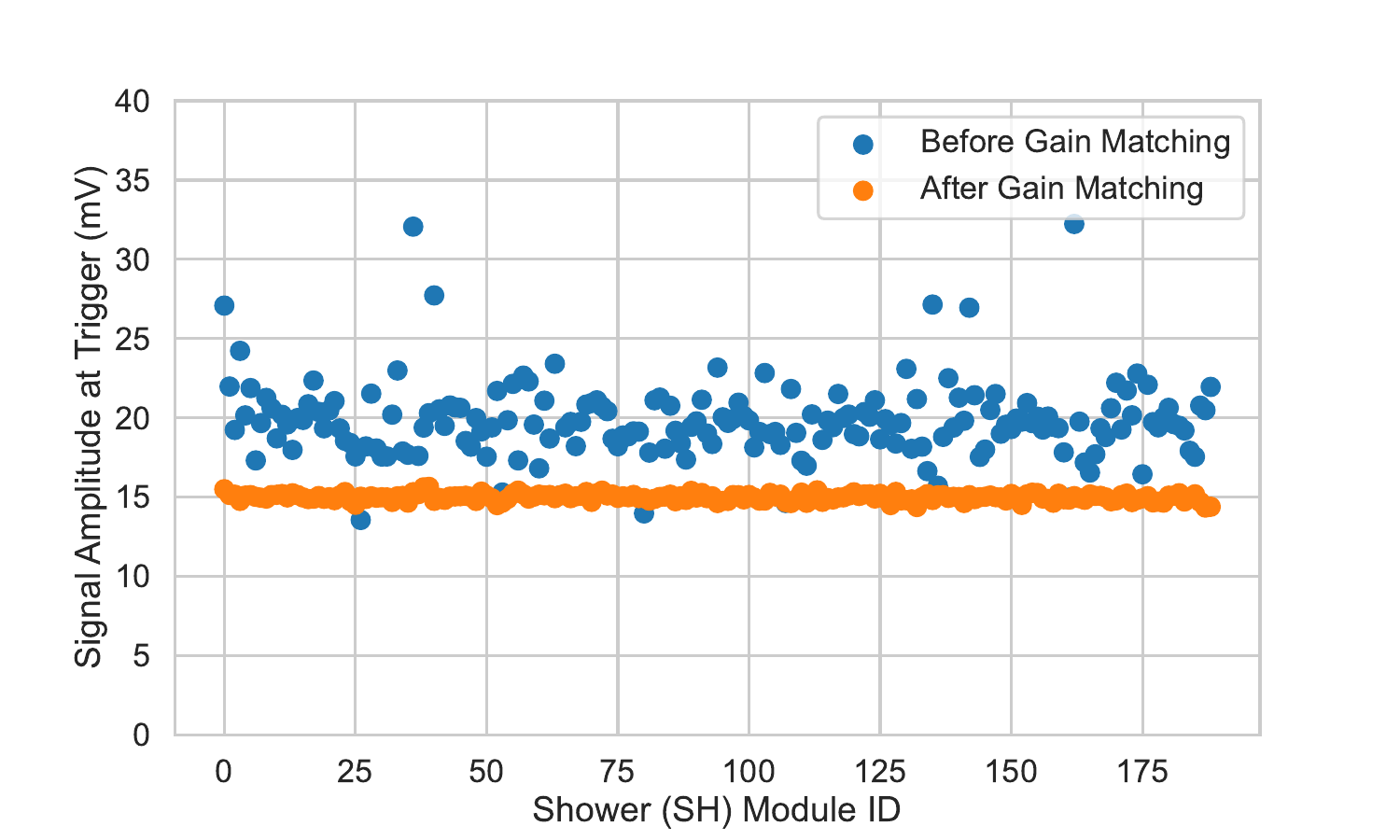}
     \end{subfigure}
     \caption{Effect of gain matching on PreShower (PS) and Shower (SH) PMTs.}
     \label{fig:ch3:bbcalgainmatch}
\end{figure}
The amplitude distributions of the good cosmic signals from individual BBCAL modules are then fitted using Gaussian functions to extract the corresponding peak positions. The desired HV to shift the cosmic signal amplitude peak position of a PMT to a target value is calculated using Equation \ref{eqn:hv}.
\begin{eqnarray}
\label{eqn:hv}
    HV_{new} = HV_{old} \left( \frac{A_{target}}{A_{old}} \right)^{\frac{1}{\alpha}}
\end{eqnarray}
where: \\
\begin{tabbing}
\hspace*{.50cm} $\alpha$ \=\hspace{1cm} PMT gain factor;\\
\hspace*{.50cm} $HV_{old}$ \>\hspace{1cm} HV before gain-matching;\\
\hspace*{.50cm} $HV_{new}$ \>\hspace{1cm} Desired HV;\\
\hspace*{.50cm} $A_{old}$ \>\hspace{1cm} Signal amplitude before gain-matching;\\
\hspace*{.50cm} $A_{target}$ \>\hspace{1cm} Target signal amplitude after gain-matching;
\end{tabbing}
\fig \ref{fig:ch3:bbcalgainmatch} shows the effect of gain matching on SH and PS PMTs. The $\alpha$ values for all BBCAL PMTs were determined through comprehensive HV scans. The approach involved collecting cosmic data at various HV settings to span the PMTs' entire operational range\footnote{$10$-$30$ mV signal amplitude ($A$) corresponding to cosmic energy deposition.}. Peak positions versus HV plots for each PMT were then fitted using a polynomial function as defined by Equation \ref{eqn:hv}, to extract the corresponding $\alpha$ parameter. Throughout \gmn, these $\alpha$ parameters remained constant for all BBCAL PMTs, confirming the stability of their gains. 

\subheading{Signal Amplitude Determination at Trigger}
As discussed above, aligning the individual BBCAL PMT signals recorded by the fADC is straightforward. However, for trigger calibration, aligning the signal amplitudes at the trigger level, specifically at the input of the LFI/O modules in the front-end, becomes necessary. Establishing a map between the signal amplitudes at the trigger level and those at the input of the fADC for each BBCAL PMT is a challenging task. The main complication arises from the fact that the Summer/Amplifier modules in the SH signal circuit, which split the SH signals into two copies for trigger formation and data acquisition, have different amplifications at the back and front output channels. Additionally, the gain of each Summer/Amplifier input channel is variable. 

\begin{figure}[ht!]
    \centering
    \includegraphics[width=0.8\columnwidth]{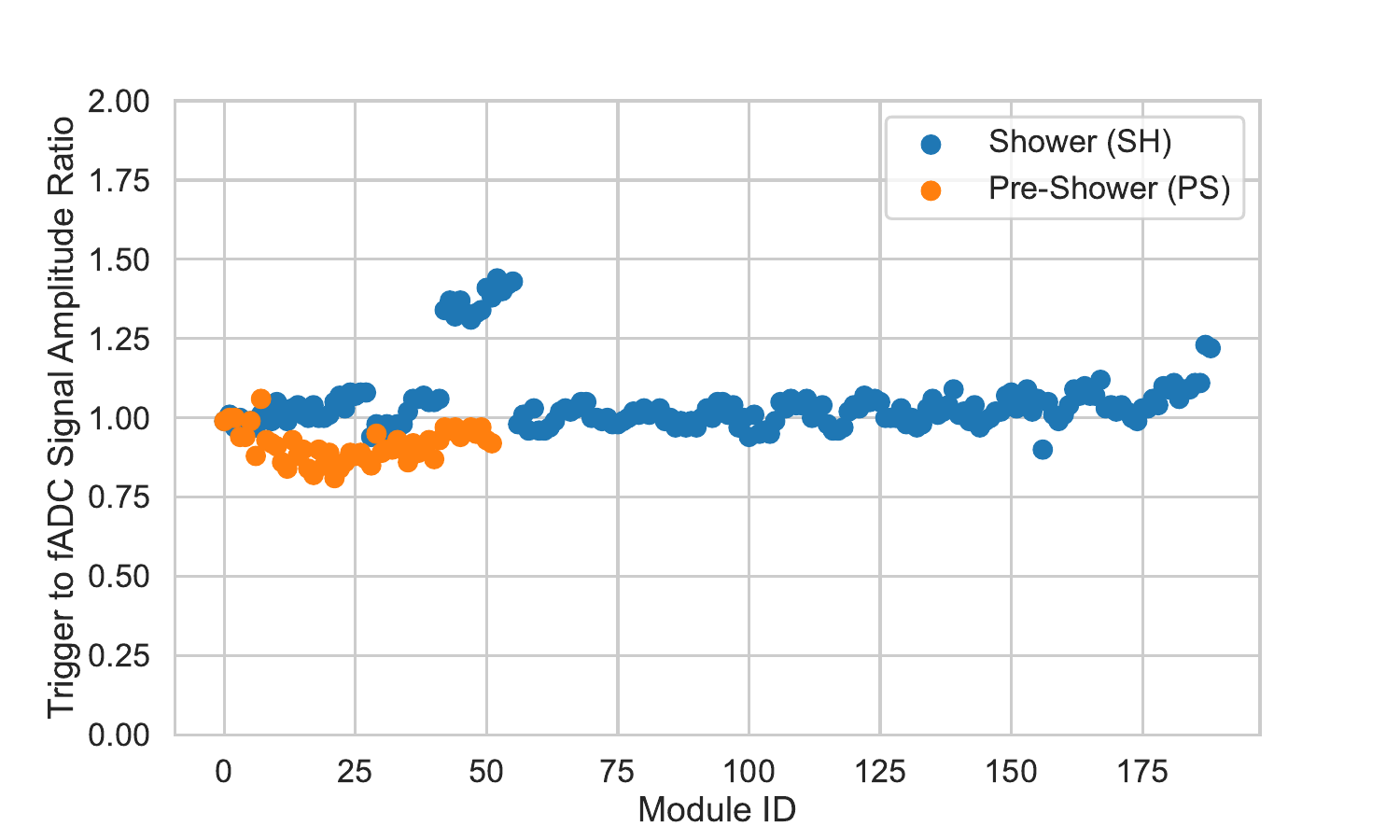}
    \caption{\label{fig:ch3:trigtofadc} Trigger to fADC signal amplitude ratios ($R_{T/F}$) across BBCAL modules.}
\end{figure}
During the commissioning phase, a test setup was established in Hall A to determine the ratio of signal amplitude at the trigger to that at the fADC for all BBCAL PMTs. The procedure had the following steps:
\begin{enumerate}
    \item SH and PS signals were replaced with pulses of known amplitude as inputs to the Summer/Amplifier modules and the PhSc 776 amplifier modules, respectively. Inputs were given to one BBCAL channel at a time.
    \item The signal amplitude at the input of the PhSc 706 discriminator was measured and recorded using an oscilloscope.
    \item A run was then started to record the corresponding signal amplitude at the fADC via the data acquisition system.
    \item Finally, the ratio of trigger to fADC signal amplitudes was computed for all BBCAL channels by taking the ratio of the values recorded in steps 2 and 3.
\end{enumerate}

Knowing these ratios for all channels, aligning the signal amplitudes at the trigger level by analyzing the fADC data is straightforward and can be achieved using Equation \ref{eqn:trigtofadc}.
\begin{eqnarray}
\label{eqn:trigtofadc}
    A^i_{Trig} = R^i_{T/F} * A^i_{fADC}
\end{eqnarray}
where:
\begin{tabbing}
\hspace*{.50cm} $i$ \=\hspace{1cm} BBCAL channel index;\\
\hspace*{.50cm} $R^i_{T/F}$ \>\hspace{1cm} Trigger to fADC signal amplitude ratio;\\
\hspace*{.50cm} $A^i_{Trig}$ \>\hspace{1cm} Signal amplitude at trigger;\\
\hspace*{.50cm} $A^i_{fADC}$ \>\hspace{1cm} Signal amplitude at FADC;
\end{tabbing}
The resulting $R_{T/F}$ values were approximately $1$ across all BBCAL modules, with some occasional outliers, as shown in \fig \ref{fig:ch3:trigtofadc}. This study allowed us to accurately calibrate the BB electron trigger using cosmic rays via BBCAL PMT gain-matching, resulting in a highly stable and efficient trigger.

%
\subsubsection{Threshold Determination}
\label{sssec:ch3:thresholddetermination}
Due to the high luminosity, over 43,000 detector channels, and significant reduction in data acquisition livetime beyond an event rate of \SI{3.5}{kHz} during the \gmn experiment, a high trigger threshold was required. This was achieved through a well-calibrated trigger system, precise threshold value determination based on physics, and reliable threshold conversion factors.

As detailed in Section \ref{sec:measurementtechnique}, the physics process of interest for \gmn is quasi-elastic (QE) electron-nucleon scattering. Thus, the trigger threshold was set according to the energy of QE electrons. The energy distribution is broad due to the large acceptance of the BB spectrometer, the lower edge of which was estimated via realistic QE Monte Carlo simulation (see Table \ref{tab:thdetermination}). The threshold was then set at least \SI{200}{MeV} below this estimate to account for calibration uncertainties and potential fluctuations in the trigger electronics.
\begin{table}[h!]
\caption{\label{tab:thdetermination} Threshold conversion factors for \gmn kinematics. {\q} is the central {\q}, $E_{beam}$ is the beam energy, $\theta_{BB}(d_{BB})$ is the BigBite central angle (target-magnet distance), $\Bar{E_{e}}$ is the average scattered electron energy,  $E_{e}$ range represents the spread of the scattered electron energy distribution, $A^{Max}_{Trig}$ is the maximum allowed trigger amplitude due to cosmic ray for the given experimental configuration to avoid saturation in the BBCAL signal circuit, $A^{Set}_{Trig}$ is the chosen trigger amplitude during experiment, and $Th_{CF}$ is the corresponding threshold conversion factor. $A^{Set}_{Trig}$ is chosen to be the nearest multiple of $5$ of $A^{Max}_{Trig}$, for convenience. A slight difference between the two is acceptable, as the estimation of $A^{Max}_{Trig}$ is conservative.}
\centering
\begin{tabular}{>{\hsptabn}c<{\hsptabn}>{\hsptabn}c<{\hsptabn}>{\hsptabn}c<{\hsptabn}>{\hsptabn}c<{\hsptabn}>{\hsptabn}c<{\hsptabn}>{\hsptabn}c<{\hsptabn}>{\hsptabn}c<{\hsptabn}>{\hsptabn}c<{\hsptabn}>{\hsptabn}c<{\hsptabn}}
\hline\hline\vspace{-1.1em} \\ 
$Q^{2}$           & $E_{beam}$& $\theta_{BB}$& $d_{BB}$& $\Bar{E_{e}}$& $E_{e}$ range (GeV)& $A^{Max}_{Trig}$& $A^{Set}_{Trig}$& $Th_{CF}$ \\ 
$(\text{GeV/c})^2$& (GeV)     & (deg)        & (m)     & (GeV)   & Low/High           & (mV)            & (mV)            & (mV/MeV)\\ \hline \vspace{-1.1em} \\
3.0  & 3.73 & 36.0 & 1.79 & 2.12 & 1.88/2.39 & 19 & 20 & 0.35 \\
4.5  & 4.03 & 49.0 & 1.55 & 1.63 & 1.43/1.86 & 24 & 25 & 0.44 \\
4.5  & 5.98 & 26.5 & 1.97 & 3.58 & 3.09/4.14 & 11 & 10 & 0.18 \\
7.4  & 5.97 & 46.5 & 1.85 & 2.00 & 1.75/2.31 & 19 & 20 & 0.35 \\
9.9  & 7.91 & 40.0 & 1.85 & 2.66 & 2.27/3.15 & 14 & 15 & 0.26 \\
13.6 & 9.86 & 42.0 & 1.55 & 2.67 & 2.29/3.25 & 14 & 15 & 0.26 \\
\hline\hline
\end{tabular}
\end{table}

This estimated threshold value is in energy units. However, as discussed in Section \ref{ssec:bbtrigimplement}, the trigger threshold is set in units of millivolts (mV), representing the minimum signal amplitude for acceptance. The conversion factor, $Th_{CF}$, between energy unit and mV is defined as:
\begin{equation}
\label{eqn:thcf}
    Th_{CF} = \frac{A_{Trig}}{57}\,\,\SI{}{mV/MeV}
\end{equation}
where $A_{Trig}$ is the signal amplitude at the trigger after gain matching, and \SI{57}{MeV} is an empirical value derived by studying the correlation between BBCAL cluster energy and the threshold value.

The choice of $A_{Trig}$ for each experimental configuration is influenced by the saturation levels of the trigger electronics, summarized in Table \ref{tab:thdetermination}. The saturation level of the Summer/Amplifier (S/A) module, at \SI{200}{mV}, is the lowest and determines the saturation of the entire BBCAL signal circuit, including the trigger system. Using this saturation value and the maximum expected scattered electron energy ($E^{Max}_{e}$) from simulation, an upper bound for $A_{Trig}$, $A_{Trig}^{Max}$, is calculated as:
\begin{equation}
\label{eqn:trigsat}
    A^{Max}_{Trig} \le \frac{5}{1.6} \times \frac{0.072}{E^{Max}_{e}} \times (200\,\text{mV}) 
\end{equation}
This calculation accounts for signal amplification at the front-end and attenuation in the long signal cable from the front-end to the DAQ bunker. $A_{Trig}^{Max}$ values for each \gmn configuration guided the selection of target trigger amplitudes for cosmic-ray gain matching of BBCAL PMTs, defining the threshold conversion factor via \eqn \ref{eqn:thcf}. A summary of the maximum trigger amplitudes and corresponding threshold conversion factors used during \gmn is presented in Table \ref{tab:thdetermination}.

\subsubsection{SBS Fringe Field Effect Mitigation}
\label{sssec:ch3:sbsfringefield}
One of the key takeaways from the above discussion is that trigger calibration is driven by the gain-matching of the BBCAL PMTs. Any instability in the gains may worsen the calibration, resulting in a highly biased and inefficient trigger. Such a situation occurred during the commissioning of the {\gmn} experiment, posing one of the biggest challenges related to data quality.

This issue was driven by the unexpectedly large fringe field effect of the SBS magnet on the BBCAL PMT gains\footnote{According to a crude field measurement performed towards the end of \gmn, the fringe field strength at the BBCAL was of the order of \SI{50}{G}. However, the BBCAL PMTs were designed to handle a maximum of \SI{35}{G}, and the mu-metal plates installed in the detector could provide shielding against an external field of up to \SI{25}{G}.}. While comparing the scattered electron energy distributions as detected by the SH and PS detectors from various runs taken at the same experimental configuration but with different SBS magnet field strengths, we noticed significant shifts in their peak positions. This would imply that the scattered electron energy is correlated with the SBS magnet field strength, which is absurd! The only explanation for this observation is that the BBCAL PMT gains are severely affected by the fringe field of the SBS magnet.

\begin{figure}[ht!]
    \centering
    \includegraphics[width=1\columnwidth]{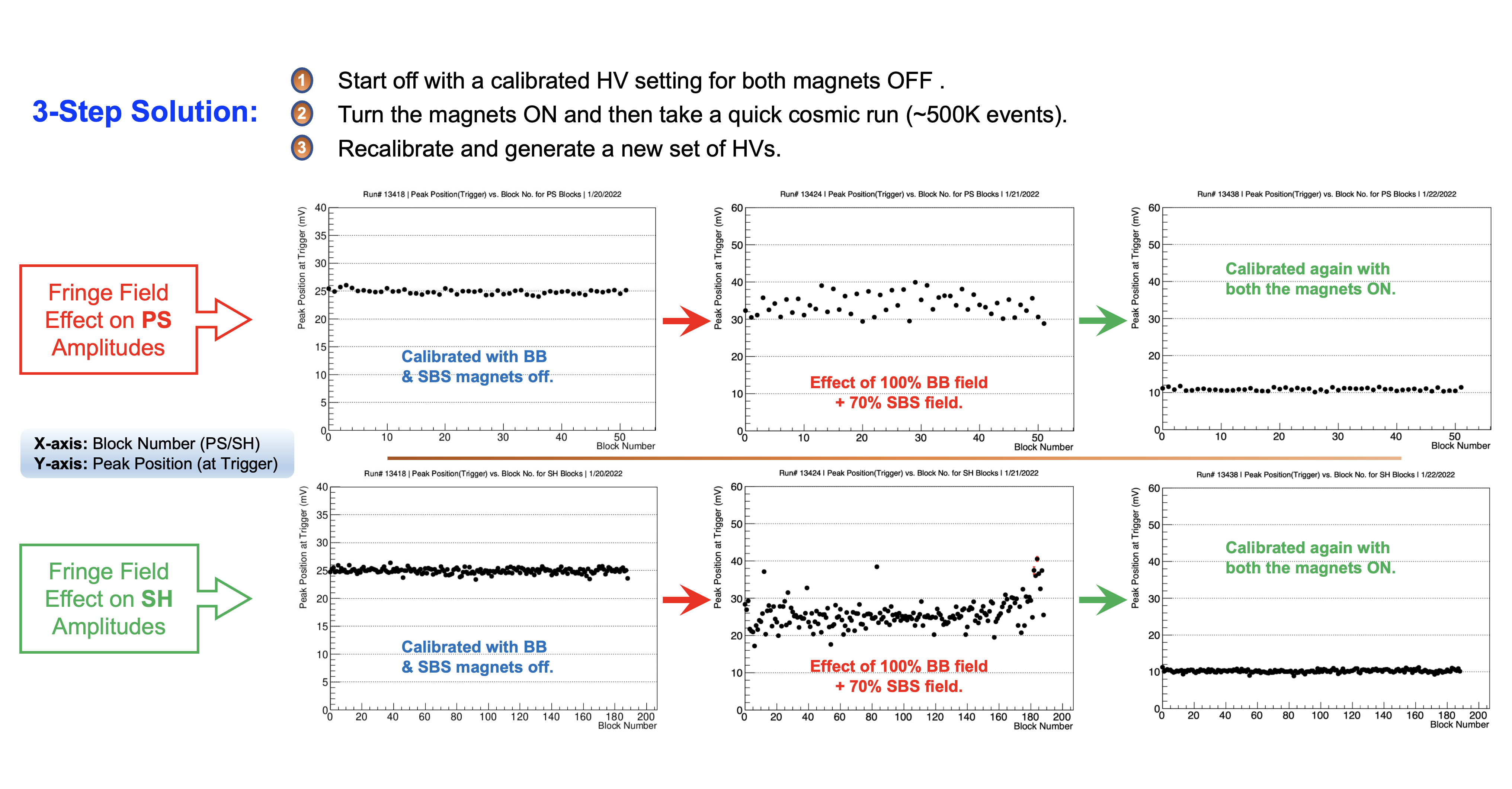}
    \caption{\label{fig:ch3:sbsfringefield}The impact of the SBS fringe field on BBCAL PMTs and the mitigation steps. Note: In step three, SH and PS signal amplitudes were aligned at a lower value (10 mV) from the initial 25 mV. Starting with a high amplitude prevents the loss of PMT signal due to the SBS fringe field, ensuring proper calibration across all BBCAL PMTs. The final low amplitude avoids the saturation of PMTs in beam. This value is kinematic-specific, as summarized in \tab \ref{tab:thdetermination}.}
\end{figure}

All BBCAL PMT gains were impacted, with some experiencing more severe effects than others. A clear correlation was observed between the PMT position in the detector and the extent of the impact. Notably, PMTs located at the edges were most affected. Overall, the PS PMTs suffered greater impact than the SH PMTs, as shown in \fig \ref{fig:ch3:sbsfringefield} likely due to the partially open and less shielded PS detector frame (see Section \ref{ssec:bbcaldetassembly}). In some cases, the impact was so severe that the signals were entirely lost! Given the proximity of the BB dipole magnet and the beam-line corrector magnets to BBCAL, we carefully assessed the effects of their fringe fields on the BBCAL PMTs as well. While the impact from the corrector magnets seemed negligible, the BB dipole magnet did affect the PMT gains, albeit to a lesser extent than the observed effects from the SBS magnet.  

As evident from the above discussion, mitigation of the SBS fringe field effect became a priority for the continuation of experimental data taking. To achieve this, a simple yet highly effective and easy-to-execute plan (see \fig \ref{fig:ch3:sbsfringefield}) was proposed, which included the following steps:
\begin{enumerate}
    \item Ensure the configuration change is fully completed. Specifically, wait until the spectrometers have been relocated to their designated positions for the {\q} point, thereby fixing the distance between BBCAL and the magnets.
    \item Take a cosmic run with the BB and SBS dipole magnets turned off. Use the data to gain-match BBCAL PMTs to a very high target signal amplitude ($\ge$ \SI{25}{mV}) at trigger, following the procedure outlined in Section \ref{ssec:trigcalib}. This ensures that the signal from any PMT is not lost when the magnets are turned on, regardless of the severity of the fringe field effect. 
    \item Turn on the SBS and BB dipole magnets to the field strengths that are required for production.
    \item Take another cosmic run, but this time with the magnets on. Upon analysis, this run provides the effective BBCAL PMT gains in the presence of the fringe fields.
    \item Use the run from the previous step to gain-match BBCAL PMTs to the target trigger amplitude, $A^{Set}_{Trig}$, as given by Table \ref{tab:thdetermination}. This ensures that the BBCAL PMTs are gain-matched in the presence of the fringe fields, confirming that the trigger is properly calibrated. 
\end{enumerate}
Once the trigger is properly calibrated in the presence of the fringe fields, we can trust the threshold values and resume normal data acquisition. This procedure was conducted whenever there was a change in the spectrometer positions or magnet field strengths during the \gmn experiment, consistently yielding the expected results.
\subsection{SBS Nucleon Trigger}
\label{ssec:sbstrig}
At the HCAL front-end, the raw PMT signals undergo amplification, splitting, and eventual summation to create an NIM-electronics-based analog nucleon trigger as outlined in Section \ref{ssec:hcalsigcircuit}. The underlying trigger logic is as follows.
\begin{enumerate}
    \item $288$ HCAL modules are grouped into eighteen $4\times4$ blocks (G1-G18). Signals from $16$ PMTs within a group are summed using nine $32$-channel UVA-120 summing modules operated in the dual sum mode.
    \item Signals from four adjacent groups are further summed together using five $4$-channel UVA-133 summing modules to form $10$ regions of $8\times8$ blocks (R1-R10) of $64$ modules. 
    \item The summed signals from $10$ regions of $8\times8$ blocks are then fed into a modified PhSc 706 discriminator. This modified discriminator module has the same features as the ones used in the BB electron trigger as described in \ref{ssec:bbtrigimplement}.  
    \item The discriminated signals are then combined to form the final trigger.  
\end{enumerate}
The poor energy resolution of HCAL makes the trigger calibration unreliable leading to ambiguity in the threshold conversion factor. Hence, the SBS nucleon trigger threshold was kept at very low value ($\approx$ \SI{-20}{mV}) throughout the experiment and the BB single-arm electron trigger was used for production data acquisition. 
\subsection{Data Acquisition}
\label{ssec:daq}
After receiving an acceptable trigger, the raw data from all the detector subsystems are recorded for the event that caused the trigger. The system governing the process of accepting and distributing triggers, transferring raw data to buffers, building events, and then recording them to file is known as the data acquisition (DAQ) system. It comprises sophisticated hardware and software tools that work in sync to accomplish these tasks. The CEBAF Online Data Acquisition (CODA) is a collection of state-of-the-art software and hardware tools necessary to implement a scalable and distributed data acquisition system as required by large-scale nuclear physics experiments at Jefferson Lab.

\begin{figure}[ht!]
    \centering
    \includegraphics[width=1\columnwidth]{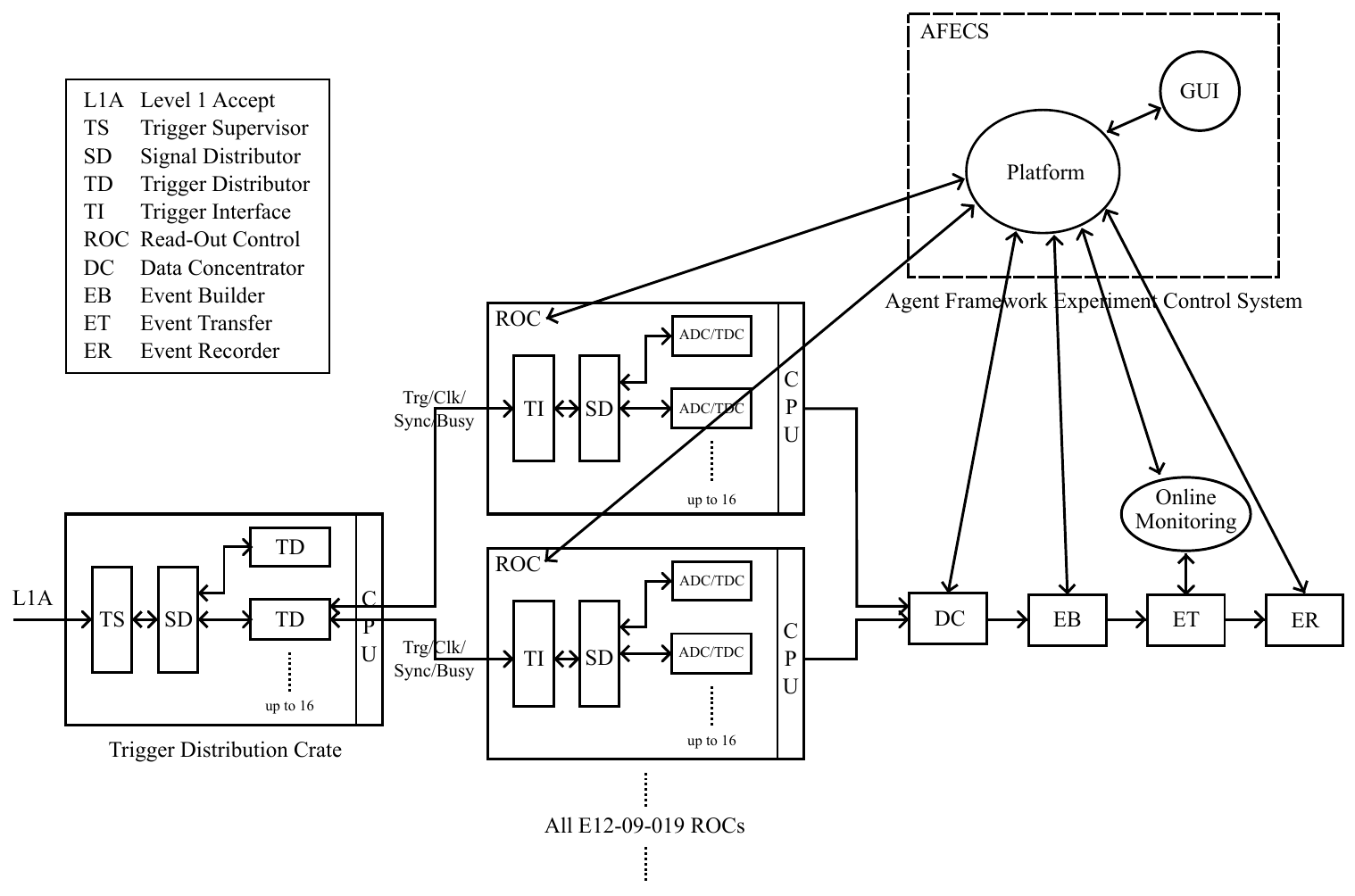}
    \caption{\label{fig:ch3:daq} High-level schematic of the SBS DAQ system. See \tab \ref{tab:ch3:daqroc} for a list of the ROCs and payload modules used during \gmn.}
\end{figure}

\subheading{The Platform in a Nutshell}
\fig \ref{fig:ch3:daq} shows a high-level schematic of the SBS DAQ system used during \gmn. It consists of multiple CODA components such as front-end electronics, including crates, ROCs, and payload modules, a Data Concentrator (DC), an Event Builder (EB), an Event Transfer (ET) system, and an Event Recorder (ER). At the front end, raw data from all detector sub-systems are processed and sent to the DC based on the accepted trigger. The DC aggregates and organizes the raw data from various sub-systems and transmits them to the EB. The EB constructs complete CODA events by consolidating the payload-specific data provided by the DC. Each payload module's raw data is preceded by a header block containing key metadata such as ROC ID, event number, event type, run number, and payload module ID. The built CODA event is then transferred to the ET system, which manages the flow of these events, distributing them to various consumers, such as online data quality monitoring or data storage via the ER. The Multi-Agent Framework for Experiment Control Systems (AFECS) provides an integrated control system environment that links the various CODA components mentioned above, ensuring coordinated operation and communication between them. A Run Control (RC) GUI is used to conveniently interact with the platform, allowing users to manage and monitor the system efficiently.

\subheading{The Trigger Supervisor System}
The DAQ front-end hardware primarily consists of electronic crates that support a highly versatile asynchronous backplane bus system, such as Versa Module Eurocard (VME) and VME-bus Switched Serial (VXS), along with their associated payload modules. The primary purpose of this hardware is to collect and process signals from the detector channels and transfer the processed data to the next stage of the DAQ system based on the acknowledged trigger signal. The Read-Out Control (ROC) system installed in each crate communicates with the Trigger Supervisor system and the CODA platform to enable this entire operation.

The trigger modules contribute to constructing the Trigger Supervisor system, which links the experiment-specific triggering system and the ROCs \cite{842691}. The system utilized during \gmn comprised a single Trigger Supervisor (TS) module, a couple of Trigger Distributor (TD) modules, and eight Trigger Interface (TI) cards, each corresponding to a front-end crate. The TS and TD modules were installed in a VME crate, specifically sbsTS21, while the TIs were installed in their respective crates.

The TS module serves as the central control point for data acquisition activity, accepting and prescaling multiple sources of triggers, including both physics and calibration types. The TD module receives trigger, clock, and synchronization signals from the TS and distributes them to up to eight crates in the front-end data acquisition system via a custom serial link to the TIs, maintaining fixed latency from the TS. It also receives status information from each crate, generating a ``busy" signal and tracking various status registers of the front-end crates and communicate those to the TS. The TS accepts new trigger only if no trigger is currently being latched and the ``busy" signal is not asserted from any of the front-end crates.

The Trigger Supervisor (TS) allows for three logical trigger levels, with Level 1 (L1) capable of processing up to twelve independent trigger streams simultaneously. During \gmn, only L1 was used, with seven inputs as listed in Table \ref{tab:gmntriggers}. Once the TS is ready to accept new triggers, any L1 trigger passing the prescale circuitry latches the TS and starts a coincidence time window. Any other L1 triggers occurring within this interval are also latched. A pattern of Level 1 Accept (L1A) signals is then generated to create ADC gates and TDC start/stop signals, dictating data transfer from the front-end payloads to the next stage of DAQ system.

\begin{table}
    \caption{Summary of sub-system-specific payload modules used during \gmn along with the available data types.}
    \label{tab:ch3:daqroc}
    \centering
    \begin{tabular}{lccc} \hline\hline
        Sub-system & ROC & Payload modules & Available data type \\ \hline
        GEM    & bbgemROC19   & APV/MPD        & ADC \\
        GRINCH & grinchROC7   & VETROC         & TDC \\
        TH     & bbhodoROC5   & V1190A         & TDC \\
        BBCAL  & bbshowerROC6 & fADC250        & ADC, ADC time\\
        HCAL   & hcalROC16/17 & fADC250, f1TDC & ADC, ADC time, TDC\\
        BB Trigger & sbsvmeROC21 & SIS3820     & Scaler counts \\ \hline\hline
    \end{tabular}
\end{table}
\tab \ref{tab:ch3:daqroc} presents a summary of the sub-system-specific payload modules employed during \gmn, along with the available data types. Since these modules are not radiation-hardened, they were housed in the DAQ bunker, located far from the spectrometers and shielded by thick concrete blocks. Raw signals from the detector front ends were transmitted to the DAQ bunker via long cables, either \SI{50}{m} or \SI{100}{m}, as discussed earlier in this chapter\footnote{The MPD modules for the GEMs, however, were placed in a smaller concrete bunker situated behind the BigBite spectrometer. The raw data from the GEMs were sent to these modules using shorter HDMI cables.}.

\chapter{Data Analysis}

Rigorous analysis of the raw data collected during \gmn, as detailed in the previous chapter, is necessary to extract meaningful physics results. Four major analysis steps precede the extraction of physics observables: event reconstruction, event selection, detector calibration, and realistic Monte Carlo simulation. The collected raw data from both spectrometers are first decoded and then combined to reconstruct the scattered electron tracks. Various cuts informed by physics and detector geometry are then applied to select physics events of interest. Several passes of detector calibrations are performed to achieve the best resolution. Only after these steps can reliable raw counts corresponding to \deen and \deep events be extracted. Realistic Monte Carlo simulations are performed in parallel, incorporating various physics and detector effects, and are then compared to the data to extract the corrected counts. In this chapter, we will discuss in detail the steps and methodology of data analysis that lead to the extraction of \gmn physics observables.
%
\section{Event Reconstruction}
The process of decoding experimental raw data event by event to extract integrated charge and/or time information from individual detector channels, and then combining this data to reconstruct relevant kinematic variables, is known as event reconstruction.\footnote{Processing the entire \gmn dataset ($\approx$ \SI{2}{PB}) takes a few hundred K-core-hours of CPU time.} In this section, we will discuss the important parts of event reconstruction in detail.  
\subsection{Software Tools}
\label{ssec:ch4:softtool}
The process of event reconstruction can become quite complex, especially with large and intricate spectrometers like BigBite and Super BigBite. To manage this complexity, sophisticated software tools are essential. A key tool in this context is the Hall A C++ analyzer, known as Podd \cite{Podd}. Podd is a modular and extensible software framework designed for event reconstruction in Hall A experiments, catering to both tracking and non-tracking detectors.

Podd is built on top of ROOT, a data analysis framework developed by CERN. It facilitates event-by-event analysis for a specific CODA run, maintaining a consistent workflow from reading the CODA raw data files (EVIO) to writing the reconstructed physics information to disk in ROOT file format. Written in an object-oriented style, Podd offers classes that can be inherited to develop detector-specific decoding and reconstruction code. A huge effort went into the development of event reconstruction software specific to the BigBite and Super BigBite spectrometers giving birth to SBS-offline \cite{SBSoffline}, a C++ library built on top of Podd. 

The analysis is configured using spectrometer- and detector-specific database (DB) files. These files contain various parameters that are crucial for defining the spectrometer and detector settings, including spectrometer position, detector geometry, analysis thresholds, and calibration coefficients. Some of these parameters remain constant throughout the experiment, while others do not. For instance, the detector geometry is expected to stay unchanged across different experimental configurations, but the spectrometer position will vary. Such changes are marked by appropriate timestamps in the DB files. When parsing the DB files, Podd matches these timestamps with the one in the EVIO file to read the correct values. The DB files used for the reconstruction of \gmn dataset are hosted in SBS-replay \cite{SBSreplay} repository.

Now that we are familiar with the software tools used for event reconstruction, let us provide a broad overview of the overall analysis flow.
%
%
\subsection{Analysis Flow}
The event reconstruction analysis of a given CODA run is executed by a C++ script, known as the replay script. Before starting the analysis, the following important parameters are set within the script:
\begin{itemize}
    \item Mode of decoding ADC and/or TDC data for all detectors.
    \item Order in which the detectors will be analyzed.
    \item Path to the directory containing database (DB) files.
    \item Format of the output file names and path to the output directory.
    \item Metadata associated with the run, including the run number and start time. 
\end{itemize}

Finally, the analysis is initialized by reading all the relevant database (DB) files and making the entries available globally. After initialization, the output ROOT file is created and a C++ object containing the run metadata is written to it. The ROOT file remains open throughout the event reconstruction process to enable writing out reconstructed data at any stage of the analysis.

Now, we are ready to begin the event-by-event analysis of all the CODA events stored in the associated EVIO file. The following tasks are carried out within the event loop in the specified order:
\begin{enumerate}
    \item One CODA event is read from the EVIO file. It includes a header block with important metadata, such as ROC ID, event number, event type, run number, and payload module ID, which precedes the payload module-specific raw data. 
    \item The appropriate analysis mode is determined based on the event type. There are two possible analysis modes: slow control analysis and physics analysis. The former is performed for scaler and EPICS events, while the latter is performed for physics events. We will focus on the physics analysis methodology going forward. The following steps are carried out during physics analysis:
    \begin{enumerate}
        \item \textbf{Event Decoding:} A map between the ROC payload module channel and the detector channel is established. Subsequently, pedestal and/or reference channel subtracted and calibrated ADC and/or TDC data from all the detector channels are read out. This process is repeated for all detectors, with tracking detectors processed before non-tracking detectors, and each group processed in the order specified in the replay script.
        \item \textbf{Coarse Reconstruction:} ADC and/or TDC data from all the channels of a detector is then combined based on a detector-specific algorithm to form a cluster of hits associated with the same scattering event. 
        \item \textbf{Track Reconstruction:} Once the clusters are formed for all detectors in the BigBite spectrometer, the scattered electron tracks are reconstructed to extract the corresponding kinematic variables.
        \item \textbf{Fine Reconstruction} At this stage, the process of event reconstruction for the associated CODA event is essentially complete. This enables the calculation of other important kinematic variables, such as the squared four-momentum transfer of the virtual photon and the squared invariant mass of the virtual photon-nucleon system.   
    \end{enumerate}
\end{enumerate}

Once the analysis of all the CODA events is completed successfully, a log file is created with important summaries related to analysis cuts and processing time. Typically, each CODA run has one associated EVIO file. However, each EVIO file can have multiple segments, as a new segment is created once the file size exceeds a user-specified limit.\footnote{During \gmn, each EVIO file segment was limited to \SI{20}{GB}, with the number of segments per run ranging from 1 to 500.} Podd is designed to handle multi-segment analysis, maintaining the correct order and tracking the global event number and time across the entire run.

As is evident from the above discussion, event decoding, cluster formation, and track reconstruction are key steps in physics event reconstruction. In the rest of this section, we will discuss these steps in greater detail. However, before delving into the event reconstruction methodology, it is necessary to familiarize ourselves with the various coordinate systems (Csys) used in this process.
\subsection{Coordinate Systems (Csys)}
\label{ssec:ch3:csys}
Events from different subsystems are reconstructed independently and then combined for physics analysis. To manage the complexity due to differences in design, position, and orientation, various coordinate systems (Csys) are used. Understanding these coordinate systems and their interconnections is crucial for performing the necessary coordinate transformations. Below is a brief description of the key coordinate systems used for event reconstruction:

\begin{itemize}
    \item \textbf{Hall/Vertex Csys:} The origin is at the center of the cryotarget cell, with $\vu{z}$ pointing downstream along the beamline, $\vu{y}$ pointing vertically upward toward the hall ceiling, and $\vu{x}$ pointing to beam left when looking downstream, forming a standard right-handed coordinate system.
    \item \textbf{Target Csys:} Derived from the Hall Csys by an anti-clockwise rotation of the BB angle ($\theta_{BB}$) about the Y-axis, followed by an additional anti-clockwise rotation of $90$ degrees about the Z-axis. Here, $\vu{z}$ points downstream along the BB spectrometer axis, $\vu{y}$ points to beam left when looking downstream, and $\vu{x}$ points vertically downward toward the hall floor.
    %
    \item \textbf{Focal Plane/Ideal Optics Csys:} Derived from the Target Csys by an anti-clockwise rotation of $10$ degrees, the tracker pitch angle, about the Y-axis, and shifting the origin downstream along the BB spectrometer axis to the ideal position of the first GEM layer\footnote{In the Focal Plane Csys, the distance from the front of the BB magnet to the first GEM layer is \SI{1.1087}{m}, indicating that the first GEM layer is approximately \SI{0.9}{m} from the BB magnet mid-plane.} as defined in the simulation.
    \item \textbf{GEM Internal Csys:} Follows the same convention as the Focal Plane Csys, but its origin and rotation angle about the Y-axis are based on the observed position and orientation of the first GEM layer rather than the ideal values. These parameters are determined from the geodetic survey and fine-tuned through BB optics calibration before being added to the database (DB) file for use in event reconstruction.
    \item \textbf{Local Detector Csys:} In this system, the X-Y plane aligns with the front face of the detector. The origin is at the physical center of the detector, with the $\vu{x}$ pointing downward toward the hall floor and the $\vu{y}$ pointing to beam left when looking downstream, following the right-handed convention. Cluster centroids obtained from event reconstruction are given in the Local Detector Csys.
    %
\end{itemize}

\subsection{Event Decoding}
The process of event decoding varies with the type of ROC payload module and the detector geometry. Even data from the same type of payload module can be decoded differently based on the data-taking mode. However, the basic algorithm is the same and has the following primary steps: 

\begin{enumerate}
    \item Begin loop over all the payload module channels based on the associated detector map. Each entry of the map contains the crate/ROC ID, slot number, start and end channel numbers, and an optional reference channel number. 
    \item Establish a map between the payload module channel and the associated detector channel based on the channel map provided in the DB file. A detector channel can be connected to multiple payload module channel. For instance, HCAL data is read by both fADC and f1TDC modules to measure ADC and TDC information, respectively.
    \item If the channel has an ADC pulse with an amplitude exceeding a specified threshold, the pedestal-subtracted amplitude, the integral of the pulse, and its time, referred to as the ADC time, are extracted\footnote{A detailed overview of the algorithm for ADC time calculation can be found at \cite{FADC250Manual}.}.
    \item If the channel has TDC pulse(s) with an amplitude greater than a specified threshold, extract the leading edge and/or trailing edge time, as well as the time over threshold, after subtracting the reference channel values. Then, convert these values to nanoseconds (ns).  
    \item Once the loop is complete, call the method to find a good TDC pulse per channel for this event based on its proximity to user defined good time cut. 
\end{enumerate}

Decoding GEM data involves additional nuances. For example, it requires two steps to establish the mapping between the GEM strips and the MPD channels. Initially, the APV map assigns the type of each GEM module across all five GEM detectors. Based on the GEM type  (see \sect \ref{sssec:sbsgemtypes}), a map between the APV card and the associated GEM strips is created. Each entry in the channel map then associates each APV card with an MPD channel, a VTP fiber/MPD module, and a VTP crate/ROC. Additionally, the presence of multiple types of APV data adds further complexity to the decoding process, with differences arising from the presence or absence of online common mode noise subtraction and/or zero suppression.

With the available good ADC and/or TDC data per detector channel for this event, we are now ready to form clusters based on the algorithms discussed below.
\subsection{Cluster Formation}
When a high-energy particle passes through the detector, it leaves a trail of hits in different modules. Clustering algorithms group these hits based on spatial and temporal proximity, enabling the reconstruction of the particle's path, energy, and other properties. Due to multiple scatterings, multiple clusters can form per CODA event. The best cluster is selected based on specific criteria for each detector. This section discusses the methodology for cluster formation across all non-tracking detectors in the BB and Super BB spectrometers. 
\subsubsection{Calorimeter Clustering}
\label{sssec:ch4:calcl}
Clusters for the BB Shower (SH) calorimeter and the hadron calorimeter (HCAL) are formed using the same algorithm. The process begins by creating a ``hit array" with the detector channels/modules with energy deposition\footnote{ADC integrals per channel are converted to energy units by applying the calibration constants. Details on the calibration process can be found in Section \ref{sec:ch4:detcalib}.} higher than the cluster hit threshold. Each element of the hit array is a C++ \textit{struct} containing the position, time, energy, and index of the corresponding module.

The cluster formation process begins by iterating through the hit array elements, sorting them by energy in descending order. The element with the highest energy is checked against the cluster seed threshold. If its energy exceeds the threshold, it is designated as the seed, added to a newly created array to hold the cluster elements, and removed from the hit array.

Using the ``Island" algorithm, a cluster is formed around this seed. A pointer is set to the seed element, and an inner loop begins to process the remaining hit array elements. Each element is added to the cluster if its center is within a specified radius of the current cluster element and if the difference between its ADC time and that of the seed is within a specified limit. Subsequently, the total energy $(E_{clus})$ and the energy-weighted centroid $(\overline{x},\overline{y})$ of the cluster are updated using the following equations:
\begin{equation}
\label{eqn:ch4:calclcentroid}
    \begin{aligned}
        E_{clus} = \sum_{i\in[1,N_{clus}]} E_i \\
        \overline{x} = \sum_{i\in[1,N_{clus}]} \frac{x_i E_i}{E_{clus}} \\
        \overline{y} = \sum_{i\in[1,N_{clus}]} \frac{y_i E_i}{E_{clus}}
    \end{aligned}
\end{equation}
In these equations, $(x_{i}, y_{i})$ are the coordinates of the center of the $i^{\text{th}}$ element in the cluster, $E_i$ is its energy, and $N_{clus}$ is the current size of the cluster. The ADC and/or TDC time of the cluster remain set to the time of its seed. 

\begin{figure}[h!]
     \centering
     \begin{subfigure}[b]{0.9\textwidth}
         \centering
         \includegraphics[width=\textwidth]{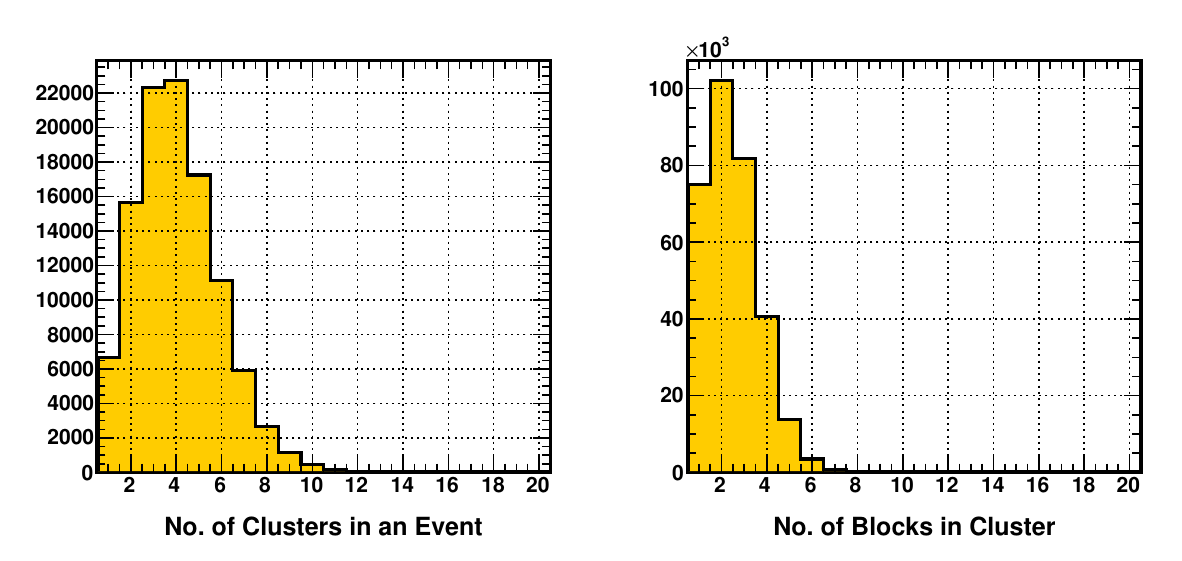}
         \caption{\qeq{3}, $\Bar{p}_{N}=$ \SI{2.4}{GeV/c}}
     \end{subfigure}
     \begin{subfigure}[b]{0.9\textwidth}
         \centering
         \includegraphics[width=\textwidth]{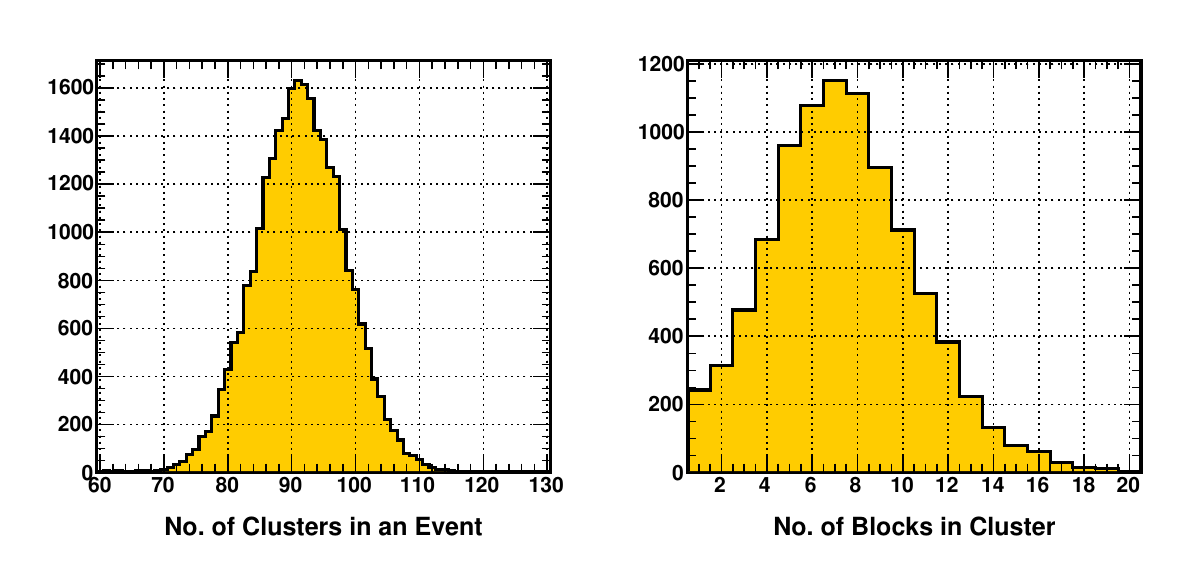}
         \caption{\qeq{13.6}, $\Bar{p}_{N}=$ \SI{8.1}{GeV/c}}
     \end{subfigure}
     \caption{Comparison of HCAL cluster multiplicity (left) and size (right) distributions using \heep events from the lowest and highest \q datasets, with average struck nucleon momentum ($\Bar{p}_{N}$) of \SI{2.4}{GeV/c} and \SI{8.1}{GeV/c}, respectively.}
     \label{fig:ch4:hclsize}
\end{figure}
Once all the eligible neighbors of the seed are added to the cluster, the pointer is incremented to point to the next element of the cluster, and the process is repeated. This continues until all eligible neighbors of all the elements already added to the cluster are included. This allows the cluster to grow in any direction within the spatial limit.

Finally, if the total energy of the cluster exceeds a specified threshold, it is added to the cluster array for the event. The process continues iteratively, looping over the remaining hit array elements to form additional clusters until the array is empty.

The ``island" algorithm allows clusters to grow in any direction within specified limits, with cluster size proportional to the energy of the incident particle. The cluster centroid $(\overline{x},\overline{y})$, formulated in Equation \ref{eqn:ch4:calclcentroid}, ensures better position resolution for larger clusters. For example, the average HCAL cluster size increases by about a factor of 4 from the lowest to the highest \q dataset, as shown in \fig \ref{fig:ch4:hclsize}, resulting in a $20\%$ improvement in position resolution.

\subheading{Pre-Shower Clustering}
At this stage, all shower (SH) clusters in this event are available, but Pre-Shower (PS) clusters have not yet been formed. Once both SH and PS clusters are available, they can be combined to form the BigBite Calorimeter (BBCAL) clusters, facilitating the energy reconstruction of the scattered electron.

PS clusters are formed by matching PS hits with available SH clusters in an event. The process begins by creating a hit array with valid PS hits, similar to what was done for SH and HCAL.

\begin{figure}[h!]
     \centering
     \begin{subfigure}[b]{0.9\textwidth}
         \centering
         \includegraphics[width=\textwidth]{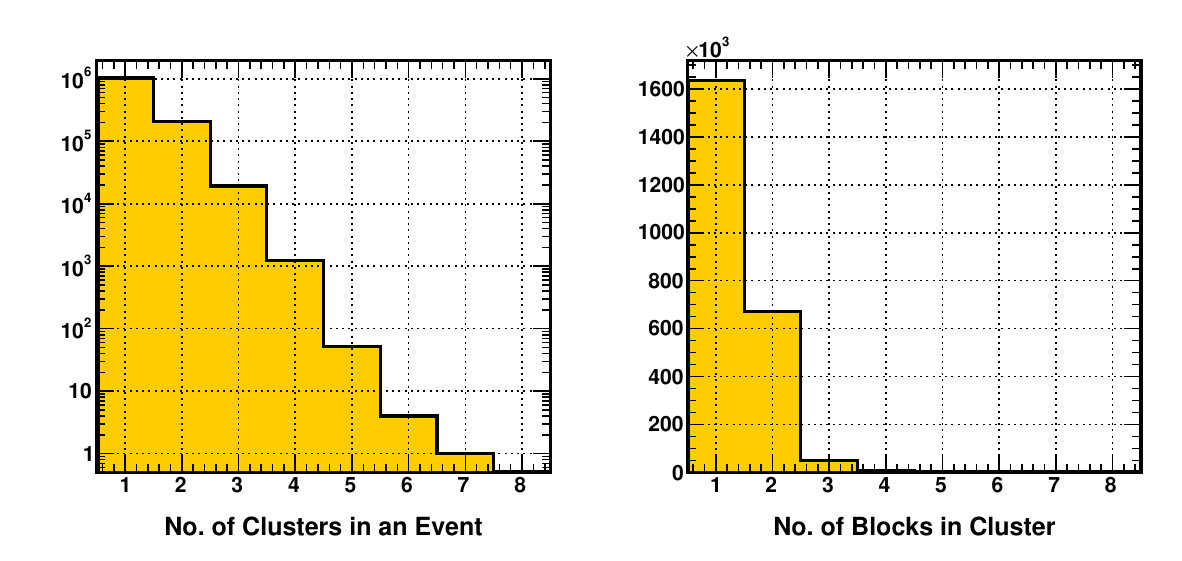}
         \caption{Pre-Shower (PS)}
     \end{subfigure}
     \begin{subfigure}[b]{0.9\textwidth}
         \centering
         \includegraphics[width=\textwidth]{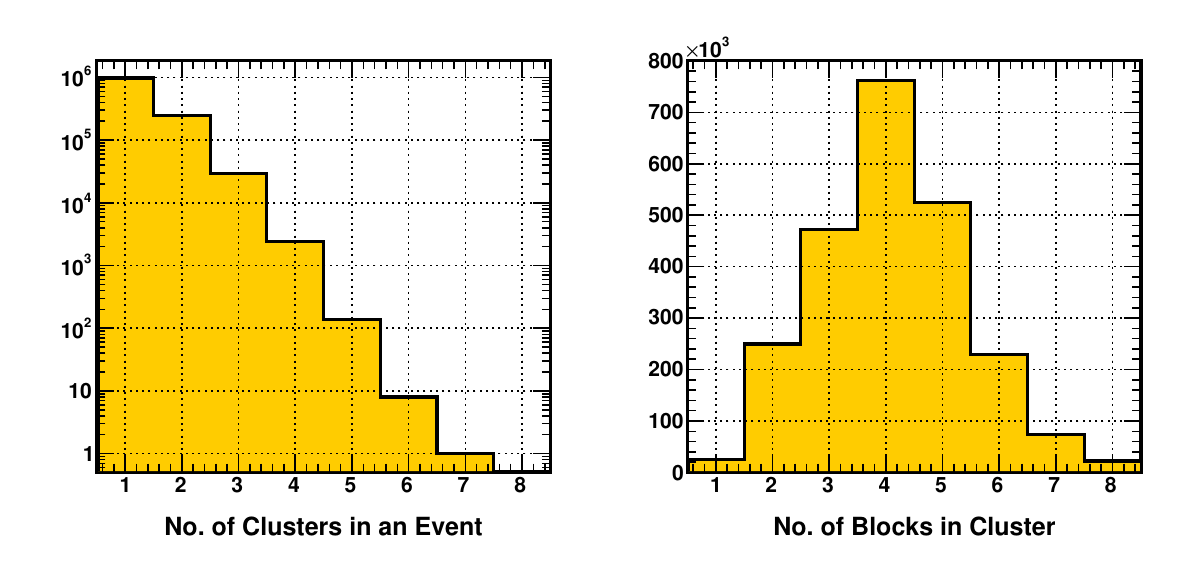}
         \caption{Shower (SH)}
     \end{subfigure}
     \caption{Comparison of Shower and Pre-Shower cluster multiplicity (left) and size (right) distributions for \heep events at \qeq{4.5} high \ep kinematics, with $\Bar{E_e}=$ \SI{3.6}{GeV}, the highest among all \gmn kinematics.}
     \label{fig:ch4:bbcalclussize}
\end{figure}
First, the algorithm loops over the SH clusters to find associated PS clusters. For each SH cluster, the position, energy, and time are retrieved. An inner loop then processes the elements of the PS hit array. If an element is not already part of any PS cluster in this event, its position and ADC time are evaluated against the current SH cluster using specified criteria. The absolute differences between the PS element center and the SH cluster centroid in both the vertical (x) and horizontal (y) directions must be within specified limits, as must the absolute difference between their ADC times.

If a PS element matches these criteria, it is marked as used to prevent it from being added to another PS cluster in the event. A PS cluster is then created with the matched PS element as the seed, and the cluster is added to the PS cluster array. If the cluster was formed in a previous iteration, the element is added to it, and the cluster centroid and energy are updated. The inner loop ends once all eligible neighbors are added to the PS cluster.

Next, the PS cluster energy is added to the energy of the current SH cluster to keep track of the matching SH and PS clusters with highest total energy in this event. This process continues iteratively, looping over the SH cluster array until all elements are checked and all possible PS clusters are formed.

\subheading{Best Calorimeter Cluster Selection}
The SH and PS cluster pair with the highest total energy in an event forms the best BBCAL cluster. However, due to HCAL's poor energy resolution ($40$-$70\%$), selecting the best cluster based solely on energy is less effective. Therefore, other methods have been explored, leveraging HCAL's high temporal ($\approx1.3$ ns) and spatial ($5$-$6$ cm) resolutions. Two particularly effective criteria are:
\begin{itemize}
    \item \textbf{In-time Criterion:} This criterion selects the highest energy HCAL cluster that is ``in-time". A cluster is considered in-time if the difference between its ADC time and that of the best SH cluster is within a specified limit.
    \item \textbf{Smallest $\theta_{pq}$ Criterion:} This criterion selects the highest energy cluster that is in-time and has the smallest $\theta_{pq}$, which is the angle between the reconstructed nucleon momentum and the momentum transfer vector.
\end{itemize}
\begin{figure}[h!]
	\centering
	\includegraphics[width=\sfig]{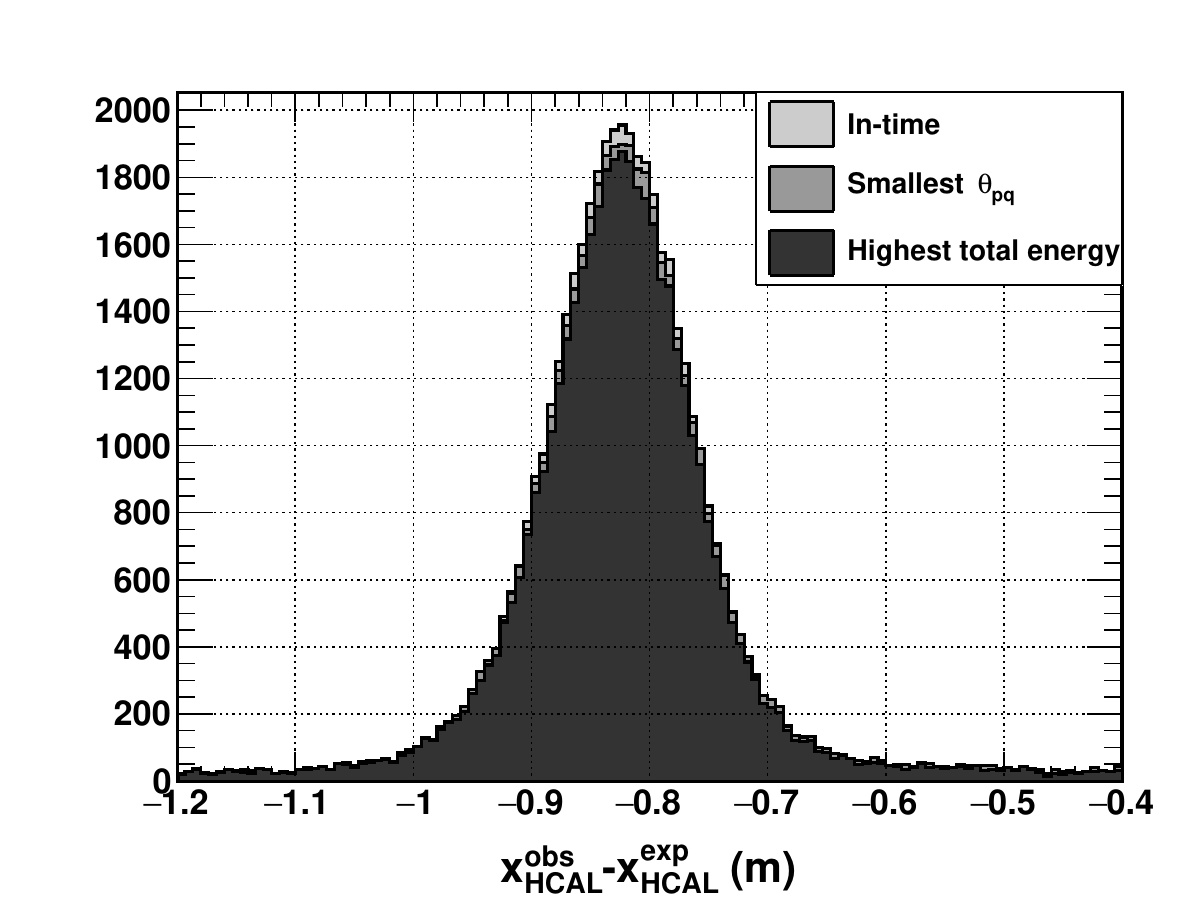}
	\caption[HCAL best cluster selection]{\label{fig:ch4:hclalgo} Comparison of various criteria for HCAL best cluster selection. The distributions represent the difference between observed (\xhob) and expected (\xhex) positions of \heep protons at HCAL in the dispersive direction at \qeq{7.4}. Notably, the ``in-time" selection criterion maximizes the yield of desired physics events.}
\end{figure}
Both criteria enhance the yields of physics events of interest compared to the highest energy cluster selection alone, as depicted in Figure \ref{fig:ch4:hclalgo}. However, the smallest $\theta_{pq}$ criterion can introduce bias due to non-negligible nucleon mis-identification probability. Consequently, the in-time criterion has been deemed superior and is used to select the best HCAL cluster in an event.

\subheading{Output Variables}
The position, energy, time, multiplicity, index, and module ID of the seed of all clusters are recorded in the Podd-generated output ROOT tree for advanced physics analysis. Additionally, the same information for all elements in the best cluster is stored primarily for calibration purposes. It is worth noting that the ``in-time" selection criterion to find the best HCAL cluster is applied in later stages of analysis. Therefore, in the Podd-generated output ROOT tree, the best HCAL cluster in an event is the one with the highest total energy.
%
\subsubsection{TH Clustering}
Timing Hodoscope (TH) bars with good TDC hits in both PMTs are used for clustering. A TDC hit is deemed good if its leading edge, trailing edge times, and Time over Threshold (ToT) fall within user-specified limits. After filtering, time walk corrections are applied.

The clustering algorithm starts by identifying local maxima from a list of good bar IDs, defining the maximum cluster size, and initializing a vector for local maxima indices. It iterates over the list of good bar IDs, calculates the ToT for each bar, and checks if the current bar’s ToT is greater than those of its neighbors, marking it as a local maximum if true. This ensures that bars with no direct neighbors or only one neighbor are also considered as potential maxima.

After identifying local maxima, the algorithm aggregates elements around each local maximum, setting minimum and maximum indices based on half-cluster size. It checks if neighboring elements are contiguous and compatible in position and time with the local maximum. If these conditions are met, the element is added to the cluster. Subsequently, the mean ToT, and the ToT-weighted time and position of the cluster are updated.

The process ensures all elements are processed and clusters are formed based on local maxima, considering proximity and compatibility in position and time.

\subheading{Best TH Cluster Selection}
The algorithm for finding the best timing hodoscope cluster involves matching a track to clusters based on their positions. It begins by obtaining the vertical (x) and horizontal (y) coordinates of the track at the TH's local coordinate system and initializes variables to track the best match and minimum x-difference.

Iterating over all available clusters, the algorithm calculates each cluster's mean x and y positions. It checks if the absolute differences between these positions and the track's coordinates fall within predefined matching cuts for x and y.

If a cluster meets these conditions, the algorithm evaluates whether the x-difference is smaller than the current minimum. If so, it updates the best match to this cluster. This process continues for all clusters, ensuring the selection of the cluster with the smallest x-difference relative to the track, provided it also satisfies the y-position condition.

\subheading{Output Variables}
The index, multiplicity, mean ToT, and ToT-weighted time and position of the best cluster are written to the output ROOT tree for advanced physics analysis. Additionally, the same information for all bars in the best cluster, along with the ToT, time, and position of the associated left and right PMTs, are recorded in the output ROOT tree primarily for calibration purposes.
\subsubsection{GRINCH Clustering}
\label{sssec:ch4:grinchclFinished}
The clustering algorithm processes a list of unused PMTs iteratively until all PMTs are assigned to clusters.

First, the algorithm selects the initial PMT from the unused PMT list. If this is the first PMT being processed and no clusters exist yet, a new cluster is created starting with this PMT. The selected PMT is then removed from the unused PMT list and added to the first cluster. Next, the algorithm examines all remaining unused PMTs, checking each one against the first PMT based on a maximum separation criterion. The separation between PMT centers is 3.1 cm in the same row and approximately 3.5 cm between adjacent rows due to column staggering. To accommodate a common event topology where several PMTs fire in a ring with a hole in the center, the optimal maximum separation is set to 7 cm. Any PMTs that meet this criterion are added to the cluster and removed from the unused PMT list. This is done in a separate pass to avoid issues with modifying the list while iterating over it.

If clusters already exist, the current unused PMT is compared against all PMTs in all existing clusters. If a match is found, the PMT is added to the corresponding cluster and removed from the unused PMT list. If no match is found, the PMT is used to seed a new cluster. It is then removed from the unused PMT list, and the algorithm searches for neighboring PMTs to add to this new cluster based on the same separation criterion.

This process ensures that at least one PMT is removed from the unused PMT list during each iteration of the main loop, and the algorithm continues until all PMTs are processed and assigned to clusters.

\subheading{Best GRINCH Cluster Selection}
The process of identifying the best cluster based on track matching begins by projecting the best track onto the GRINCH entry window. The correlation between the cluster's mean x and y coordinates and the track projection is then analyzed.

\begin{figure}[h!]
	\centering
	\includegraphics[width=\sfig]{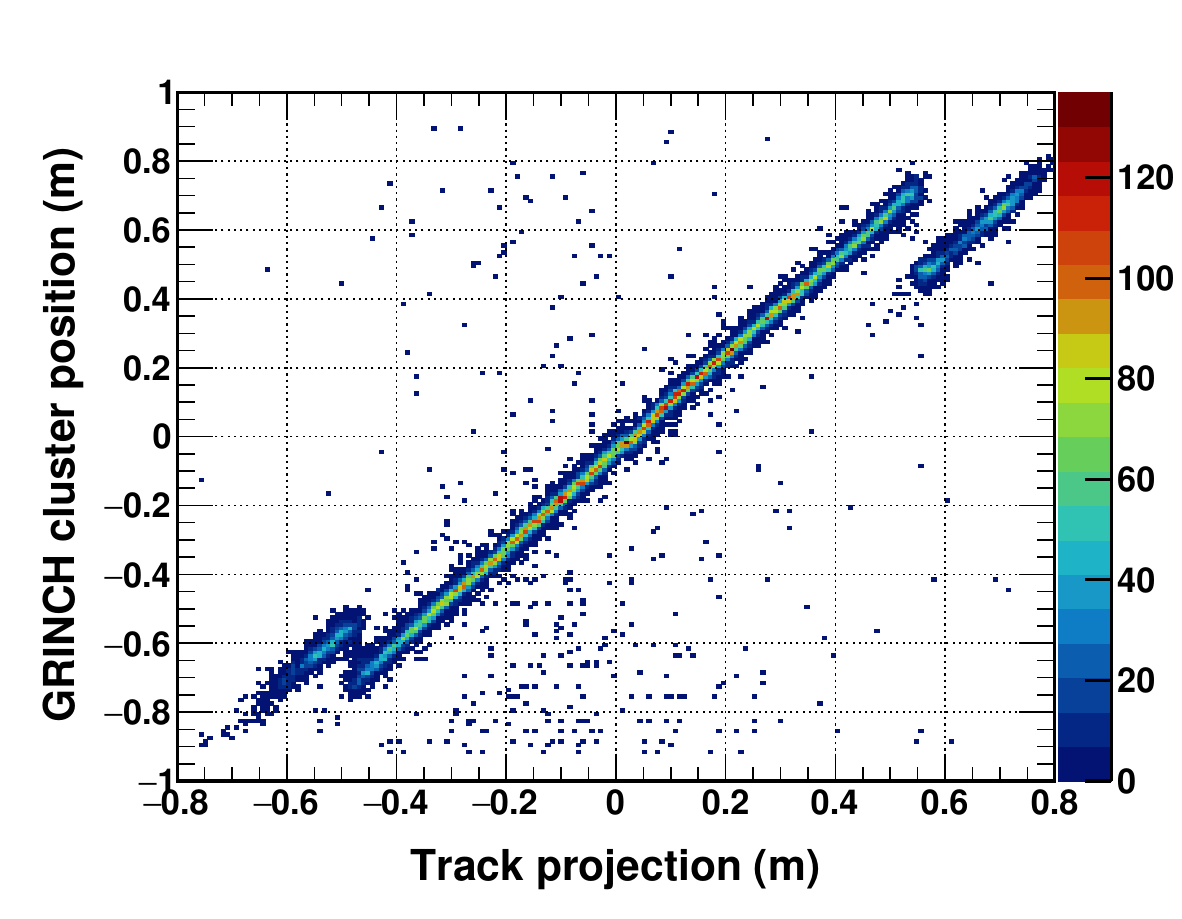}
	\caption[Correlation between GRINCH cluster position and track projection to GRINCH]{\label{fig:ch4:grinchcl} Correlation between the best GRINCH cluster mean position and the best track projection to the GRINCH entry window in the dispersive ($x$) direction. The discontinuities at the edges are due to slightly different orientation of the mirrors to maximize light collection efficiency. \heep data from the \qeq{4.5}, low \ep configuration was used to generate this plot.}
\end{figure}
To achieve this, specific matching cuts are defined for each of the four mirrors. These mirrors have designated ranges for allowed x track projections, along with a defined slope for the linear correlation between track and cluster positions, an offset, and a standard deviation. This precise definition ensures that the matching process is accurate and reliable.

When a cluster passes the track match cut for at least one mirror, it is considered a ``match". The corresponding track index is then assigned to that cluster and all the hits it contains, ensuring that the track and cluster data are properly correlated.

Additionally, matching parameters are defined for the y direction. While the correlations in the y direction are not perfectly linear and the cuts are broader, they still contribute significantly to the matching process, helping to refine the accuracy of the cluster identification.

If any track-matched clusters are found, the one with the largest number of PMTs fired is selected as the best cluster. If no track-matched clusters are found, the largest cluster from among all clusters is chosen. This approach ensures that the most significant cluster, whether track-matched or not, is identified for further analysis.

\subheading{Output Variables}
The index, multiplicity, mirror ID, mean position, time, and ToT of the best cluster are written to the output ROOT tree for advanced physics analysis. As discussed in \sect \ref{ssec:ch3:grinch}, GRINCH data is only usable/useful for \qeq{4.5} kinematics. 
\subsection{Track Reconstruction}
\label{ssec:trackreconst}
Cluster formation in the non-tracking detectors of the BB spectrometer sets the stage for reconstructing the scattered electron tracks using GEM hits. Initially, 1D clusters are formed along each GEM axis and then combined to create 2D hits. These hits across all five GEM layers are matched to form a track. However, due to very high luminosity ($\approx 10^{38}$ cm$^{-2}$s$^{-1}$), a large number of GEM strips\footnote{During \gmn, raw GEM occupancies reached as high as approximately 30\% in the worst-case scenario.} receive signals in a given CODA event, resulting in an overwhelming number of combinatorial possibilities, making track reconstruction impractical.

\begin{figure}[h!]
	\centering
	\fboxsep=0.75mm
    \fboxrule=1pt
	\includegraphics[width=0.75\columnwidth]{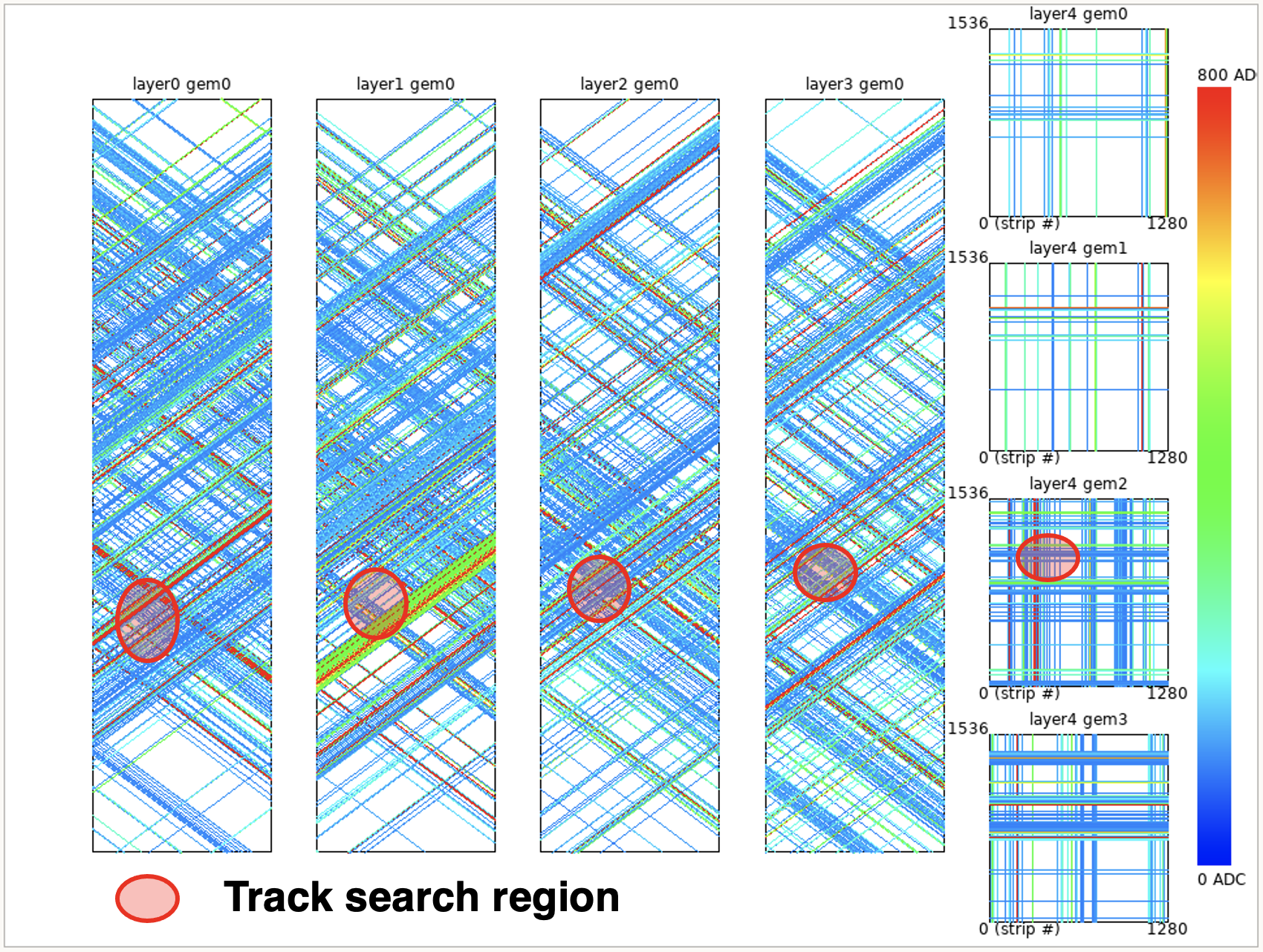}
    \caption{\label{fig:ch4:trsearch} Event display showing hits on all five GEM layers in a single triggered event for \qeq{4.5} low \ep dataset. GEM event display courtesy of Xinzhan Bai, UVA.}
\end{figure}

To address this challenge, the position and energy of the best BBCAL cluster, along with the position of the matched TH cluster, are used to define constraint regions on each of the GEM layers as shown in Figure \ref{fig:ch4:trsearch}. The track search is then performed only within these regions, reducing the search area to just $2-3\%$ of the GEM active area, thereby making reconstruction feasible. This section will discuss the methods of constraint point calculation and GEM clustering, followed by the algorithm for finding tracks.
\subsubsection{Back and Front Constraint Points Calculation}
\label{sssec:ch4:trackconstraint}
The track reconstruction process in the BB spectrometer is performed in reverse order. First, the centroid of the constraint point for the back tracker ($x_{bcp},y_{bcp},z_{bcp}$) is calculated using the following equations:
\begin{equation}
    \begin{aligned}
        x_{bcp} &= \frac{\sum_d x_{d}w_d^x}{\sum_{d} w_d^x}, \,\,\,\, d\in\{\text{SH,PS,TH}\} \\
        y_{bcp} &= \frac{\sum_{d} y_{d}w_d^y}{\sum_{d} w_d^y} \\
        z_{bcp} &= \frac{\sum_{d,i} z_{d}w_{d}^i}{\sum_{d,i} w_d^i}, \,\,\,\, i\in\{x,y\}
    \end{aligned}
\end{equation}
Here, $(x,y,z)$ are the cluster centroids of the associated detector in the transport coordinate system, and the weight factors, $w^i_d$, are the inverse squares of the corresponding resolution parameters. The SH calorimeter, with its high position resolutions ($11-$\SI{14}{mm}) in both vertical (dispersive) and horizontal directions, has the strongest contribution. In contrast, the PS's poor horizontal position resolution makes its contribution to $y_{bcp}$ negligible. The TH's contribution is optional and only considered if a TH cluster is found that matches the best BBCAL cluster.

Next, the back constraint point is projected to the upstream GEM layers by estimating the track slopes in both the vertical ($x'_{bcp}$) and horizontal ($y'_{bcp}$) directions. Assuming that the track originated from the center of the target, $y'_{bcp}$ can be expressed in terms of the back constraint points and the first-order optics coefficients as:
\begin{equation}
    y'_{bcp} = \frac{y_{bcp}}{z_{bcp} - \frac{\mathcal{M}^{y'_{tg}}_{0001}}{\mathcal{M}^{y_{tg}}_{0100}}}
\end{equation}
where $\mathcal{M}$ are the elements of optical matrix defined in Equation \ref{eqn:ch4:optics}. Estimating the vertical slope is more complex due to the momentum-dependent dispersion of tracks introduced by the BB dipole. However, BBCAL enables this by reconstructing the scattered electron's energy (equivalently, the track momentum) with high resolution ($5.4-6.5\%$). The estimated track energy, back constraint points, and the first-order optics coefficients provide enough constraints to define $x'_{bcp}$ as:
\begin{equation}
    x'_{bcp} = \frac{ \ebbcal (10^{\circ}+\mathcal{M}^{x'_{tg}}_{1000}x_{bcp}) - A(1+B\mathcal{M}^{x'_{tg}}_{1000}x_{bcp})  }{ AB(\mathcal{M}^{x'_{tg}}_{0010}-\mathcal{M}^{x'_{tg}}_{0010}z_{bcp}) + \ebbcal (1-\mathcal{M}^{x'_{tg}}_{0010}+\mathcal{M}^{x'_{tg}}_{0010}z_{bcp})}
\end{equation}
where $A$ and $B$ are the momentum reconstruction coefficients defined in Equation \ref{eqn:ch4:momrecon}, $\ebbcal$ is the BBCAL cluster energy, and $10^{\circ}$ is the pitch angle of the focal plane coordinate system relative to the target coordinate system.

Once the slopes are determined, the back constraint point is projected to the first GEM layer to calculate the front constraint point ($x_{fcp},y_{fcp},z_{fcp}$) using the following equations:
\begin{equation}
    \begin{aligned}
        z_{fcp} &= 0 \\
        x_{fcp} &= x_{bcp} + x'_{bcp} (z_{fcp} - z_{bcp}) \\
        y_{fcp} &= y_{bcp} + y'_{bcp} (z_{fcp} - z_{bcp}) 
    \end{aligned}
\end{equation}
Projection at any intermediate GEM layer is also straightforward since their positions relative to the first GEM layer are known with sub-millimeter accuracy from the geodetic survey.

Finally, the search regions are defined around the back and front constraint points based on user-specified widths. Typically, around the back constraint point, widths of approximately \SI{5}{cm} in the dispersive direction and \SI{7}{cm} in the non-dispersive direction are chosen. For the front constraint point, the width in the dispersive direction is set to be roughly twice that of the back constraint point to account for estimation uncertainty.
\subsubsection{GEM Clustering}
The availability of pedestal and common mode noise-subtracted, zero-suppressed signals from individual GEM strips during the event decoding phase, combined with predefined search regions, sets the stage for GEM clustering. Key parameters for clustering include the ADC sum, which represents the total signal strength; the ADC sum cut, a threshold for considering a strip; and the weighted strip time and position, calculated from the ADC sums. These parameters are calibrated using runs with lower occupancies to ensure accurate and efficient clustering.

The algorithm begins by performing clustering separately on each axis before combining the results into 2D clusters. Initially, individual strips along each axis are examined to identify local maxima, which are strips where the signal strength is higher than that of adjacent strips. To identify a local maximum, the signal strength is compared to adjacent strips, ensuring the signal exceeds a predefined threshold, and confirming that the strip timing is close to the expected mean times.

\begin{figure}[h!]
     \centering
     \begin{subfigure}[b]{0.496\textwidth}
         \centering
         \includegraphics[width=\textwidth]{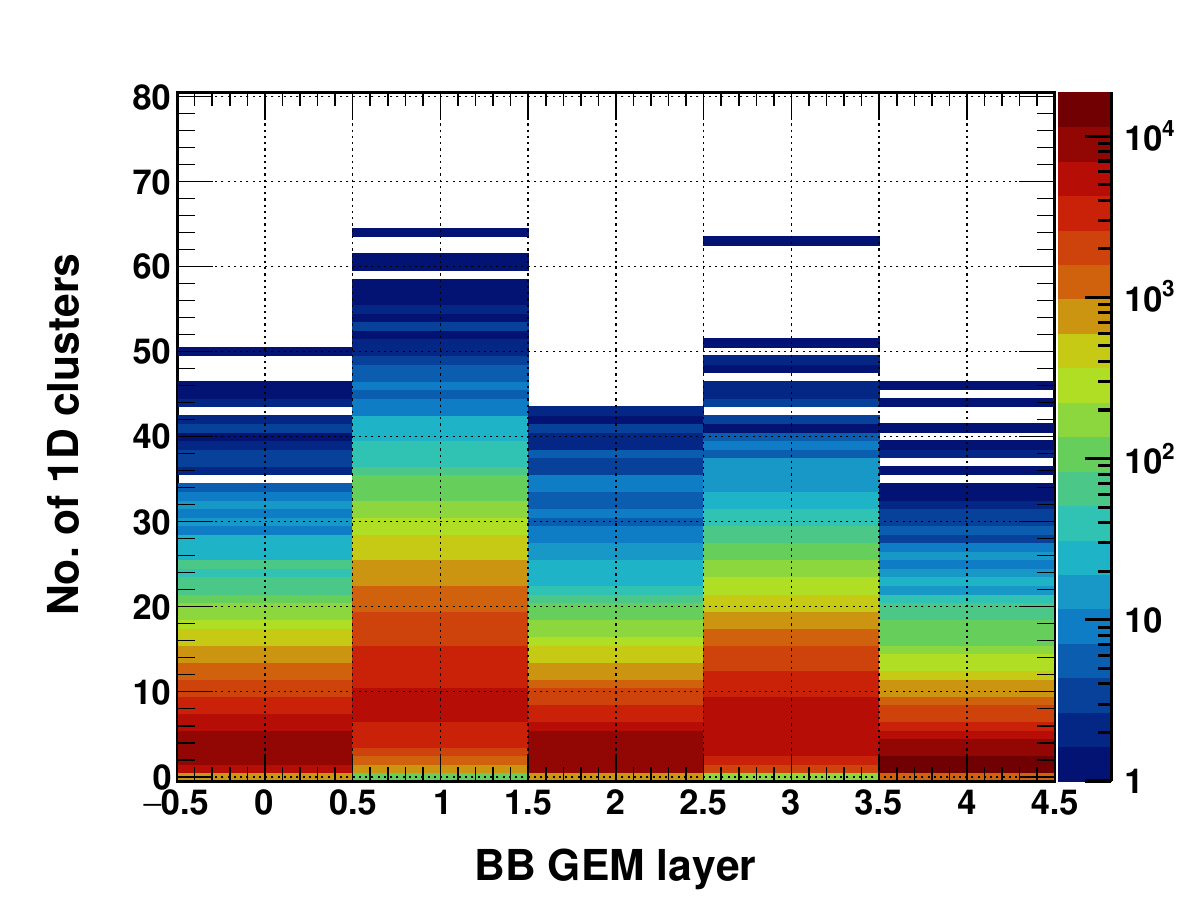}
     \end{subfigure}
     \begin{subfigure}[b]{0.496\textwidth}
         \centering
         \includegraphics[width=\textwidth]{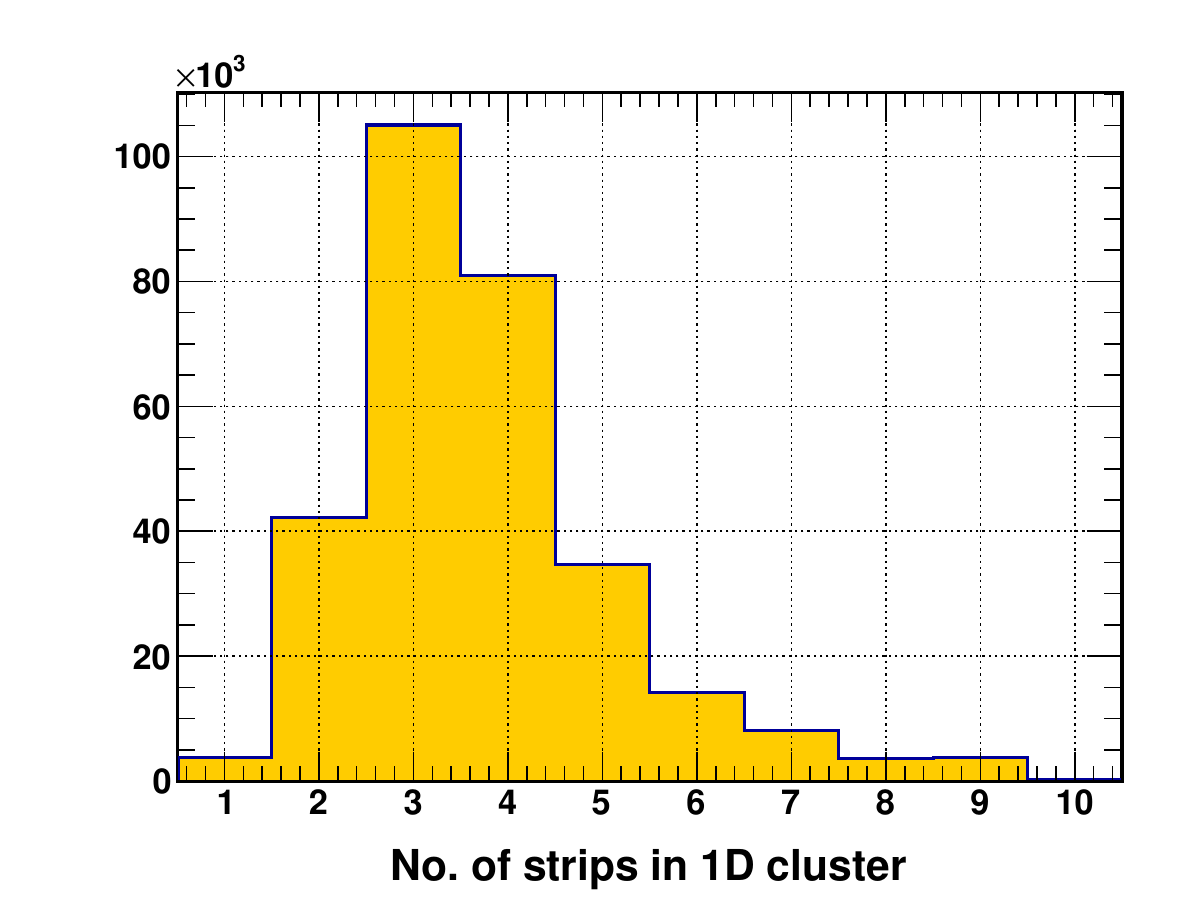}
     \end{subfigure}
     \caption{BB GEM 1D cluster multiplicity (left) and average cluster size (right) distributions using data from \qeq{7.4} on the \lh target at a beam current of $10,,\mu$A. Notably, the cluster multiplicity shown here is limited to the track search region, not the entire detector, which would exhibit a much higher multiplicity depending on the run conditions.}
     \label{fig:ch4:gemclus}
\end{figure}
Once local maxima are identified, the algorithm evaluates these peaks in relation to their neighboring strips. If two maxima are within eight strips of each other, the algorithm examines the minimum signal between them—referred to as the ``valley"—to determine peak prominence. Prominence, calculated as the signal difference between a peak and the valley, is compared against noise thresholds. Peaks with prominence below these thresholds are not considered separate clusters.

After establishing the local maxima, the algorithm expands each cluster by incorporating adjacent strips (up to four on each side) that meet specific criteria. These criteria ensure that the timing of adjacent strips aligns closely with that of the local maximum and that the signals are sufficiently correlated. For each GEM strip, six time samples per event are recorded at a 24-ns interval. These samples are used to evaluate GEM timing, calculated as the ADC-weighted mean sample time, and to compute a correlation coefficient. This coefficient, which measures the correlation of two sets of six APV25 samples in time, approaches 1 when the samples originate from the same hit. 

The iterative process continues until all qualifying strips are added, forming a comprehensive 1D cluster for each axis. The properties of these clusters—such as total signal strength, timing, and position—are calculated based on the weighted contributions of individual strips and time samples.


Next, the algorithm forms 2D clusters, or ``hits", by combining 1D clusters from both axes. Each possible combination undergoes further validation to ensure the cluster positions are within the search region, the time difference between the two clusters is within an acceptable range, the correlation between the signals is strong, and the signal asymmetry between the two axes is within an acceptable range. 2D-hit candidates passing these validation checks are compiled to identify tracks corresponding to real particles.
%
%
\subsubsection{Track Finding}
With a list of potential 2D hits for each layer, the track-finding algorithm aims to identify straight-line trajectories indicative of particle tracks. This process involves several steps:

First, each tracking layer is divided into a uniform 2D rectangular grid with a bin width of $1 \times 1$ \SI{}{cm^2}, approximately 100 times the spatial resolution along each dimension. The algorithm accumulates a list of 2D-hit candidates within each grid bin.

Next, the algorithm ensures there are at least three GEM detectors with 2D hits to form a viable track. It then iterates through all possible combinations of 2D hits from the two outermost layers (within the search region) to form straight-line projections extending to the inner layers. Each proposed track’s consistency is verified by projecting it to the target and then back to the focal plane using a forward optics model based on simulation estimates.

\begin{figure}[h!]
	\centering
	\includegraphics[width=\sfig]{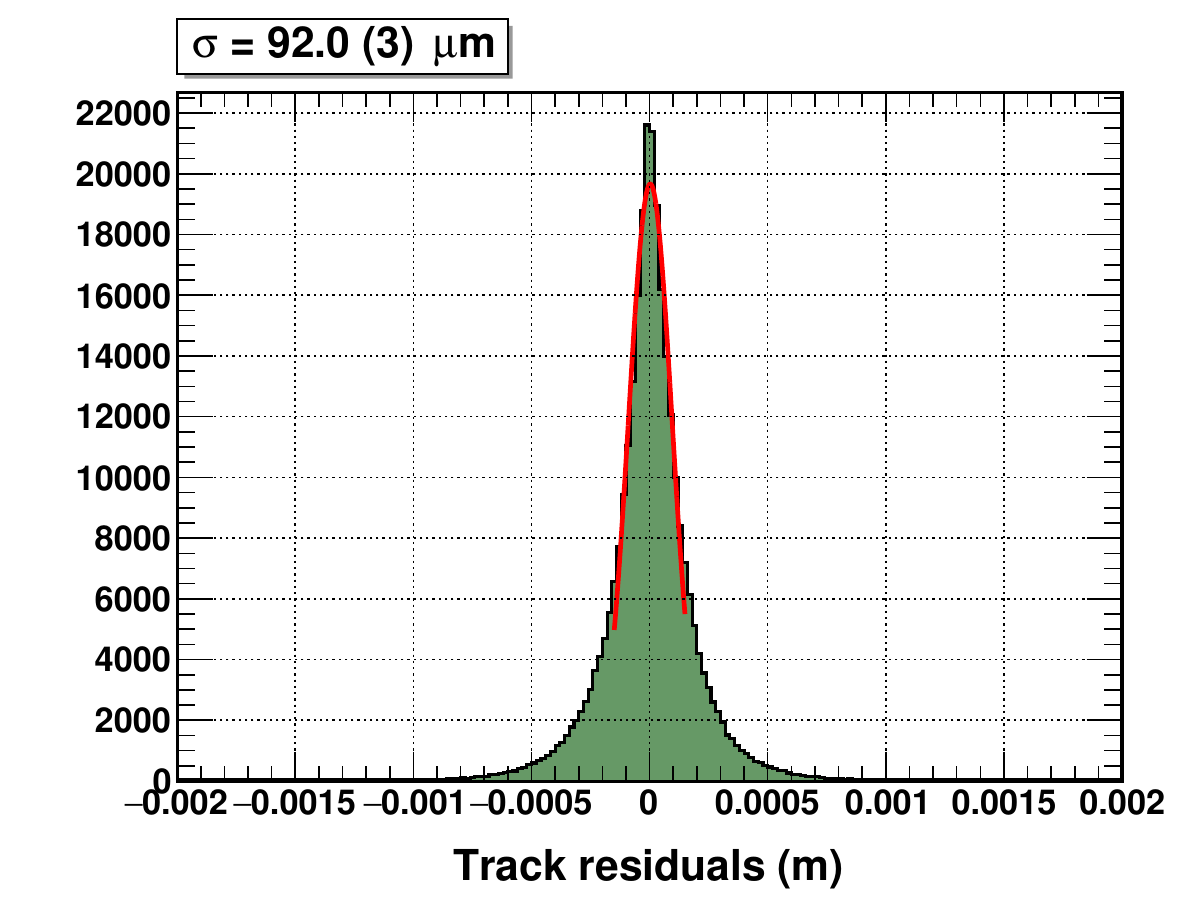}
	\caption{\label{fig:ch4:trresid} BB GEM tracking residuals, averaged over all planes.}
\end{figure}
For each combination, the algorithm evaluates every 2D hit on the intermediate layers near the projected tracks, looping over all potential 2D-hit combinations within grid bins that align with the straight-line projection from the outer layers. It identifies the 2D-hit combination with the best $\chi^2$ per degree of freedom (NDF). Basic track quality, track slope, and optics-based constraints are imposed on track candidates to eliminate poorly matched hit combinations within the search region.

In each iteration, tracks with the largest number of GEM layers with hits are prioritized. Tracks with hits on all five GEM layers are identified first, followed by those with a minimum of four hits, and so on down to three-hit tracks. This process continues until no new tracks are found, and all remaining combinations of 2D-hit candidates are exhausted.

Finally, the algorithm compiles a list of tracks that pass all criteria, designating the one with the largest number of GEM layers with hits and the lowest $\chi^2/\text{NDF}$ as the best track in the event. Figure \ref{fig:ch4:trresid} shows the tracking residuals, averaged across all planes, using \heep data from the \qeq{7.4} dataset. The final resolution achieved is approximately \SI{92}{\micro\meter} ($1\sigma$).

\subheading{Output Variables}
Arrays containing the position and slope of the track in the GEM internal coordinate system, total number of tracks in the event, and $\chi^2/\text{NDF}$ values for all the tracks are written to the output ROOT tree. The first elements of these arrays carry information regarding the best track in the event. 

\subsection{BB Optics}
\label{ssec:ch4:bboptics}
Reconstructing the track position and slope in the GEM internal coordinate system fully defines the scattered electron's trajectory within the detector stack. However, the task of reconstructing the trajectory back to the interaction vertex is still remaining. Once the trajectory bend angle is known, the track momentum can be reconstructed, leading to a full kinematic characterization of the scattered electron. The primary purpose of the BB optics program is to enable this task.

A trajectory originating from the target and passing through the BB dipole magnet is defined by its position in the dispersive ($x_{tg}$) and non-dispersive ($y_{tg}$) directions at the origin, the angles ($x'{tg}, y'{tg}$), and the deviation of its momentum from the central value ($\delta \equiv \frac{p - p_0}{p_0}$)\footnote{The effectively infinite momentum bite of BB invalidates momentum reconstruction in terms of $\delta$, requiring an alternative approach as discussed in \sect \ref{sssec:ch4:momrecon}}. Given the small raster size (\SI{2}{mm} $\times$ \SI{2}{mm}) and the relatively thin production target, assuming $x_{tg} = 0$ is reasonable. The task then becomes determining the remaining variables based on the known track position and slope at the BB detector stack.

Although analytical calculation is possible, it has several shortcomings. The transport matrix formalism is preferred, where the target variables are expressed as polynomials of the known detector variables:
\begin{equation}
\label{eqn:ch4:optics}
    \Xi_{tg} = \sum_{i,j,k,l=0}^{i+j+k+l\leq N} \mathcal{M}^{\Xi_{tg}}_{ijkl} (x_{fp})^i (y_{fp})^j (x'_{fp})^k (y'_{fp})^l, \,\,\,\, \Xi_{tg} \in{\{x'_{tg},y'_{tg},y_{tg},\delta_{tg}\}}  
\end{equation}
where ($x_{fp}, y_{fp}, x'_{fp}, y'_{fp}$) are the focal-plane trajectory coordinates and $N$ is the order of expansion. In this formalism, the BB optics is determined by the optical ``matrix" formed by the expansion coefficients $\mathcal{M}^{\Xi_{tg}}_{ijkl}$.

BB optics is sensitive to the distance between the target and the BB dipole magnet. Consequently, dedicated optics data was taken at every \gmn configuration. Although the target-to-magnet distance was the same between the \qeq{9.9\,\,\&\,\,7.4} configurations, the replacement of GEM layers at the beginning of the latter necessitated dedicated optics runs. Data was taken on very thin \ce{C} foil targets with known $z$-positions, as listed in Table \ref{tab:tgtsolid}, while a sieve slit collimator was installed in front of the BB dipole magnet. The sieve slit is a \SI{1.5}{in} thick rectangular lead plate with several circular and a few rectangular holes, as depicted in Figure \ref{sfig:ch4:bbsieve}. The rectangular holes are included to avoid orientation ambiguity. The positions and dimensions of these holes are known from surveys. The known $z$-position of the thin foil target, along with the sieve slit collimator, provides enough constraints for the electron tracks passing through the sieve holes to optimize BigBite optics with the desired resolution.

The program consists of three major steps: zero-field alignment of the GEM layers, angle and vertex reconstruction, and momentum reconstruction. Each of these steps will be discussed in greater detail in the remainder of this section. As evident, the optics program starts with fragmented information from various trajectory locations, each provided in different coordinate systems. Combining these pieces for overall reconstruction requires multiple coordinate transformations, specifically using the GEM Internal, Focal Plane, Target, and Hall/Vertex coordinate systems. Refer to \sect \ref{ssec:ch3:csys} for their definitions.
\subsubsection{Zero-Field Alignment}
\label{sssec:ch4:zeroalign}
The purpose of zero-field alignment is to align the GEM stack with the focal plane coordinate system and to determine the distance from the target to the sieve slit with greater accuracy.

\begin{figure}[ht!]
     \centering
     \begin{subfigure}[b]{0.59\columnwidth}
         \centering
         \includegraphics[width=\textwidth]{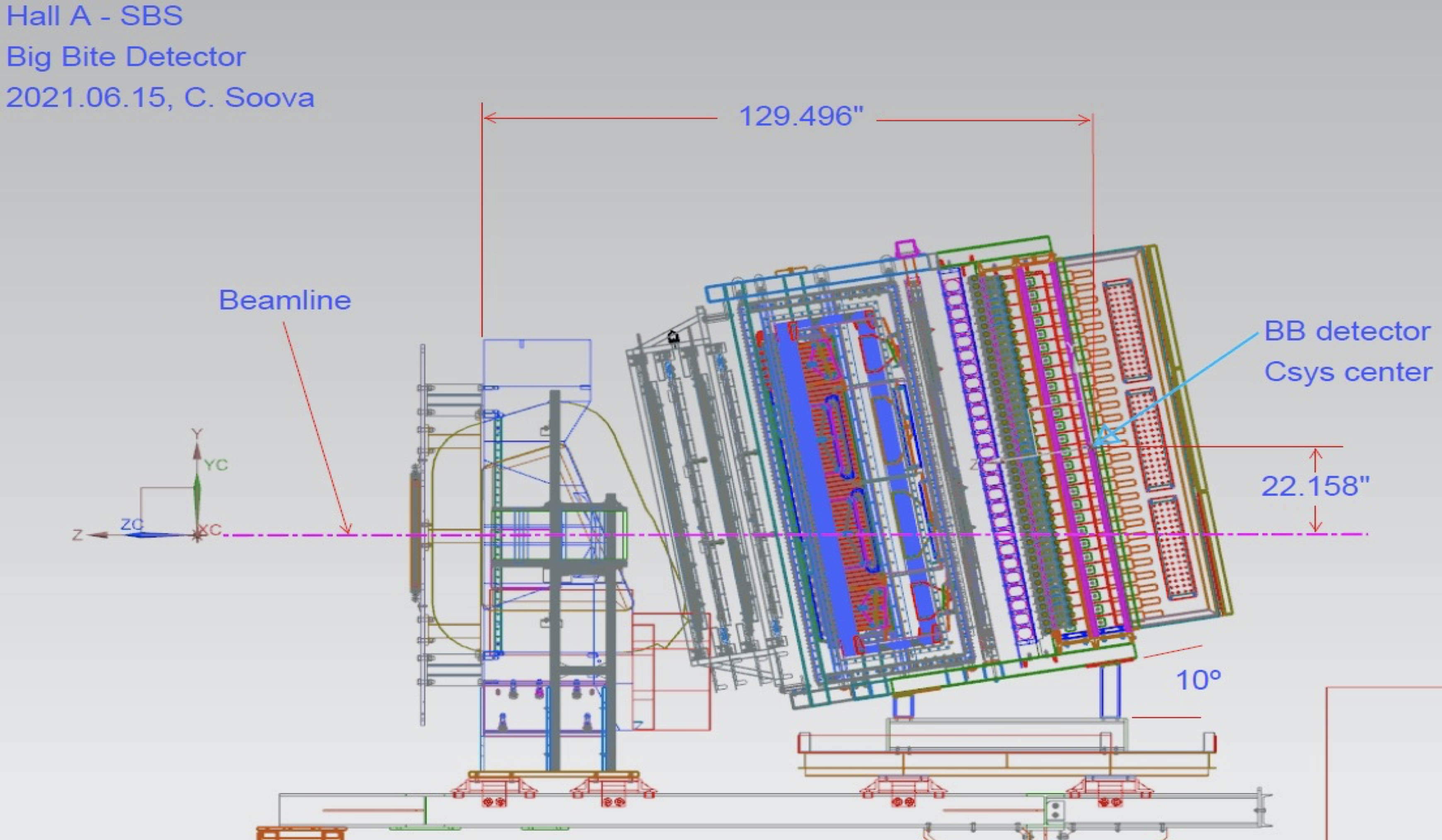}
         \caption{}
         \label{sfig:ch4:bbspec}
     \end{subfigure}
     \hfill
     \begin{subfigure}[b]{0.64\columnwidth}
         \centering
         \includegraphics[width=\textwidth]{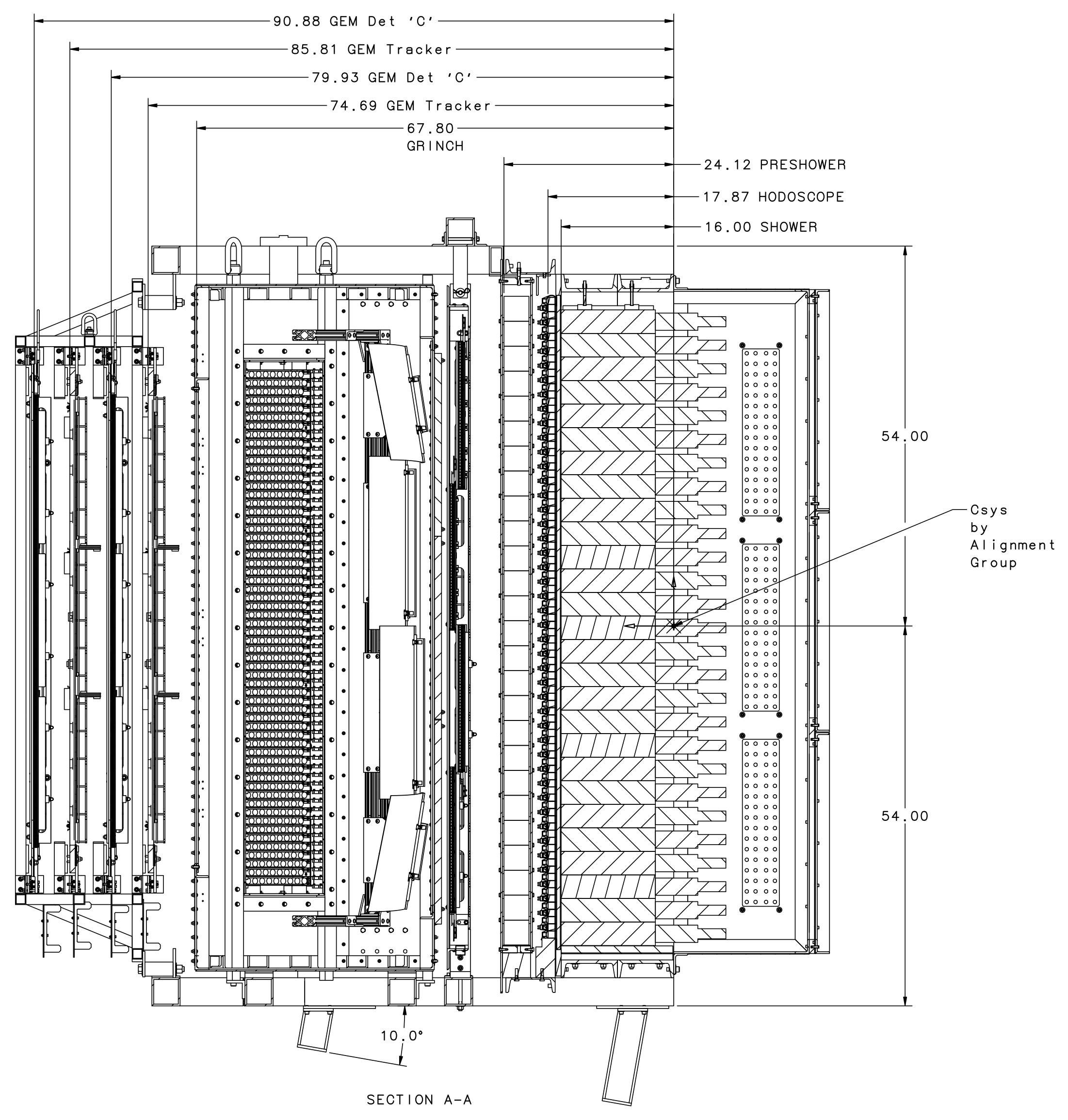}
         \caption{}
         \label{sfig:ch4:bbdetstack}
     \end{subfigure}
     \caption{Engineering drawings of the BB spectrometer. (a) BB position and dimensions relative to the beamline. (b) Relative $z$ distances of the detectors in the BB detector coordinate system (Csys).}
\end{figure}

Data from a single foil \ce{C} target is used for this purpose. The BB and SBS magnets were kept off during data collection to obtain straight tracks, which are then used for GEM alignment — hence the name ``zero-field" alignment.\footnote{During \qeq{7.4} single foil data taking, the SBS magnet was mistakenly kept on at \SI{1470}{A}. This could have influenced track directions due to the fringe field, but the effect is considered negligible.} The process relies on precise knowledge of the following parameters:
\begin{itemize}
    \item The \ce{C} foil is placed at $z_{hall}\equiv z_{tg}=0$. Due to its negligible thickness, it acts as a point source of straight-line rays originating from the global origin and passing through the sieve holes into the GEMs.
    \item The position and orientation of the magnet relative to the target.
    \item The position of the sieve slit relative to the magnet, along with the positions and dimensions of the holes in the sieve slit.
    \item The positions and orientation of the BB detector stack relative to the magnet and the positions of the detectors relative to the first GEM detector, as shown in Figure \ref{sfig:ch4:bbdetstack}.
    \item The dimensions of the GEM layers, which are internally aligned using cosmic rays before data taking.
\end{itemize}
The analysis begins by selecting good electron tracks using the following cuts:
\begin{itemize}
    \item The track should have hits on four or more GEM layers.
    \item The track's $\chi^2/NDF$ should be less than $30$.
    \item The position of the track projected on the SH should be within a \SI{3}{cm} radius of the SH cluster centroid.
    \item PS cluster energy should be greater than \SI{200}{MeV}.
    \item BBCAL cluster energy should be above a threshold chosen based on the central electron momentum.
\end{itemize}
The positions and slopes of these tracks are known in the GEM internal coordinate system which can then be transformed into the focal plane coordinate system by estimating the origin ($x^0_{fp},y^0_{fp},z^0_{fp}$) and angles ($x'_{fp},y'_{fp}$) of the first GEM layer in that coordinate system. Simultaneously, the expected positions and angles of the same tracks in the same coordinate system can be calculated by associating each track with the respective sieve hole it passed through and projecting a straight line from the point target to the first GEM layer. Discrepancies between the observed and expected values in the focal plane coordinate system are then minimized by tuning the parameters $x^0_{fp},y^0_{fp},z^0_{fp},x'_{fp},y'_{fp},$ and the target-to-sieve distance. The $\chi^2$ has the following form:
\begin{equation}
    \chi^2 = \frac{(x^{exp}_{fp}-x^{obs}_{fp})^2 + (y^{exp}_{fp} - y^{obs}_{fp})^2}{\sigma^2_{xy}} + \frac{(x'^{exp}_{fp}-x'^{obs}_{fp})^2 + (y'^{exp}_{fp} - y'^{obs}_{fp})^2}{\sigma^2_{x'y'}}
\end{equation}
where ($x^{obs}_{fp}, y^{obs}_{fp}, x'^{obs}_{fp}, y'^{obs}_{fp}$) and ($x^{exp}_{fp}, y^{exp}_{fp}, x'^{exp}_{fp}, y'^{exp}_{fp}$) are the observed and expected focal-plane coordinates, respectively. The position ($\sigma_{xy}$) and slope ($\sigma_{x'y'}$) uncertainty parameters are assumed to be \SI{0.1}{mm} and $6\times10^{-5}$, respectively, for this analysis.

Precise knowledge of the GEM position and orientation in the focal plane coordinate system, along with the target-to-sieve distance obtained from zero-field alignment, sets the stage for angle and vertex reconstruction.




%
\subsubsection{Angle and Vertex Reconstruction}
\label{sssec:ch4:angnverrecon}
Angle and vertex reconstruction involves optimizing the expansions of the $x'_{tg}$, $y'_{tg}$, and $y_{tg}$ parameters, as described in Equation \ref{eqn:ch4:optics}.
%
%

%
\begin{figure}[ht!]
     \centering
     \begin{subfigure}[b]{0.25\textwidth}
         \centering
         \includegraphics[width=\textwidth]{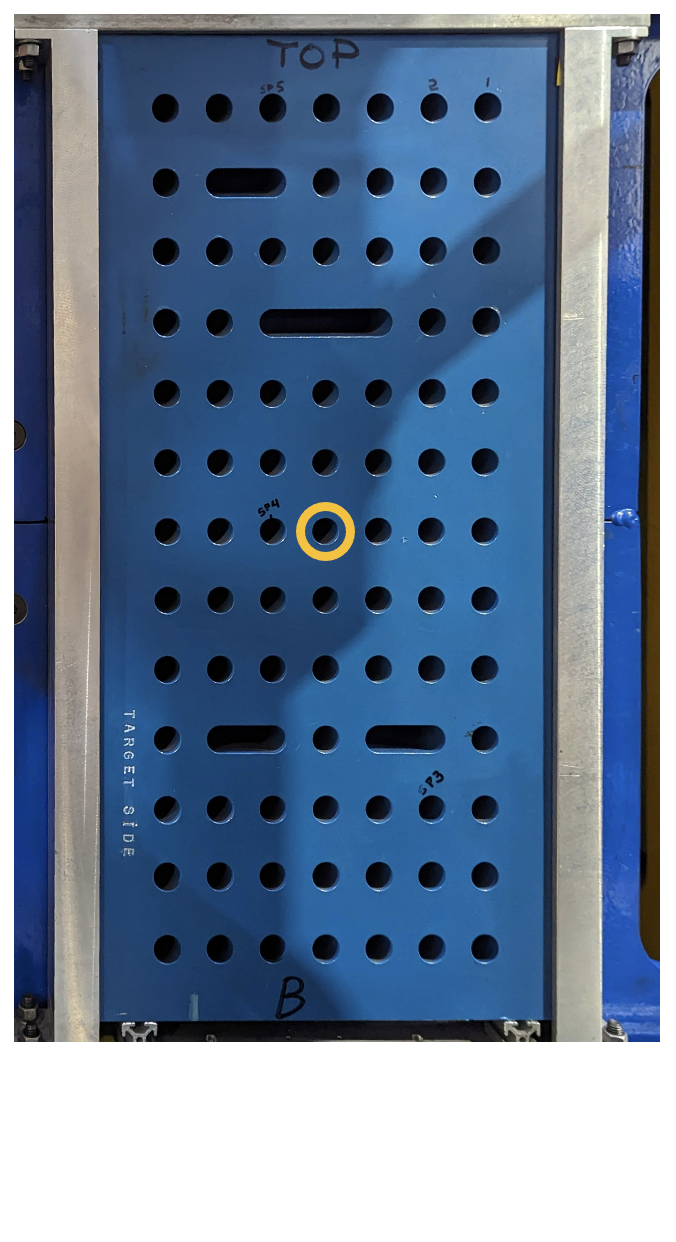}
         \caption{}
         \label{sfig:ch4:bbsieve}
     \end{subfigure}
     \hfill
     \begin{subfigure}[b]{0.65\textwidth}
         \centering
         \includegraphics[width=\textwidth]{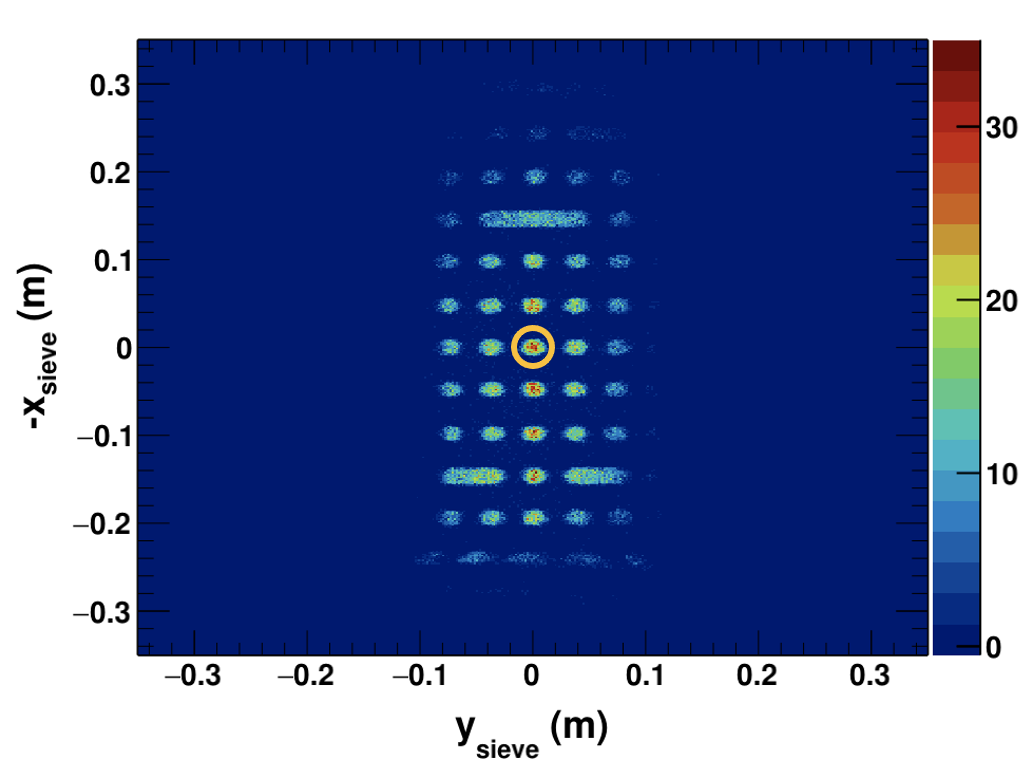}
         \caption{}
         \label{sfig:ch4:bbsieve_recon}
     \end{subfigure}
     \caption{Hole pattern of the BB sieve slit collimator. (a) The actual sieve slit installed in front of the BigBite dipole magnet. (b) Reconstructed hole patterns from projecting tracks onto the sieve slit. Electron tracks with a central angle of $49^{\circ}$, originating from five \ce{C}-foil targets, were used. The yellow circle indicates the central hole in both images for reference.}
\end{figure}
For this purpose, data is collected from four- and five-\ce{C} foil targets with the BB and SBS \footnote{The non-negligible effect of SBS magnet's fringe field on trajectory direction is accounted for in the optical matrix elements by keeping the magnet on at its production settings during the optimization process.} magnets turned on at production settings. These foils provide nine interaction points at $z_{hall} = 0$, $\pm$\SI{2.5}{cm}, $\pm$\SI{5}{cm}, $\pm$\SI{7.5}{cm}, and $\pm$\SI{10}{cm}, covering more than the entire \SI{15}{cm} thickness of the production target. In addition to the precise locations of these foils, the exact positions and orientations of the GEM layers in the focal plane coordinate system, along with the target-to-sieve distance, are known from zero-field alignment. 
%

%
\begin{figure}[h!]
	\centering
	\includegraphics[width=\sfig]{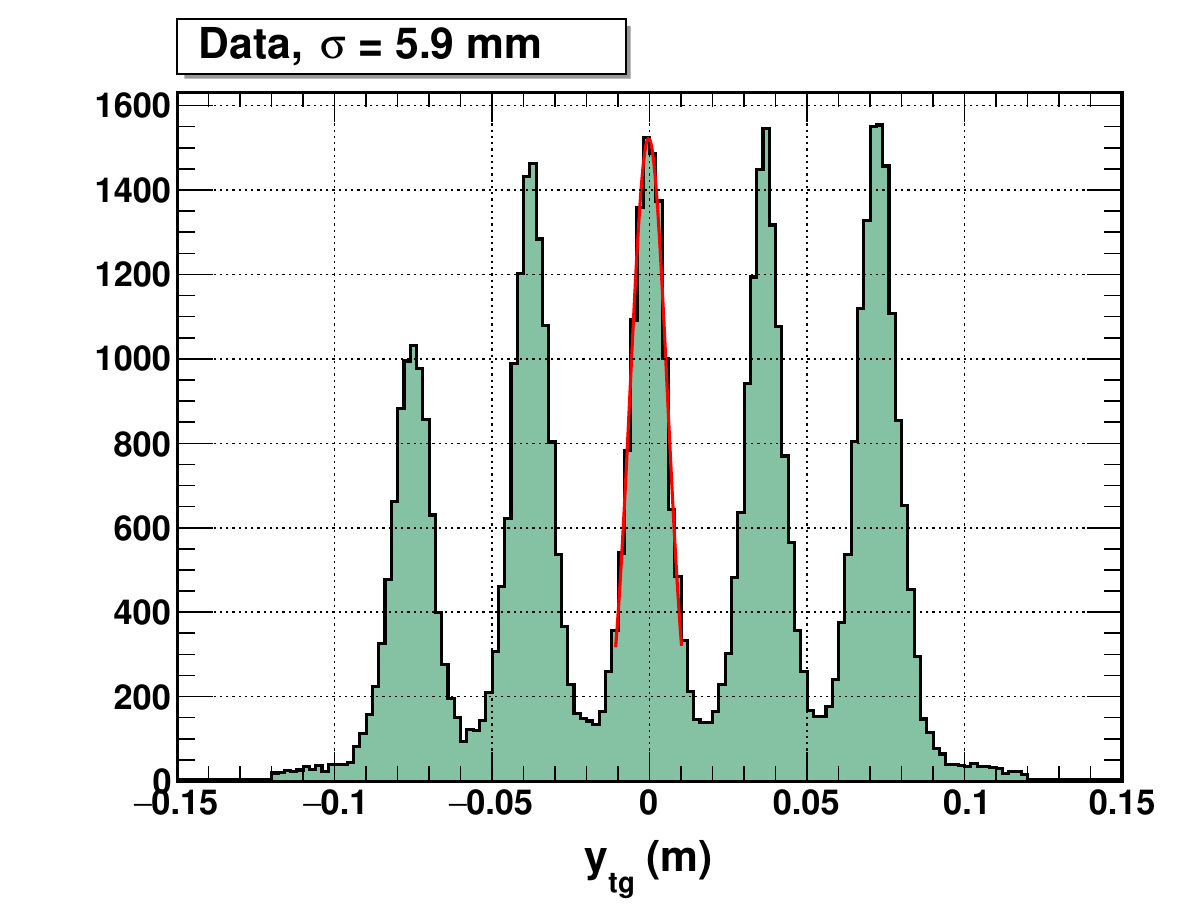}
	\caption{\label{fig:ch4:vz} Reconstructed $y_{tg}$ distribution from optics data taken during \qeq{4.5} low-\ep kinematics, resolving all five \ce{C} foil targets.} 
\end{figure}
The same analysis cuts used to select good electron tracks during zero-field alignment are used here as well. The focal-plane coordinates ($x_{fp}, y_{fp}, x'_{fp}, y'_{fp}$) of the tracks, obtained from track reconstruction, constitute the right-hand side of the expansion in Equation \ref{eqn:ch4:optics}. These coordinates are compared to initial track slopes ($x'_{tg},y'_{tg}$) and position ($y_{tg}$) in the target coordinate system, obtained by associating the known position of the interaction point with the known position of the small sieve hole it passed through. Multiple such events, covering most of the BB acceptance in $x'_{tg}$, $y'_{tg}$, and $y_{tg}$ are then combined to calculate the respective expansion coefficients, $\mathcal{M}^{\Xi_{tg}}_{ijkl}$, corresponding to the most optimized solution. Expansions up to the second order ($N=2$) for each target variable have been deemed optimal to balance between overfitting and underfitting.

Figure \ref{sfig:ch4:bbsieve_recon} shows the reconstructed sieve hole patterns obtained by projecting tracks onto the sieve slit. Displayed are electron tracks with a central angle of $49^{\circ}$, originating from five \ce{C}-foil targets positioned at $z_{hall} = 0, \pm$\SI{5}{cm}, and $\pm$\SI{10}{cm}. The reconstructed $z$ positions of these targets in the hall/vertex coordinate system, denoted as $v_z$, can be calculated from the reconstructed $y_{tg}$ values using the following equation:
\begin{equation}
    v_{z} = - \frac{y_{tg}}{\sin{\theta_{BB}} + y'_{tg}\cos{\theta_{BB}}}
\end{equation}
where $\theta_{BB}$ is the BB central angle, and the negative sign arises because the BB is located on the left side of the beam line when looking downstream.

\fig \ref{fig:ch4:vz} shows the $y_{tg}$ distribution, resolving all five \ce{C} foil targets with an approximate resolution of \SI{6}{mm}. This resolution is lower than expected, primarily due to multiple scattering from air molecules between the scattering chamber and the first GEM layer. The ``true" vertex resolution from simulation produces a significantly better result of approximately \SI{2}{mm}, as anticipated.

%
\subsubsection{Momentum Reconstruction}
\label{sssec:ch4:momrecon}
The momentum reconstruction in terms of $\delta$, as formulated in Equation \ref{eqn:ch4:optics}, has proven highly successful for Jefferson Lab's high-precision spectrometers including HRSs, HMS, and SHMS. However, an effectively infinite momentum bite of the BB spectrometer makes this method less effective. Therefore, a slightly different approach has been adopted, which will be discussed in this section. 
\begin{figure}[h!]
     \centering
     \begin{subfigure}[b]{0.496\textwidth}
         \centering
         \includegraphics[width=\textwidth]{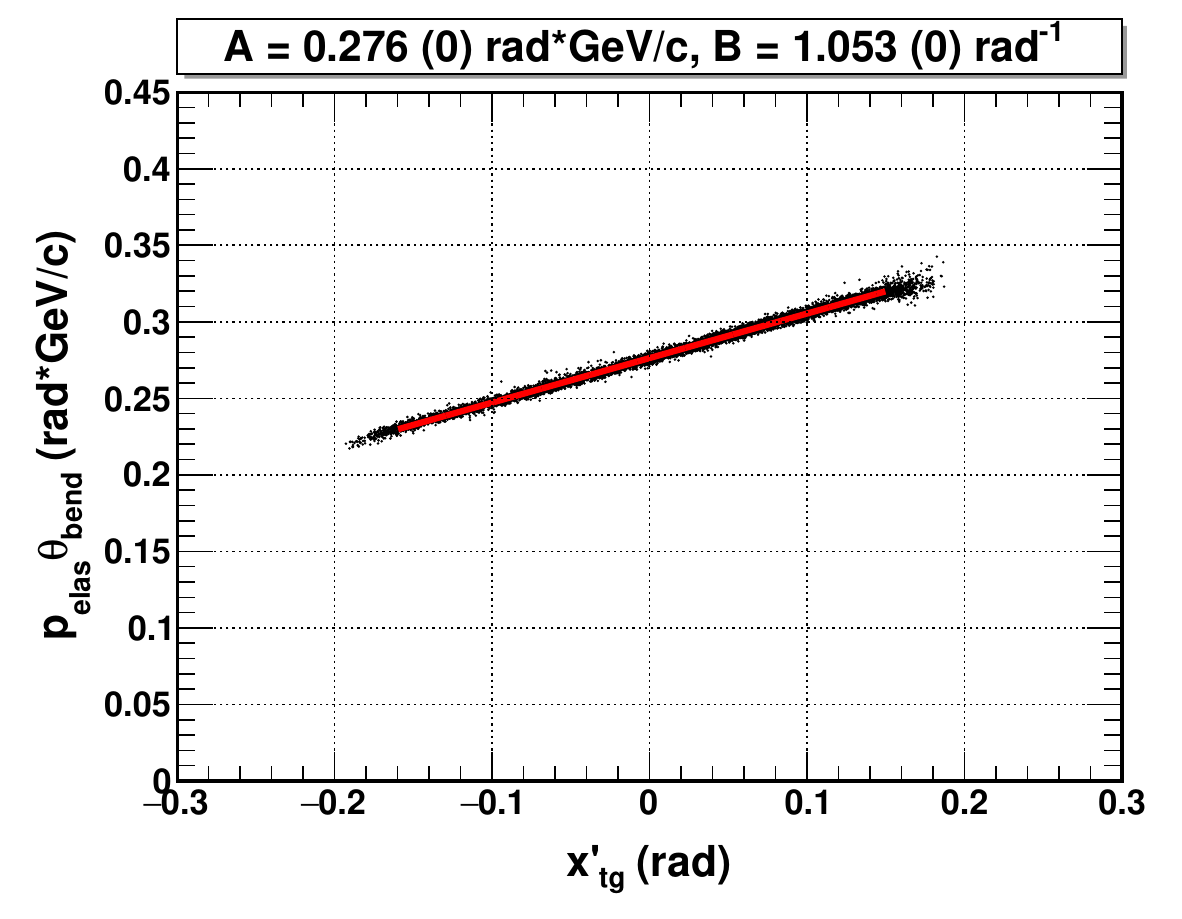}
         \caption{Simulation}
         \label{fig:ch4:momrecon_pthbnd_simu}
     \end{subfigure}
     \begin{subfigure}[b]{0.496\textwidth}
         \centering
         \includegraphics[width=\textwidth]{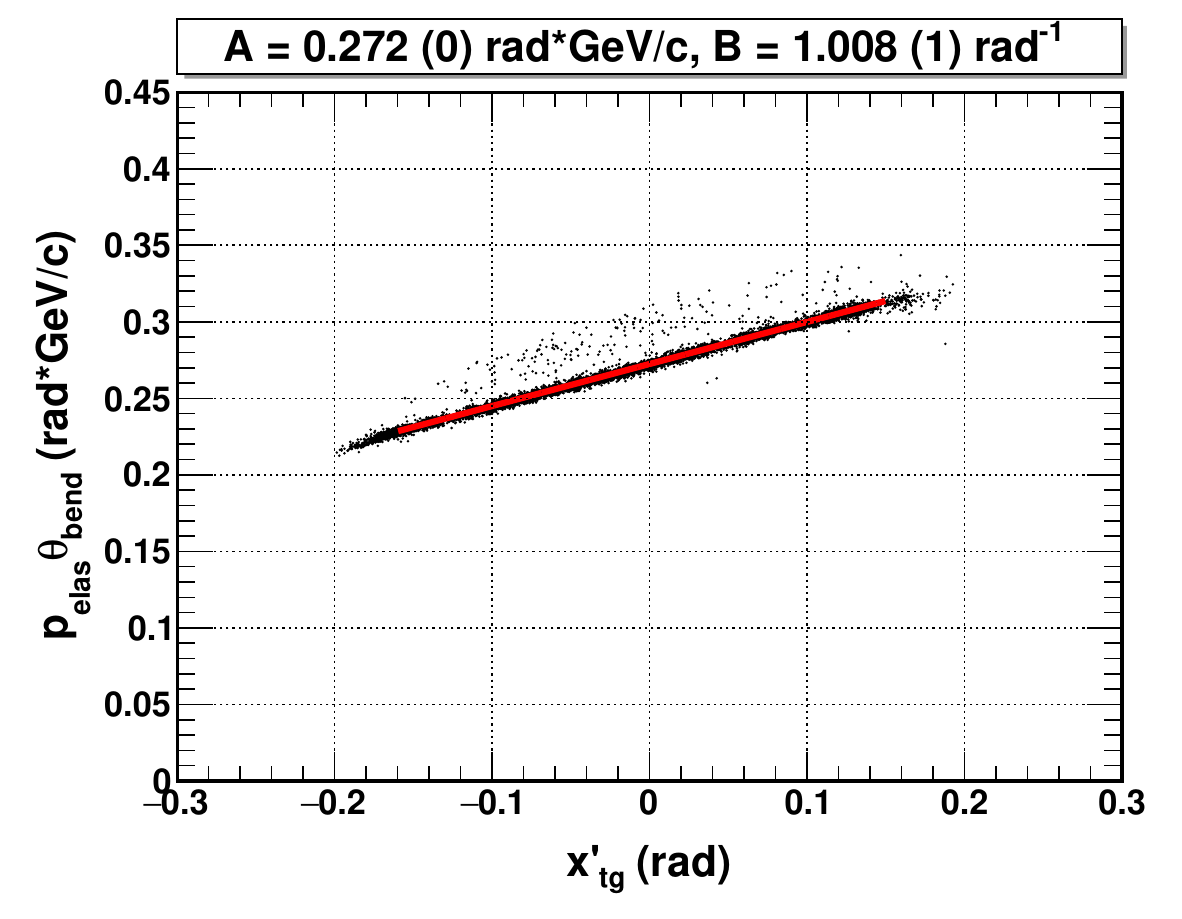}
         \caption{Data}
         \label{fig:ch4:momrecon_pthbnd_data}
     \end{subfigure}
     \caption{The product of the calculated track momentum and the trajectory bend angle as a function of the dispersive plane angle, $x'_{tg}$, from (a) simulation and (b) data. \heep events at  \qeq{3} were used to generate these plots. $A$ and $B$ are first-order momentum reconstruction parameters defined in Equation \ref{eqn:ch4:momrecon}.}
     \label{fig:ch4:momrecon_pthbnd}
\end{figure}

Studies using the simulation to track electrons through the BigBite magnetic field have revealed the following findings \cite{SUPPMAT}:
\begin{itemize}
    \item The product of the track momentum ($p$) and the trajectory bend angle ($\theta_{bend}$) is primarily sensitive to the dispersive plane angle of the trajectory, $x'_{tg}$, given the angle ($\theta_{BB}$), distance ($d_{BB}$), and field strength ($H_{BB}$) of the BB dipole magnet.
    \item The relationship between $p\theta_{bend}$ and $x'_{tg}$ is linear (see Figure \ref{fig:ch4:momrecon_pthbnd_simu}) and exhibits the following properties:
    \begin{itemize}
        \item The intercept depends solely on $H_{BB}$ and is proportional to it.
        \item The ratio of the slope to the intercept depends only on $d_{BB}$.
    \end{itemize}
    \item To first order, $p\theta_{bend}$ is not sensitive to $\theta_{BB}$ for constant $d_{BB}$ and $H_{BB}$.
\end{itemize}

$H_{BB}$ was kept constant throughout the experiment by operating the BB magnet only at \SI{750}{A}. However, $\theta_{BB}$ and $d_{BB}$ varied significantly across configurations, as listed in Table \ref{tab:sbsconfig}. Combining these facts with the above findings, $p\theta_{bend}$ can be expanded to first order in $x'_{tg}$ in the following form:
\begin{equation}
\label{eqn:ch4:momrecon}
    p\theta_{bend} = A(1 + B(1 + Cd_{BB}) x'_{tg})
\end{equation}
However, the considerations of point-to-point systematic uncertainties related to the estimations of beam energy, $\theta_{BB}$, and absolute $\theta_{bend}$ while combining data from different configurations were deemed inefficient. Consequently, $C$ was set to zero, simplifying the task of momentum reconstruction to optimizing $A$ and $B$ for each configuration\footnote{The SBS fringe field invalidated the assumption of universality for the $A$ and $B$ parameters within a configuration, necessitating separate momentum calibration for datasets recorded at different SBS field strengths.}. 

\begin{figure}[h!]
     \centering
     \begin{subfigure}[b]{0.496\textwidth}
         \centering
         \includegraphics[width=\textwidth]{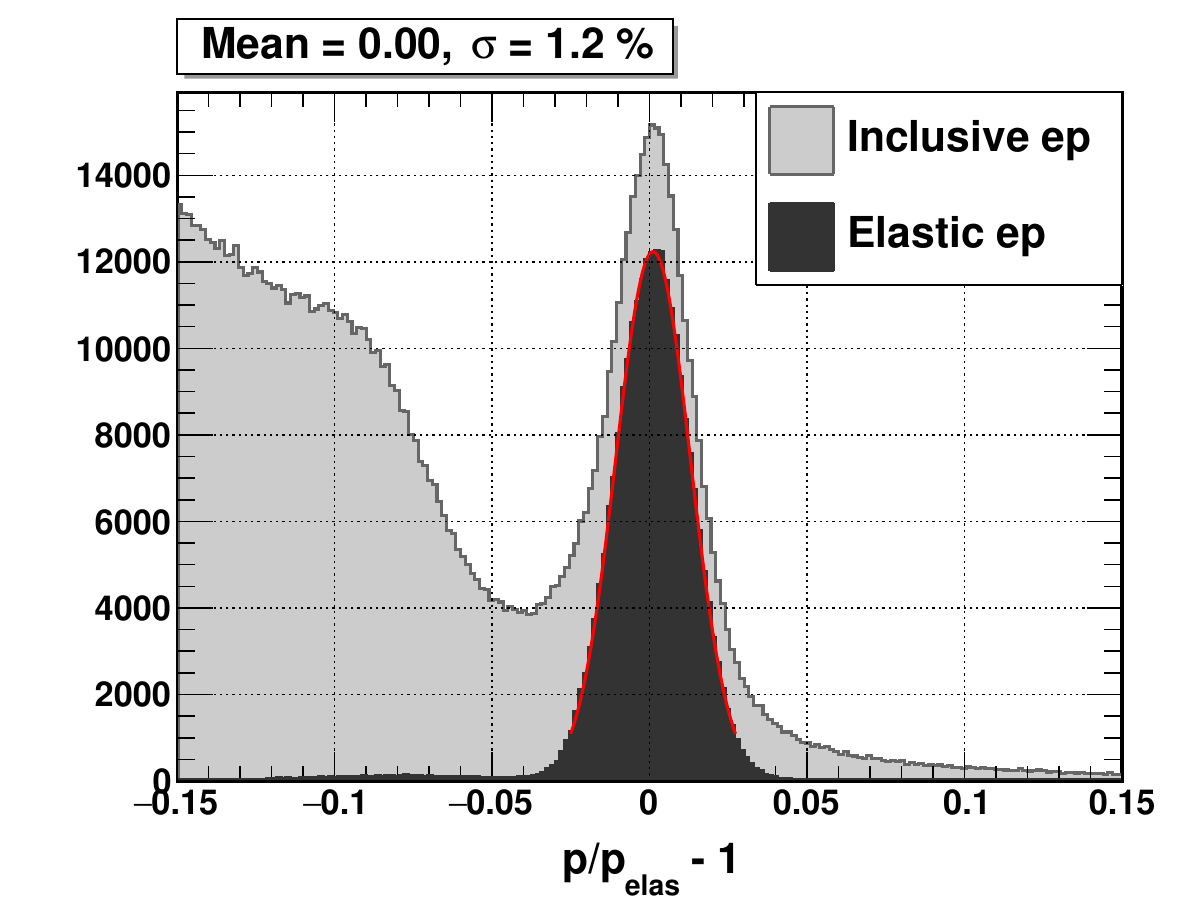}
         \caption{\qeq{3} \vspace{0.8em}}
     \end{subfigure}
     \begin{subfigure}[b]{0.496\textwidth}
         \centering
         \includegraphics[width=\textwidth]{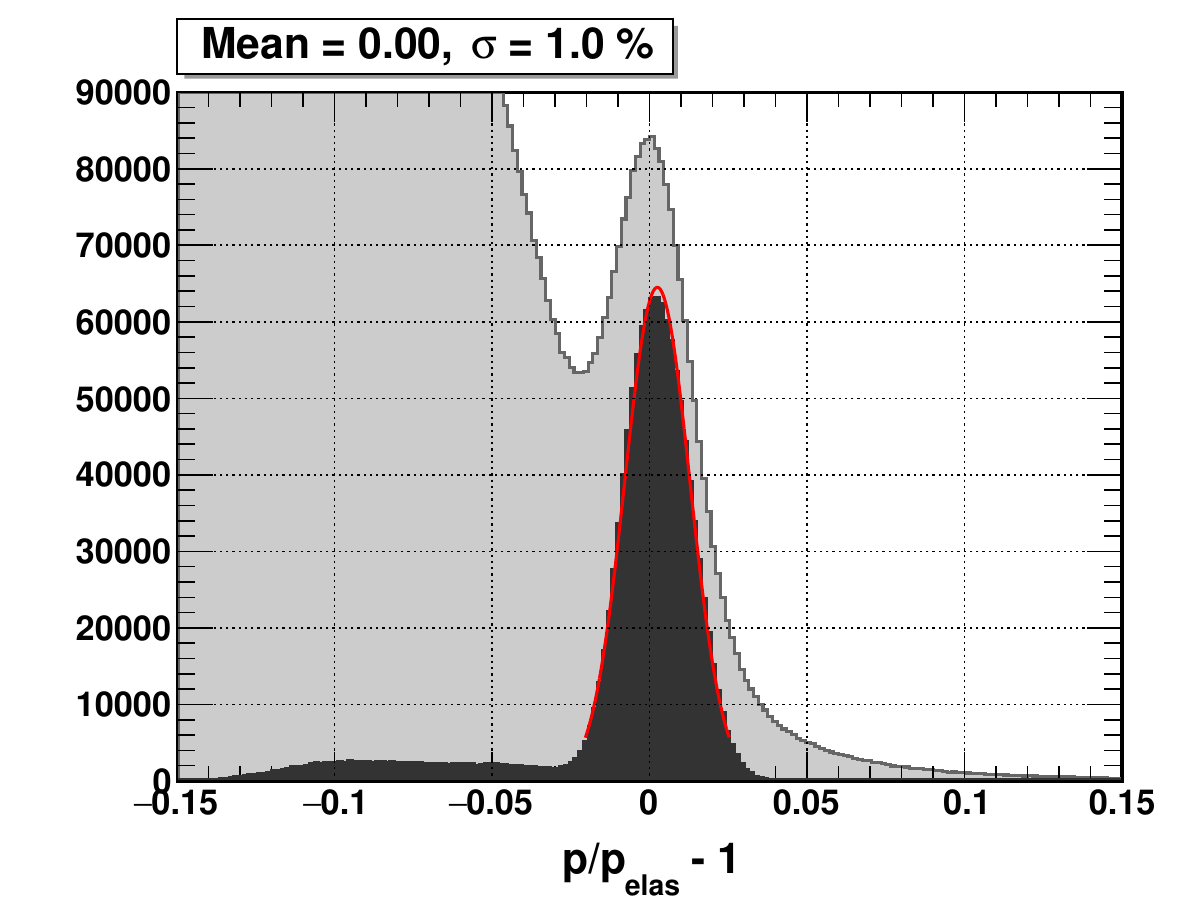}
         \caption{\qeq{4.5}, high \ep \vspace{0.8em}}
     \end{subfigure}
     \begin{subfigure}[b]{0.496\textwidth}
         \centering
         \includegraphics[width=\textwidth]{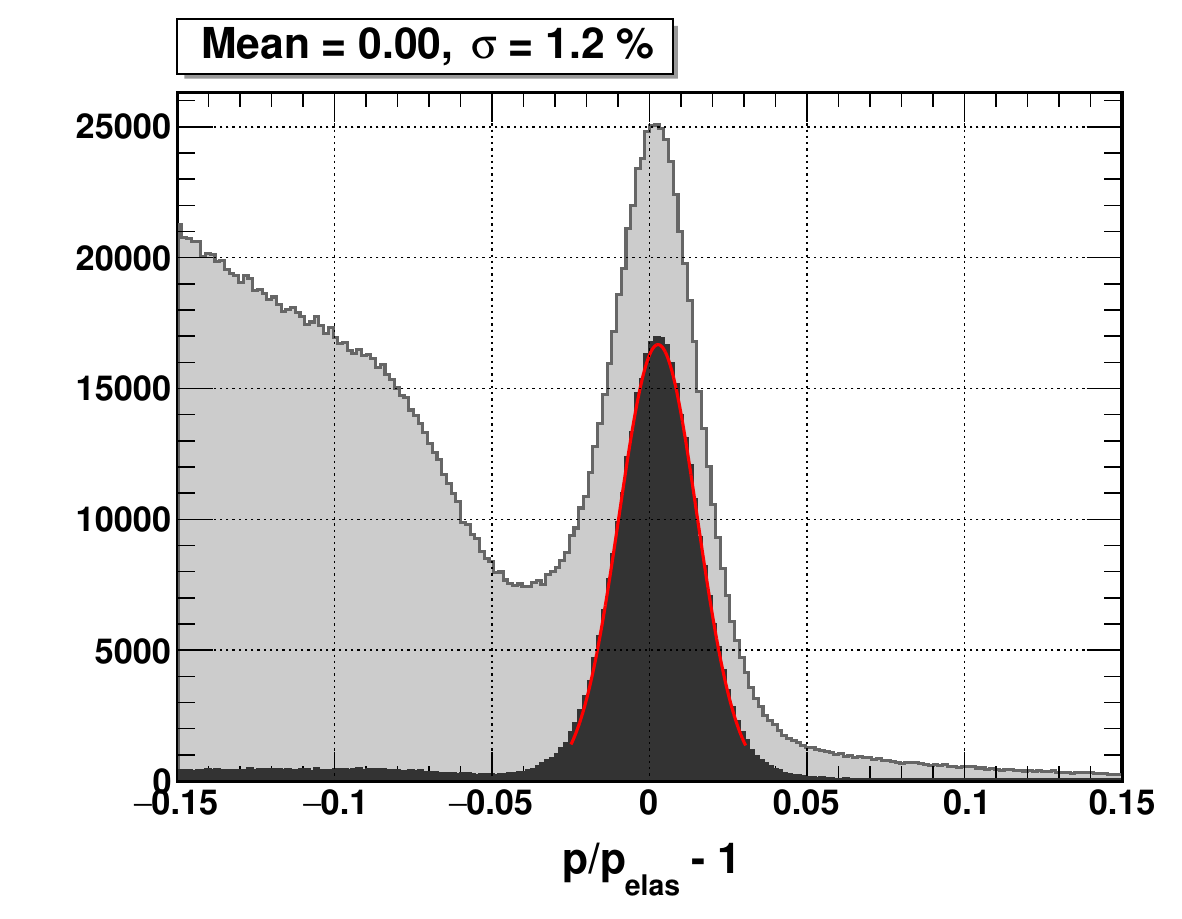}
         \caption{\qeq{4.5}, low \ep}
     \end{subfigure}
     \begin{subfigure}[b]{0.496\textwidth}
         \centering
         \includegraphics[width=\textwidth]{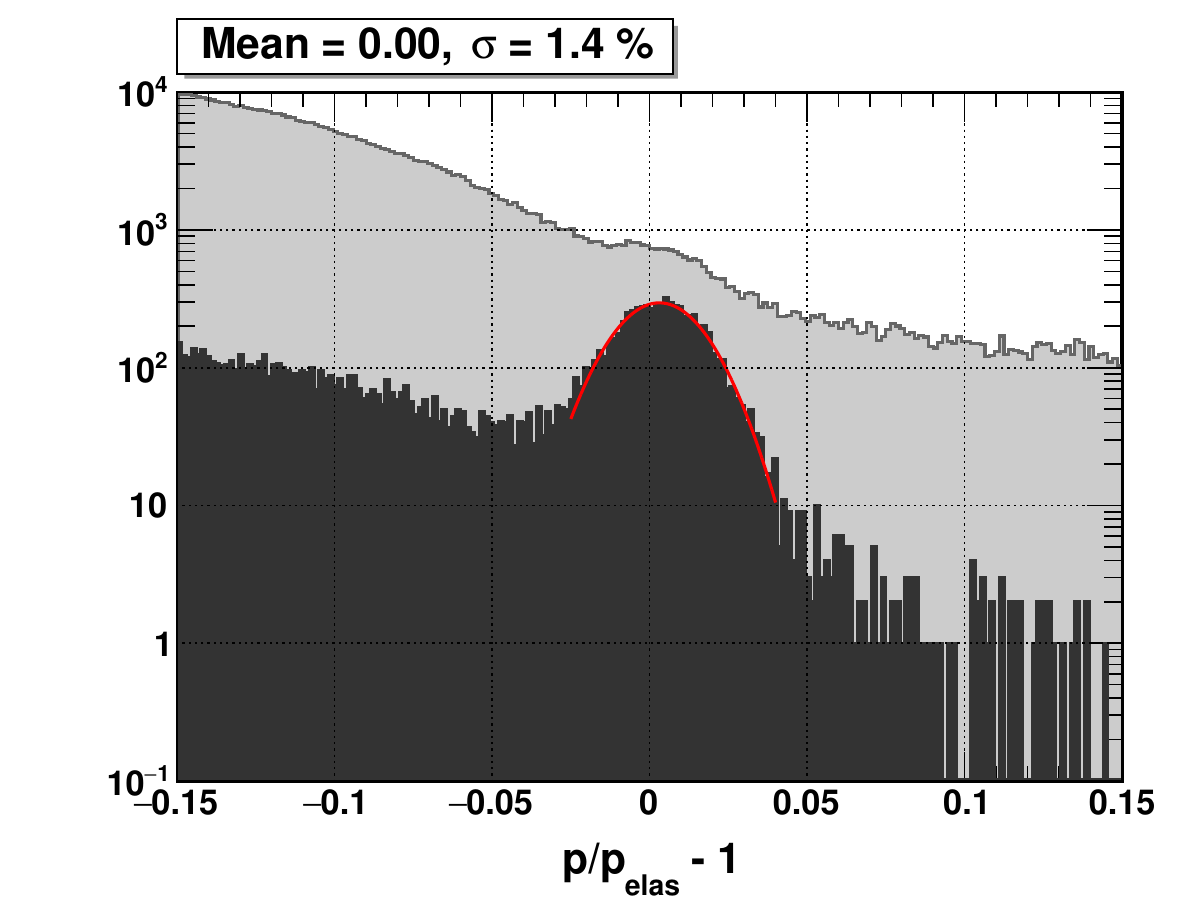}
         \caption{\qeq{13.6}}
     \end{subfigure}
     \caption{Results of momentum reconstruction from various kinematics with target-to-BB magnet distances ($d_{BB}$) and average scattered electron energies ($\Bar{E_{e}}$): (a) $d_{BB}=$ \SI{1.79}{m}, $\Bar{E_{e}}=$ \SI{2.12}{GeV}, (b) $d_{BB}=$ \SI{1.97}{m}, $\Bar{E_{e}}=$ \SI{3.58}{GeV}, (c) $d_{BB}=$ \SI{1.55}{m}, $\Bar{E_{e}}=$ \SI{1.63}{GeV}, (d) $d_{BB}=$ \SI{1.55}{m}, $\Bar{E_{e}}=$ \SI{2.67}{GeV}. Refer to the text for more details.}
     \label{fig:ch4:momrecon}
\end{figure}
As is evident, it is crucial to select clean \heep events from data for optimal results. The data collected on the \lh target at each configuration provides sufficient statistics for this optimization process. The clean selection of \heep events is ensured by applying stringent electron and exclusivity cuts, some of which are kinematic-dependent as detailed in Section \ref{sec:ch4:evselect}. The optimization process for a given configuration involves the following steps:
\begin{enumerate}
    \item \textbf{Calculation of Initial Momentum:} The initial momentum ($p_{elas}$) of the elastically scattered electron at the reconstructed polar scattering angle ($\theta$) is calculated for each event using:
    \begin{equation}
    \label{eqn:ch4:pelas}
        p_{elas} =  \frac{E_{beam}}{1 + \frac{E_{beam}}{M_{p}}(1 - \cos{\theta})}
    \end{equation}
    where $E_{beam}$ is the beam energy corrected for the energy loss in the target before scattering. Subsequently, the incident momentum at the BB magnet is obtained by subtracting the post-scattering energy loss from $p_{elas}$.
    \item \textbf{Determination of True Bend Angle:} The ``true" bend angle ($\theta_{bend}$) of the corresponding trajectory is then calculated from the reconstructed track directions at the target ($\vu{e}_{tg}$) and at the focal plane ($\vu{e}_{fp}$) using:
    \begin{equation}
        \theta_{bend} = \arccos(\vu{e}_{fp} \vdot \vu{e}_{tg})
    \end{equation}
    \item \textbf{Optimization of Parameters:} The plot of $p_{elas}\theta_{bend}$ vs. $x'_{tg}$ for all elastic events are fitted with the equation \ref{eqn:ch4:momrecon} to extract optimized parameters $A$ and $B$, enabling momentum reconstruction for events in this configuration. Figure \ref{fig:ch4:momrecon_pthbnd_data} shows an example of such optimization for \qeq{3} dataset.
\end{enumerate}

Figure \ref{fig:ch4:momrecon} shows the deviation of reconstructed momentum from $p_{elas}$ across four configurations, revealing a BB momentum resolution of $1-1.4\%$. Notably, the resolution worsens significantly at the highest \q, likely due to increased survival of inelastic events through exclusivity cuts, biasing the optimization. Interestingly, the high \ep dataset at \qeq{4.5} shows better momentum resolution than the low \ep dataset. Although higher $\theta_{bend}$ at lower $\Bar{E_{e}}$ should theoretically improve resolution, increased multiple scattering from air molecules at lower $\Bar{E_{e}}$ worsens it, evidently dominating the result.

\section{Event Selection}
\label{sec:ch4:evselect}
The reconstructed events include both signal and background, and event selection aims to separate the two. \gmn measurements rely on isolating quasi-elastic electron-nucleon scattering events from the \ld target to extract $G_M^n$. Additionally, selecting elastic electron-proton scattering events from the \lh target is crucial for calibration. However, various sources of background—such as inelastic scattering, multiple scattering, target end-window scattering, and accidental two-arm coincidences—make this separation challenging. Among these sources, inelastic scattering generates the most abundant background events, with their number increasing with rising \q. The primary processes contributing to this background include pion electroproduction, photoproduction, and deep inelastic scattering.

The innovative design of the BB and Super BB spectrometers enables robust background rejection. The upward tilt of the BB detector stack optimizes acceptance for up-bending particles while excluding most down-bending $\pi^+$ tracks. The Pre-Shower and GRINCH detectors, optimized for particle identification, effectively reject $\pi^-$ tracks. On the hadron arm, negatively charged pions are deflected from HCAL acceptance by the strong SBS dipole magnet field. Additionally, simultaneous detection of scattered electrons and nucleons allows for coincidence timing cuts, effectively reducing accidental background.

These features allow for the application of several analysis cuts to select good electron and nucleon events detected by BB and Super BB spectrometers, respectively. Quasi-elastic and elastic event selection cuts, based on kinematic constraints and $eN$ kinematic correlations, are then applied to isolate the physics events of interest. The details and definitions of these cuts will be discussed in this section.

\subsection{Good Electron Event Selection}
The cuts to select good electron events originating from the target consist of various track quality and PID cuts discussed below.
\subsubsection{Track Quality Cuts}
The quality of the reconstructed tracks are assured by the following cuts:
\begin{itemize}
    \item \textbf{Minimum Number of GEM Layers with Hits $(N_{hit}^{GEM})$} --- Tracks with hits in more GEM layers are considered more reliable. Typically, we select tracks with hits in at least four out of five BB GEM layers. However, for the highest \q, this requirement is reduced to three due to the failure of the first GEM layer halfway through data collection.
    \item \textbf{Cut on Track $\chi^2/NDF$} --- This ensures the quality of the straight-line fit for track reconstruction. A loose cut of $\chi^2/NDF<50$ is used for tracks with hits on four or more GEM layers. However, when three hit tracks are also accepted, as in the case of the highest \q dataset, using a much stricter cut of $\chi^2/NDF<15$ dramatic reduces the number of fake tracks. 
    \item \textbf{\bm{$z$}-Vertex $(v_z)$ Cut} --- A cut of $|v_z|\leq0.075$, on the reconstructed $z$-position of the interaction vertex ($v_z$) ensures that the corresponding track originates from the target, reflecting the cryotargets' thickness of \SI{15}{cm}. Events beyond this range are not necessarily background, as the resolution of BB vertex reconstruction is finite at approximately \SI{6}{mm} (discussed in \sect \ref{sssec:ch4:angnverrecon}). Due to this finite resolution, some contamination from the target's entrance and exit windows in the selected events is also expected.

    \begin{figure}[h!]
	   \centering
	   \includegraphics[width=\sfig]{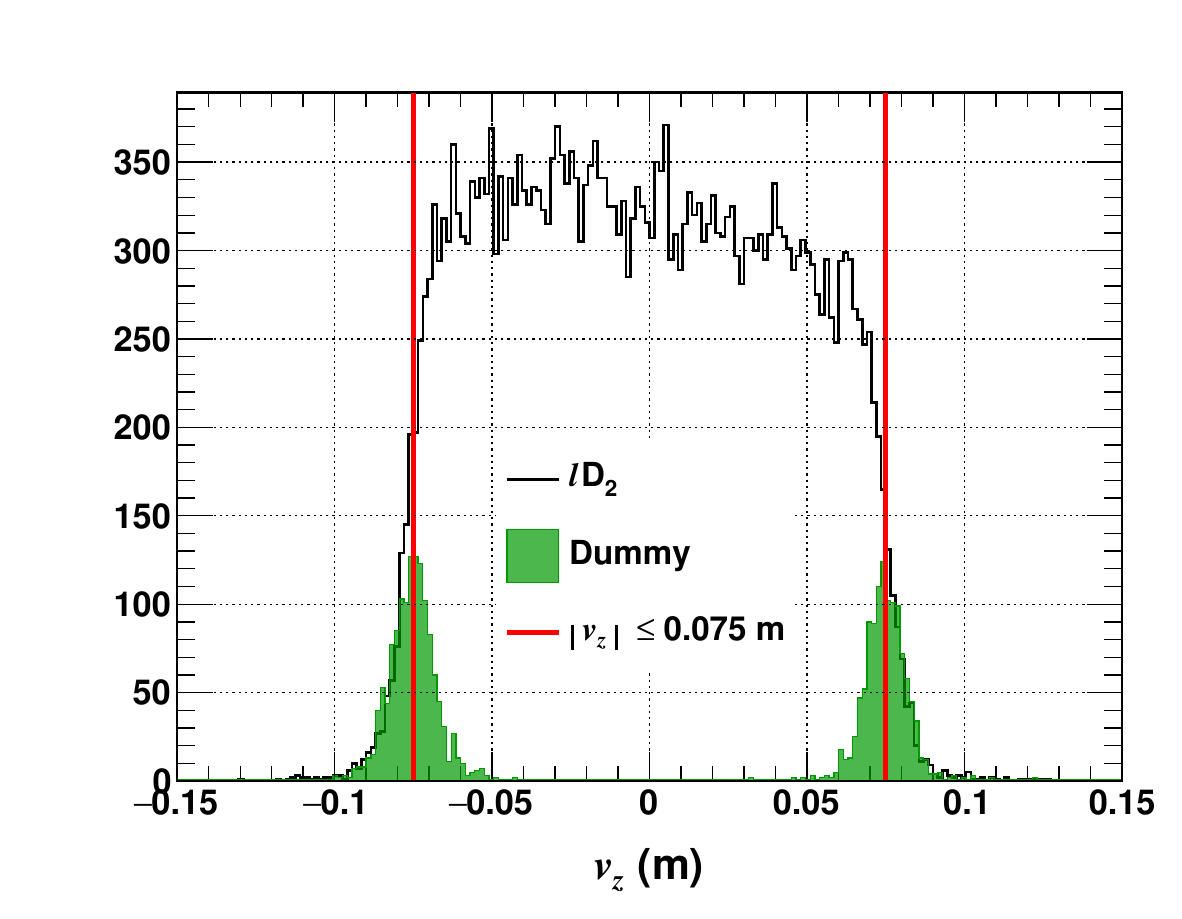}
	   \caption[Distribution of the reconstructed $z$-position of the interaction vertex ($v_z$)]{\label{fig:ch4:vzcut} The reconstructed $z$-position of the interaction vertex ($v_z$) for \ld and dummy targets at \qeq{3}. Only events passing the exclusivity cuts are shown. The relative normalization between \ld and dummy data is arbitrary.}
    \end{figure}
    \hspace{1em}\fig \ref{fig:ch4:vzcut} shows the $v_z$ distributions for \ld and dummy targets from the \qeq{3} dataset, plotting only events passing the exclusivity cuts. It qualitatively shows the severity of contamination from target end-window scattering in quasi-elastic event selection. Reducing or extending the cut range by one $\sigma$ of the dummy target $v_z$ distribution from both sides changes the statistics by approximately $3\%$. Given these caveats, determining a fixed cut range is non-trivial. Therefore, we choose the optimal $v_z$ cut range at the later stages of analysis based on the stability of the physics observable studied for each \gmn configuration individually.
    \item \textbf{Optics Validity Cut} --- This cut selects the region of uniform field within the BB magnet, as the performance of the BB optics model degrades with field non-uniformity. Although a higher-order optics model can improve performance, it is computationally expensive and increases the risk of overfitting. The track coordinates projected to the BB magnet mid-plane,  ($x_{BB}, y_{BB}$), are used as the cut variables, which can be calculated as follows:
    \begin{equation}
        \begin{aligned}
            x_{BB} &= x_{fp} - 0.9 x'_{fp} \\
            y_{BB} &= y_{fp} - 0.9 y'_{fp}
        \end{aligned}
    \end{equation}        
    where ($x_{fp}, y_{fp}, x'_{fp}, y'_{fp}$) are the focal plane coordinates of the track, and \SI{-0.9}{m} is the $z$-position of the BB magnet mid-plane relative to the focal plane coordinate system. 

    \begin{figure}[h!]
	   \centering
	   \includegraphics[width=1\columnwidth]{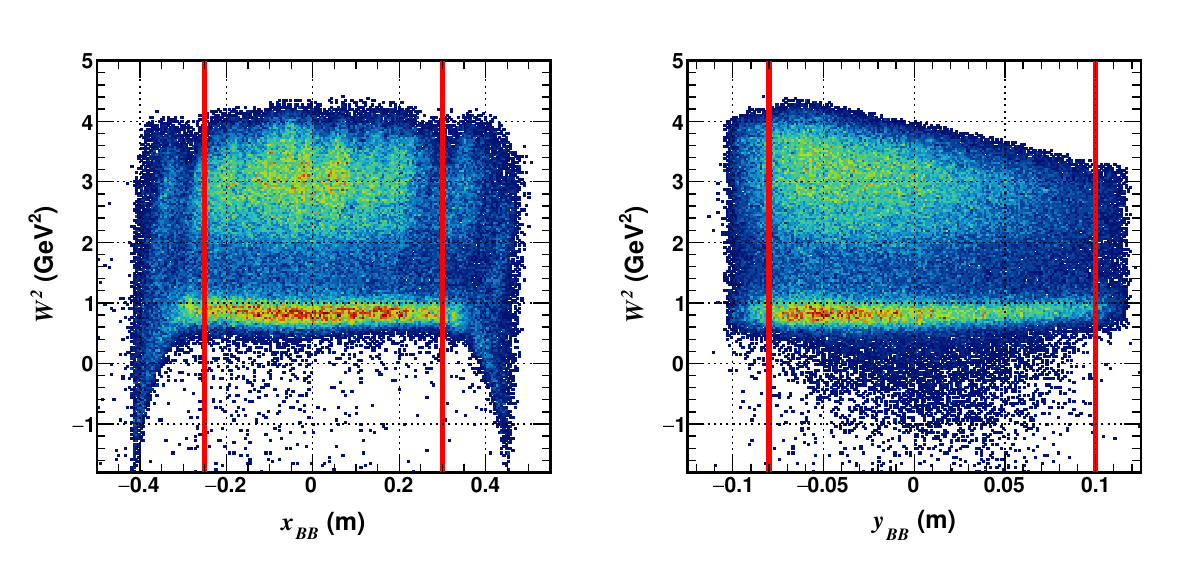}
	   \caption{\label{fig:ch4:opvcut} The evaluation of BB optics validity cut region for \heep events at \qeq{7.4}. Good electron events passing a loose \thpq cut are shown. $x_{BB}$ and $y_{BB}$ are the reconstructed track coordinates projected to the mid-plane of the BB magnet. Refer to the text for more information.}
    \end{figure}
    %
    \hspace{1em}\fig \ref{fig:ch4:opvcut} shows the correlations of these variables with \w, highlighting the regions of non-uniformity in both the dispersive and transverse directions, which guided the determination of the cut ranges. Notably, the acceptance of the BB magnet varies with kinematics, necessitating the optimization of the optics validity cut region for each experimental configuration.     
\end{itemize}

\subsubsection{PID Cuts}
Tracks passing quality assurance cuts are not guaranteed to be associated with electrons. The largest contamination comes from negatively charged pions generated due to inelastic scattering, the rate of which increases with rising \q. The following cuts help select electron tracks by rejecting pions:
\begin{itemize}
    \begin{figure}[h!]
	   \centering
	   \includegraphics[width=\sfig]{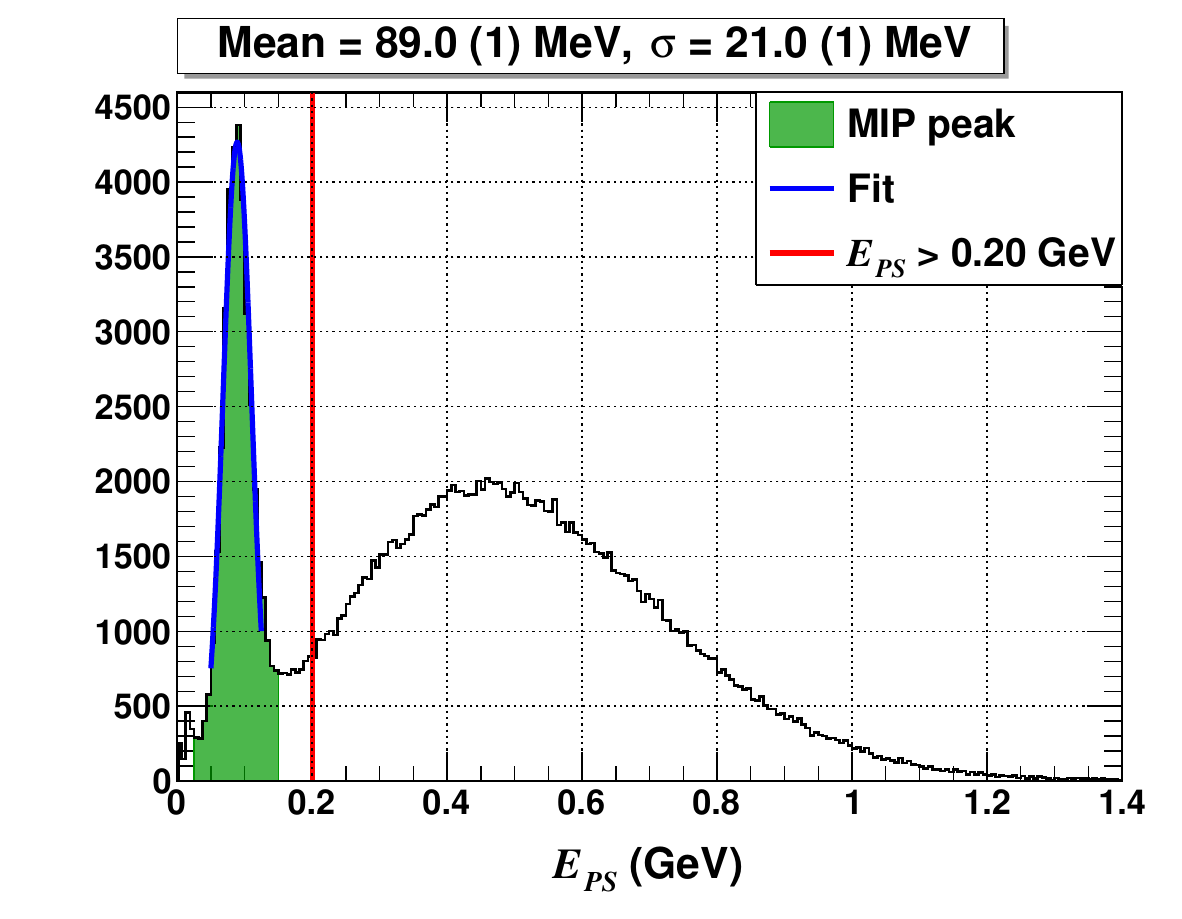}
	   \caption{\label{fig:ch4:psecut} The evaluation of Pre-Shower (PS) cluster energy ($E_{PS}$) threshold based on the position of MIP peak using high-quality tracks from the \qeq{4.5}, low \ep dataset.}
    \end{figure}
    \item \textbf{PS Energy ($E_{PS}$) Cut} --- As discussed in Section \ref{ssec:bbcaldetassembly}, the segmented design of BBCAL effectively rejects pions. High-energy scattered electrons deposit only a fraction of their energy in the PS blocks through electromagnetic showers, while hadrons (mainly pions) behave as minimum ionizing particles (MIPs) in this momentum range. This results in two distinct peaks in the PS cluster energy distribution: one for electrons and one for MIPs. Therefore, a simple threshold cut on the PS energy to exclude the MIP peak effectively filters out much of the pion background. \fig \ref{fig:ch4:psecut} shows the PS energy distribution using high-quality tracks from the \qeq{4.5}, low \ep dataset. A threshold of \SI{200}{MeV}, chosen as the cut limit, clearly removes the MIP peak contribution. 

    \item \textbf{\eovpb Cut} --- Assuming the electron's rest mass is negligible, the reconstructed scattered electron energy measured by BBCAL ($E_{BBCAL}$) should match its momentum ($p$) from momentum reconstruction. This assumption doesn't hold for pions. Therefore, selecting events with \eovp values close to unity helps reject pions and reduces fake tracks.
    \begin{figure}[h!]
	   \centering
	   \includegraphics[width=\sfig]{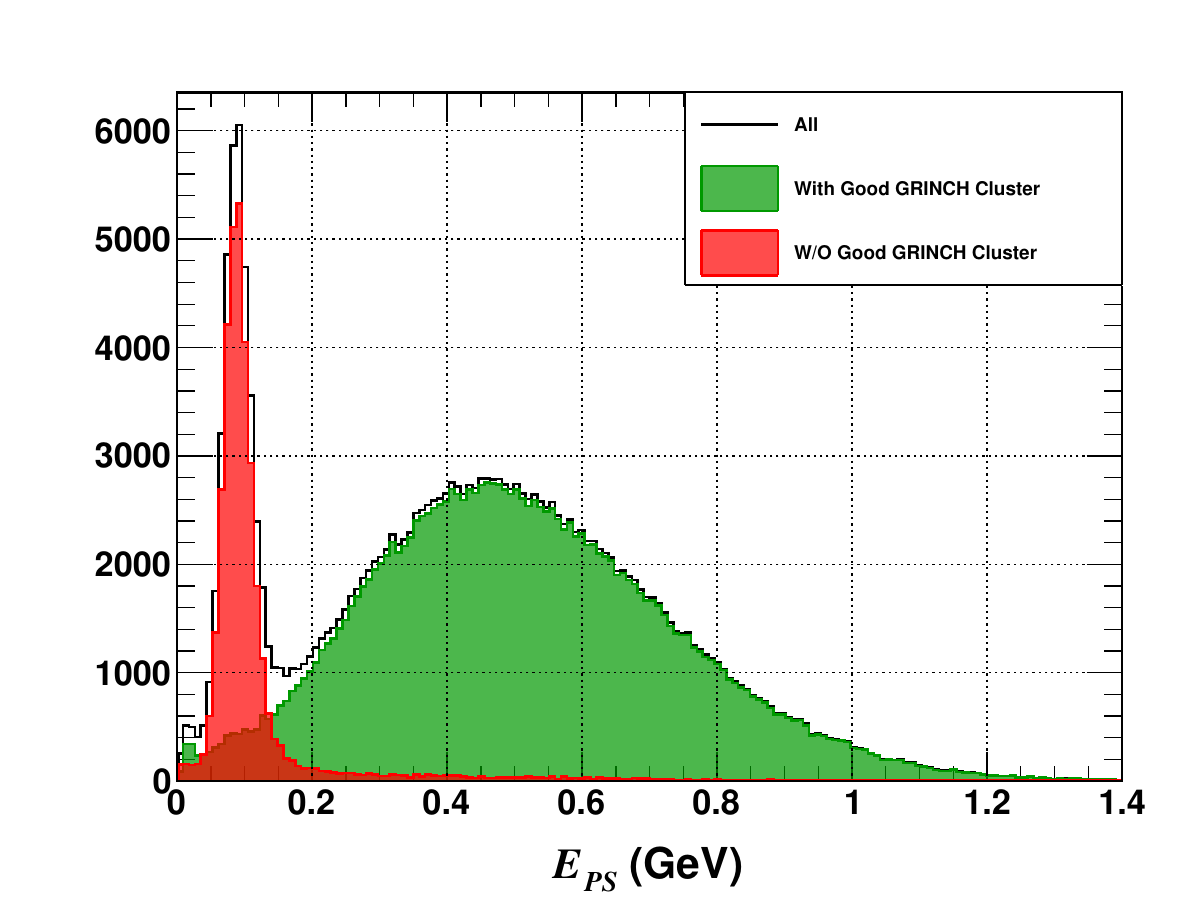}
	   \caption{\label{fig:ch4:grinchcut} PID by GRINCH at \qeq{4.5}, low \ep configuration. High-quality tracks with (without) an associated GRINCH cluster correspond to the electrons (MIPs) in the PS cluster energy ($E_{PS}$) distribution, as expected. A GRINCH cluster is considered good if it is track-matched (see \sect \ref{sssec:ch4:grinchclFinished}) and consists of more than two PMTs.}
    \end{figure}
    \item \textbf{GRINCH Cluster Cut} --- According to its design, only electron tracks should form clusters in the GRINCH. Thus, selecting events with a good GRINCH cluster effectively rejects pions. Figure \ref{fig:ch4:grinchcut} shows GRINCH's PID capability using high-quality tracks from the \qeq{4.5}, low \ep dataset. However, it is important to note that GRINCH data is only usable for the \qeq{4.5} datasets. Even among \qeq{4.5} datasets, its performance is less effective for the high \ep point, as the high-energy pions generated in this configuration often have energy higher than GRINCH's pion threshold of \SI{2.7}{GeV}. Refer to \sect \ref{ssec:ch3:grinch} for more details.

\end{itemize}
%
%
\subsection{Good Nucleon Event Selection}
Unlike BB, the Super BB spectrometer lacks high-precision particle trackers and PID detectors to cleanly select nucleon tracks. However, the detected HCAL events can still be cleaned up significantly using the cuts discussed below.
%
%
\subsubsection{HCAL Cluster Energy Cut}
\begin{figure}[h!]
    \centering
    \includegraphics[width=\sfig]{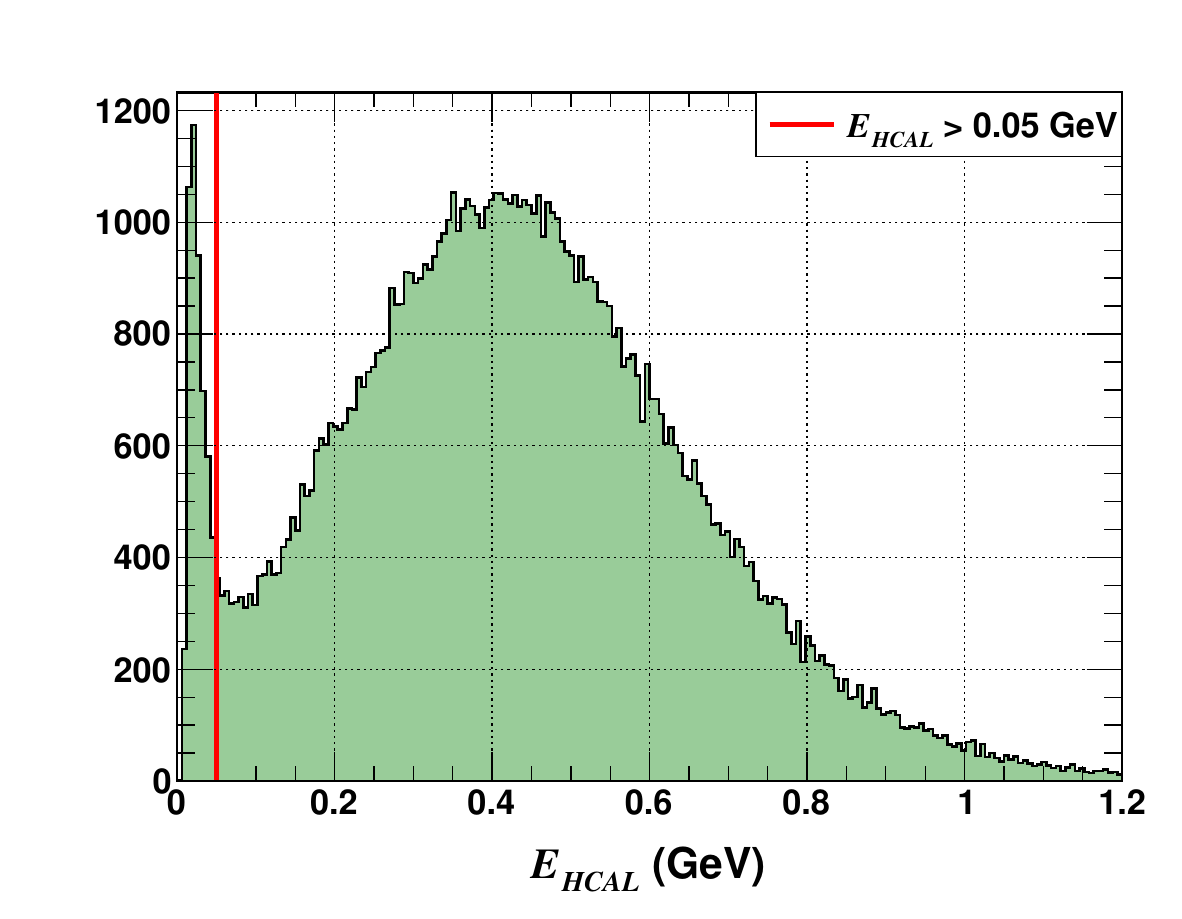}
    \caption{\label{fig:ch4:ehcal} HCAL cluster energy ($E_{HCAL}$) distribution using quasi-elastic \deeN events at \qeq{7.4}. The red vertical line indicates a typical threshold cut.}
\end{figure}

Rejecting events with very low HCAL cluster energy ($E_{HCAL}$) significantly reduces background from inelastic scattering, as shown in \fig \ref{fig:ch4:ehcal}. However, determining the cut limit is challenging due to its high sensitivity to the concurrent loss of quasi-elastic scattering events. Additional complications arise from discrepancies observed in the $E_{HCAL}$ distributions for neutron and proton events. The poor energy resolution of HCAL ($40-70\%$) further complicates the situation. Therefore, as described in \sect \ref{ssec:ch4:cutoptim}, cut sensitivity study is performed for each experimental configuration to determine the optimal $E_{HCAL}$ cut limit.   
\subsubsection{Electron-Nucleon Coincidence Time $(t_{coin})$ Cut}
Electron and nucleon tracks from the same scattering events should be in sync. Thus, rejecting out-of-time events effectively reduces background from accidental hits. One can use the time difference between HCAL and the timing hodoscope (TH) or HCAL and Shower (SH) to form the electron-nucleon coincidence time distribution. The goal is to achieve the best possible resolution with minimal loss of statistics.

In BB, the best time resolution is expected from TH TDC data, but in reality, it appears to be much worse, as discussed in \sect \ref{ssec:ch4:thcalib}. Additionally, HCAL TDC data is missing for a significant portion of good elastic events, especially for high-\q kinematics where we cannot afford to lose any statistics (see \sect \ref{ssec:ch4:hcalcalib}). Due to these issues, we are currently unable to use TH and HCAL TDC data for \gmn analysis, despite their potential suitability. However, we are hopeful that a more concentrated effort in improving the timing analysis will make this data useful for at least some of the kinematics.

\begin{figure}[h!]
    \centering
    \includegraphics[width=\sfig]{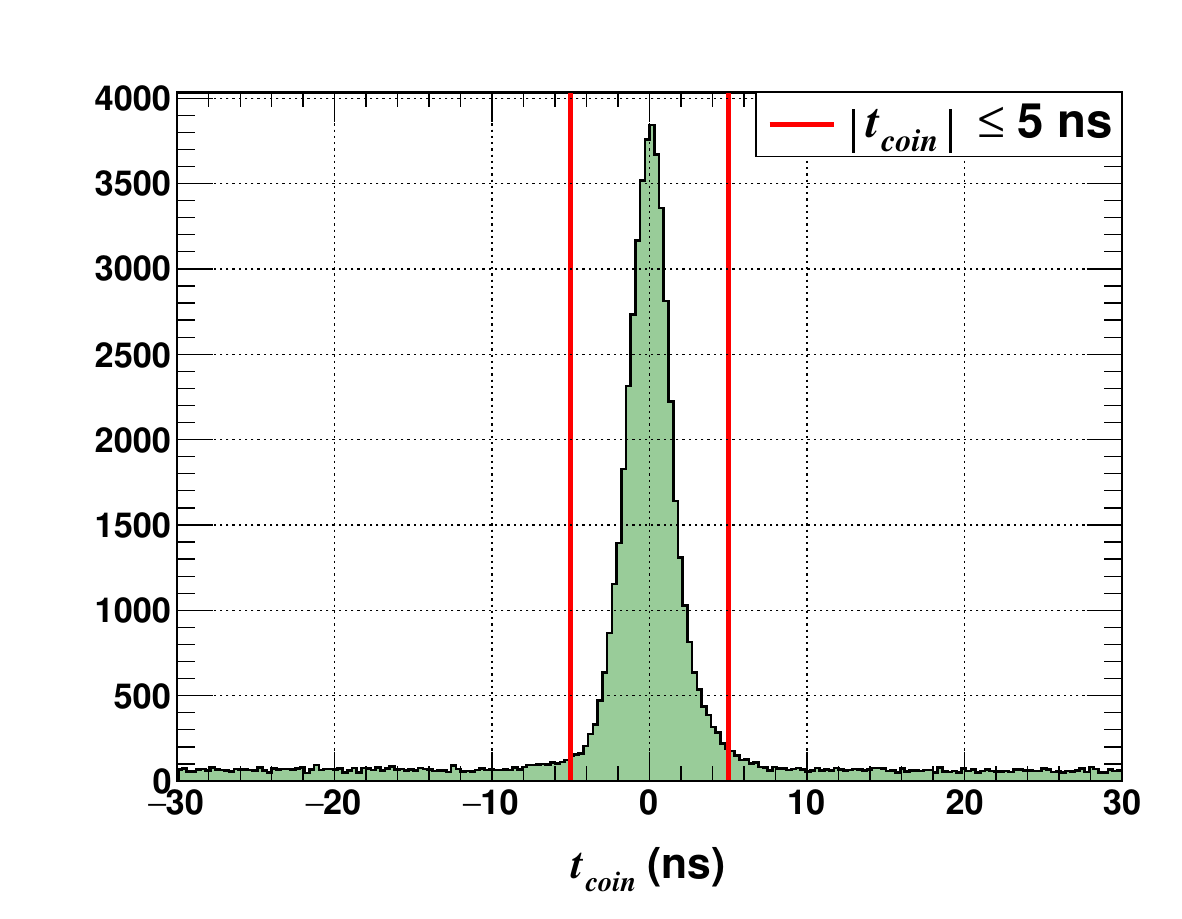}
    \caption{\label{fig:ch4:coint} HCAL-SH ADC coincidence time distribution for \qeq{4.5} low \ep dataset.}
\end{figure}
Using TH and HCAL TDC data would be beneficial but is not strictly necessary for \gmn analysis. This is because the coincidence between HCAL and SH ADC time, which can be calculated for the entire experimental dataset, provides excellent resolution in the range of $1.2-1.4$ ns. \fig \ref{fig:ch4:coint} shows the distribution of HCAL-SH ADC coincidence time ($t_{coin}$) using good electron events from \qeq{4.5}, low \ep dataset. The signal sits on a flat background attributable to accidental hits, with a significant portion of this background surviving the exclusivity cuts.

Typically, selecting events within $3.5$ $\sigma$ around the mean of the coincidence peak effectively eliminates accidental background. However, similar to the $E_{HCAL}$ cut, a kinematic-specific cut sensitivity study is performed to determine the optimal $t_{coin}$ cut range (see \sect \ref{ssec:ch4:cutoptim}. In the rest of this document, a coincidence time cut refers to a cut on $t_{coin}$, which is the difference between HCAL and SH ADC time.
%
\subsection{$eN$ Kinematic Correlation: HCAL \dx \& \dy Distributions}
\label{ssec:ch4:enkinecorr}
In Born approximation, the four-momentum of the virtual photon ($q$) in elastic electron-nucleon scattering, \enscat, is defined as follows:
\begin{equation}
\label{eqn:ch4:qvect}
    k - k' = q \equiv (\nu,\vb{q})
\end{equation}
where $k$ ($k'$) is the four-momentum of the incident (scattered) electron and $\nu$ ($\vb{q}$) is the energy (momentum) transfer in the scattering process. The deviation of the scattered nucleon momentum from the direction $\vu{q} = \vb{q}/|\vb{q}|$ indicates the inelasticity of a scattering event. Therefore, selecting events with forward $\theta_{pq}$ angles, defined by the angle between the \qvect and the reconstructed nucleon momentum $\vb{p}$, effectively isolates quasi-elastic (and elastic) electron-nucleon scattering events, as desired by \gmn. Equivalently, selecting events with minimal difference between the observed nucleon position at HCAL and the expected value, calculated based on $\vu{q}$, should serve the same purpose.
%

The observed nucleon position at HCAL, (\xhob,\yhob), is the HCAL best cluster centroid, available directly from event reconstruction, as discussed in \sect \ref{sssec:ch4:calcl}. Calculating the expected nucleon position, (\xhex,\yhex), is more complex and based solely on the kinematics of the scattered electrons detected by BB. This involves calculating $\vu{q}$ and projecting it onto the face of HCAL to determine the expected nucleon position coordinates. This section will provide a detailed discussion of this process.


%
\subsubsection{Calculation of \qvect}
\label{sssec:ch4:qvectcalculation}
\subheading{4-Vector Method}
In the lab frame, the components of $k$ and $k'$ can be expressed in terms of known parameters as follows:
\begin{equation}
\label{eqn:ch4:4vectmethod}
    \begin{aligned}
        k^\mu &\equiv (E_e,\vb{k}) = (E^{corr}_{beam},0,0,E^{corr}_{beam}) \\
        k'^{\mu} &\equiv (E'_e,\vb{k'}) = (E^{corr}_{e},E^{corr}_{e}\sin\theta_e\cos\phi_e,E^{corr}_{e}\sin\theta_e\sin\phi_e,E^{corr}_{e}\cos\theta_e)
    \end{aligned}
\end{equation}
where $E_{beam}^{corr}$ is the beam energy corrected for energy loss in the target before scattering, $E^{corr}_{e}$ is the reconstructed scattered electron momentum corrected for energy loss after scattering\footnote{Energy loss in the target material, target wall, scattering chamber window, and any external \ce{Al} or polyethylene shield is considered in the calculation.}, and $\theta_e$ and $\phi_e$ are the reconstructed polar and azimuthal scattering angles, respectively. Now, substituting \eqn \ref{eqn:ch4:4vectmethod} into \ref{eqn:ch4:qvect}, we obtain:
\begin{align}
    \vb{q} = (-E^{corr}_{e}\sin\theta_e\cos\phi_e,-E^{corr}_{e}\sin\theta_e\sin\phi_e,E^{corr}_{beam}-E^{corr}_{e}\cos\theta_e)
\end{align}

Notably, \qvect, as calculated here, utilizes both angle and momentum reconstructions. Alternatively, \qvect can be derived using either reconstructed angles or momentum alone, as one can be expressed in terms of the other assuming elastic scattering. The calculation using reconstructed angles as the independent variable, known as the ``angles-only" method, is important and will be discussed below.

\begin{figure}[h!]
	\centering
	\fboxsep=0.75mm
    \fboxrule=1pt
	\includegraphics[width=0.75\columnwidth]{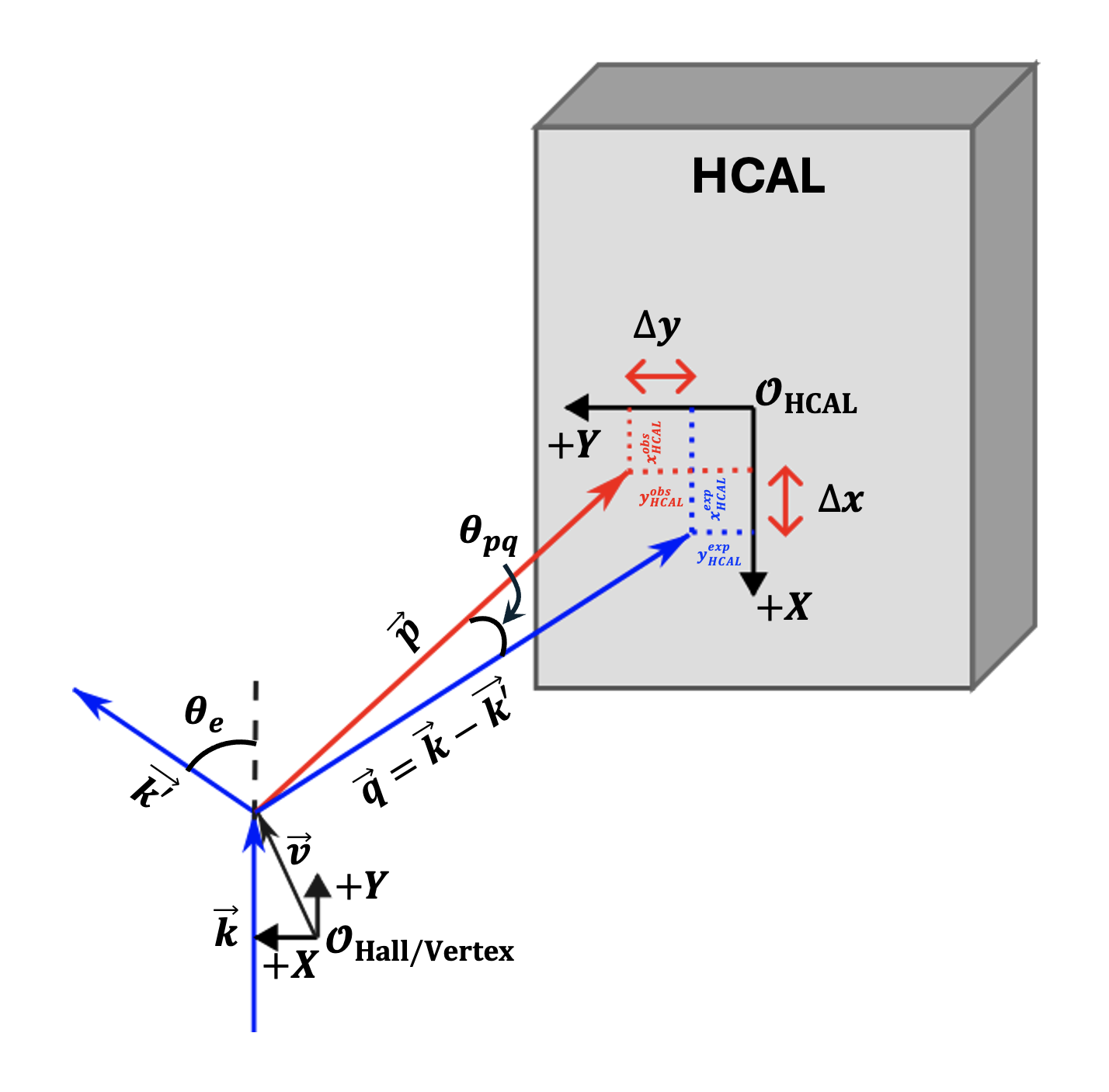}
	\caption{\label{fig:ch4:sketchdxdy} A conceptual and exaggerated diagram displaying the formulation of HCAL \dx and \dy variables as a difference between the reconstructed (\xhob,\yhob) and the expected (\xhex,\yhex) nucleon position at HCAL. The reconstructed positions are available from HCAL clustering while the expected ones are obtained by projecting the estimated \qvect to HCAL's front face. See text for more details. NOTE: The presence of the SBS magnet has been ignored here.}
\end{figure}
\subheading{Angles-Only Method}
%
In this approach, the components of $k$ and $k'$ in the lab frame are expressed as follows:
\begin{equation}
    \begin{aligned}
        k^\mu &\equiv (E_e,\vb{k}) = (E^{corr}_{beam},0,0,E^{corr}_{beam}) \\
        k'^{\mu} &\equiv (E'_e,\vb{k'}) = (p_{elas},p_{elas}\sin\theta_e\cos\phi_e,p_{elas}\sin\theta_e\sin\phi_e,p_{elas}\cos\theta_e)
    \end{aligned}
\end{equation}
where $p_{elas}$ is the momentum of the elastically scattered electron calculated based on $\theta_e$ using \eqn \ref{eqn:ch4:pelas}. Additionally, the four-momentum of the incident and scattered nucleons can be defined as:
\begin{equation}
    \begin{aligned}
        p &\equiv (p^{0},\vb{p}) = (M_{N},\vb{0}) \\
        p' &\equiv (p'^{0},\vb{q}) = (M_N+\nu,-|\vb{q}|\sin\theta_N\cos\phi_N,-|\vb{q}|\sin\theta_N\sin\phi_N,|\vb{q}|\cos\theta_N)        
    \end{aligned}
\end{equation}
where $\nu=E_{beam}^{corr}-p_{elas}$ is the energy transfer in the scattering process. Clearly, in this approach, the calculation of \qvect comes down to the estimation of $|\vb{q}|$ and the nucleon scattering angles $\theta_N$ and $\phi_N$. 

Squaring the four-momentum of the virtual photon, $q \equiv (\nu,\vb{q})$, yields:
\begin{equation}
\label{eqn:ch4:qsq1}
    q^2 = \nu^2 - |\vb{q}|^2  
\end{equation}

Alternatively, from four-momentum conservation:
\begin{equation}
\label{eqn:ch4:qsq2}
    \begin{aligned}
        q^{2} &= (p' - p)^2\\
              &= p'^2 - 2\vb{p'}\vdot\vb{p} + M_N^{2}\\
              &= 2(M_N^2 - (M_N+\nu)M_N), \,\,\,\\\
              &= -2M_N\nu
    \end{aligned}
\end{equation}

Combining \eqn \ref{eqn:ch4:qsq1} and \ref{eqn:ch4:qsq2}:
\begin{equation}
\label{eqn:ch4:qvectmod}
    |\vb{q}| = \sqrt{\nu^2 + 2M_N\nu}
\end{equation}

Coplanarity requirement of the elastic scattering yields:
\begin{equation}
\label{eqn:ch4:phin}
    \phi_N = 2\pi - \phi_e      
\end{equation}

Conserving the $z$-component of the momentum in the scattering process, we obtain:
\begin{align}
\label{eqn:ch4:thetan}
    &|\vb{q}|\cos\theta_N = E^{corr}_{beam} - p_{elas}\cos{\theta_e} \notag\\
    \Rightarrow\,\, &\theta_N = \arccos{\frac{E^{corr}_{beam}-p_{elas}\cos{\theta_e}}{|\vb{q}|}}
\end{align}

Therefore, the \qvect calculated using ``angles only" method can be written as:
\begin{equation}
\label{eqn:ch4:qvectanglesonly}
    \vb{q} = (-|\vb{q}|\sin\theta_N\cos\phi_N,-|\vb{q}|\sin\theta_N\sin\phi_N,|\vb{q}|\cos\theta_N)
\end{equation}
where $|\vb{q}|$, $\theta_N$, and $\phi_N$ are given by Equations \ref{eqn:ch4:qvectmod}, \ref{eqn:ch4:phin}, and \ref{eqn:ch4:thetan}, respectively.

The ``4-vector" method is optimal for \deeN events as it offers better separation of inelastic background. In contrast, the ``angles-only" method provides better resolution for \heep events, as these events are free from nuclear effects and uncertainties regarding the struck nucleon type. Consequently, the ``4-vector" method will be the default for calculating \qvect for \deeN events, while the ``angles-only" method will be used for \heep events in the analyses presented here, unless stated otherwise.

\subsubsection{Calculation of \xhex and \yhex} 
The expected nucleon coordinates at HCAL in the dispersive (\xhex) and transverse (\yhex) directions can be determined by projecting a straight line from the interaction vertex to the HCAL face along the direction of $\vu{q}$ using the following steps:
\begin{enumerate}
    \item First, the parametric distance between the interaction vertex and the expected nucleon position at HCAL, $s$, is calculated in the following form:
    \begin{equation*}
        s = \frac{(\vb{O} - \vb{v})\vdot\vu{z}}{\vu{q}\vdot\vu{z}}
    \end{equation*}
    where $\vb{O}$ and $\vb{v}$ are vectors to the HCAL origin and the interaction vertex from the Hall center, respectively, and $\vu{z}$ is a unit vector normal to the HCAL front surface pointing downstream. All these parameters are known from event reconstruction and geodetic survey. 
    \item The vector from the Hall origin to the expected nucleon position, $\vb{h}$, can then be calculated as follows:
    \begin{equation*}
        \vb{h} = \vb{v} + s\vu{q}
    \end{equation*}
    \item Finally, the coordinates of the expected nucleon position in the dispersive (\xhex) and horizontal (\yhex) directions in the HCAL coordinate system can be calculated as follows:
    \begin{equation*}
        \begin{aligned}
            x_{HCAL}^{exp} &= (\vb{h} - \vb{O}) \vdot \vu{x} \\
            y_{HCAL}^{exp} &= (\vb{h} - \vb{O}) \vdot \vu{y} 
        \end{aligned}
    \end{equation*}
    where $\vu{x}$ and $\vu{y}$ are directions along the respective HCAL coordinate axes.
\end{enumerate}

Notably, this calculation, based on straight-line projection, does not account for the deflection of charged particle tracks by the SBS dipole. Therefore, it is exact only for neutral particles like neutrons. The expected proton positions calculated this way will be offset by the magnitude of the SBS kick, as discussed below.
%

\subsubsection{HCAL \dxb and \dyb Distributions}
The difference between the observed and expected nucleon positions at HCAL in the dispersive (\dx) and horizontal (\dy) directions are defined as:
\begin{equation}
    \begin{aligned}
        \Delta{x} &= x_{HCAL}^{obs} - x_{HCAL}^{exp} \\
        \Delta{y} &= y_{HCAL}^{obs} - y_{HCAL}^{exp} 
    \end{aligned}
\end{equation}
The distributions of each of these variables reveal very interesting features and are crucial for the extraction of physics observables from \gmn data.

\fig \ref{fig:ch4:dxcomp} shows \dx distributions for (quasi) elastically scattered electron events, obtained with a strict cut on \w (see \sect \ref{sssec:ch4:w2cut}), from both \lh and \ld targets. Key observations from these distributions include:
\begin{figure}[h!]
    \centering
    \includegraphics[width=\sfig]{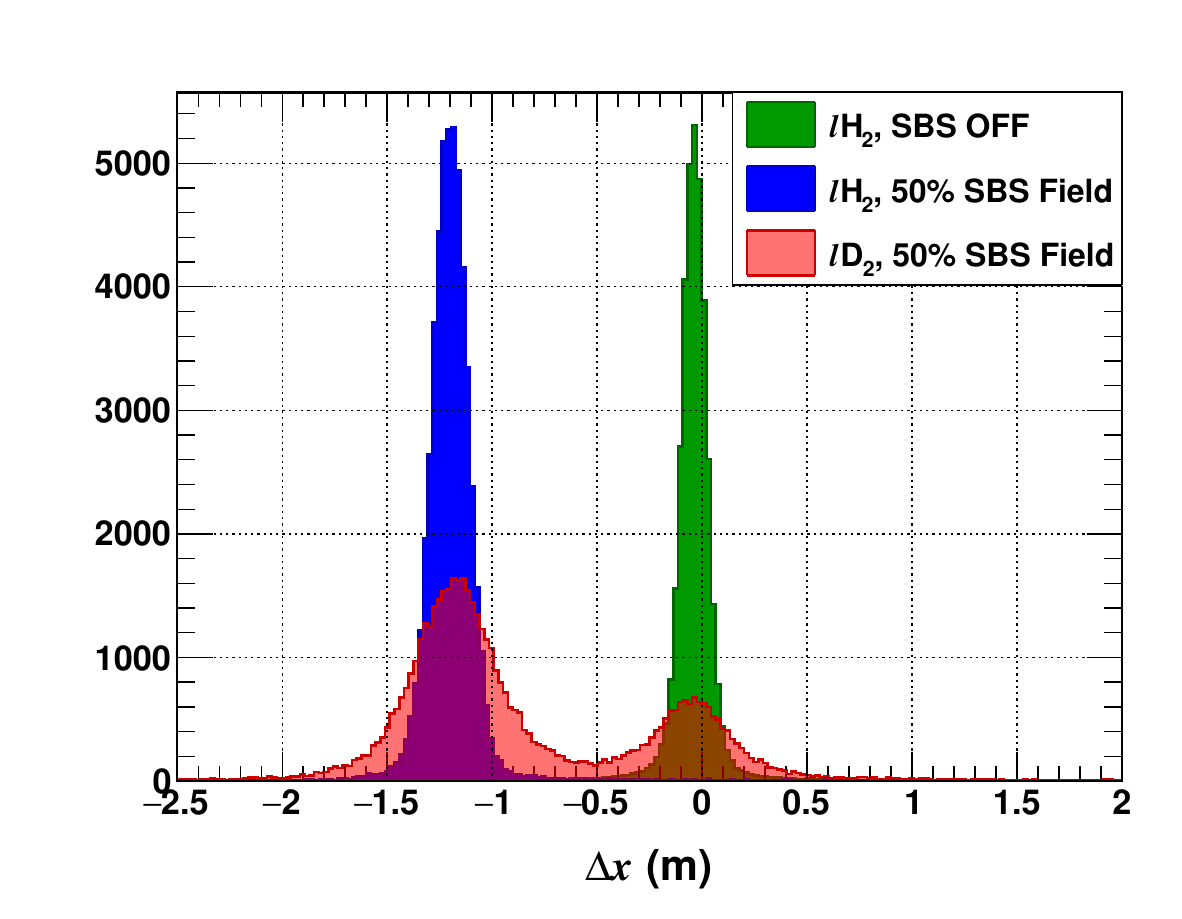}
    \caption{\label{fig:ch4:dxcomp} The comparison of \dx distributions generated using elastic and quasi-elastic events from \qeq{3} taken with multiple SBS field strengths. The scaling between different datasets is arbitrary.}
\end{figure}
\begin{enumerate}
    \item The \lh target distributions, resulting from \heep events, display a sharp peak at the origin or shifted left, depending on whether the SBS magnet is off or on. In contrast, the \deeN event distribution from the \ld target, with the SBS magnet on at the same field strength, features two peaks that align with the \heep event peaks. This trend is explained by the effect of the SBS dipole magnet on the scattered nucleon trajectories.

    \hspace{1em}It is crucial to note that the deflection of charged particles by the SBS dipole magnet is not accounted for in the calculation of \xhex, the expected nucleon position at HCAL in the dispersive direction. While neutron tracks pass straight through the dipole, proton tracks are deflected upwards by the field, with the magnitude of deflection determined by the SBS field strength, nucleon momentum, and SBS magnet-to-HCAL distance. Consequently, for scattered proton tracks, \xhob is shifted upwards from \xhex by the deflection introduced by the SBS magnet, causing a leftward shift in the corresponding \dx distribution. In contrast, scattered neutron tracks remain unaffected by the SBS field, causing their \dx distribution to peak at the origin. However, when the SBS magnet is turned off, the \dx distribution for scattered proton events should also peak at the origin, which is observed in \fig \ref{fig:ch4:dxcomp}.

    \hspace{1em}For the \ld target, the peak at the origin is due to \deen events, while the other peak is due to \deep events. The widths of these peaks are comparable, but their heights are not, indicating a much higher probability of quasi-elastic electron scattering off the proton within the deuteron than off the neutron at a given \q. This discrepancy in scattering cross-sections is due to fundamental differences in the internal structures of the nucleons.
\begin{figure}[h!]
    \centering
    \includegraphics[width=\sfig]{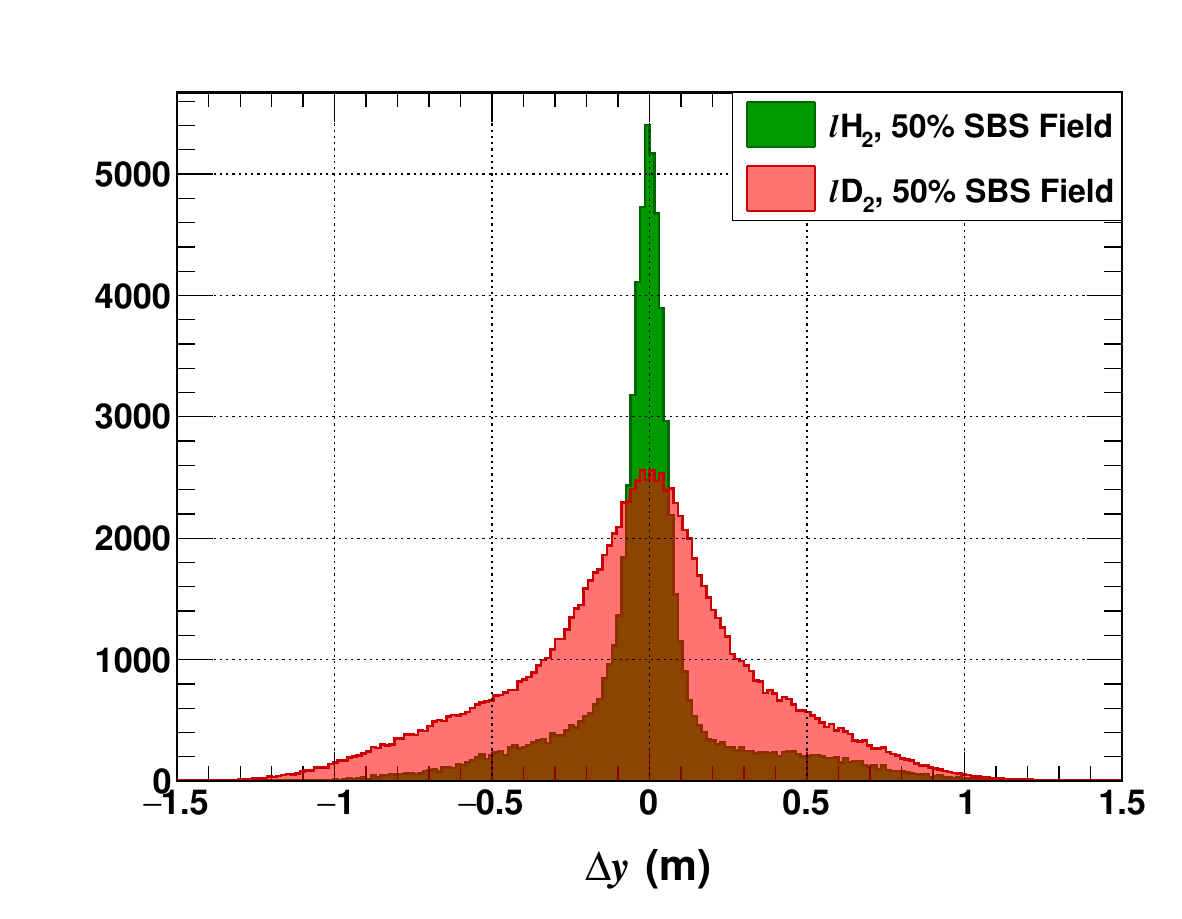}
    \caption{\label{fig:ch4:dycomp} Comparison of \dy distributions obtained using elastic and quasi-elastic events from \qeq{7.4} dataset. The former is calculated using the 4-vector method, while the latter uses the angles-only method. The scaling between datasets is arbitrary.}
\end{figure}
    \item The distributions for \deeN events are significantly broader than those for \heep events, primarily due to nuclear effects in the deuteron nucleus, including the Fermi motion of nucleons. For a given target, the peaks sharpen with increasing \q, driven by kinematic focusing and the improved position resolution of HCAL at higher nucleon momenta, resulting from the larger cluster size.
    %
\end{enumerate}   
Such clear separation of \deen and \deep events in the \dx distribution allows for the extraction of corresponding event counts to form the ratio \rqe, the direct experimental observable of \gmn, defined in \sect \ref{sec:measurementtechnique}. Refer to \sect \ref{sec:ch4:extrationofexpobservable} for a detailed discussion on the methodology of experimental observable extraction from \dx distribution.

The \dy distribution, being the nucleon position difference in the non-dispersive (horizontal) direction, features only one peak for both \heep and \deeN events, as shown in \fig \ref{fig:ch4:dycomp}. Good $e$-$N$ coincidence events passing a strict \w cut (see \sect \ref{sssec:ch4:w2cut}) at \qeq{7.4} have been used. Like \dx, the resolution of \dy is worse for \deeN events, largely due to the Fermi motion of the nucleons in the deuteron nucleus.

\subsection{Exclusive Event Selection}
Selecting good electron and nucleon events does not restrict the underlying scattering process. However, to extract physics observables from \gmn, we need to exclusively select (quasi) elastic scattering events. Exclusivity cuts, based on energy-momentum conservation, are necessary for this purpose. This section will discuss the definition and implementation of these cuts in detail.
%

%
\subsubsection{Invariant Mass Cut}
\label{sssec:ch4:w2cut}
The invariant mass squared, \w, of the virtual photon - struck nucleon system is a crucial kinematic variable for identifying elastic scattering events. It is a Lorentz invariant quantity, defined as follows:
\begin{align}
\label{eqn:ch4:w2}
    W^2 &= (q + p)^2
\end{align}
where $q$ and $p$ are the four-momenta of the virtual photon and the struck nucleon, respectively. Assuming the struck nucleon is at rest, $W^2$ can be reconstructed for each event using the known beam parameters and the measured scattered electron kinematics by BB.
    
Applying four-momentum conservation, $q = p' - p$, to \eqn \ref{eqn:ch4:w2} yields: 
\begin{equation}
    W^2 = {p'}^2
\end{equation}
where $p'$ is the four-momentum of the scattered nucleon. Therefore, $W^2$ should equal the mass squared of the nucleon, $M_N^2$, for elastic $eN$ scattering events. $M_N$ is defined as follows based on the type of the target:
\begin{equation}
\label{eqn:ch4:mn}
    M_N =
    \begin{cases}
        M_p,      & \text{for \lh target} \\
        \frac{1}{2}(M_p + M_n), & \text{for \ld target}
    \end{cases}
\end{equation}
For \ld target, the average nucleon mass is assumed  to account for the fact that the type of the struck nucleon is not known a priori.  

\begin{figure}[h!]
    \centering
    \begin{subfigure}[b]{0.496\textwidth}
         \centering
         \includegraphics[width=\textwidth]{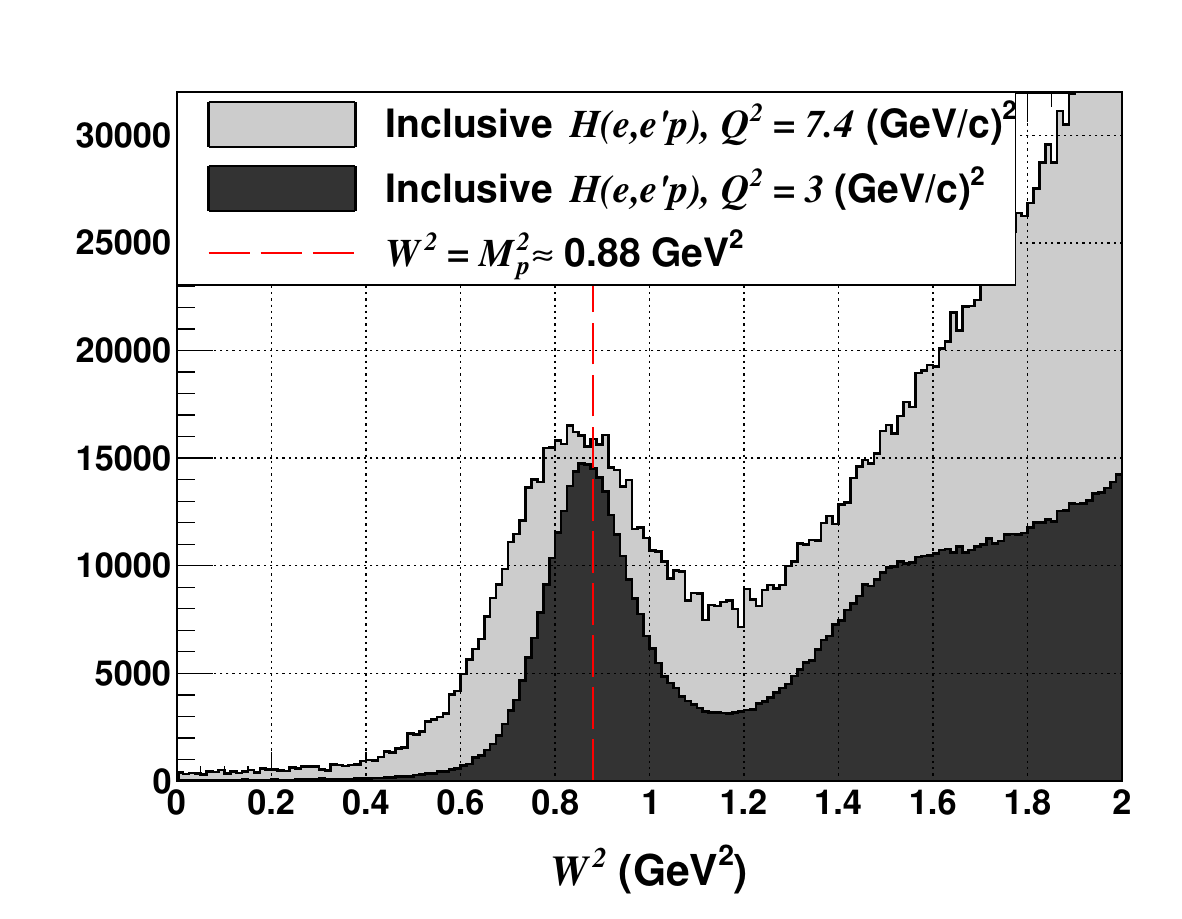}
         \caption{}
         \label{sfig:ch4:w2cut1}
    \end{subfigure}
    \hfill
    \begin{subfigure}[b]{0.496\textwidth}
        \centering
        \includegraphics[width=\textwidth]{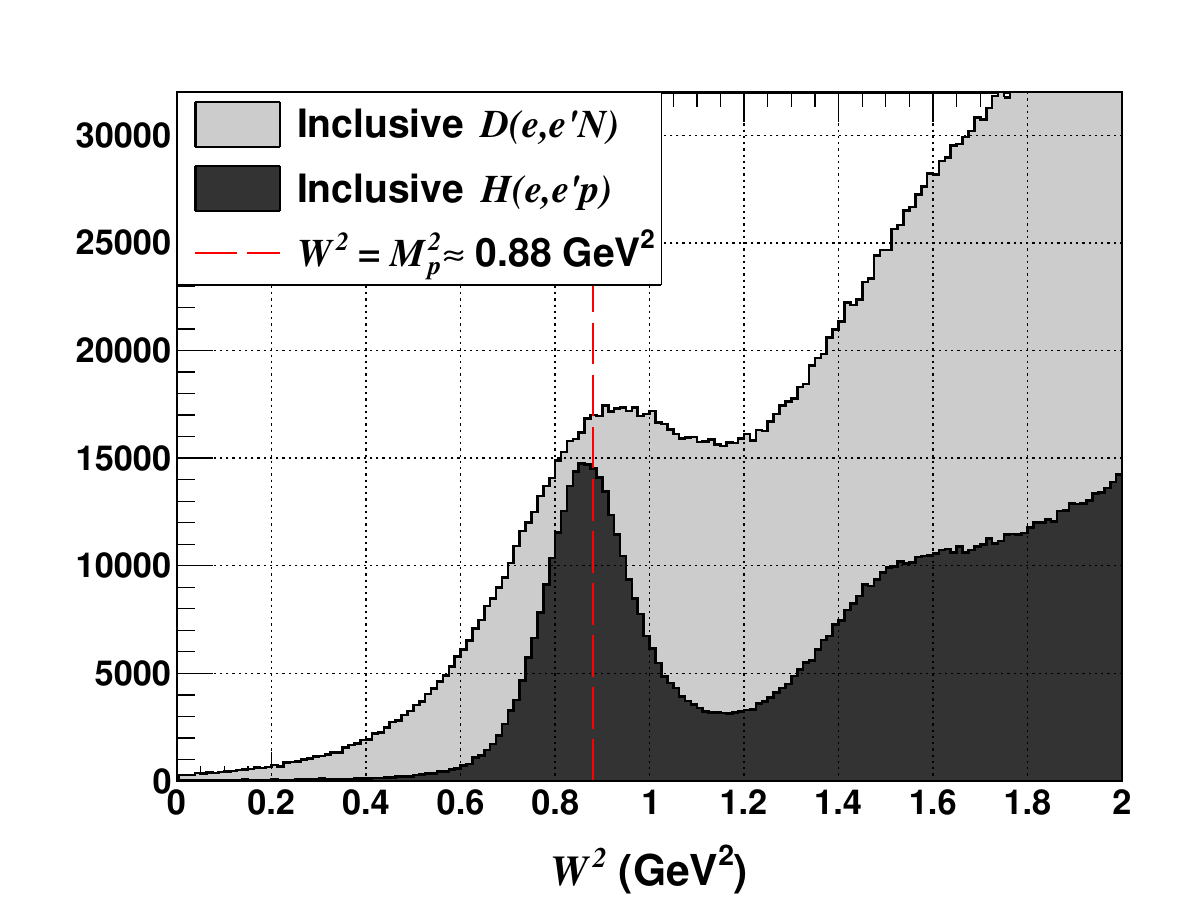}
        \caption{}
        \label{sfig:ch4:w2cut2}
    \end{subfigure}
    \caption{Reconstructed \w distributions for inclusive electron-nucleon scattering from \lh and \ld targets. The peak near the proton mass squared ($M_p^2$) is due to elastic scattering events. Comparison of \w resolution for elastic scattering events: (a) with increasing \q and (b) between \lh and \ld targets at \qeq{3}.}
    \label{fig:ch4:w2cut}
\end{figure}
\fig \ref{sfig:ch4:w2cut1} shows the reconstructed \w distributions for inclusive electron-proton scattering from \lh at \qeq{3\,\,\&\,\,7.4}. The peak near the proton mass is attributed to elastic scattering. The spread of this peak depends on various factors, including detector resolution, uncertainties in beam parameters, and radiative effects. Notably, \w resolution worsens with increasing \q due to the sensitivity of the measured scattered electron kinematic variables to initial conditions, a phenomenon known as ``kinematic broadening". Additionally, for a \ld target, various nuclear effects become significant. For instance, the Fermi motion of the target nucleon within the deuteron increases sensitivity to initial conditions by violating the assumption that the target nucleon is at rest, resulting in further broadening of the \w distribution at a given \q, as shown in \fig \ref{sfig:ch4:w2cut2}. 

\begin{figure}[h!]
    \centering
    \includegraphics[width=\sfig]{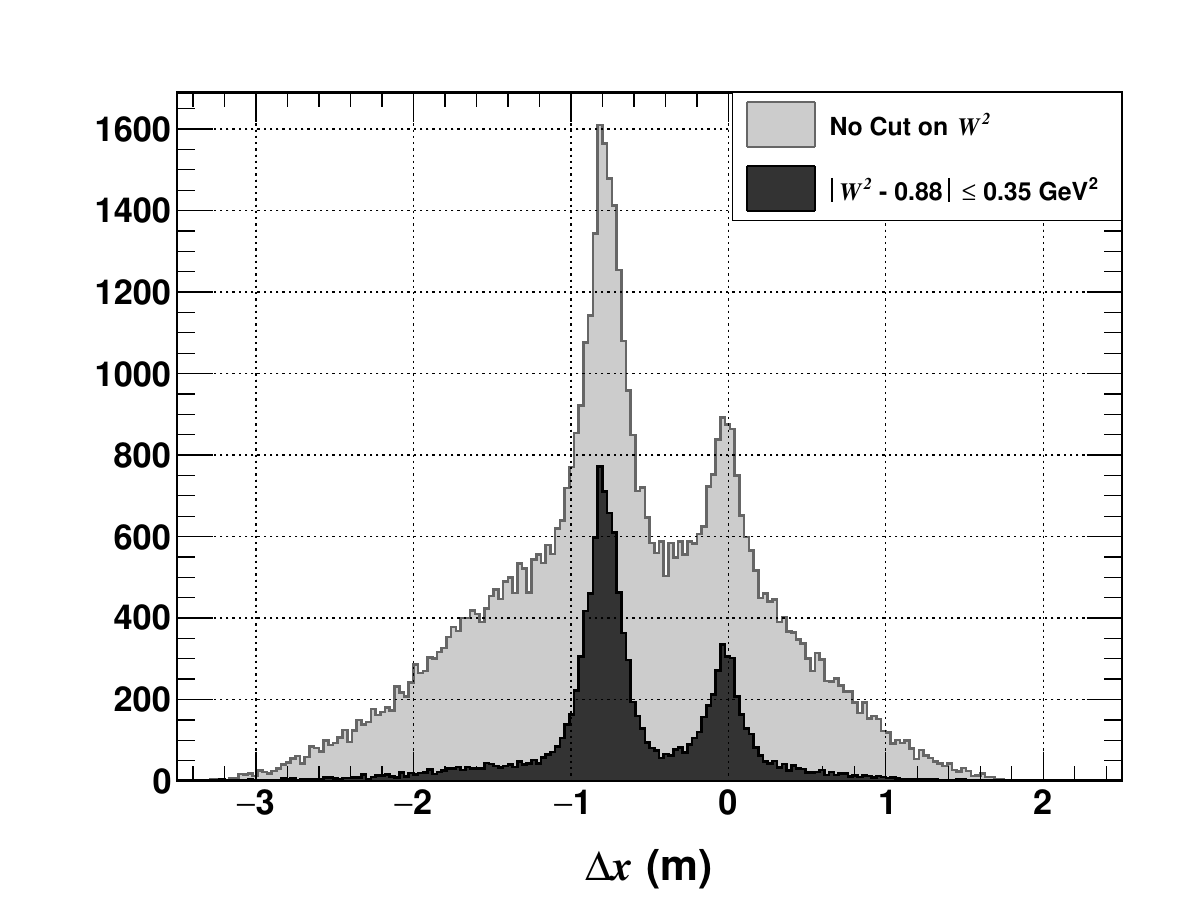}
    \caption{\label{fig:ch4:w2cuteffect}Effect of \w cut on quasi-elastic event selection at \qeq{9.9}.}
\end{figure}
Elastic electron-nucleon scattering events can be isolated by applying cuts around the elastic peak in the reconstructed \w distribution, as shown in \fig \ref{fig:ch4:w2cuteffect}. This cut region slightly varies with configuration to account for kinematic-dependent smearing. The optimal \w cut region for each configuration is determined based on realistic MC simulation and the stability of physics observables, a process discussed later in this chapter. Additionally, smearing of the \w distribution introduces background from inelastic scattering under the elastic peak, with this background rate increasing with \q, making elastic event selection more challenging. The following exclusivity cuts, based on electron-nucleon coincidence, greatly reduce this background.
%
%
\begin{figure}[] 
  \centering
  \includegraphics[width=1\columnwidth]{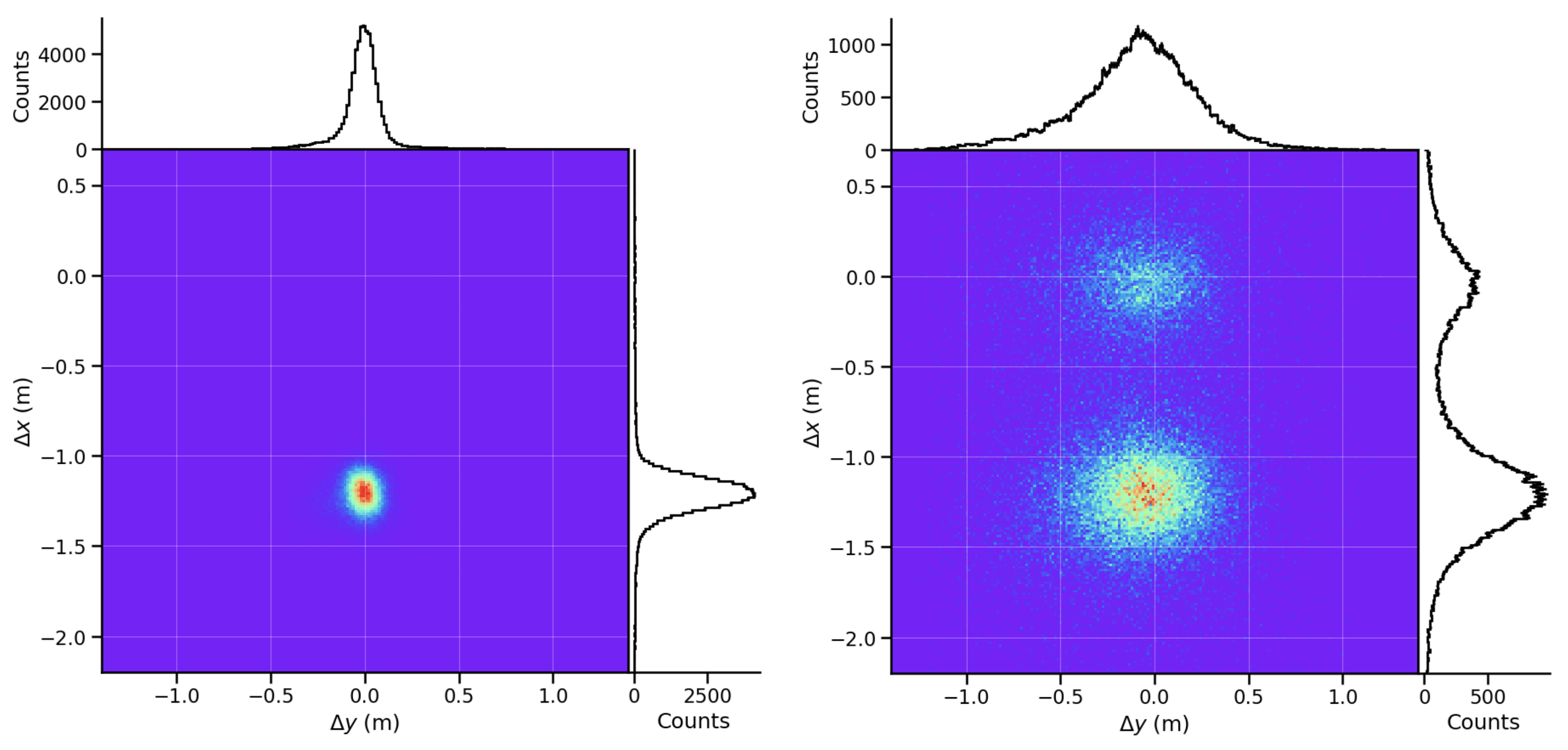}
  \caption{Comparison of HCAL \dx-\dy correlation between elastic \heep (left) and quasi-elastic \deeN (right) events using the \qeq{3} SBS $50\%$ field strength dataset. Good electron-nucleon coincidence events passing a strict \w cut (identical in both cases) are shown. Expected nucleon positions have been calculated using the ``angles only" method for \heep events and the ``4-vector" method for \deeN events. The elastic spot is very sharp and shifted away from the origin in \dx, indicating proton deflection by the SBS field. The quasi-elastic spots, while not as sharp due to nuclear effects, are clearly visible and well-separated. The spot from \deen events is centered at $(0,0)$, whereas the spot from \deep events is shifted in \dx by the same amount observed in the \heep events, as expected.}
  \label{fig:ch4:dxdycorr}
\end{figure}
%
\subsubsection{$eN$ Angular Correlation/\dx-\dy Correlation/\thpq Cuts}
\label{sssec:ch4:dxdycorrcut}
Simultaneous detection of scattered electrons and nucleons allows for the formation of exclusivity cuts based on their kinematic correlation. The \dx and \dy variables, defined in \sect \ref{ssec:ch4:enkinecorr}, measure the deviation of the nucleon position at HCAL from the predicted values, calculated under the assumption of (quasi) elastic scattering. Selecting events where these deviations fall within a user-specified tolerance effectively isolates (quasi) elastic scattering.
%

%
\begin{figure}[h!]
    \centering
    \includegraphics[width=\sfig]{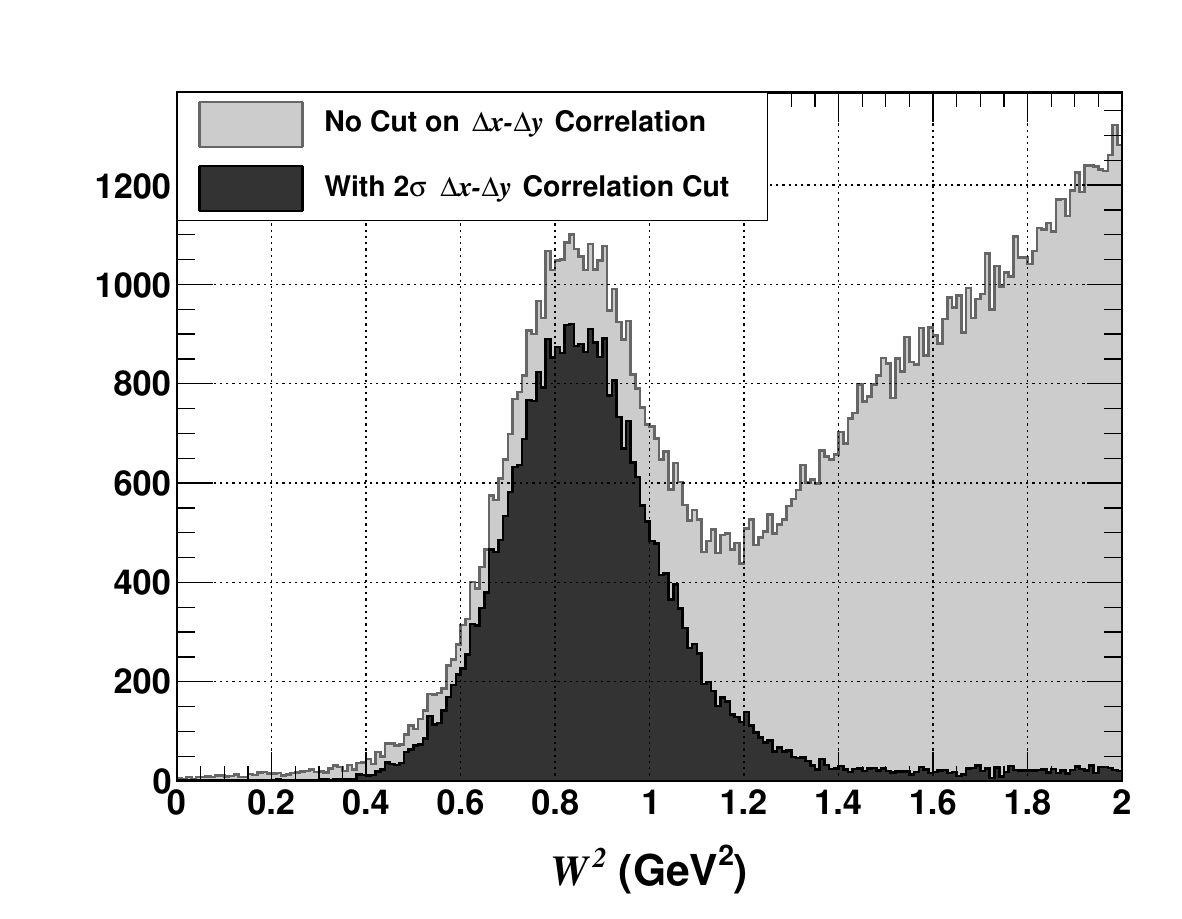}
    \caption{\label{fig:ch4:dxdycuteffect}Effect of \dx-\dy correlation cut on elastic event selection at \qeq{7.4}.}
\end{figure}
\fig \ref{fig:ch4:dxdycorr} shows the correlation between the \dx and \dy variables for \deeN events at \qeq{3}. Only good electron-nucleon coincidence events passing the \w cut are plotted. The elliptical spots represent quasi-elastic events: the one at \dx $=0$ corresponds to \deen events, while the other corresponds to \deep events. By forming elliptical cuts using the \dx and \dy variables, one can exclusively isolate \deen (within the red circle) and \deep events (within the blue circle). The cut ranges along \dx and \dy are guided by the widths of the corresponding distributions. A typical $2\sigma$ cut around the peak in both directions effectively suppresses inelastic background at \qeq{7.4}, as shown in \fig \ref{fig:ch4:dxdycuteffect}. However, there is room for optimization.

\subheading{\thpq Cut Formation and Its Equivalence to the \dx-\dy Correlation Cuts}
Instead of cutting on the \dx-\dy correlation, which can be inconvenient due to its elliptical nature, a cut on the angle between the \qvect and the reconstructed nucleon momentum $\vb{p}$, \thpq, can be imposed with an equivalent outcome. The concept behind calculating \thpq is similar to determining the expected nucleon coordinates at HCAL, but the process is reversed. For expected nucleon coordinates, a ray is projected from the interaction vertex to HCAL along the $\vb{q}$ direction, while for \thpq, a ray is projected from the observed nucleon positions at HCAL back to the interaction vertex. The angle between this ray and the \qvect gives the desired angle. This calculation is straightforward for neutron tracks, which travel directly through the SBS dipole magnet. However, complexity arises because the observed proton coordinates include the SBS kick. Therefore, making the \thpq calculation meaningful for protons requires accurately estimating their deflection by the SBS dipole event-by-event.

\begin{figure}[h!]
    \centering
    \includegraphics[width=1\columnwidth]{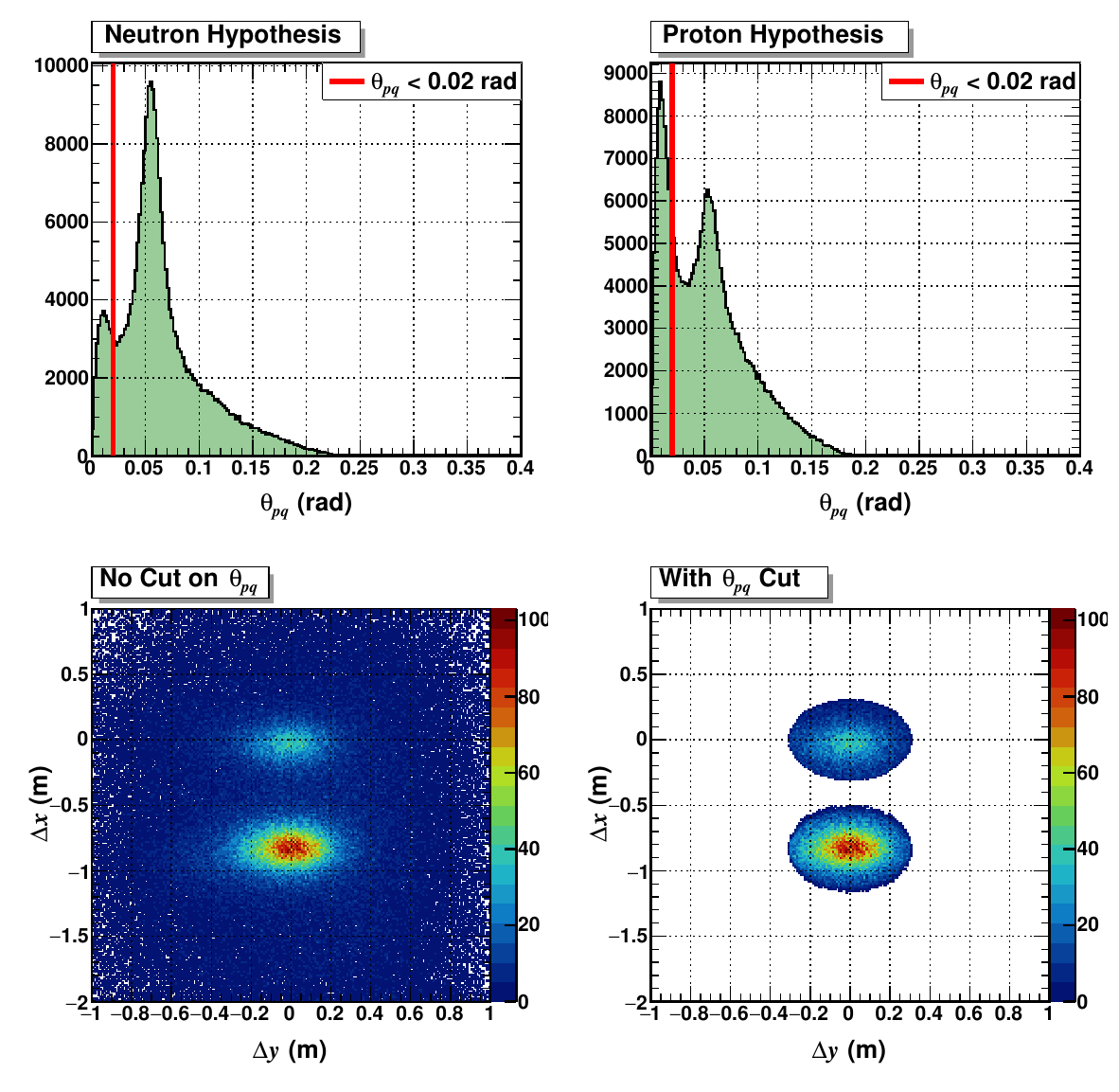}
    \caption{\label{fig:ch4:thpqcut}Effect of \thpq cuts on quasi-elastic event selection at \qeq{7.4}. The top row shows the \thpq distributions calculated using the neutron hypothesis (left) and proton hypothesis (right), with the red lines indicating typical \thpq cut regions for quasi-elastic event selection. The bottom row displays the \dx-\dy correlation without (left) and with (right) the \thpq cuts applied. The elliptical spots for \deen and \deep events resulting from these cuts replicate the \dx-\dy correlation cuts shown in \fig \ref{fig:ch4:dxdycorr}, confirming their equivalence. A $\theta_{pq} < 0.02$ rad cut is approximately equivalent to a $2\sigma$ \dx-\dy correlation cut at this \q.}
\end{figure}
Fortunately, SBS is a simple dipole magnet with a nearly uniform field within the gap, allowing a straightforward yet sufficiently accurate event-by-event calculation of the proton track deflection ($\delta x_{SBS}$) at HCAL using the equation:
\begin{equation}
\label{eqn:ch4:protondeflection}
    \delta x_{SBS} = \tan{\theta_{bend}} \, (d_{HCAL} - (d_{SBS} + d_{SBS}^{gap}/2))
\end{equation}
where $d_{HCAL}$ and $d_{SBS}$ are the distances of HCAL and the SBS magnet front-face from the target, $d_{SBS}^{gap} \approx 1.22$ m is the SBS dipole gap, and $\theta_{bend} = 0.3 \times BdL/|\vb{p}|$ is the proton track deflection angle. Here, $B$, representing the SBS field strength, is estimated by analyzing the correlation between \dx and $|\vb{p}|$ across \lh runs taken with different SBS field settings for a given kinematics. Ideally, the maximum SBS field ($B_{max}$) should be consistent across all kinematics, however, three distinct values were observed. At the start of \gmn, during the lowest \q kinematics, $B_{max}$ was approximately \SI{1.78}{T}. It was subsequently reduced to \SI{1.275}{T} by deactivating certain coils to mitigate stray fields. This value remained consistent for other kinematics, except for the high \ep dataset at \qeq{4.5}, where $B_{max}$ appeared to be \SI{1.23}{T}, roughly $3.5\%$ lower than expected. This discrepancy also affected the BBCAL cluster energy, as discussed in \sect \ref{sssec:ch4:bbcalibeng}, and is suspected to be related to hysteresis, though this remains unconfirmed.

The \thpq angle calculated using the proton hypothesis, accounting for the mentioned proton deflection, can be used to isolate quasi-elastic (and elastic, for \lh data) $ep$ events. In contrast, the same angle calculated with the neutron hypothesis, without accounting for the SBS kick, isolates quasi-elastic $en$ events. Together, these cuts effectively select quasi-elastic \deeN events from \ld data, as required by \gmn. The equivalence of these cuts to the \dx-\dy elliptical cuts is shown in \fig \ref{fig:ch4:thpqcut}.
%
%
%
\subsubsection{\dyb Cut}
\begin{figure}[h!]
    \centering
    \includegraphics[width=\sfig]{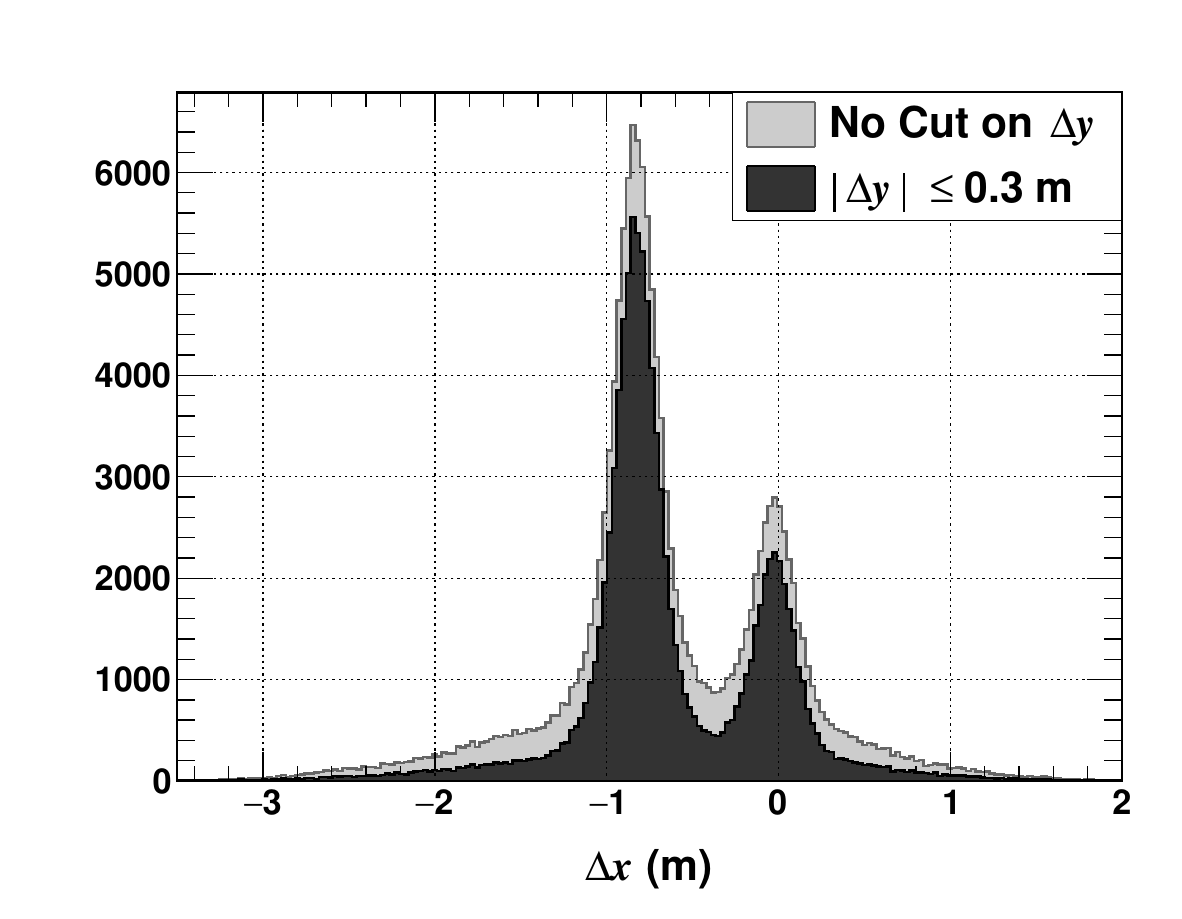}
    \caption{\label{fig:ch4:dycuteffect}Effect of \dy cut on the reduction of inelastic background in the \dx distribution at \qeq{7.4}.}
\end{figure}
As discussed above, the clear separation of \deen and \deep events in the \dx distribution is used to extract quasi-elastic yields from \gmn data. Therefore, the \dx-\dy correlation cut cannot be used for this purpose. Instead, cutting only on the \dy variable can help reduce background in the \dx distribution, leading to a cleaner extraction of physics observables. \fig \ref{fig:ch4:dycuteffect} shows the effect of the \dy cut on background reduction in the \dx distribution for good electron-nucleon coincidence events passing the \w cut at \qeq{7.4}. Similar to other cuts discussed, the optimal \dy cut range is determined via a cut sensitivity study, as discussed in \sect \ref{ssec:ch4:cutoptim}.

In summary, the exclusivity cuts used to isolate quasi-elastic scattering events for extracting physics observables from \gmn include the \w cut and the \dy cut. The optimal ranges of these cuts vary slightly across kinematics and are determined via cut sensitivity studies. It is worth noting that the \w and \dy variables are correlated, so the cut on one variable should be relaxed during the optimization of the other. Refer to \sect \ref{ssec:ch4:cutoptim} for a detailed overview of this process.
\subsection{HCAL Fiducial Cut}
A fiducial cut, based on the reconstructed electron kinematics detected by the BB, can be applied to the direction of $\vb{q}$ (as defined in \sect \ref{ssec:ch4:enkinecorr}) to restrict the trajectory of elastically scattered nucleons within a desired phase space. Applying the same fiducial region for both \deen and \deep events minimizes relative acceptance and efficiency differences, thereby significantly reducing systematic errors associated with the measurement of the ratio \rqe.

%
\begin{figure}[h!]
     \centering
     \begin{subfigure}[b]{0.65\columnwidth} 
         \centering
         \includegraphics[width=\textwidth]{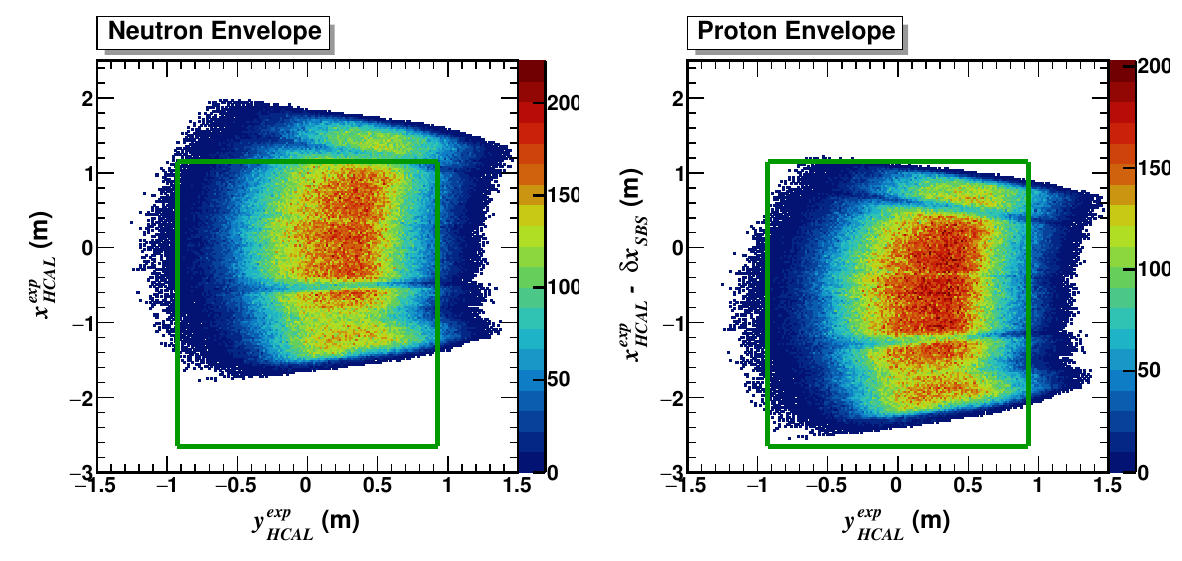}
         \caption{No fiducial cut.}
         \label{fig:ch4:fidu1}
     \end{subfigure}
     \begin{subfigure}[b]{0.65\columnwidth}
         \centering
         \includegraphics[width=\textwidth]{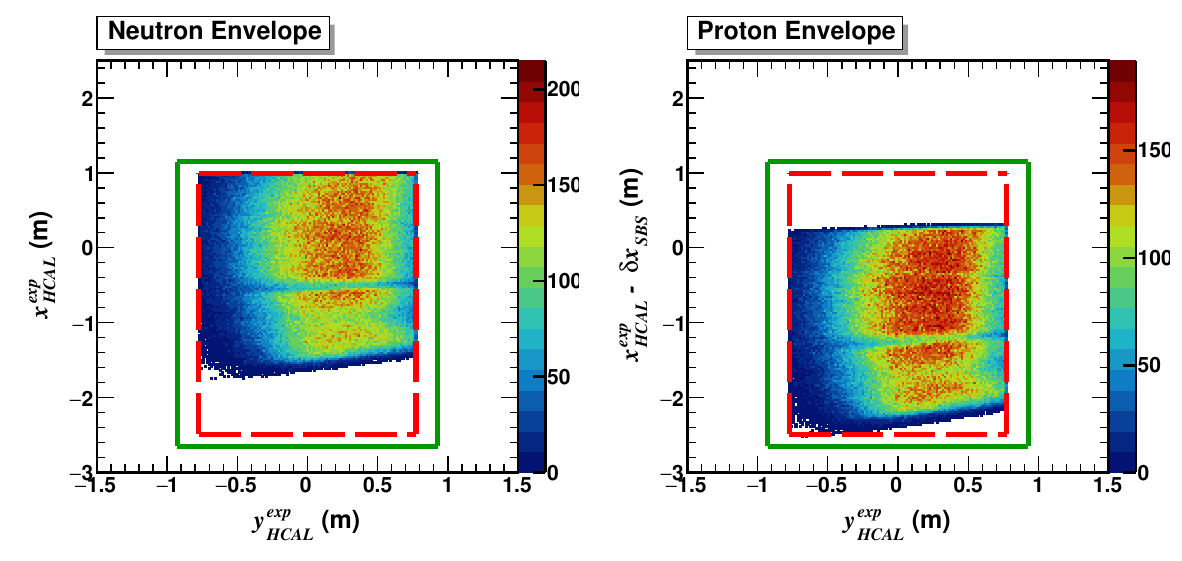}
         \caption{Fiducial region contained within HCAL active area.}
         \label{fig:ch4:fidu2}
     \end{subfigure}
     \begin{subfigure}[b]{0.65\columnwidth}
         \centering
         \includegraphics[width=\textwidth]{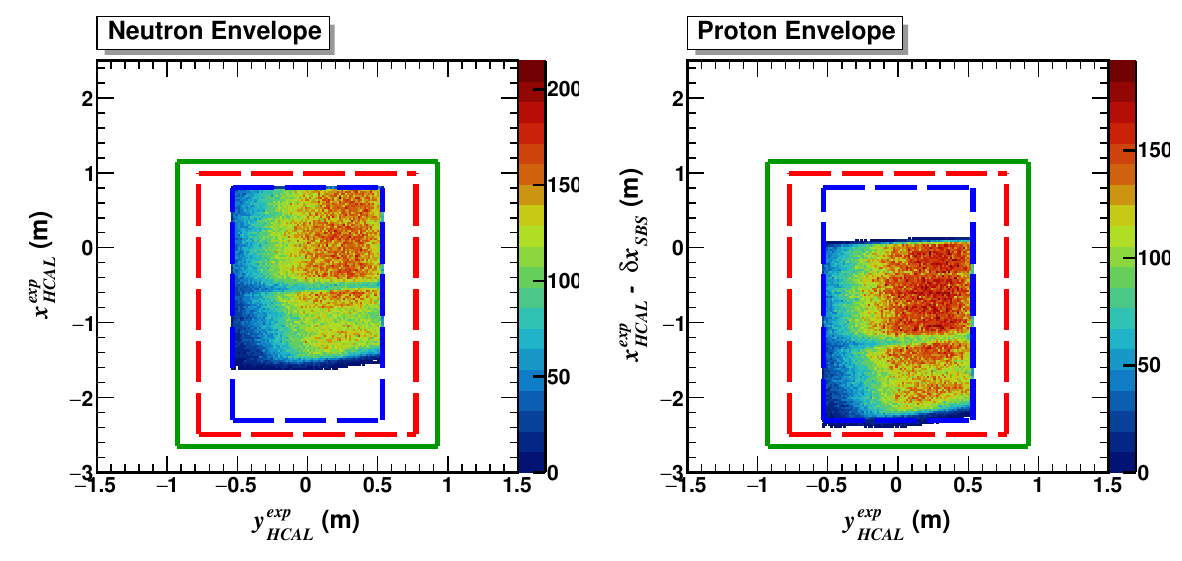}
         \caption{Fiducial region contained within additional safety margin.}
         \label{fig:ch4:fidu3}
     \end{subfigure}     
     \caption[Effect of fiducial cut on quasi-elastic events at \qeq{3}]{Effect of fiducial cut on quasi-elastic events at \qeq{3}. Green and red boundaries represent HCAL physical and active area, respectively. The gap between the red and blue boundaries defines the safety margin with a width of $1\sigma_{\Delta{x}}$ ($1\sigma_{\Delta{y}}$) in the dispersive (transverse) directions. With the fiducial cut applied, the effective acceptance matching between the proton and neutron envelopes is clearly visible.} 
     \label{fig:ch4:fiducut}
\end{figure}

As discussed in \sect \ref{ssec:ch4:enkinecorr}, the reconstructed $\vb{q}$ can be mapped to the expected nucleon position at HCAL in terms of \xhex and \yhex. The 2D correlation between these variables forms an ``envelope" of the expected nucleon position at HCAL, as shown in \fig \ref{fig:ch4:fidu1}. The proton envelope accounts for the deflection of proton tracks by the SBS dipole magnet ($\delta x_{SBS}$), shifting it vertically towards the top of HCAL. To determine if a specific electron-deuteron scattering event falls within the fiducial region, we calculate \xhex and \yhex for both proton and neutron events with the same underlying kinematics. The event is considered within the fiducial region if both particles fall within HCAL's active area, delineated by the red dashed lines in \fig \ref{fig:ch4:fidu2}. The HCAL active area is generally defined as the full physical area, excluding the outermost rows and columns.

Fermi motion in quasi-elastic events causes additional smearing of the observed nucleon position, introducing further acceptance loss. To mitigate this, a safety margin is established between the fiducial region and HCAL’s active area on all four sides, represented by the blue dashed lines in \fig \ref{fig:ch4:fidu3}. The width of this margin in the dispersive (transverse) directions is based on the widths of the \dx and \dy distributions, typically set at $1.5\sigma$ around their peaks. However, the cut range is optimized for each kinematic setting via cut-sensitivity studies.

Equipped with all the event selection cuts discussed in this section we are now ready to extract yields for \deep and \deen events to form the ratio \rqe from which $G_M^n$ can be extracted. However, it is of utmost importance to ensure the best possible calibration for all the detector sub-systems before the extraction of experimental observables to minimize both statistical and systematic uncertainties. 

\section{Detector Calibration and Performance}
\label{sec:ch4:detcalib}
Detector calibration is one of the most arduous yet crucial steps in data analysis. It is essential to ensure accurate and efficient event reconstruction and optimal signal to background ratio. Each detector sub-system is calibrated separately according to the experimental configuration, undergoing multiple rounds of calibration. Initial calibrations are performed using cosmic rays to ensure high-quality data collection. Subsequently, they are fine-tuned using elastic \heep events. The most up-to-date calibration parameters from all the detector sub-systems are then used to reconstruct the entire experimental dataset. The second pass of fine-tuning the detector calibration and reconstructing the entire \gmn dataset was completed in February 2024, and the resulting data have been used for the analysis presented in this thesis. In this section, we will briefly discuss the calibration steps, resulting detector performance, and any potential areas for improvement for all the detector sub-systems.

\subsection{Gas Electron Multipliers}
\label{ssec:ch4:gemcalib}
The calibration of the BB Gas Electron Multiplier (GEM) detectors involve several steps as listed below \cite{SUPPMAT}:
\begin{enumerate}
    \item \textbf{Alignment:} The initial step of GEM detector calibration involves aligning individual GEM modules relative to the position of the first GEM layer. The resulting positions and angles of the individual GEM modules are recorded in the database (DB) file (see \sect \ref{ssec:ch4:softtool}) for use during event reconstruction. This analysis is conducted using low current data runs taken throughout \gmn. Dedicated alignment is necessary for data taken in-between events that could potentially cause misalignment of the GEM layers, such as experimental configuration changes and the replacement of a faulty GEM layer. The special resolution averaged over all five GEM layers after alignment is observed to be approximately 100 $\mu$m.
    %
    \item \textbf{Gain Matching:} The gains of individual APV channels (see \sect \ref{ssec:ch3:bbgem}) within a GEM module are matched to zero ADC asymmetries across all GEM detectors. The resulting gain coefficients are added to the DB file. This calibration requires a lot of statistics, so both \lh and \ld runs were used. New gain matching is necessary whenever an APV card is replaced.
    %
    \item \textbf{ADC Threshold Calibration:} This is the immediate next step after ADC gain matching. Various ADC thresholds are calibrated to cleanly select good GEM hits. This includes thresholds for individual ADC samples, 1D clusters, and 2D clusters. The calibrated threshold values are added to the DB file to ensure clean and efficient track reconstruction.
    \item \textbf{Timing Calibration:} Several time cuts are implemented at the level of GEM strips to cleanly select good GEM hits that are in time. Fine-tuning these cut ranges using low current runs is necessary to avoid bias. Similar to ADC threshold calibration, timing calibration also requires the successful completion of gain matching. 
    \item \textbf{Track Search Region Constraint Calibration:} The track search region constraints, defined in \sect \ref{sssec:ch4:trackconstraint}, need to be fine-tuned whenever there is an update in the BBCAL energy calibration and/or optics calibration to avoid bias. 
\end{enumerate}

Following the second pass of fine-tuning GEM calibration, the performance of track reconstruction has been deemed sufficient to extract preliminary results.
\subsection{Gas Ring Imaging Cherenkov}
\label{ssec:ch4:grinchcalib}
\begin{figure}[h!]
    \centering
    \begin{subfigure}[b]{0.496\textwidth}
         \centering
         \includegraphics[width=\textwidth]{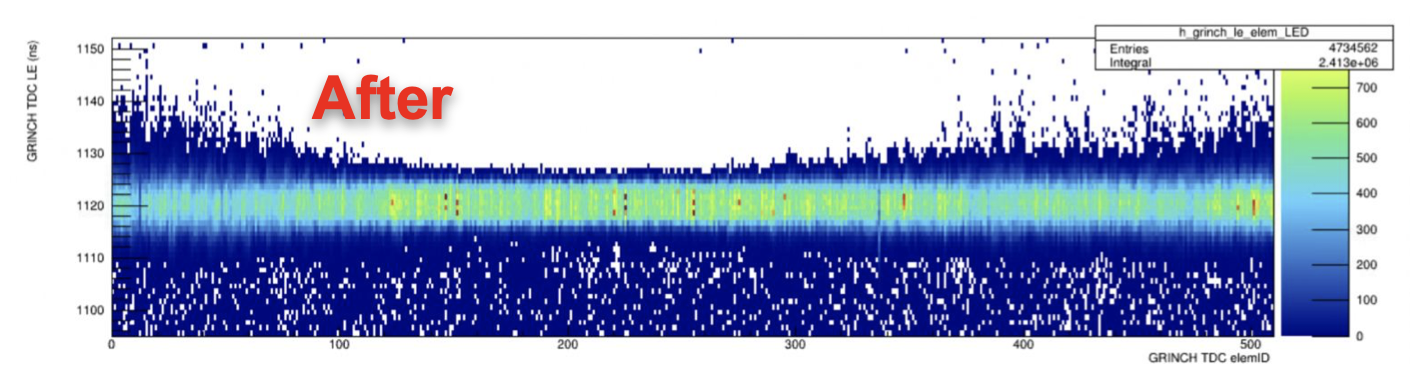}
         \caption{}
         \label{sfig:ch4:grinchcalib1}
    \end{subfigure}
    \hfill
    \begin{subfigure}[b]{0.496\textwidth}
        \centering
        \includegraphics[width=\textwidth]{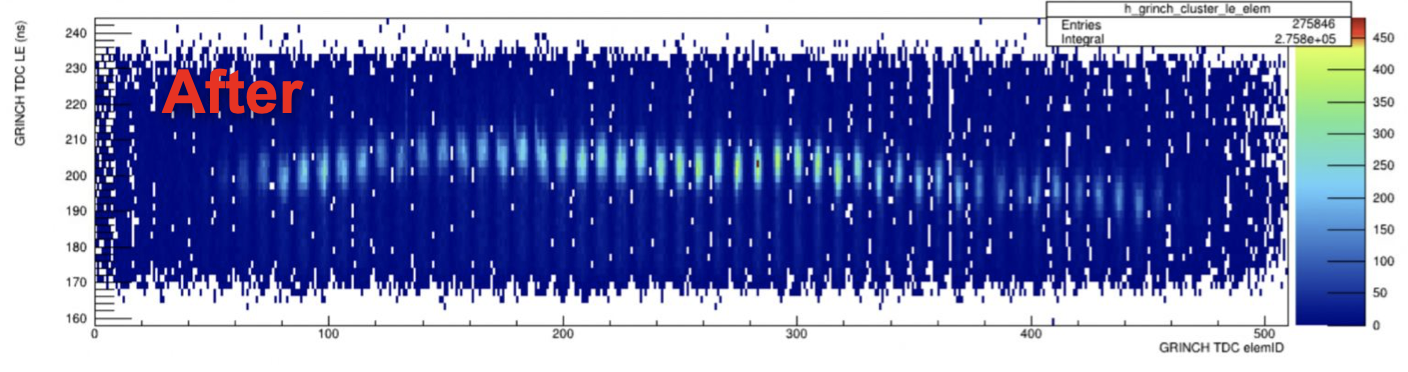}
        \caption{}
        \label{sfig:ch4:grinchcalib2}
    \end{subfigure}
    \caption{Result of leading edge time alignment for GRINCH PMTs \cite{SUPPMAT}.}
    \label{fig:ch4:grinchcalib}
\end{figure}

The primary calibration of Gas Ring Imaging Cherenkov (GRINCH) TDC data involves aligning the leading edge (LE) times of all the PMTs relative to a common reference. This alignment accounts for relative offsets among detector channels due to differences in the electronic circuitry, cable lengths, and PMT response. Achieving uniform calibration requires illuminating the entire detector. However, this cannot be done using cosmic or electron beam data due to GRINCH's unique design relative to other detector subsystems, as discussed in \sect \ref{ssec:ch3:grinch}. Therefore, LEDs installed in the frame of the mirrors, capable of illuminating the entire detector, are used. Signals with known characteristics (\SI{1}{ns} FWHM, \SI{100}{Hz}) generated from a signal generator were used to light up the LEDs, allowing for precise determination of offsets in the LE time for all the detector channels \cite{SUPPMAT}. \fig \ref{sfig:ch4:grinchcalib1} shows the aligned LE TDC times from all the detector channels using LED data. The effect of such alignment on elastic \heep events at \qeq{4.5} (low \ep) dataset is depicted in \fig \ref{sfig:ch4:grinchcalib2}.   

The correlation between the GRINCH cluster position and the projected electron track position at the GRINCH entry window varies slightly between kinematics. This variation likely arises because electrons from different kinematics strike a given position on the GRINCH mirror at different angles. These electrons are deflected by varying amounts by the BB dipole due to momentum differences, which alters the position of the associated Cherenkov light cone on the PMT array. To address this, a combined analysis of elastic \heep events from \qeq{4.5} low- and high-\ep kinematics was conducted to determine a universal set of parameters \cite{SUPPMAT}, enabling accurate matching of the GRINCH cluster with the track for optimal cluster selection.
\subsection{Timing Hodoscope}
\label{ssec:ch4:thcalib}
The calibration of the timing hodoscope (TH) leading edge (LE) TDC data involves fine-tuning several correction factors. The corrected TDC time for a TH PMT, $t_{corr}$, can be expressed as:
\begin{equation}
\label{eqn:ch4:thcalib}
    t_{corr} = t_{raw} - t_{0} - t_{TW} - t_{prop}
\end{equation}
where $t_{raw}$ is the raw PMT signal and $t_{0}$, $t_{TW}$, and $t_{prop}$ are correction factors accounting for unknown zero offset (including PMT response and cable length differences), time-walk in the NINO ASIC discriminator, and variation in light propagation time in the scintillator bar, respectively. The calibration procedure includes the following steps:
\begin{enumerate}
    \item \textbf{Initial Offset Alignment:} Align the TDC times for each PMT to an arbitrary value, initially set to zero. This involves aligning the time difference distribution for each bar to have a mean of zero, providing initial offsets for the left and right PMTs. Then, adjust the mean time to zero, further refining the offsets for each PMT. Record the resulting offsets in the database (DB) file (see \sect \ref{ssec:ch4:softtool}).
    \item \textbf{Time-Walk Correction:} Correct for time-walk in the NINO ASIC discriminator, indicated by the correlation between the LE TDC time and the Time Over Threshold (TOT)\footnote{The TOT is calibrated using the cosmic signal amplitude available via fADC data for a handful of PMTs. This calibration is crucial since the production data lacks fADC information for all PMTs, making TOT essential for time walk correction.} for each PMT, which is linear in nature. Fit a straight line to the distribution, and record the intercept ($a$) and slope ($b$) in the DB file.
    \begin{figure}[h!]
    \centering
        \includegraphics[width=0.95\columnwidth]{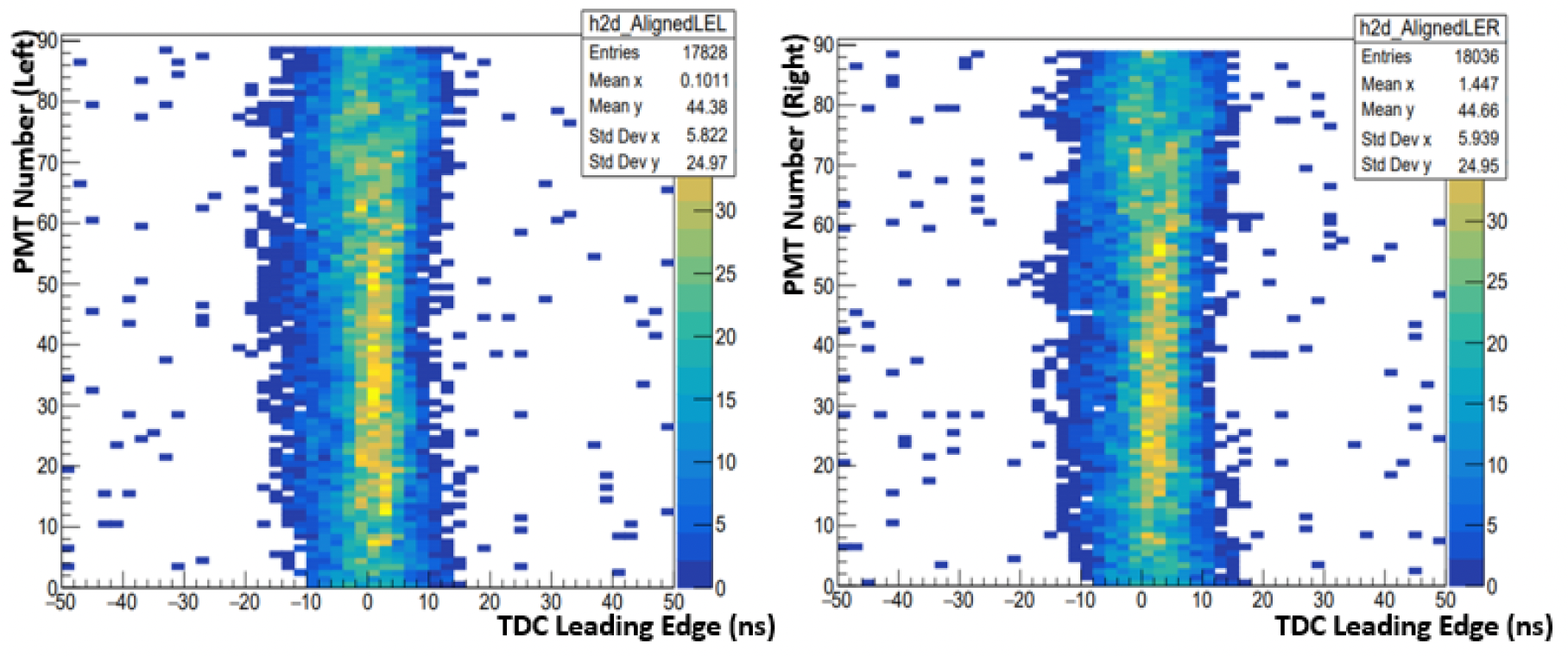}
        \caption{\label{fig:ch4:thalignment}Result of leading edge time alignment for timing hodoscope PMTs \cite{thesisRalphM}.}
    \end{figure}
    \item \textbf{GEM Track Matching:} Match GEM tracks to the hodoscope plane. Fit the correlation between the time-walk corrected left and right PMT time difference ($t_{diff}$) for each bar and the GEM track hit position on the bar ($y_{tr}$) with a straight line:
    \begin{align*}
        t_{diff} = t_{0} + \frac{y_{tr}}{v_{scint}}
    \end{align*}
    The slope provides the effective light propagation speed ($v_{scint}$) in the bar, accounting for physical non-uniformity, and the intercept aligns the $t_{diff}$ of the bar with the GEM tracks. Record these values in the DB file.
\end{enumerate}
Repeat the above steps several times to achieve the best possible calibration. \fig \ref{fig:ch4:thalignment} shows the alignment of LE times of all the TH PMTs after calibration.

\subheading{Performance of TH during \gmn}
The intrinsic time resolution of a TH bar, $\sigma_t$, is determined by calculating the difference in mean time between that bar and an adjacent bar within the same cluster, which cancels out common uncertainties, such as the trigger time resolution. The average intrinsic resolution across all TH bars ($\Bar{\sigma_t}$) serves as a key benchmark for performance, with an expected value of approximately \SI{200}{ps} by design \cite{SUPPMAT}.

\begin{figure}[h!]
    \centering
    \begin{subfigure}[b]{0.496\textwidth}
         \centering
         \includegraphics[width=\textwidth]{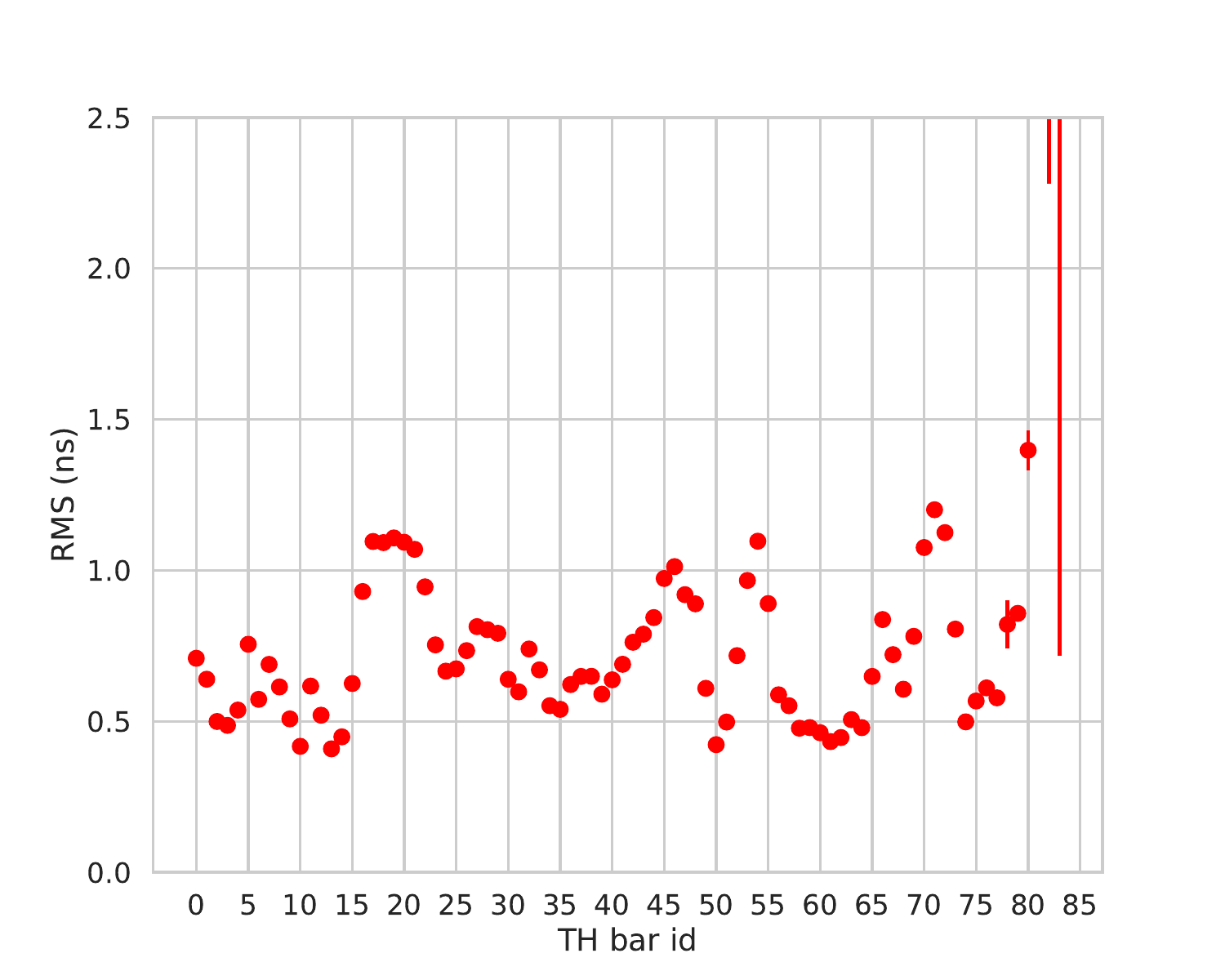}
         \caption{}
         \label{sfig:ch4:thresolution1}
    \end{subfigure}
    \hfill
    \begin{subfigure}[b]{0.496\textwidth}
        \centering
        \includegraphics[width=\textwidth]{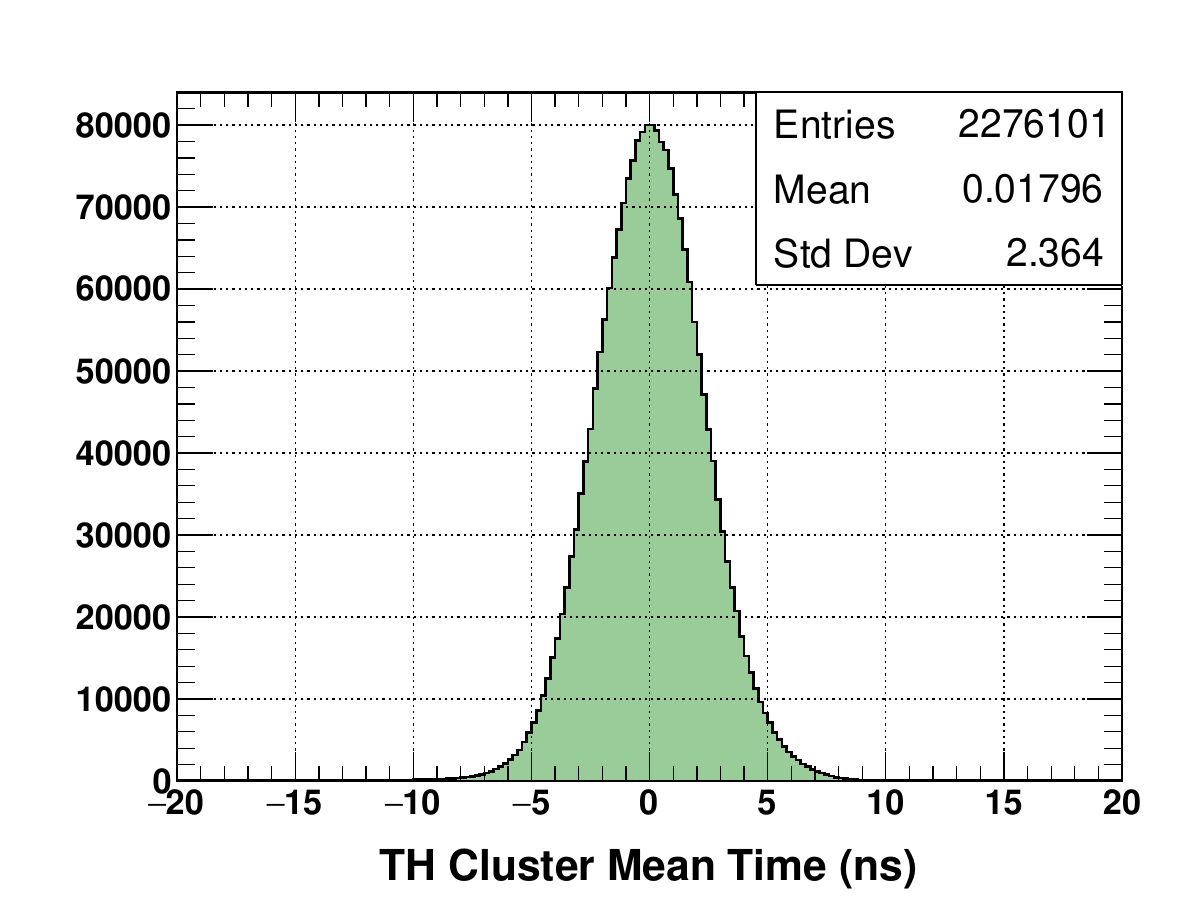}
        \caption{}
        \label{sfig:ch4:thresolution2}
    \end{subfigure}
    \caption{Current status of the Timing Hodoscope (TH) calibration for \gmn. (a) Profile of the RMS of the mean time difference between two consecutive TH bars in a cluster. (b) Distribution of the cluster mean time. These plots were generated using \lh data from the \qeq{3} kinematics.}
    \label{fig:ch4:thresolution}
\end{figure}
Analysis using cosmic data from the \gmn experiment shows a $\Bar{\sigma_{t}}$ of approximately \SI{300}{ps}, closely aligning with expectations. However, analysis using good electron events at \qeq{3} yields a significantly worse $\Bar{\sigma_{t}}$ of approximately \SI{500}{ps} \cite{SUPPMAT}. The cause of this degradation in resolution with electron beam data is not fully understood and is expected to improve with better calibration.

Additionally, random variations in $\sigma_t$ across TH bars, as shown in \fig \ref{sfig:ch4:thresolution1}, exceed the expected intrinsic resolution and do not correlate with physics observables like electron time-of-flight (TOF). This suggests that the current zero offsets for TH PMTs ($t_0$) and bars are suboptimal. Further analysis across all \gmn kinematic points shows a variation in $\Bar{\sigma_t}$ from \SI{450}{ps} to \SI{750}{ps}, depending on the scattered electron momentum. This trend of degraded resolution with increasing momentum is likely due to the spatial spread of hits on the TH bars, caused by the electromagnetic shower from the electron track in the Pre-Shower calorimeter located just upstream of the TH, as supported by simulation \cite{SUPPMAT}. However, the magnitude of this degradation is expected to be reduced with improved calibration.
%

For \gmn analysis, the goal is to achieve the best possible resolution for the two-arm coincidence time, defined as the difference between the HCAL cluster time and the TH cluster mean time. However, several challenges must be addressed: 
\begin{itemize}
    \item With the current calibration, the width of the TH cluster mean time distribution is approximately $2-3$ ns, as shown in \fig \ref{sfig:ch4:thresolution2}, which is about ten times the intrinsic resolution of TH. This poor resolution, primarily dominated by trigger time resolution and its variation within the acceptance, limits the effectiveness of additional background rejection beyond the existing event selection cuts.
    %
    \item The optimal coincidence time resolution should come from the difference between TH and HCAL TDC times. However, as discussed in \sect \ref{sssec:ch4:hcaltimingcalib}, HCAL TDC data is missing for many quasi-elastic events, particularly at high-\q points, necessitating the use of HCAL ADC time, which has worse resolution than HCAL TDC time.
\begin{figure}[h!]
    \centering
    \begin{subfigure}[b]{0.496\textwidth}
         \centering
         \includegraphics[width=\textwidth]{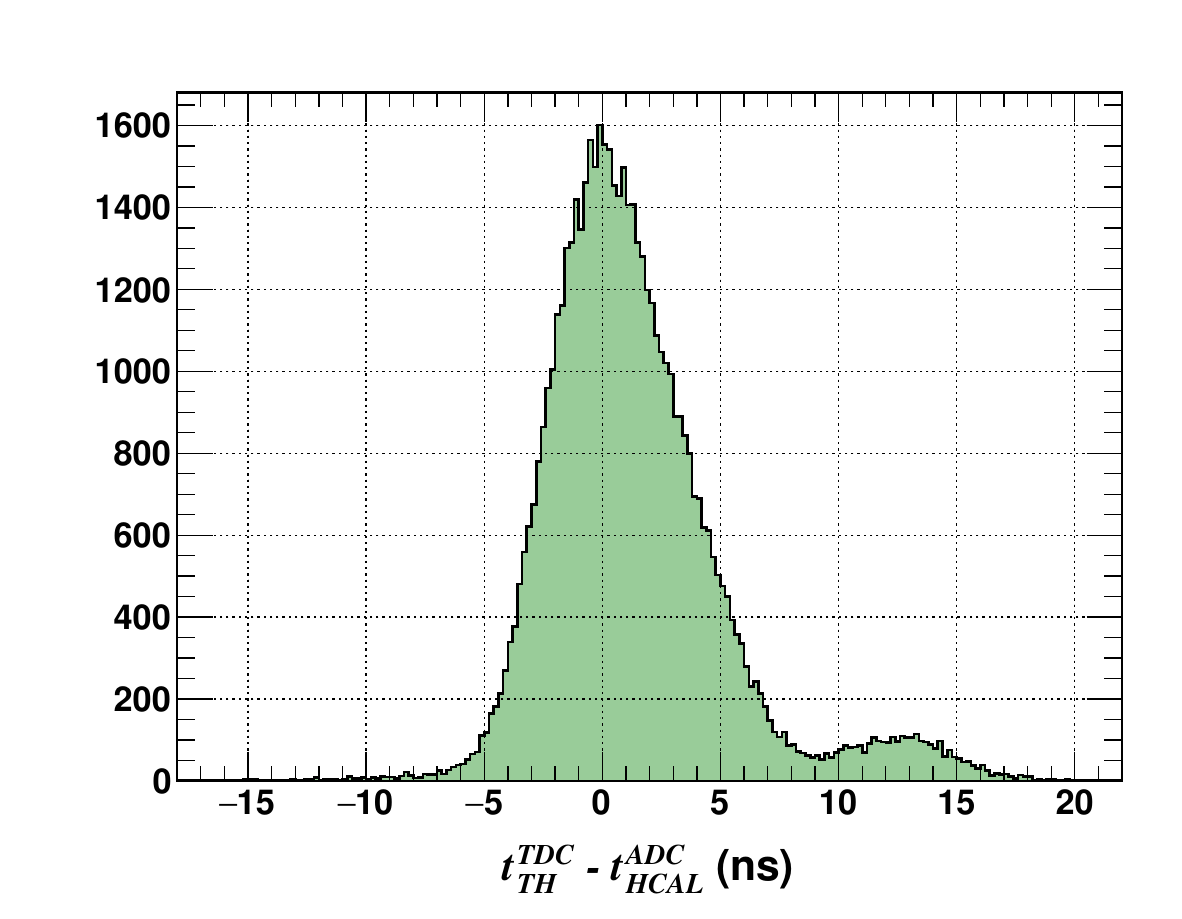}
         \caption{}
         \label{sfig:ch4:thissue1}
    \end{subfigure}
    \hfill
    \begin{subfigure}[b]{0.496\textwidth}
        \centering
        \includegraphics[width=\textwidth]{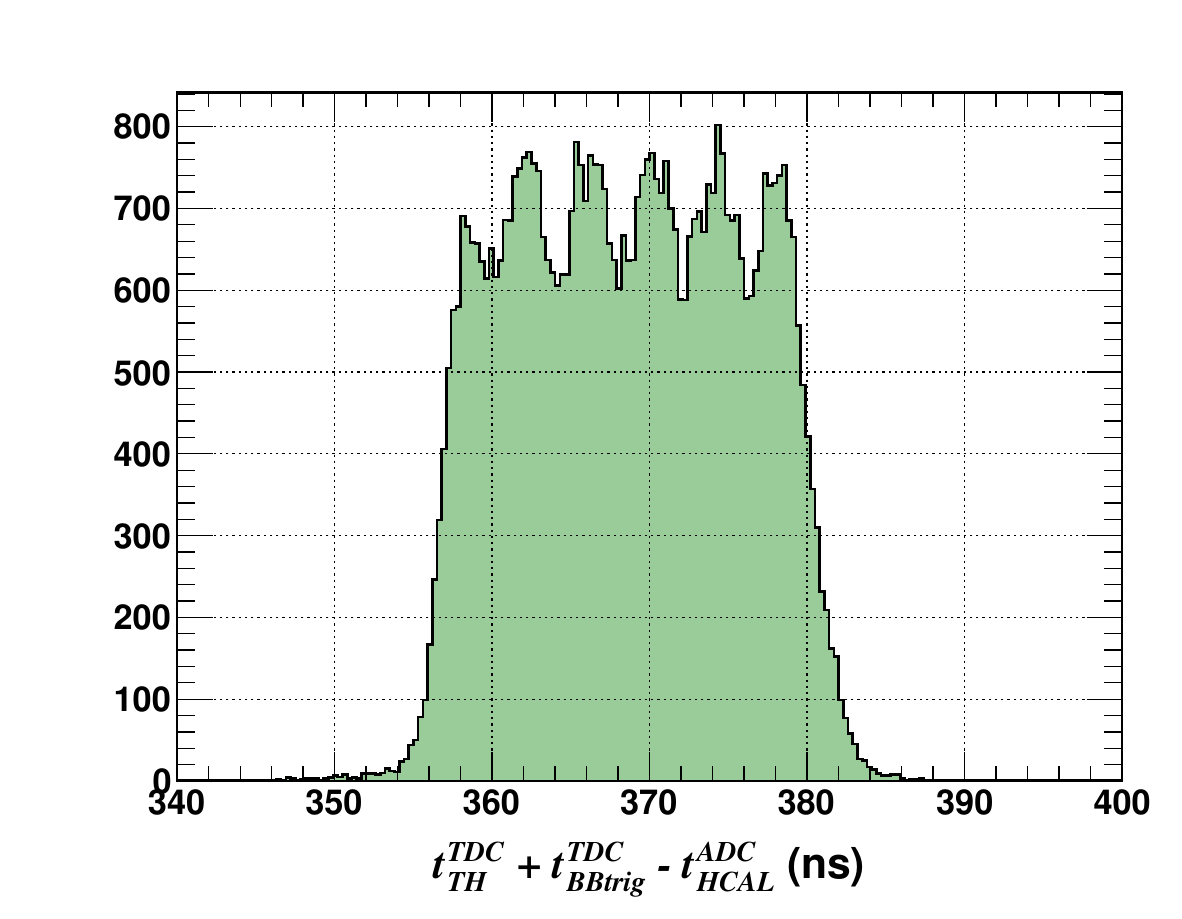}
        \caption{}
        \label{sfig:ch4:thissue2}
    \end{subfigure}
    \caption{Issues related to the computation of the time difference between the Timing Hodoscope (TH) TDC and HCAL ADC. Data shown is from \lh at \qeq{3} kinematics. See text for details.}
    \label{fig:ch4:thissue}
\end{figure}
    \item TH TDC data and HCAL ADC data are not directly comparable. TH TDC data is reference time subtracted, while HCAL ADC data is not. The trigger time, used as the reference, has exhibited random shifts of approximately \SI{12}{ns} throughout \gmn, likely originating from the Trigger Supervisor and affecting all detector subsystems equally. This shift is removed from TH TDC data but remains in HCAL ADC data, causing a double-peaking structure in the coincidence time distribution, as shown in \fig \ref{sfig:ch4:thissue1}. This double peak complicates the selection of good coincidence events. 
    
    \hspace{1em}Although adding the trigger time back to TH TDC data before subtracting HCAL ADC time addresses the above-mentioned issue, it reveals multiple peaks separated by \SI{4}{ns}, as shown in \fig \ref{sfig:ch4:thissue2}. This structure is suspected to be an artifact of the \SI{4}{ns} sampling of the fADC, but further investigation is needed before it can be corrected and used for physics analysis.

    \hspace{1em}Additionally, the different internal clocks associated with the CAEN 1190 TDCs (24 ns) and fADC 250s (4 ns) introduce a random 24 ns jitter in the former relative to the latter. Corrections to remove this jitter have yet to be developed.
\end{itemize}

Due to these issues, TH data has not been used to calculate the two-arm coincidence time for the analysis presented in this thesis. However, the individual TH bar mean times are used as a common reference to align the times of BBCAL and HCAL modules.  

\subheading{Thoughts on Future Improvements}
The existing TH calibration was conducted using cosmic data from early in the \gmn run. Recalibration with elastic \heep events is expected to yield improved results. Furthermore, implementing kinematic-specific calibration should help address the correlation with scattered electron momentum, as previously mentioned.

The particle arrival time measured by the TH can be biased due to variations in the trajectory path length from the interaction vertex to different bars of the detector. Although this variation is small (ranging from $14.8$ to $15.6$ ns within acceptance at $\Bar{E_{e}}=1.63$ GeV), it matters at the scale of our desired resolution of $0.1-0.2$ ns \cite{SUPPMAT}. The current TH calibration framework, as expressed in \eqn \ref{eqn:ch4:thcalib}, does not account for this correction factor, known as the time-of-flight (TOF) correction. Recalibrating TH data with the TOF correction implemented is expected to significantly improve $\sigma_t$ observed during \gmn.
\begin{figure}[h!]
    \centering
    \includegraphics[width=\sfig]{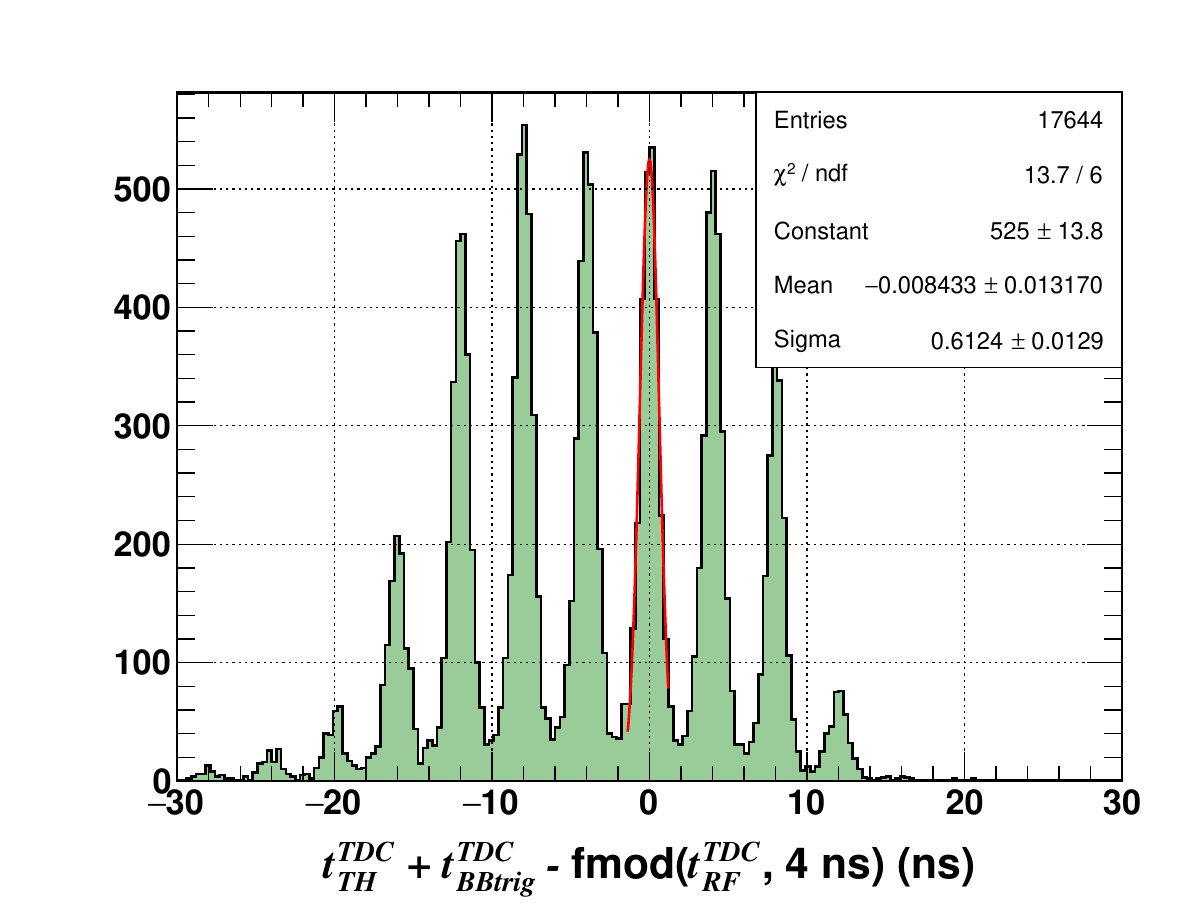}
    \caption{\label{fig:ch4:rfstructure} RF time structure obtained by subtracting the RF time ($t^{TDC}_{RF}$) from the raw Timing Hodoscope (TH) time ($t^{TDC}_{TH} + t^{TDC}_{BBtrig}$) recorded during \qeq{4.5} high-\ep kinematics. Only \lh events where TH bar 44 is the primary bar in the cluster are shown.}
\end{figure}

Proper analysis of the accelerator Radio Frequency (RF) time data recorded during \gmn allowed us to reconstruct the electron beam bunch spacing of \SI{4}{ns} with a very high resolution of approximately \SI{0.6}{ns}, as shown in \fig \ref{fig:ch4:rfstructure}. This is expected to mitigate the effect of poor trigger time resolution, which tends to dominate the individual bar mean time resolution as mentioned above. Implementation of such an analysis is a work-in-progress, but once available, it should significantly improve the current situation.

In summary, TH performed very well during \gmn, though there is room for improvement in its calibration, which is currently being addressed. However, it is important to note that further enhancement of TH performance, while beneficial, is not essential for \gmn analysis. The resolution achieved from the BBCAL-HCAL ADC coincidence time is already sufficient to effectively suppress accidental background across all kinematic points and still has potential for further improvement.
\subsection{BigBite Calorimeter}
\label{ssec:ch4:bbcalcalib}
The BigBite calorimeter (BBCAL) measures both the energy and arrival time of scattered electrons. To enhance its overall performance, both energy and time calibrations are performed. As discussed in \sect \ref{ssec:ch3:bbcal}, BBCAL has a segmented design, consisting of Pre-Shower (PS) and Shower (SH) calorimeters. Therefore, optimal performance can only be achieved by calibrating all channels associated with both SH and PS.

\subsubsection{BBCAL Timing Calibration}
\label{sssec:ch4:bbcalcalibtime}
The BBCAL data, unlike TH and GRINCH, is not read out by TDCs; instead, timing information is derived from the ADC time, which is essentially the leading-edge (LE) time calculated from the ADC pulse recorded by the fADC. For optimal resolution, channel-specific offsets in the ADC time must be aligned across all detector channels, similar to TDC alignment. Given that TH is designed with better timing resolution than BBCAL, the ADC time offsets for all BBCAL channels are determined relative to TH. This alignment is carried out separately for each SH and PS channel through the following steps:
\begin{figure}[h!]
    \centering
    \begin{subfigure}[b]{0.496\textwidth}
         \centering
         \includegraphics[width=\textwidth]{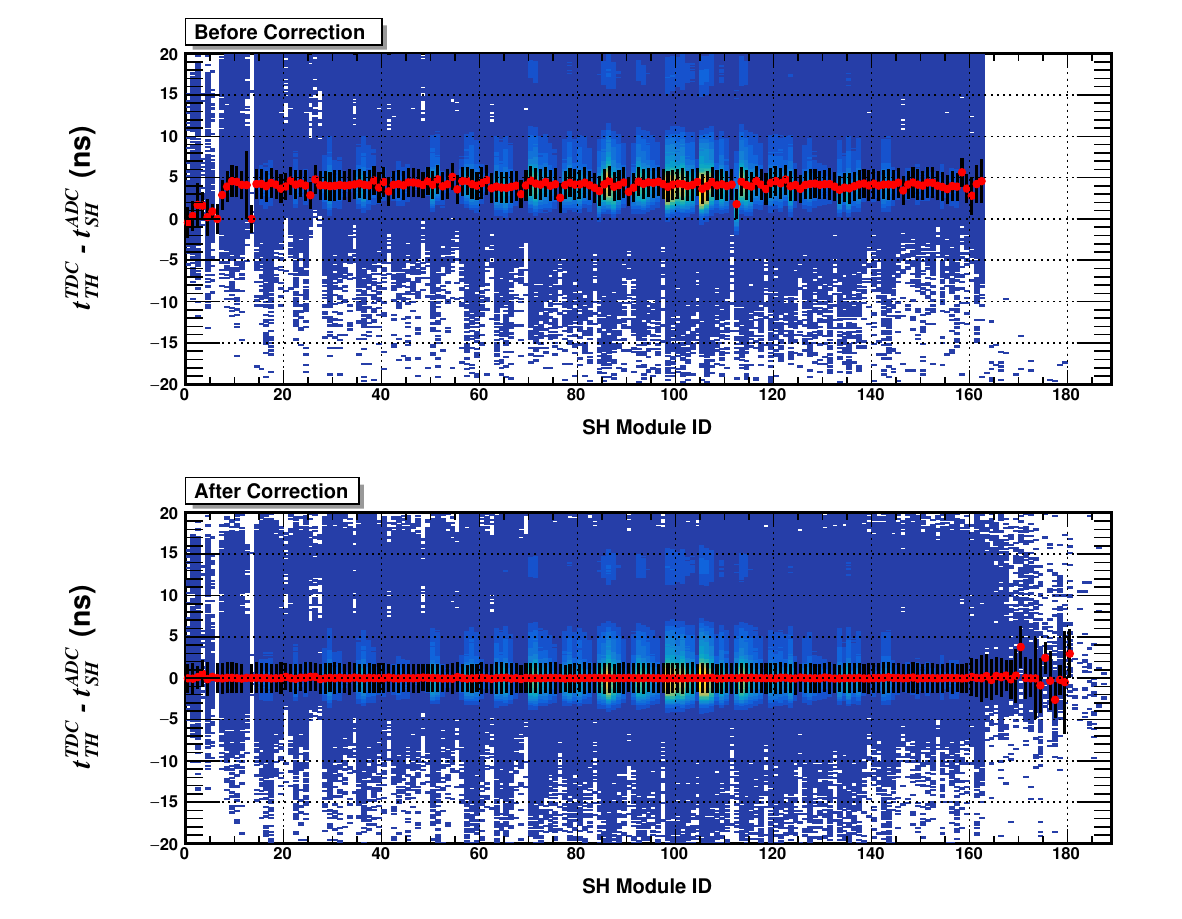}
         \caption{Shower (SH)}
    \end{subfigure}
    \hfill
    \begin{subfigure}[b]{0.496\textwidth}
        \centering
        \includegraphics[width=\textwidth]{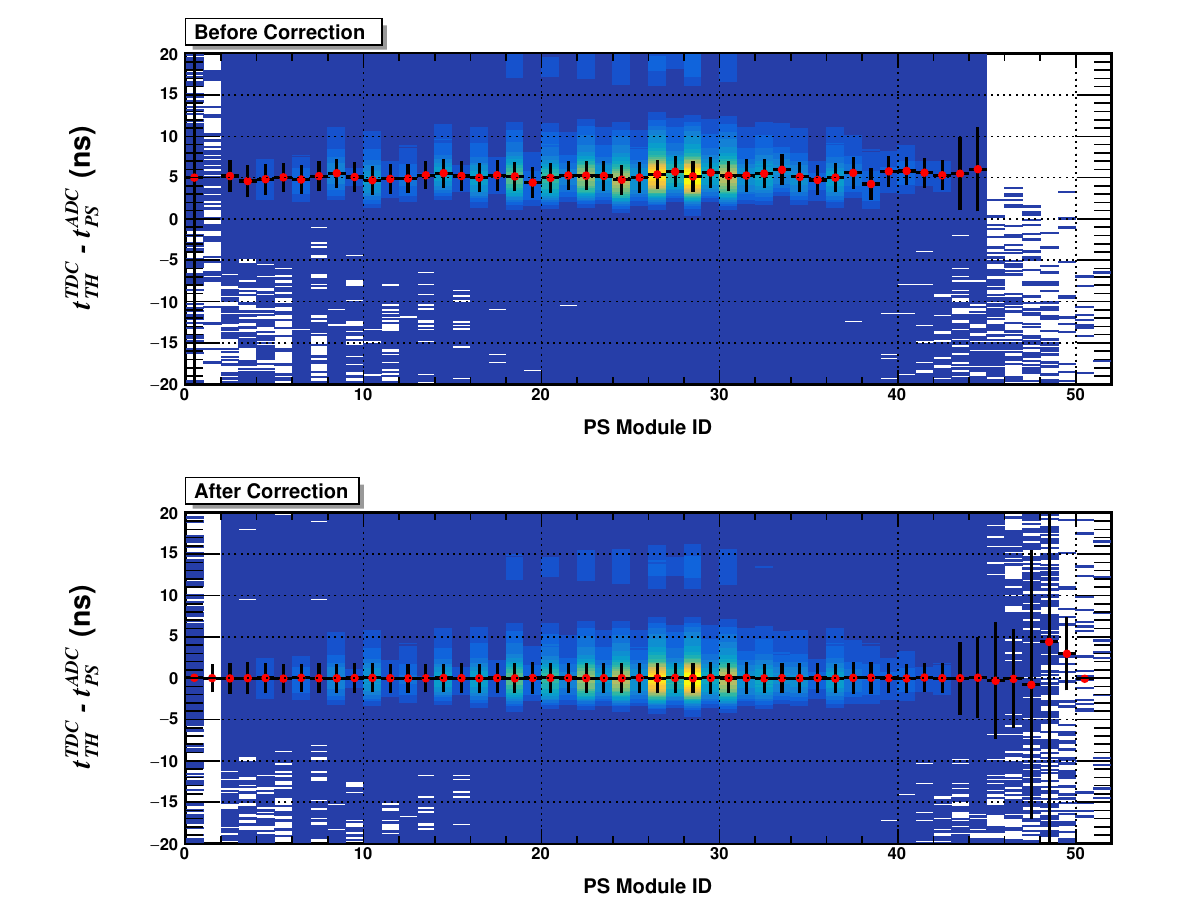}
        \caption{Pre-Shower (PS)}
    \end{subfigure}
    \caption{Effect of ADC time alignment on (a) Shower and (b) Pre-Shower PMTs. Plots generated using all \lh and \ld data taken at \qeq{9.9} kinematics.}
    \label{fig:ch4:bbcalatimealign}
\end{figure}
\begin{enumerate}
    \item For each event that passes good electron event selection cuts, two histograms are filled. The first histogram records the difference between the TH best cluster mean time and the ADC time of the highest energy module in the SH cluster. The second histogram records the same difference for the highest energy module in the PS cluster.
    \item After processing all events, the distributions for each SH and PS channel are fitted, and their mean values are extracted. Channels outside of the acceptance may have insufficient statistics, but this is not considered problematic.
    \item The extracted means are used to calculate the ADC time offsets for each SH and PS channel, which are then updated in the database files. During event reconstruction, these offsets are subtracted from the uncorrected ADC times of the corresponding channels, resulting in well-aligned ADC times across the detector.
\end{enumerate}

Alignment has been performed separately for each kinematic point. All the \lh and \ld runs were used in combination to ensure enough statistics for calibration for channels near the edge of the acceptance to maximize uniformity. \fig \ref{fig:ch4:bbcalatimealign} shows the improvement of ADC time alignment across all SH and PS channels with calibration at \qeq{9.9}.

\subheading{Notable Issues and Mitigation}
Several issues related to BBCAL ADC time observed during \gmn, their underlying causes, and the mitigation strategies employed are discussed below.
\begin{itemize}
    \item \textbf{ADC Time Shift Within a Configuration:} An overall shift of the same magnitude in the ADC times of all SH and PS channels was observed in certain runs taken during the \qeq{7.4\,\,\&\,\,13.6} kinematics, as shown in \fig \ref{sfig:ch4:bbcalatimeissue1}. Specifically, the \qeq{7.4} data showed an approximate \SI{-12}{ns} shift in a few initial runs, which then returned to normal and remained stable. In contrast, a \SI{-5}{ns} shift occurred midway through the \qeq{13.6} production. 
    
    \hspace{1em}The cause of these abrupt time shifts is unclear, but identical shifts observed in HCAL ADC times rule out a detector-specific issue. The fact that payload modules across multiple ROCs were affected similarly points to a possible DAQ-related problem. Although these shifts followed experimental downtimes, the reasons behind those downtimes do not appear correlated, suggesting that DAQ system adjustments during these periods might have caused the shifts, which is plausible given the DAQ system's volatility during \gmn.
\begin{figure}[h!]
    \centering
    \begin{subfigure}[b]{0.496\textwidth}
         \centering
         \includegraphics[width=\textwidth]{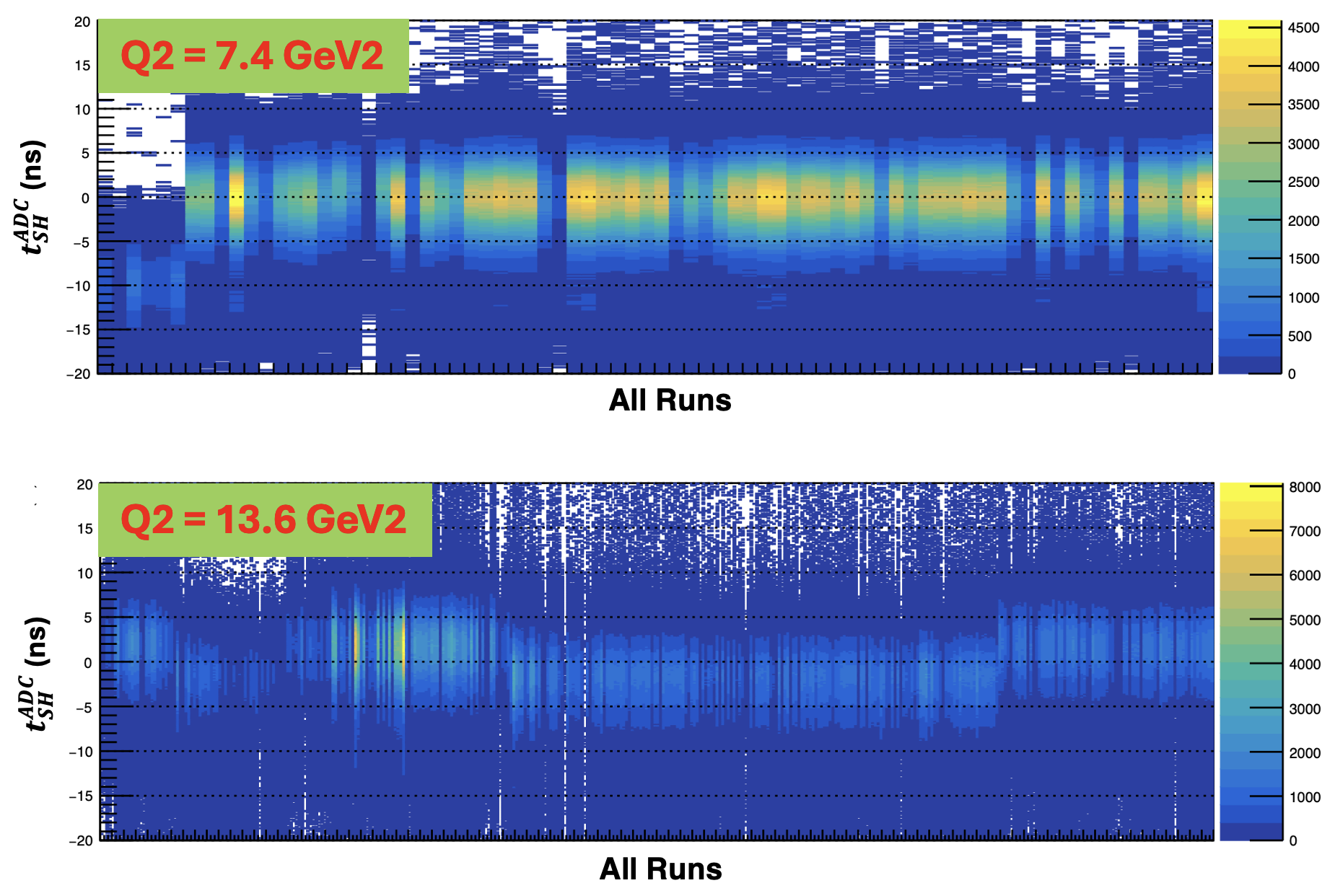}
         \caption{Before Correction}
         \label{sfig:ch4:bbcalatimeissue1}
    \end{subfigure}
    \hfill
    \begin{subfigure}[b]{0.496\textwidth}
        \centering
        \includegraphics[width=\textwidth]{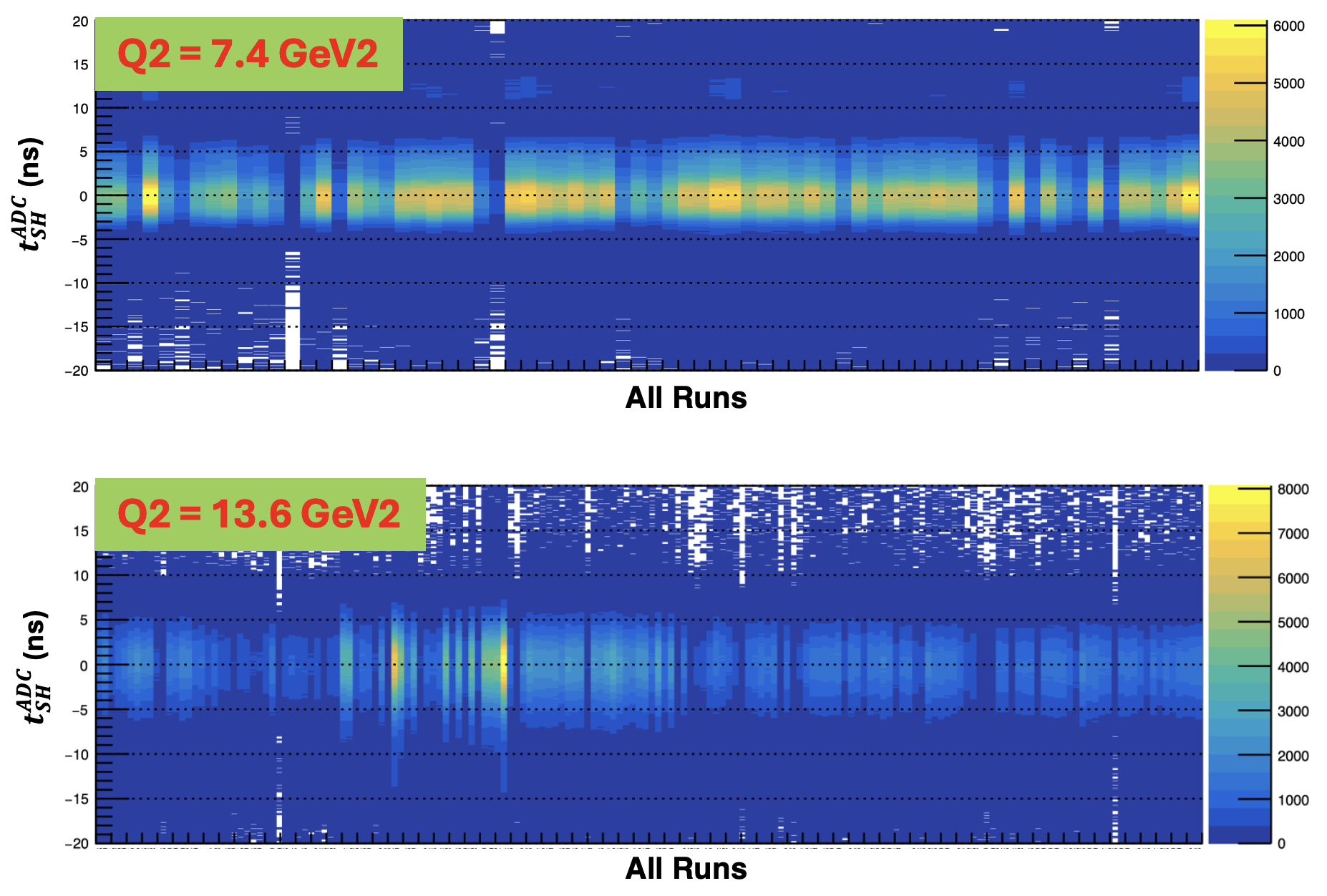}
        \caption{After Correction}
        \label{sfig:ch4:bbcalatimeissue1solved}
    \end{subfigure}
    \caption{Mitigation of ADC time shift between runs observed during \qeq{7.4\,\,\&\,\,13.6} kinematics. See text for details.}
    \label{fig:ch4:bbcalatimeissue1}
\end{figure}

    \hspace{1em}Fortunately, correcting these shifts with calibration is straightforward. The process involves grouping runs with the same latency within a configuration and performing timing calibration for each group separately. This ensures that the resulting offsets align the ADC times from all runs to zero, regardless of the shifts. \fig \ref{sfig:ch4:bbcalatimeissue1solved} demonstrates the successful mitigation of SH ADC time shifts through calibration across all affected runs, with similar results observed for PS ADC times as well \cite{SUPPMAT}.
    \item \textbf{ADC Time Shift Within Runs:} The aforementioned ADC time shifts occurred across runs, but random shifts of approximately \SI{-12}{ns} were also observed for a fraction of events within runs throughout \gmn production, leading to a double-peaking structure in the ADC time distribution, as shown in \fig \ref{sfig:ch4:bbcalatimeissue2}. This shift is evident in all the plots in Figures \ref{fig:ch4:bbcalatimealign} and \ref{fig:ch4:bbcalatimeissue1}. ADC times from all the detector subsystems, including SH, PS, and HCAL, were affected. However, this shift cancels out in the decoded TDC data due to reference time subtraction. 
\begin{figure}[h!]
    \centering
    \begin{subfigure}[b]{0.496\textwidth}
         \centering
         \includegraphics[width=\textwidth]{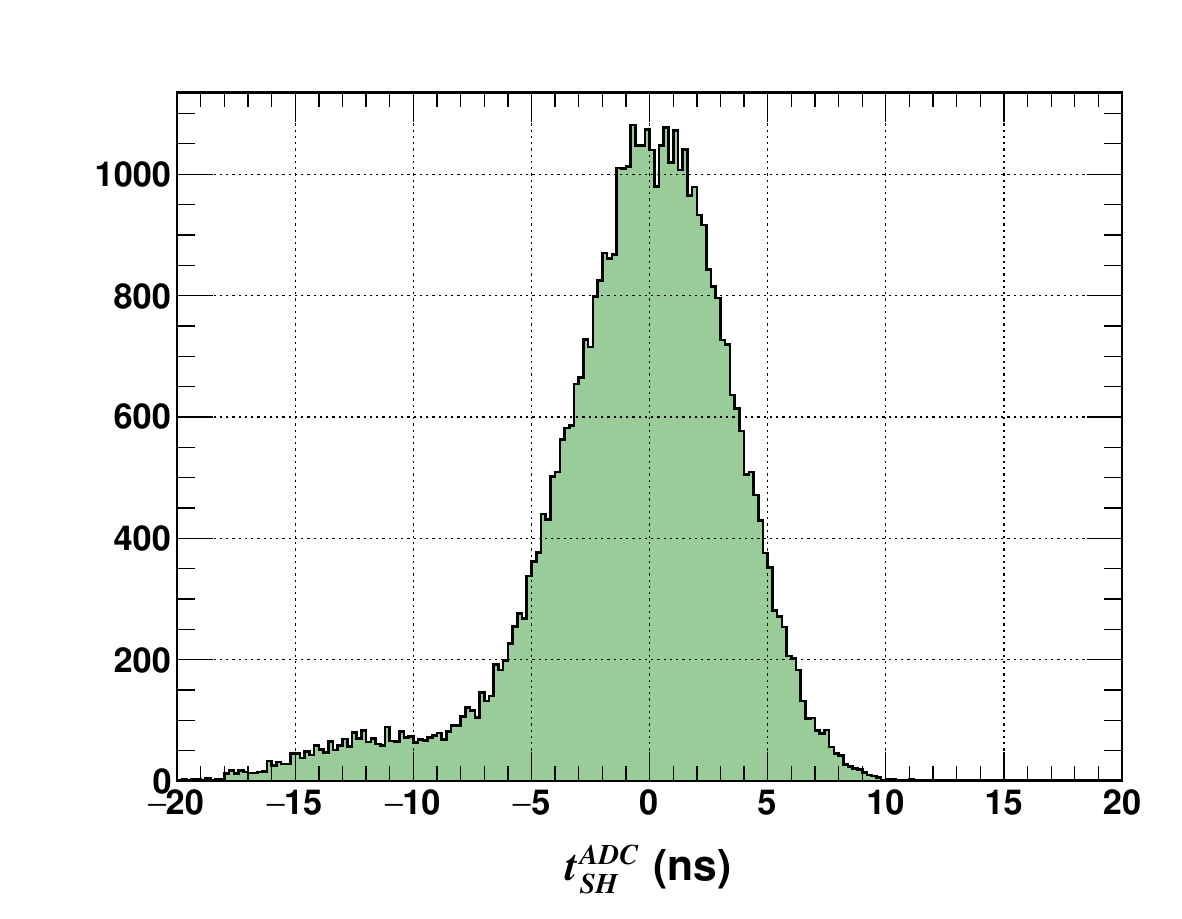}
         \caption{}
         \label{sfig:ch4:bbcalatimeissue2}
    \end{subfigure}
    \hfill
    \begin{subfigure}[b]{0.496\textwidth}
        \centering
        \includegraphics[width=\textwidth]{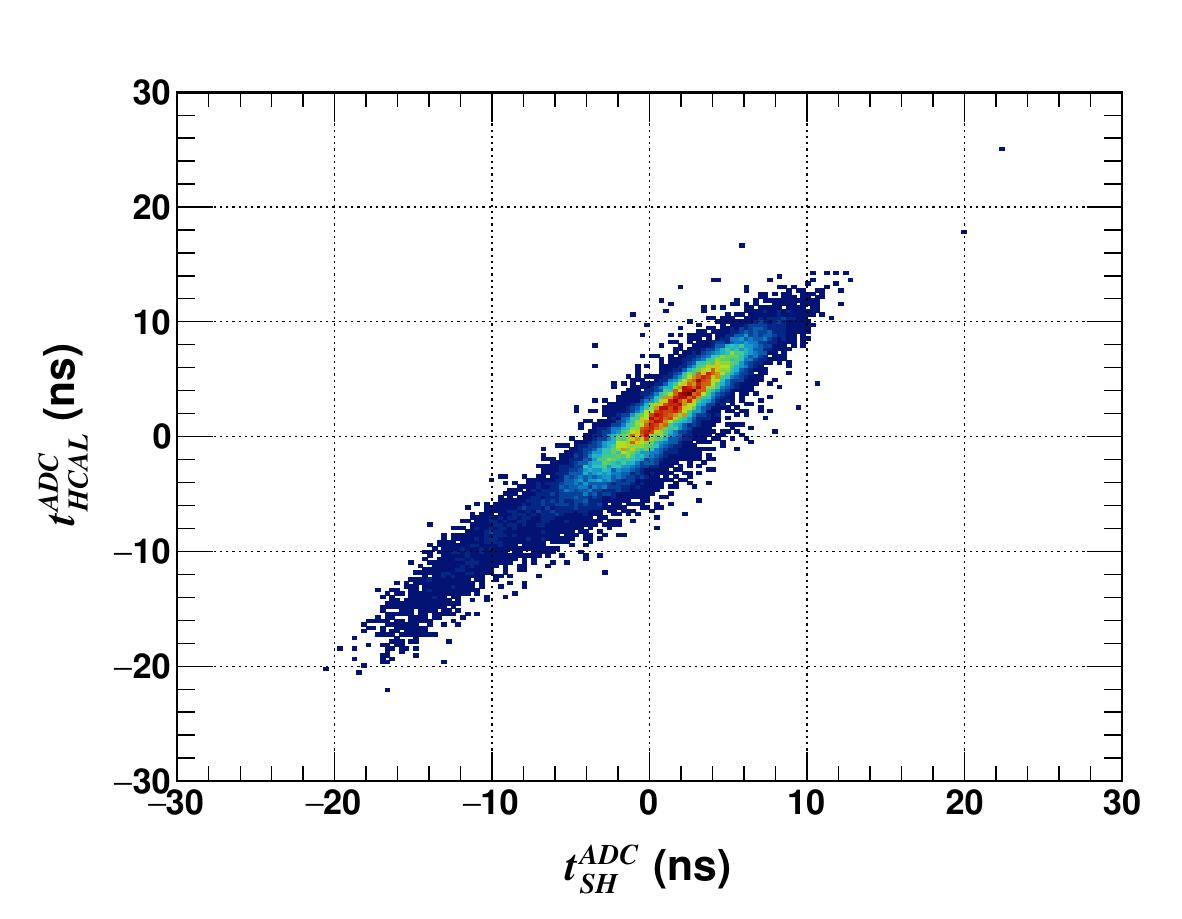}
        \caption{}
        \label{sfig:ch4:bbcalatimeissue2solved}
    \end{subfigure}
    \caption{Effect and scope of the ADC time shift observed within a given CODA run throughout \gmn. (a) The Shower (SH) ADC time distribution, where the secondary peak at \SI{-12}{ns} is clearly visible. (b) The correlation between the SH and HCAL times, showing that the perfect correlation observed for events in the secondary peak implies the shift affected both detectors equally.}
    \label{fig:ch4:bbcalatimeissue2}
\end{figure}

    \hspace{1em}As shown in \fig \ref{sfig:ch4:bbcalatimeissue2solved}, the SH and HCAL ADC times are similarly correlated for both the shifted and normal events. The same holds true between PS and HCAL as well. This suggests that the data from different subsystems recorded by different ROCs were similarly affected, indicating that the shift is related to the trigger time, which is common to all ROCs in the DAQ system. The Trigger Supervisor (TS) distributes the BBCAL singles trigger, primary trigger for the \gmn experiment, to all the ROCs. A copy of the BBCAL singles trigger, taken from upstream of the TS in the DAQ circuit, is recorded by a CAEN V1190 TDC module based on the trigger received from the TS. If the \SI{-12}{ns} shift originated from any electronics module upstream of the TS, the BBCAL singles trigger time recorded by the V1190 should not exhibit it. However, the same shift is observed in the raw BBCAL singles trigger time signal, perfectly correlating with the shift seen in other detectors. This strongly suggests that the random shift in question was introduced by the TS itself.

    \hspace{1em}Since events within a given run are randomly affected by this shift, it cannot be straightforwardly corrected through calibration as in the previously mentioned case. Fortunately, only a fraction of events are impacted by the shift, allowing us to use the primary peak for timing calibration. Additionally, because HCAL and SH ADC times remain perfectly correlated despite the shift, the double-peaking structure vanishes in the SH-HCAL coincidence time, resulting in a sharp Gaussian distribution. This enables the formulation of a strict two-arm coincidence time cut to effectively suppress accidental background as necessary for the physics analysis.
    \item \textbf{Out-of-Time Hits in SH Clusters:} The ADC time difference between the primary and secondary modules in a SH cluster shows a double-peaking structure, with a small secondary bump peaking approximately \SI{17}{ns} away from the primary peak. However, no such shift is observed in PS data. 
\begin{figure}[h!]
    \centering
    \begin{subfigure}[b]{0.496\textwidth}
         \centering
         \includegraphics[width=\textwidth]{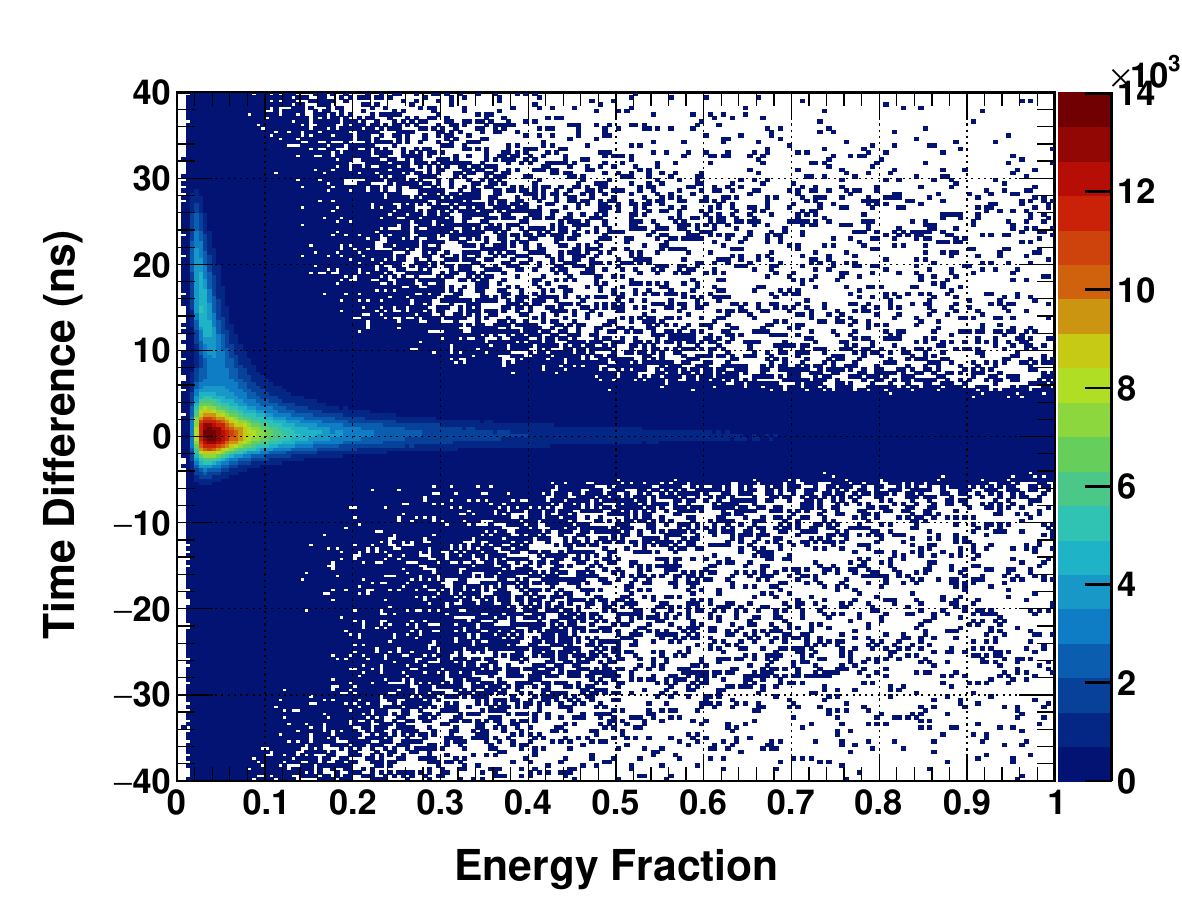}
         \caption{}
         \label{sfig:ch4:bbcalatimeissue3ev1}
    \end{subfigure}
    \hfill
    \begin{subfigure}[b]{0.496\textwidth}
        \centering
        \includegraphics[width=\textwidth]{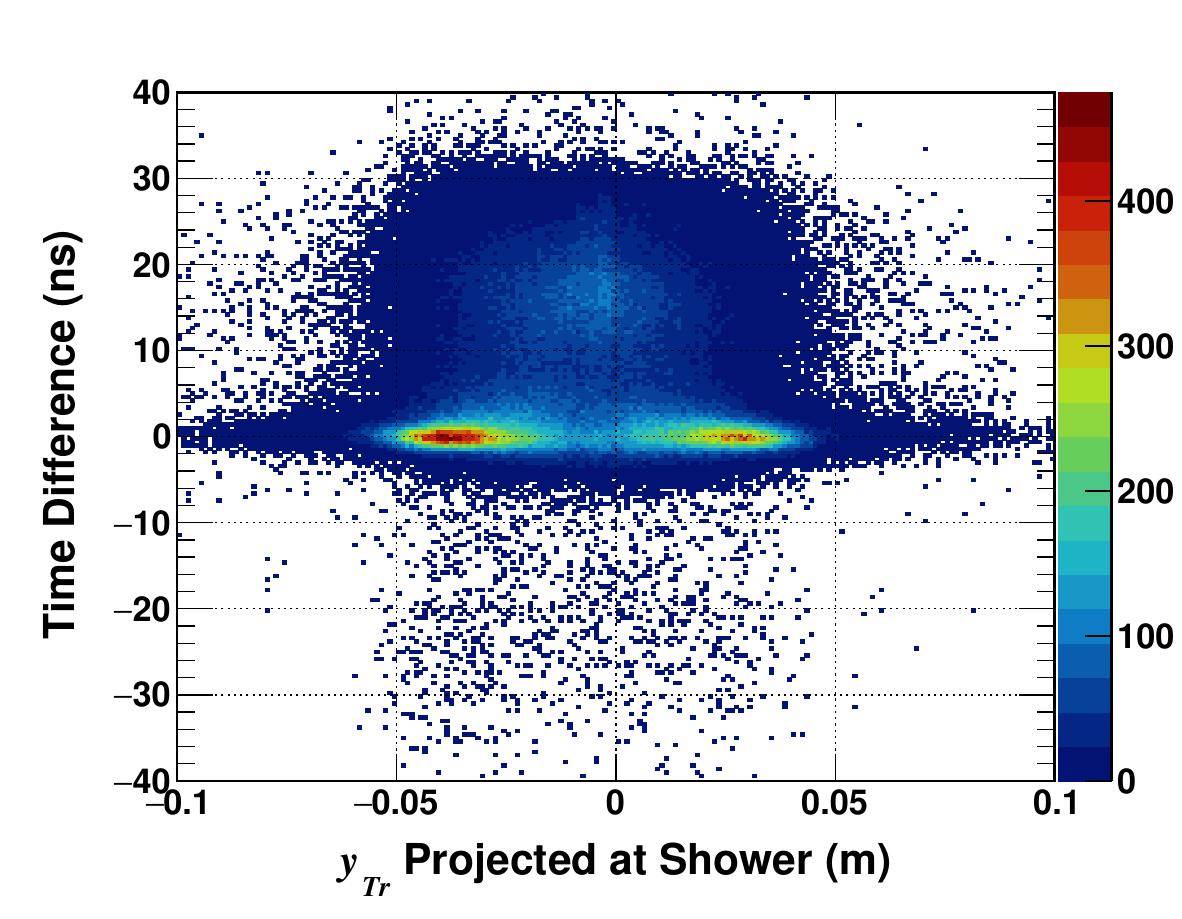}
        \caption{}
        \label{sfig:ch4:bbcalatimeissue3ev2}
    \end{subfigure}
    \caption{Out-of-time hits within a Shower (SH) cluster. (a) Correlation between the time difference of the secondary and primary modules in a shower cluster and the ratio of the secondary module’s energy to that of the primary, with low-energy out-of-time hits clearly visible. (b) Time difference as a function of the transverse track position ($y_{Tr}$) projected onto the shower, where secondary modules are in the same row as the primary and the primary modules lie exclusively in the fourth SH column. Out-of-time hits predominantly occur when the track hits the center of a SH module.}
    \label{fig:ch4:bbcalatimeissue3}
\end{figure}

    \hspace{1em}The out-of-time hits contributing to the secondary bump have very low energy, about $15\%$ or less than that of the primary module in the cluster, as illustrated in \fig \ref{sfig:ch4:bbcalatimeissue3ev1}. Further analysis reveals that these out-of-time hits predominantly occur when the electron track hits near the center of a SH module (see \fig \ref{sfig:ch4:bbcalatimeissue3ev2}), and the secondary modules are in a different row than the primary module. These observations suggest that the low-energy out-of-time hits originate from multiple scattering of low-energy events, which are remnants of the main electromagnetic shower. The row-dependent bias is likely due to the presence of Mu-metal plates installed between SH rows, which may enhance the probability of multiple scattering. Preliminary studies with full optical photon simulations support these hypotheses. However, a more in-depth analysis is necessary for any conclusive remarks.

    \hspace{1em}For the purpose of \gmn analysis, understanding whether to include these modules in the cluster or not is sufficient. Rigorous analysis across several kinematics has consistently shown that excluding these events from SH clustering yields better energy resolution. Additionally, an optimal cut range of $\pm10$ ns on the difference between the ADC time of the cluster seed and the secondary module has been implemented for both SH and PS clustering. Even though PS events don't exhibit a secondary peak, implementing the aforementioned cut helps reduce some background.
\end{itemize}

\begin{figure}[h!]
    \centering
    \begin{subfigure}[b]{0.496\textwidth}
         \centering
         \includegraphics[width=\textwidth]{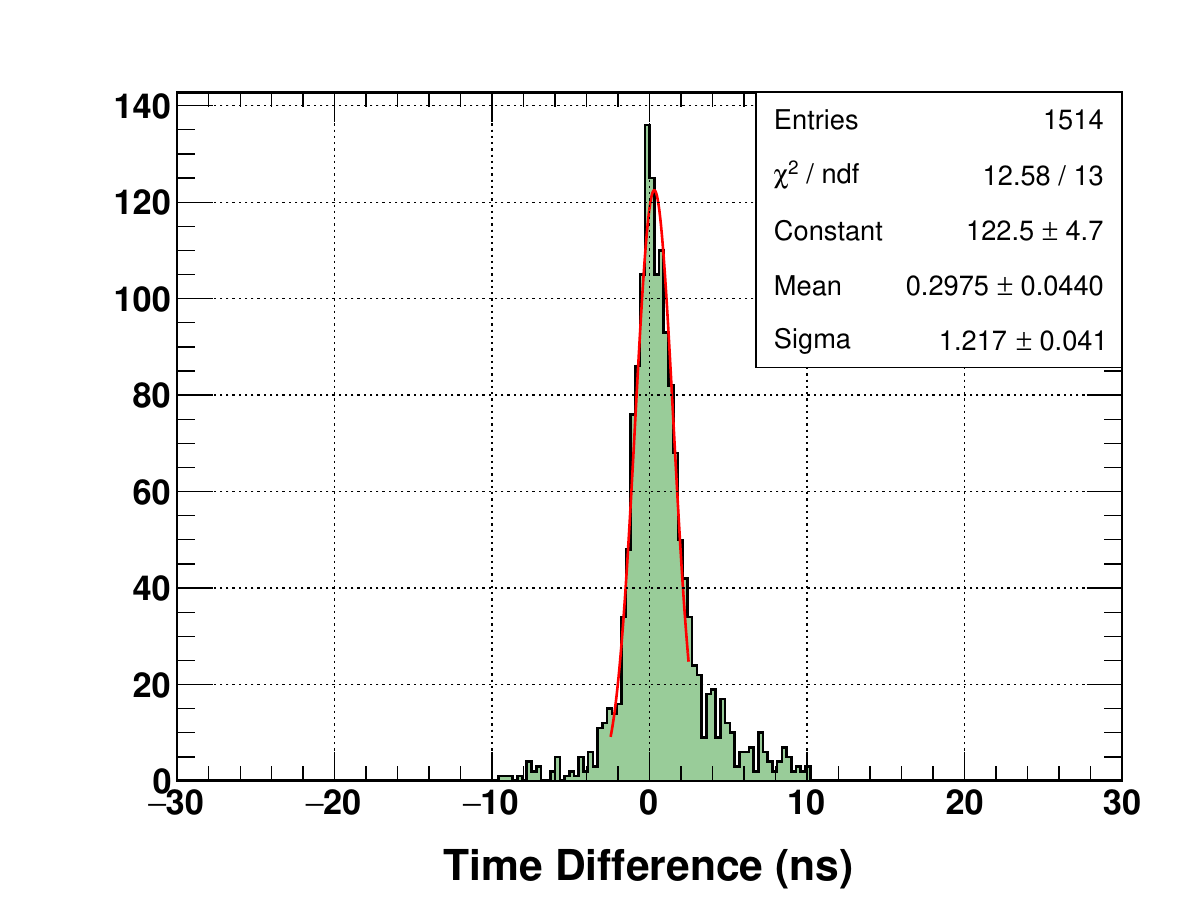}
         \caption{Shower (SH)}
    \end{subfigure}
    \hfill
    \begin{subfigure}[b]{0.496\textwidth}
        \centering
        \includegraphics[width=\textwidth]{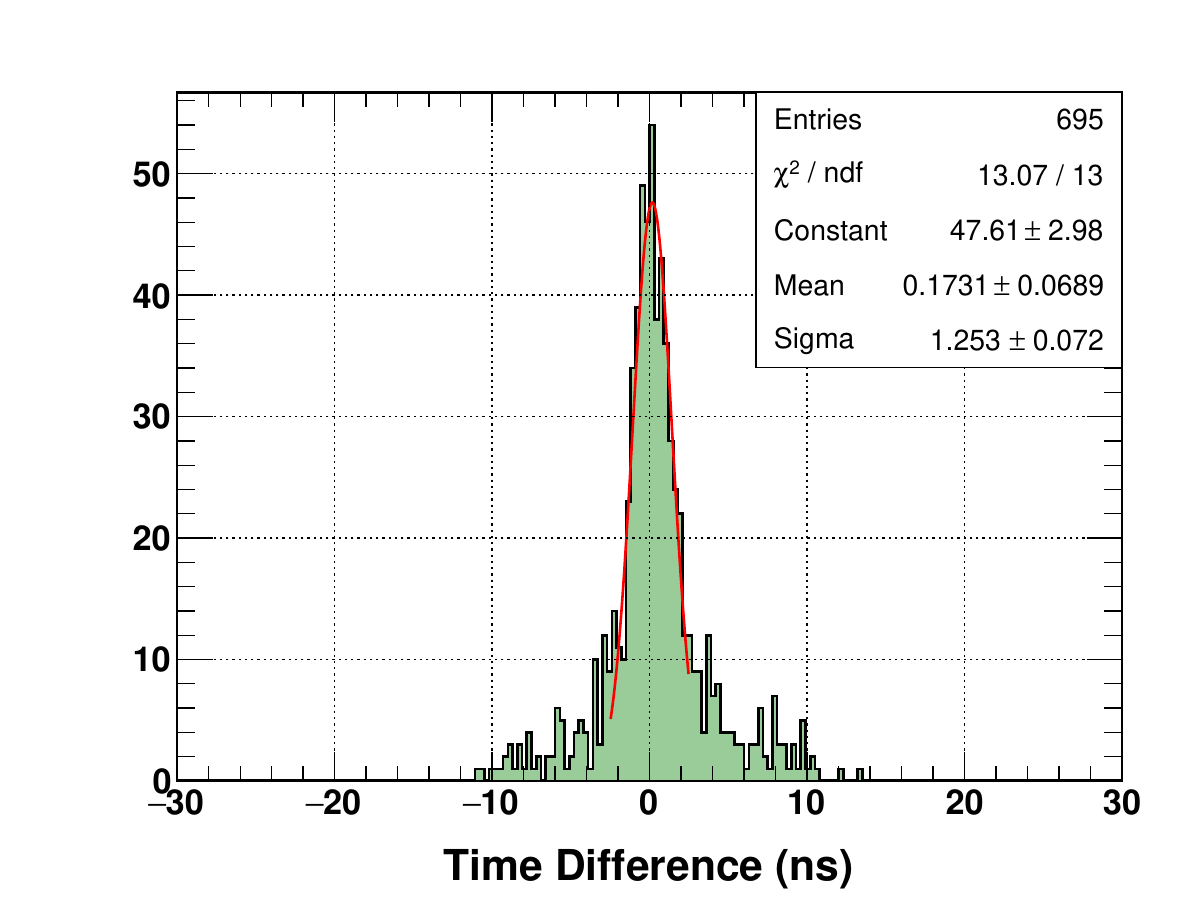}
        \caption{Pre-Shower (PS)}
    \end{subfigure}
    \caption{Intrinsic resolutions of (a) Shower (SH) and (b) Pre-Shower (PS) ADC times, obtained by taking the difference between the ADC times of the primary and secondary modules in SH and PS clusters, respectively. These plots were generated using \lh events passing good electron and \w cuts at \qeq{4.5} (low \ep). Note: The hard cut-offs at $\pm10$ ns are artifacts of the ADC time cut implemented during BBCAL cluster formation to filter out ``out-of-time hits in SH clusters", as discussed in the text.}
    \label{fig:ch4:bbcalatimeresolution}
\end{figure}
\subheading{Performance Parameters}
After second pass fine-tuning of BBCAL timing calibration, the intrinsic ADC time resolutions of SH and PS, obtained by taking the ADC time differences between the primary and secondary modules in a cluster, are observed to be approximately $1.2-1.3$ ns across \gmn kinematics (see \fig \ref{fig:ch4:bbcalatimeresolution}). Achieving such high resolution using ADC time meets or even exceeds our expectations. 
\subsubsection{BBCAL Energy Calibration}
\label{sssec:ch4:bbcalibeng}
The amplitude and integral of the signal generated by a BBCAL PMT are proportional to the energy deposited in the corresponding lead-glass (LG) module. The primary goal of BBCAL energy calibration is to determine the proportionality constant, known as the ADC gain coefficient, for all Shower (SH) and Pre-Shower (PS) channels with high precision.

For the \gmn analysis, BBCAL energy calibration involved two stages: an initial cosmic calibration followed by in-beam calibration using \lh data. 

\vspace{1em}
\hspace{{-1em}}\textbf{Cosmic calibration} was conducted during commissioning to ensure high-quality data acquisition. It began with gain-matching all BBCAL PMTs by adjusting their operating high voltages (HVs), as described in \sect \ref{ssec:trigcalib}. This process leverages the well-defined energy deposition of cosmic ray muons in the LG as minimum ionizing particles (MIPs). After gain matching, the ADC gain coefficients ($c_{i}$) for all BBCAL channels were calculated as follows:
\begin{equation}
    c_{i} = R_{i}^{A/I} \frac{E_{cosmic}}{A_{Trig}^{Set}} 
\end{equation}
where $R_{i}^{A/I}$ is the ratio of signal amplitude to integral of the $i^{th}$ BBCAL module, $E_{cosmic}$ is the estimated average cosmic energy deposition in each BBCAL module, and $A_{Trig}^{Set}$ is the signal amplitude corresponding to each BBCAL module for $E_{cosmic}$ energy deposition after gain matching. Multiplying the ADC integral of a given BBCAL channel by its corresponding $c_i$ value, conventionally expressed in GeV/pc, yields the associated energy deposition, which is meaningful for physics analysis.

Cosmic calibration was performed at the start of each \gmn kinematic configuration, with updated $A_{Trig}^{Set}$ values based on \tab \ref{tab:thdetermination}. Additionally, the effect of the SBS fringe field on BBCAL PMT gains, as detailed in \sect \ref{sssec:ch3:sbsfringefield}, required updated calibrations within a kinematic point whenever the SBS magnet field strength was adjusted. With cosmic calibration, the energy resolution of BBCAL, evaluated using elastic \heep events by considering the scattered electron energy obtained from BB optics as the true energy deposition, was observed to be about $10-12\%$, which is suboptimal. This is expected due to various uncertainties associated with cosmic calibration, including the ambiguity in the precise knowledge of $E_{cosmic}$. However, the quality of the cosmic calibration was sufficient to establish unbiased track search region constraints, facilitating proper track reconstruction during data collection and enabling effective online data quality monitoring.

\vspace{1em}
\hspace{-1em}\textbf{In-Beam calibration} using \lh data employs a more statistically rigorous approach to fine-tune the ADC gain coefficients obtained from the initial cosmic calibration. This process involves minimizing a $\chi^2$ function, defined by the difference between the ``true" and the reconstructed scattered electron energy:
\begin{equation}
\label{eqn:ch4:bbcalibchi2}
    \chi^2 = \sum_{i=1}^{N} \left( E_{e}^{i} - \sum_{j=0}^{M} c_{j}A^{i}_{j} \right)^{2}
\end{equation}
Here, $N$ is the number of events and $E^{i}_{e}$ represents the track momentum (effectively energy, as the electrons are relativistic) of the $i^{th}$ event, as determined by BB optics, and is considered the ``true" scattered electron energy. The term $\sum_{j=0}^{M} c_{j}A^{i}_{j}$ denotes the BBCAL cluster energy ($E_{BBCAL}^i$) for the same event, representing the reconstructed scattered electron energy. This cluster includes $M$ SH and PS modules, where $c_{j}$ is the ADC gain coefficient and $A^{i}_{j}$ is the ADC pulse integral for the $j^{th}$ module. Now, minimizing $\chi^2$ with respect to $c_{j}$ gives: 
\begin{equation}
    \frac{\partial \chi^2}{\partial c_{j}} = 0 
\Rightarrow \sum_{i=1}^{N}  \left( A_{j}^{i} - \sum_{k=0}^{M} \frac{A_{j}^{i} A_{k}^{i}}{E_{e}^{i}}c_{k}^{i} \right) = 0
\end{equation}
This results in a system of 241 linear equations representing all $189$ SH and $52$ PS modules, which can be trivially solved using standard linear algebra libraries to obtain the calibrated ADC gain coefficients.

\begin{figure}[h!]
    \centering
    \begin{subfigure}[b]{0.9\textwidth}
         \centering
         \includegraphics[width=\textwidth]{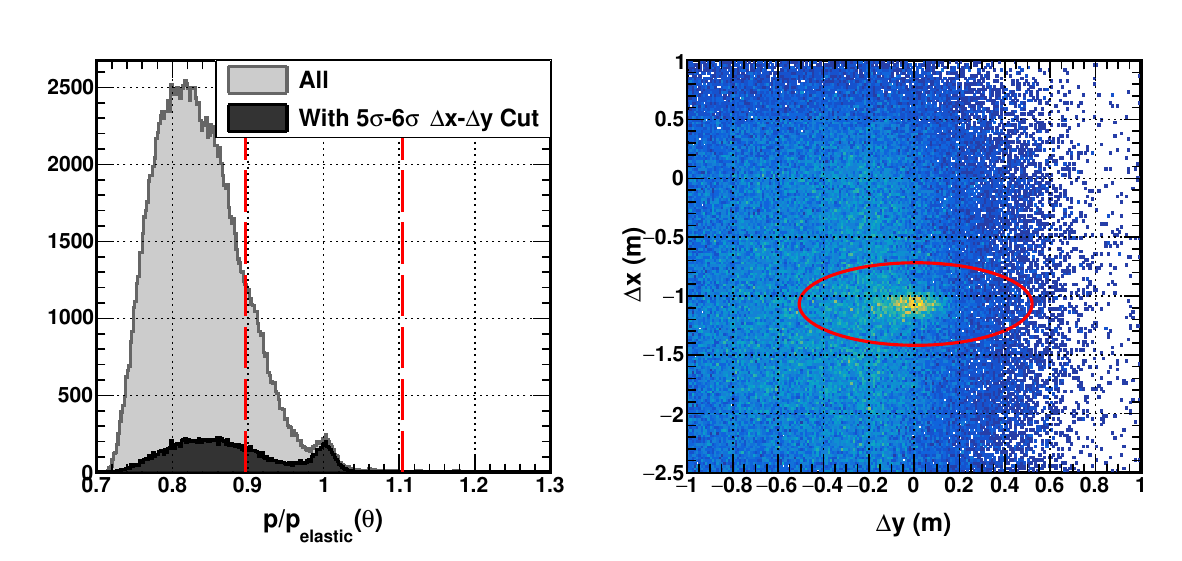}
         \caption{\qeq{9.9}}
    \end{subfigure}
    \hfill
    \begin{subfigure}[b]{0.9\textwidth}
        \centering
        \includegraphics[width=\textwidth]{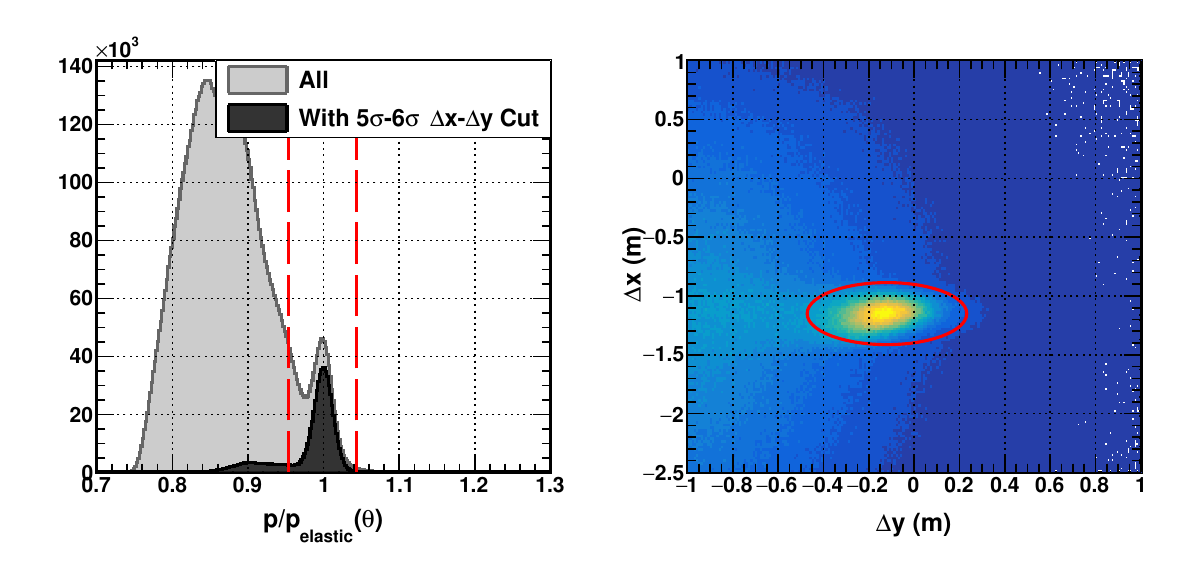}
        \caption{\qeq{4.5}, high \ep}
    \end{subfigure}
    \caption{Elastic \heep event selection cut ranges for BBCAL energy calibration. (a) A very loose $6\sigma$ $p/p_{elas}$ cut, combined with a $5\sigma$-$6\sigma$ \dx-\dy correlation cut, was used for \qeq{9.9}, the most statistically challenged setting. (b) Much stricter cuts, including a $4\sigma$ $p/p_{elas}$ cut and a $2\sigma$ \dx-\dy correlation cut, were applied for the \qeq{4.5} high \ep setting, where statistics were most abundant.}
    \label{fig:ch4:bbcalengelcut}
\end{figure}
This method depends heavily on the quality of momentum reconstruction from BB optics, making it crucial to use events with accurate momentum reconstruction for calibration to avoid bias. Although cleanly selected good electron events should suffice, BB momentum reconstruction is less reliable for inelastic scattering events, as discussed in \sect \ref{sssec:ch4:momrecon}. However, using only events passing strict elastic event selection cuts can lead to low statistics per BBCAL module, potentially resulting in poor calibration. To balance this, loosely selected elastic \heep events were used for BBCAL in-beam calibration. The selection criteria included all good electron event selection cuts defined in \sect \ref{sec:ch4:evselect} except for the \eovp cut. Loose cuts on \dx-\dy correlation and $p/p_{elas}$\footnote{The ratio of reconstructed track momentum to the elastically scattered electron momentum (calculated using the reconstructed scattering angle), $p/p_{elas}$, is strongly correlated with \w. While cutting on one effectively impacts the other, $p/p_{elas}$ more clearly identifies regions of good momentum reconstruction, making it preferable for BBCAL calibration.}, as shown in \fig \ref{fig:ch4:bbcalengelcut}, were applied to isolate elastic events. An active area cut was also used to exclude events with clusters centered in the outermost rows and columns of the SH detector to prevent bias introduced by potential energy leakage.

In-beam calibration was conducted separately for each kinematic point and for data taken with different SBS fields within a kinematic, following the same rationale as for cosmic calibration. For some kinematic points, separate calibrations were required even within datasets recorded with the same SBS field strength to avoid biases observed across runs, as discussed later in this section. 

\subheading{Quality Assurance of BBCAL In-Beam Calibration}
The quality of each BBCAL in-beam calibration, after the second pass of fine-tuning, was assessed using several criteria discussed below. Example plots from one or two settings are shown here due to space constraints, but representative plots from all settings can be found at \cite{SUPPMAT}.
\begin{itemize}
    \item \textbf{Energy Resolution:} The ratio of reconstructed ($E_{BBCAL}$) to ``true" ($E'_e$) scattered electron energy for elastic \heep events serves as a measure of BBCAL energy calibration quality. Ideally, this ratio should be 1 for all events, but in practice, it forms a distribution with finite width. Better calibration shifts the mean of this distribution closer to 1 while simultaneously reducing its width, which defines the energy resolution of BBCAL. A significant improvement in energy calibration using \lh data over cosmic data for the \qeq{4.5} high \ep dataset is clearly visible in \fig \ref{sfig:ch4:bbcalengcalib1}.  

\begin{figure}[h!]
    \centering
    \begin{subfigure}[b]{0.496\textwidth}
         \centering
         \includegraphics[width=\textwidth]{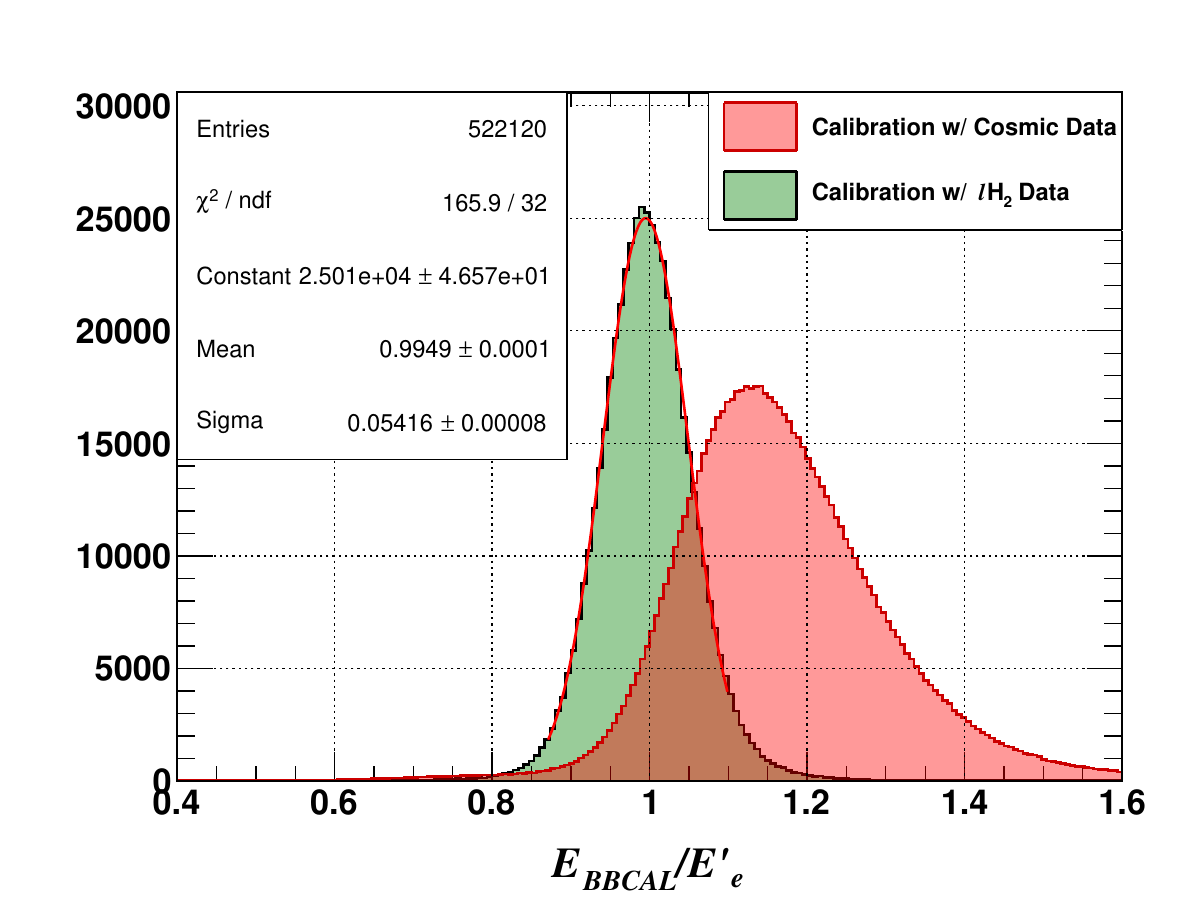}
         \caption{$\Bar{E_{e}}=3.6$ GeV}
         \label{sfig:ch4:bbcalengcalib1}
    \end{subfigure}
    \hfill
    \begin{subfigure}[b]{0.496\textwidth}
        \centering
        \includegraphics[width=\textwidth]{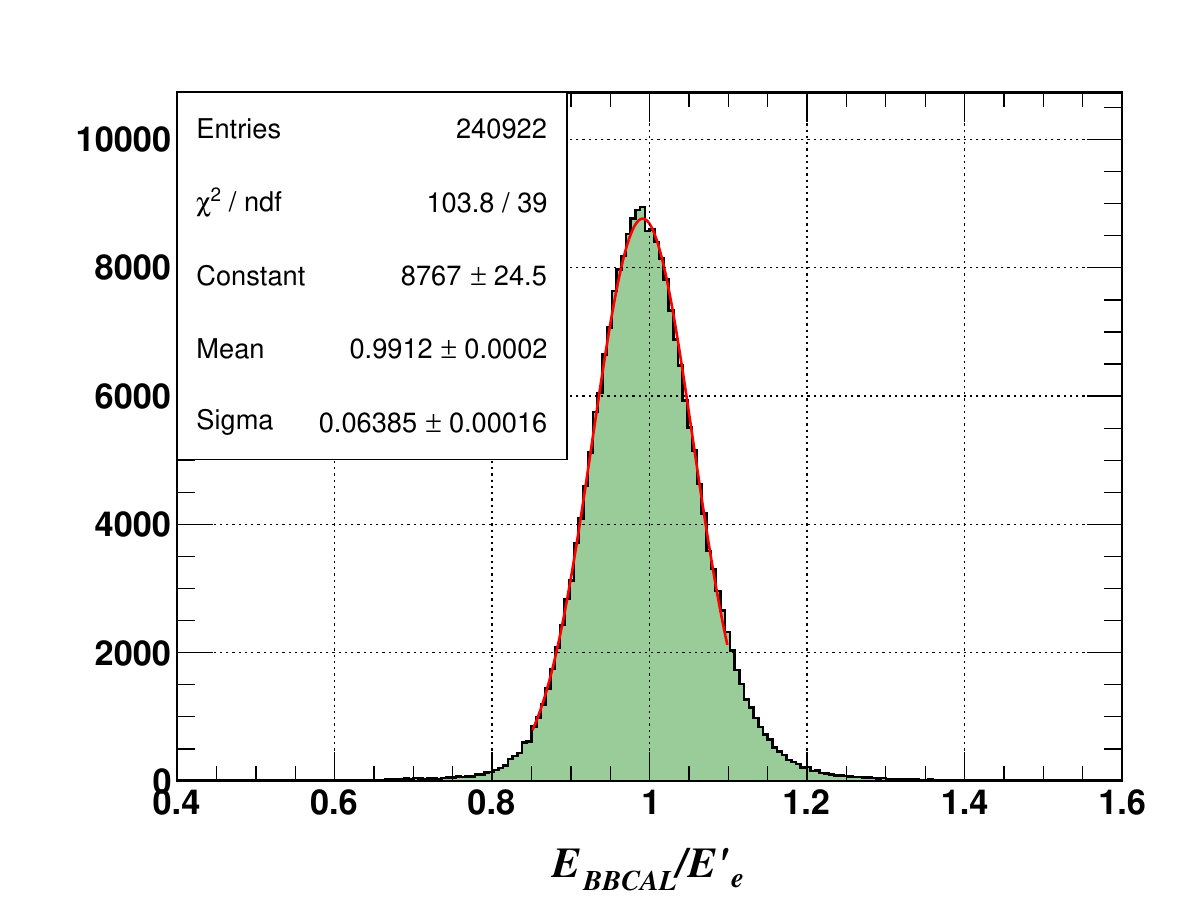}
        \caption{$\Bar{E_{e}}=1.6$ GeV}
    \end{subfigure}
    \caption{\eove distributions after the second pass of BBCAL energy calibration fine-tuning for \heep data at \qeq{4.5}: (a) high-\ep and (b) low-\ep datasets. The former plot clearly shows the significant improvement in calibration using \lh data compared to cosmic data.}
    \label{fig:ch4:bbcalengcalib}
\end{figure}
    \hspace{1em}The improved resolutions observed for the high and low \ep datasets are $5.4\%$ and $6.4\%$, respectively. BBCAL energy resolution is expected to be better for higher average scattered electron energy ($\Bar{E_{e}}$) due to larger cluster size, which explains the difference between these resolution values. As summarized in \tab \ref{tab:sbsconfig}, $\Bar{E_{e}}$ associated with the \qeq{4.5} high (low) \ep dataset is the highest (lowest) across \gmn kinematics. Therefore, \fig \ref{fig:ch4:bbcalengcalib} illustrates the upper and lower bounds of BBCAL energy resolution from the second pass of fine-tuning, which meets or even exceeds our expectations.
    %
    %
%
\begin{figure}[h!]
    \centering
    \includegraphics[width=\sfig]{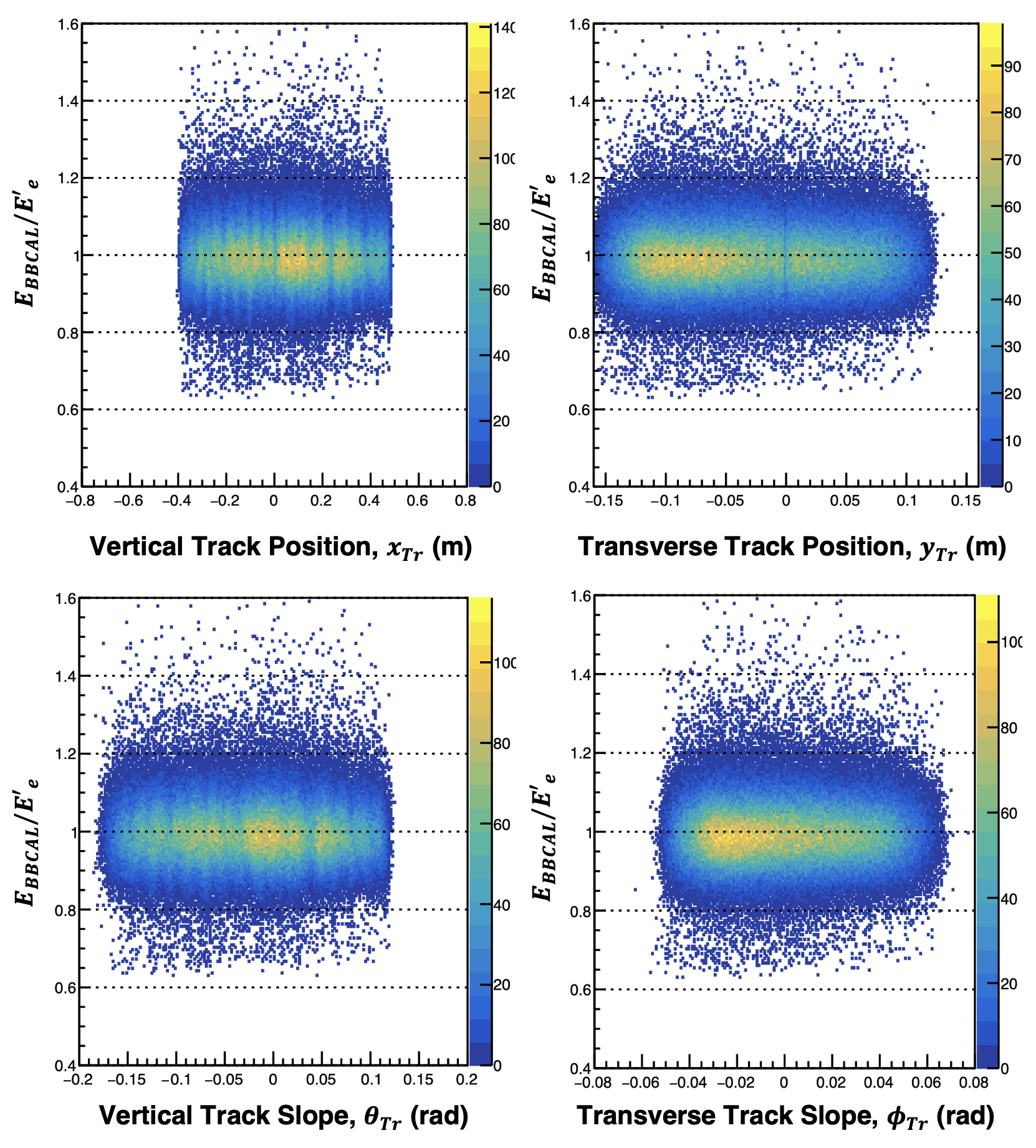}
    \caption{\label{fig:ch4:eovpvstr} Uniformity of BBCAL energy calibration within the BB acceptance after the second pass of fine-tuning. Elastic \heep events from the \qeq{4.5} low-\ep dataset are shown.}
\end{figure}
    \item \textbf{Uniformity Within Acceptance:} Uniform BBCAL energy calibration within the acceptance is crucial to avoid bias during event reconstruction. Any position-dependent non-uniformity in calibration would appear in the correlation between \eove and track positions and angles. \fig \ref{fig:ch4:eovpvstr} shows these correlations using elastic \heep events from the \qeq{4.5} low \ep dataset, demonstrating excellent uniformity of calibration across the entire acceptance.
\begin{figure}[h!]
    \centering
    \includegraphics[width=1\columnwidth]{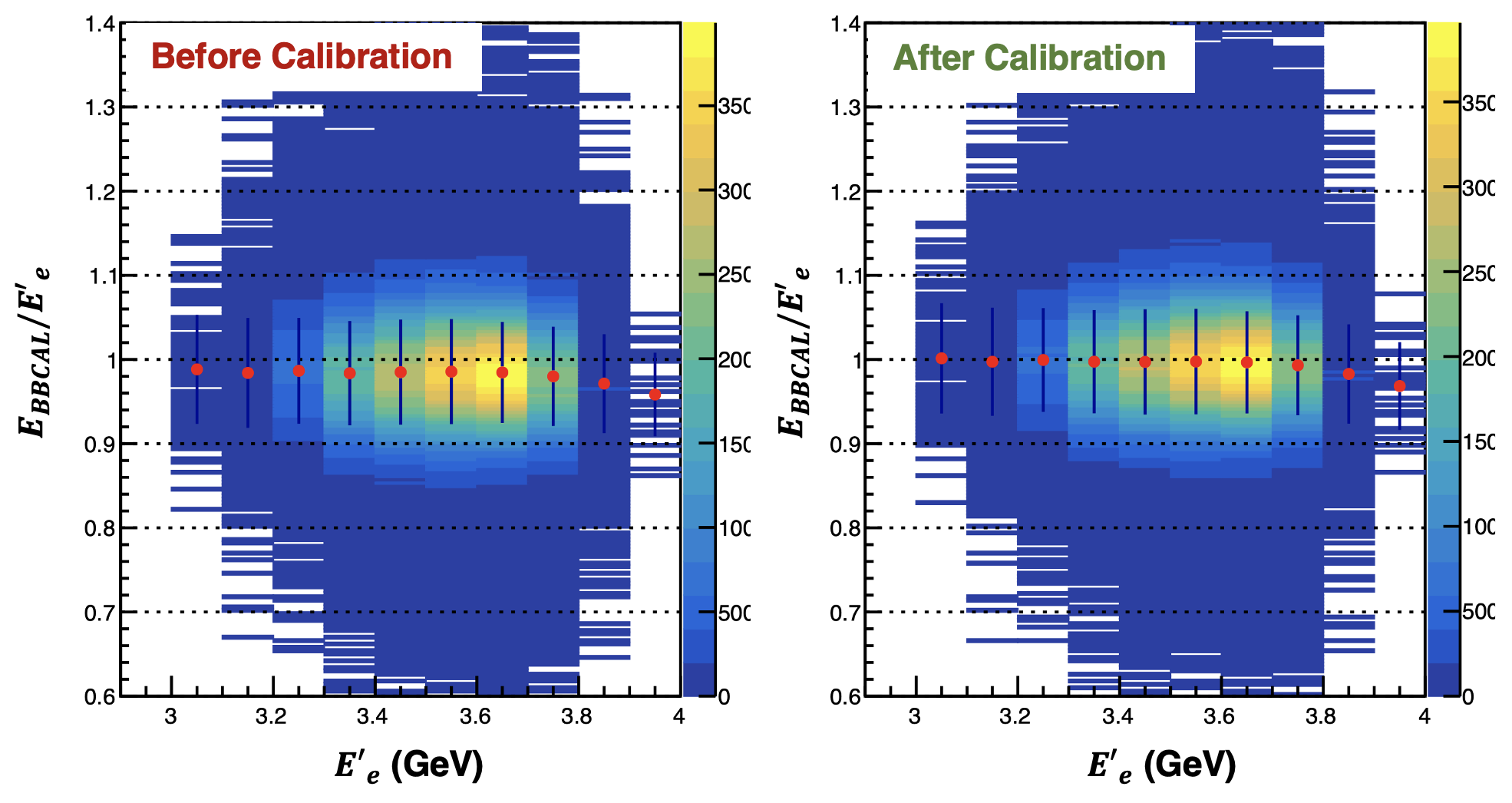}
    \caption{\label{fig:ch4:eovpvsp} Improved uniformity of \eove vs track momentum with calibration for \qeq{4.5} high \ep dataset.}
\end{figure}
    \item \textbf{Correlation with \bm{$p$}:} The BB spectrometer has a large momentum bite, making uniform BBCAL energy calibration across the entire range essential. \fig \ref{fig:ch4:eovpvsp} illustrates the improved uniformity of \eove with respect to track momentum after calibration for elastic \heep events in the \qeq{4.5} high \ep dataset. The apparent non-uniformity at large $p$ is due to lack of statistics.
    %
    %
%
\begin{figure}[h!]
    \centering
    \includegraphics[width=1\columnwidth]{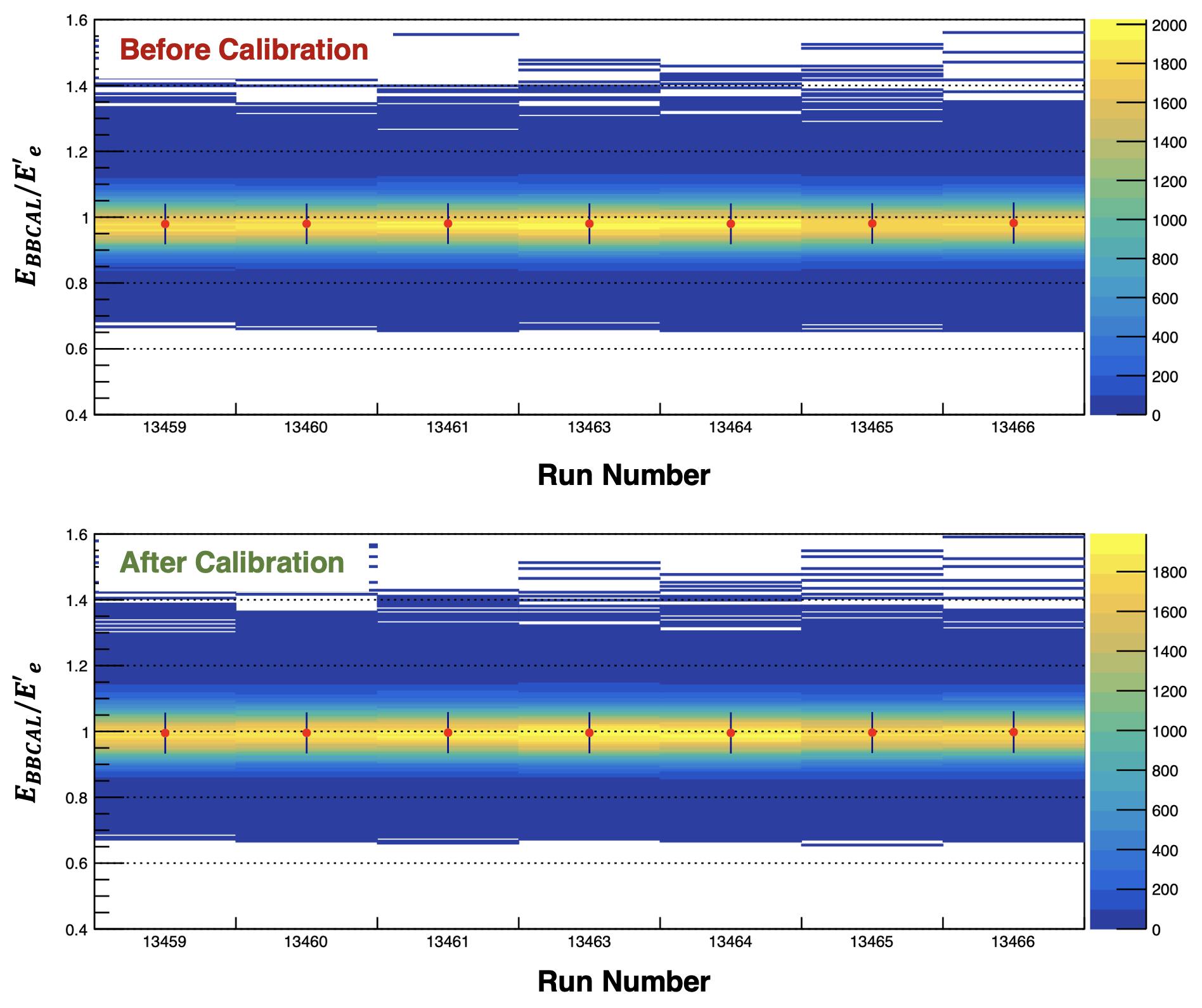}
    \caption{\label{fig:ch4:bbcaleovpvsrun} Uniformity of BBCAL energy calibration across runs for \qeq{4.5} high-\ep \lh data recorded with the SBS magnet turned off.}
\end{figure}
    \item \textbf{Uniformity Across Runs:} Calibration for a given experimental configuration should affect all CODA runs taken under the same conditions similarly. Therefore, the mean and $\sigma$ of the \eove distribution should remain uniform across all runs in a dataset sharing the same BBCAL energy calibration constants. This uniformity was ensured for all calibration settings. For example, \fig \ref{fig:ch4:bbcaleovpvsrun} demonstrates near-perfect uniformity in the mean and $\sigma$ of the \eove distribution across all \lh runs recorded during \qeq{4.5} high \ep data with the SBS magnet turned off. As in previous examples, these plots were generated using elastic \heep events.
\end{itemize}

\subheading{Notable Issues and Mitigation:}
In a few settings, significant non-uniformity of \eove across runs was observed, with varying trends:
\begin{enumerate}
\begin{figure}[h!]
    \centering
    \begin{subfigure}[b]{1\textwidth}
         \centering
         \includegraphics[width=\textwidth]{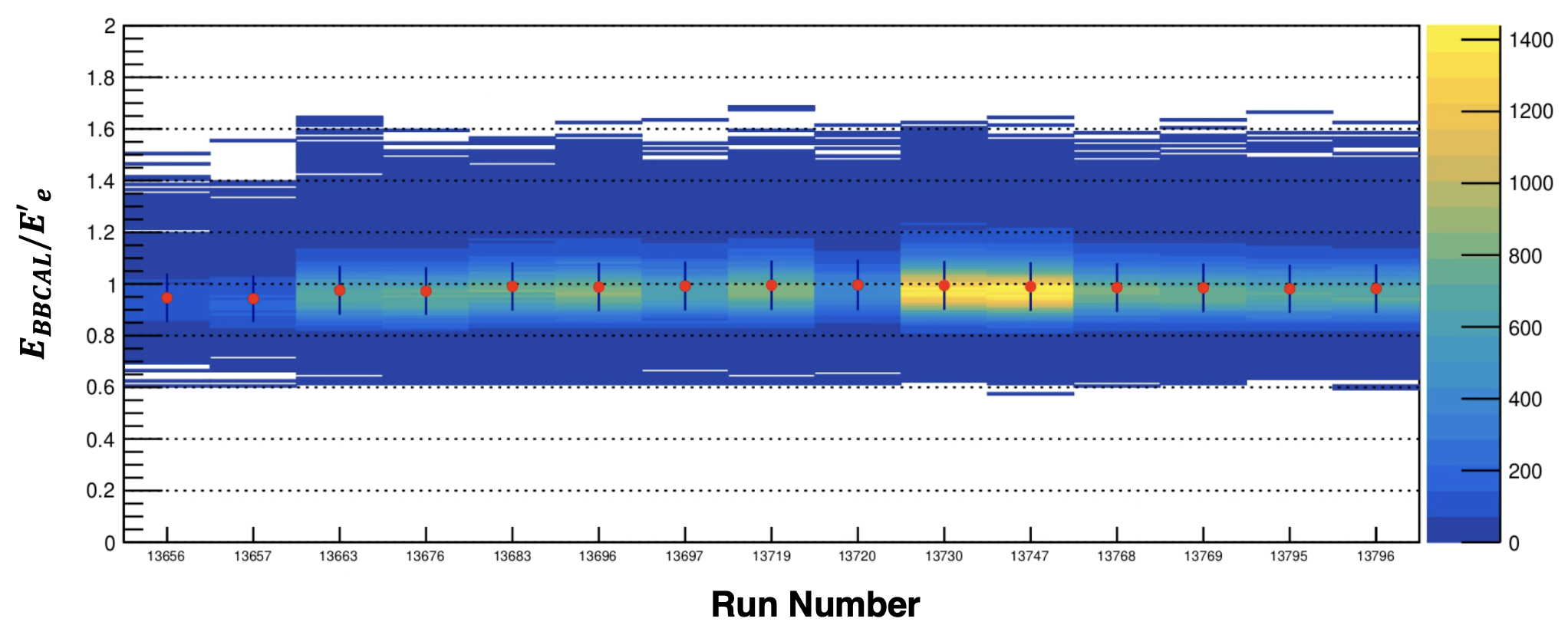}
         \caption{}
         \label{sfig:ch4:bbcalengissue1a}
    \end{subfigure}
    \hfill
    \begin{subfigure}[b]{1\textwidth}
        \centering
        \includegraphics[width=\textwidth]{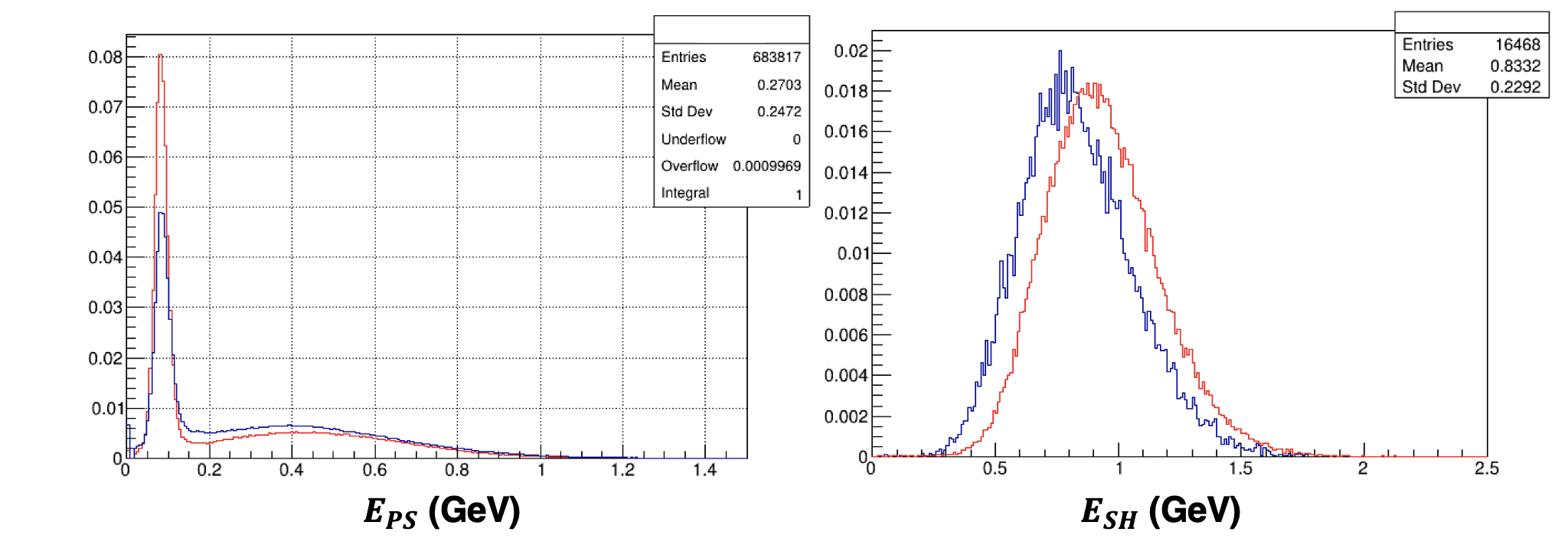}
        \caption{}
        \label{sfig:ch4:bbcalengissue1b}
    \end{subfigure}
    \caption{Non-uniformity in the \eove distribution across runs. Elastic \heep events at \qeq{4.5} low \ep kinematics are shown. (a) \eove vs. \lh runs, where a handful of runs exhibit an approximately $5\%$ shift in the \eove peak position from nominal. (b) Comparison of Pre-Shower (left) and Shower (right) cluster energy distributions between the affected (blue) and unaffected (red) runs reveals that the shift predominantly originates from the Shower.}
    \label{fig:ch4:bbcalengissue1}
\end{figure}
    \item One or two CODA runs exhibited a significant shift in the \eove distribution. For instance, \lh run 13240 from the \qeq{7.4} $70\%$ SBS field settings and runs 13656 and 16657 from the \qeq{4.5} low \ep dataset (see \fig \ref{sfig:ch4:bbcalengissue1a}) showed a leftward shift of $5–6\%$ in the \eove distribution, while the rest of the \lh runs recorded under the same settings peaked at 1. A similar but less pronounced shift was observed in a few \lh runs from the high-\q production settings as well. These runs were conducted with a much lower beam current than the rest, and in some cases, a different BBCAL trigger threshold was used.

    \hspace{1em}No significant variation in the distribution of $p$ between high and low current runs was observed, confirming that the shift originated from changes in the $E_{BBCAL}$ distribution. Further analysis revealed that this shift was primarily due to changes in the SH cluster energy distribution, as no significant differences were observed in the PS distribution (see \fig \ref{sfig:ch4:bbcalengissue1b}).

    \hspace{1em}These observations strongly suggest a rate-dependent effect in BBCAL energy calibration. However, various studies conducted to confirm this hypothesis yielded negative results. The causes of these shifts remain unclear and uncorrected. This should have negligible effect on \gmn analysis given only a handful of runs were affected.
    \item For \lh runs taken during the \qeq{7.4\,\,\&\,\,9.9} production settings, a slight shift ($\approx 2\%$) in the \eove peak position was observed, separating the runs into two distinct groups. Further investigation revealed that a long experimental downtime separated the recording of these group of runs for \qeq{7.4}. However, this alone does not fully explain the shift.

    \hspace{1em}Since multiple runs in each group were affected similarly, individual calibrations were performed for each group. This approach successfully removed the discrepancy and improved the overall energy resolution for the corresponding settings.
    \item The same trend discussed previously was also observed in \lh runs recorded during the \qeq{4.5} high \ep production settings. However, in this case, the shift in the \eove peak position between the two groups of runs ($\approx 5\%$) was nearly three times greater in magnitude (see \fig \ref{sfig:ch4:bbcalengissue3a}). According to the logbook, during the short experimental downtime (about 4 hours) between the recordings of these two groups, the beam energy pass was changed for Hall B, and multiple cosmic runs were taken for BBCAL with varying SBS field strengths.

\begin{figure}[h!]
    \centering
    \begin{subfigure}[b]{1\textwidth}
         \centering
         \includegraphics[width=\textwidth]{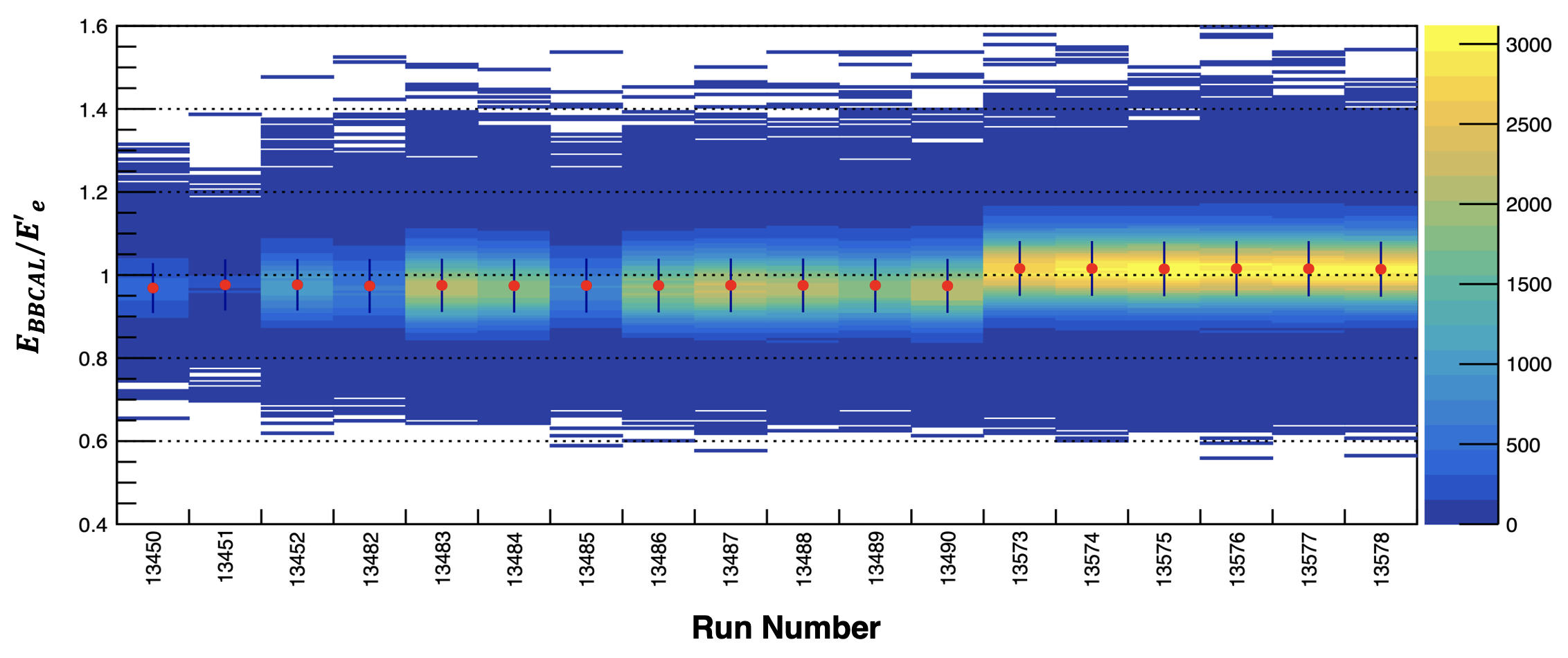}
         \caption{}
         \label{sfig:ch4:bbcalengissue3a}
    \end{subfigure}
    \hfill
    \begin{subfigure}[b]{1\textwidth}
        \centering
        \includegraphics[width=\textwidth]{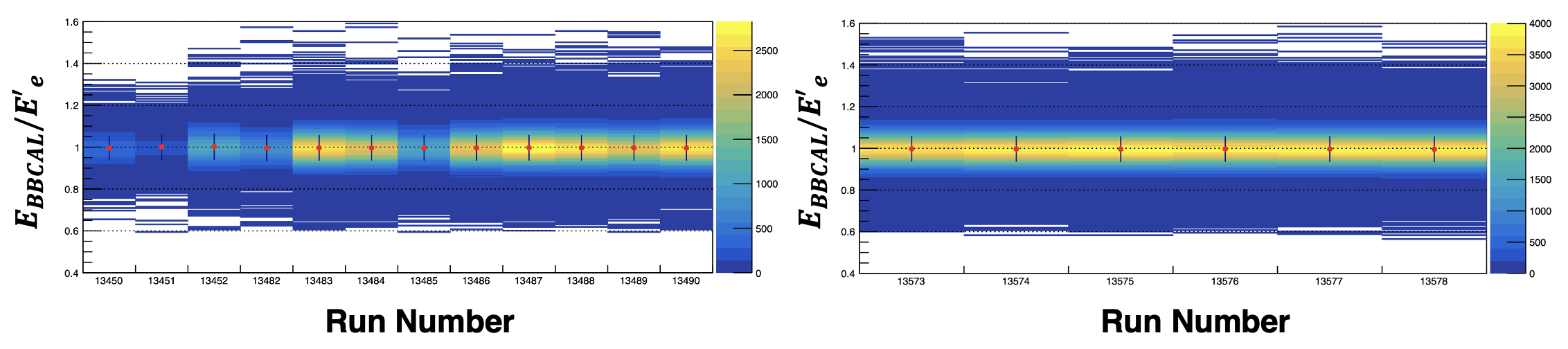}
        \caption{}
        \label{sfig:ch4:bbcalengissue3b}
    \end{subfigure}
    \caption{Issue and mitigation of the \eove shift observed across runs taken at \qeq{4.5} high \ep kinematics. Only elastic \heep events are shown. (a) \eove vs. \lh runs, clearly showing a sudden jump in the \eove peak position. (b) After calibration, \eove is perfectly aligned at $1$ for both sets of runs.}
    \label{fig:ch4:bbcalengissue3}
\end{figure}
    \hspace{1em}To rule out the possibility that the Hall B pass change affected the beam energy in Hall A and caused the observed shift, the average beam energy for all runs before and after the shift was compared, revealing no significant discrepancies. However, an interesting observation was made when studying the variation in SBS magnet current read-back during this period, suggesting that the recording of BBCAL cosmic runs with various SBS field strengths might have caused the shift. Before the shift, the SBS magnet current was almost constant at the set value, but after the shift, a significant variation of about \SI{16}{A} was observed. The reason for this sudden instability is unknown, but hysteresis resulting from the frequent changes in SBS field strength during the BBCAL cosmic runs immediately before the shift could be a possible explanation.

    \hspace{1em}Given the strong presence of the SBS fringe field at the BBCAL's location, even a slight difference in SBS field strength could affect the BBCAL PMT gains, resulting in a shift in the \eove peak position. Further investigation is necessary to confirm this hypothesis.

    \hspace{1em}Since the trend was similar, the same solution discussed previously was applied here, leading to an overall improvement in the uniformity (see \fig \ref{sfig:ch4:bbcalengissue3b}) and BBCAL energy resolution.
\end{enumerate}
These issues have been properly mitigated for all \gmn kinematic points, resulting in sixteen different calibration settings. 
%

%
\begin{table}[h!]
\caption[]{\label{tab:ch4:bbcalibsummary}Summary of BBCAL energy calibration performance parameters across all settings after the second pass of fine-tuning. Here, \q represents the central {\q}, \ep denotes the longitudinal polarization of the virtual photon, $\Bar{E_{e}}$ is the average scattered electron energy, $I_{BB}$ ($I_{SBS}$) represents the BB (SBS) magnet current as a percentage of \SI{750}{A} (\SI{2100}{A}), and Set \# is the index for different calibration sets within a setting, used to address \eove non-uniformity across runs.}
\centering
\begin{tabular}{cccccccc}
\hline\hline\vspace{-1.1em} \\ 
$Q^{2}$          &\multirow{2}{*}{\ep} &$\Bar{E_{e}}$ &\multirow{2}{*}{$I_{BB}$ ($\%$)} &\multirow{2}{*}{$I_{SBS}$ ($\%$)} & \multirow{2}{*}{Set \#} &\multicolumn{2}{c}{\eove} \\ \vspace{-1.15em} \\ \cline{7-8} \vspace{-1.15em} \\ 
\SI{}{(GeV/c)^2} &                     &GeV           &                                 &             &           & Mean              & Sigma ($\%$)         \vspace{0.2em} \\ \hline \vspace{-1.15em} \\ 
\multirow{3}{*}{3.0}  & \multirow{3}{*}{0.72} & \multirow{3}{*}{2.12} & \multirow{3}{*}{100} & 0 & \multirow{3}{*}{N/A} & 1.00 & 5.9\\ 
  & & &  & 30 & & 1.00 & 6.0\\ 
  & & &  & 50 & & 1.00 & 6.1\\ \hline 
  4.5  & 0.51 & 1.63 & 100 & 70 & N/A & 0.99 & 6.4 \\ \hline
\multirow{5}{*}{4.5}  & \multirow{5}{*}{0.80} & \multirow{5}{*}{3.58} & \multirow{5}{*}{100} & 0 & \multirow{2}{*}{N/A} & 0.99 & 5.4 \\ 
  & & &  & 50 & & 1.00 & 5.4 \\ \cline{5-8}
  & & &  & \multirow{2}{*}{70} & 1 & 0.99 & 5.4 \\ 
  & & &  &                     & 2 & 0.99 & 5.4 \\ \cline{5-8}
  & & &  & 100 & N/A & 0.99 & 5.5 \\ \hline
\multirow{3}{*}{7.4}  & \multirow{3}{*}{0.46} & \multirow{3}{*}{2.00} & \multirow{3}{*}{100} & 0 & N/A & 1.00 & 6.3 \\ \cline{5-8}
  & & &  & \multirow{2}{*}{70} & 1 & 1.00 & 6.0 \\
  & & &  &                     & 2 & 1.00 & 6.3 \\ \hline
\multirow{2}{*}{9.9}  & \multirow{2}{*}{0.50} & \multirow{2}{*}{2.66} & \multirow{2}{*}{100} & \multirow{2}{*}{85} & 1 & 1.00 & 5.7\\ 
  & & &  & & 2 & 1.00 & 5.8 \\ \hline
\multirow{2}{*}{13.6} & \multirow{2}{*}{0.41} & \multirow{2}{*}{2.67} & \multirow{2}{*}{100} & 0 & \multirow{2}{*}{N/A} & 0.98 & 5.8 \\ 
  & & &  & 100 &  & 0.99 & 6.5 \\ 
\hline\hline
\end{tabular}
\end{table}
\begin{figure}[h!]
    \centering
    \includegraphics[width=1\columnwidth]{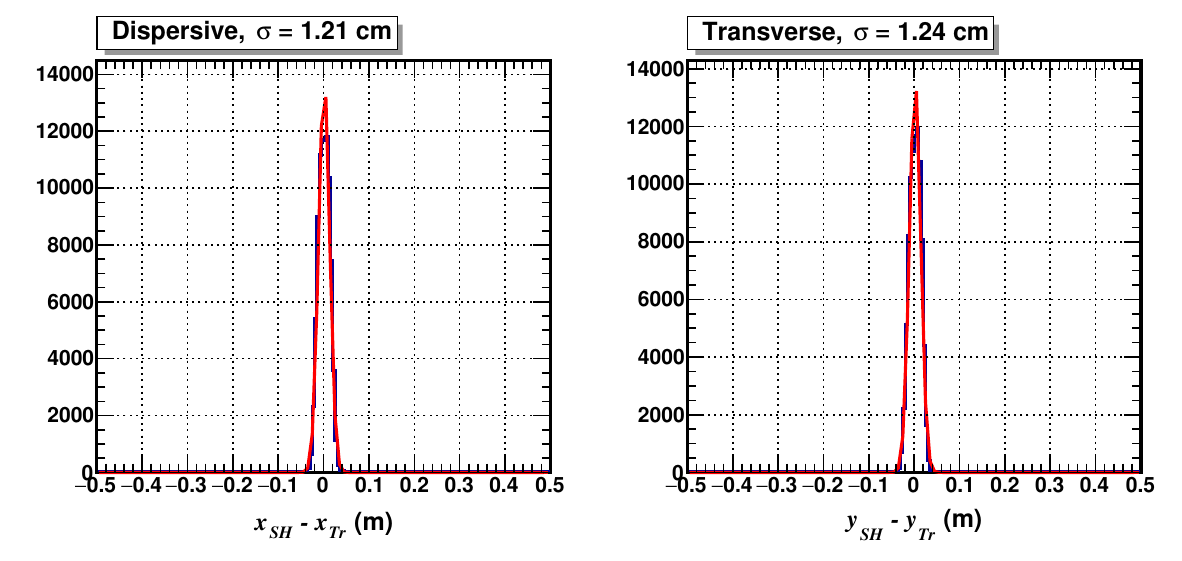}
    \caption{\label{fig:ch4:bbcalibposres} Position resolution of Shower (SH) clusters in (left) dispersive and (right) transverse directions, defined by the difference between the reconstructed SH cluster position and the position of track projected to the face of SH. Elastic \heep events from the \qeq{4.5} high-\ep dataset are shown.}
\end{figure}
\subheading{Performance Parameters}
\tab \ref{tab:ch4:bbcalibsummary} summarizes the performance of the BBCAL energy calibration after the second pass of fine-tuning across all sixteen calibration settings throughout \gmn. The observed energy resolutions range from $5.4\%$ to $6.4\%$, with the average \eove converging to $1$ in all cases. Rigorous quality assurance, based on the criteria discussed above, has ensured optimal uniformity across the acceptance at each setting. As a result of the improved BBCAL energy calibration, the energy-weighted cluster centroids of SH (and PS), as defined in \eqn \ref{eqn:ch4:calclcentroid}, have also improved, leading to a position resolution of approximately $1.2-1.4$ cm in both the dispersive and horizontal directions (see \fig \ref{fig:ch4:bbcalibposres}).

\subsection{Hadron Calorimeter}
\label{ssec:ch4:hcalcalib}
Despite significant differences in the design of the hadron calorimeter (HCAL) and the BigBite calorimeter (BBCAL), their calibration procedures are largely similar. Like BBCAL, HCAL measures both the arrival time and energy of detected particles, requiring calibration for both timing and energy.

\subsubsection{HCAL Timing Calibration}
\label{sssec:ch4:hcaltimingcalib}
The HCAL data is read out by fADCs and f1TDCs, providing both ADC and TDC times for each HCAL channel. These times are calibrated separately for all HCAL channels.

\begin{figure}[h!]
    \centering
    \includegraphics[width=1\columnwidth]{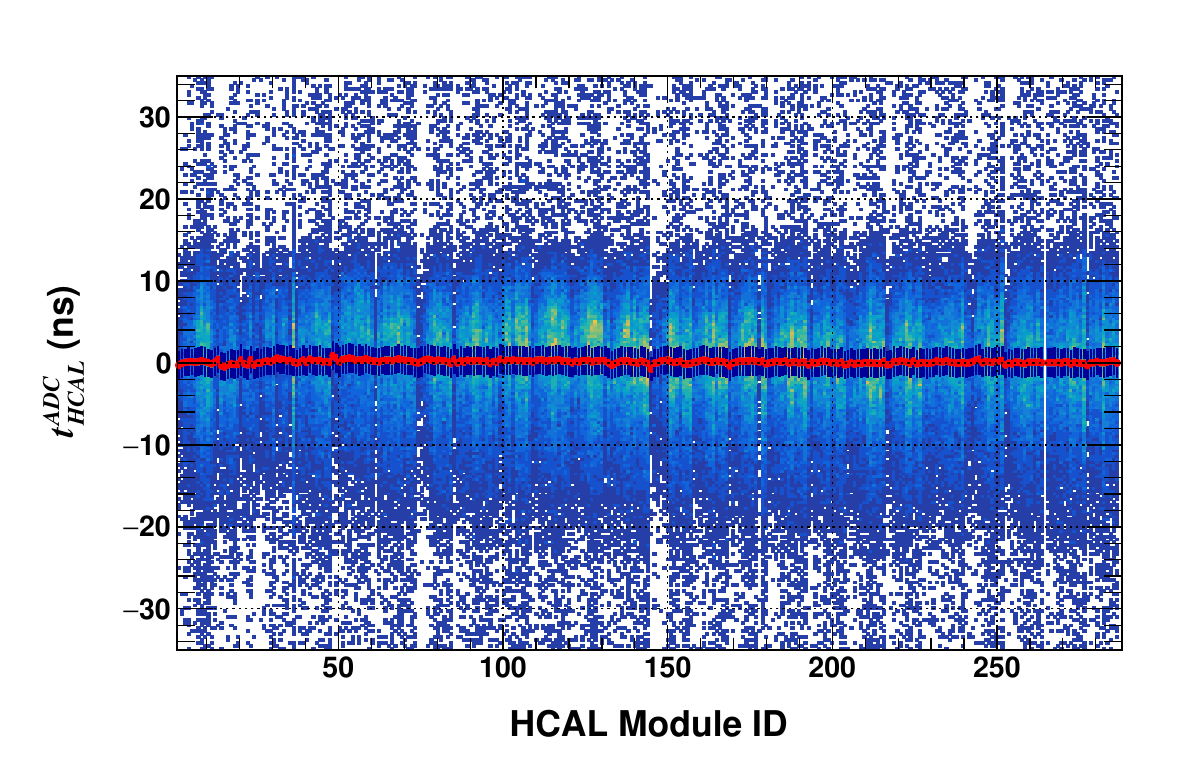}
    \caption{\label{fig:ch4:hcalatime} HCAL ADC time alignment across modules for \qeq{9.9} kinematics using \lh data.}
\end{figure}

\subheading{ADC Time Calibration}
HCAL ADC time is calibrated in the same manner as the BBCAL ADC time, as detailed in \sect \ref{sssec:ch4:bbcalcalibtime}. This process involves aligning the ADC time of each HCAL channel with respect to the timing hodoscope (TH) cluster mean time. The resulting channel-specific offsets are recorded in the database (DB) file (see \sect \ref{ssec:ch4:softtool}) and subtracted from the raw ADC time of the corresponding HCAL channel during event reconstruction. As expected, the reconstructed ADC times from all HCAL channels align nicely at zero, as shown in \fig \ref{fig:ch4:hcalatime}.

\lh and \ld events that pass good electron event selection cuts, a very low threshold HCAL cluster energy cut, and HCAL fiducial cuts are used for calibration. Separate calibrations were performed for each \gmn configuration to avoid biases associated with kinematic-specific uncertainties. Additionally, multiple calibration sets within the \qeq{7.4\,\,\&\,\,13.6} configurations were necessary to mitigate the ADC time shift observed across runs. This same shift was also observed in BBCAL ADC times, as discussed in \sect \ref{sssec:ch4:bbcalcalibtime}.

It is worth noting again that the ADC time shift of about \SI{-12}{ns} observed within the same CODA run for BBCAL also affects HCAL, as clearly demonstrated by their correlation in \fig \ref{sfig:ch4:bbcalatimeissue2solved}. Although there is no straightforward way to correct for these out-of-time events, the shift cancels out perfectly in the HCAL-BBCAL ADC coincidence time, enabling effective suppression of accidental background, crucial for the \gmn analysis.

\subheading{TDC Time Calibration}
The offsets to align the TDC times across all HCAL channels are determined in the same way as for HCAL ADC time. Additionally, a time-walk correction of the form:
\begin{equation}
\label{eqn:ch4:hcaltw}
    t_{HCAL} = a + b/(E_{HCAL})^c
\end{equation}
is applied to remove any energy-dependent bias. In \eqn \ref{eqn:ch4:hcaltw}, $a$, $b$, and $c$ are fit parameters, while $t_{HCAL}$ and $E_{HCAL}$ represent the TDC time and the energy of the best HCAL cluster. The same type of events used for ADC time calibration are also used here. The resulting channel-specific offsets and the parameters for time-walk correction are recorded in the DB file for event-by-event correction of the raw TDC time for all HCAL channels during event reconstruction.

\begin{figure}[h!]
    \centering
    \includegraphics[width=\sfig]{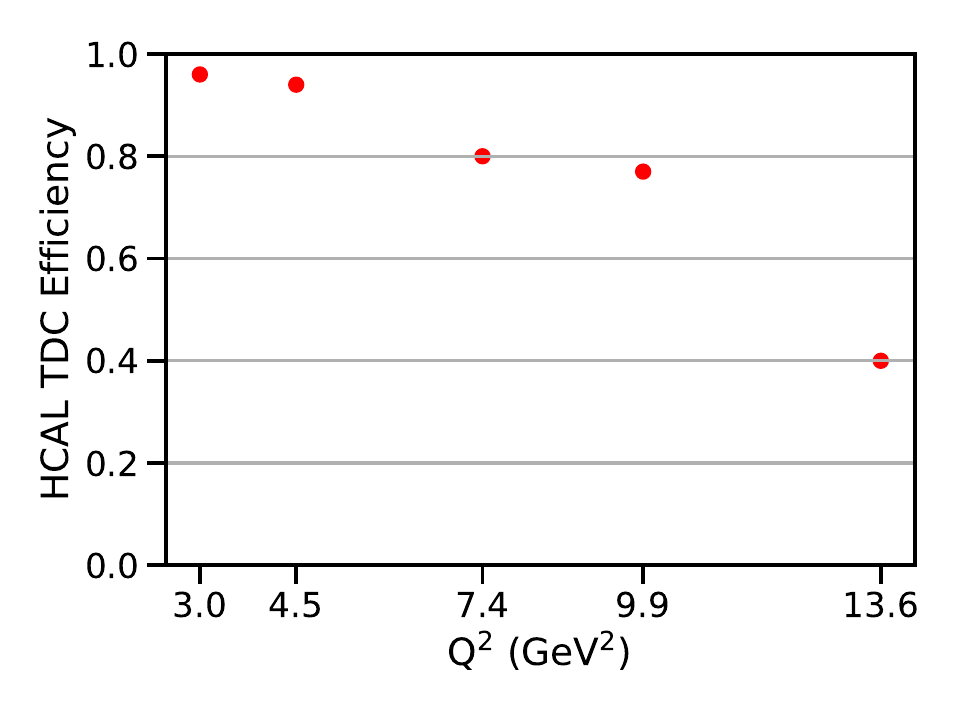}
    \caption{\label{fig:ch4:hcaltdceffi} Severity of missing HCAL TDC data across \gmn production kinematics. HCAL TDC efficiency is defined as the ratio of detected to expected good TDC events. The total number of elastic \heep events represents the expected number of events, while the detected number of events includes an additional cut requiring a good TDC signal in the primary block of the HCAL cluster.}
\end{figure}
Separate calibrations were performed for each kinematic point. In a handful of runs recorded toward the end of the highest-\q, a sudden shift of about \SI{30}{ns} in the TDC time was observed. The latency shift, observed at the same kinematics in HCAL ADC times, as mentioned above, affected a different set of runs. The reasons behind these shifts are yet to be confirmed. However, similar to the ADC time shift, the shift observed in TDC time was corrected by calibrating the affected and unaffected runs separately.

During offline analysis, it was discovered that HCAL TDC data were missing for a significant fraction of elastic events, which are the key physics events of interest, across all \gmn kinematics. The high \q kinematics, where statistics are limited, were the most affected (see \fig \ref{fig:ch4:hcaltdceffi}). Further investigation confirmed that the issue was caused by very high occupancy in the f1TDCs under high rate conditions, resulting from a lower than expected threshold applied to the discriminators upstream of the f1TDCs in the HCAL signal circuit. Unfortunately, this meant that the missing HCAL TDC data are unrecoverable. 

\begin{figure}[h!]
    \centering
    \includegraphics[width=\sfig]{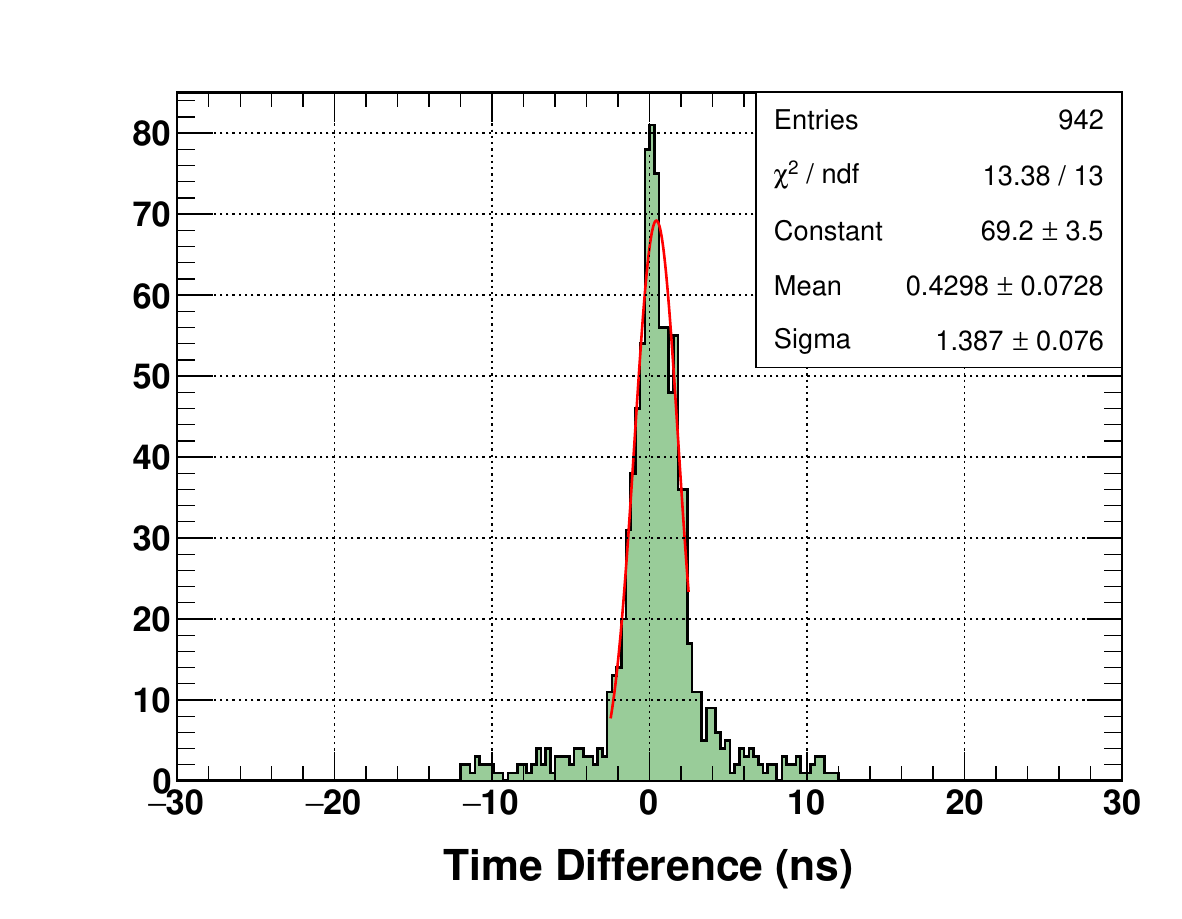}
    \caption{\label{fig:ch4:hcaltimeresolution} Intrinsic resolution of HCAL ADC time, determined by the time difference between the ADC times of primary and secondary modules in HCAL clusters. \lh events passing good electron and \w cuts at \qeq{4.5} (low-\ep) are shown.}
\end{figure}
\subheading{Performance Parameters}
With the second pass of fine-tuning, the intrinsic resolution of HCAL ADC time, defined by the ADC time difference between the primary and secondary modules in a cluster, was observed to be in the range of $1.2$ to $1.4$ ns across \gmn kinematics (see \fig \ref{fig:ch4:hcaltimeresolution}). Similar results were observed for HCAL TDC time as well. Although more advanced calibration of TDC time, including nucleon time-of-flight (TOF) and RF time corrections, could potentially improve performance, the previously mentioned missing TDC data issue renders such efforts impractical. Nevertheless, the observed ADC time resolution is highly encouraging and provides high-resolution two-arm coincidence time resolution when subtracted from BBCAL ADC time, as required for the \gmn analysis.

\subsubsection{HCAL Energy Calibration}
\label{sssec:ch4:hcalengcalib}
Similar to timing calibration, HCAL energy calibration shares many similarities with BBCAL's. At the start of \gmn, HCAL PMTs were gain-matched using cosmic data, following the same procedure used for BBCAL (see \sect \ref{ssec:trigcalib}). The high voltage (HV) settings from this gain matching were used for data acquisition during the lowest \q kinematics and remained constant until CODA run number $11581$. Subsequently, the operating HVs for the HCAL PMTs were updated based on in-beam calibration and then kept constant for the remainder of \gmn. Since the HCAL HVs were updated only once, two sets of in-beam energy calibrations were sufficient for the entire \gmn dataset \cite{SUPPMAT}.

\begin{figure}[h!]
    \centering
    \includegraphics[width=1\columnwidth]{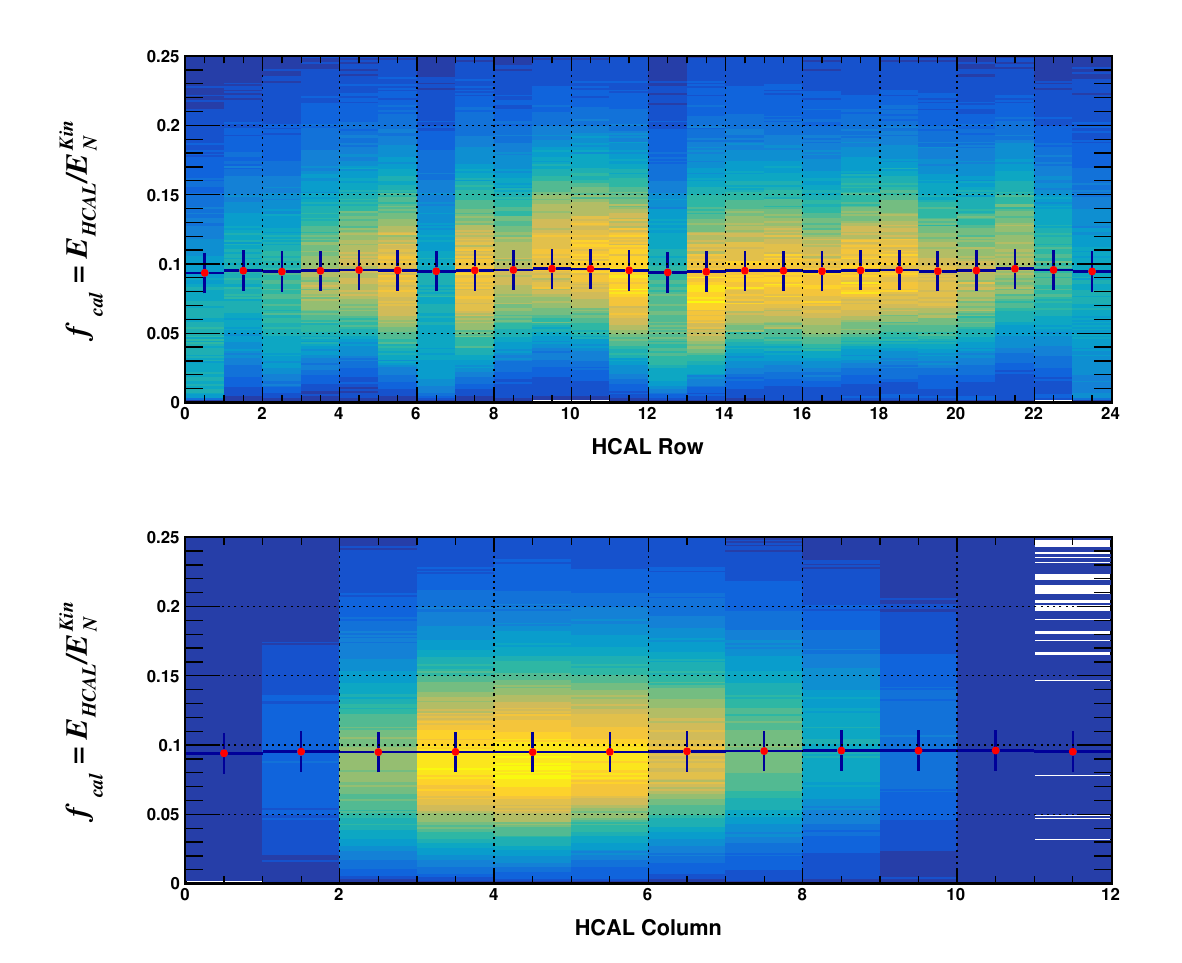}
    \caption{\label{fig:ch4:hcalenguniformity} Uniformity of HCAL energy calibration within the acceptance for the \qeq{4.5} high-\ep dataset. The top plot shows the HCAL energy sampling fraction $f_{cal}$ as a function of HCAL rows, while the bottom plot shows $f_{cal}$ as a function of HCAL columns. \lh events passing strict elastic cuts are shown.}
\end{figure}
The same chi-squared minimization formalism used for BBCAL in-beam energy calibration is applied here as well. However, to account for the fact that HCAL is a sampling rather than a homogeneous calorimeter, the $\chi^2$ function is slightly modified as follows:
\begin{equation}
\label{eqn:ch4:hcalibchi2}
    \chi^2 = \sum_{i=1}^{N} \left( \Bar{f}_{cal}\nu^{i} - \sum_{j=0}^{M} c_{j}A^{i}_{j} \right)^{2}
\end{equation}
Here, $\nu^i$, the kinetic energy of the struck nucleon in the $i^{th}$ event, is defined by the difference between the beam energy and the scattered electron energy reconstructed by BB optics. When multiplied by $\Bar{f}_{cal}$, the average energy sampling fraction of HCAL, it gives the ``true" energy deposition of an elastically scattered nucleon in HCAL for the same event. The rest of the variables have the same meanings as in \eqn \ref{eqn:ch4:bbcalibchi2}, with BBCAL replaced by HCAL. $\Bar{f}_{cal}$ is obtained from realistic \textit{Geant4} simulation.

Naturally, elastic \heep events are used for HCAL energy calibration. \lh runs taken with various SBS field settings during the \qeq{4.5} high \ep configuration (see \tab \ref{tab:sbsconfig2}) are combined to calibrate HCAL energy for data recorded after the HCAL HV change, which includes most of the \gmn dataset, except for some runs from the lowest \q point. The ample statistics recorded for each SBS field setting effectively cover the entire active area of HCAL, which is crucial to avoid position-dependent bias in the calibration. \fig \ref{fig:ch4:hcalenguniformity} illustrates the uniformity of HCAL energy calibration within acceptance in both the dispersive and transverse directions for the \qeq{4.5} high \ep dataset.

\begin{figure}[h!]
    \centering
    \begin{subfigure}[b]{0.496\textwidth}
         \centering
         \includegraphics[width=\textwidth]{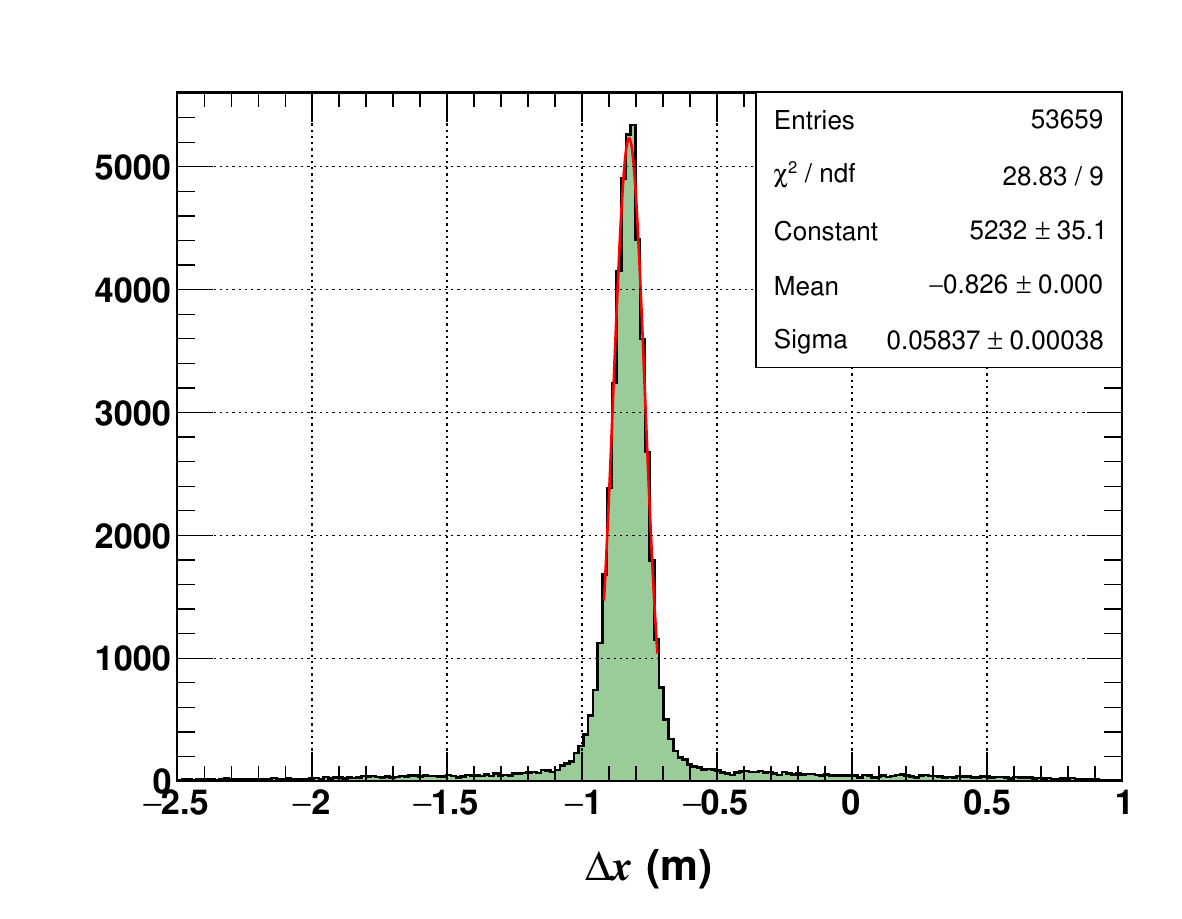}
    \end{subfigure}
    \hfill
    \begin{subfigure}[b]{0.496\textwidth}
        \centering
        \includegraphics[width=\textwidth]{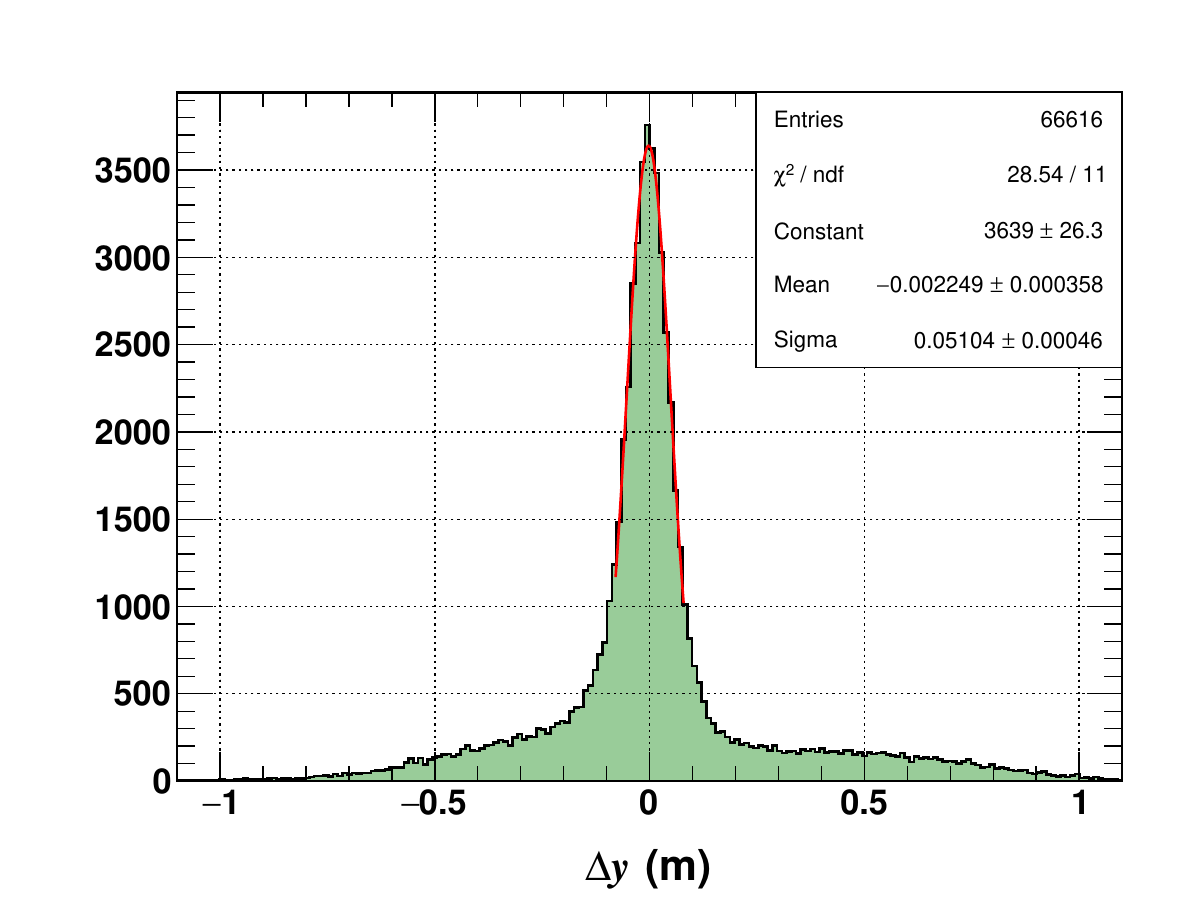}
    \end{subfigure}
    \caption{\label{fig:ch4:hcalengposresolution} Position resolution of HCAL in the (left) dispersive and (right) transverse directions for \qeq{7.4}. Elastic \heep events are shown.}
\end{figure}
%
\subheading{Performance Parameters}
After the second pass of fine-tuning HCAL energy calibration, the energy resolution has been observed to be between $41\%$ and $67\%$ across \gmn kinematics \cite{SUPPMAT}. Although HCAL, being a sampling calorimeter, is not expected to have high energy resolution, its ability to accurately reconstruct the position of detected nucleons is invaluable for \gmn analysis. Improved energy calibration is crucial for this, as the reconstructed cluster centroids are energy-weighted, as defined in \eqn \ref{eqn:ch4:calclcentroid}. With the latest calibration, the position resolution of HCAL in both the dispersive and transverse directions was observed to be approximately $5-6$ cm across \gmn configurations, (see \fig \ref{fig:ch4:hcalengposresolution}), which meets the design expectation.

\section{Monte Carlo Simulation}
\label{sec:ch4:mcsimulation}
Monte Carlo (MC) simulation plays a crucial role in the \gmn data analysis. It enables the extraction of the elastic electron-neutron ($en$) to electron-proton ($ep$) cross-section ratio, $R$, from \rqe, the quasi-elastic $en$ to $ep$ cross-section ratio, by accurately accounting for the relevant physics and detector effects. The process involves realistic simulations of the distributions of key kinematic variables, which can then be directly compared to those obtained from real data on an equal basis. The simulated distribution is fitted to the data, along with an estimated background shape, to extract the corrected signal. Achieving such a realistic simulation of physics and detector effects is highly challenging and requires sophisticated computational tools. This section provides an overview of the MC framework used in the \gmn analysis, highlighting its key components and validating its accuracy through data/MC comparisons of essential kinematic variables.
\subsection{Software Tools and Analysis Flow}
\gmn MC machinery consists of four different software libraries: \simc \cite{SIMC}, \gfsbs \cite{G4SBS}, libsbsdig \cite{LIBSBSDIG}, and SBS-offline \cite{SBSoffline}. Below is a brief overview of their roles:
    \begin{itemize}
        \item \textbf{\simc:} A FORTRAN-based MC simulation library primarily used for coincidence reactions in Jefferson Lab Hall C experiments. SIMC provides realistic quasi-elastic and elastic event generators that incorporate sophisticated radiative and nuclear correction models, rigorously validated over time. These generators were naturally adapted for the \gmn analysis, necessitating several upgrades to the standard \simc library:
        \begin{itemize}[label=$\circ$]
            \item Integration of box detectors to replicate the acceptances of BB and Super BB spectrometers.
            \item Incorporation of realistic cryotarget geometry.
            \item Self-consistent implementation of the \deen process alongside the existing \deep process.
            \item Introduction of rejection sampling to distribute events according to the cross-section model, boosting event generation efficiency by approximately $99\%$, allowing simulations with the desired statistical precision across \gmn kinematics.
            \item Enhancement of the \ce{D2} model, extending the missing momentum range from \SI{490}{MeV} to \SI{1.2}{GeV}, making Fermi motion effects more realistic in quasi-elastic analysis.
        \end{itemize}
    A more detailed discussion of the models included in the quasi-elastic event generator is presented in the following section.
    \item \textbf{\gfsbs:} A \geantf-based library built exclusively for the Super BigBite Spectrometer collaboration experiments, including \gmn. It contains detailed geometry of the spectrometers and the experimental setup, essential for conducting realistic detector simulations.
    \item \textbf{libsbsdig:} A C++ library built on top of \gfsbs to digitize its output, generating pseudoraw data.
    \item \textbf{SBS-offline:} A C++ library built on top of Podd for event reconstruction of \gmn data, as discussed in \sect \ref{ssec:ch4:softtool}. This library can also reconstruct the pseudoraw data generated by libsbsdig, ensuring equivalent reconstruction uncertainties between data and MC.
\end{itemize}

\begin{figure}[h!]
    \centering
    \includegraphics[width=1\columnwidth]{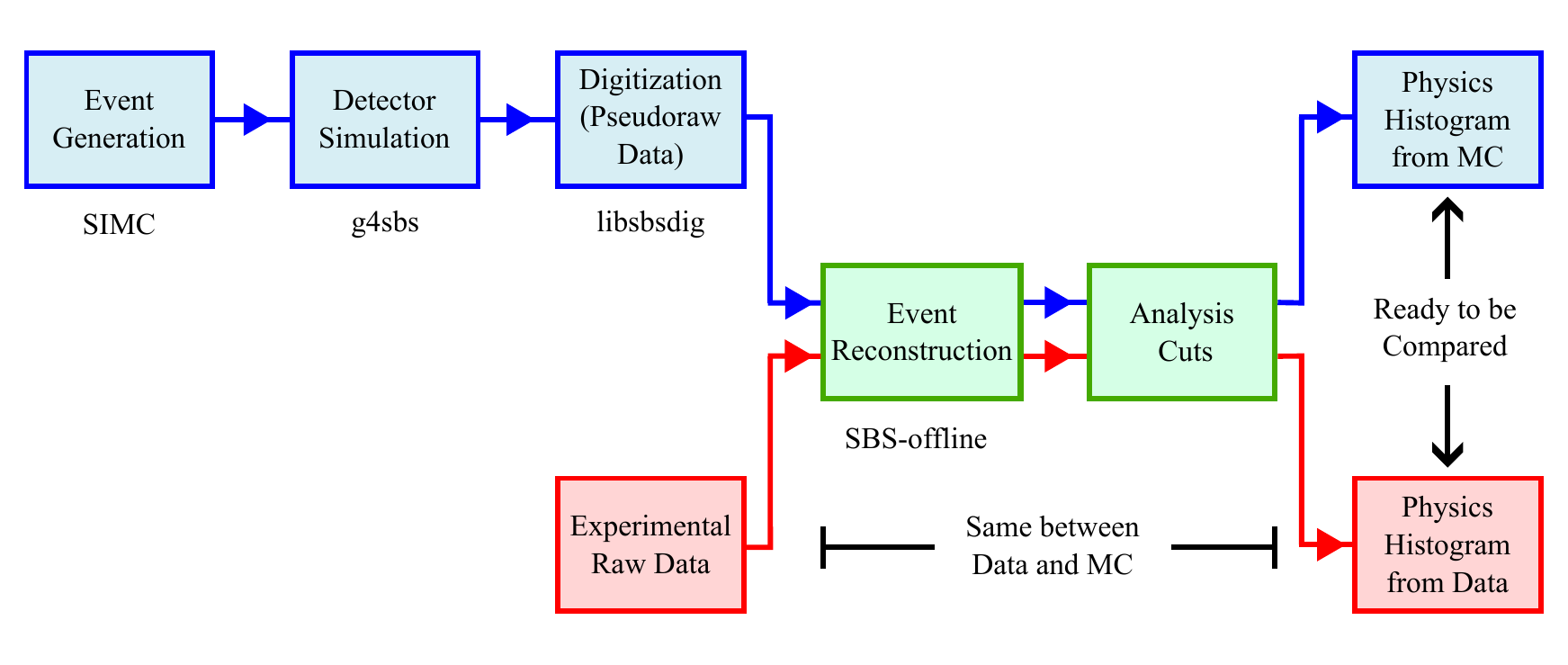}
    \caption{\label{fig:ch4:mcdmcsteps} Steps to perform realistic data/MC comparison for \gmn.}
\end{figure}
As illustrated in \fig \ref{fig:ch4:mcdmcsteps}, the libraries mentioned above work in sync to produce realistic physics histograms from MC, which can be directly compared to data. The process begins by generating quasi-elastic (or elastic, in the case of \lh data) events via the \simc generator, incorporating realistic radiative and nuclear effects. These generated events are then processed through the BigBite and Super BigBite spectrometers using \gfsbs to simulate realistic detector effects, followed by digitization with libsbsdig to create pseudoraw data. Finally, the pseudoraw data is reconstructed by SBS-offline using the same event reconstruction algorithms applied to real data. Physics histograms generated from the reconstructed MC events, which pass the same analysis cuts as the real data, are then ready for direct comparison. 

\subsection{Event Generation}
\label{ssec:mceventgeneration}
The scattering processes of interest for \gmn analysis include \deen, \deep, and \heep. The first two processes, representing quasi-elastic $eN$ scattering from \ce{D2}, are used for extracting experimental observables, while the last one, representing elastic $ep$ scattering from \ce{H2}, is essential for detector calibration and performance characterization. These events are generated by \simc-based event generators that incorporate realistic radiative and nuclear effects. 

Additionally, inclusive inelastic $eN$ events are generated for background estimation using an inelastic event generator based on \gfsbs, which includes a realistic cross-section model for inclusive inelastic $ep$ scattering but lacks radiative corrections. This is acceptable for the \gmn analysis, as the approach focuses on generating signal shapes as realistically as possible and then using reasonable models to estimate the background shape, as discussed in \sect \ref{ssec:ch4:inelbg}.

A brief description of the key features of these generators is provided below.
\subsubsection{Quasi-Elastic Event Generation}
\label{sssec:qeeventgeneration}
The quasi-elastic (QE) event generator in \simc separately generates \deep and \deen events, but the underlying mechanism remains the same. Scattered electron and nucleon angles are generated uniformly within user-defined $x'_{tg}$ and $y'_{tg}$ limits, followed by the generation of scattered electron energy based on the desired energy range. The polar and azimuthal scattering angles are then computed using the $x'_{tg}$, $y'_{tg}$ values and the spectrometer central angles.

Next, the scattered nucleon energy ($E_N$) is determined by fixing the missing energy ($E_{miss}$) to the deuteron binding energy, and a Jacobian of the form $|\dv{E_N}{E_{miss}}|$ is computed to correctly weight the event. Events are then radiated according to the built-in radiative correction model \cite{PhysRevC.64.054610}. Spectrometer cuts are applied to ensure the event is accepted only if both the nucleon and electron fall within the respective spectrometer acceptances\footnote{Toy box-shaped spectrometer models are implemented in \simc to simulate the acceptances of the BB and Super BB spectrometers.}.

User-specified beam energy, corrected for target energy loss, and beam position, smeared based on the raster size and pattern, are incorporated. Combining these with the already determined kinematic variables and scattering angles, remaining physics quantities such as $\nu$, \q, and the missing momentum vector ($\va{p}_{\textbf{miss}}$) are calculated. The spectral function weight is then determined based on the missing momentum, followed by the calculation of the cross section.

The final event weight is the product of the Jacobian, spectral function weight, and cross section weight. The event is accepted if the ratio of this weight to a predefined maximum weight is greater than or equal to a random number drawn from a uniform distribution between 0 and 1. This rejection sampling process ensures that the accepted events are distributed based on the cross-section model, significantly increasing  (approximately $99\%$) the event reconstruction efficiency\footnote{The maximum weight is determined from a large sample of events without rejection sampling.}.

In post-analysis, the weight for each event is calculated as:
\begin{equation}
    \text{Weight} = \frac{\text{Maximum Weight} \times \text{Luminosity} \times \text{Generation Volume}}{\text{Total Number of Tries}}
\end{equation}

A brief description of the \ce{D2} theory and the cross-section model is provided below.
%
\begin{figure}[h!]
    \centering
    \begin{subfigure}[b]{0.496\textwidth}
         \centering
         \includegraphics[width=\textwidth]{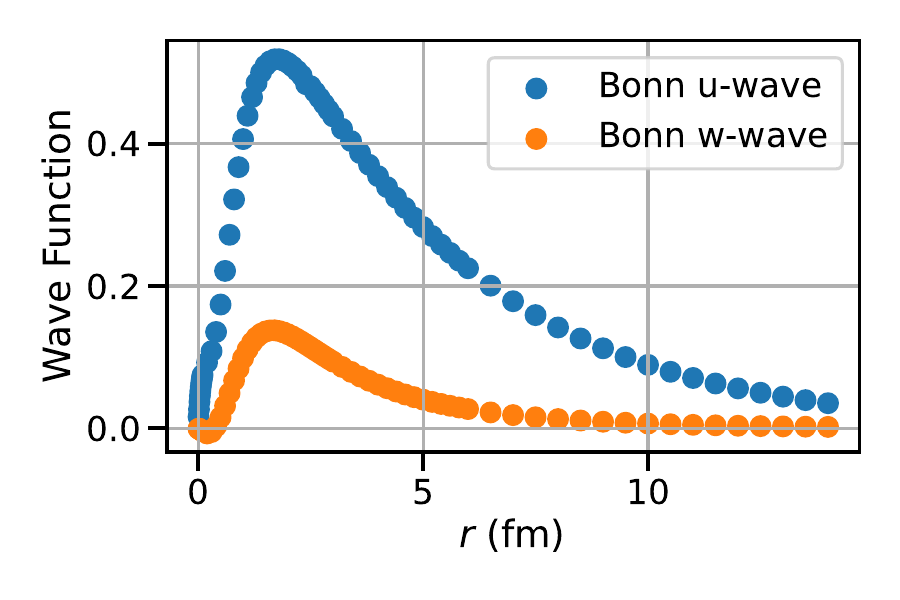}
    \end{subfigure}
    \hfill
    \begin{subfigure}[b]{0.496\textwidth}
        \centering
        \includegraphics[width=\textwidth]{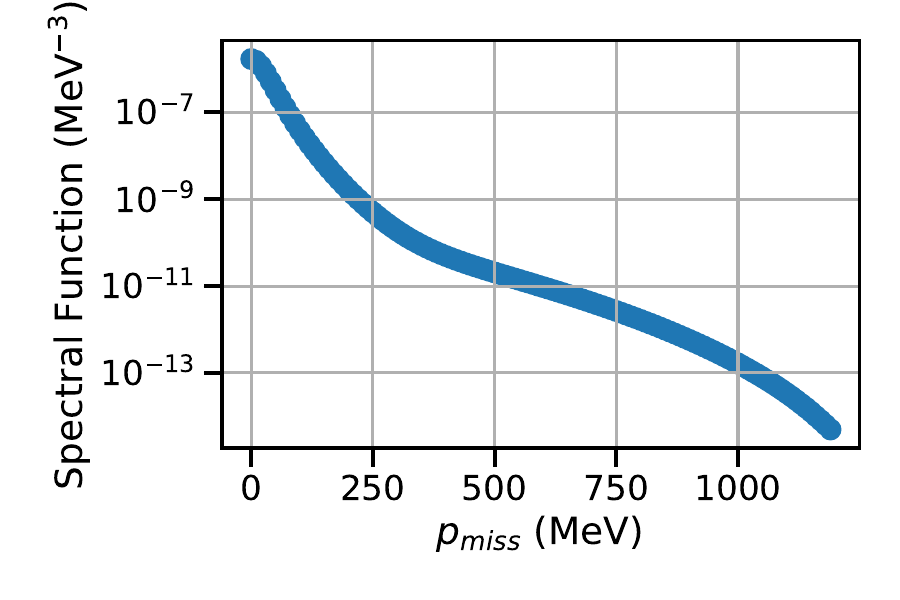}
    \end{subfigure}
    \caption{\label{fig:ch4:h2theory} Coordinate-space deuteron wave functions, $u(r)$ and $w(r)$, obtained using the Bonn meson-exchange potential (left). Spectral function distribution in missing momentum ($p_{miss}$) included in \simc (right).}
\end{figure}

\subheading{\ce{D2} Theory}
The nucleon-nucleon ($NN$) interaction within the deuteron is modeled using the Bonn meson-exchange potential \cite{MACHLEIDT19871}. The corresponding coordinate-space wave functions, $u(r)$ and $w(r)$, are shown in \fig \ref{fig:ch4:h2theory}. The spectral function $(S)$, computed using this model at a fixed missing energy equal to the deuteron binding energy, is used to estimate the spectral function weight for a given missing momentum ($p_{miss}$) in the event generator. The existing spectral function calculation in \simc spans the $p_{miss}$ range from $0$ to $1.2$ GeV, as shown in \fig \ref{fig:ch4:h2theory}, adequately capturing the high-momentum tail of the deuteron wave function for quasi-elastic scattering analysis.

\subheading{Cross Section Model}
The quasi-elastic scattering cross section is calculated using the de Forest model \cite{DEFOREST1983232}, which incorporates an off-shell extrapolation of the Rosenbluth cross section within the general framework of $(e,e'N)$ scattering in the Plane Wave Impulse Approximation (PWIA). The off-shell cross section can be expressed as:
\begin{equation}
    \frac{d^4\sigma}{d\Omega_{E'_e}d\Omega_{|\vb{k'}|}d\Omega_{E'_p}d\Omega_{|\vb{p'}|}} = E'_e E'_p \sigma_1^{cc} S(p,E_p)
\end{equation}
where $k'=(E'_e,\vb{k'})$ is the final electron four-momentum, $p=(E_p,\vb{p})$ ($p'=(E'_p,\vb{p'})$) is the initial (final) nucleon four-momentum,  $S(p,E_p)$ is the above-discussed spectral function, and $\sigma_1^{cc}$ is defined as:
\begin{equation}
    \begin{aligned}    
        \sigma_1^{cc} = \sigma_{\text{Mott}}\frac{E_e}{E'_e} &\left[ \frac{q^4}{|\vb{q}|^4} W_C 
        + \left( \frac{q^2}{2|\vb{q}|^2} + \tan^2\frac{\theta_e }{2} \right) W_T \right. \\
        &\quad + \frac{q^2}{|\vb{q}|^2} \left( \frac{q^2}{|\vb{q}|^2} + \tan^2\frac{\theta_e }{2} \right)^{\frac{1}{2}} W_I \cos\phi_e \\
        &\left. \quad + \left( \frac{q^2}{|\vb{q}|^2}\cos^2\phi_e + \tan^2\frac{\theta_e }{2} \right) W_S \right]
    \end{aligned}
\end{equation}
where $\sigma_{\text{Mott}}$ is the Mott cross section (see \eqn \ref{eqn:ch1:mott}), $\theta_e$ ($\phi_e$) is the polar (azimuthal) scattering angle, and the $W$ terms are given by:
\begin{equation}
    \begin{aligned}
        W_C &= \frac{1}{4E_pE'_p}\left[ (E_p + E'_p)^2 \left( \frac{G_E^2 + \tau_N G_M^2}{1+\tau_N} \right) - |\vb{q}|^2 G_M^2 \right] \\
        W_T &= \frac{Q^2}{2E_pE'_p} G_M^2 \\
        W_I &= \frac{{E'_p}^2 \sin^2 \theta_{pq}}{E_pE'_p} \left( \frac{G_E^2 + \tau_N G_M^2}{1+\tau_N} \right) \\
        W_S &= - \frac{E'_p \sin \theta_{pq}}{E_pE'_p} (E_p + E'_p) \left( \frac{G_E^2 + \tau_N G_M^2}{1+\tau_N} \right)
    \end{aligned}
\end{equation}
where $Q^2=-q^2$, $\tau_N = \frac{Q^2}{4M_N^2}$, \thpq is the angle between $\vb{p'}$ and $\vb{q}$, and $G_E$ and $G_M$ are the nucleon electromagnetic form factors. The Kelly parametrization \cite{PhysRevC.70.068202} is used to calculate $G_E^p$, $G_M^p$, and $G_M^n$, while the Riordan parametrization \cite{Riordan:2010id} is used for $G_E^n$, in the following form:
\begin{equation}
\label{eqn:ch4:emffparametrization}
    \begin{aligned}
        G_E^n &= \frac{(1.520\tau_n + 2.629\tau_n^2 + 3.055\tau_n^3)G_D}{1.0+5.222\tau_n+0.040\tau_n^2+11.438\tau_n^3} \\
        G_M^n &= \frac{-1.913 (1.0+2.33\tau_n)}{1.0 + 14.72\tau_n + 24.20\tau_n^2 + 84.1\tau_n^3} \\
        G_E^p &= \frac{1.0-0.24\tau_p}{1.0 + 10.98\tau_p + 12.82\tau_p^2 + 21.97\tau_p^3} \\
        G_M^p &= \frac{2.79 (1.0+0.12\tau_p)}{1.0 + 10.97\tau_p + 18.86\tau_p^2 + 6.55\tau_p^3}
    \end{aligned}
\end{equation}
where $G_D$ is the dipole form factor.

\begin{figure}[h!]
    \centering
    \includegraphics[width=0.7\columnwidth]{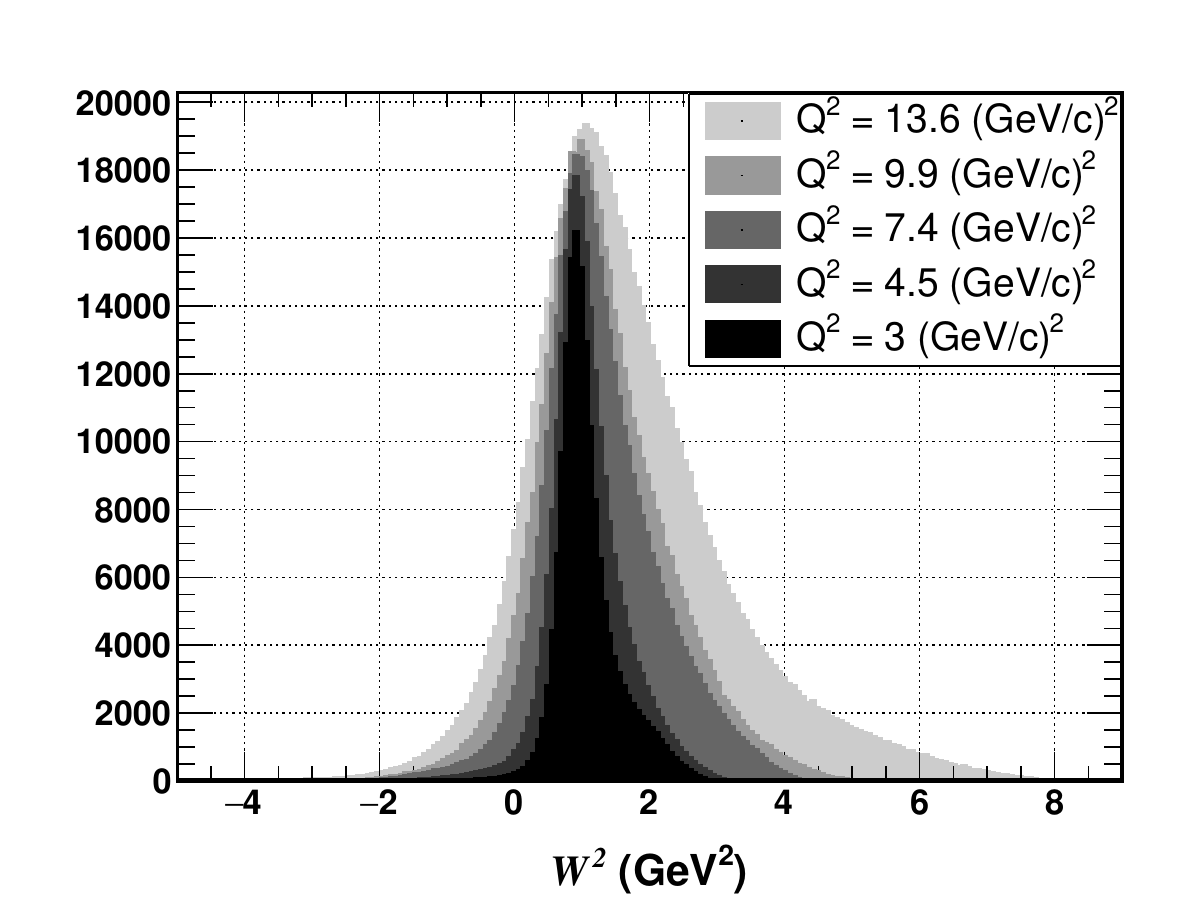}
    \caption{\label{fig:ch4:qegenw2distall} Comparison of \w distribution generated using quasi-elastic event generator across \gmn kinematics. The kinematic broadening of the \w distribution with rising \q is clearly visible.}
\end{figure}
\fig \ref{fig:ch4:qegenw2distall} shows a comparison of \w, the squared invariant mass of the virtual photon-nucleon system, distribution generated using the quasi-elastic event generator across \gmn kinematics. The expected kinematic broadening of the \w distribution with rising \q is clearly increasing in the plot.

\subsubsection{Inelastic Event Generation}
\label{sssec:ineleventgeneration}
Inelastic events are simulated using an event generator based on \gfsbs. Currently, it produces only single pion-nucleon ($\pi N$) final states and does not include radiative or nuclear corrections. The cross-section model relies on empirical fits to inclusive inelastic electron-proton cross-section data \cite{Christy_2010}, as well as electron-deuteron and electron-neutron transverse cross-section data \cite{PhysRevC.77.065206} in the resonance region. This model covers the kinematic range $0 \leq Q^2 < 8$ (GeV/c)$^2$ and $1.1 < W < 3.1$ GeV.

The event generation process for inelastic scattering begins by sampling initial kinematic variables, $k$ and $p$, using a parametrization specific to the target type. This is followed by generating the scattered electron’s polar angle, $\theta_e$, and energy, $E'_e$, from a uniform distribution. Using these values, the momentum transfer $q$, the squared four-momentum transfer \q, the Bjorken scaling variable $x_{bjk}$, and the squared invariant mass \w are calculated.

\begin{figure}[h!]
    \centering
    \begin{subfigure}[b]{0.496\textwidth}
         \centering
         \includegraphics[width=\textwidth]{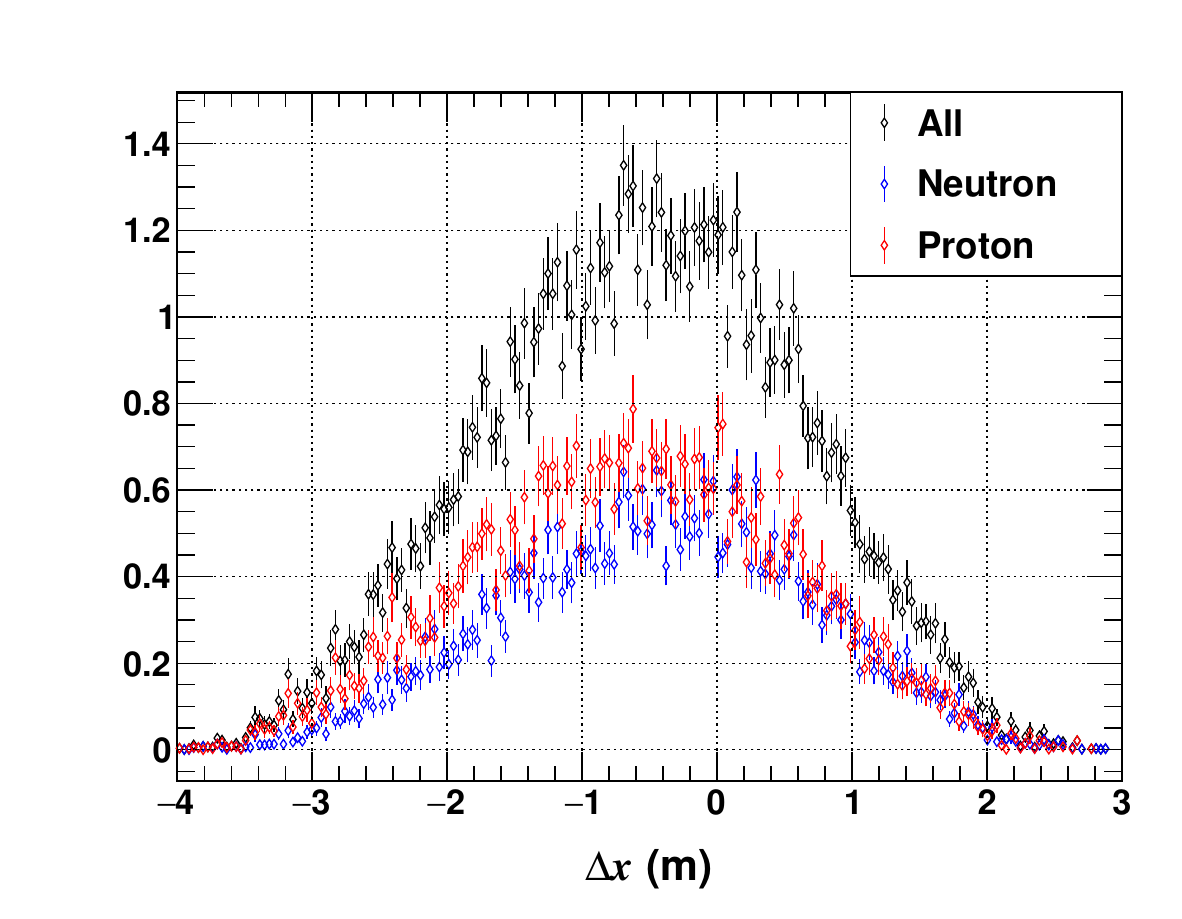}
         \caption{\qeq{3}}
    \end{subfigure}
    \hfill
    \begin{subfigure}[b]{0.496\textwidth}
        \centering
        \includegraphics[width=\textwidth]{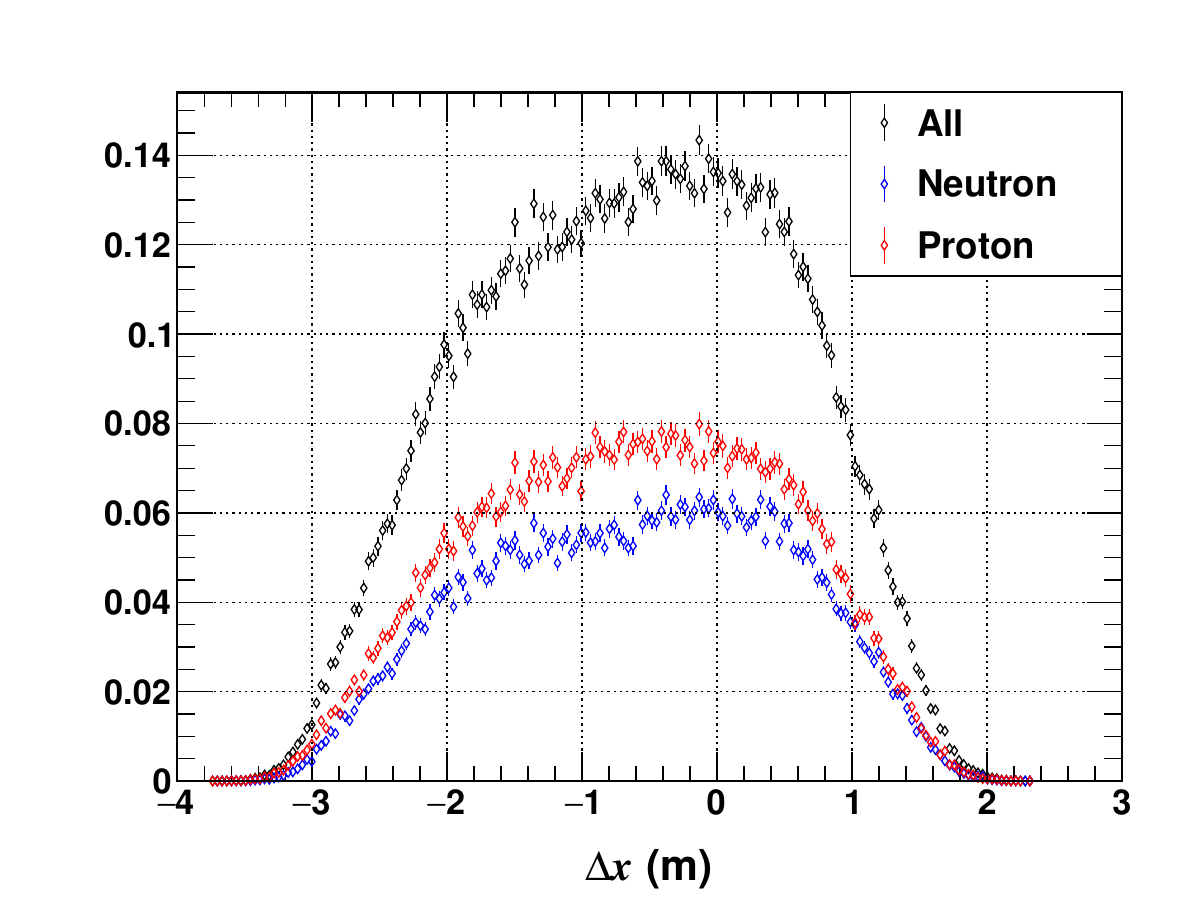}
        \caption{\qeq{9.9}}
    \end{subfigure}
    \caption{\label{fig:ch4:mcinelshape} Shape of the inelastic background in the \dx distribution generated from simulation for (a) the lowest and (b) the second highest \q points. In both cases, the background shape is smooth and flat compared to the quasi-elastic distributions (see Appendix \ref{appen:qeyield}). The simulated events have been digitized and reconstructed to produce these plots.}
\end{figure}
The event proceeds if \w exceeds the single pion production threshold. In such cases, the available energy in the system is determined, and the decay angle is generated isotropically, assuming that the pion and recoiling nucleon are emitted back-to-back in the virtual-photon nucleon center-of-mass (CM) frame. The charge of the final-state nucleon is assigned with a probability of $\frac{2}{3}$ for the same charge as the initial nucleon and $\frac{1}{3}$ for the opposite charge. Once the final nucleon’s charge is established, charge conservation is used to determine the type of the final-state pion.

The cross-section for the event is calculated based on the energies of the incident and scattered electrons and the scattering angle. For neutron cross-sections, the approximation $\sigma_n = 2\sigma_D - \sigma_p$ is applied, where $\sigma_D$ denotes the electron-deuteron scattering cross-section, assumed to represent the average cross section of a free proton and a free neutron.

Examples of \dx and \w distributions generated with the inelastic generator are shown in Figures \ref{fig:ch4:mcinelshape} and \ref{fig:ch4:mccompw2}, respectively. 



\subsection{Comparison to Data: Introducing Ad-Hoc Corrections}
\label{ssec:ch4:dmccomp}
Physics histograms generated from digitized and reconstructed MC events can be directly compared to those obtained from data, as they incorporate both realistic detector and physics effects. This section presents such comparisons for key variables in the electron and hadron arms, obtained by following the analysis flow depicted in \fig \ref{fig:ch4:mcdmcsteps}, to validate the accuracy of the MC.

\subsubsection{\w Distribution}
\begin{figure}[h!]
    \centering
    \includegraphics[width=0.8\columnwidth]{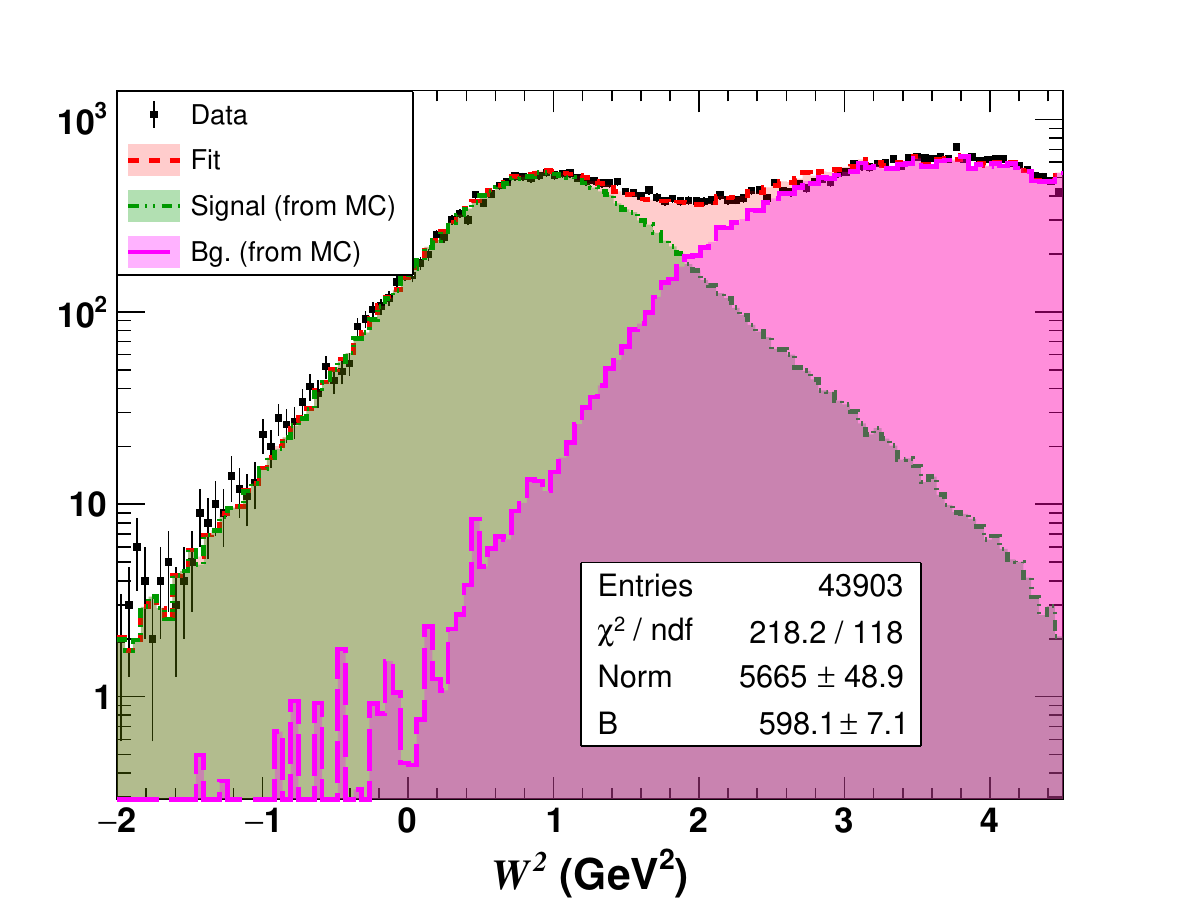}
    \caption{\label{fig:ch4:mccompw2} Data/MC fit to the \w distribution at \qeq{13.6}, using both quasi-elastic signal and inelastic background shapes generated from MC. \ld events selected with very strict \thpq cut have been used.}
\end{figure}

\fig \ref{fig:ch4:mccompw2} shows a data/MC fit to the \w distribution using both quasi-elastic signal and inelastic background generated from MC. \ld events from the highest-\q kinematics were selected with strict \thpq cuts, and identical cuts were applied to the MC-generated events. The resulting fit matches the data closely across the entire range of interest, confirming that the MC event generators realistically produce both quasi-elastic signal and inelastic background shapes.

To achieve the best possible $\chi^2/NDF$, the MC \w distributions were allowed to float horizontally. An overall shift of approximately \SI{0.1}{GeV^2} (see \tab \ref{tab:adhoccorrection}) was observed for the signal distribution. Although this discrepancy is small, it could affect the equivalence of cut regions between data and MC, which is crucial for extracting \rsf, the experimental observable (see \sect \ref{ssec:ch4:rsfintro}). Consequently, these offsets were carefully determined for each kinematic using \heep events, which yields better resolution, and applied as ad-hoc corrections to the MC events. This ensured equivalence during data/MC comparison for the extraction of the physics observable.

\subsubsection{HCAL Cluster Energy}
\begin{figure}[h!]
    \centering
    \begin{subfigure}[b]{0.496\textwidth}
         \centering
         \includegraphics[width=\textwidth]{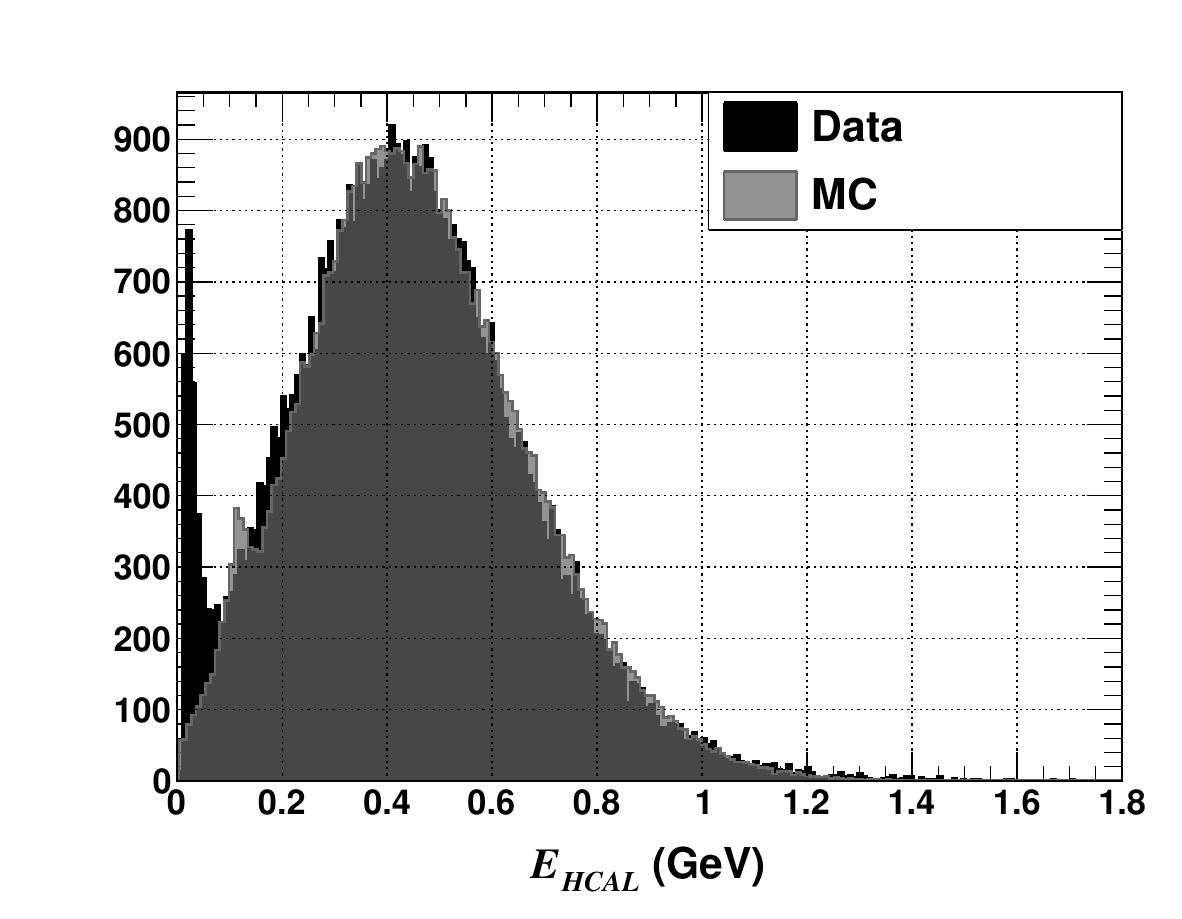}
    \end{subfigure}
    \hfill
    \begin{subfigure}[b]{0.496\textwidth}
        \centering
        \includegraphics[width=\textwidth]{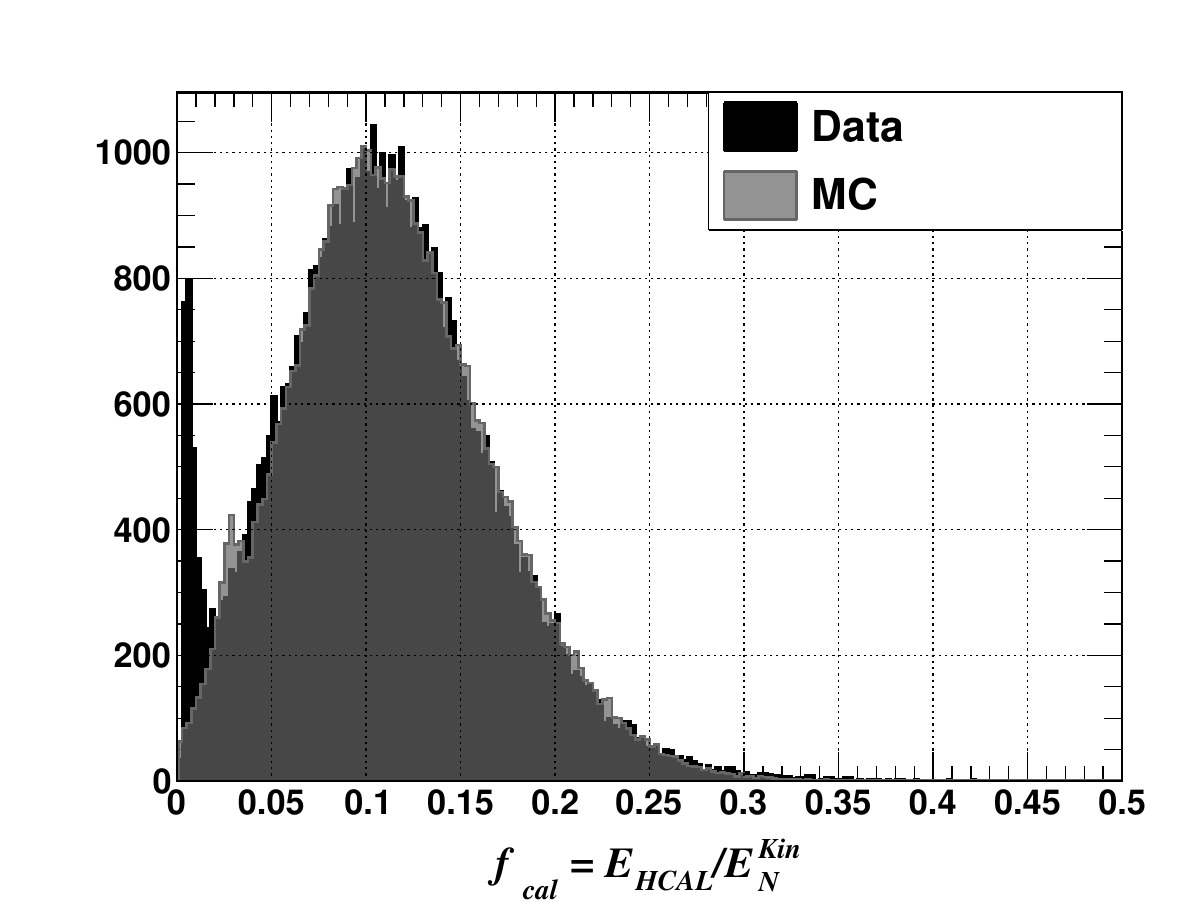}
    \end{subfigure}
    \caption{\label{fig:ch4:mccompehcal} Data/MC comparison of the (left) HCAL cluster energy and (right) energy sampling fraction distributions for \qeq{7.4} kinematics, showing excellent agreement. Only good electron-nucleon events passing strict quasi-elastic selection cuts are shown.}
\end{figure}

MC events effectively reproduce the shape of the HCAL cluster energy and sampling fraction distributions observed in real data across \gmn kinematics, as shown in \fig \ref{fig:ch4:mccompehcal}. Good electron-nucleon coincidence events that passed strict quasi-elastic event selection cuts were used to generate these plots.
%

\subsubsection{HCAL \dx Distribution}
\begin{figure}[h!]
    \centering
    \includegraphics[width=1\columnwidth]{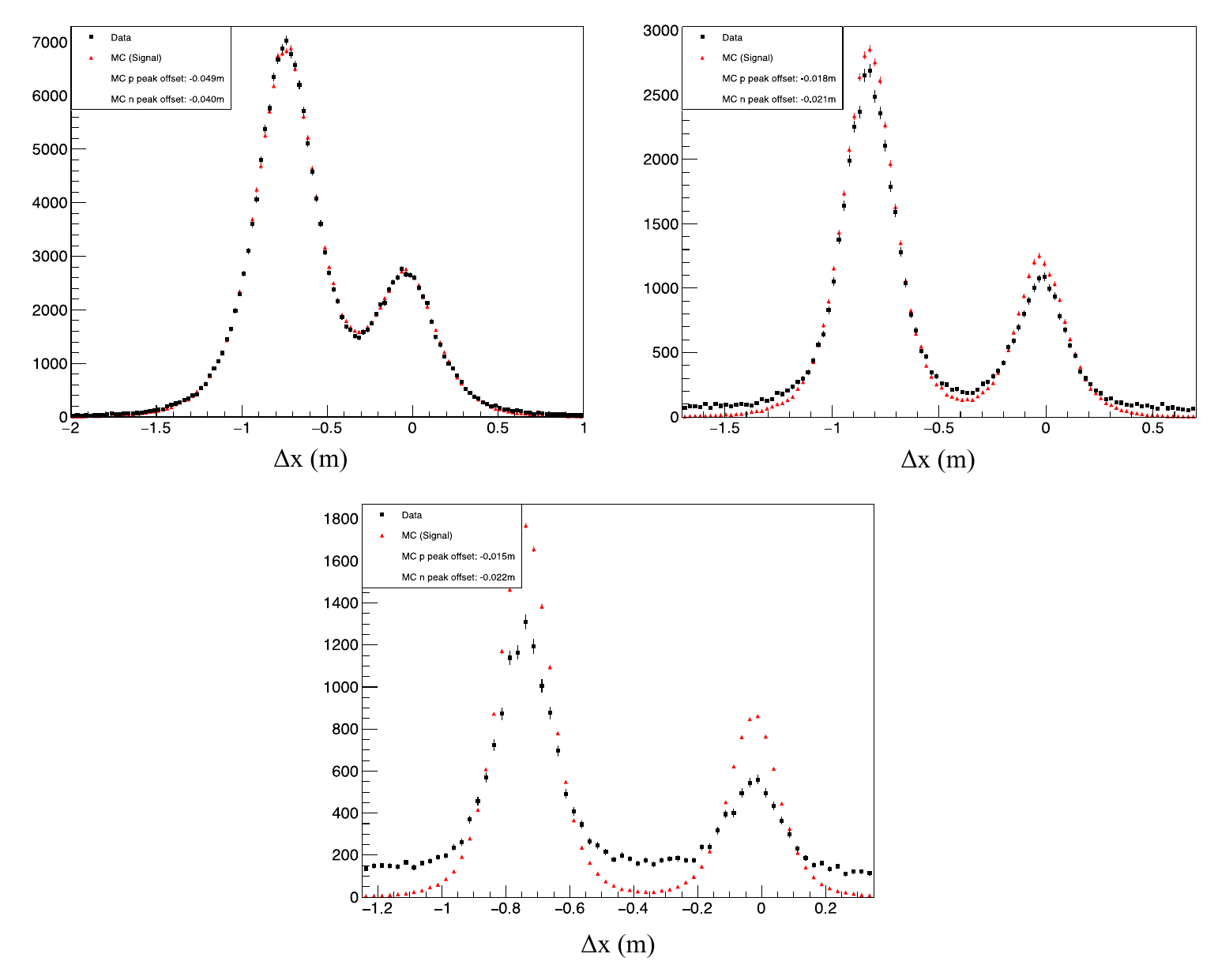}
    \caption{\label{fig:ch4:mccompedx} Data/MC comparison of the HCAL \dx distribution for \qeq{3} (top left), \qeq{7.4} (top right), and \qeq{13.6} (middle) production datasets. Only good electron-nucleon coincidence events passing \w and \dy cuts are shown, with identical cuts applied to both data and MC. The MC signal (red) represents pure quasi-elastic events generated using the SIMC generator. A clear increase in the number of background events passing \w and \dy cuts is observed with rising \q. The MC \deen and \deep signal distributions have been shifted in X by the amounts indicated to align with the data. The net shift between the \deen and \deep distributions is within a few millimeters in each case, which is well within the HCAL’s position resolution.}
\end{figure}

\fig \ref{fig:ch4:mccompedx} presents the data/MC comparison of the HCAL \dx distribution using production data from several \gmn kinematics. The MC signal represents pure quasi-elastic events generated with the SIMC generator, and identical analysis cuts, including \w and \dy, have been applied to both data and MC. At the lowest-\q, where minimal background events survive these cuts, the comparison appears nearly perfect. However, at higher-\q, adding a background model to the MC signal becomes necessary to achieve a similar level of comparison. 

While the distribution of \deen events from MC is well aligned to zero, the corresponding data shows a slight offset from zero in all cases, primarily due to calibration uncertainties. Although these offsets are within HCAL's position resolution, they can influence the data/MC fit to the \dx distribution used for extracting the physics observable from \gmn, as discussed in \sect \ref{ssec:ch4:rsfintro}. Consequently, these offsets are carefully evaluated and applied to the MC \dx distribution for data/MC comparison for each kinematic setting. Additionally, the SBS field strength in MC for each setting is fine-tuned to closely match the proton deflection observed in the data.

\subsubsection{\dx $-$ \xhob Correlation: Effective HCAL Distance}
\begin{figure}[h!]
     \centering
     \begin{subfigure}[b]{1\textwidth} 
         \centering
         \includegraphics[width=\textwidth]{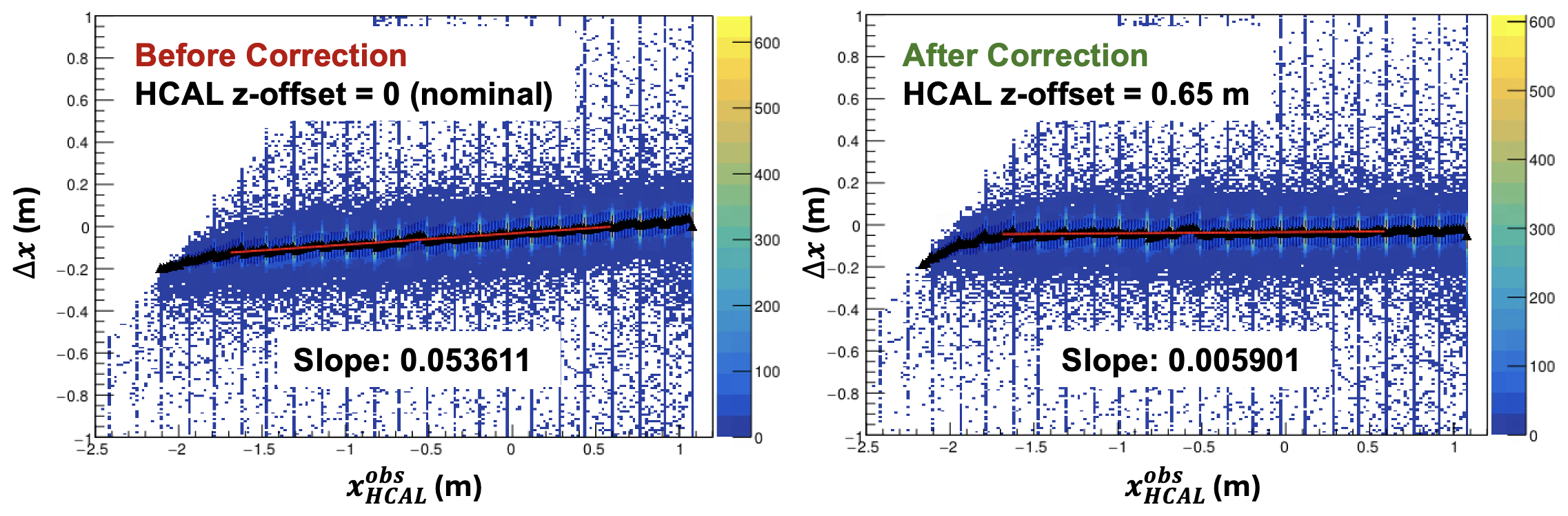}
         \caption{Data}
     \end{subfigure}
     \begin{subfigure}[b]{1\textwidth}
         \centering
         \includegraphics[width=\textwidth]{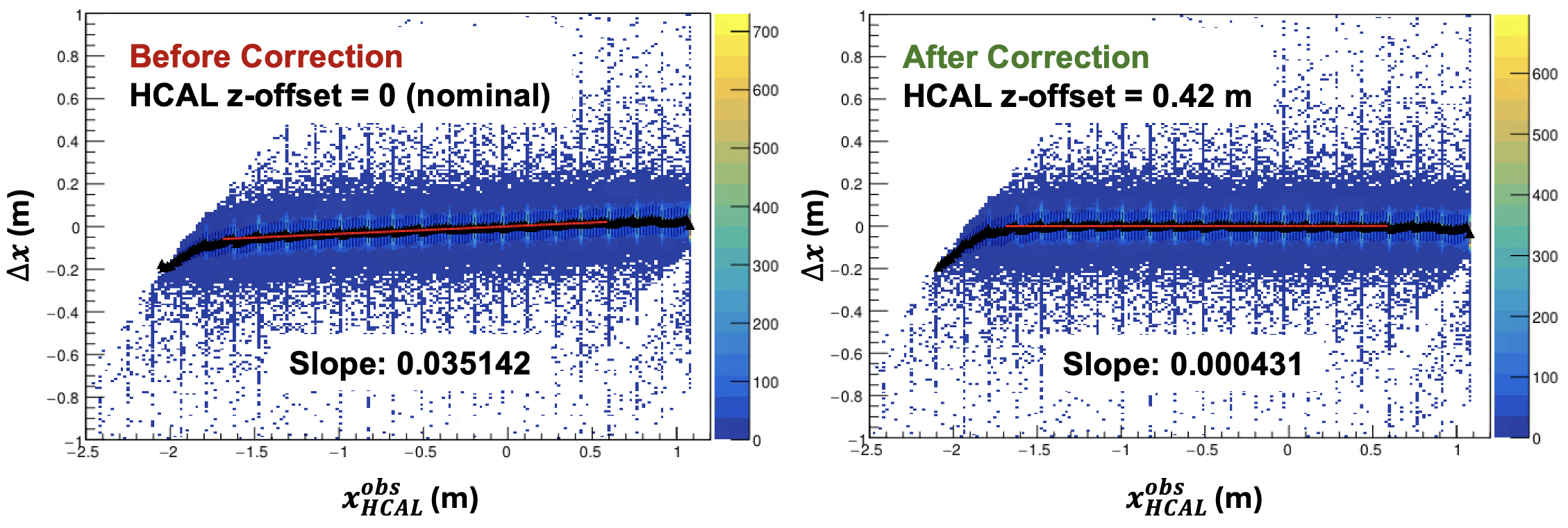}
         \caption{MC}
     \end{subfigure}
     \caption{Correlation between \dx and \xhob, the dispersive component of the observed nucleon position at HCAL, from both data and MC. Elastic \heep events from \qeq{3} SBS zero field data were used to generate this plot. The applied HCAL z-offset effectively eliminates the correlation in both data and MC.} 
     \label{fig:ch4:hcalzoff}
\end{figure}

\dx, calculated using the nominal target-to-HCAL distance listed in \tab \ref{tab:sbsconfig}, appears positively correlated with the dispersive component of the observed nucleon position at HCAL (\xhob) for both data and MC, as shown in \fig \ref{fig:ch4:hcalzoff}. This correlation likely arises because the depth of the HCAL cluster centroid increases with the incoming nucleon's angle of incidence. The correlation is consistent across all kinematics, with its magnitude increasing as nucleon momentum rises, further supporting this explanation.

Correcting this correlation is essential to eliminate position-dependent bias in calculating the \dx variable. This is done by introducing an ad-hoc offset to HCAL's z-coordinate, effectively increasing its distance from the target. These kinematic-dependent correction factors are carefully evaluated for both data and MC using elastic \heep events, with SBS zero field data utilized when available.
\subsubsection{Summary of Ad-Hoc Corrections}
\tab \ref{tab:adhoccorrection} summarizes the ad-hoc correction factors discussed in this section, used to address calibration uncertainties and optimize agreement between data and MC. Detailed discussions on the HCAL nucleon detection efficiency comparison between data and MC, as well as the data/MC fit to the \dx distribution for experimental observable extraction using these corrections, are provided in the following sections.

\begin{table}[h!]
    \centering
    \caption{Summary of ad-hoc correction factors used to address calibration uncertainties and achieve optimal agreement between MC and data. $z_{HCAL}$ ($y_{HCAL}$) denotes the longitudinal (transverse) component of the HCAL origin in the Hall Csys, with the remaining variables representing their usual meanings. The offsets shown are added to the respective variables. All offsets, except those explicitly mentioned, are applied only to MC events.}
    \label{tab:adhoccorrection}
    \begin{tabular}{cccccccc} \hline\hline\vspace{-1em} \\
\multirow{3}{*}{\q} & \multirow{3}{*}{\ep} & \multicolumn{6}{c}{Ad-Hoc Offset for} \\ \cline{3-8} \vspace{-1em} \\
         &  & \multirow{2}{*}{\w (GeV$^2$)} & \multicolumn{2}{c}{$z_{HCAL}$ (cm)} & \multirow{2}{*}{$y_{HCAL}$ (cm)} & \multicolumn{2}{c}{\dx (cm)} \\ \vspace{-0.95em} \\ \cline{4-5} \cline{7-8} \vspace{-0.95em} \\
         &  &  & Data & MC &  & \deen & \deep \vspace{0.2em} \\ \hline \vspace{-1em} \\
        $3$    & $0.72$ & $-0.018$ & $65$   & $42$ & $3$  & $4$   & $4.9$\\
        $4.5$  & $0.51$ & $-0.058$ & $70$   & $41$ & $-6$ & $1.8$ & $4$\\
        $7.4$  & $0.46$ & $-0.076$ & $113$  & $48$ & $3$  & $2.1$ & $1.8$\\
        $9.9$  & $0.50$ & $-0.060$ & $92.5$ & $55$ & $2$  & $2.5$ & $2.5$\\
        $13.6$ & $0.41$ & $-0.109$ & $127$  & $59$ & $3$  & $2.2$ & $1.5$\\
        \hline\hline
    \end{tabular}
\end{table}

\section{HCAL Detection Efficiency}
\label{sec:ch4:hcalnde}
The measurement of quasi-elastic electron-neutron (\deen) to electron-proton (\deep) scattering cross-section ratio \rqe, which is the direct observable of \gmn, relies heavily on the hadron calorimeter's (HCAL) ability to detect both nucleons with high and comparable efficiencies. Any non-uniformity in the relative detection efficiencies could alter the counts of the corresponding scattering events leading to systematic uncertainty in \rqe. Addressing this uncertainty requires a thorough evaluation of HCAL's nucleon detection efficiency (NDE) and its uniformity across the acceptance, as detailed in this section.

The HCAL NDE is defined as:
\begin{equation}
\label{eqn:ch4:hcalnde}
    \epsilon^{p(n)}_{HCAL} = \frac{N^{det}_{HCAL}}{N^{exp}_{HCAL}} = \frac{\text{Number of proton (neutron) events detected by HCAL}}{\text{Number of proton (neutron) events expected to hit HCAL}},
\end{equation}
where $\epsilon^{p(n)}$ represents the proton (neutron) detection efficiency. Since the expected and detected events are drawn from the same underlying distribution, the statistical error on the efficiency is calculated using the binomial method:
\begin{equation}
\label{eqn:ch4:hcalndeerror}
    \sigma_{\epsilon_{HCAL}} = \left[\frac{\epsilon_{HCAL}(1 - \epsilon_{HCAL})}{N^{exp}_{HCAL}}\right]^{\frac{1}{2}}
\end{equation}
In this section, different methods for HCAL NDE estimation will be presented. However, the definition of NDE and the statistical error calculation method, as discussed above, are consistent across all approaches.
\subsection{``True" NDE from MC}
The realistic HCAL simulation (see \sect \ref{sec:ch4:mcsimulation}) was used to estimate its design-based ``true" NDE across the relevant nucleon momentum range. The method involved simulating nucleons with known energies and angles within HCAL’s full active area and then determining both the expected and detected event counts. 

\begin{figure}[h!]
    \centering
    \includegraphics[width=\sfig]{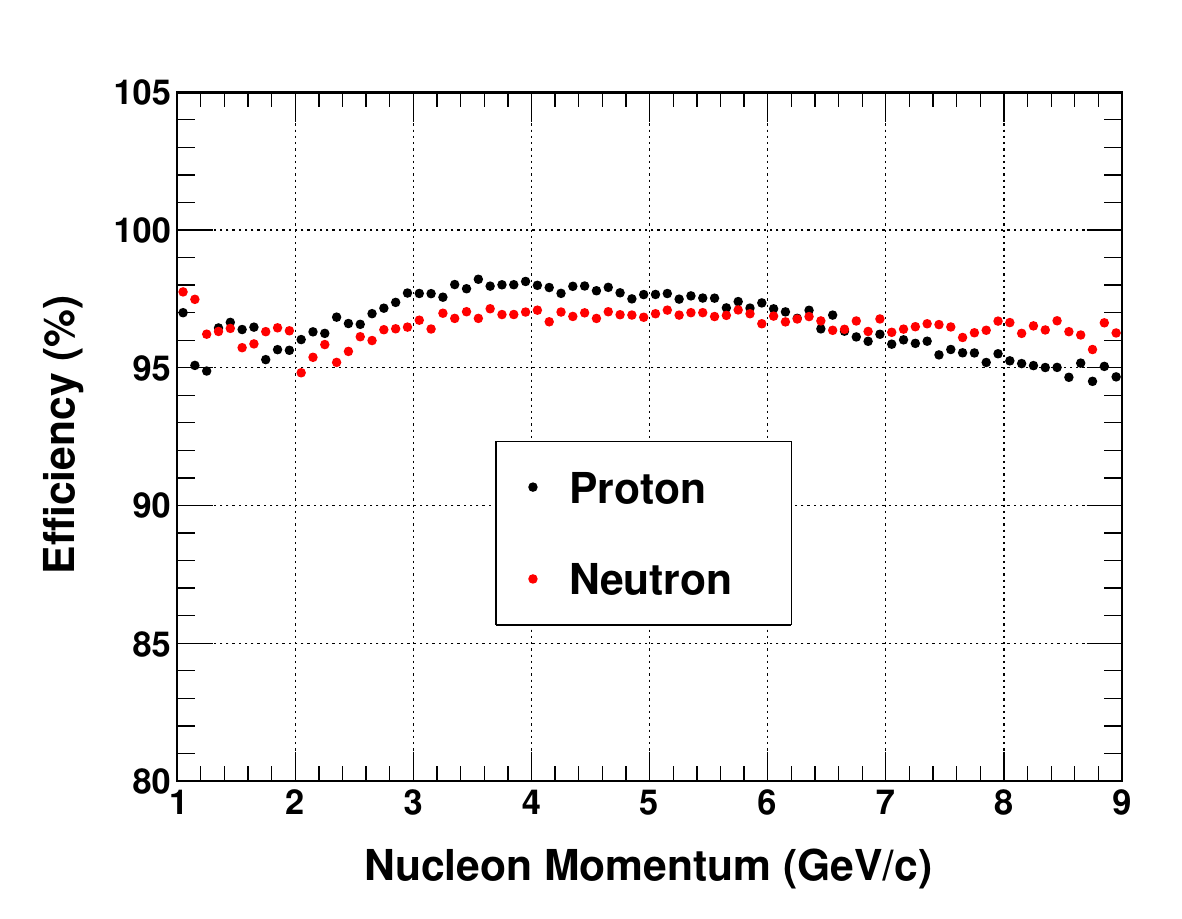}
    \caption{\label{fig:ch4:hcalndetrue1} HCAL ``true" nucleon detection efficiency (NDE) as a function of nucleon momentum, estimated from simulation.}
\end{figure}
Events generated by the \gfsbs ``particle gun" generator were used for this study. This generic event generator can produce various particle types, including nucleons, with user-specified energy and angular limits. The generation limits were carefully chosen to populate HCAL’s entire active area while ensuring nucleon tracks remained within the SBS dipole magnet acceptance to avoid false negatives. Additionally, the SBS field was turned off to match the acceptance for both nucleons. The simulated events were digitized and reconstructed to account for detector resolution and uncertainties in event reconstruction.

%
\begin{figure}[h!]
    \centering
    \includegraphics[width=1\columnwidth]{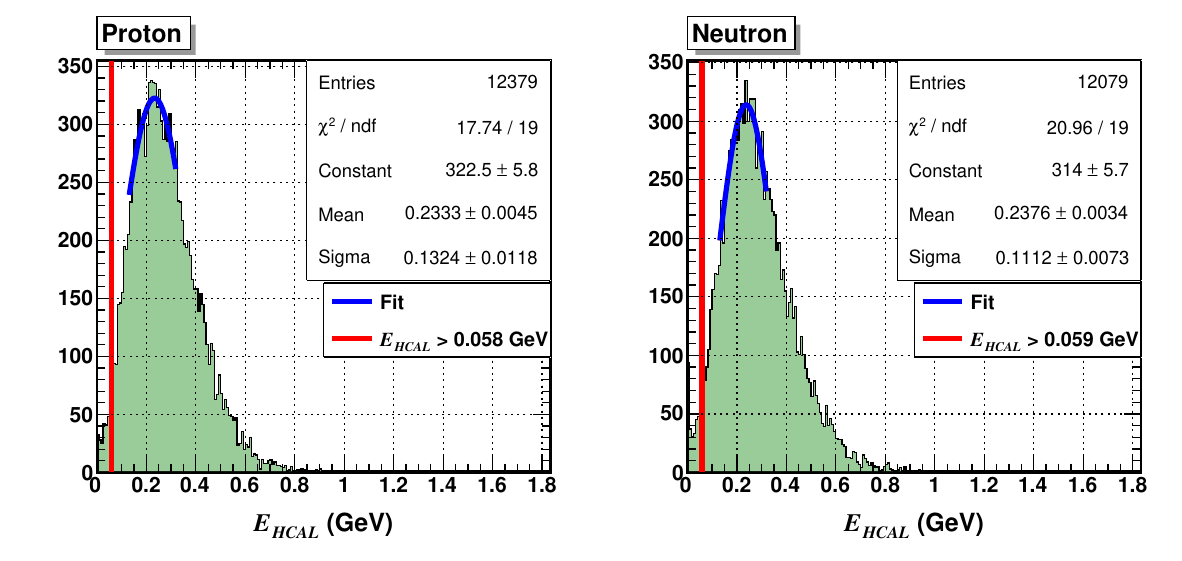}
    \caption{\label{fig:ch4:hcalndetrue2} Digitized and reconstructed HCAL energy distributions for proton and neutron events generated by the \gfsbs ``particle gun" generator for a single momentum slice, highlighting the threshold value used in the ``true" HCAL nucleon detection efficiency analysis.}
\end{figure}
Only events expected to land within HCAL’s active area\footnote{The active area of HCAL excludes the outermost rows and columns.} were analyzed. An event was considered detected or missed based on its energy deposition. An energy threshold ($E_{Th}$) of $E_{HCAL}^{peak}/4$, where $E_{HCAL}^{peak}$ is the most probable energy deposition by nucleons for a given momentum range, was applied. The total event count in the energy distribution formed the denominator of \eqn \ref{eqn:ch4:hcalnde}, while events with energy above $E_{Th}$ formed the numerator, yielding the corresponding detection efficiency. \fig \ref{fig:ch4:hcalndetrue1} shows the resulting HCAL proton and neutron detection efficiencies as functions of momentum. The momentum range covers the entire scope of \gmn (see \tab \ref{tab:sbsconfig}). The detection efficiencies for both protons and neutrons are consistently high and comparable across this range, aligning with HCAL’s design expectations.

It is important to note that the absolute efficiency value depends on the choice of $E_{Th}$. The threshold $E_{Th}=E_{HCAL}^{peak}/4$ was chosen as an educated guess based on the shape of the HCAL energy distribution, which is non-Gaussian, as shown in Figure \ref{fig:ch4:hcalndetrue2}. Raising $E_{Th}$ by a factor of two reduces the NDE by approximately $3\%$, though the relative NDE between neutrons and protons remains nearly unchanged.
\subsection{Proton DE: Data/MC Comparison}
\label{ssec:ch4:hcalprotonde}
The ``true” HCAL NDE obtained from MC, which aligns with design expectations, is reassuring; however, it is essential to validate these results against the NDE observed in real data. Elastic events selected from \lh data can be used to unambiguously select proton events in the absence of nuclear effects, providing a basis for estimating the corresponding detection efficiency, which is the focus of this section. In contrast, directly measuring neutron detection efficiency is more complex due to the absence of a free neutron target, as discussed later in Section \ref{ssec:ch4:hcalneutronde}.  

\subheading{Procedure}
To compare the proton detection efficiencies (pDE) between real data and MC, a straightforward ``cut-based” analysis was employed. In this approach, elastic \heep events were selected using strict event selection criteria to calculate pDE as a function of the expected proton positions in the dispersive (\xhexp)\footnote{$[]^p$ denotes calculations made using the proton hypothesis, which accounts for the deflection of proton tracks by the SBS field in each event. Therefore, \begin{equation*}    
{\left[x_{HCAL}^{exp}\right]}^p = x_{HCAL}^{exp} - \delta x_{SBS},\end{equation*} where $\delta x_{SBS}$ represents the event-by-event deflection of the proton track by the SBS magnet (see \eqn \ref{eqn:ch4:protondeflection}). Naturally, $\delta x_{SBS}$ is zero for neutrons.} and transverse (\yhex) directions. An weighted average of the efficiencies in each direction was then computed to determine the acceptance-averaged pDE. The analysis procedure, kept consistent between data and MC, followed these steps:

\begin{enumerate}
    \item \lh data from a given \gmn kinematic setting recorded with the same SBS field strength was used. MC events were generated with matching kinematic variables and SBS field strength.
    \item Strict track quality cuts and PID cuts, as described in \sect \ref{sec:ch4:evselect}, were applied to select good electron events. Identical cut ranges were used for both data and MC.
    \item A very strict cut on \w was applied to select a region with minimal inelastic contamination, eliminating the need for statistical background subtraction to a good approximation. The same cut was applied to MC events, even though they are purely elastic, to maintain consistency. 
    \item The total number of events in each slice of \xhexp (\yhex) was selected from the chosen elastic events, with a fiducial cut on \yhex (\xhexp) to include only those expected to land within the HCAL fiducial region for efficiency estimation, avoiding bias from false negatives. These counts represent the events expected to hit HCAL.  
    \item To determine the number of detected events for each slice of \xhexp and \yhex, a sufficiently loose cut on \thpq, calculated using the proton hypothesis, was applied.
    \item The ratio of detected to expected events for each \xhexp and \yhex slice yielded the corresponding pDE (see \eqn \ref{eqn:ch4:hcalnde}).  
\end{enumerate} 

It’s important to note that the \gfsbs particle gun generator could not be used for this purpose, as it does not replicate the event distribution found in real data. Instead, \heep events generated using the \simc elastic event generator, which includes realistic radiative corrections as described in \sect \ref{sssec:qeeventgeneration}, were used. The simulated events were digitized and reconstructed using the same procedures applied to real data to account for realistic detector effects. 

\subheading{Implementation}
The "cut-based" method relies on the assumption that inelastic contamination within the tight \w cut region is statistically negligible, which becomes less valid at higher \q. Consequently, this method is expected to yield reliable pDE estimates only for low-\q data points, specifically \qeq{3\,\,\&\,\,4.5}. Analysis performed on the lowest-\q dataset recorded with $30\%$ SBS field strength is discussed in detail below.

\begin{figure}[h!]
    \centering
    \begin{subfigure}[b]{1\textwidth}
         \centering
         \includegraphics[width=\textwidth]{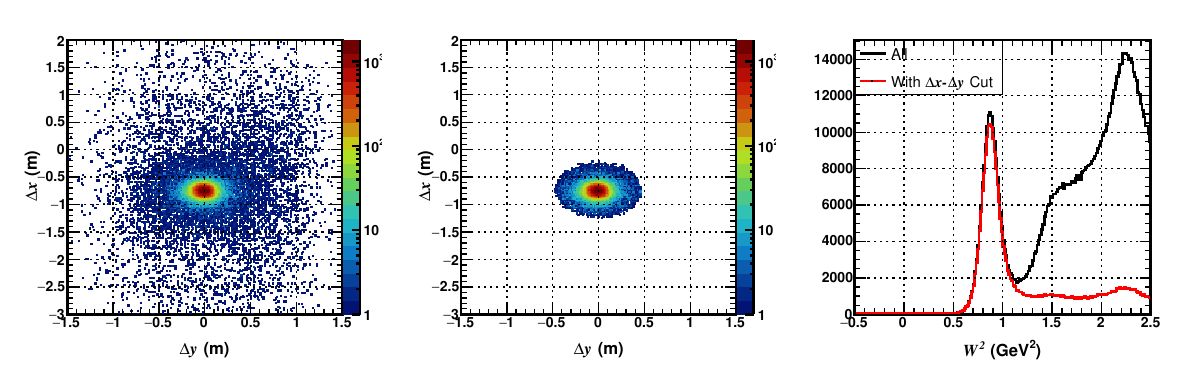}
         \caption{Data}
    \end{subfigure}
    \hfill
    \begin{subfigure}[b]{1\textwidth}
        \centering
        \includegraphics[width=\textwidth]{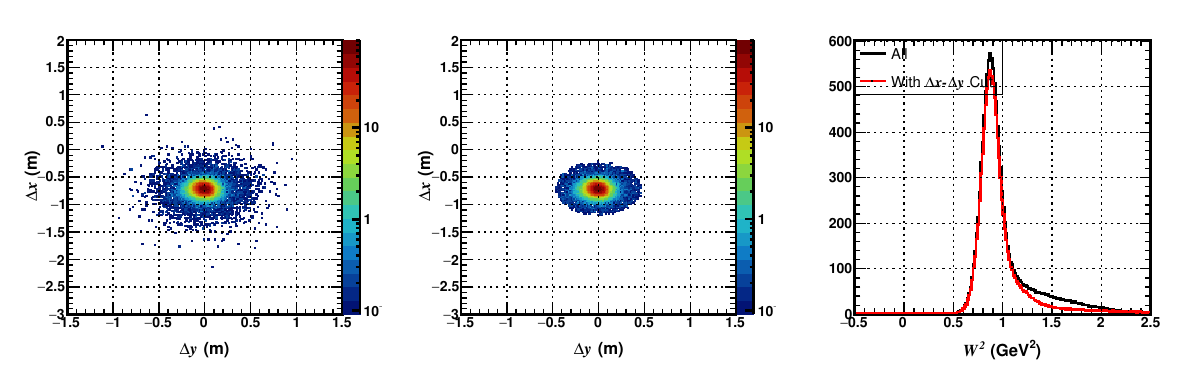}
        \caption{MC}
    \end{subfigure}
    \caption{\label{fig:ch4:hcalndedatamccut} Relevant analysis cut regions for the proton detection efficiency comparison between data and MC for \qeq{3} SBS $30\%$ field settings. The leftmost column shows \dx-\dy correlation plots without a \thpq cut, while the middle column includes the \thpq cut. In the rightmost column, \w distributions are displayed with and without the \thpq cut.}
\end{figure}
The following electron arm cuts were used to determine the number of events expected to hit HCAL (refer to \sect \ref{sec:ch4:evselect} for a detailed description of the cut variables):
\begin{itemize}
    \item Good electron event selection cuts
    \begin{itemize}[label=$\circ$]
        \item Minimum number of GEM layers with hits $>3$.
        \item Track $\chi^2/NDF<15$
        \item $|v_z|<0.065$ m
        \item $|x_{BB}|<0.25$ m
        \item $E_{PS}>0.2$ GeV
    \end{itemize}
    \item $|W^2-0.8|<0.25$ GeV$^2$
    \item Fiducial cuts
    \begin{itemize}[label=$\circ$]
        \item $-2.4<$ \xhexp $<0.9$ m
        \item $-0.5<$ \yhex $<0.5$ m
    \end{itemize}
\end{itemize}

Additional HCAL-specific cuts were applied to determine the number of events detected by HCAL:
\begin{itemize}
    \item $E_{HCAL}>0$
    \item $\theta_{pq}<0.04$ rad
\end{itemize}

The \w and \thpq cut regions from both data and MC are shown in \fig \ref{fig:ch4:hcalndedatamccut} for reference.

\begin{figure}[h!]
    \centering
    \begin{subfigure}[b]{0.8\textwidth}
         \centering
         \includegraphics[width=\textwidth]{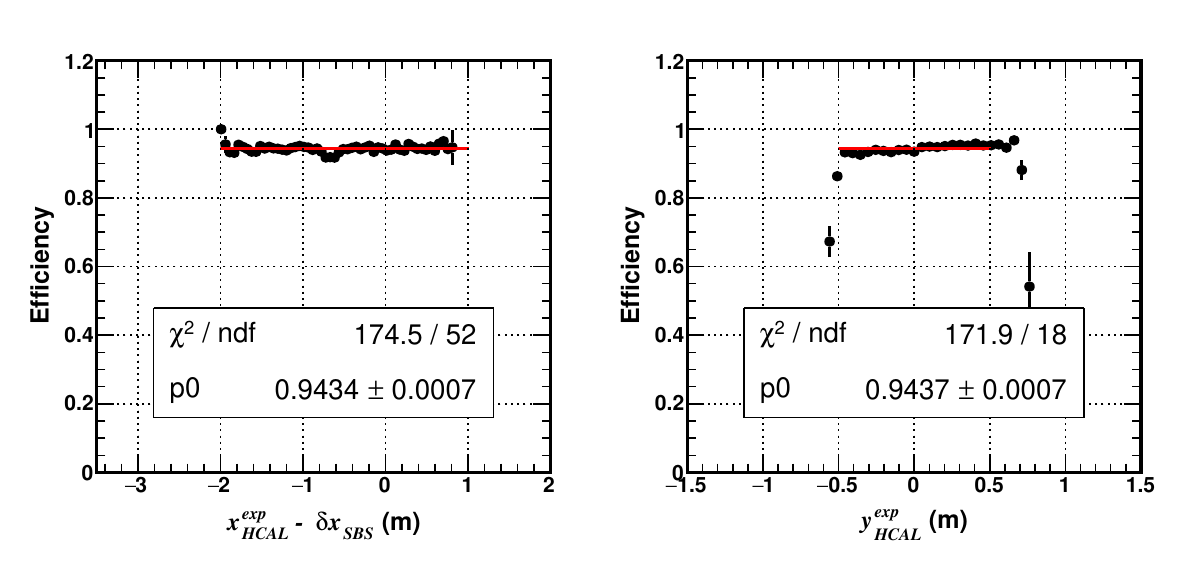}
         \caption{\label{sfig:ch4:hcalndexyeffiwocorrdata}Data}
    \end{subfigure}
    \hfill
    \begin{subfigure}[b]{0.8\textwidth}
        \centering
        \includegraphics[width=\textwidth]{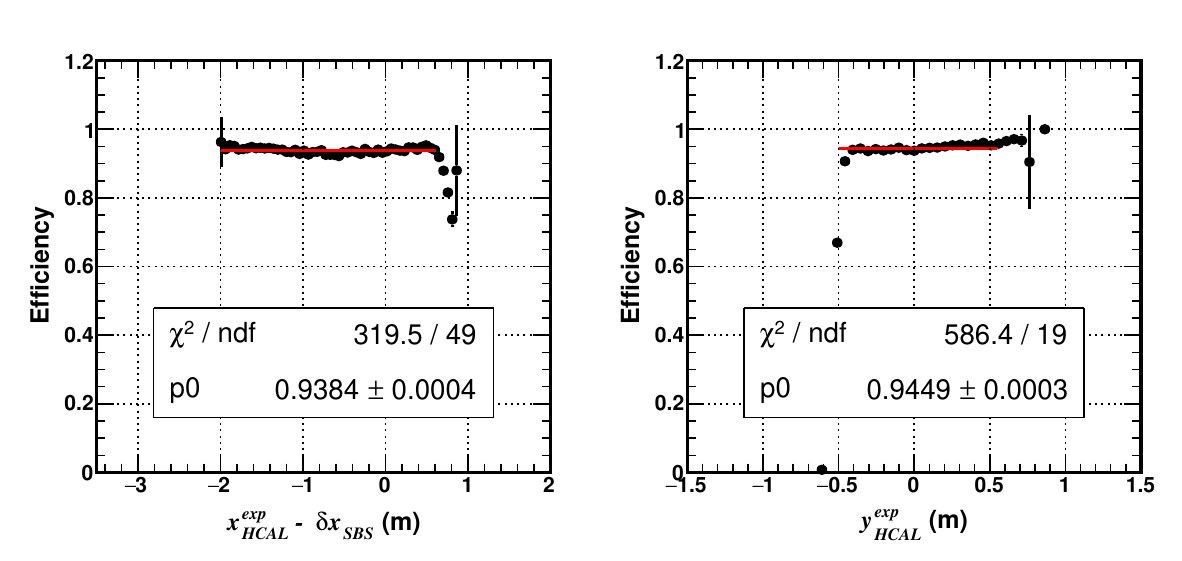}
        \caption{MC}
    \end{subfigure}
    \caption{\label{fig:ch4:hcalndexyeffiwocorr} HCAL proton detection efficiency (pDE) profile comparison between data and MC, showing excellent agreement. The plots in the left column show the pDE profile in the dispersive direction, while those in the right column show it in the transverse direction. There is a slight indication of non-uniform efficiency around -0.6 m in the dispersive direction for the data, which is absent in the MC.}
\end{figure}
\subheading{Results}
The resulting pDEs in the dispersive ($x$) and transverse ($y$) directions from both data and MC are shown in \fig \ref{fig:ch4:hcalndexyeffiwocorr}, leading to the following observations:
\begin{itemize}
    \item The acceptance-averaged pDE values between data and MC agree within $0.7\%$, which is highly encouraging. Such a small discrepancy is expected given that the simulation is realistic but not exact.
    \item The acceptance-averaged pDE values in the $x$ and $y$ directions align closely for both data and MC, as expected.
    \item The trend in the $y$ direction matches almost perfectly between data and MC.
    \item In the $x$ direction, a dip in efficiency around \xhexp $= -0.6$ m is observed only in real data. This dip, occurring near the center of HCAL's acceptance, suggests position-dependent non-uniformity in HCAL detection efficiency, which could introduce bias in the measurement of \rqe. Addressing this non-uniformity is critical and will be discussed in the next section.
\end{itemize}

The strong overall agreement of pDE estimates between data and MC in both the dispersive and transverse directions indicates that the MC provides a realistic simulation of HCAL detection efficiencies.

\subsection{Addressing Non-Uniformity: Efficiency Map}
\label{ssec:ch4:hcalefficiencymap}
\begin{figure}[h!]
    \centering
    \begin{subfigure}[b]{0.7\textwidth}
         \centering
         \includegraphics[width=\textwidth]{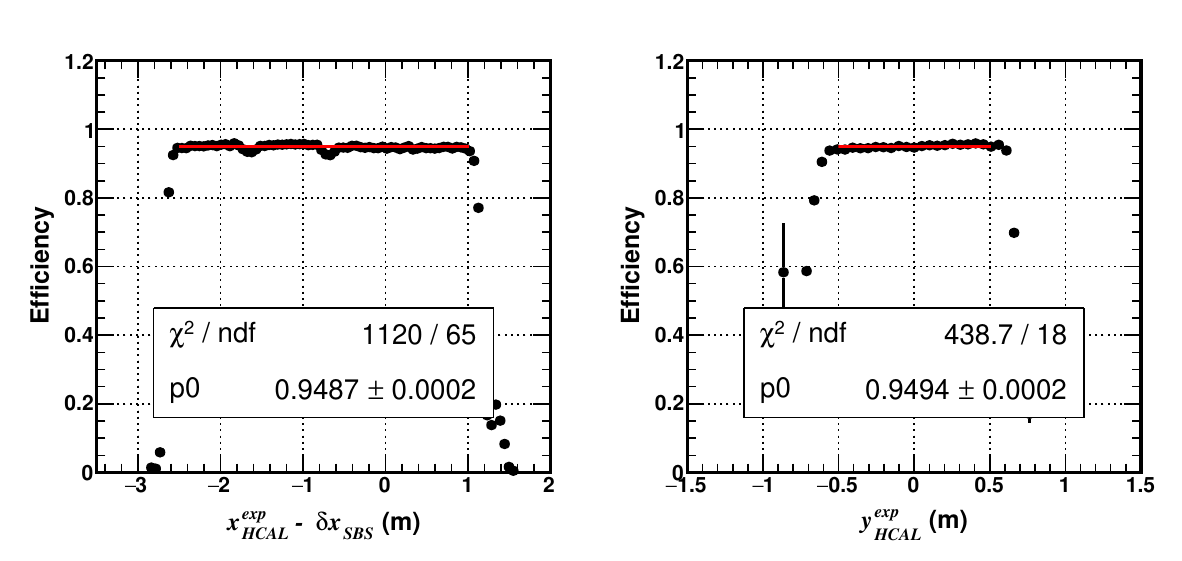}
         \caption{}
    \end{subfigure}
    \hfill
    \begin{subfigure}[b]{0.7\textwidth}
        \centering
        \includegraphics[width=\textwidth]{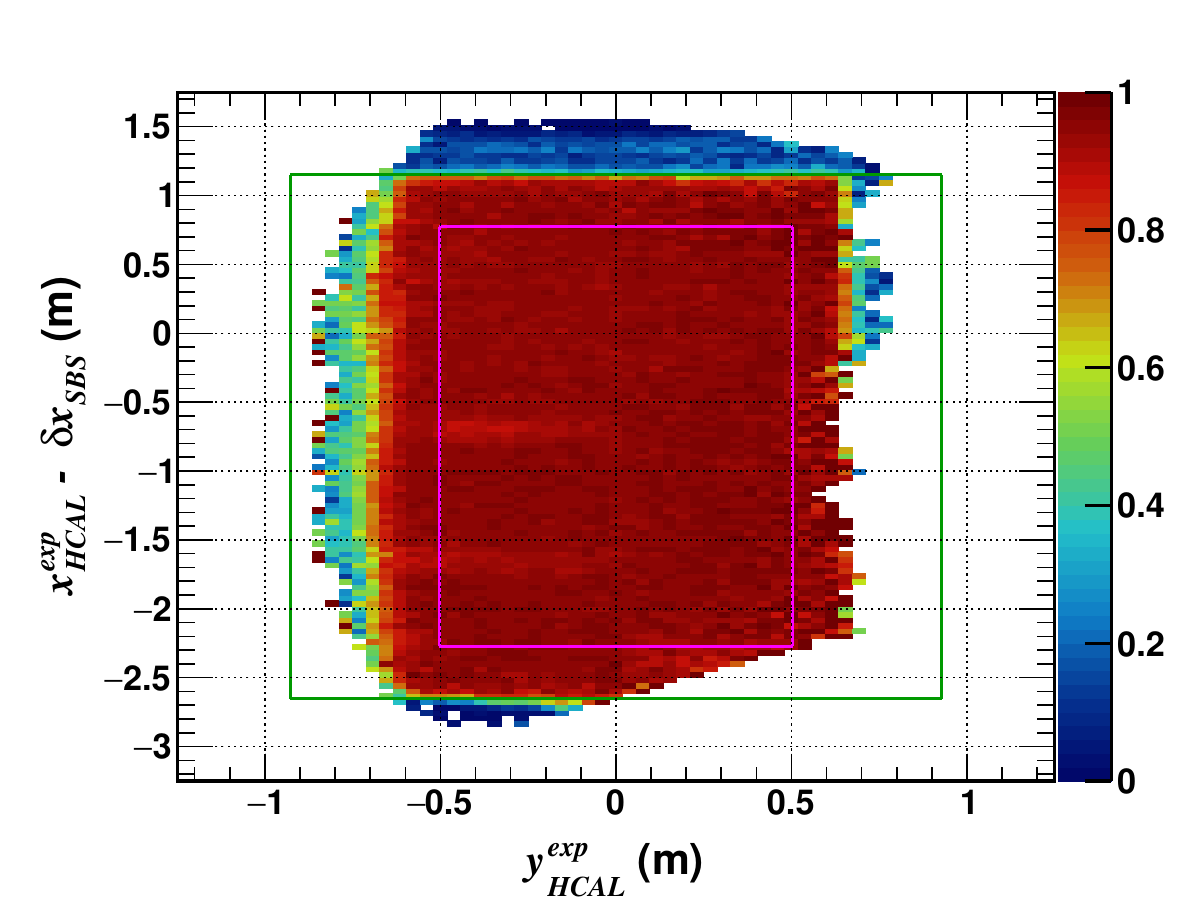}
        \caption{}
    \end{subfigure}
    \caption{\label{fig:ch4:hcalndeeffimap} HCAL proton detection efficiency (pDE) results from a combined analysis of \lh data recorded with four different SBS field strengths ($0\%$, $50\%$, $70\%$, and $100\%$) at \qeq{4.5} high-\ep kinematics. (a) Efficiency profile in both dispersive and transverse directions, with a larger HCAL acceptance range revealing an additional energy dip around -1.7 m in the dispersive direction. (b) 2D efficiency map, effectively capturing the efficiency profile across the HCAL active area. The green rectangle represents the HCAL physical area, while the magenta rectangle shows an example of a fiducial region typically selected for analysis.}
\end{figure}

The position-dependent non-uniformity in the dispersive direction, observed in the proton detection efficiency (pDE) estimation from the lowest-\q dataset (see \fig \ref{sfig:ch4:hcalndexyeffiwocorrdata}), was confirmed through independent analyses using data recorded with different SBS field settings at the same \q and at \qeq{4.5}. These analyses, covering a broader HCAL acceptance than the \qeq{3} SBS $30\%$ field dataset, also revealed another efficiency dip near the top of HCAL at around $-1.7$ m, as shown in \fig \ref{fig:ch4:hcalndeeffimap}. Correcting these non-uniformities is crucial to avoid bias in the measurement of \rqe, the direct observable of \gmn. 

Since $eN$ correlation varies with SBS field strength, the consistent efficiency profiles observed across different field settings at a given kinematic configuration rule out any influence from non-uniformities in the electron arm. This reinforces that the efficiency dips in HCAL are likely caused by hardware issues. Further investigation confirmed this, revealing inefficient modules associated with the channels located in the problematic regions.

\subheading{Possible Approaches to Handle NDE Non-Uniformity}
Although the issue appears to be hardware-related, several approaches can be taken to address it at the software level, as outlined below:
\begin{itemize}
    \item Improve HCAL energy calibration to minimize the non-uniformity observed in real data. This solution directly targets the problem at its source and is ideal; however, significant improvements beyond the second pass of fine-tuning may be difficult to achieve.
    \item Model the efficiency loss mechanism within the MC. One possible method is to reduce the gains of the affected HCAL channels during the digitization of simulated events. However, accurately mapping the problematic channels and their gain variations from real data is challenging, potentially leading to overcorrection or undercorrection of the efficiencies.
    \item Apply position-dependent efficiency corrections to MC events before comparing them to data. This approach is the most balanced in terms of simplicity, correctness, and feasibility for implementation and has been adopted in this analysis.
\end{itemize}

\begin{figure}[h!]
    \centering
    \includegraphics[width=0.8\columnwidth]{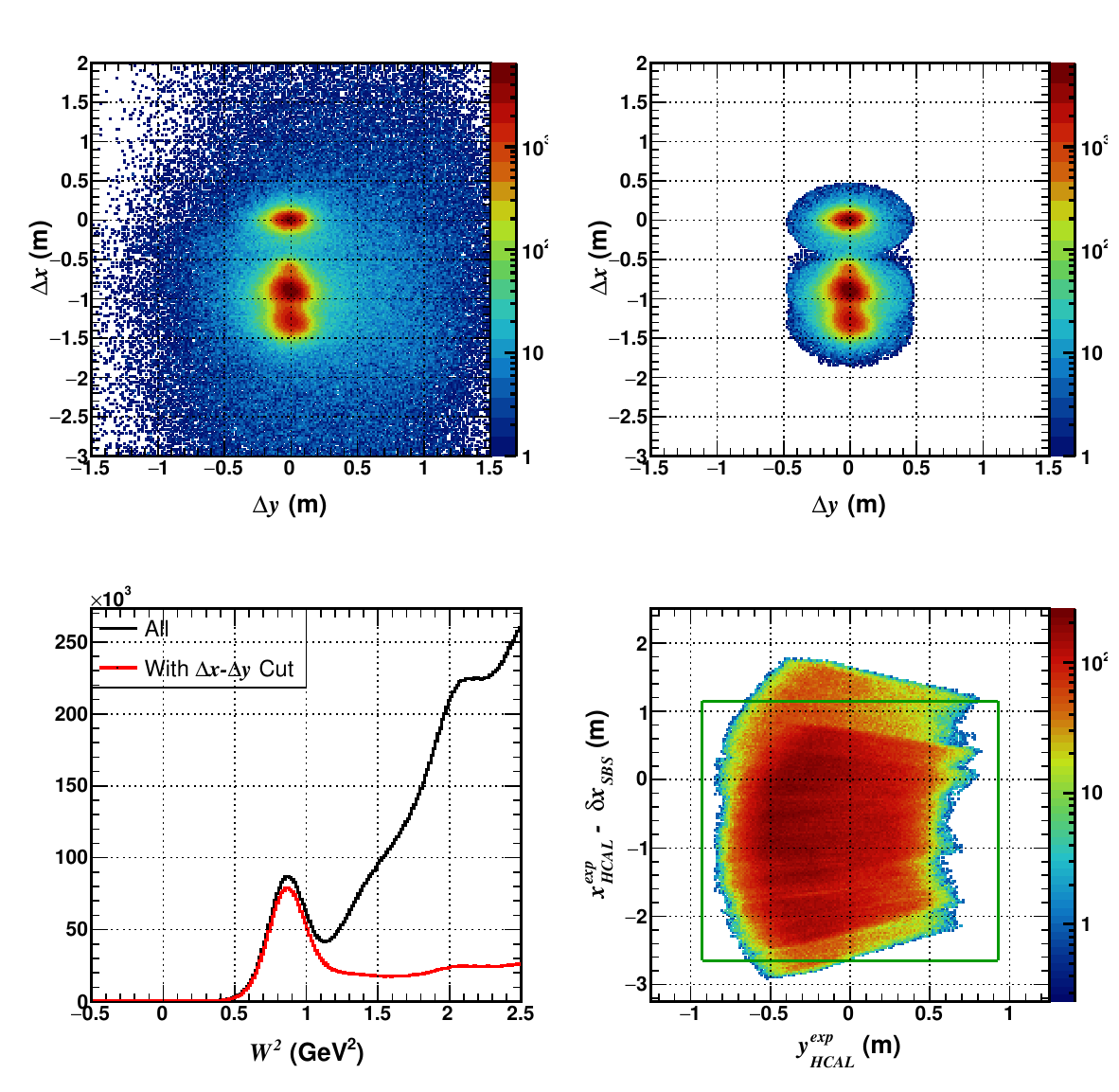}
    \caption{\label{fig:ch4:hcalndecutscombined} The effect of analysis cuts and the resulting acceptance for the combined \lh dataset recorded with four different SBS field strengths ($0\%$, $50\%$, $70\%$, and $100\%$) at \qeq{4.5} high-\ep kinematics. The top-left plot shows the combined \dx-\dy distributions without the \thpq cut, while the top-right plot includes the \thpq cut, demonstrating self-consistency of the \thpq cuts across data from different SBS field settings. The bottom-left plot displays the combined \w distributions with and without the \thpq cuts. The bottom-right plot shows the envelope of expected nucleon positions, ensuring effective HCAL acceptance coverage. The green rectangle represents the HCAL's physical area for reference.}
\end{figure}

\subheading{Implementation of Position Dependent Efficiency Corrections to MC Events}
The core idea behind this approach is to introduce artificial position-dependent non-uniformity in the HCAL NDE within MC, which is uniform initially (as shown in \fig \ref{fig:ch4:hcalndecutscombined}), to better reflect what is observed in real data. This approach assumes that efficiency varies only with position and that this variation affects protons and neutrons equally.

The simplest and most effective way to apply this correction is by adjusting the weights for MC events. Events where the nucleon is expected to land in a region with reduced (or enhanced) efficiency relative to the acceptance-averaged value should have their weights reduced (or increased) accordingly. The efficiency correction factor $c(x,y)$ for a given position at HCAL can be defined as follows \cite{SUPPMAT}:
\begin{equation}
c(x,y) = \frac{\epsilon^{data}_{HCAL}(x,y)}{\langle \epsilon^{data}_{HCAL} \rangle}
\end{equation}
Here, $\epsilon^{data}_{HCAL}(x,y)$ represents the interpolated NDE value at position $(x,y)$ derived from data, while the denominator is the acceptance-averaged value of the same observable.

%
%
The position $(x,y)$ for a given event, used to select the appropriate correction factor $c$, should be based on the expected nucleon positions rather than the observed ones, thereby avoiding bias due to detector resolution. These expected positions are calculated using MC truth information to accurately account for nuclear and radiative effects. Additionally, the deflection due to the SBS magnet ($\delta x_{SBS}$) is properly accounted for based on the nucleon type, which is also known \textit{a priori} from the MC truth information.

\begin{figure}[h!]
    \centering
    \begin{subfigure}[b]{0.8\textwidth}
         \centering
         \includegraphics[width=\textwidth]{ch4/figs/hcalnde_xyeffi_data_4.pdf}
         \caption{Data}
    \end{subfigure}
    \hfill
    \begin{subfigure}[b]{0.8\textwidth}
        \centering
        \includegraphics[width=\textwidth]{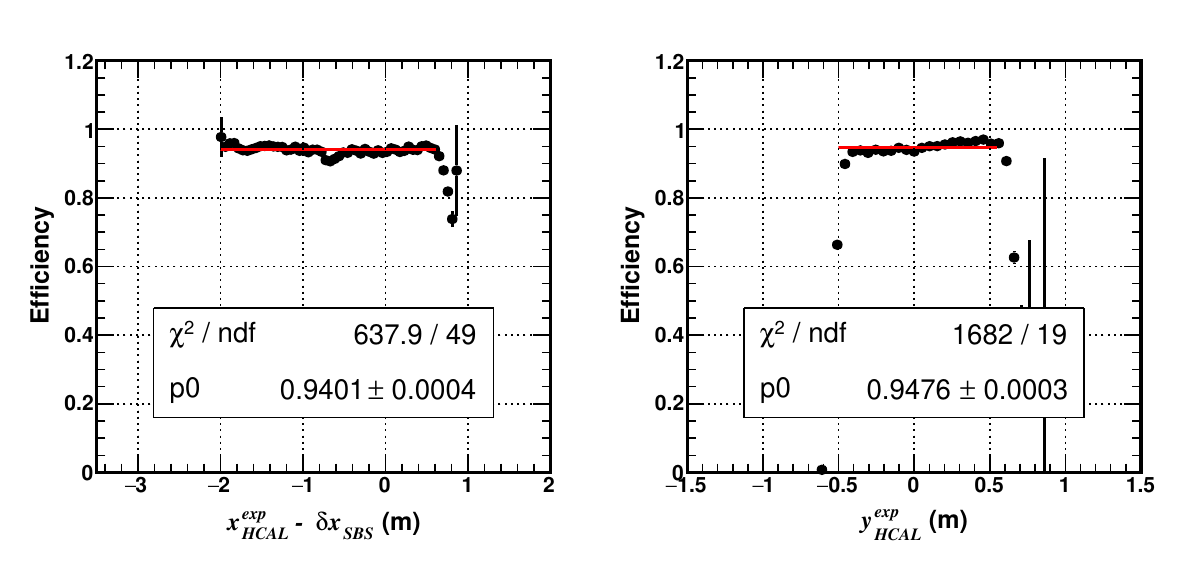}
        \caption{MC}
    \end{subfigure}
    \caption{\label{fig:ch4:hcalndexyeffiwcorr} HCAL proton detection efficiency profile comparison between data and MC after applying efficiency corrections. The efficiency dip observed in the dispersive direction in the data has been successfully reproduced in the MC.}
\end{figure}
Implementing this method requires a detailed efficiency map covering the entire HCAL acceptance, evaluated from data. To generate this map, reliable pDE estimates with sufficient statistical precision across the entire active area are necessary. Consequently, pDE estimates obtained using datasets recorded with different SBS field strengths at \qeq{4.5} high \ep kinematics were combined self-consistently. The cut regions and the resulting proton envelope based on their expected HCAL positions, effectively covering the HCAL acceptance, are shown in \fig \ref{fig:ch4:hcalndecutscombined}.

The resulting efficiency map from this combined analysis is shown in \fig \ref{fig:ch4:hcalndeeffimap}, clearly highlighting regions of non-uniformity. A coarse binning size, set to one-quarter of an HCAL module, was chosen to maintain a balance between statistical precision and resolution.

%
\subheading{Results}
The implementation of efficiency corrections in MC, using the above-mentioned efficiency map, produced expected outcomes. As shown in \fig \ref{fig:ch4:hcalndexyeffiwcorr}, the comparison of efficiency profiles between data and MC for the \qeq{3} SBS $30\%$ field dataset reveals that the efficiency dip in MC aligns well with the location observed in real data. 
\subsection{Notes on Neutron DE}
\label{ssec:ch4:hcalneutronde}
Direct measurement of neutron detection efficiency (nDE) is inherently more challenging than proton detection efficiency (pDE) due to the lack of stable free neutron targets. The \gmn data offers multiple options for measuring nDE, though each comes with its own limitations, as discussed below:

\begin{itemize}
    \item \textbf{Using \ld data:} Quasi-elastic events from the \ld target are a natural source of neutrons in the \gmn data. However, identifying the scattered nucleon type without any ambiguity is only possible if the spectator nucleon is also detected, which is impossible in the \gmn setup.
    \item \textbf{Using \lh data:} Exclusively selected $\gamma p \rightarrow \pi^+ n$ events from \lh data could provide a clean sample of tagged neutrons. This requires detecting the $\pi^+$ in BB, which is possible but not ideal. BB is optimized for detecting scattered electrons, which bend upward due to the BB dipole magnet. Detecting downward-bending $\pi^+$ with the same setup limits BB acceptance, resulting in tagged neutrons populating only a small portion of the bottom right corner of HCAL. Additionally, BB’s momentum resolution is insufficient for effectively suppressing background from processes like multi-pion production.\footnote{The initial plan was to collect dedicated tagged neutron data using the Left HRS, which offers significantly better momentum resolution than the BB. However, it could not be implemented due to time constraints caused by unexpected delays before and during data collection.}
    \item \textbf{Using both:} It is theoretically possible to extract nDE by comparing absolute yields from \lh and \ld data for a given configuration. However, the associated systematic uncertainty is likely to exceed the precision needed.
\end{itemize}

While a direct nDE measurement from \gmn data to validate MC would be valuable, it is not strictly necessary for extracting the physics observable. The strong agreement between pDE from MC and real data, coupled with the "true" nucleon detection efficiency (NDE) from MC showing high and comparable efficiencies for neutrons and protons largely independent of momentum, provides confidence that MC reliably represents nDE within the required error margin. 

%

\section{Extraction of Experimental Observables}
\label{sec:ch4:extrationofexpobservable}
The \gmn experiment directly measures the ratio of neutron-tagged ($N_{D(e,e'n)}$) to proton-tagged ($N_{D(e,e'p)}$) quasi-elastic scattering events from \ce{D2}, forming the ratio \rqe, defined as:
\begin{equation}
\label{eqn:ch4:expobsrpp}
    \frac{N_{D(e,e'n)}}{N_{D(e,e'p)}} = R^{QE} = \frac{\dv{\sigma}{\Omega}|_{D(e,e'n)}}{\dv{\sigma}{\Omega}|_{D(e,e'p)}}.
\end{equation}
From this, the elastic neutron-to-proton cross-section ratio $R$ and, subsequently, $G_M^n$ can be extracted, as discussed in \sect \ref{sec:measurementtechnique}.

The \dx distribution, which offers the cleanest separation between \deen and \deep events, is used to extract the counts $N_{D(e,e'n)}$ and $N_{D(e,e'p)}$. This involves fitting the \deen and \deep signal peaks in the \dx distribution using realistic models for both the signal shapes and background. The background-subtracted signals resulting from the fit provide the desired counts.

\begin{figure}[h!]
	\centering
	\fboxsep=0.4mm
    \fboxrule=0.9pt
	\fcolorbox{gray}{lightgray}{\includegraphics[width=0.98\columnwidth]{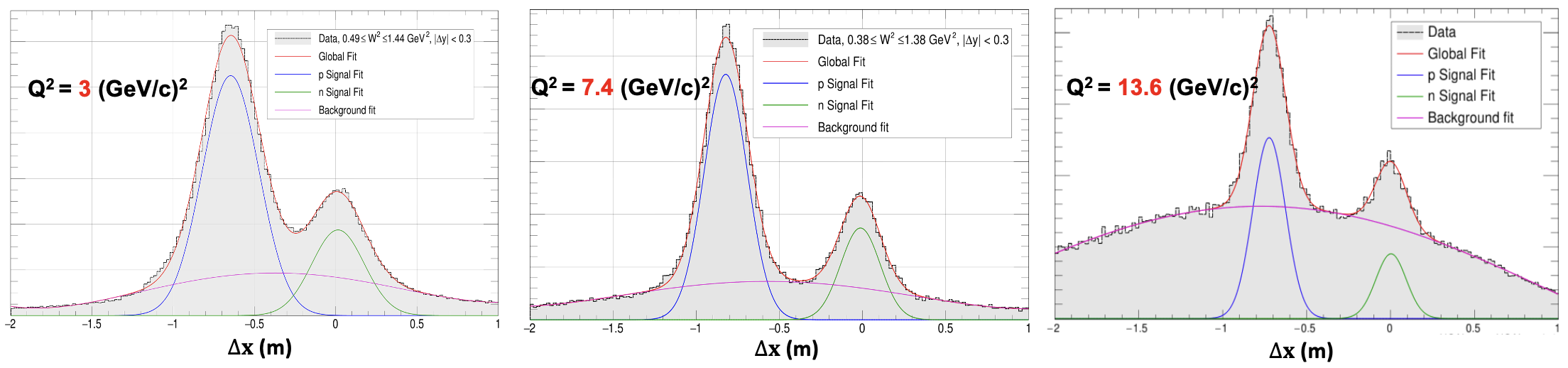}}
	\caption{\label{fig:ch4:dxfitbasic}Fitting HCal $\Delta x$ distributions using two Gaussian functions to model \deen and \deep signals and a $4^{th}$ order polynomial to model the background.}
\end{figure}
\fig \ref{fig:ch4:dxfitbasic} shows a basic fit to the \dx distribution across multiple \q points, assuming a pure Gaussian shape for the signals and using a polynomial function to model the background. While the overall fit is encouraging, discrepancies near the peak and tail regions of the \deep distribution indicate that it does not fully capture the signal shape, as expected from such a simplistic approach. 

A more accurate method would involve obtaining the signal shapes from quasi-elastic Monte Carlo simulations that incorporate all relevant physics and detector effects, as outlined in \sect \ref{sec:ch4:mcsimulation}. Additionally, the unknown true background shape can be better handled by using multiple models to estimate the background, with the resulting variation in the extracted observable quoted as a systematic uncertainty. This approach has been adopted in the analysis presented here and will be detailed in the following sections.
\subsection{Background Shape Estimation}
\label{ssec:ch4:inelbg}
Before delving into the discussion of realistic fits to the \dx distribution using MC-generated signals, it's essential to understand the background shape in the \dx distribution, how it correlates with the \w distribution, and what strategies can be employed to model it.

\begin{figure}[h!]
     \centering
     \begin{subfigure}[b]{0.9\columnwidth} 
         \centering
         \includegraphics[width=\textwidth]{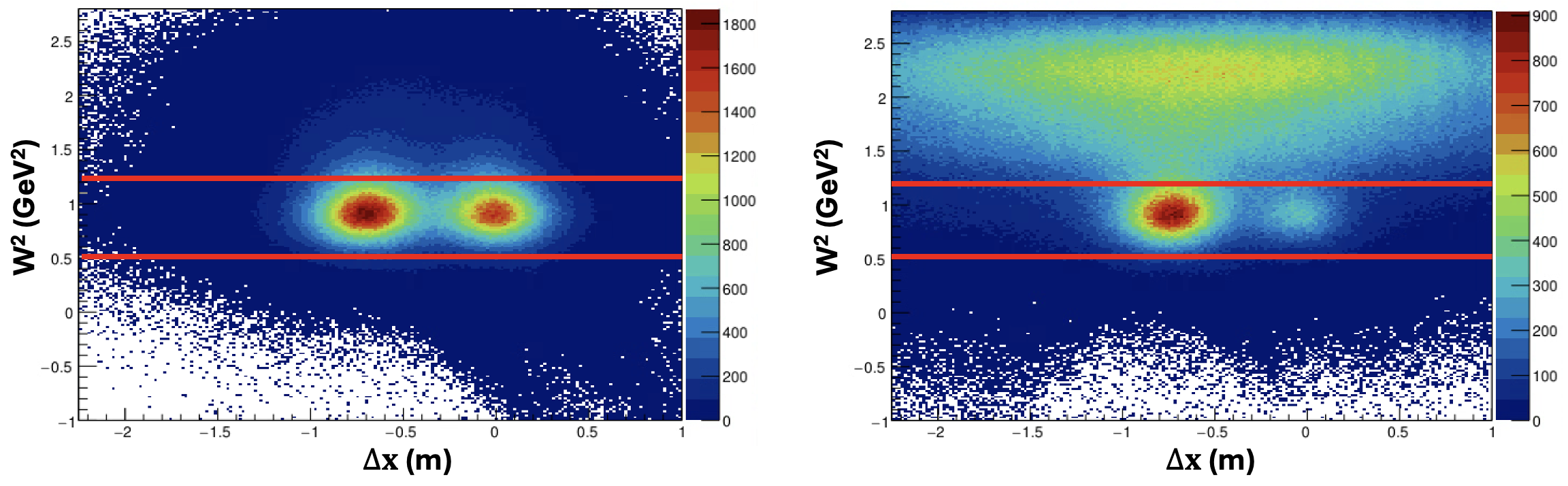}
         \caption{\qeq{3}}
     \end{subfigure}
     \begin{subfigure}[b]{0.9\columnwidth}
         \centering
         \includegraphics[width=\textwidth]{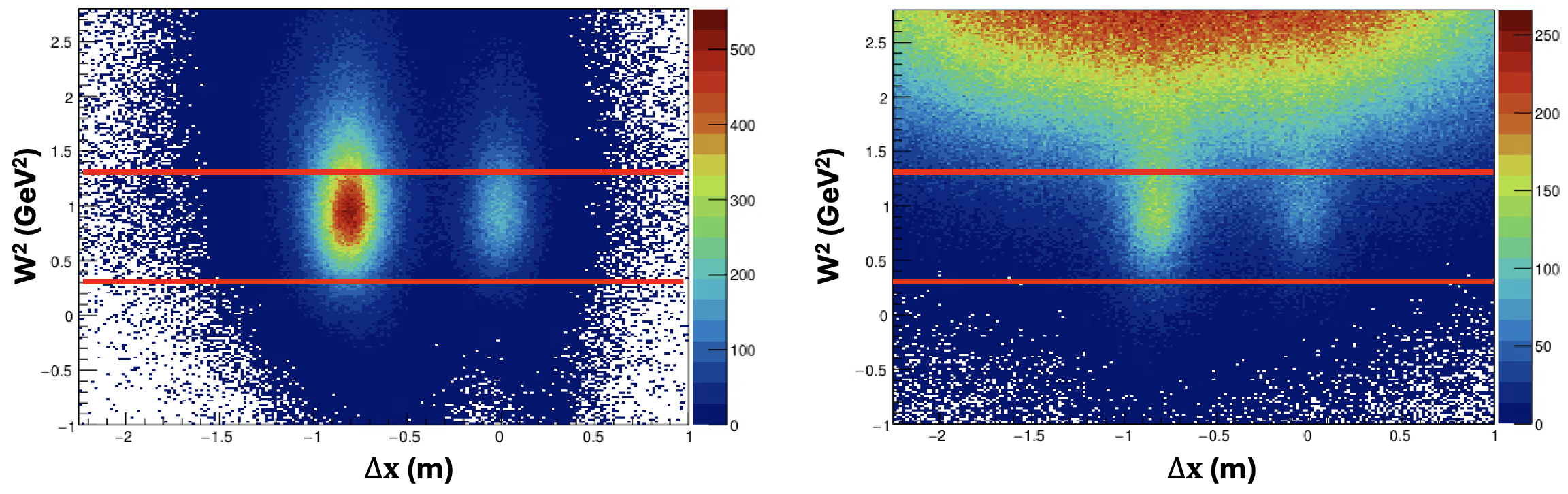}
         \caption{\qeq{7.4}}
     \end{subfigure}
     \begin{subfigure}[b]{0.9\columnwidth}
         \centering
         \includegraphics[width=\textwidth]{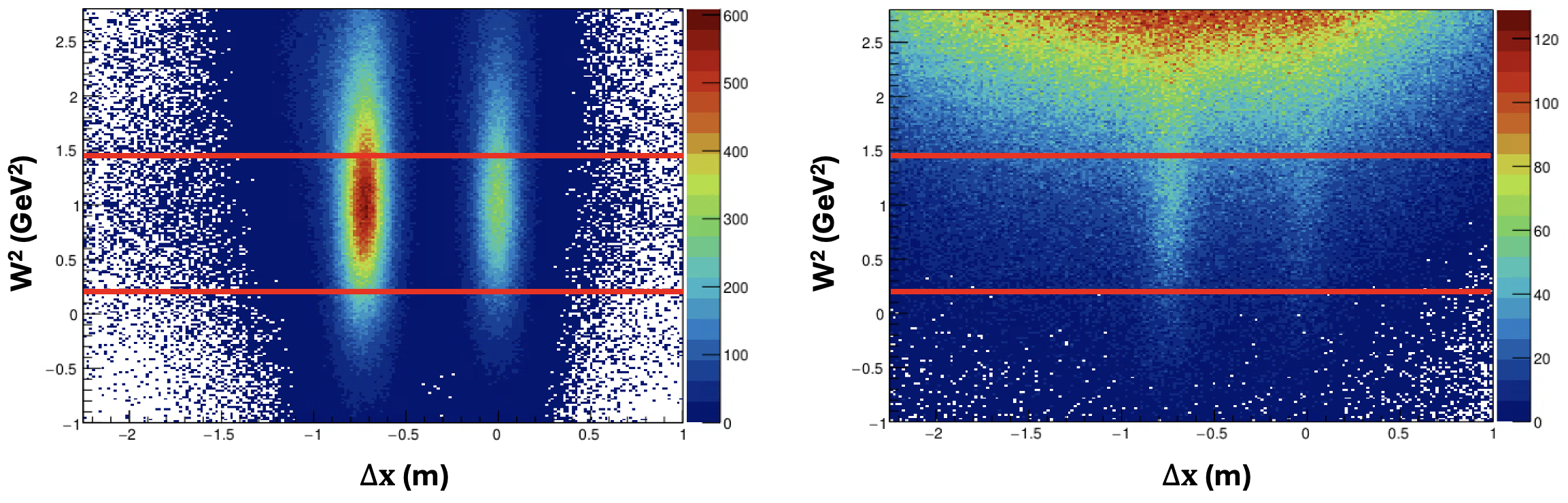}
         \caption{\qeq{13.6}}
     \end{subfigure}     
     \caption{Correlation of \w and \dx across \gmn kinematics using pure quasi-elastic signals from MC (left) and production data (right). The region within the red vertical lines represents the optimized cut region used for \rsf extraction. In this region, the signal-to-background ratio is sufficiently high, and the background shape in the \dx distribution appears smooth and symmetric, even at the highest-\q point.} 
     \label{fig:ch4:bgshapew2dx}
\end{figure}
\fig \ref{fig:ch4:bgshapew2dx} illustrates the correlation between the \w and \dx distributions for three different kinematic settings, covering the low to high \q regions. The plots on the left are generated using quasi-elastic MC, while the ones on the right are from \ld data. These plots include only events passing good electron event selection cuts. As expected, there is a noticeable increase in background and a broadening of the \w distribution as \q increases. Nevertheless, the quasi-elastic spots corresponding to \deep and \deen events remain distinct, even at the highest \q point. Importantly, the background shape in the \dx distribution appears smooth and symmetric within the \w cut region across all cases, which is reassuring.

Given the smooth and symmetric nature of the background shape, simple parameterizations can be used to model it. Alternatively, background shapes can be extracted directly from data by applying anti-cuts to select regions far from the quasi-elastic signals. Inelastic events generated from MC simulation provide another reliable method for generating the background shape. The following list outlines the background models/shapes used in the analysis presented here, incorporating all these approaches:
\begin{enumerate}
    \item \textbf{Anti-dy:} \dx distribution generated from the same dataset as the signal using good electron events that pass a loose \w cut but fail the \dy cut
    \item \textbf{Anti-dt:} \dx distribution generated from the same dataset as the signal using good electron events that fail the coincidence time cut
    \item \textbf{Inel-MC:} Background shape generated by the simulation of inelastic events.
    \item \textbf{Gaussian:} Gaussian distribution
    \item \textbf{Poly2:} Second-order polynomial
    \item \textbf{Poly3:} Third-order polynomial
\end{enumerate}

Each of the background models mentioned above can be combined with quasi-elastic signal shapes derived from MC to perform a realistic fit to the \dx distribution from data. The details of this process are discussed in the following section. 

\subsection{Data/MC Fit to \dx: $R_{n/p}^{sf}$ Extraction}
\label{ssec:ch4:rsfintro}
The quasi-elastic event generator based on \simc was used to generate \deen and \deep signal shapes for this analysis. As detailed in \sect \ref{sssec:qeeventgeneration}, this generator includes all relevant physics effects. The subsequent digitization and reconstruction of these events incorporate detector effects and reconstruction uncertainties, allowing for direct comparisons to data.

\begin{figure}[h!]
    \centering
    \includegraphics[width=1\columnwidth]{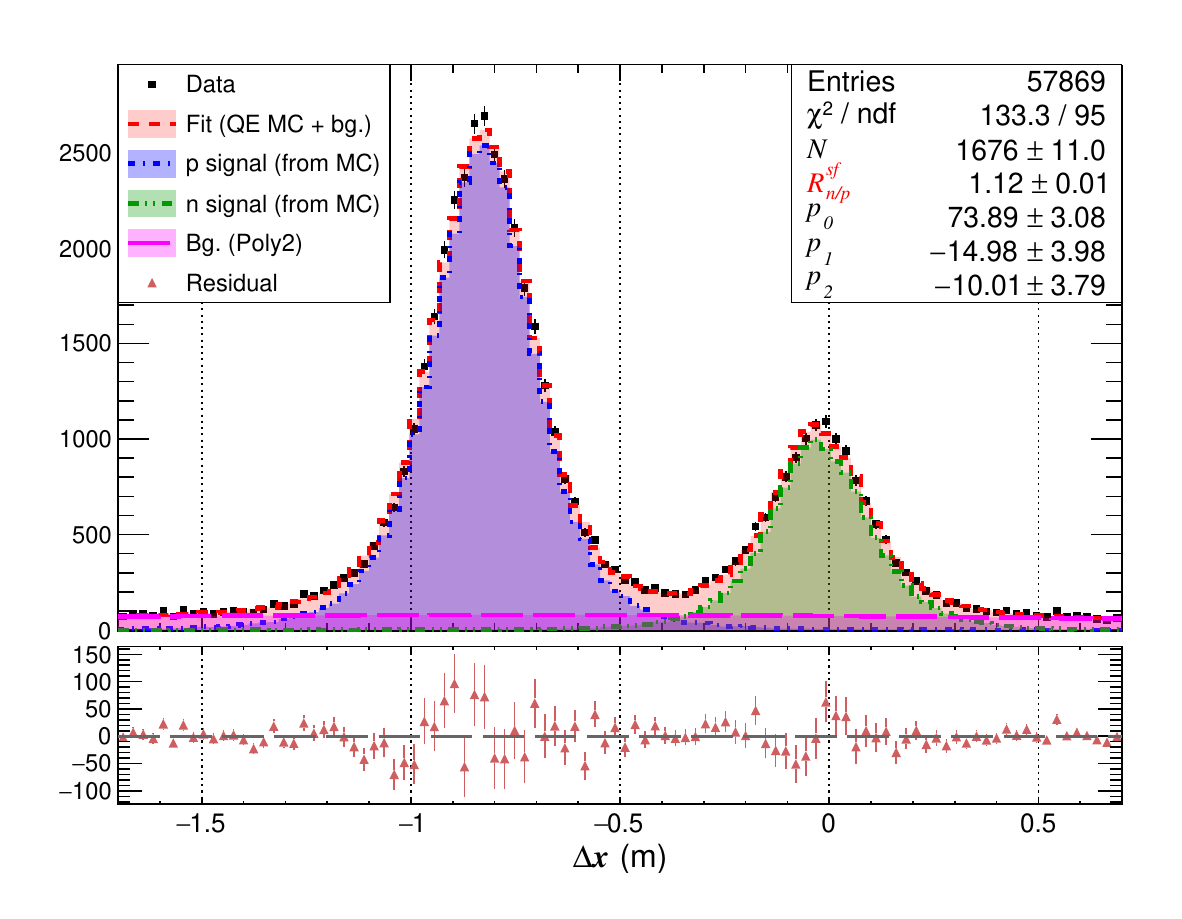}
    \caption{\label{fig:ch4:dxfitdmc1} Example of data/MC fit to the \dx distribution for \qeq{7.4} kinematics using signal shapes from MC and second-order polynomial to model the background. Good electron events passing \w and \dy cuts are shown.}
\end{figure}
The approach combines the signal histograms with the background using relative normalizations as free parameters to fit the \dx histogram from the data. The content of each bin in the combined histogram, $h_{i}^{est}$, is defined as:
\begin{equation}
    h_{i}^{est} = N (h_i^p + R_{n/p}^{sf} h_i^n) + B h_i^{bg}
\end{equation}
where $h_i^p$, $h_i^n$ and $h_i^{bg}$ represent the contents of the $i^{th}$ bin for the estimated \deep, \deen, and background histograms, respectively. Here, $N$, $R_{n/p}^{sf}$, and $B$ are the parameters for overall proton normalization, relative neutron to proton normalization, and overall background normalization, respectively, and are allowed to float freely during minimization. For a background model based on parameterization, $B h_i^{bg}$ is replaced with a functional form evaluated at the corresponding bin center. For instance, for a second-order polynomial background, $h_{i}^{est}$ is defined as:
\begin{equation}
    h_{i}^{est} = N (h_i^p + R_{n/p}^{sf} h_i^n) + p_0 + p_1 x_i^{bg} + p_2 (x_i^{bg})^2
\end{equation}
where $p_0$, $p_1$ and $p_2$ are the parameters, and $x_i^{bg}$ is the center of the $i^{th}$ bin.

In this approach, the fit parameter $R_{n/p}^{sf}$ serves as the key experimental observable, reflecting the relative discrepancy between the modeling of \deep and \deen events in MC, and can be used to extract $G_M^n$, as detailed in \sect \ref{sec:ch4:gmnextraction}. The ratio of quasi-elastic \deen to \deep events, \rqe (see \eqn \ref{eqn:ch4:expobsrpp}), can also be obtained by integrating the fitted signal distributions, but it includes detector effects like HCAL efficiency, requiring further corrections to extract $G_M^n$. In contrast, $R_{n/p}^{sf}$ inherently accounts for these corrections, making it more suitable for such extraction.

\begin{figure}[h!]
    \centering
    \includegraphics[width=0.85\columnwidth]{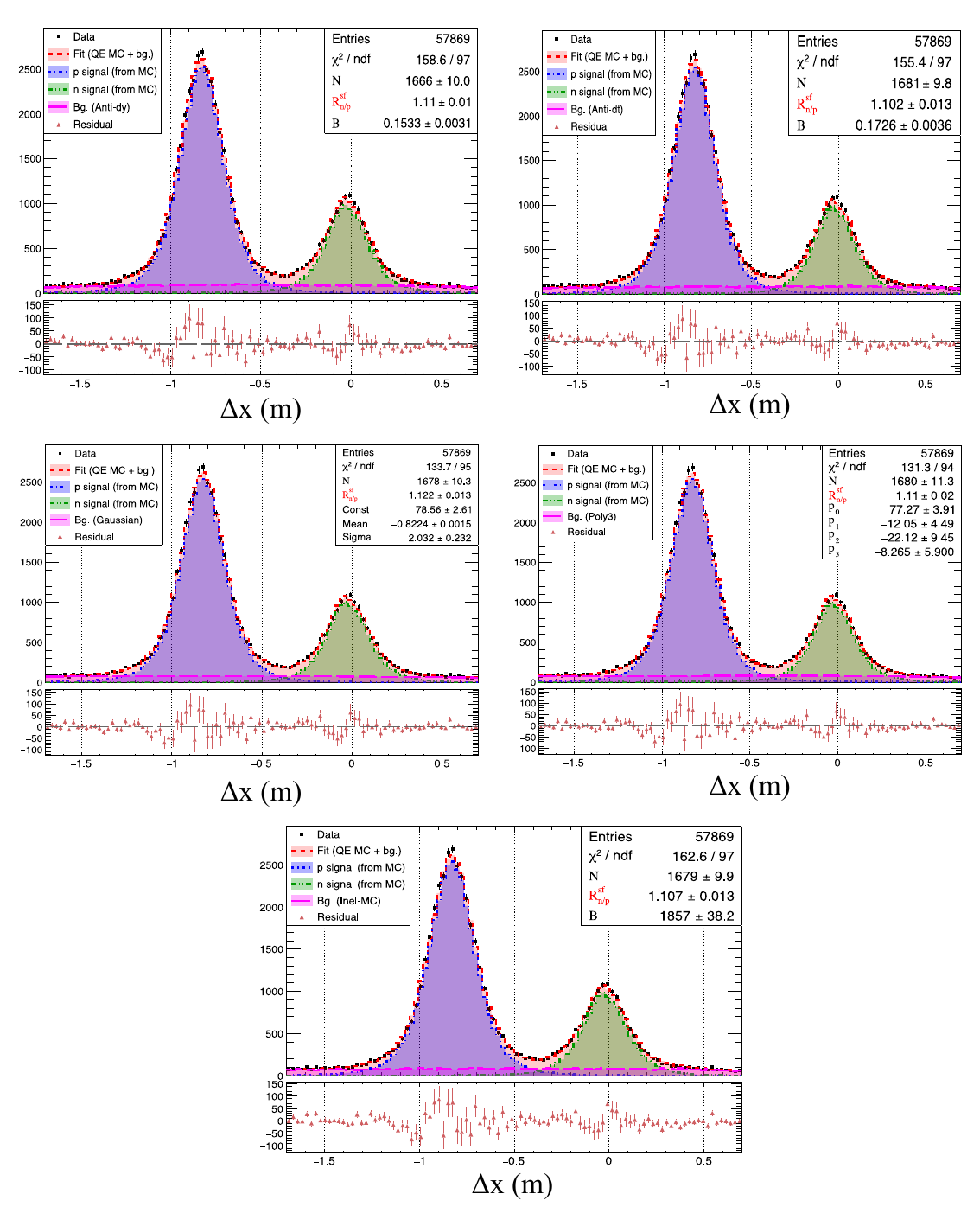}
    \caption{\label{fig:ch4:dxfitdmc2} Comparison of data/MC fit to the \dx distribution for \qeq{7.4} kinematics between various estimations of background models. Good electron events passing \w and \dy cuts are shown.}
\end{figure}
The associated $\chi^2$ is defined as:
\begin{equation}
    \chi^2 = \sum_{i=1}^{N_{bin}^{tot}} \left( \frac{h_i^{data} - h_i^{est}}{\sigma_i^{data}} \right)^2
\end{equation}
where $h_i^{data}$ and $\sigma_i^{data}$ are the content and statistical error of the $i^{th}$ bin in the data histogram, and $N_{bin}^{tot}$ represents the total number of bins. Sufficient MC statistics (at least $3$ to $5$ times more than data) were generated to ensure the statistical error of the MC sample has a negligible effect on the $\chi^2$ value. CERN ROOT’s built-in optimization process, which uses the \textit{Minuit} package at its back-end, was employed to perform the $\chi^2$ minimization.

\fig \ref{fig:ch4:dxfitdmc1} shows an example of such a fit using \ld data from \qeq{7.4} kinematics\footnote{Refer to Appendix \ref{appen:qeyield} for similar plots from the remaining \gmn kinematics.}. Good electron events passing the \w and \dy cuts were used to populate the data histogram and the pure quasi-elastic signals histograms from MC, ensuring consistent cut ranges between them\footnote{The ad-hoc correction factors summarized in \tab \ref{tab:adhoccorrection} were applied to address minor offsets between data and MC.}. A second-order polynomial models the background. The fit quality is solid, with no statistically significant trends observed in the residuals across the fit range, suggesting that the models accurately capture all key features present in the data.

Repeating the same fit with different background models listed in the previous section yields similar results, as shown in \fig \ref{fig:ch4:dxfitdmc2}\footnote{Refer to Appendix \ref{appen:qeyield} for similar plots from the remaining \gmn kinematics.}. The extracted fit parameter $R_{n/p}^{sf}$, the experimental observable, remains consistent within $1\%$ across these models.  This stability suggests that the signal extraction is robust and largely independent of the background assumptions, implying minimal interference between the signal and background contributions. 

\subsection{Cut Optimization}
\label{ssec:ch4:cutoptim}
Optimizing analysis cut ranges is essential for the reliable extraction of \rsf. Loose cut ranges risk excess background contamination, complicating subtraction, while overly tight cuts increase statistical uncertainty and may differently impact \deen and \deep events, introducing bias into the fit.

The optimal cut range is identified by analyzing the stability of the experimental observable (\rqe and/or \rsf) as a function of the cut variables, with this process performed separately for each kinematic setting. Particular attention is given to quasi-elastic event selection cuts (\w and \dy), as they have the biggest impact on the extraction. The other event selection cuts (discussed in \sect \ref{sec:ch4:evselect}) remain largely consistent across kinematics. The approach includes the following steps: 

\begin{figure}[h!]
    \centering
    \includegraphics[width=0.88\columnwidth]{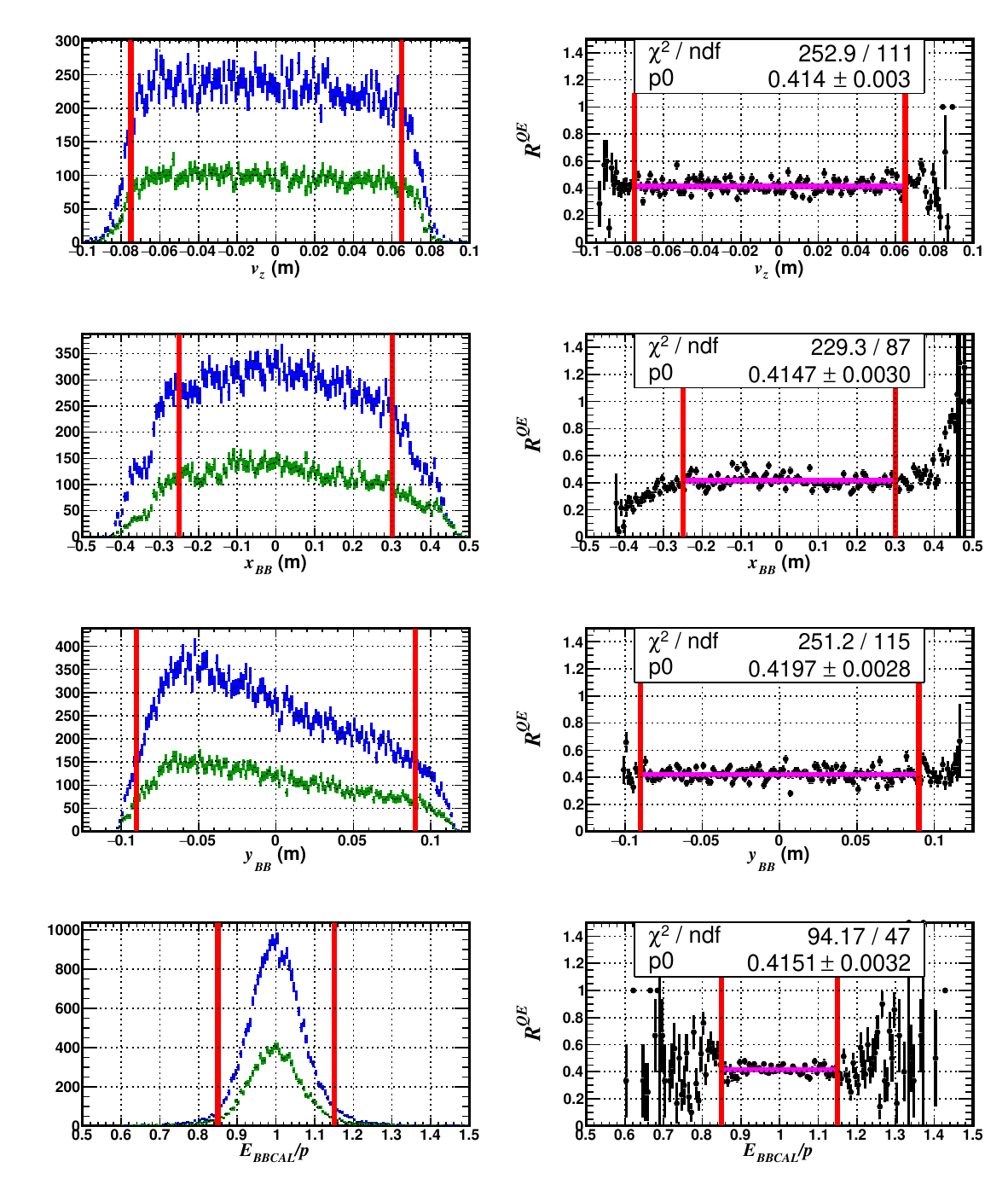}
    \caption{\label{fig:ch4:cutvarall1} Stability study of \rqe as a function of various cut variables for \qeq{7.4} production data. The left column displays the distribution of cut variables for \deep (blue) and \deen (green) events selected by \thpq cuts, while the right column presents their ratio, \rqe, plotted against the corresponding cut variable. The region within the vertical red lines represents the accepted cut range.}
\end{figure}
\begin{figure}[h!]
    \centering
    \includegraphics[width=0.9\columnwidth]{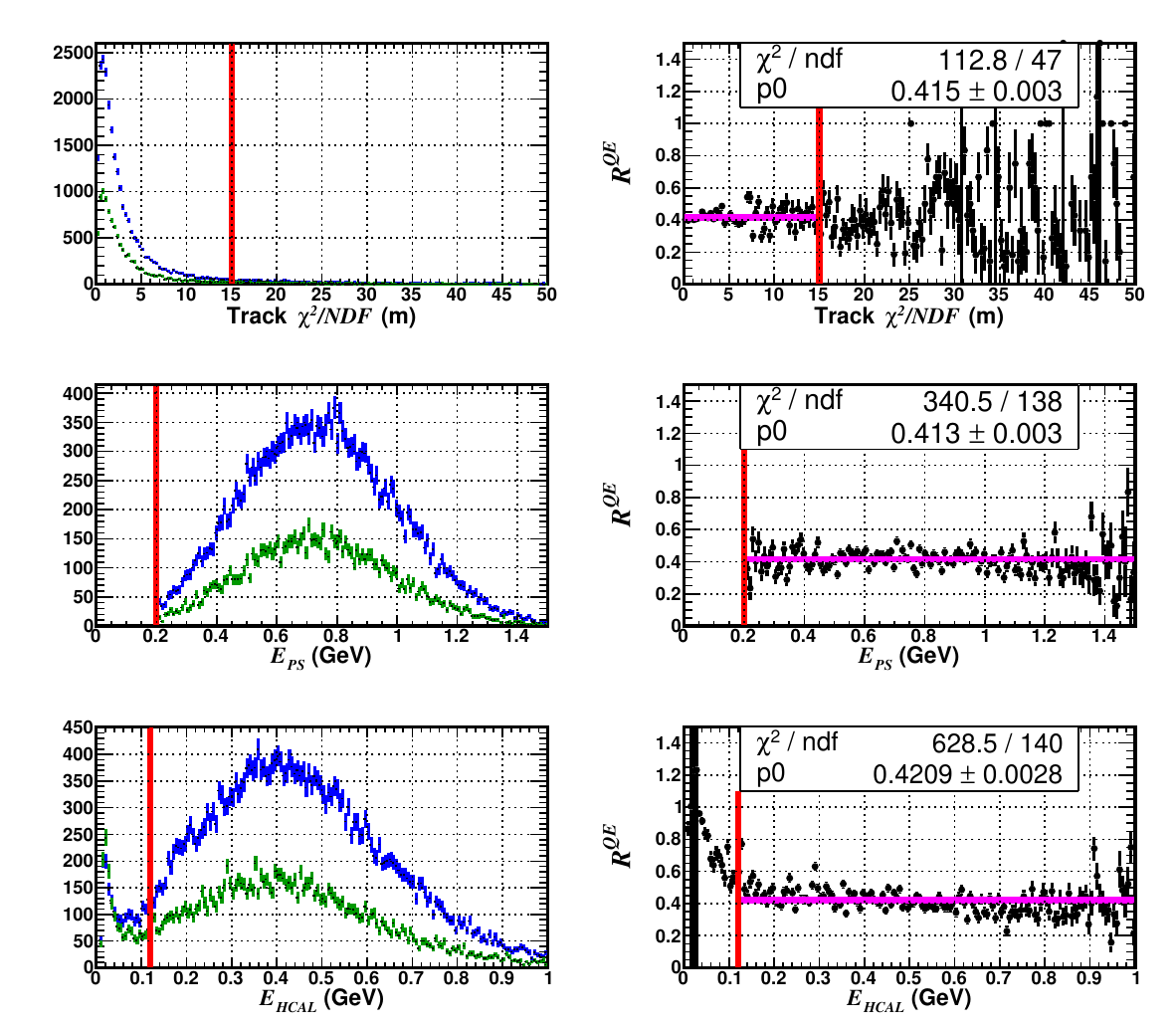}
    \caption{\label{fig:ch4:cutvarall2} Stability study of \rqe as a function of various cut variables for \qeq{7.4} production data. The left column displays the distribution of cut variables for \deep (blue) and \deen (green) events selected by \thpq cuts, while the right column presents their ratio, \rqe, plotted against the corresponding cut variable. The vertical red line represents the accepted cut threshold. For $E_{PS}$ and $E_{HCAL}$, values higher than the threshold are accepted, while for track $\chi^2/NDF$, values lower than the specified threshold are accepted.}
\end{figure}
\begin{enumerate} 
    \item \rqe, the ratio of quasi-elastic \deen to \deep events obtained from data, is plotted against the good electron and nucleon event selection cut variables, with fairly strict \w and \thpq cuts. The stability region of \rqe is evaluated for each case, guiding the cut ranges for the corresponding variable. Figures \ref{fig:ch4:cutvarall1} and \ref{fig:ch4:cutvarall2} show example plots from the \qeq{7.4} dataset, the intermediate \q point. The accepted cut region based on the stability of \rqe lies within the red vertical lines. For some variables shown in \fig \ref{fig:ch4:cutvarall2}, threshold cuts are used instead of ranges. For $E_{PS}$ and $E_{HCAL}$, values higher than the threshold are accepted, while for track $\chi^2/NDF$, values lower than the specified threshold are accepted. 
\begin{figure}[h!]
     \centering
     \begin{subfigure}[b]{0.9\textwidth} 
         \centering
         \includegraphics[width=\textwidth]{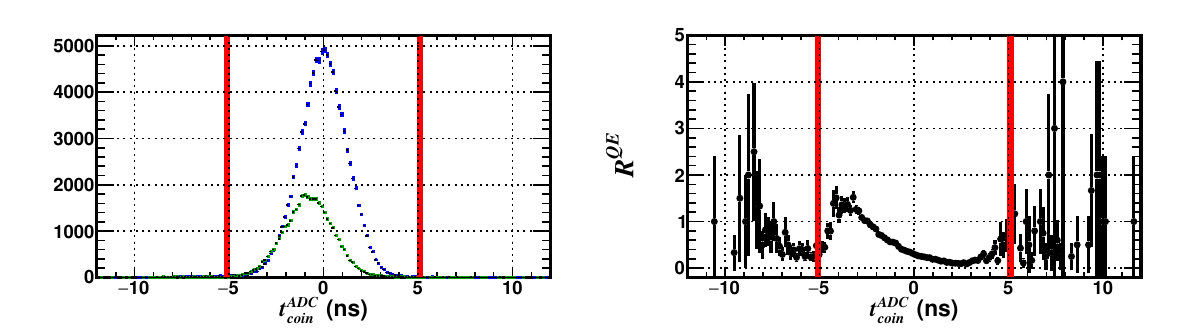}
         \caption{\qeq{3}}
     \end{subfigure}
     \begin{subfigure}[b]{0.9\textwidth}
         \centering
         \includegraphics[width=\textwidth]{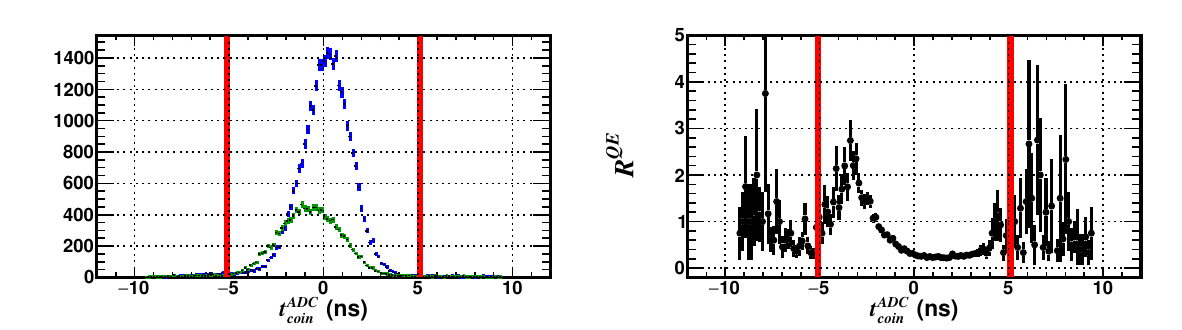}
         \caption{\qeq{7.4}}
     \end{subfigure}
     \begin{subfigure}[b]{0.9\textwidth}
         \centering
         \includegraphics[width=\textwidth]{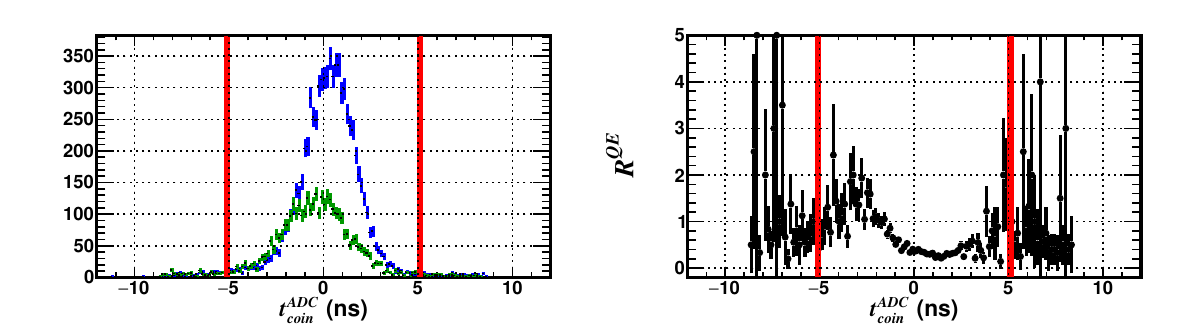}
         \caption{\qeq{13.6}}
     \end{subfigure}     
     \caption{Stability of \rqe relative to the Shower-HCAL ADC coincidence time ($t_{coin}^{ADC}$) cut across different \gmn kinematics. The left column shows the distribution of cut variables for \deep (blue) and \deen (green) events selected by \thpq cuts, while the right column presents their ratio, \rqe, plotted against $t_{coin}^{ADC}$. The misalignment between the \deep and \deen peak positions leads to significant non-uniformity in the region of interest. A cut region of $\pm5.1$ ns, indicated by red vertical lines, effectively avoids the region of instability in all cases.} 
     \label{fig:ch4:cutoptimcointime}
\end{figure}
\begin{figure}[h!]
     \centering
     \begin{subfigure}[b]{0.58\textwidth} 
         \centering
         \includegraphics[width=\textwidth]{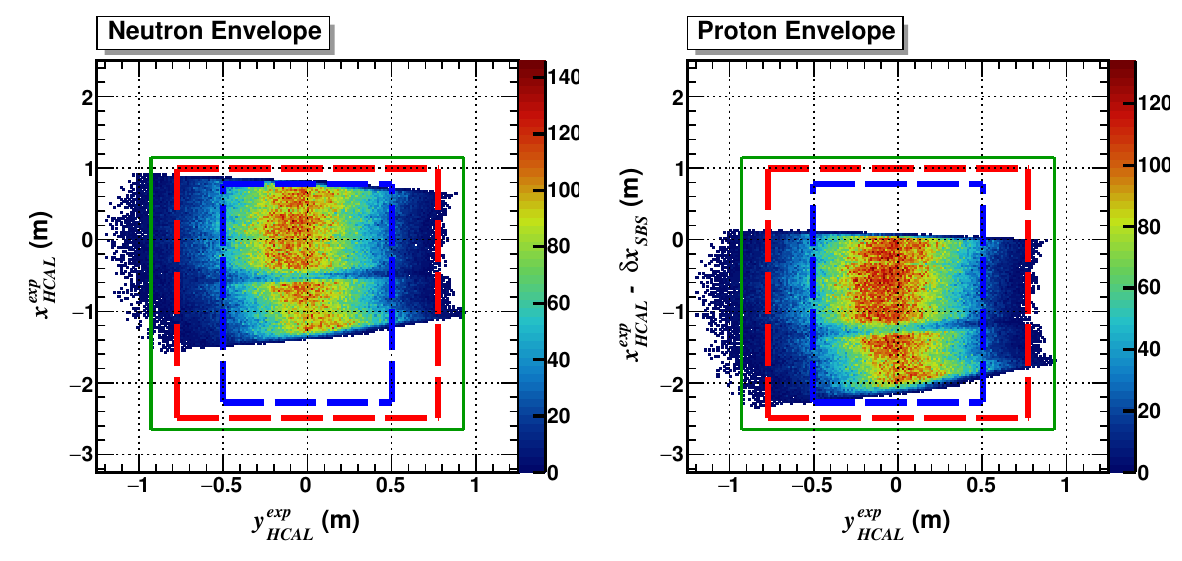}
         \caption{\qeq{3}}
     \end{subfigure}
     \begin{subfigure}[b]{0.58\textwidth}
         \centering
         \includegraphics[width=\textwidth]{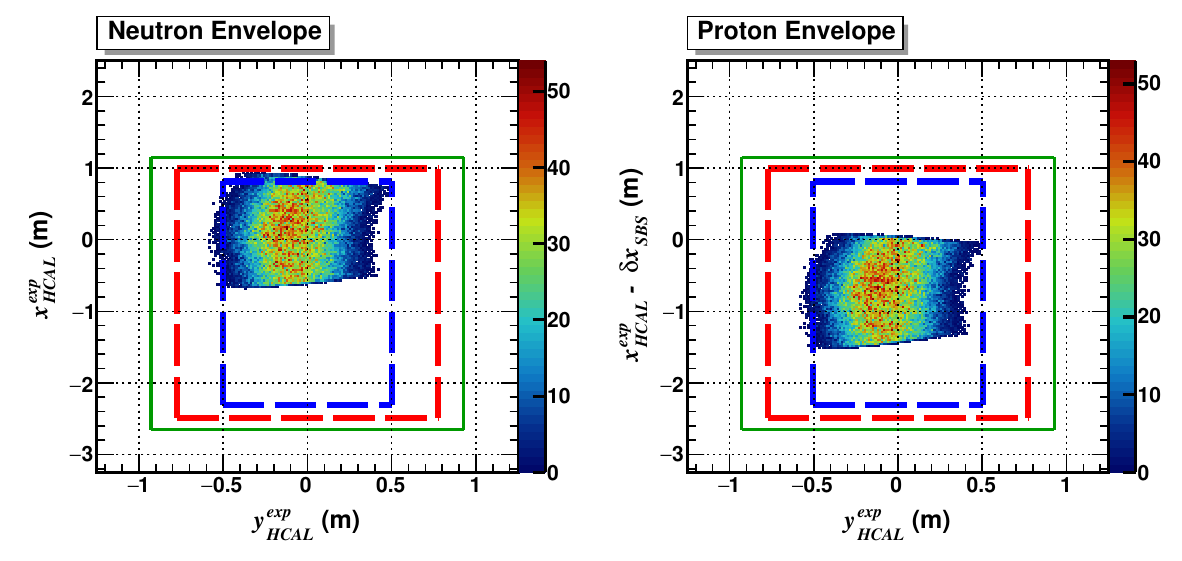}
         \caption{\qeq{7.4}}
     \end{subfigure}
     \begin{subfigure}[b]{0.58\textwidth}
         \centering
         \includegraphics[width=\textwidth]{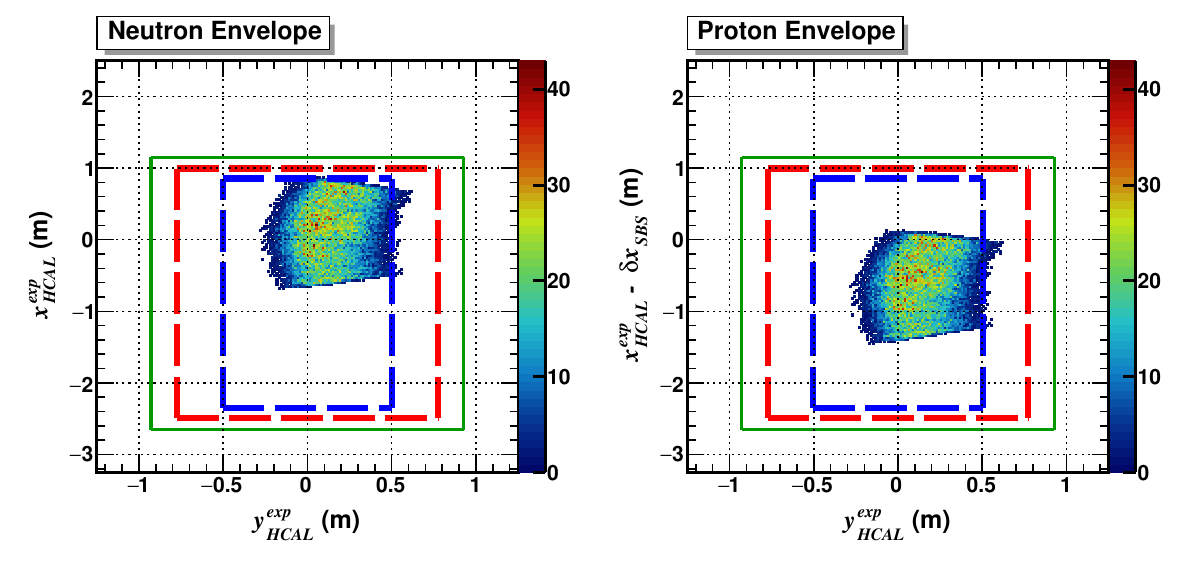}
         \caption{\qeq{13.6}}
     \end{subfigure}     
     \caption{Optimized HCAL fiducial cut regions superposed on the envelope of expected nucleon positions across different \gmn kinematics. The green (red) rectangle represents the HCAL physical (active) area, while the blue rectangle denotes the safety margin. The boundaries in the transverse direction are kept constant at $\pm0.5$ m, and in the dispersive direction, they maintain a gap of $1.5\sigma_{\Delta{x}}$ from the HCAL active area. Here, $\Delta x_{SBS}$ refers to the proton deflection due to the SBS magnet. Events passing all optimized cuts, except for the HCAL fiducial, are shown. The optics validity cut ensures matched proton and neutron acceptances, as clearly observed in the lowest-\q dataset.} 
     \label{fig:ch4:cutoptimfiducial}
\end{figure}
   \item Variables showing significant non-uniformity within the expected cut region are further analyzed relative to \rsf to ensure the cut range is broad enough to avoid separate effects on \deen and \deep events, which could bias the fit. Such behavior has only been observed with the Shower-HCAL ADC coincidence time variable, as shown in \fig \ref{fig:ch4:cutoptimcointime}. The difference in the $t_{coin}^{ADC}$ peak position between \deen and \deep events is clearly visible (plots on the left column), with the magnitude being consistent across kinematics. A cut range of $\pm$ \SI{5.1}{ns}, indicated by the vertical red lines, effectively avoids the region of instability. This discrepancy primarily arises from the non-uniformity in HCAL ADC time alignment existing after the second calibration pass. Efforts for improvement are ongoing.
\begin{figure}[h!]
    \centering
    \begin{subfigure}[b]{0.496\textwidth}
         \centering
         \includegraphics[width=\textwidth]{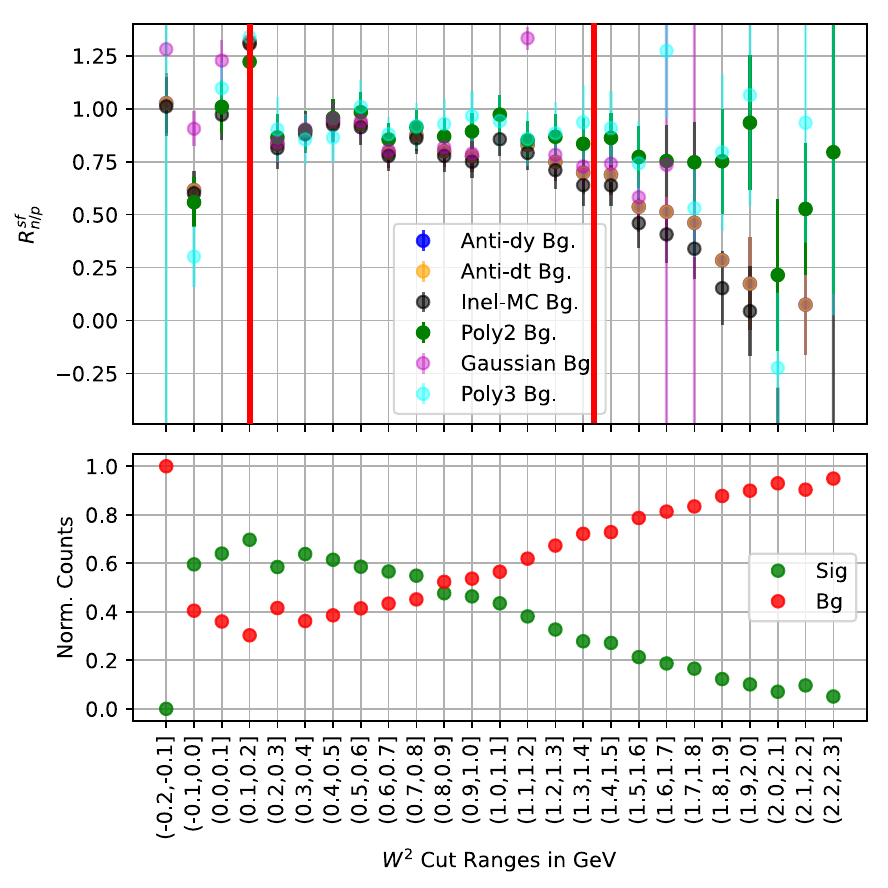}
    \end{subfigure}
    \hfill
    \begin{subfigure}[b]{0.496\textwidth}
        \centering
        \includegraphics[width=\textwidth]{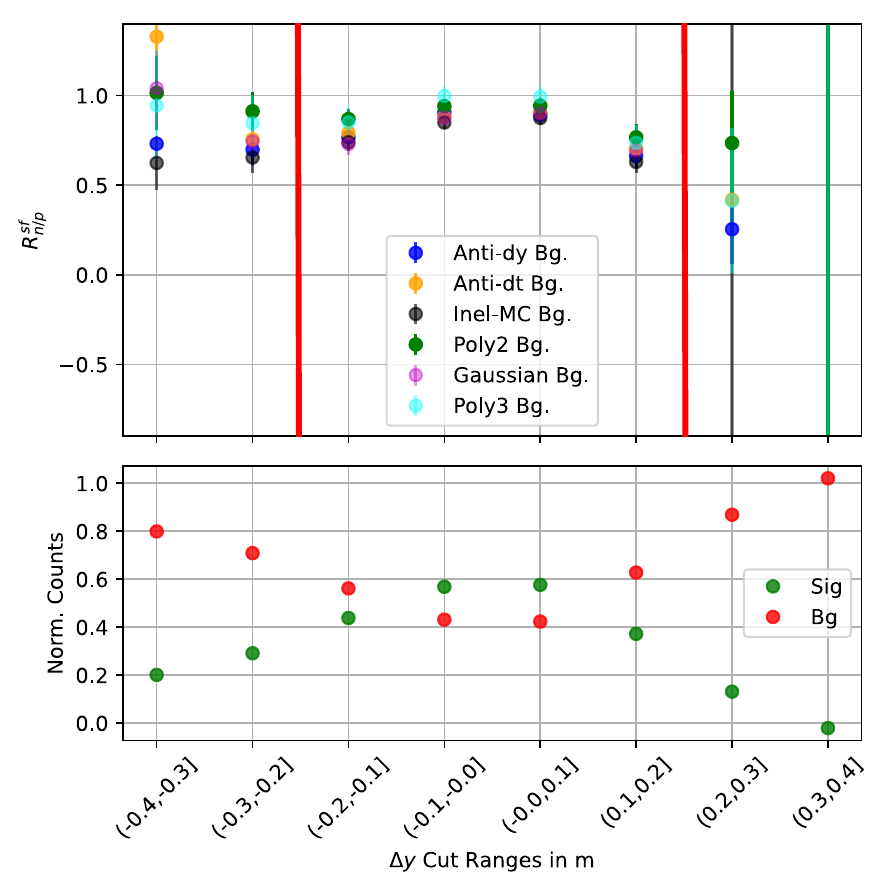}
    \end{subfigure}
    \caption{\label{fig:ch4:cutoptimdxdy} Stability study of \rsf as a function of the \w (left) and \dy (right) cuts, the quasi-elastic event selection cuts, for \qeq{13.6} production data. In each plot, the top pad shows \rsf extracted using various background models, while the bottom pad presents the normalized counts of signal and background events obtained using the ``Poly2" background model. The region within the vertical red lines represents the accepted cut range. The optimized \w cut region superposed on the \w-\dx correlation plot obtained from both data and MC can be found in \fig \ref{fig:ch4:bgshapew2dx}.}
\end{figure}
    \item The HCAL fiducial cut region is evaluated using guidance from the HCAL nucleon detection efficiency (NDE) study, considering the Fermi motion of nucleons within the deuteron nucleus. The HCAL NDE map obtained from elastic \heep events data analysis, as shown in \fig \ref{fig:ch4:hcalndeeffimap}, reveals low-efficiency regions resulting from acceptance cutoff beyond pm 0.5 m in the transverse direction. This sets the HCAL fiducial region's transverse range for all kinematics. In the dispersive direction, the fiducial region's width is determined by $1.5\sigma$ of the \dx distribution width for a given kinematic. This accounts for potential nucleon displacement due to the high momentum tail of the deuteron wavefunction, reducing uncertainty related to acceptance losses. \fig \ref{fig:ch4:cutoptimfiducial} shows the accepted fiducial region superimposed on the expected nucleon envelope across different kinematics. 
    %

%
\begin{table}[h!]
    \centering
    \caption{\label{tab:ch4:optcut}Summary of the optimized set of cuts used for the final analysis. Cut variables marked with * are applied only to data. The variables \xhex and \yhex are the expected nucleon positions at HCAL. The notation {[]}$^{(p,n)}$ indicates calculations using both the proton and neutron hypotheses, where the former accounts for the SBS kick and the latter does not. Cut on these variables forms the HCAL fiducial region. A detailed description of the remaining cut variables is provided in \sect \ref{sec:ch4:evselect}.} 
    \begin{tabular}{>{\hsptab}l<{\hsptab}>{\hsptab}c<{\hsptab}>{\hsptab}c<{\hsptab}>{\hsptab}c<{\hsptab}>{\hsptab}c<{\hsptab}>{\hsptab}c<{\hsptab}} \hline\hline\vspace{-1em} \\
        \multirow{2}{*}{Cut Variable} & \multicolumn{5}{c}{\q (\ep)} \vspace{0.2em} \\ \cline{2-6} \vspace{-1em} \\
                                  & 3 (0.72) & 4.5 (0.51) & 7.4 (0.46) & 9.9 (0.50) & 13.6 (0.41) \vspace{0.2em} \\ \hline \vspace{-1em} \\
        $N_{hit}^{GEM}$           & $>3$           & $>3$         & $>3$         & $>3$         & $>2$         \\ \vspace{-0.9em} \\
        Track $\chi^2/NDF$        & $<15$          & $<15$        & $<15$        & $<15$        & $<15$        \\ \vspace{-0.9em} \\
        $v_{z}$ (cm)              & $(-7,7)$       & $(-7,7)$     & $(-7.5,6.5)$ & $(-7,7)$     & $(-7.5,7.5)$     \\ \vspace{-0.9em} \\
        $x_{BB}$ (cm)             & $(-12,30)$     & $(-20,35)$   & $(-25,25)$   & $(-20,30)$   & $(-25,25)$      \\ \vspace{-0.9em} \\
        $y_{BB}$ (cm)             & $(-9,9)$       & $(-9,10)$    & $(-9,9)$     & $(-9,9)$     & $(-9,9)$      \\ \vspace{-0.9em} \\
        $E_{PS}$ (GeV)            & $>0.2$         & $>0.2$       & $>0.2$       & $>0.2$       & $>0.2$       \\ \vspace{-0.9em} \\
        $E_{BBCAL}/p$             & $(0.8,1.2)$    & $(0.7,1.3)$  & $(0.85,1.15)$& $(0.8,1.2)$  & $(0.8,1.2)$ \\ \vspace{-0.9em} \\
        $Size_{clus}^{GRINCH}$ *  & -              & $>2$         & -            & -            & -            \\ \vspace{-0.9em} \\
        $E_{HCAL}$ (GeV)          & $>0.025$       & $0.1$        & $>0.12$      & $>0.2$       & $>0.2$         \\ \vspace{-0.9em} \\
        $|t_{coin}^{ADC}|$ (ns) * & $<5.1$         & $<5.1$       & $<5.1$       & $<5.1$       & $<5.1$         \\ \vspace{-0.9em} \\
        \w (GeV$^2$)              & $(0.5,1.2)$    & $(0.25,1.2)$ & $(0.3,1.3)$  & $(0.3,1.3)$  & $(0.2,1.45)$   \\ \vspace{-0.9em} \\
        \dy (m)                   & $(-0.3,0.3)$   & $(-0.3,0.3)$ & $(-0.3,0.3)$ & $(-0.3,0.3)$ & $(-0.25,0.25)$ \\ \vspace{-0.9em} \\
$\left[x_{HCAL}^{exp}\right]^{(p,n)}$ (m)& $(-2.22,0.72)$ & $(-2.28,0.78)$ & $(-2.32,0.82)$ & $(-2.36,0.86)$ & $(-2.36,0.86)$\\ \vspace{-0.9em} \\
        \yhex (m)                 & $(-0.5,0.5)$   & $(-0.5,0.5)$ & $(-0.5,0.5)$ & $(-0.5,0.5)$ & $(-0.5,0.5)$\\ \vspace{-0.9em} \\
    \hline\hline
    \end{tabular}
\end{table}
    \item Finally, quasi-elastic event selection cuts (\w and \dy) are studied separately using the optimized cut regions obtained in the previous steps. The process involves dividing these distributions into small slices covering the entire quasi-elastic signal range, guided by MC, and then extracting \rsf for each slice by performing a data/MC fit to the \dx distribution. The range where \rsf shows no statistically significant fluctuation is chosen as the optimal cut range. Care is taken to vary the cut regions equivalently between data and MC signals. Since \w and \dy are correlated, the cut on one variable is removed or sufficiently relaxed during variation for the other. \fig \ref{fig:ch4:cutoptimdxdy} shows example plots from the highest \q dataset, the most challenging \gmn kinematic for these studies. 

\end{enumerate}

This process is iterative, and the steps are repeated multiple times before finalizing the optimal set of analysis cuts for a given kinematic. \tab \ref{tab:ch4:finalrsf} presents a summary of the optimized set of cuts across all kinematics used for the final extraction of \rsf \ref{tab:ch4:optcut}.
\subsection{Final $R_{n/p}^{sf}$ Values}
\label{ssec:ch4:finalrsf} 
The final \rsf values obtained with the optimized set of cuts are summarized in \tab \ref{ssec:ch4:cutoptim}. Production datasets from all \gmn kinematics, as listed in \tab \ref{tab:sbsconfig2}, were used for these extractions\footnote{The \qeq{4.5} high-\ep kinematics dataset is excluded as it primarily belongs to the \ntpe experiment.}. \rsf values from all six background shapes are listed; among them, the result from the ``Poly2" background shape was selected for the extraction of final physics results presented in the next chapter, as it provides the best fit to the data without overfitting.

\begin{table}[h!]
    \centering
    \caption{Final \rsf values obtained from \gmn production data using the optimized set of cuts listed in \tab \ref{tab:ch4:optcut}. ``Bg. model" refers to the background model used for \rsf extraction, detailed in \sect \ref{ssec:ch4:inelbg}. The model marked with * indicates the one selected for the final physics extraction discussed in the next chapter.}
    \label{tab:ch4:finalrsf}
    \begin{tabular}{>{\hsptab}l<{\hsptab}>{\hsptab}c<{\hsptab}>{\hsptab}c<{\hsptab}>{\hsptab}c<{\hsptab}>{\hsptab}c<{\hsptab}>{\hsptab}c<{\hsptab}} \hline\hline\vspace{-1em} \\
        \multirow{2}{*}{Bg. Model} & \multicolumn{5}{c}{$R_{n/p}^{sf} \pm \Delta(R_{n/p}^{sf})_{stat}$ at \q (\ep)}  \vspace{0.2em} \\ \cline{2-6} \vspace{-1em} \\
                                & 3 (0.72) & 4.5 (0.51) & 7.4 (0.46) & 9.9 (0.50) & 13.6 (0.41) \vspace{0.2em} \\ \hline \vspace{-1em} \\
        Anti-dy  & $0.9536\pm0.0060$ & $1.087\pm0.003$ & $1.110\pm0.010$ & $1.055\pm0.040$ & $0.8605\pm0.0224$ \\ \vspace{-0.9em} \\
        Anti-dt  & $0.9566\pm0.0059$ & $1.097\pm0.003$ & $1.102\pm0.013$ & $1.076\pm0.040$ & $0.8764\pm0.0224$ \\ \vspace{-0.9em} \\
        Inel-MC  & $0.9559\pm0.0060$ & $1.070\pm0.000$ & $1.107\pm0.013$ & $1.044\pm0.040$ & $0.8446\pm0.0223$ \\ \vspace{-0.9em} \\
        Gaussian & $0.9628\pm0.0059$ & $1.115\pm0.004$ & $1.125\pm0.013$ & $1.057\pm0.042$ & $0.9095\pm0.0249$ \\ \vspace{-0.9em} \\
        Poly2 *  & $0.9615\pm0.0062$ & $1.105\pm0.003$ & $1.120\pm0.010$ & $1.055\pm0.042$ & $0.9174\pm0.0254$ \\ \vspace{-0.9em} \\
        Poly3    & $0.9591\pm0.0067$ & $1.093\pm0.004$ & $1.110\pm0.020$ & $1.073\pm0.049$ & $0.9244\pm0.0323$ \\ \vspace{-0.9em} \\
    \hline\hline
    \end{tabular}
\end{table}

\chapter{Results}

\section{Extraction of $G_M^n$ from $R_{n/p}^{sf}$}
\label{sec:ch4:gmnextraction}
The first step in extracting $G_M^n$ from $R_{n/p}^{sf}$ is determining the elastic electron-neutron to electron-proton scattering cross-section ratio, known as the Born cross-section ratio $R$. From $R$, $G_M^n$ can be obtained using the Rosenbluth formula. It's important to note that extracting $G_M^n$ from $R$ depends on the models chosen to estimate the proton cross-section and $G_E^n$. As better models become available, the extraction of $G_M^n$ can be refined using the measured $R$ values, which are model-independent and represent the most fundamental physics observables of the \gmn experiment. Therefore, in this work, the values and associated uncertainties of $R$ will be presented alongside $G_M^n$. 

\subheading{$R$ from $R_{n/p}^{sf}$}
The deviation from unity of the extracted \rsf (see \sect \ref{ssec:ch4:rsfintro}) indicates discrepancies in the modeling of \deen and \deep events in the MC. These differences must stem from limitations in the simulation of parameters that affect these event types differently. The radiative and nuclear effects included in the MC are realistic and influence neutrons and protons similarly, with good approximation. Simulated detector effects have also been validated through data/MC comparisons across various physics variables. Among these, HCAL nucleon detection efficiency is particularly crucial for \rsf. Rigorous analysis in \sect \ref{sec:ch4:hcalnde} shows that the proton detection efficiency in MC aligns closely with data. Although a direct measurement of neutron detection efficiency (nDE) was not possible, it is reasonable to assume that the simulated nDE is consistent with data within the desired margin of error. Since the MC reasonably captures all relevant physics and detector effects, the discrepancy in modeling \deen and \deep events, represented by $R_{n/p}^{sf}$, must be due to differences in the neutron-to-proton Born cross-section ratio between MC and data. Therefore, the the ratio $R$ can be extracted from $R_{n/p}^{sf}$ as:
\begin{equation}
    R(Q^2,\epsilon) = R_{n/p}^{sf} \times R^{Born}_{MC}(Q^2,\epsilon)
\end{equation}
where $R_{MC}^{Born}$ is the neutron-to-proton Born cross section ratio assumed in MC using the nucleon electromagnetic form factor (EMFF) parametrizations listed in \eqn \ref{eqn:ch4:emffparametrization}. The same parametrizations are utilized to retrieve the value of $R^{Born}_{MC}$ for a give \q and \ep.

\subheading{$G_M^n$ from $R$}
$G_M^n$ can be expressed in terms of $R$ via \eqn \ref{eqn:ch3:gmnfromr}, which has the following form:
\begin{equation}
    G_M^n (Q^2) = -\left[ \frac{1}{\tau_n}\frac{\epsilon_{n}(1+\tau_{n})}{\epsilon_{p}(1+\tau_{p})}\sigma^p_{Red}(Q^2,\epsilon)R(Q^2,\epsilon) - \frac{\epsilon_n}{\tau_n}{G_E^n}^2(Q^2) \right]^{\frac{1}{2}}
\end{equation}
where $\sigma_{Red}^p (Q^2,\epsilon) = \tau_p {G_M^p}^2 (Q^2) + \epsilon_p {G_E^p}^2 (Q^2)$ is the proton reduced cross section and the other variables carry their usual meanings. EMFF parametrization by Ye \textit{et al} \cite{YE20188} is used to estimate $\sigma_{Red}^p$ and $G_E^n$ for the extraction of $G_M^n$ at a given \q and \ep.  

\subheading{Calculation of $\langle Q^2\rangle$ and $\langle \epsilon\rangle$}
The \q and \ep distributions for a given \gmn kinematics are broad due to the large acceptance of BB. To account for this, acceptance-averaged values are used to extract $R$ and apply the appropriate proton cross-section and $G_E^n$ corrections to $G_M^n$. For this purpose, the ``true" particle four-momentum at the vertex is taken from MC, free from any smearing from post-vertex kinematics or detector resolutions. The \q and \ep calculated from these variables are then averaged over all events passing the analysis cuts to determine the acceptance-averaged values. A summary of these values, along with the spread of the corresponding distributions across all \gmn kinematics, is provided in \tab \ref{tab:ch5:accavgq2ep}.

\begin{table}
    \caption{Acceptance averaged \q and \ep values along with the spread of the range of the corresponding distribution for all \gmn kinematics. The \q values are quoted in units of (GeV/c)$^2$.}
    \centering
    \begin{tabular}{cccccc} \hline\hline\vspace{-1em} \\ 
        $\langle Q^2 \rangle$ & $Q^2_{min}$ & $Q^2_{max}$ & $\langle \epsilon \rangle$ & $\epsilon_{min}$ & $\epsilon_{max}$ \\ \hline \vspace{-1.1em} \\
        2.989 & 2.560 & 3.520 & 0.722 & 0.660 & 0.770 \\
        4.488 & 4.000 & 5.067 & 0.515 & 0.450 & 0.575 \\
        7.464 & 7.147 & 7.787 & 0.469 & 0.435 & 0.505 \\
        9.834 & 9.013 & 10.773 & 0.502 & 0.430 & 0.565 \\
        13.465 & 12.587 & 14.400 & 0.417 & 0.355 & 0.475
         \\ 
         \hline\hline
    \end{tabular}
    \label{tab:ch5:accavgq2ep}
\end{table}
\section{Statistical Uncertainties}
\label{sec:ch5:staterror}
The statistical uncertainty on $R_{n/p}^{sf}$ is directly taken from the fit parameter error provided by CERN ROOT. This error is estimated using the covariance matrix derived from chi-squared minimization, which accounts for any correlations between the fit parameters. It is then propagated to $R$ as follows:
\begin{equation}
\label{eqn:ch5:errorproprsftor}
    \Delta(R)_{stat} = |R_{MC}^{Born}| \, \Delta (R_{n/p}^{sf})_{stat}
\end{equation}
The statistical uncertainty on  $R_{MC}^{Born}$ is zero, as it is known exactly. Finally, the statistical error on $G_M^n$ is evaluated using:
\begin{align}
\label{eqn:ch5:errorproprtogmn}
    \Delta(G_M^n)_{stat} = \frac{1}{2}\left| \frac{1}{G_M^n} \frac{1}{\tau_n}\frac{\epsilon_{n}(1+\tau_{n})}{\epsilon_{p}(1+\tau_{p})}\sigma^p_{Red} \right| \Delta(R)_{stat}
\end{align}
with the assumptions: $\Delta(\sigma^p_{Red})_{stat} = 0$ and $\Delta(G_E^n)_{stat} = 0$.

\section{Systematic Uncertainties}
\label{sec:ch5:syserror}
Given the availability of sufficient statistics, the overall precision of the \gmn measurements will be primarily determined by systematic uncertainties. Fortunately, the ratio method of measurement minimizes many systematic errors, including DAQ live time, trigger efficiency, electron track reconstruction efficiency, and target density. The remaining significant sources of systematic uncertainty and the methods used to quantify them are discussed in this section.
\subsection{Inelastic Contamination}
The contamination from inelastic scattering within the quasi-elastic event sample is significant, even for the lowest-\q dataset, making it the largest source of systematic uncertainty in \gmn measurements. Since the ``true” shape of this background in the \dx distribution is unknown, it cannot be precisely subtracted from the data. However, as the underlying distribution is smooth and symmetric, it can be effectively modeled and subtracted, as detailed in Sections \ref{ssec:ch4:inelbg} and \ref{ssec:ch4:rsfintro}. For each dataset, the data/MC fit to extract $R_{n/p}^{sf}$ is repeated multiple times, using consistent signal shapes but varying background models. Specifically, five different background models, listed in the above-mentioned sections, are employed. The consistency of the extracted $R_{n/p}^{sf}$ values across these fits reflects how accurately the inelastic background was modeled. Therefore, the standard deviation of these values is quoted as the systematic uncertainty due to inelastic contamination. The propagation of this error to $R$ and $G_M^n$ is straightforward and follows Equations \ref{eqn:ch5:errorproprsftor} and \ref{eqn:ch5:errorproprtogmn}, respectively. The error numbers are summarized in \tab \ref{tab:ch5:systematicerror}.
\subsection{HCAL Detection Efficiency}
Any discrepancy in the description of HCAL nucleon detection efficiency (NDE) between data and MC affects the extraction of \rsf, necessitating the quantification of associated uncertainty as part of the systematic error estimation. As discussed in \sect \ref{sec:ch4:hcalnde}, a rigorous analysis was performed to determine the uniformity of proton detection efficiency within HCAL acceptance using elastic \heep events. Based on these findings, an efficiency map was created to capture efficiency differences within HCAL acceptance effectively. Using this map, the relative weights of the MC events were adjusted to replicate the efficiency variation observed in the data. With this efficiency correction, the efficiency profile observed in the data could be effectively reproduced in the otherwise uniform MC. The difference in \rsf values extracted with and without the efficiency correction provides an upper bound on the uncertainty related to HCAL NDE uniformity and is thus quoted as the corresponding systematic uncertainty.

The efficiency map used to address HCAL NDE non-uniformity is based solely on proton detection efficiency (pDE) estimated from data, as no straightforward method exists to directly measure neutron detection efficiency (nDE) from \gmn data. Although the strong agreement between MC and data for pDE provides confidence that MC reliably represents nDE within the required error margin, it is still necessary to establish an upper bound on the uncertainty of HCAL nDE. Additionally, the inability to calculate HCAL proton detection efficiency reliably at higher \q kinematics due to significant inelastic background introduces further uncertainty that also requires careful quantification. In this work, a 2$\%$ systematic error on \rsf has been assigned across all kinematic settings to account for these uncertainties based on informed estimates, though dedicated analysis efforts are ongoing for a more precise evaluation.

\subsection{Cut Stability}
As discussed in \sect \ref{ssec:ch4:cutoptim}, the optimal set of analysis cuts was determined by examining the stability of the experimental observable as a function of the cut variables. A systematic approach, guided by both data and MC, was employed to identify the stability region. However, this region is more distinct for some cut variables than for others. The diminishing signal-to-background ratio at higher-\q further complicates this determination, making it more uncertain. Quantifying the associated uncertainty is crucial as part of the systematic error in measuring \rsf.

\begin{figure}[h!]
    \centering
    \includegraphics[width=1\columnwidth]{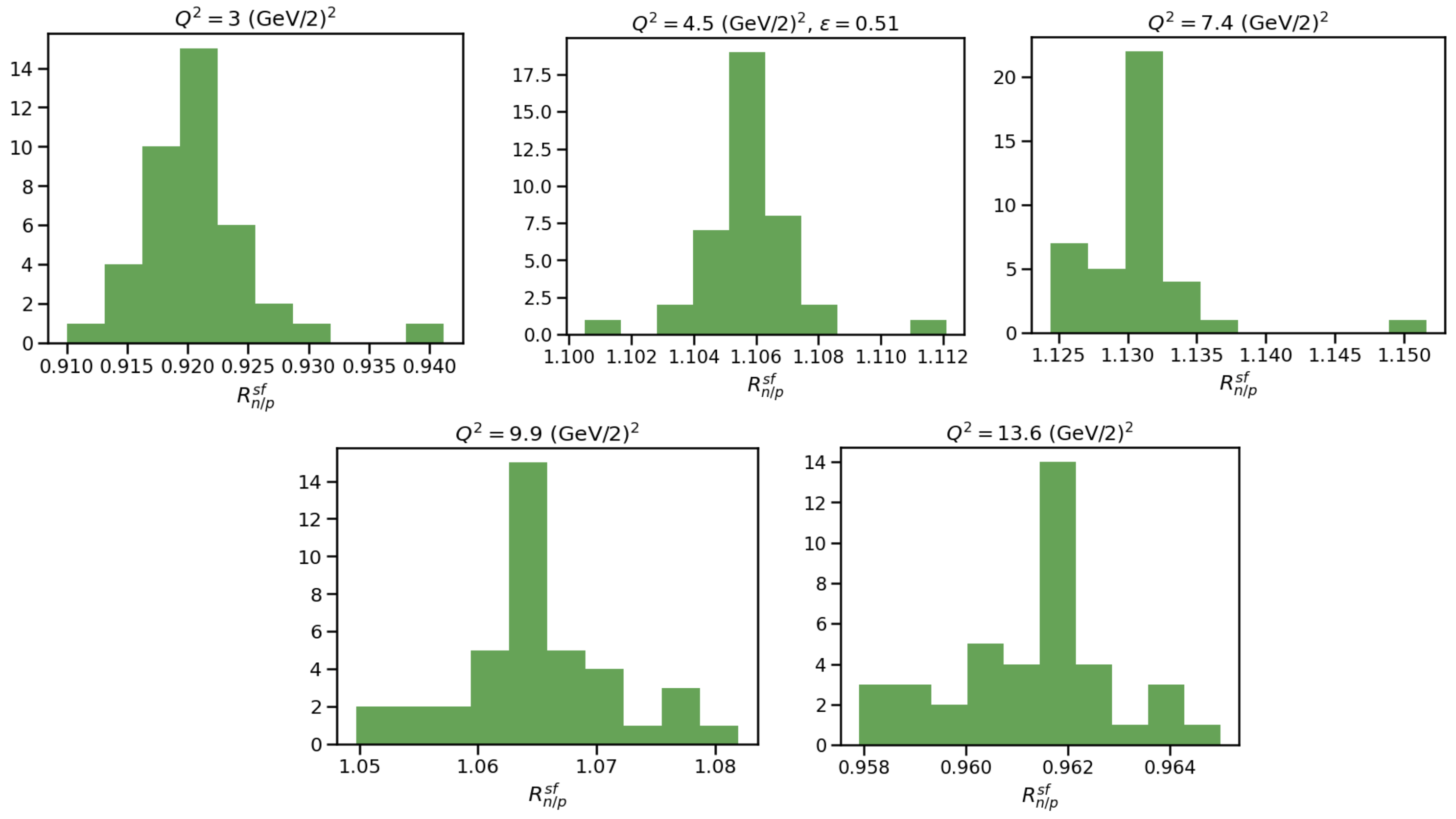}
    \caption{\label{fig:ch5:cutstability} Distribution of \rsf for all \gmn kinematics, resulting from the cut stability study. ``Poly2" background model has been used to perform these extractions.}
\end{figure}
The approach involves varying the range (both lower and upper separately, if applicable) of each cut variable by $\pm10\%$ while keeping the other cuts at their optimized values. For each variation, a data/MC fit to the \dx distribution is performed to extract \rsf\footnote{Care was taken to ensure that the cuts were varied equivalently between data and MC signals.}. The standard deviation of the resulting \rsf distribution reflects the spread of the data due to sensitivity to the analysis cuts, thereby quantifying the associated uncertainty. All cut variables except for $N_{hit}^{GEM}$, $t_{coin}^{ADC}$, and $Size_{clus}^{GRINCH}$ listed in \tab \ref{tab:ch4:optcut} were varied for a given kinematic setting to determine the systematic uncertainty due to cut stability. \fig \ref{fig:ch5:cutstability} shows the distribution of \rsf values, determined with the ``Poly2" background model, resulting from the cut stability studies across all \gmn kinematics. In each case, the distributions approximately follow a Gaussian shape, as expected.
\subsection{Final State Interaction}
In quasi-elastic electron-deuteron scattering, the struck nucleon can interact with the spectator nucleon, a phenomenon known as final state interaction (FSI). FSIs lead to significant modifications in the momentum and angular distributions of outgoing nucleons, affecting the cross-section. However, because this effect is largely similar for neutrons and protons, it nearly cancels out in the cross-section ratio. Nevertheless, it is necessary to calculate the associated correction factor—larger at lower momentum transfers due to stronger inter-nucleon interactions—using theoretical models and apply it to the extracted ratio.

SIMC, the quasi-elastic event generator used for \gmn analysis, does not include models for FSIs; therefore, such corrections must be calculated and applied separately. Initial discussions with theorists suggest a correction factor of approximately $0.5\%$ on $R$ for low-\q kinematics. Further efforts are ongoing to accurately compute correction factors for high-\q data points. In this work, a $0.5\%$ systematic error on $R$ has been assigned across all kinematic settings to account for FSI-related corrections.    

\subsection{$\sigma_{Red}^p$ Estimation}
Proton reduced cross section ($\sigma_{Red}^p$) is evaluated using multiple models for a given kinematics to estimate the associated systematic uncertainty. This process involves calculating $G_M^n$ from $R$ using three different models for $\sigma_{Red}^p$ while keeping the $G_E^n$ value fixed to the one obtained from Ye \textit{et al.}. The models used are:

\begin{enumerate} 
    \item Ye \textit{et al.} \cite{YE20188}, 
    \item Kelly \textit{et al.} \cite{PhysRevC.70.068202}, and 
    \item Arrington \textit{et al.} \cite{PhysRevC.76.035205} 
\end{enumerate}
The standard deviation of the resulting $G_M^n$ values is then quoted as the associated systematic uncertainty for $\sigma_{Red}^p$ estimation.

It is worth noting that the models used in this work include corrections for the two-photon exchange (TPE) cross-section in $ep$ scattering. The ideal approach, however, would be to use a TPE-uncorrected model to estimate $\sigma_{Red}^{p}$ and then multiply it by $R$ to obtain the TPE-uncorrected Born cross-section for $en$ scattering. This result could then be corrected for the TPE contribution to yield the neutron Born cross-section. Finally, the ``true" $G_M^n$ value can be extracted from the neutron Born cross-section by applying the $G_E^n$ correction. Efforts are ongoing to implement this approach.

\subsection{$G_E^n$ Estimation}
The same procedure used to estimate systematic uncertainly on $\sigma_{Red}^p$ is used here as well. $G_M^n$ is calculated from $R$ using the following three models for $G_E^n$ while keeping the $\sigma_{Red}^p$ value fixed to the one obtained from Ye \textit{et al.}:
\begin{enumerate}
    \item Ye \textit{et al.} \cite{YE20188}
    \item Kelly \textit{et al.} \cite{PhysRevC.70.068202}, and 
    \item Galster \textit{et al.} \cite{GALSTER1971221}
\end{enumerate}
The standard deviation of the resulting $G_M^n$ values is then quoted as the associated systematic uncertainty for $G_E^n$ estimation.
\subsection{Total Systematic Error Budget}
\tab \ref{tab:ch5:systematicerror} provides a summary of the systematic errors from all the sources discussed above for each \gmn kinematics. These values are quoted for the $R$ value extracted using the ``Poly2" background model and the $\frac{G_M^n}{\mu_nG_D}$ value calculated from this $R$ value, with the reduced proton cross section and $G_E^n$ values estimated using the Ye parametrization. \cite{YE20188}

\begin{table}[h!]
    \centering
    \caption{Total systematic error budget for \gmn kinematics. The \q and \ep values are central values, with \q quoted in (GeV/c)$^2$. Among the sources of systematic error, Inel. represents inelastic contamination, NDE1 refers to HCAL nucleon detection efficiency non-uniformity, NDE2 pertains to HCAL neutron detection efficiency and the inability to reliably estimate proton detection efficiency for high-\q kinematics, Cut S. indicates cut stability, and FSI represents final state interactions. Errors from individual sources have been combined in quadrature to calculate the total error. Refer to the text for further details.}
    \label{tab:ch5:systematicerror}
    \begin{tabular}{ccccccc} \hline\hline\vspace{-1em} \\
         & Error            & \multicolumn{5}{c}{\q (\ep)} \\ \cline{3-7} \vspace{-1em} \\
         & Sources          & 3 (0.72) & 4.5 (0.51) & 7.4 (0.46) & 9.9 (0.50) & 13.6 (0.41) \\ \hline \vspace{-1em} \\        
         \multirow{6}{*}{$\Delta(R)_{sys}$}  & Inel.            & 0.0014 & 0.0056 & 0.0030 & 0.0045 & 0.0130\\
         & NDE1             & 0.0004 & 0.0007 & 0.0011 & 0.0011 & 0.0040\\
         & NDE2             & 0.0076 & 0.0081 & 0.0079 & 0.0077 & 0.0072 \\
         & Cut S.           & 0.0006 & 0.0006 & 0.0015 & 0.0024 & 0.0020\\
         & FSI              & 0.0019 & 0.0020 & 0.0020 & 0.0019 & 0.0018\\\cline{2-7}
         & Total            & 0.0080 & 0.0101 & 0.0089 & 0.0095 & 0.0156\\
         \hline\hline\vspace{-1em} \\
         \multirow{7}{*}{$\Delta(\frac{G_M^n}{\mu_nG_D})_{sys}$}& Inel.            & 0.0019 & 0.0068 & 0.0035 & 0.0049 & 0.0139\\
         & NDE1             & 0.0005 & 0.0009 & 0.0012 & 0.0012 & 0.0043\\
         & NDE2             & 0.0098 & 0.0098 & 0.0091 & 0.0085 & 0.0076\\
         & Cut S.           & 0.0008 & 0.0007 & 0.0018 & 0.0027 & 0.0022\\
         & FSI              & 0.0024 & 0.0024 & 0.0023 & 0.0021 & 0.0019\\         
         & $\sigma_{Red}^p$ & 0.0080 & 0.0090 & 0.0123 & 0.0129 & 0.0102\\
         & $G_E^n$          & 0.0053 & 0.0061 & 0.0054 & 0.0052 & 0.0038\\\cline{2-7}
         & Total            & 0.0141 & 0.0163 & 0.0169 & 0.0174 & 0.0199\\
        \hline\hline
    \end{tabular}
\end{table}


\section{Discussion of the Results}
\tab \ref{tab:ch5:results} presents the preliminary $R$ and $G_M^n$ values, along with associated statistical and systematic uncertainties, for all \gmn kinematics. The systematic errors on $R$ dominate the overall uncertainty across all kinematics except for \qeq{9.9}, where statistical error prevails due to relatively low statistics. This lower data yield was caused by a significant loss of beam time during the experimental run. Notably, the quoted systematic uncertainties due to HCAL neutron detection efficiency and final state interactions (FSIs) are currently based on educated estimates. Efforts are underway to refine the quantification of these effects. 
\begin{table}[]
    \caption[Preliminary results]{Preliminary results. The $\langle Q^2 \rangle$ and $\langle \epsilon \rangle$) values are acceptance averaged values, with \q quoted in (GeV/c)$^2$.}
    \label{tab:ch5:results}
    \centering
    \begin{tabular}{cccc} \hline\hline\vspace{-1em} \\
       $\langle Q^2 \rangle$  & $\langle \epsilon \rangle$ & $R\pm\Delta(R)_{stat}\pm\Delta(R)_{sys}$ & $\frac{G_M^n}{\mu_nG_D}\pm\Delta(\frac{G_M^n}{\mu_nG_D})_{stat}\pm\Delta(\frac{G_M^n}{\mu_nG_D})_{sys}$ \\ \vspace{-1em} \\ \hline \vspace{-1em} \\ 
        2.989  & 0.722 & $0.3808\pm0.0025\pm0.0080$ & $0.9774\pm0.0033\pm0.0141$\\
        4.488  & 0.515 & $0.4037\pm0.0011\pm0.0101$ & $0.9763\pm0.0014\pm0.0163$\\
        7.464  & 0.469 & $0.3974\pm0.0035\pm0.0089$ & $0.9071\pm0.0042\pm0.0169$\\
        9.834  & 0.502 & $0.3868\pm0.0154\pm0.0095$ & $0.8473\pm0.0173\pm0.0174$\\
        13.465 & 0.417 & $0.3615\pm0.0100\pm0.0156$ & $0.7582\pm0.0107\pm0.0199$\\
        \hline\hline
    \end{tabular}
\end{table}
\begin{figure}[]
    \centering
    \includegraphics[width=1\columnwidth]{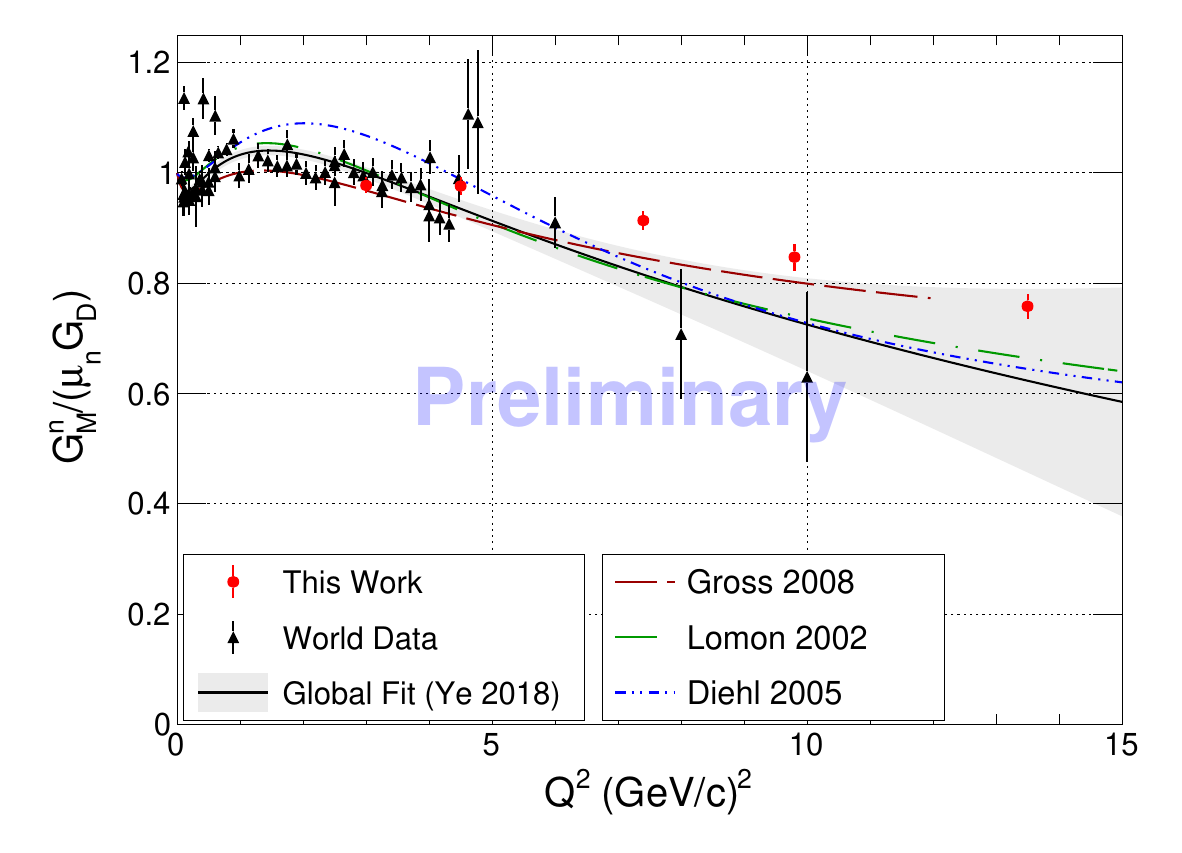}
    \caption{\label{fig:ch5:gmn} $G_M^n$ world data, including the preliminary results from this work. The error bars represent the total error, calculated by combining statistical and systematic errors in quadrature. The global fit is sourced from \cite{YE20188}. Among the theory curves, the one by Gross et al. \cite{PhysRevC.77.015202} is based on the covariant spectator model, Lomon et al. on the vector meson dominance (VMD) model, and Diehl on GPD calculations. See \sect \ref{ssec:ch2:neutronff} for a detailed overview of the $G_M^n$ world data.}
\end{figure}
\hspace{-0.5em}Additionally, the two-photon exchange (TPE) contribution to the $eD$ scattering cross-section has been omitted in the calculation of the $G_M^n$ values presented here, but efforts are ongoing to quantify these corrections, which are expected to be non-negligible.

When plotting the obtained $G_M^n$ values alongside existing world data, as shown in \fig \ref{fig:ch5:gmn}, interesting insights emerge. The preliminary results at lower \q points closely align with existing world data, while the results at higher \q greatly extend the range in which $G_M^n$ is precisely known. The observed trend suggests a slower falloff of $G_M^n/\mu_nG_D$ than previously anticipated, implying a more compact neutron structure. The model by Gross et al. \cite{PhysRevC.77.015202}, published in 2008, appears to most closely capture this observed trend. Their calculation is based on the covariant spectator model, where the virtual photon interacts with a single quark while the remaining quarks are treated as an on-shell diquark with a definite mass. A comprehensive interpretation of these results will likely necessitate significant refinement of existing nucleon models and parametrizations, thereby enhancing our understanding of the neutron's internal structure.   

\appendix
\chapter{Supplementary Derivations for the Rosenbluth Formula}
\label{appen:rosenbluth}
This section will provide detailed intermediate steps in the derivation of the Rosenbluth formula presented in \sect \ref{sec:ch1:eNscattering}.
\section{Useful Identities}
\label{sec:appena:identities}
Minkowski Metric Tensor:
\begin{equation}
    \begin{aligned}
        g = \text{diag}(1,&-1,-1,-1) \\
        g^{\mu\nu}g_{\mu\nu} &= 4                    
    \end{aligned}    
\end{equation}
Gamma Matrices:
\begin{equation}
    \begin{aligned}
        \{\gamma^\mu,\gamma^\nu\} = 2g^{\mu\nu}
    \end{aligned}
\end{equation}
Traces:
\begin{equation}
\label{eqn:appena:traceidentities}
    \begin{aligned}
        \text{Tr}(A+B) &= \text{Tr}(A) + \text{Tr}(B)   \\
        \text{Tr}(\alpha A) &= \alpha\text{Tr}(A) \\
        \text{Tr}(AB) &= \text{Tr}(BA) \\
        \text{Tr}(\gamma^\mu\gamma^\nu) &= g^{\mu\nu} \\
        \text{Tr}(\gamma^\mu\gamma^\nu\gamma^\lambda\gamma^\sigma) &= 4\left(g^{\mu\nu}g^{\lambda\sigma} - g^{\mu\lambda}g^{\nu\sigma} + g^{\mu\sigma}g^{\nu\lambda}\right) \\
        \text{The trace of the product of an}\,&\,\text{odd number of gamma matrices in zero.}
    \end{aligned}
\end{equation}
Casimir's Trick:
\begin{equation}
\label{eqn:appena:casimirstrick}
    \sum_{\text{all spins}} [\Bar{u}(a)\Gamma_1u(b)][\Bar{u}(a)\Gamma_2u(b)] = \text{Tr}[\Gamma_1(\slashed{p_b}+m_b)\Bar{\Gamma}_2(\slashed{p_a}+m_a)], \quad \text{where}\,\,\Bar{\Gamma}_2=\gamma^0\Gamma_2^\dagger \gamma^0
\end{equation}
Lab Frame Kinematics (see \sect \ref{ssec:ch1:labframekine}):
\begin{equation}
\label{eqn:appena:labframe}
    \begin{aligned}
        q^2 &= (k-k')^2 = -2k\cdot k' = -2M(E_e-E'_e) = -4E_eE'_e\sin^2\frac{\theta_e}{2} \\
        &= (p'-p)^2 = 2M^2 - 2p\cdot p' \\
        k\cdot q  &= -k'\cdot q = \frac{q^2}{2}\\
        k\cdot p  & = k'\cdot p' = ME_e \\
        k\cdot p' &= k'\cdot p = ME_e + \frac{q^2}{2} 
    \end{aligned}
\end{equation}
\section{Hadronic Tensor}
\label{sec:appena:derivehadronictensor}
The hadronic tensor ($W^N_{\mu\nu}$) is defined as follows (see \eqn \ref{eqn:ch1:hadronictensor}):
\begin{alignat}{2}
    W^N_{\mu\nu} &= \frac{1}{2} \sum_{r',r} [\Bar{u}^{r'}(p')(-ie\Gamma_\mu)u^r(p)] [\Bar{u}^{r'}(p')(-ie\Gamma_\nu)u^r(p)]^* \notag \\
        &= \frac{1}{2}\,\text{Tr}[\Gamma_\mu(\slashed{p} + M)\Gamma_\nu(\slashed{p'} + M)] \qquad [\text{using}\,\,\eqref{eqn:appena:casimirstrick}] \notag \\
        &= \frac{1}{2}\,\text{Tr} \left[ \left(\gamma_\mu (F_1+F_2) - \frac{p'_{\mu} + p_\mu}{2M} F_2\right) (\slashed{p} + M)  \left(\gamma_\nu (F_1+F_2) - \frac{p'_{\nu} + p_\nu}{2M} F_2\right) (\slashed{p'} + M) \right] \notag\\
        &= \frac{1}{2}\,\text{Tr}\left[ (F_1+F_2)^2\gamma_\mu(\slashed{p}+M)\gamma_\nu(\slashed{p'}+M) - (F_1+F_2)F_2\frac{p'_\mu+p_\mu}{2M}(\slashed{p}+M)\gamma_\nu(\slashed{p'}+M) \right. \notag\\
        &\qquad\qquad \left.- (F_1+F_2)F_2\frac{p'_\nu+p_\nu}{2M}\gamma_\mu(\slashed{p}+M)(\slashed{p'}+M) + F_2^2\frac{p'_\nu+p_\nu}{2M}\frac{p'_\mu+p_\mu}{2M}(\slashed{p}+M)(\slashed{p'}+M) \right] \notag\\
        &= \frac{1}{2}\,\text{Tr}\left[ (F_1+F_2)^2\gamma_\mu(\slashed{p}+M)\gamma_\nu(\slashed{p'}+M) - 2(F_1+F_2)F_2\frac{p'_\mu+p_\mu}{2M}(\slashed{p}+M)\gamma_\nu(\slashed{p'}+M) \right. \notag\\
        \displaybreak         
        &\qquad\qquad \left. + F_2^2\frac{p'_\nu+p_\nu}{2M}\frac{p'_\mu+p_\mu}{2M}(\slashed{p}+M)(\slashed{p'}+M) \right] \notag\\       
        &= (F_1+F_2)^2 \frac{1}{2}\,\text{Tr}[\gamma_\mu(\slashed{p}+M)\gamma_\nu(\slashed{p'}+M)] - 2(F_1+F_2)F_2\frac{p'_\mu+p_\mu}{2M} \frac{1}{2}\,\text{Tr}[(\slashed{p}+M)\gamma_\nu(\slashed{p'}+M)] \notag\\
        &\quad + F_2^2\frac{p'_\nu+p_\nu}{2M}\frac{p'_\mu+p_\mu}{2M}\frac{1}{2}\,\text{Tr}[(\slashed{p}+M)(\slashed{p'}+M)] \notag\\
        &= (F_1+F_2)^2\, 2\,[p_\mu p'_\nu + p'_\mu p_\nu - g_{\mu\nu}(p\cdot p' - M^2)] - 2(F_1+F_2)F_2\frac{p'_\mu+p_\mu}{2M} 2M(p'_\nu + p_\nu) \notag\\
        &\quad + F_2^2\frac{p'_\nu+p_\nu}{2M}\frac{p'_\mu+p_\mu}{2M}2(p\cdot p' + M^2) \qquad [\text{using}\,\,\eqref{eqn:appena:traceidentities}] \notag\\
        &= 2\, \bigg\{\bigg.(F_1+F_2)^2[p_\mu p'_\nu + p'_\mu p_\nu - g_{\mu\nu}(p\cdot p' - M^2)] - (F_1+F_2)F_2 (p'_\mu+p_\mu) (p'_\nu+p_\nu)] \notag\\
        &\qquad\quad + \frac{F_2^2}{4M^2} (p'_\mu+p_\mu) (p'_\nu+p_\nu) (p\cdot p' + M^2) \bigg.\bigg\} \notag\\
        &= 2\, \left[(F_1+F_2)^2 \mathcal{A} - (F_1F_2 + F_2^2)\mathcal{B} + \frac{F_2^2}{4M^2}\mathcal{B}(2M^2-\frac{q^2}{2})\right]
\end{alignat}
where $\mathcal{A}=p_\mu p'_\nu + p'_\mu p_\nu - g_{\mu\nu}(p\cdot p' - M^2)$ and $\mathcal{B}=(p'_\mu+p_\mu) (p'_\nu+p_\nu)$. In the last line, we have used the relation $p\cdot p' + M^2=2M^2-\frac{q^2}{2}$, assuming lab frame kinematics (see $\eqref{eqn:appena:labframe}$). 

Next, we replace $F_1F_2$ with $\frac{1}{2}[(F_1+F_2)^2 - F_1^2 - F_2^2]$, followed by some straightforward rearrangements. This leads to the simplified expression for $W^N_{\mu\nu}$ in the lab frame, as follows:
\begin{equation}
\begin{aligned}
    W_{\mu\nu}^N &= (p'_\mu + p_\mu)(p'_\nu + p_\nu)\left(F_1^2 - \frac{q^2}{4M^2}F_2^2\right) \\
    &\qquad - [(p'_\mu - p_\mu)(p'_\nu - p_\nu) + 2g_{\mu\nu} (p\cdot p' - M^2)] (F_1 + F_2)^2
\end{aligned}
\end{equation}
\section{Differential Phase Space}
\label{sec:appena:derivephasespace}
The general expression for the differential cross section for two-body scattering is given by (see \eqn \ref{seqn:dphi}):
\begin{align}
    d\Phi &= (2\pi)^4 \delta^4 (k + p - k' - p') \frac{d^3\vb{k'}}{(2\pi)^32E'_e} \frac{d^3\vb{p'}}{(2\pi)^32E'_N} \notag\\
          &= \frac{1}{16\pi^2} \frac{\delta^0 (E_e+M-E'_e-E'_N)}{E'_eE'_N} \delta^3(\vb{k}-\vb{k'}-\vb{p'})d^3\vb{p'}d^3\vb{k'} \notag\\
          &= \frac{1}{16\pi^2} \frac{\delta^0 \left(E_e+M-E'_e-\sqrt{(\vb{k}-\vb{k'})^2+M^2}\right)}{E'_e \sqrt{(\vb{k}-\vb{k'})^2+M^2}} {E'_e}^2dE'_ed\Omega \label{eqn:appena:dphi0}
\end{align}
The last line has been obtained by integrating over the outgoing nucleon momentum ($\vb{p'}$). Here, $|\vb{k}|=E_e$ and $|\vb{k'}|=E'_e$\footnote{Neglecting the rest mass of the electron.}, therefore,
\begin{equation}
\label{eqn:appena:dphi1}
    (\vb{k}-\vb{k'})^2 = E^2_e + {E'_e}^2 - 2E_eE_e'\cos\theta_e
\end{equation}
Now, considering
\begin{equation}
\label{eqn:appena:dphi2}
    z = E'_e + \sqrt{E^2_e + {E'_e}^2 - 2E_eE_e'\cos\theta_e + M^2},
\end{equation}
we obtain,
\begin{equation}
\label{eqn:appena:dphi13}
    \frac{dz}{dE'_e} = \frac{z-E_e\cos\theta_e}{\sqrt{z-E'_e}}
\end{equation}
Substituting \eqref{eqn:appena:dphi1}, \eqref{eqn:appena:dphi2}, and \eqref{eqn:appena:dphi13} into \eqref{eqn:appena:dphi0}, we obtain,
\begin{align}
    d\Phi &= \frac{1}{16\pi^2} \frac{\delta^0 (E_e+M-z)}{E'_e(z-E_e\cos\theta_e)}{E'_e}^2dzd\Omega \notag\\
    &= \frac{1}{16\pi^2} \frac{{E'_e}^2}{E'_e(E_e+M-E_e\cos\theta_e)} d\Omega \notag\\
    &= \frac{1}{16\pi^2M} \frac{E'_e}{E_e} \left(\frac{E_e}{1+\frac{E_e}{M}(1-\cos\theta_e)}\right) d\Omega \notag\\
    &= \frac{1}{16\pi^2M} \frac{E'^{2}_e}{E_e} d\Omega
\end{align}
This is the desired expression for the differential phase space in the lab frame for two-body scattering.
\section{Spin Averaged Squared Amplitude}
\label{sec:appena:deriveamplitude}
The expression for the spin averaged squared amplitude $\langle |\mathcal{M}|^2 \rangle$ is given by \eqn \ref{eqn:ch1:squaredamplitudeintermediate} as:
\begin{align}
\label{eqn:ch1:squaredamplitudeintermediateappena}
    \langle |\mathcal{M}|^2 \rangle &= \frac{e^4}{q^4} L^{\mu\nu}_e W^N_{\mu\nu} \notag\\
    &= \frac{2e^4}{q^4} (k^\mu k'^{\nu} + k'^{\mu} k^{\nu} - g^{\mu\nu}k\cdot k') \notag\\
    &\quad\qquad \times \bigg\{\bigg.(p'_\mu + p_\mu)(p'_\nu + p_\nu)\left(F_1^2 - \frac{q^2}{4M^2}F_2^2\right) \notag\\
    &\quad\qquad\qquad - [(p'_\mu - p_\mu)(p'_\nu - p_\nu) + 2g_{\mu\nu} (p\cdot p' - M^2)] (F_1 + F_2)^2 \bigg.\bigg\} \notag\\
    &= \frac{2e^4}{q^4} \left[ \mathcal{A} \left(F_1^2 - \frac{q^2}{4M^2}F_2^2\right) - \mathcal{B} (F_1 + F_2)^2 \right]
\end{align}
where
\begin{equation}
    \begin{aligned}
        \mathcal{A} &= (k^\mu k'^{\nu} + k'^{\mu} k^{\nu} - g^{\mu\nu}k\cdot k') (p'_\mu + p_\mu)(p'_\nu + p_\nu) \\
        \mathcal{B} &= (k^\mu k'^{\nu} + k'^{\mu} k^{\nu} - g^{\mu\nu}k\cdot k') [(p'_\mu - p_\mu)(p'_\nu - p_\nu) + 2g_{\mu\nu} (p\cdot p' - M^2)] 
    \end{aligned}
\end{equation}
We will evaluate these coefficients separately for convenience. 

\subheading{Evaluating $\mathcal{A}$}
\vspace{-3em}
\begin{equation}
\label{eqn:appena:mathcala}
    \mathcal{A} = (\underbrace{k^\mu k'^{\nu}}_{(1)} + \underbrace{k'^{\mu} k^{\nu}}_{(2)} - \underbrace{g^{\mu\nu}k\cdot k'}_{(3)}) \underbrace{(p'_\mu + p_\mu)(p'_\nu + p_\nu)}_{(4)}
\end{equation}
$(1)\times(4)$:
\begin{equation}
\label{eqn:appena:mathcala1}
    \begin{aligned}
        k^\mu k'^{\nu} (p'_\mu + p_\mu)(p'_\nu + p_\nu) &= (k\cdot p' + k\cdot p) (k'\cdot p' + k'\cdot p) \\
        &= \left(2ME_e + \frac{q^2}{2}\right)\left(2ME_e + \frac{q^2}{2}\right) \qquad[\text{using}\,\,\eqref{eqn:appena:labframe}]
    \end{aligned}
\end{equation}
$(2)\times(4)$:
\begin{equation}
\label{eqn:appena:mathcala2}
    \begin{aligned}
        k'^{\mu} k^{\nu} (p'_\mu + p_\mu)(p'_\nu + p_\nu) &= (k'\cdot p' + k'\cdot p) (k\cdot p' + k\cdot p) \\
        &= \left(2ME_e + \frac{q^2}{2}\right)\left(2ME_e + \frac{q^2}{2}\right) \qquad[\text{using}\,\,\eqref{eqn:appena:labframe}]
    \end{aligned}
\end{equation}
$(3)\times(4)$:
\begin{equation}
\label{eqn:appena:mathcala3}
    \begin{aligned}
        g^{\mu\nu}k\cdot k'(p'_\mu + p_\mu)(p'_\nu + p_\nu) &= k\cdot k'(p'^\nu + p^\nu)(p'_\nu + p_\nu) \\
        &= k\cdot k' (p' + p)^2 \\
        &= k\cdot k' (2M^2 + 2p\cdot p') \\
        &= -\frac{q^2}{2}(4M^2-q^2) \qquad[\text{using}\,\,\eqref{eqn:appena:labframe}]
    \end{aligned}
\end{equation}
Now, substituting \eqref{eqn:appena:mathcala1}, \eqref{eqn:appena:mathcala2}, and \eqref{eqn:appena:mathcala3} into \eqref{eqn:appena:mathcala}, we obtain,
\begin{align}
    \mathcal{A} &= (1)\times(4) + (2)\times(4) - (3)\times(4) \notag\\
    &= 2 \left(2ME_e + \frac{q^2}{2}\right)^2 + \frac{q^2}{2}(4M^2-q^2) \notag\\
    &= 8M^2E_e^2 + 4ME_e(-2ME_e+2ME'_e) + 2M^2 \left(-4E_eE'_e\sin^2\frac{\theta_e}{2}\right) \qquad[\text{using}\,\,\eqref{eqn:appena:labframe}] \notag\\
    &= 8M^2E_eE'_e\cos^2\frac{\theta_e}{2} \notag\\
    &= -2M^2\frac{\cos^2\frac{\theta_e}{2}}{\sin^2\frac{\theta_e}{2}} \left(-4E_eE'_e\sin^2\frac{\theta_e}{2}\right) \notag\\
    &= -2M^2q^2\frac{\cos^2\frac{\theta_e}{2}}{\sin^2\frac{\theta_e}{2}} \label{eqn:appena:mathcalafinal}
\end{align}

\subheading{Evaluating $\mathcal{B}$}
\vspace{-3em}
\begin{equation}
\label{eqn:appena:mathcalb}
    \mathcal{B} = (\underbrace{k^\mu k'^{\nu}}_{(1)} + \underbrace{k'^{\mu} k^{\nu}}_{(2)} - \underbrace{g^{\mu\nu}k\cdot k'}_{(3)}) [\underbrace{(p'_\mu - p_\mu)(p'_\nu - p_\nu)}_{(4)} + \underbrace{2g_{\mu\nu} (p\cdot p' - M^2)}_{(5)}]
\end{equation}
$(1)\times(4)$:
\begin{equation}
\label{eqn:appena:mathcalb1}
    \begin{aligned}
        k^\mu k'^{\nu} (p'_\mu - p_\mu)(p'_\nu - p_\nu) &= (k\cdot p' - k\cdot p) (k'\cdot p' - k'\cdot p) 
    \end{aligned}
\end{equation}
$(2)\times(4)$:
\begin{equation}
\label{eqn:appena:mathcalb2}
    \begin{aligned}
        k'^{\mu} k^{\nu} (p'_\mu - p_\mu)(p'_\nu - p_\nu) &= (k'\cdot p' - k'\cdot p) (k\cdot p' - k\cdot p)
    \end{aligned}
\end{equation}
$(3)\times(4)$:
\begin{equation}
\label{eqn:appena:mathcalb3}
    \begin{aligned}
        g^{\mu\nu}k\cdot k'(p'_\mu - p_\mu)(p'_\nu - p_\nu) &= k\cdot k' (p'-p)^2 
    \end{aligned}
\end{equation}
$(1)\times(5)$:
\begin{equation}
\label{eqn:appena:mathcalb4}
    k^\mu k'^{\mu} \, 2 g_{\mu\nu} (p\cdot p' - M^2) = 2 k\cdot k' (p\cdot p' - M^2)
\end{equation}
$(2)\times(5)$:
\begin{equation}
\label{eqn:appena:mathcalb5}
    k'^{\mu} k^{\nu} \, 2 g_{\mu\nu} (p\cdot p' - M^2) = 2 k\cdot k' (p\cdot p' - M^2)
\end{equation}
$(3)\times(5)$:
\begin{equation}
\label{eqn:appena:mathcalb6}
    g^{\mu\nu}k\cdot k' \, 2 g_{\mu\nu} (p\cdot p' - M^2) = 8 k\cdot k' (p\cdot p' - M^2)
\end{equation}
Now, substituting \eqref{eqn:appena:mathcalb1}, \eqref{eqn:appena:mathcalb2}, \eqref{eqn:appena:mathcalb3}, \eqref{eqn:appena:mathcalb4}, \eqref{eqn:appena:mathcalb5}, and \eqref{eqn:appena:mathcalb6} into \eqref{eqn:appena:mathcalb}, we obtain,
\begin{align}
    \mathcal{B} &= (1)\times(4) + (2)\times(4) - (3)\times(4) + (1)\times(5) + (2)\times(5) - (3)\times(5) \notag\\
    &= 2 (k'\cdot p' - k'\cdot p) (k\cdot p' - k\cdot p) - k\cdot k' (p'-p)^2 - 4 k\cdot k' (p\cdot p' - M^2) \notag\\
    &= 2 k'\cdot (p'-p)\, k\cdot (p'-p) - k\cdot k' (p'-p)^2 - 4 k\cdot k' (p\cdot p' - M^2) \notag\\
    &= 2 \left(-\frac{q^2}{2}\right) \left(\frac{q^2}{2}\right) - \left(-\frac{q^2}{2}\right) (q^2) - 4 \left(-\frac{q^2}{2}\right) \left(-\frac{q^2}{2}\right)  \qquad[\text{using}\,\,\eqref{eqn:appena:labframe}] \notag\\
    &= - q^4 \label{eqn:appena:mathcalbfinal}
\end{align}

\subheading{Coming Back to $\langle |\mathcal{M}|^2 \rangle$}
Substituting \eqref{eqn:appena:mathcalafinal} and \eqref{eqn:appena:mathcalbfinal} into \eqref{eqn:ch1:squaredamplitudeintermediateappena}, we obtain,
\begin{align}
    \langle |\mathcal{M}|^2 \rangle &= \frac{e^4}{q^4} L^{\mu\nu}_e W^N_{\mu\nu} \notag\\
    &= \frac{2e^4}{q^4} \left[ -2M^2q^2\frac{\cos^2\frac{\theta_e}{2}}{\sin^2\frac{\theta_e}{2}} \left(F_1^2 - \frac{q^2}{4M^2}F_2^2\right) + q^4 (F_1 + F_2)^2 \right] \notag\\
    &= \frac{e^4M^2 \cos^2\frac{\theta_e}{2}}{E_eE'_e\sin^4\frac{\theta_e}{2}} \left[ \left(F_1^2 - \frac{q^2}{4M^2}F_2^2\right) - \frac{q^2}{2M^2} (F_1 + F_2)^2 \tan^2\frac{\theta_e}{2}\right] \qquad[\text{using}\,\,\eqref{eqn:appena:labframe}]
\end{align}
This is the desired expression for the spin averaged squared amplitude in the lab frame for the unpolarized elastic $eN$ scattering in one-photon exchange (OPE) approximation.
\chapter{Supplementary Plots for Data/MC Fits to \dx Distribution}
\label{appen:qeyield}
\section{\qeq{3}}
\begin{figure}[h!]
    \centering
    \includegraphics[width=0.8\columnwidth]{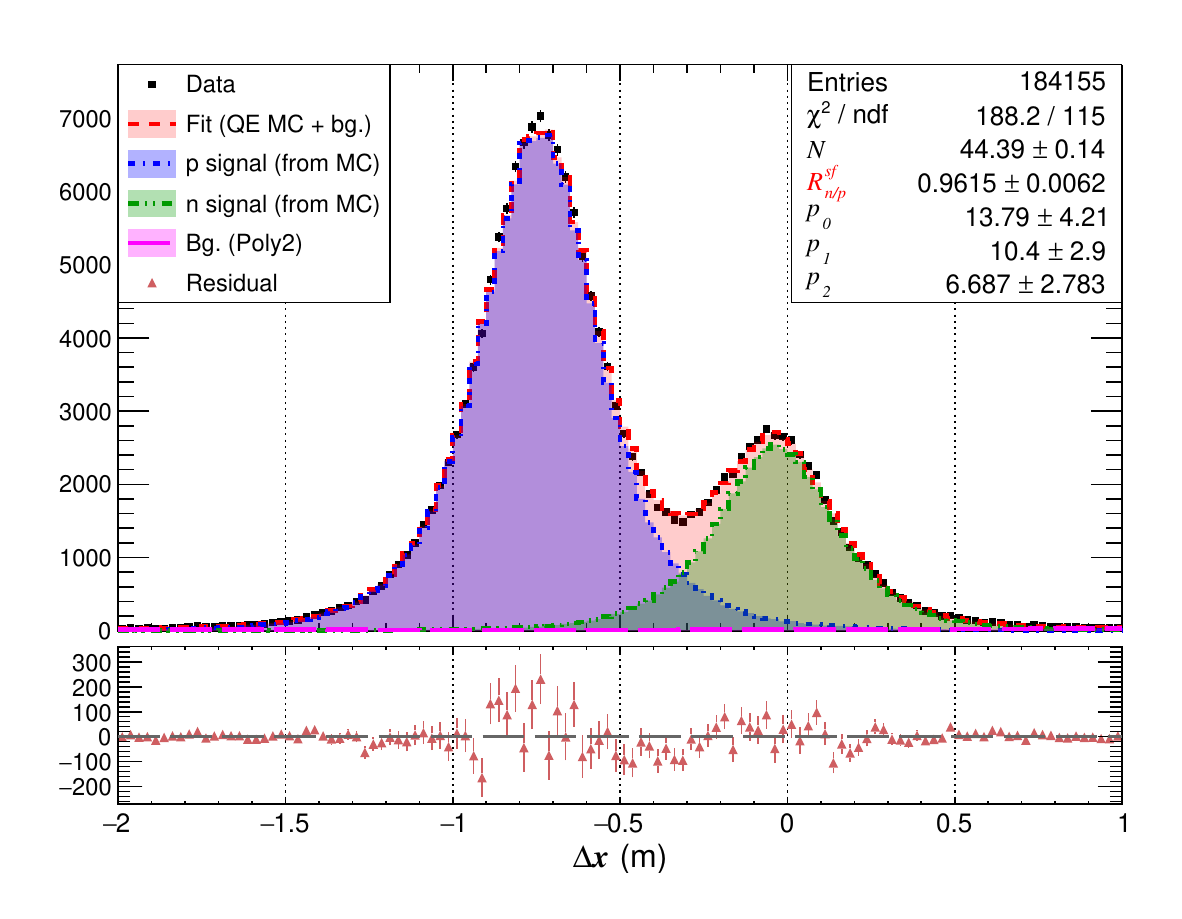}
    \caption{\label{fig:appen:dxfitsbs4} Example of data/MC fit to the \dx distribution for \qeq{3} kinematics using signal shapes from MC and second-order polynomial to model the background. Good electron events passing \w and \dy cuts are shown.}
\end{figure}
\begin{figure}[h!]
    \centering
    \includegraphics[width=0.9\columnwidth]{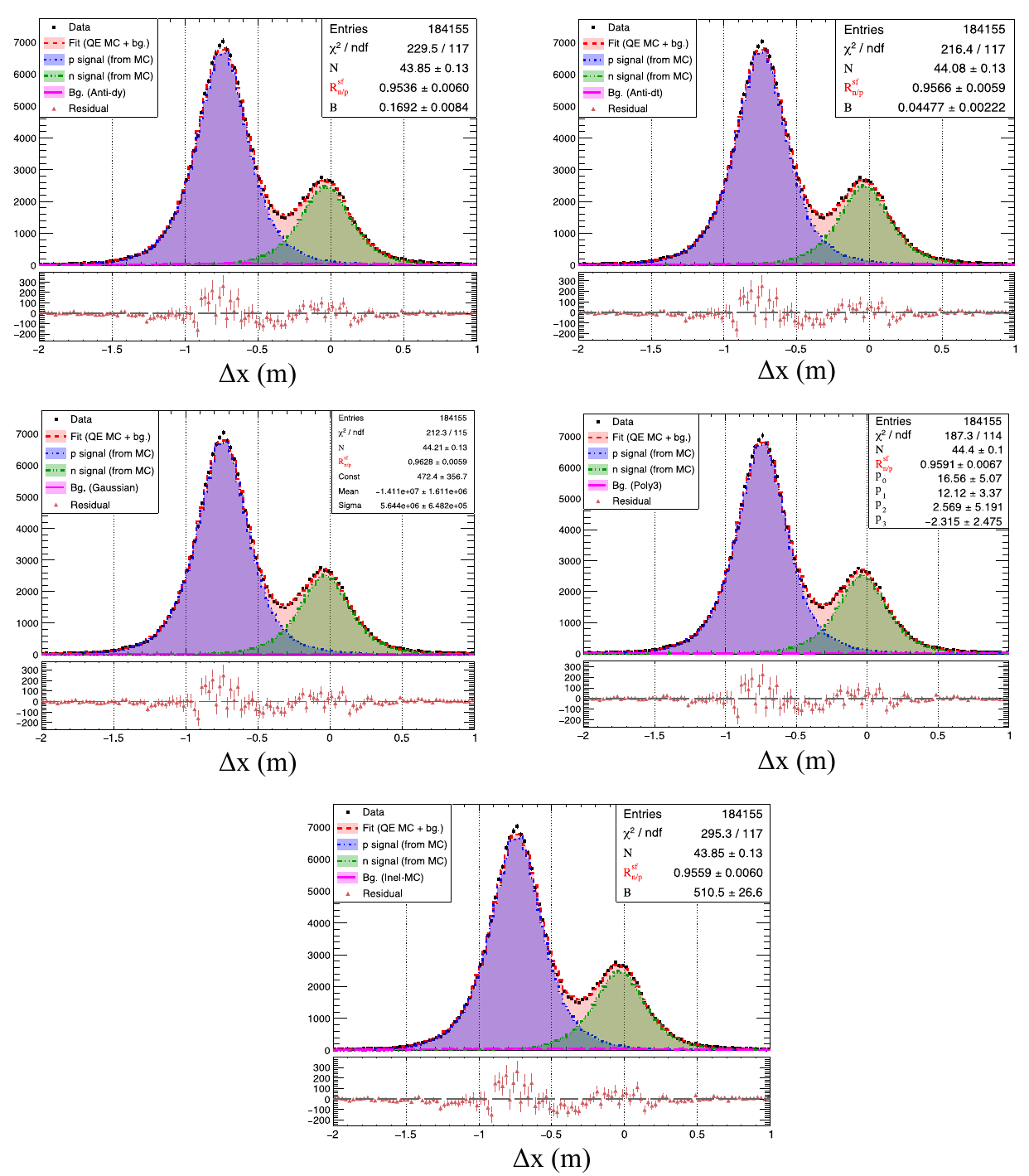}
    \caption{\label{fig:appen:dxfitsbs4_2} Comparison of data/MC fit to the \dx distribution for \qeq{3} kinematics between various estimations of background models. Good electron events passing \w and \dy cuts are shown.}
\end{figure}
\section{\qeq{4.5}}
\begin{figure}[h!]
    \centering
    \includegraphics[width=1\columnwidth]{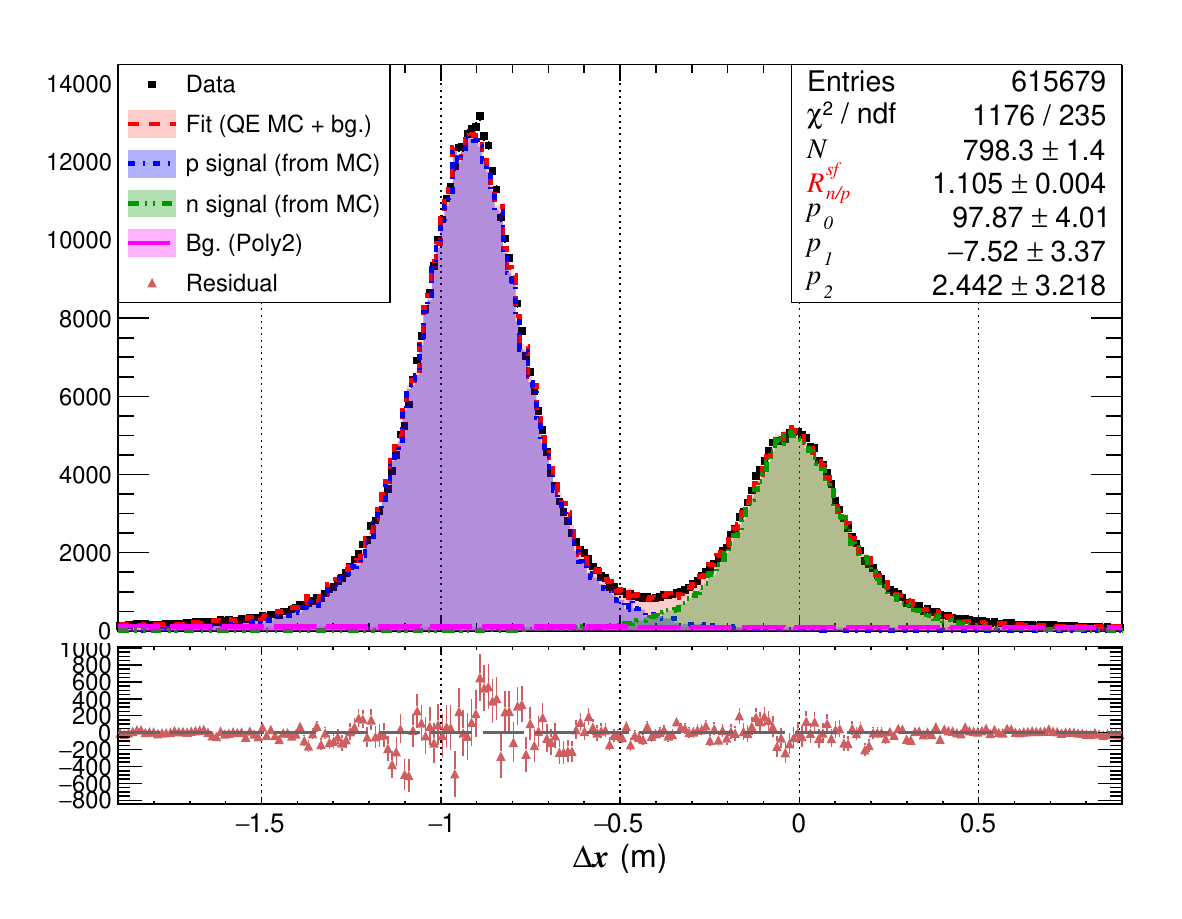}
    \caption{\label{fig:appen:dxfitsbs9} Example of data/MC fit to the \dx distribution for \qeq{4.5} kinematics using signal shapes from MC and second-order polynomial to model the background. Good electron events passing \w and \dy cuts are shown.}
\end{figure}
\begin{figure}[h!]
    \centering
    \includegraphics[width=0.9\columnwidth]{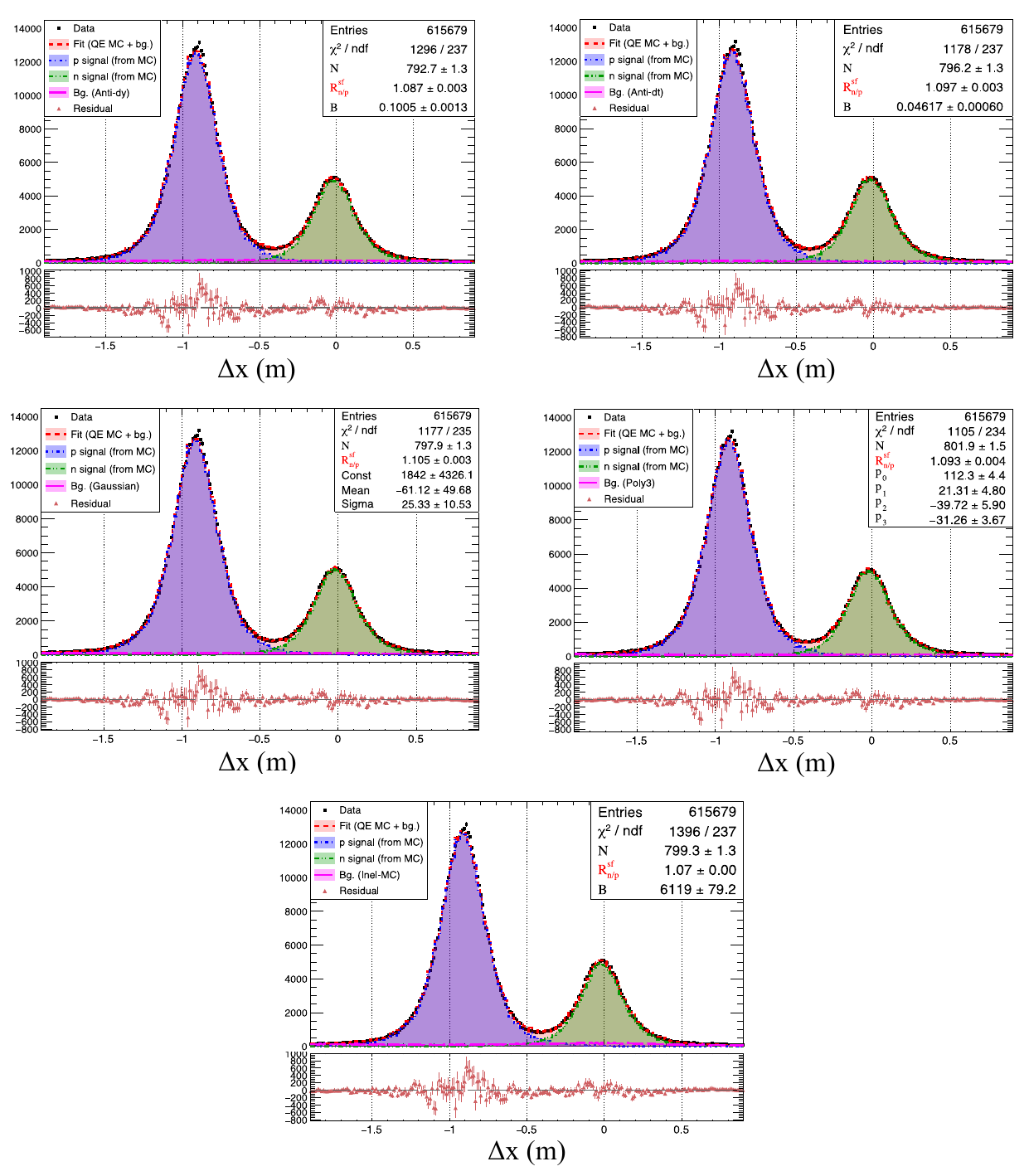}
    \caption{\label{fig:appen:dxfitsbs9_2} Comparison of data/MC fit to the \dx distribution for \qeq{4.5} kinematics between various estimations of background models. Good electron events passing \w and \dy cuts are shown.}
\end{figure}
\newpage

\section{\qeq{9.9}}
\begin{figure}[h!]
    \centering
    \includegraphics[width=1\columnwidth]{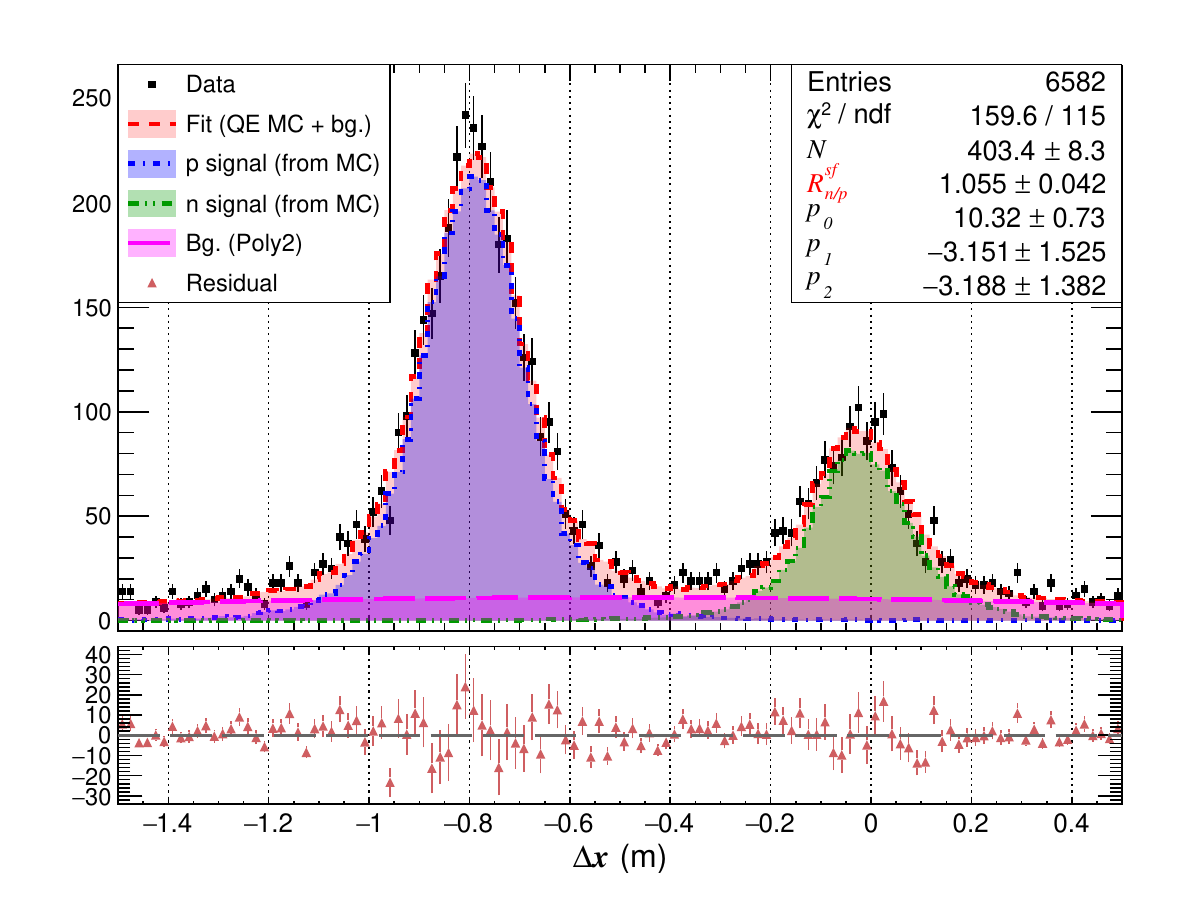}
    \caption{\label{fig:appen:dxfitsbs7} Example of data/MC fit to the \dx distribution for \qeq{9.9} kinematics using signal shapes from MC and second-order polynomial to model the background. Good electron events passing \w and \dy cuts are shown.}
\end{figure}
\begin{figure}[h!]
    \centering
    \includegraphics[width=0.9\columnwidth]{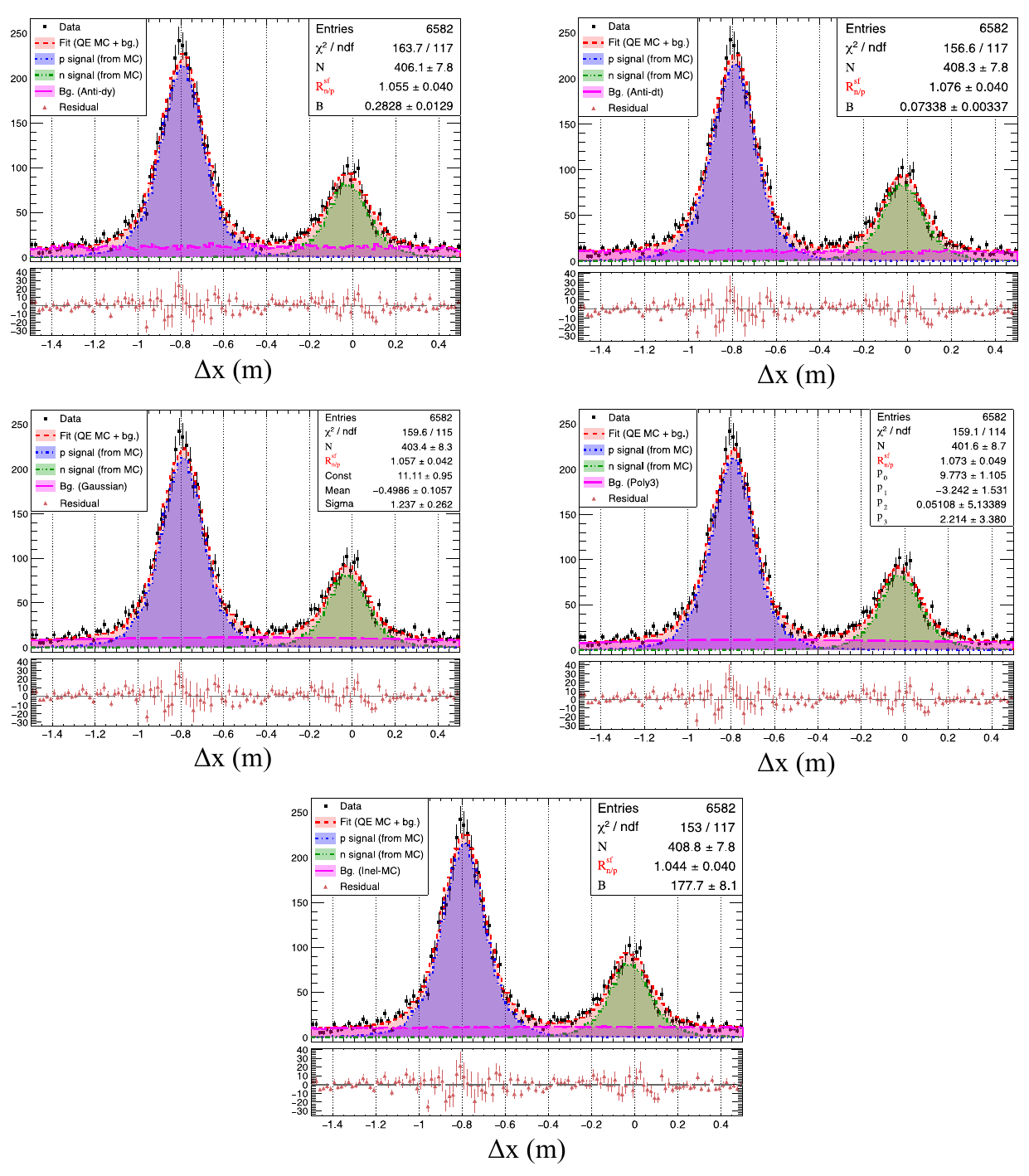}
    \caption{\label{fig:appen:dxfitsbs7_2} Comparison of data/MC fit to the \dx distribution for \qeq{9.9} kinematics between various estimations of background models. Good electron events passing \w and \dy cuts are shown.}
\end{figure}
\newpage

\section{\qeq{13.6}}
\begin{figure}[h!]
    \centering
    \includegraphics[width=1\columnwidth]{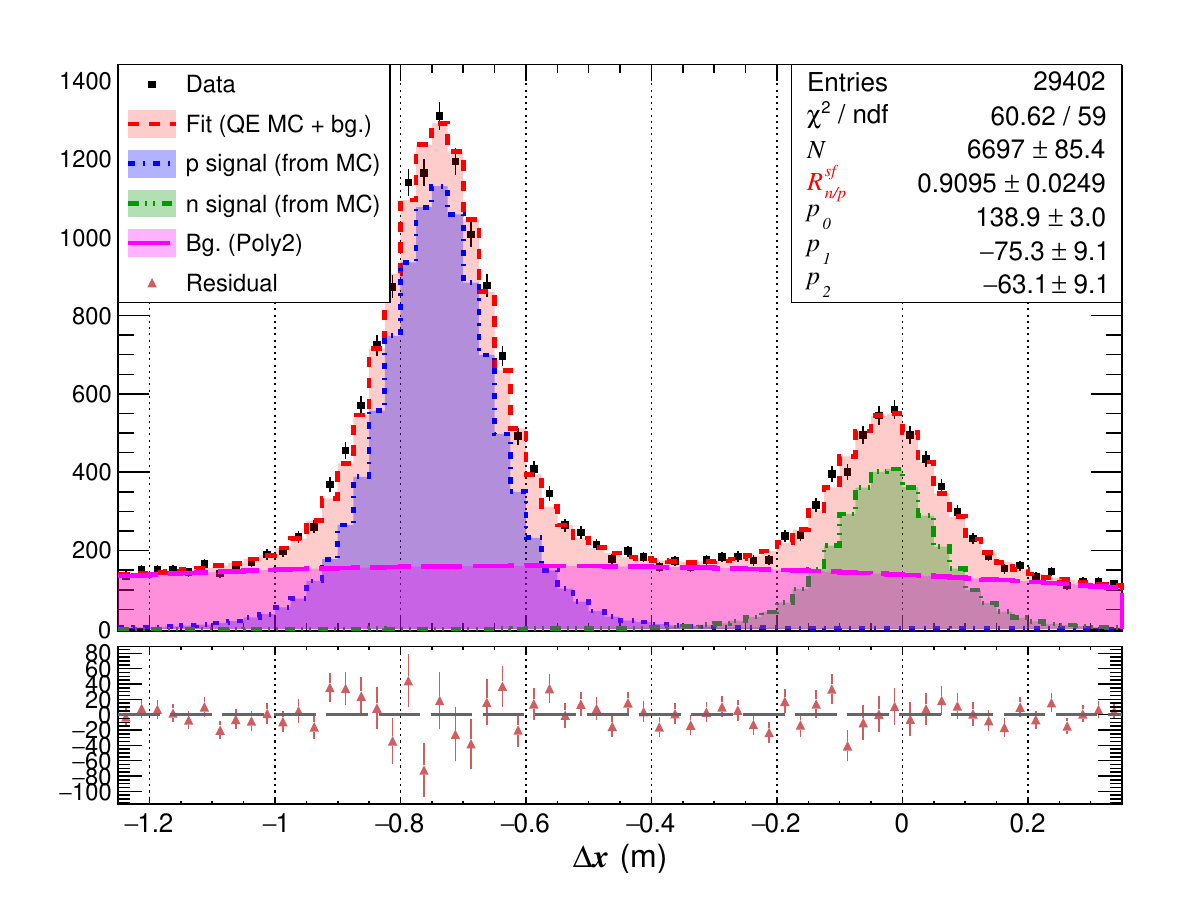}
    \caption{\label{fig:appen:dxfitsbs11} Example of data/MC fit to the \dx distribution for \qeq{13.6} kinematics using signal shapes from MC and second-order polynomial to model the background. Good electron events passing \w and \dy cuts are shown.}
\end{figure}
\begin{figure}[h!]
    \centering
    \includegraphics[width=0.9\columnwidth]{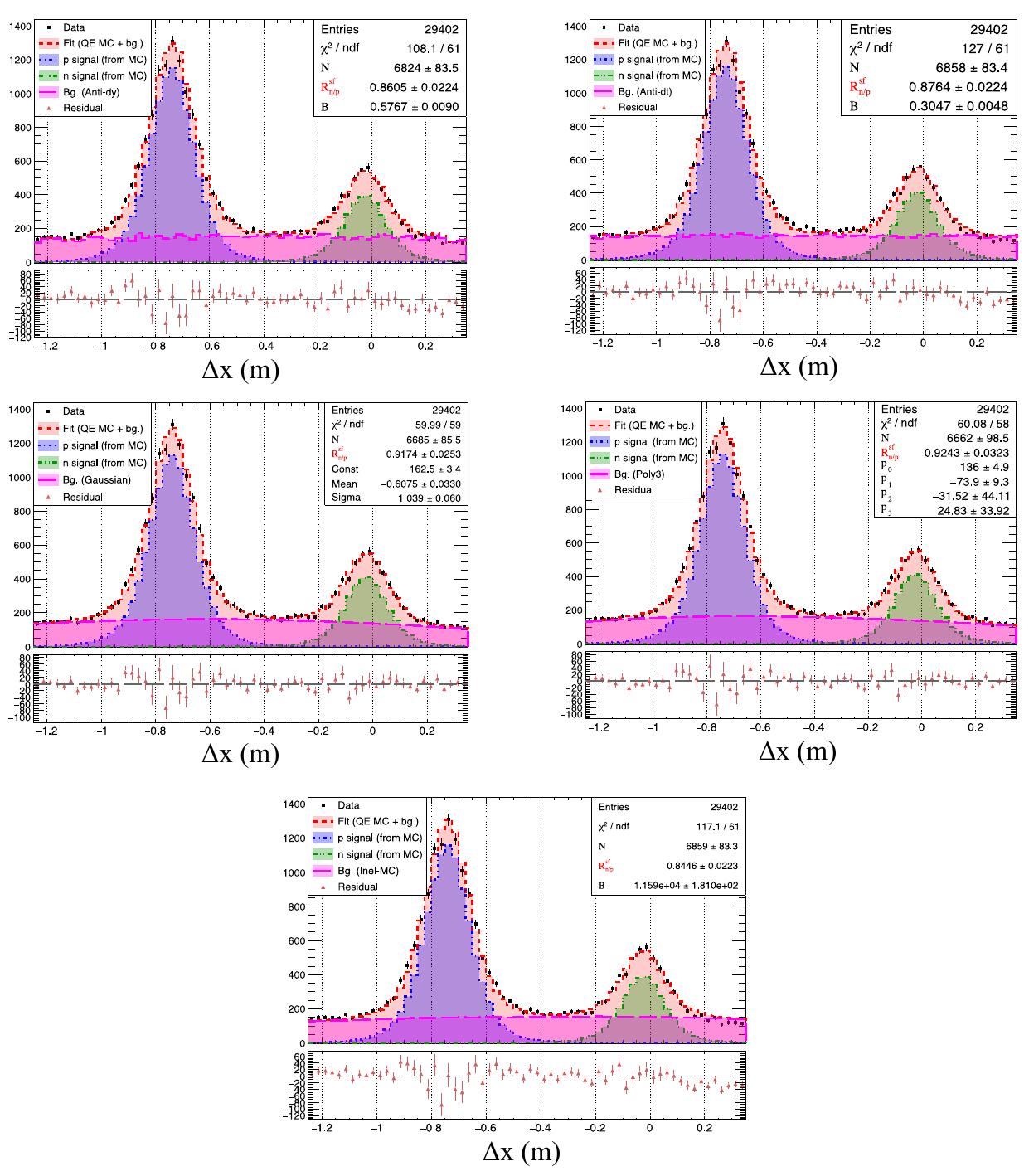}
    \caption{\label{fig:appen:dxfitsbs11_2} Comparison of data/MC fit to the \dx distribution for \qeq{13.6} kinematics between various estimations of background models. Good electron events passing \w and \dy cuts are shown.}
\end{figure}
%

\clearpage
\phantomsection
\addcontentsline{toc}{chapter}{Bibliography}
\printbibliography
\end{document}